\newif\ifpdf
	\ifx\pdfoutput\undefined
	   \pdffalse
	\else
	   \pdfoutput=1
	   \pdftrue
	\fi
\documentclass[12pt,dvips,a4paper,final]{bookdom}
\usepackage{a4p}
\usepackage{cite,mcite}
\ifpdf
\usepackage[pdftex]{graphicx}
\usepackage[pdftex,bookmarks,bookmarksopen,
bookmarksopenlevel={0},
pdfauthor={Dominique J. Mangeol},
pdftitle={Correlations in the Charged-Particle Multiplicity Distribution},
pdfkeywords={correlations, moments, QCD, pQCD, HEP, factorial, cumulant}]
{hyperref}
\usepackage{thumbpdf}
\else
\usepackage{graphicx}
\usepackage[bookmarks,
pdfauthor={Dominique J. Mangeol},
pdftitle={Correlations in the Charged-Particle Multiplicity Distribution},
pdfkeywords={correlations, moments, QCD, pQCD, HEP, factorial, cumulant}]
{hyperref}
\fi
\usepackage{physics}
\usepackage{amsmath2000}
\usepackage{amssymb}
\usepackage{stmaryrd}
\usepackage{pennames}
\usepackage{multicol}
\usepackage{subfigdom}
\usepackage{feynmf}
\usepackage{rotating}
\usepackage{epic}
\usepackage{eepic}
\usepackage{ifthen}
\graphicspath{{figures/chapter1/}
{figures/chapter2/}
{figures/chapter3/}
{figures/chapter4/}
{figures/chapter5/}
{figures/chapter6/}
{figures/chapter7/}
{figures/chapter8/}
{figures/chapter9/}
}
\ifpdf
\DeclareGraphicsExtensions{.pdf,.jpg}
\else
\DeclareGraphicsExtensions{.eps,.ps}
\fi
\newlength{\capindent}
\newlength{\capwidth}
\newlength{\figwidth}
\newcommand{\icaption}[2][!*!,!]{\hspace*{\capindent}%
  \begin{minipage}{\capwidth}
    \ifthenelse{\equal{#1}{!*!,!}}%
      {\caption{#2}}%
      {\caption[#1]{#2}}
  \end{minipage}}
\newcommand{\Ecal}{E^\mathrm{cal}_\mathrm{vis}}
\newcommand{\Ecperp}{E^{\mathrm{cal}}_{\perp}}
\newcommand{\Ecpar}{E^{\mathrm{cal}}_{\sslash}}

\newcommand{\Z}{$\mathrm{Z}^0$}

\newcommand{\toto}{$\tau^+\tau^-$}
\newcommand{\mumu}{$\mu^+\mu^-$}
\newcommand{\qq}{$\text{q}\bar{\text{q}}$}

\newcommand{\kl}{$\text{K}^0_\text{s}\text{ and }\Lambda$}
\newcommand{\cpmd}{charged-particle multiplicity distribution}
\newcommand{\mcpm}{mean charged-particle multiplicity}
\newcommand{\Cpmd}{Charged-particle multiplicity distribution}
\newcommand{\jmd}{jet multiplicity distribution}
\newcommand{\Ycut}{$y_\text{cut}$}
\newcommand{\Ejet}{$E^\text{jet}_\text{t}$}

\newcommand{\eeh}{$\Pep\Pem \longrightarrow \text{hadrons}$}

\newcommand{\Zto}{$\text{Z}\rightarrow $}
\newcommand{\scaption}[1]{\caption{\small{#1}}}

\begin{document}
\title{Correlations in the\\
Charged-Particle Multiplicity Distribution}
\author{Dominique J. Mangeol}
\maketitle
  
\newpage{\pagestyle{empty}\cleardoublepage}

\clearpage
\thispagestyle{empty}

\begin{center}

$$
$$


{\Large\bf{Correlations in the\\
\vspace{0.4cm}
Charged-Particle Multiplicity Distribution}}

\vspace{2.5cm}

{{}}

\vspace{4cm}

{\large{Doctoral dissertation}}

\vspace{2cm}

{{to obtain the degree of doctor\\
from the University of Nijmegen\\
according to the decision of the Council of Deans\\
to be defended in public\\
on Monday, 21 January 2002\\
at 3:30 pm precisely\\
by}}

\vspace{2cm}

{\large\bf{Dominique Jean-Marie Jacques Mangeol}}

\vspace{1.5cm}

{{born in Epinal, France\\
on 23 June 1971}}

\end{center}

\newpage{\thispagestyle{empty}}

\clearpage
\thispagestyle{empty}

\begin{center}

$$
$$


{\Large\bf{Correlaties in de\\
\vspace{0.4cm}
Multipliciteitsverdeling van Geladen Deeltjes}}

\vspace{1.5cm}

{{Een wetenschappelijke proeve op het gebied van de\\ 
Natuurwetenschappen, Wiskunde en Informatica}}

\vspace{4cm}

{\large{Proefschrift}}

\vspace{2cm}

{{ter verkrijging van de graad van doctor\\
aan de Katholieke Universiteit Nijmegen\\
volgens besluit van het college van Decanen\\
in het openbaar te verdedigen op\\
maandag 21 januari 2002\\
des namiddags om 3.30 uur precies\\
door}}

\vspace{2cm}

{\large\bf{Dominique Jean-Marie Jacques Mangeol}}

\vspace{1.5cm}

{{geboren op 23 juni 1971 te Epinal, Frankrijk}}

\end{center}

\newpage{\thispagestyle{empty}}

\vspace{0.5cm}

Supervisor: \hspace{2.25cm}{\bf{Prof.~Dr. E.W.~Kittel}}

\vspace{0.5cm}

Co-supervisor: \hspace{1.68cm}{\bf{Dr. W.J.~Metzger}}

\vspace{0.5cm}

Manuscript Committee: \hspace{-0.02cm}{\bf{Prof.~Dr. A.~Giovannini (University of Torino)}}

\hspace{4.45cm}{\bf{Prof.~Dr. E.~Longo (University of Roma)}}

\hspace{4.45cm}{\bf{Dr. J.~Field (University of Geneva)}}

\vspace{18.5cm}

\noindent The work described in this thesis is part of the research programme of the 
``Nationaal Instituut voor Kernfysica en Hoge-Energie Fysica'' (NIKHEF). The author 
was financially supported by the ``Stichtig voor Fundamenteel Onderzoek der Materie'' 
(FOM).

\begin{center}
{{ISBN 90-9015317-9}}
\end{center}

\clearpage

\newpage{\pagestyle{empty}}

\thispagestyle{empty}

$$
$$

\vspace{10cm}

\hspace{3cm}{\large\it La chose la plus difficile n'est pas de 
s'attaquer \`a une de ces}

\hspace{3cm}{\large\it grandes questions insolubles mais bien d'adresser 
\`a quelqu'un}

\hspace{3cm}{\large\it un petit mot d\'elicat o\`u tout est dit et rien.}

\hspace{12cm}{\large E.M.~Cioran}

\clearpage
\newpage{\pagestyle{empty}}
\thispagestyle{empty}
\cleardoublepage

\pagenumbering{roman}
\tableofcontents

\cleardoublepage
\pagenumbering{arabic}

\chapter*{Introduction}
\addcontentsline{toc}{chapter}{\numberline{}Introduction}
\markboth{\large{Introductions}}{\large{Introductions}}

What can we learn about the dynamics of particle production 
in \ee collision by just looking at the number of particles 
produced in the final state ?

This is the main question which is addressed in this thesis.

It is remarkable that the study of a static variable, such 
as the number of particles found in the final state, might 
reveal anything about the chain of processes which have 
occurred in the early stage of the reaction. 

By analogy, one can imagine an extra-terrestrial being  
wanting to understand Humanity from outside the Earth 
by using a giant telescope and looking at the distribution of 
light density at the surface of the Earth, from the 
behavior of this light density trying to access information
about Humanity and its evolution.

What he would see is that the light density is not uniform, 
that in some areas there is practically nothing, while at other places 
huge clusters of light density occur.  
He might also see that large light density areas are connected 
to middle range light density areas by thin light density lines. 
He would then find an obvious hierarchy in the distribution of these 
lights. In order to understand the mechanism which governs 
the apparition and the evolution of the light density, 
he would try to understand how all these lights are connected 
to each other. Therefore, he would study 
the correlations between these light densities.

He would certainly not understand the human spirit, 
but he might be able to understand some important points, both about the 
present and the past of Humanity and also about its evolution. 
Quickly, he would understand that these areas of high light 
densities have a ``capital'' importance for the whole light density 
distribution and that they, somehow, govern the evolution of 
smaller light density areas. 
He may argue that this importance dates back to some early 
stage in Humanity evolution when the light population 
was smaller and that by some iterative migration waves, 
the light population has increased. He may even argue 
about the reason why these areas have been favored above the 
others. He may decompose the dynamics of this light 
density evolution in this area into two stages, a first 
stage dating back to the origin of the apparition of the light 
in this area and a second one coming from the fact that the 
growing importance of this area itself might have 
led this area to grow in importance (and in light 
density) even faster.
He may not realize that Humanity (at least what he 
thinks Humanity is) has anything to do with some living 
creatures crawling at the surface of Earth. However, he will 
always be able to manage some understanding concerning  
the main points in the evolution of Humanity.

Luckily, particle physics is rather well understood 
(so we think). Main stream theories collected under the 
common name of Standard Model of particle physics 
describe in great detail many aspects of particle 
physics. Therefore, we should be able to learn more 
about the dynamics of particle production than our 
E.T. scientist can about Humanity.

The main-stream theory of primary concern in this thesis is is 
Quantum ChromoDynamics (QCD), the theory of the 
strong interaction which insures at its smallest scale the cohesion 
of matter, explaining the way in which quarks and gluons are bound  
together within protons and neutrons, which are themselves held within 
the atomic nucleus.  
QCD was born in the 1970's. So it is still a rather young theory and 
many questions have yet to be answered. 
One of them concerns the evolution of partons into hadrons.
Currently, no theory is able to describe the entire process.

This thesis describes research that takes place at the interface between 
soft and hard QCD (the sector of QCD which describes hadrons and that 
which describes partons, respectively) and it is intended to help to 
understand the general picture of the evolution of partons to hadrons.

By studying the \cpmd{}, we are able to access 
the dynamics of the process. The way partons  
evolve into hadrons leaves footprints in the \cpmd{} of 
the final state. These footprints 
are the correlations which exist between the particles. 
By studying these correlations, one might be able to 
pinpoint the various processes responsible for the particles  
we record in our detector. 

Therefore, to give a short answer to the question asked 
at the beginning of this introduction, the study of the 
\cpmd{} will help us to better understand  
the chain of processes and their hierarchical importance
in the production of final state particles.

The layout of the thesis is organized as follows:
In Chapter 1, a short theoretical introduction to multiparticle 
production is given. LEP and the L3 detector, the source of  
the data used in our analysis, are  described in Chapter 2.  
In Chapter 3, the selection, \ie{}, the necessary step to isolate a pure 
sample of hadronic \Z{} decays, is described. The measurement of our 
main variable, the charged-particle multiplicity, is described in Chapter 4. 
The inclusive charged-particle spectrum is measured in Chapter 5. 
Chapter 6 is the first chapter dedicated to the study of the shape 
of the \cpmd{}. In this chapter the study is limited to the full event sample 
and to a comparison with theoretical predictions. In Chapter 7, the 
shape of the jet multiplicity distribution is studied in order to 
enable direct comparison with QCD and jets obtained at perturbative energy
 scales. In Chapter 8, the \cpmd{} of the full sample is subdivided into 
that of 2-jet and 3-jet samples in order to compare to phenomenological models.
The \cpmd{} is also studied in restricted rapidity intervals in Chapter 9.
Finally, the conclusions are summarized.

\chapter{Theory}\label{chap:theo}

We present in this chapter a short introduction to the theoretical 
knowledge and understanding of the processes occurring during 
an \ee{} collision and responsible for the properties of 
the observables such as the \cpmd{} and reduced momentum distribution 
studied in this thesis.

\section[Multiparticle production in {\ee{}} collisions]
{Multiparticle production in \boldmath{\ee{}} collisions}

Multiparticle production is studied in a wide range 
of processes, from heavy-ion to \ee{} interactions.
Unlike other types of collisions, 
the \ee{} interaction has the advantage of offering 
a clean framework for this study. Only one collision 
per bunch crossing occurs, furthermore 
all the available center-of-mass energy is used in the 
interaction. Electrons and positrons being 
point-like and massless particles, and interacting only {\it{via}} 
the Electroweak interaction, their interaction is  
well understood and described by the Standard Model.

At the center-of-mass energy of $\sqrt{s}=91.2~\GeV{}$, 
\ie{} at the \Z{} resonance, the \ee{} interaction is dominated 
by the production of a quark and anti-quark pair (and hence 
of hadrons) {\it via} the formation of a \Z{} boson.  
The cross section of this process is  
about $69.9\%$~\cite{pdg} of the total \ee{} cross section. 
From a theoretical point of view, the process \eeh{} is understood 
as a succession of phases, each of them being described 
by a different theory.
The main phases in the evolution
of an \ee{} multi-hadronic event are shown in 
Fig.~\ref{fig:shower} together with its structure.

\subsubsection{The electroweak phase}

The first phase of the process concerns the collision itself. 
The electron-positron pair annihilates into a virtual 
photon or a \Z{} boson. This phenomenon may be accompanied 
by the emission of photons (initial-state radiation) prior
to the annihilation.
Following its creation, the vector boson decays 
into a quark-antiquark (\qq{}) pair. As for the annihilation, the 
creation of the fermion pair may be accompanied 
by the emission of one or more photons (final-state radiation). 
Both initial- and final-state radiation affect 
the system by reducing the energy available for 
hadron production. However, both initial- and final-state 
radiations are not very common at the energy of the \Z.
All these phenomena are described in the framework of 
the very successful electroweak model.

\subsubsection{The perturbative QCD phase}

\begin{figure}[htbp]
  \begin{center}
    \includegraphics[width=10cm]{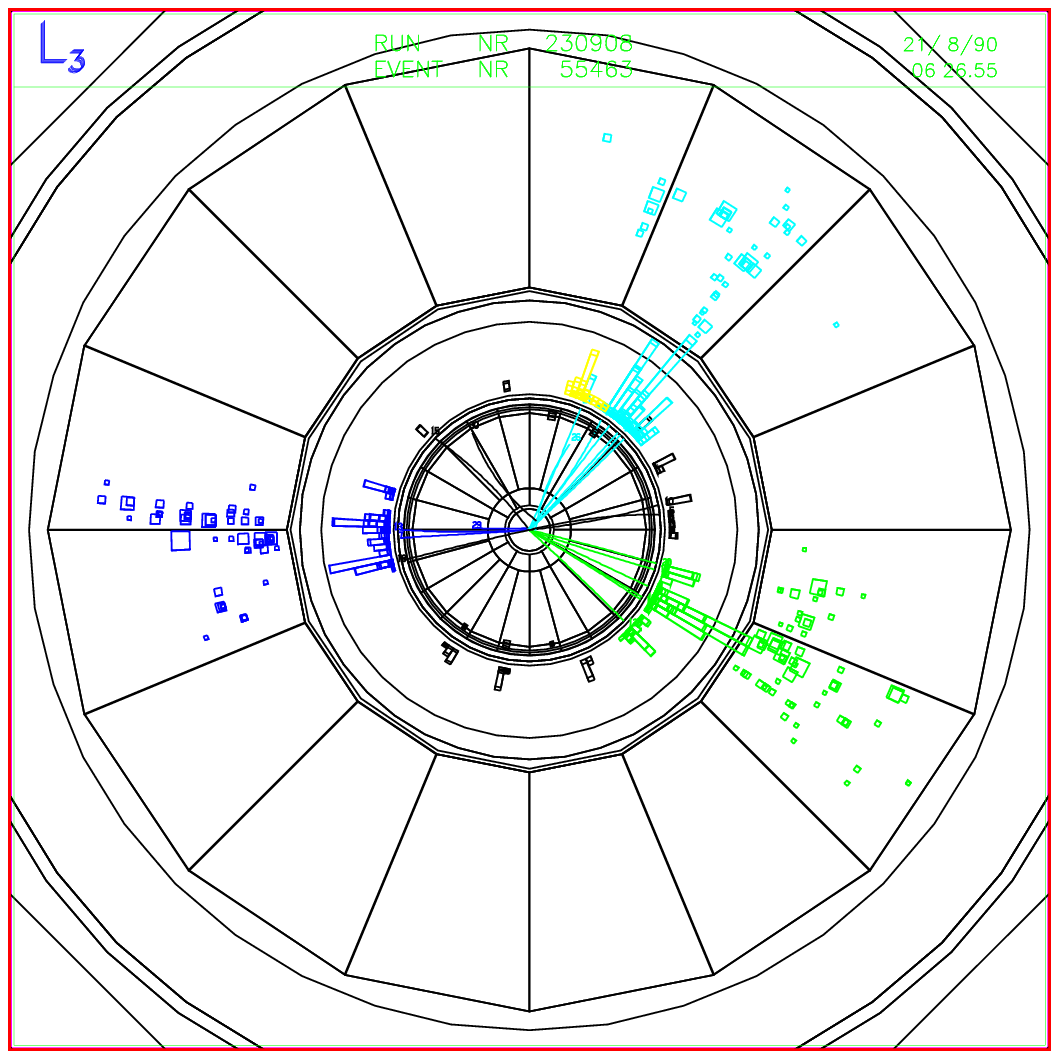}
  \end{center}
\scaption{A hadronic event as seen in the L3 detector.}
  \label{fig:l3event}  
\end{figure}

From the \qq{} pair created in the previous phase, 
a large number of particles are produced in the final-state 
in a very characteristic jet structure, as seen 
in Fig.~\ref{fig:l3event} for a 3-jet event as it appears 
in the L3 detector. Quantum ChromoDynamics (QCD) gives, in principle, 
a description and explanation for both the large number 
of particles produced in the final-state and its jet structure. 

\begin{figure}[htbp]
\hspace{0.5cm}
\begin{fmffile}{fmftempl}
\unitlength=1mm
\begin{fmfchar*}(100,65)
  \fmfpen{thin}
  \fmfset{curly_len}{1.5mm}
  \fmfset{wiggly_len}{5mm}
  \fmfleftn{a}{10} \fmfrightn{o}{18}
  \fmf{fermion}{a4,i1,a7}
  \fmflabel{{\Large $\text{e}^+$}}{a7}
  \fmflabel{{\Large $\text{e}^-$}}{a4}
  \fmf{photon,label={\Large $\gamma^\star,,\text{Z}$}}{i1,v8}
  \fmf{plain}{o1,v1,v2,v3,v4,v5,v6,v7,v8,v9,v10,v11,v12,v13,v14,v15,o18}
  \fmflabel{{\Large q}}{v10}
  \fmflabel{{\Large $\bar{\text{q}}$}}{v6}
  \fmffreeze
  \fmf{curly}{v5,g2}
  \fmf{plain}{p1,g2,p2}
  \fmf{curly}{o2,p1,o3}
  \fmf{curly}{o4,p2,o5}

  \fmf{curly}{v7,g4}
  \fmf{plain}{o6,g5,g4,o9}
  \fmffreeze
  \fmf{curly}{o7,g5,o8}

  \fmf{curly}{v11,g1,p4}
  \fmf{curly}{g1,p3}
  \fmf{curly}{o12,p3,o13}
  \fmf{plain}{o10,p4,o11}

  \fmf{curly}{v13,g3}
  \fmf{curly}{p5,g3,p6}
  \fmf{curly}{o14,p5,o15}
  \fmf{plain}{o16,p6,o17}

  \fmf{zigzag}{o1,o2}
  \fmf{zigzag}{o2,o3}
  \fmf{zigzag}{o3,o4}
  \fmf{zigzag}{o4,o5}
  \fmf{zigzag}{o5,o6}
  \fmf{zigzag}{o6,o7}
  \fmf{zigzag}{o7,o8}
  \fmf{zigzag}{o8,o9}
  \fmf{zigzag}{o9,o10}
  \fmf{zigzag}{o10,o11}
  \fmf{zigzag}{o11,o12}
  \fmf{zigzag}{o12,o13}
  \fmf{zigzag}{o13,o14}
  \fmf{zigzag}{o14,o15}
  \fmf{zigzag}{o15,o16}
  \fmf{zigzag}{o16,o17}
  \fmf{zigzag}{o17,o18}
\end{fmfchar*}
\hspace{0.25cm}
\begin{rotate}{90}
{\hspace{1.25cm} \bf HADRONIZATION}
\end{rotate}
\hspace{0.15cm}
\begin{fmfchar*}(50,65)
  \fmfpen{thin}
  \fmfset{curly_len}{1.5mm}
  \fmfset{wiggly_len}{2mm}
  \fmfstraight
  \fmfleftn{a}{10} \fmfrightn{o}{14}
  \fmf{zigzag}{a1,a2}
  \fmf{zigzag}{a2,a3}
  \fmf{zigzag}{a3,a4}
  \fmf{zigzag}{a4,a5}
  \fmf{zigzag}{a5,a6}
  \fmf{zigzag}{a6,a7}
  \fmf{zigzag}{a7,a8}
  \fmf{zigzag}{a8,a9}
  \fmf{zigzag}{a9,a10}

  \fmf{plain}{a2,p2}
  \fmf{plain}{o1,p2,p3}
  \fmf{plain}{o2,p3,o4}
  \fmf{plain}{a3,o5}
  \fmf{plain}{a4,j1}
  \fmf{plain}{o6,j1,o7}

  \fmf{plain}{a7,p4,p6,o11}
  \fmffreeze
  \fmf{plain}{p4,p5}

  \fmf{plain}{p6,o10}

  \fmf{plain}{o9,p5,o8}
  \fmf{plain}{a8,j2}
  \fmf{plain}{o12,j2,o14}
\end{fmfchar*}
\vspace{-0.6cm}
\setlength{\unitlength}{18mm}
\begin{picture}(0,2)(300,0)
\linethickness{1pt}
\put(300,1){\vector(1,0){10}}
\thicklines

\multiputlist(300,0.75)(.2,0){,,,,,,,,,,$91.2\GeV$,
,,,,$10\GeV$,,,,,,,,,,,,,$\approx 1\GeV{}$,
,,,,,,,,,,,,,,$\approx 0.1\GeV$,,,,,,}
\multiputlist(300,1)(.2,0){,,,,,,,,,,{\dashline{3}[0.7](0,0.2)(0,0)},
,,,,{\dashline{3}[0.7](0,0.2)(0,0)},,,,,,,,,,,,,
{\dashline{3}[0.7](0,0.2)(0,0)},,,,,,,,,,,,,,,
{\dashline{3}[0.7](0,0.2)(0,0)},,,,,,}

\multiputlist(300,1.5)(.2,0){,,,,,{\large Electroweak},,,,,
{\dashline{3}[0.7](0,0.2)(0,0)},
,,,,,,,{\large Perturbative QCD},,,,,,,,,,{\dashline{3}[0.7](0,0.2)(0,0)},
,,,,,,,,,{\large Non-perturbative QCD},,,,,,,,,,,}

\multiputlist(300,1.2)(.2,0){,,,,,,,,,,{\dashline{3}[0.7](0,0.2)(0,0)},
,,{M.E.},,{\dashline{3}[0.7](0,0.2)(0,0)},,,,,,{Parton Shower},,,,,,,
{\dashline{3}[0.7](0,0.2)(0,0)},,,,,,,,,,,,,,,,,,,,,}
\end{picture}
\end{fmffile}

\vspace{-6.6cm}
{\hspace{13cm} \bf DECAY}
\vspace{6.6cm}

\vspace{-1cm}
 \scaption{Schematic representation of a hadronic \ee{} event.}
   \label{fig:shower}  
\end{figure}

In Quantum ChromoDynamics, multiparticle production arises
from the interactions of quarks and gluons~\cite{gary}. 
Because of their properties,
these interactions are responsible for the 
creation of additional quark-antiquark pairs and gluons (\ie{} partons) 
in a cascade process near the 
direction of the primary partons (\ie{} the initial 
quarks and gluons), thus giving 
this typical jet structure to the events. 
Ultimately, hadronization gives birth 
to a large number of hadrons arranged into jets.

Two approaches may be used to described the 
production of partons. 

The first one, known as the matrix element method (M.E.), 
consists of performing the exact calculation   
perturbatively at each order of the strong coupling 
constant $\alpha_\text{s}$, 
taking into account all Feynman diagrams. 
Unfortunately, the difficulty increases sharply 
with the order considered and such a calculation has not yet
been performed to more than the second order in $\alpha_\text{s}$.
Therefore, this method cannot account for more than 
4 partons in the final-state.

Instead of using the exact calculation, 
the second approach, known as the parton shower 
approach, attempts to reproduce the cascade 
process responsible for the jet structure of
the event.
This is achieved by making iterative use
of the three basic branchings 
allowed by QCD, $\text{q}\rightarrow \text{qg}$, 
$\text{g}\rightarrow \text{q}\bar{\text{q}}$ and 
$\text{g}\rightarrow \text{gg}$. The probabilities 
governing the occurrence of these branchings are 
obtained from the 
Dokshitzer-Gribov-Lipatov-Altarelli-Parisi (DGLAP) 
evolution equations~\cite{dglap} as a function of the 
transverse momentum of the partons.
These equations are calculated using perturbative theory, 
in the Leading-Logarithm Approximation (LLA)
by taking into account in the expansion only the 
leading terms in $(\alpha_\text{s}\ln^2(s))^n$, 
the so-called leading logarithms. 
Extensions to this model such as Double LLA (DLLA), 
Modified LLA (MLLA), Next to LLA (NLLA) and even
Next to Next to LLA (NNLLA) have been investigated. 
These approximations take into account subleading terms 
ignored in the LLA, which allows them to account 
for effects such as gluon coherence (responsible 
for angular ordering which causes each subsequent gluon 
to be radiated within a smaller cone 
than its parent) and which better 
incorporate energy-momentum conservation.

The parton shower approach makes use of the running property 
of $\alpha_\text{s}$, which decreases to 0
at large energy scales (asymptotic freedom),
 enabling perturbative calculations to be carried 
out. On the other hand, at small energy scales  
$\alpha_\text{s}$ becomes large, thus forbidding the use 
a of power expansion in $\alpha_\text{s}$. 
This imposes a limit on the perturbative calculation of the 
development in the cascade process to energy scales larger 
than about 1~\GeV{}, defining what is often called the perturbative 
region (illustrated in Fig~\ref{fig:shower}) of QCD. 

\subsubsection{The non-perturbative phase}

At small energy scales, where $\alpha_\text{s}$ 
is large, the use of perturbation theory can not be justified.
Therefore, this phase is called the non-perturbative phase.

This third phase may itself be decomposed 
into two parts.
In the first part, the hadronization, colored partons  
fragment into colorless hadrons.
In the second part, these hadrons, which are for the most part 
unstable, decay into the stable particles 
which constitute the final-state particles 
observed in the detector (Fig.~\ref{fig:l3event}). 

In order to make final-state particles accessible 
to theoretical predictions, two approaches are often used:

The Analytical Perturbative Approach 
widely used to extrapolate analytical QCD predictions
to final-state particles assumes 
Local Parton-Hadron Duality (LPHD)~\cite{lphd}. 
The LPHD hypothesis relies on the idea of 
pre-confinement~\cite{preconf}, which implies that before 
hadronization colored partons are locally (in phase space) 
grouped into colorless clusters keeping 
the main properties of the partonic final-state. Consequently, 
the hadronic final state can be directly compared to the   
analytical QCD predictions for partons.  
The use of this method is limited to infrared safe variables, 
which are, in principle, not distorted by the hadronization 
phase. For such variables, partons and hadrons differ 
by only a proportionality factor, 

\begin{equation}
\label{eq:lphd}
\text{q}_\text{hadron} \propto \text{q}_\text{parton}.
\end{equation}

This method does not describe the final-state particles, 
but offers a rather good description of the behavior of some 
of the quantities characterizing the final-state particles,  
such as the energy evolution of the average number of charged 
particles and of their momentum.

Since the description of the final-state particles cannot be 
accessed analytically, 
a second approach is to use phenomenological models. 
Such models can provide a more complete 
description of the final-state. 
The various models available to describe 
the hadronization are Monte Carlo based 
models. They are described in the following 
section. 

For completeness, we also mention lattice  
QCD~\cite{lattice}, which has enjoyed large success in describing 
non-perturbative effects, but which has not yet 
been applied to hadronization.

\section{QCD generators}

Monte Carlo generators are essential for our study. 
They are able to generate a complete 
particle final-state which can then be compared to 
the final-state particles resulting 
from the \ee{} interaction, from which the quantities 
relevant to our analysis are extracted.

Each event is generated independently of 
the others. 
For each event, the whole chain of processes 
leading to the hadronic final-state is generated.
Each property, such as quark flavor, particle directions, 
energy they carry, the way they decay, 
is randomly generated according to its probability 
of occurrence determined by the physics of the process.
The generator also takes into account 
all the constraints and limitations imposed by the dynamics  
and the kinematics imposed by the whole chain of processes.

The main Monte Carlo models used to simulate 
hadronic Z decays in this thesis are the 
JETSET 7.4~\cite{jetset},
HERWIG 5.9~\cite{herwig} and ARIADNE 4.08~\cite{ariadne,ariad} 
Monte Carlo programs.

The generation of hadronic events proceeds in two 
main stages. The parton generation, which 
implements perturbative QCD to produce partons, 
and the fragmentation, which treats in a phenomenological way the 
hadronization as well as the decay of the particles. 
The main approaches used in the Monte Carlo generators 
to describe these stages and their specification are  
reviewed in the following two sub-sections.
Furthermore, these models incorporate special treatment to 
take into account the weak decay of heavy quarks.

\subsection{Parton generation}

Both Matrix Element and Parton Shower QCD approaches to 
parton generation have been implemented in 
Monte Carlo generators. 

\subsubsection{The Matrix Element approach}

The Matrix Element approach is found as an option 
in JETSET. It implements matrix element 
calculations up to  
$\mathcal{O}(\alpha_\text{s}^2)$ allowing to choose 
between a maximum of 2, 3 or 4 partons in the final-state.

\subsubsection{The parton shower approach}

Parton shower models are implemented 
in JETSET, HERWIG, and ARIADNE.
An important advantage over the analytical perturbative 
QCD calculations is that full energy-momentum conservation 
is imposed at each branching in the Monte Carlo models. 
Thus, Monte Carlo parton shower models implement,  
intrinsically, some higher-order corrections ignored by 
their analytical counterparts.

The JETSET parton shower implementation generates 
partons according to the LLA framework.
This model does not 
take into account subleading effects responsible 
for gluon interference, but an option  
exists to force it by requiring angular ordering 
explicitly. This makes the parton shower
equivalent to an MLLA parton shower.

HERWIG and ARIADNE with its color dipole 
cascade~\cite{dipole} use different approaches 
for their parton shower. Both of them take 
intrinsically into account coherence effects, 
which also makes them equivalent to a MLLA parton shower.

\subsection{Fragmentation models}

Since hadronization cannot be described analytically, 
phenomenological models are used. There are mainly three 
different models, the Lund string model~\cite{string} 
implemented in JETSET,  
which is the most popular (and also the most successful 
in describing the data), the cluster model~\cite{cluster} 
implemented in the HERWIG generator, and the independent 
fragmentation model~\cite{if2,if}, an older model still implemented
in JETSET as an option.

\subsubsection{Independent fragmentation model}

In the independent fragmentation model,
each final-state parton fragments 
independently from the others.
It fragments into a mesonic cascade 
until no energy is left to allow further splitting.

This model, whose origin dates back to the beginning 
of the seventies, gives only a poor description of the 
data. It has been supplanted by more recent 
models, such as the string and cluster models discussed 
below. Therefore, it is practically not used anymore.
Nevertheless, since all partons fragment 
independently, 
we will find it useful as a toy model to investigate the 
origin of certain correlations.
It may help understanding part of  
the effects brought in by the hadronization. 
But, because of the very approximative description of the 
hadronization, we cannot make detailed comparisons of 
this model to the data.  

\subsubsection{Cluster fragmentation model}

The cluster fragmentation model is an implementation 
of the pre-confinement property, from which originates 
the LPHD hypothesis. It assumes that after the parton 
shower, partons are locally aggregated into colorless hadrons.
Therefore, in the cluster model, all the gluons 
resulting from the parton shower are split into light (u or d) 
quark-antiquark or diquark-antidiquark pairs. 
These clusters are then fragmented into hadrons.

\subsubsection{The Lund string model}

The Lund string model is certainly the most 
popular and successful fragmentation model.
In the string model, a color string is stretched between 
quark and anti-quark.
The quark and antiquark moving apart along this string 
lose energy. This causes the string to break into two 
new quark-antiquark systems, resulting in two new strings 
which will break up similarly. The breaking process eventually 
stops when the mass of the string pieces has fallen to the hadron mass.
These string pieces form the hadrons. 
In this model, gluons are treated as kinks on the string.

This model appears to give a good description of the 
data in the final-state.

\subsection{Final-state particles}

Most of the hadrons produced by the fragmentation models described  
above are unstable and must decay into stable particles.
The quantitative knowledge we have about decays 
is mainly experimental. Therefore, at this stage, most of the 
masses and branching 
ratios governing the decays of these particles are taken from 
experimental results, with subsequent tuning in order to optimize 
the description of the data by the Monte Carlo models.

Furthermore, all these Monte Carlo models have adjustable parameters 
and switches, whose values are chosen in order to ensure a 
good description of the experimental data.

\section[{$\xi$} spectrum]{\boldmath{$\xi$} spectrum}

The single-particle inclusive momentum spectrum in $\xi=-\ln(x)$, where $x$ is 
the scaled momentum (\ie{} $x=2p/\sqrt{s}$, where $p$ is the 
particle momentum and $\sqrt{s}$ the center of mass energy),   
is very sensitive to soft gluon radiation. Its description, therefore, 
constitutes  an important test of perturbative QCD, in 
particular of the MLLA, which takes into account subleading terms introducing 
soft gluon radiation corrections. 
These corrections take into account color coherence, which leads   
to soft gluon suppression at large angles and hence to 
hadron depletion at low momentum.
This effect is also characterized by a strong 
angular ordering of gluon production, each gluon being 
emitted with a smaller angle than its parent. 
The effect of color coherence can be seen in 
the $\xi$ distribution, where it results in the so-called 
MLLA hump-backed plateau~\cite{basics,ochs} shape of the 
$\xi$ distribution. 
 
In the DLLA approximation which contains the premise of 
the color coherence effect~\cite{xidok} 
(and only a rough estimate of the angular ordering), 
the $\xi$ spectrum is described by a Gaussian.
Applying a next to leading order correction to the DLLA prediction, 
corresponding to MLLA, causes the $\xi$ distribution to 
deviate from the Gaussian shape, becoming a  
platykurtic shape~\cite{fw}, which 
has the appearance of a skewed and flattened Gaussian.
This is also characterized by a shift to lower momentum of 
the peak position, $\xi^\star$, of the $\xi$ distribution. 

Under the LPHD assumption, this behavior is not 
distorted by hadronization and is therefore identical 
for partons and hadrons. 

Furthermore, as a confirmation of the MLLA (or as a need
to take into account angular ordering), the evolution 
of the peak position $\xi$ with the center-of-mass 
energy has been found to be described by the MLLA, 
while the DLA has failed to describe it~\cite{ksi-l3,ksi-tasso,
ksi-aleph,ksi-delphi,ksi-opal}.

\section{The \cpmd{}}\label{sec:tcpmd}

One of the most fundamental observables in any high-energy collision 
process is the total number of particles produced in the 
final-state and by extension the number of charged particles 
which detection are easier. 
Even if this is only a global measure of the characteristics  
of the final-state, it is an important parameter in the understanding 
of hadron production. Independent emission of single particles leads 
to a Poissonian multiplicity distribution. Deviations from the 
Poissonian shape reveal correlations~\cite{kittel}. 
Therefore, these correlations are the signatures of the 
mechanisms involved from the early stage of the 
interaction with the appearance of the primary partons 
to the production of the particles in the final-state.

Using appropriate  
tools, it is, therefore, possible to extract information  
about the dynamics of particle production from the shape of the \cpmd{}. 

The usual way of studying the \cpmd{} and its shape, is to 
calculate its moments. 
General characteristics of the \cpmd{} are obtained using 
low-order moments, such as the mean, $\mu_1$, 
the dispersion, $D$, which estimates the width of the distribution,
the skewness, $S$, which measures how symmetric the distribution is,   
and the kurtosis, $K$, which measures how  
sharply peaked the distribution is. 
With $P(n)$ the \cpmd{} and $\langle q\rangle$ symbolizing  
the average of a quantity $q$, these moments are defined by 

\begin{equation}
\label{eq:moment1}
\mu_1=\langle n\rangle=\overset{\infty}{\underset{n=1}{\sum}} n P(n)
\text{ , } 
%
D=\sqrt{\langle(n-\mu_1 )^2\rangle}\text{ , }
S=\frac{\langle(n-\mu_1 )^3\rangle}{D^3}\text{ , }
K=\frac{\langle(n-\mu_1)^4\rangle}{D^4}-3.
\end{equation}

However, these moments only give information about the 
main properties of the distribution. 
A more detailed study of the \cpmd{} and of its shape, 
and in particular the study of correlations (and hence, 
the study of particle production) requires 
high-order moments~\cite{kittel}. 
A way, often used, of studying the correlations between 
particles in the \cpmd{} is to measure the normalized factorial 
moments of order $q$, 

\begin{equation}
\label{eq:fq}
F_q=\frac{\overset{\infty}{\underset{n=q}{\sum}}n(n-1)....(n-q+1)P(n)}
{\left (\overset{\infty}{\underset{n=1}{\sum}} n P(n)\right)^q}.
\end{equation}

\begin{figure}[htbp]
  \begin{center}
    \includegraphics[width=8.4cm]{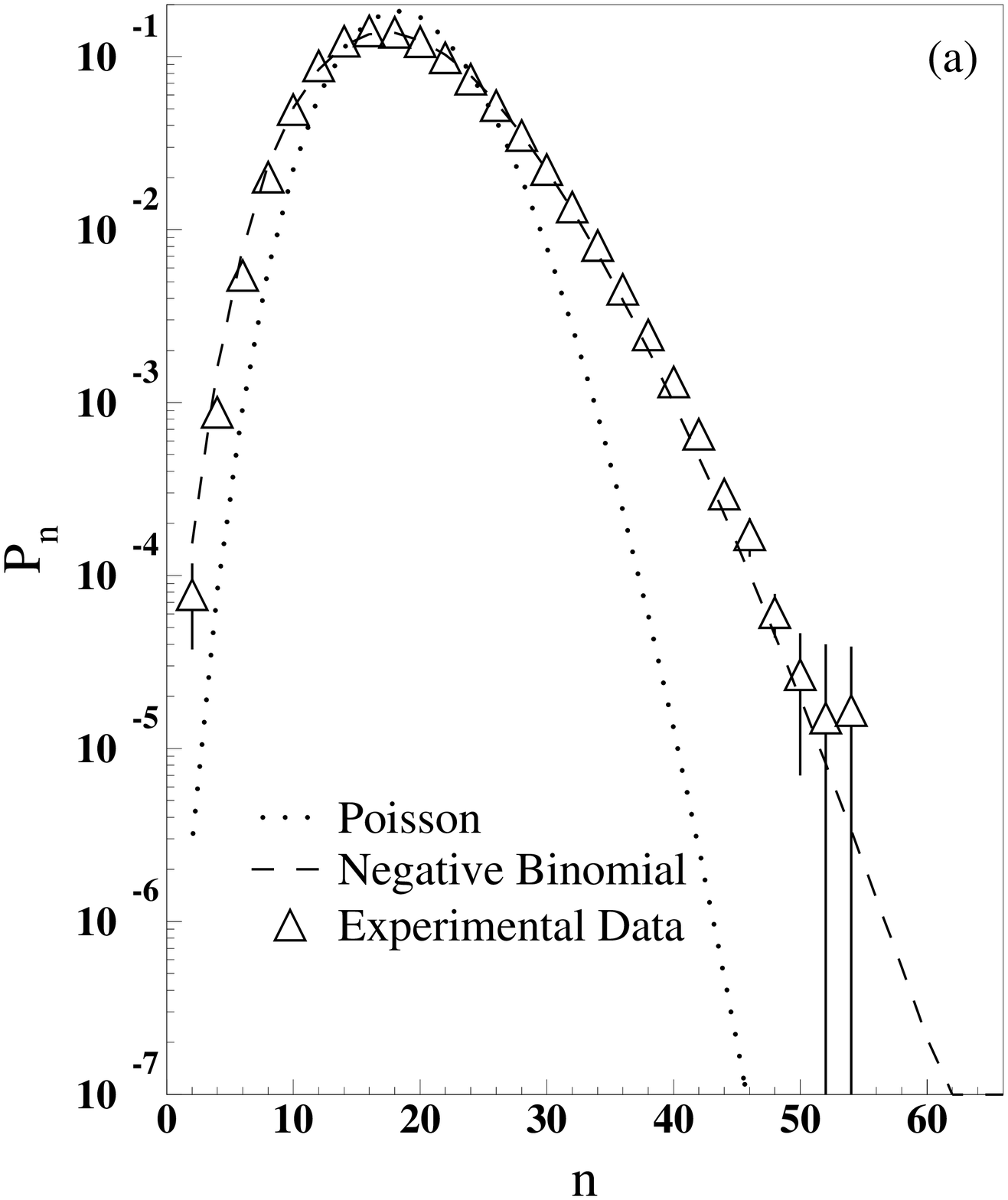}
    \includegraphics[width=8.4cm]{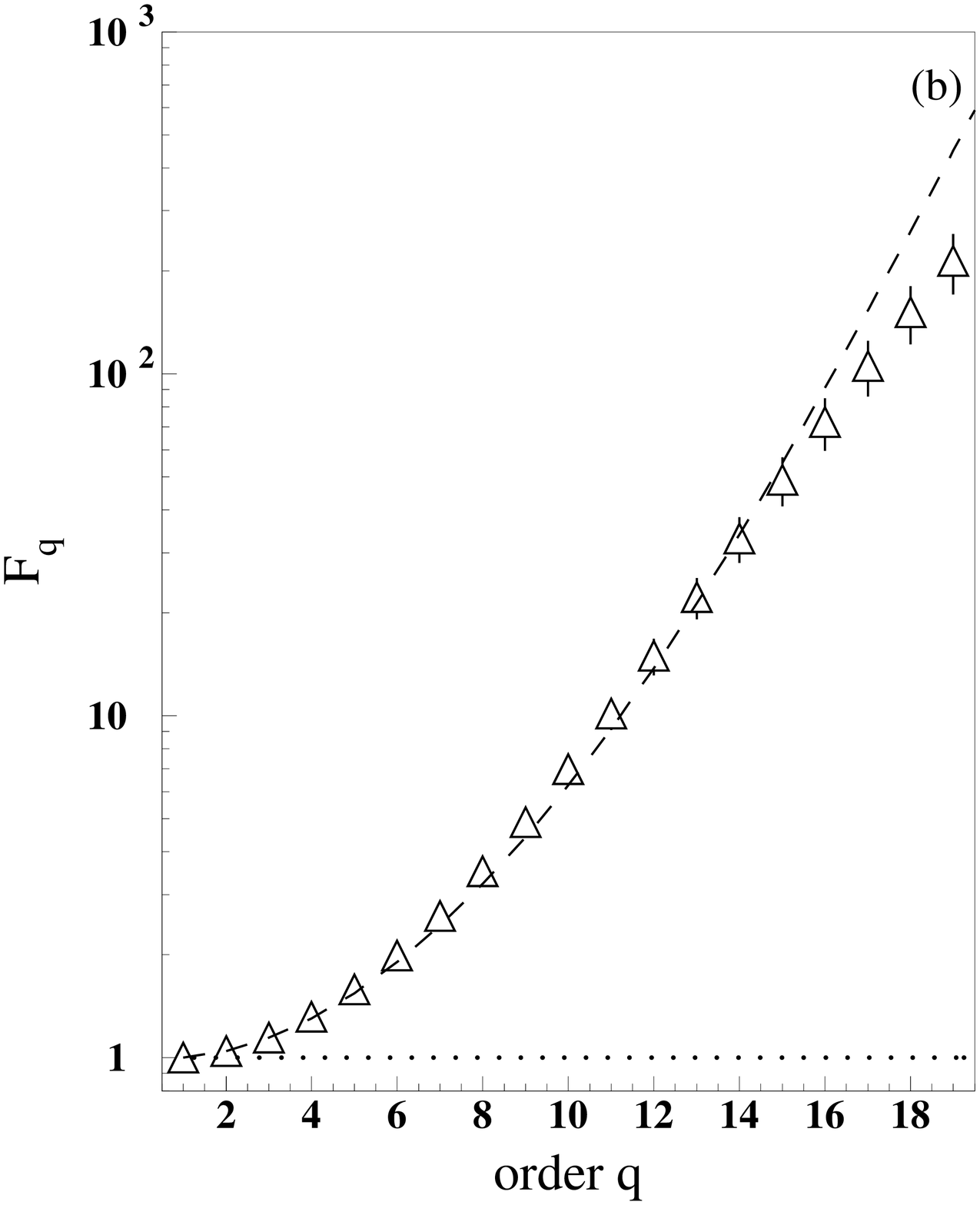}
  \end{center}
\vspace{-1cm}
\scaption{Poisson and negative binomial distributions ,
together with the \cpmd{} of the experimental data (a).
$F_q$ {\it vs} $q$ obtained from 
Poisson and negative binomial distributions,  
together with the $F_q$ measured from the experimental data (b).}
  \label{fig:pnfqp}  
\end{figure}

The factorial moment of order $q$ corresponds to an integral over the 
$q$-particle density and reflects correlations in particle 
production. If particles are produced independently the multiplicity  
distribution is Poissonian (see Fig.~\ref{fig:pnfqp}(a)), 
and all the $F_q$ are equal to unity.
If the particles are correlated, the distribution is broader than 
Poisson and the $F_q$ are larger than unity (see example of the 
negative binomial distribution in Fig.~\ref{fig:pnfqp}(a)). In the opposite case, 
if the particles are anti-correlated, the distribution is narrower 
than Poisson, and the $F_q$ are smaller than unity.
Two examples of $F_q$ plotted as a function of the order $q$ 
are shown in Fig.~\ref{fig:pnfqp}(b), for distributions 
such as the Poisson distribution (no correlation) 
and for a negative binomial distribution (positive correlation).

\begin{figure}[htbp]
  \begin{center}
    \includegraphics[width=8.4cm]{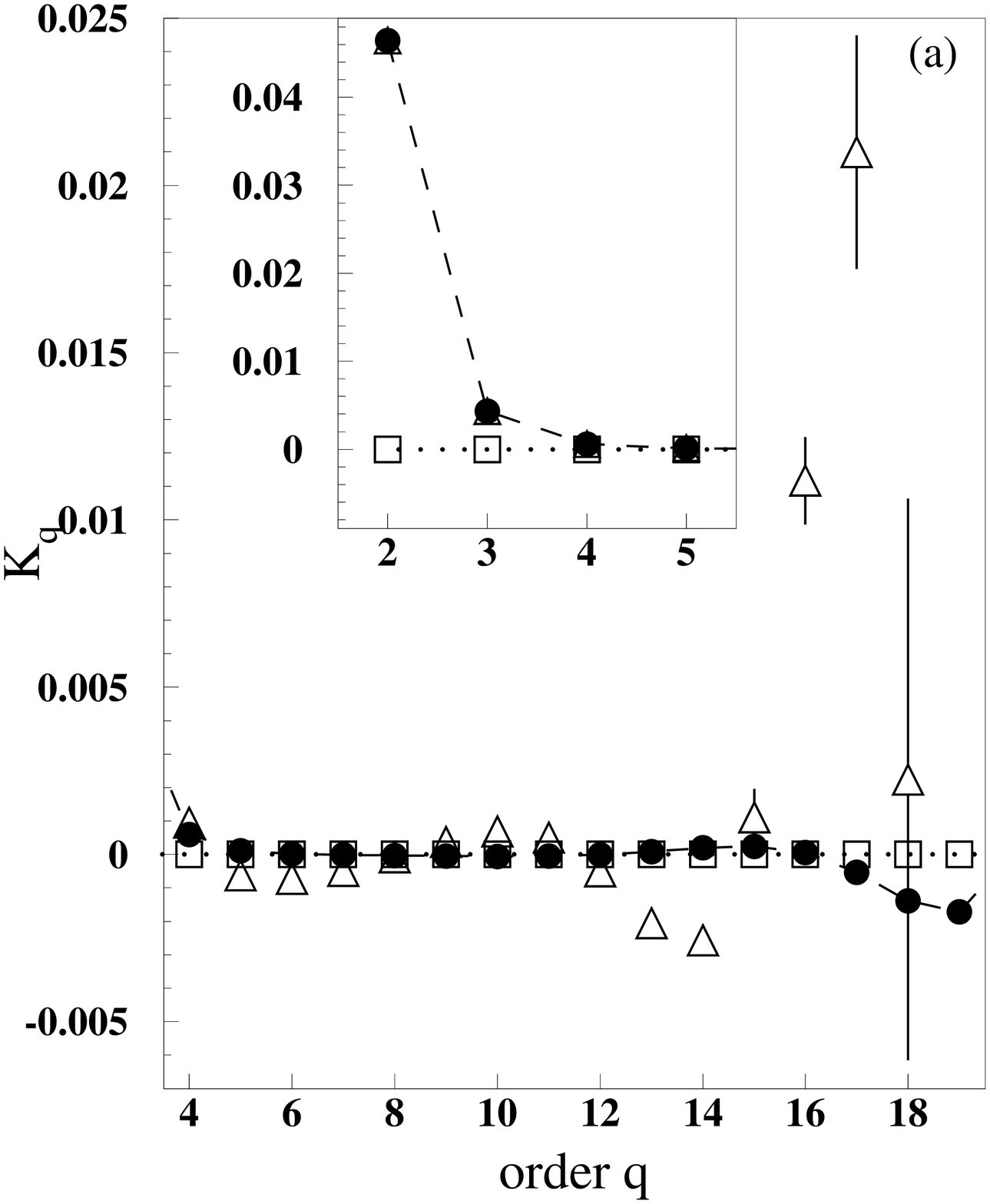}
    \includegraphics[width=8.4cm]{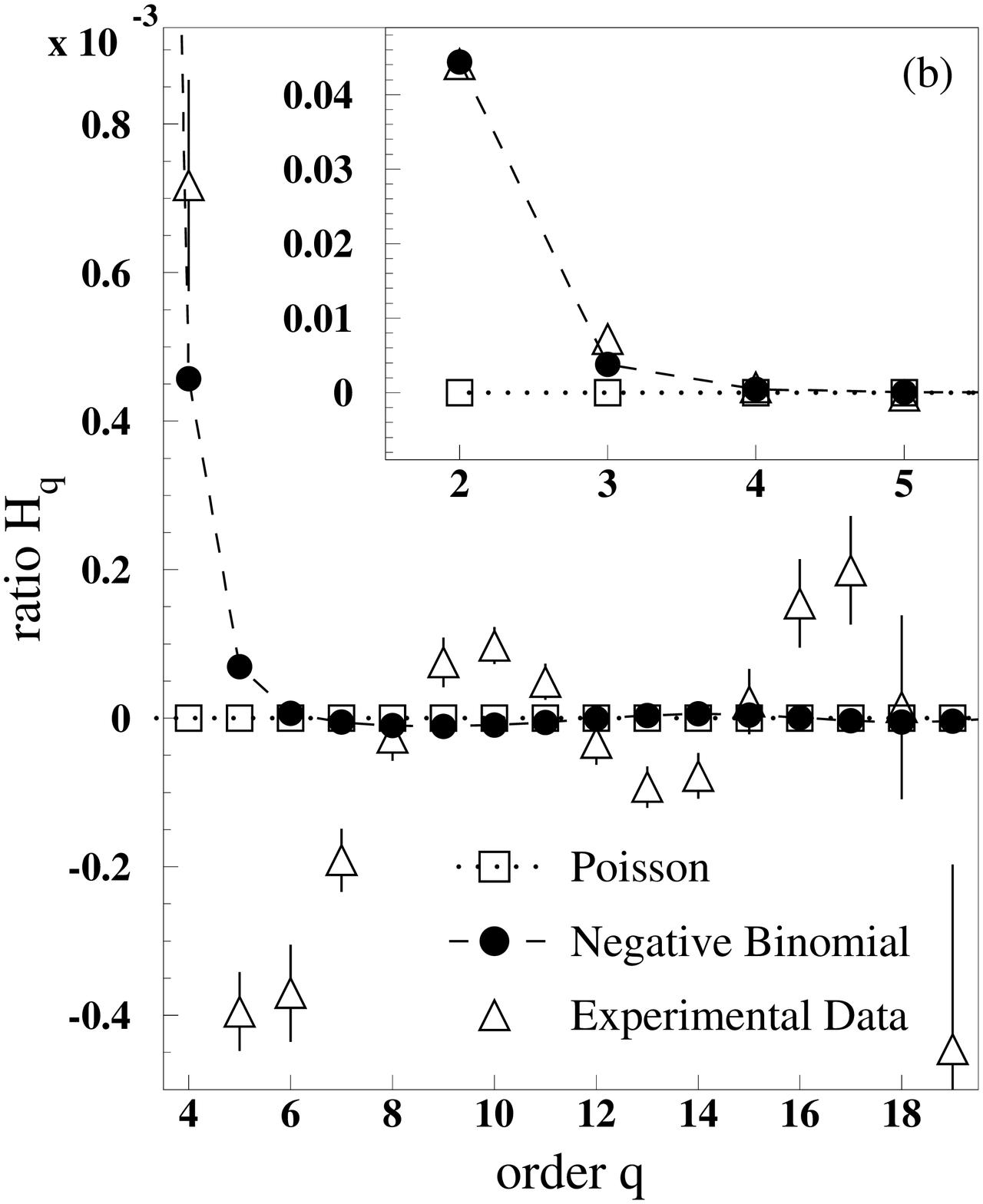}
  \end{center}
\vspace{-1cm}
\scaption{$K_q$ {\it vs} $q$ (a) and $H_q$ {\it vs} $q$ (b),  
obtained from Poisson and negative binomial distributions,  
together with those measured from the experimental data.}
  \label{fig:hkqp}  
\end{figure}

However, with the $F_q$ we only access the sum of all 
correlations existing among $q$ or fewer particles.  
It is a combination which takes into account all possible 
correlations between any number of particles,    
smaller or equal to the order $q$. Therefore, 
one can access  the genuine $q$-particle correlation by the 
use the normalized cumulant factorial moments, $K_q$,  
which are obtained from the normalized factorial moments by 
\begin{equation}
\label{eq:kq}
K_q=F_q-\overset{q-1}{\underset{m=1}{\sum}}
\frac{(q-1)!}{m!(q-m-1)!}K_{q-m}F_m.
\end{equation}
These $K_q$ correspond to the phase-space integral over the genuine 
$q$-particle 
correlation. If the particles result from independent emission 
(Poissonian behavior), the $K_q$ are equal to 0. The $K_q$ are 
positive if the particles are correlated and negative if 
the particles are anti-correlated. 
As examples, the $K_q$ plotted versus the order $q$ are given 
in Fig.~\ref{fig:hkqp}(a), for the Poisson (no correlation) and 
for the negative binomial distribution (positive correlation) 
and also for the experimental \cpmd{}, which will be studied 
in details in chapter~\ref{chap:cpmd}.

Since $F_q$ and $|K_q|$ both increase with the order $q$, it is 
useful to define the ratio $H_q$:

\begin{equation}
\label{eq:hqt}
H_q=\frac{K_q}{F_q}=1-
\overset{q-1}{\underset{m=1}{\sum}}\frac{(q-1)!}{m!(q-m-1)!}H_{q-m}
\frac{F_mF_{q-m}}{F_q}.
\end{equation}
The $H_q$ moments reflect the genuine $q$-particle correlation 
integrals relative to the density integrals. They characterize the 
weight of the genuine $q$-particle correlations with respect to   
the whole spectrum of correlations between $q$ particles. 
Furthermore, the $H_q$ moments have the advantage over 
the $F_q$ and $K_q$ of being of the same order of magnitude for 
a large range of $q$. Examples of $H_q$ plotted versus the 
order $q$ are given in Fig.~\ref{fig:hkqp}(b), for the Poisson and 
the negative binomial distribution, together with the one measured 
from the experimental \cpmd{}. 

More astonishing than from Poisson and negative binomial 
are the $H_q$ moments obtained 
from the experimental \cpmd{s} (Fig.~\ref{fig:hkqp}(b)), 
exhibiting an oscillatory behavior when 
plotted versus the order $q$.  
Furthermore, the same qualitative oscillatory behavior 
has been observed not only in \ee{} collisions, but also in 
hadron-hadron, hadron-ion and even ion-ion 
collisions~\cite{dremin2,hqhadron}.

The usual way to interpret this oscillatory behavior 
is to refer to perturbative QCD, which provides us with 
calculations for the $H_q$ 
of the parton multiplicity distribution~\cite{dremin1,dremin2}. 
Under the local parton-hadron duality hypothesis, 
which assumes that the shape of the parton multiplicity 
distribution is not distorted by hadronization, 
perturbative QCD prediction may be valid for hadrons, 
thereby allowing the extension of perturbative QCD predictions 
to the shape of the \cpmd{}.

However, this result can also be interpreted in a
more phenomenological way by viewing the shape of the 
charged-particle multiplicity distribution as the result 
of the fact that different types of events,  
such as 2-jet or 3-jets events, compose the total \cpmd{}~\cite{23jet}.

\subsection[{$H_q$} moments and analytical QCD predictions]{\boldmath{$H_q$} moments and analytical QCD predictions}

\begin{figure}[htbp]
  \begin{center}
    \includegraphics[width=10cm]{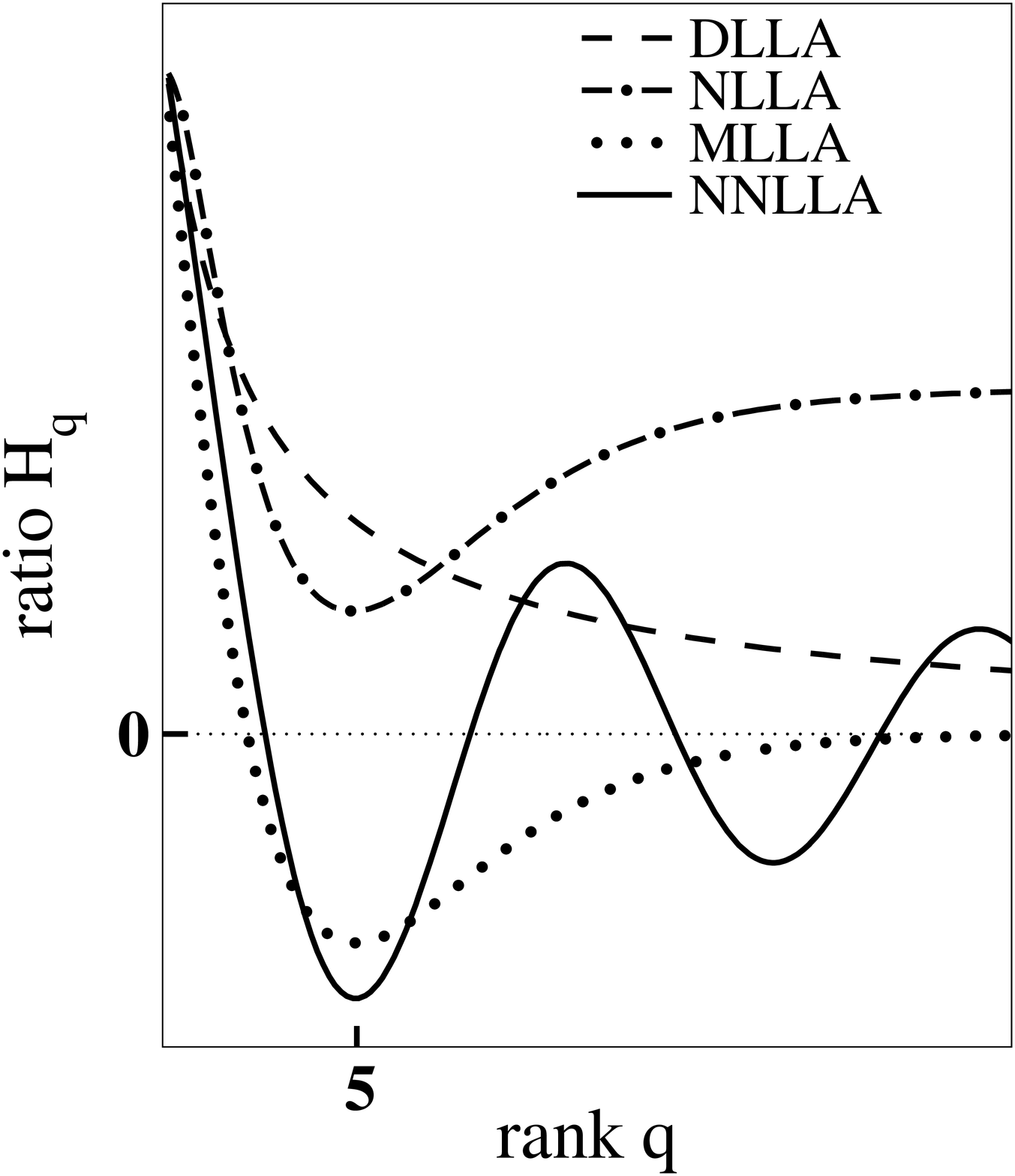}
  \end{center}
\scaption{Qualitative behavior of $H_q$ {\it vs.} $q$ for
various perturbative QCD approximations.}
  \label{fig:pred}  
\end{figure}

Since the evolution equations of QCD can be described in 
probabilistic terms using generating functions, it is, 
in principle, possible to describe analytically the 
parton multiplicity distribution. 
Nevertheless, even an approximate solution to this 
equation cannot be obtained easily. However, it has turned out to 
be a relatively easy problem to solve for the moments 
of the multiplicity distribution.  
Therefore, the $H_q$ moments have been calculated up to 
the next-to-next-to-leading logarithm approximation~\cite{dremin1}. 
The expected behavior of $H_q$ for various approximations  
is qualitatively plotted as a function of $q$ in Fig.~\ref{fig:pred}.

\begin{itemize}

\item  For the Double Leading Logarithm Approximation (DLLA), 
       $H_q$ decreases to 0 as $q^{-2}$.

\item  For the Modified Leading Logarithm Approximation (MLLA), 
       $H_q$ decreases to a negative minimum at $q=5$, and then 
       rises to approach 0 asymptotically.

\item  For the Next-to-Leading Logarithm Approximation (NLLA), 
       $H_q$ decreases to a positive minimum at $q=5$ and then 
       increases to a positive constant value for large moment rank.

\item  For the Next-to-Next to Leading Logarithm Approximation (NNLLA), 
       $H_q$ decreases to a negative first minimum for $q=5$, and 
       for $q>5$, $H_q$ shows quasi-oscillations about 0. 

\end{itemize}

The main difference between all these approximations lies in how 
energy momentum conservation is incorporated. 
The most accurate treatment is given by the NNLLA.
 
Similar behaviors as those predcited 
are expected for the charged-particle multiplicity distribution 
under the Local Parton-Hadron Duality hypothesis. 
The oscillatory behavior observed in 
Fig.~\ref{fig:hkqp}(b) is often 
interpretated as a confirmation of NNLLA and LPHD.

\subsection{Phenomenological approaches}

However, the $H_q$ oscillatory behavior may also be interpreted 
in a more phenomenological way making use of different 
classes of events, which themselves do not necessarly have 
$H_q$ oscillations. 

These approaches are based on the
idea that the main features of the shape of 
the \cpmd{} and the oscillatory behavior of the $H_q$ could 
originate from the superposition of different types of events, 
as the 2-jet and 3-jet events~\cite{23jet} or the light- and 
b-quark samples~\cite{lbquark}, which compose the full sample.

Under this hypothesis, assuming we are able to describe  
individually the \cpmd{s} of these different types of events 
using suitable parametrizations, the \cpmd{} of the full sample 
could then be described by a weighted sum of all the 
individual parametrizations, the weight being related to
the rates of the various type of events.

If the full sample can be resolved into  
various classes of events, we can express its 
\cpmd{} as a sum of the various contributions:

\begin{equation}
\label{eq:multi}
P(n)=R_\alpha P_\alpha(n)+
               R_\beta P_\beta(n)+
               R_\gamma P_\gamma(n)+\ldots,
\end{equation}
where the $P_\alpha(n)$, 
          $P_\beta(n)$ and 
          $P_\gamma(n)$ 
are the \cpmd{s} of events of type $\alpha, \beta \text{ and }
\gamma$, while 
$R_{\alpha,\beta,\gamma}$ are their respective rates.

Assuming these \cpmd{s} are described by the parametrizations 
$f_\alpha(n)$,
$f_\beta(n)$ and 
$f_\gamma(n)$,  
the \cpmd{} of the full sample,  
$P(n)$, will then be described by $f_\text{full}(n)$,  
the weighted sum of all the parametrizations:

\begin{equation}
\label{eq:fits}
f_\text{full}(n)=R_\alpha f_\alpha(n)+
                 R_\beta  f_\beta(n)+
                 R_\gamma f_\gamma(n)+\ldots.
\end{equation}

In our analysis, we make the choice to use the Negative 
Binomial Distribution (NBD) as parametrization. 
The NBD has been already used, with more or less success, 
in many types of interactions to describe their \cpmd{s}~\cite{nbd}. 
The use of the NBD, as for the phenomenological approach, 
in multiparticle dynamics is intimately associated to 
the clan concept~\cite{clan,clan2}. This concept 
was used to explain the apparent NBD behavior of the 
multiplicity distribution 
observed in many experiments, processes and energies 
in both full and restricted phase space.
A clan is defined as a group of particles originating  
from the same parent particle~\cite{ugo}. 
While the particle distribution within a clan is assumed to be 
logarithmic, its composition with other clans (which are assumed to be 
independent of each other) leads to the NBD.

In \ee{} annihilation at the \Z{} energy, it has been found 
that the \cpmd{} cannot be described by a single NBD~\cite{delfull}. 
Therefore, it is interesting to try combinations of NBDs~\cite{23jet,lbquark}. 
The NBD parametrization is given by
\begin{equation}
f^\text{NB}(n,\langle n\rangle,k)=\frac{\Gamma(k+n)}{\Gamma(n+1)\Gamma(k)}
\left(\frac{k}{\langle n\rangle+k}\right)^k \left(\frac{\langle n\rangle}
{\langle n\rangle+k}\right)^n,
\end{equation}
where $\langle n\rangle$ is the mean of the distribution and $k$ is given 
by 
\begin{equation}
\label{eq:park}
\frac{D^2}{\langle n\rangle^2}=\frac{1}{\langle n\rangle}+\frac{1}{k},
\end{equation}
$D$ being the dispersion.
Using the means and dispersions from the experimental 
distributions, we can then have fully constrained 
parametrizations of the \cpmd{s} of the various classes 
of events and subsequently of the full sample. 
In Chapter~\ref{chap:hq23j}, several phenomenological 
approaches based on this concept will be examined and 
confronted with the experiment.

\chapter{Experimental apparatus}

\section{The LEP collider at CERN}

\begin{figure}[htbp]
   \begin{center}
     \includegraphics[width=16cm]{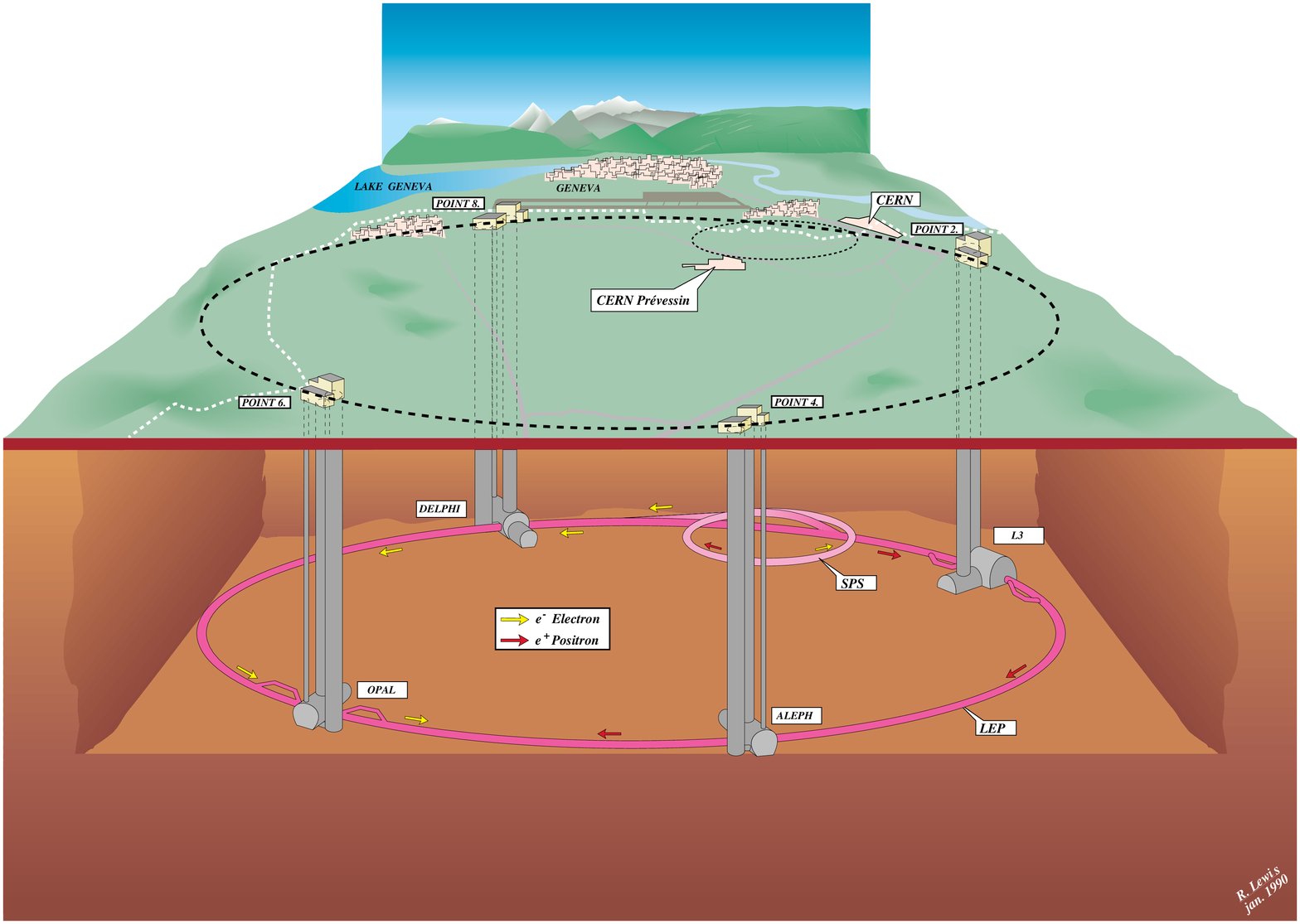}
   \end{center}
 \vspace{-0.5cm}
\scaption{Underground and aerial view of LEP and its surrounding area.}
  \label{fig:lep}  
  \begin{center}
\ifpdf
  \includegraphics[width=16cm]{lucomp_00}
\else
  \includegraphics[width=16cm]{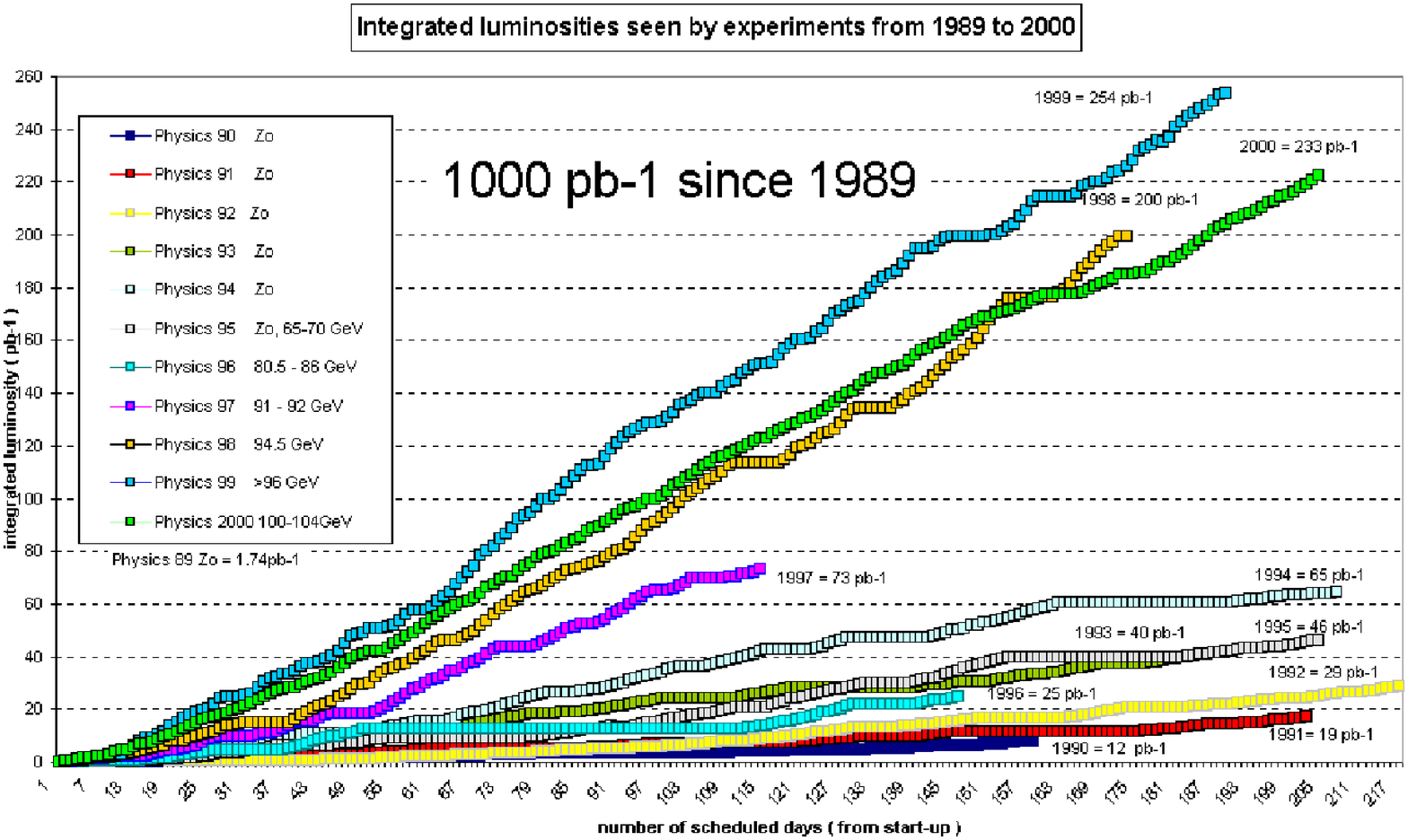}
\fi
  \end{center}
\vspace{-0.5cm}
 \scaption{Integrated luminosity per LEP experiment.}
   \label{fig:lulep} 
 \end{figure}

Located near Geneva, between the Alps and the Jura, 
the Large Electron Positron collider, LEP (Fig.~\ref{fig:lep}),
commissioned and operated by CERN, straddles the French-Swiss border 
at an average depth of about 100 meters. 

LEP, with a circumference of 27 km, is the largest (electron-positron) 
collider built so far. 
It was designed to store and accelerate electrons and positrons, which it did
up to the energy of $104.5 \GeV$ per beam reached during the last  
data taking period in 2000.
The electrons and positrons are produced and accelerated up to 
$20\GeV$ by lower energy CERN accelerators.
They are then injected into LEP and concentrated into equidistant
bunches circulating in opposite direction. Finally, they are 
accelerated to their final energy.
Once they have reached that energy, they are allowed to collide 
at four of the eight equidistant crossing points.
At these four crossing points are positioned 
the four LEP experiments: ALEPH, OPAL, DELPHI and in particular 
L3, the source of the data used in this analysis.

During more than 10 years, LEP has exploited to its limit 
the current technology and design. The limitation for such a 
design is synchrotron radiation, 
which depends on the curvature of the collider and on the energy of the 
electron or positron beam. Higher energies than those reached by LEP would 
require more power to replace the energy radiated or  
even larger rings to reduce the curvature and hence the synchrotron radiation,  
both of which are expensive.
Therefore, the future of electron-positron colliding machines
lies in new technologies which explore linear collider design (\cf{} TESLA at DESY, 
CLIC at CERN, JLC in Japan and NLC 
in the U.S., where already the first collider of this kind, the SLC at SLAC, 
has successfully fulfilled its goals).

The LEP collider was in operation from August 1989 to November 2000. 
A summary of the whole LEP activity is given in Fig.~\ref{fig:lulep} in 
terms of integrated luminosities per LEP experiment.
From 1989 to 1995, the LEP I period, its working energy was around the \Z{} mass, 
near $91.2\GeV$. This period was dedicated to the extensive study of 
the parameters of the Z boson. About 4 million \Z{} events were collected 
during this period. 
During 1995, 
a major upgrade took place in order 
to increase the LEP working energy, to enable the production of 
$\text{W}^\pm$ bosons and to continue the search 
for Higgs bosons and for supersymmetry already started 
at LEP I. During this new era, 
called LEP II, the LEP energy was gradually increased up to 
$209\GeV$ in 2000. 
The year 2000 also saw the report by ALEPH and L3 of 
events compatible with a Higgs signal. 
However, too few events were reported to confirm a discovery, but 
too many to be rejected as a simple statistical fluctuation. Nevertheless, 
this was the motivation for an extension of the data taking period  
from September to November 2000. However the additional data collected 
during this period  did not settle the issue. 
It was finally decided
to definitely close LEP on this {\it status quo}, 
leaving this question unanswered, but open to the higher energy collider  
TEVATRON at FNAL as well as to the next generation 
of colliders  and, in particular, for the Large Hadron Collider (LHC), 
whose construction has already started in the LEP tunnel. 

\section{The L3 detector}

\begin{figure}[htbp]
  \begin{center}
    \includegraphics[width=15cm]{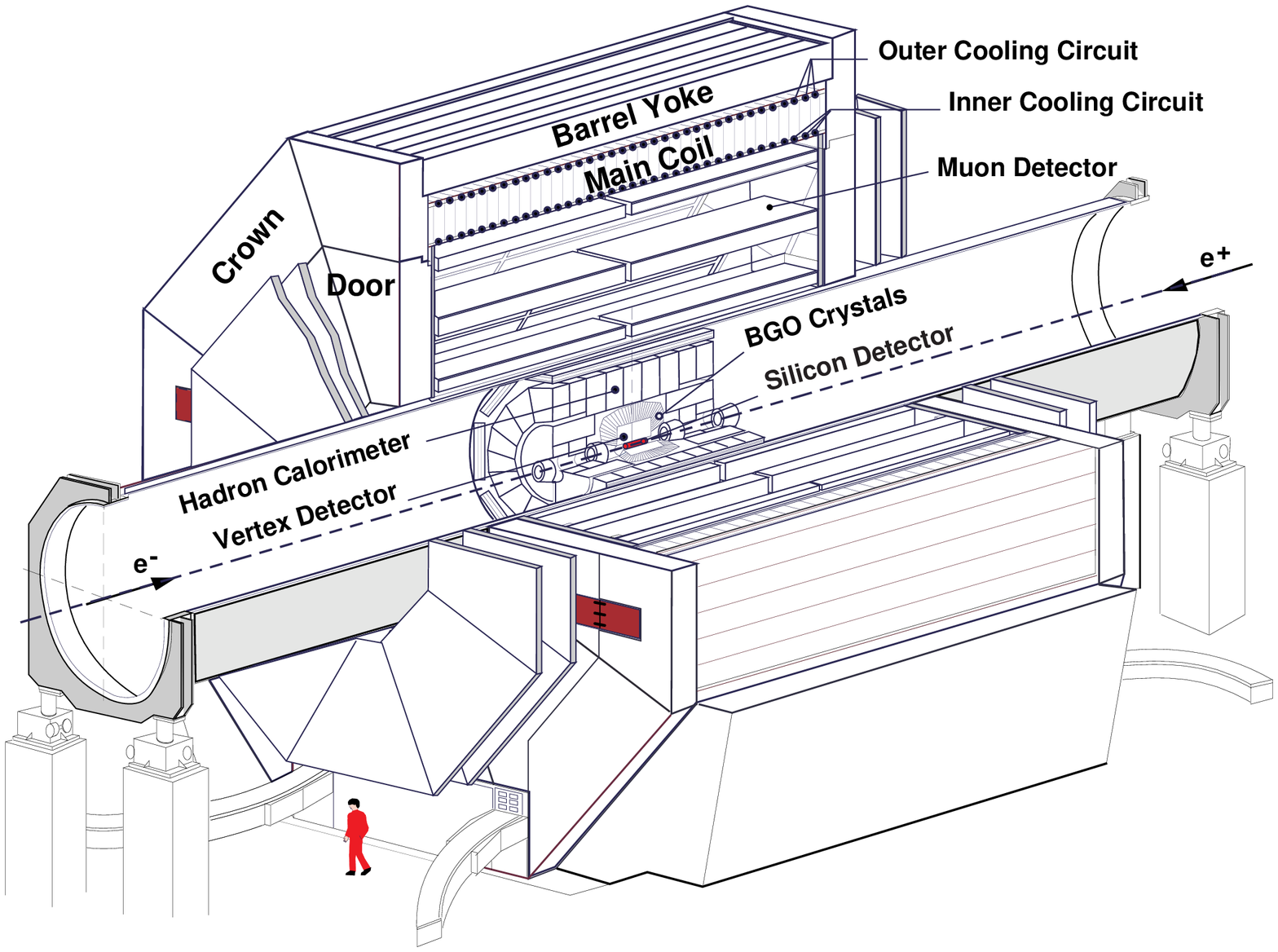}
  \end{center}
\vspace{-0.5cm}
\scaption{Perspective view of the L3 detector.}
  \label{fig:l3d}  
  \begin{center}
    \includegraphics[width=16cm]{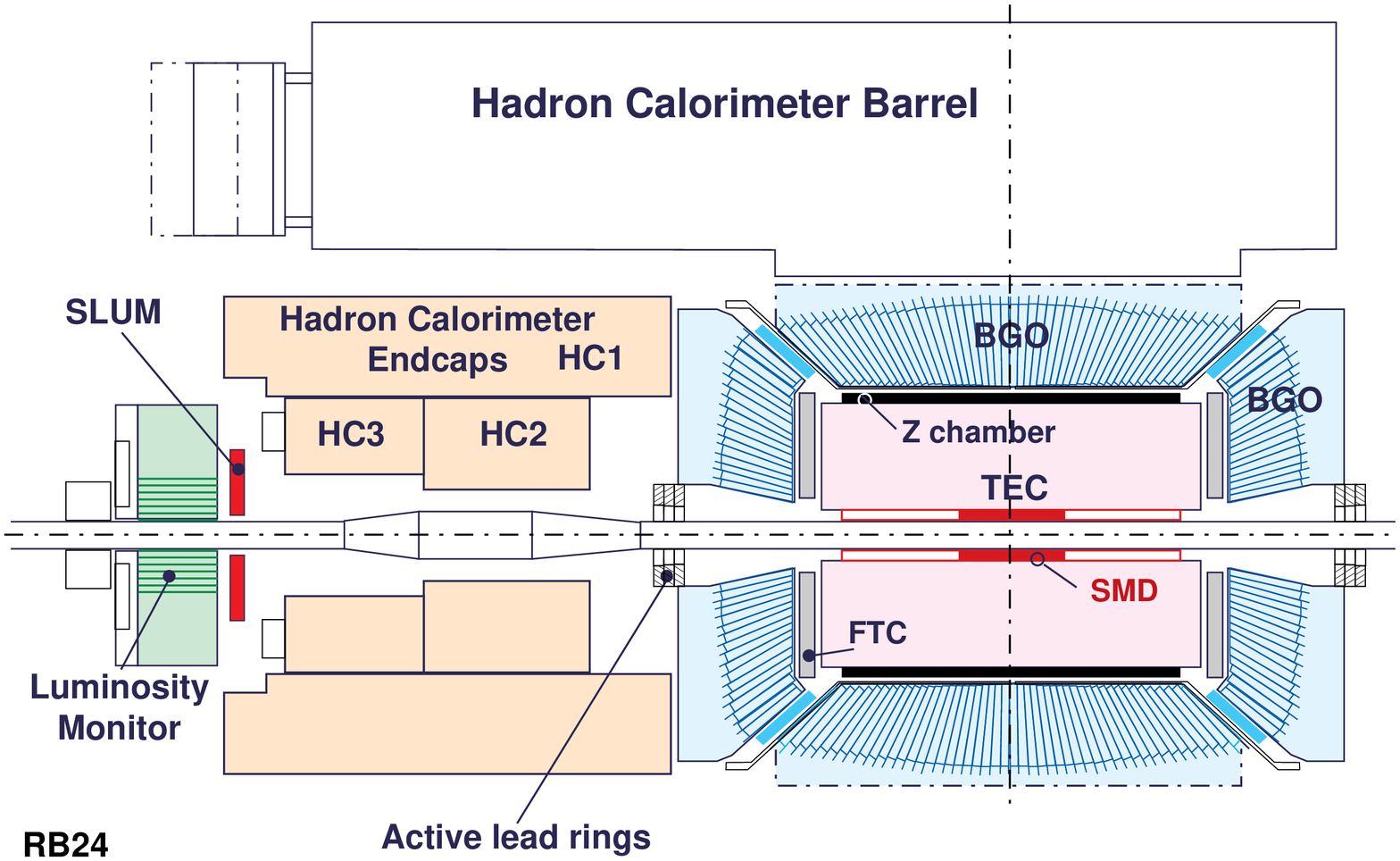}
  \end{center}
\vspace{-1.cm}
\scaption{Longitudinal view of the inner part of the L3 detector.}
  \label{fig:inl3}  
\end{figure}

Fig.~\ref{fig:l3d} shows a perspective view of the L3 detector. 
The basic orientation of the detector is defined from the 
interaction point (at the center of the detector), which is  
the origin of the coordinate system in which the analysis takes 
place. 
Using the interaction point as origin, we define   
the coordinate system in the following way. The $x$ axis 
is perpendicular to the beam pipe, toward  
the center of LEP ring, 
the $y$ axis perpendicular to the beam pipe, pointing 
towards the top of the detector,  
the $z$ axis along the beam pipe, in the 
direction of the electron beam. It is also useful to define
this system in spherical coordinates, where $r$ is the distance
taken from the origin, $\theta$ is the angle
between $\vec{r}$ and the $z$ axis, and $\phi$ the angle between
the $x$ axis and $\vec{s}$, the projection of $\vec{r}$ onto  
the $xy$ plane.

The L3 detector is, in fact, composed of several subdetectors, which 
are fully described in \cite{det1}. 
These subdetectors are all inside a huge octagonal iron magnet 
of $12\text{m}\times 12\text{m}\times12\text{m}$, which delivers 
a uniform magnetic field of 0.5 T along the $z$ axis.

From the magnet wall to the interaction point, and 
by increasing order of importance for this analysis, we have: 

\subsubsection{Muon Chambers (MUCH)}

Between the magnet and the inner part of the detector lies  
the muon chamber system. It is located far away from the interaction 
point, so that only energetic muons (with momentum larger than $3\GeV$)
can reach it and be detected, other particles being totally absorbed by 
the material between the interaction point and the muon chambers. 
The system consists of 3 layers of drift chamber grouped in 8 octants 
covering the region around the beam pipe 
(\ie{} $45^\circ<\theta<135^\circ$). 
It gives a measure of the momentum of a muon track in the 
$xy$ plane. In addition, the measurement of the $z$ 
coordinate is given by Z-chambers located 
on top and bottom of the first and third layers of 
drift chambers.

\subsubsection{Hadron Calorimeter (HCAL)}

The hadron calorimeter (Fig.~\ref{fig:inl3}) 
is made of 5 mm thick depleted
uranium plates ($\text{U}_3\text{O}_8$) interleaved
with proportional wire chambers. The uranium plates act 
as an absorber while the proportional chambers enable us to record
the position of the hadron along its path 
through the calorimeter and to measure its energy  
by the total absorption technique.
Such a measurement is only effective if the hadron 
is totally absorbed in the calorimeter.  
Therefore, a high density material is required as an absorber and  
Uranium 238 fulfills this requirement.  
Furthermore, its natural radioactivity is an advantage for 
 the calibration of the calorimeter.

With components both in the 
barrel and in the endcap, this detector has a geometrical 
coverage of the interaction point close to $4\pi\text{ sr}$ 
($99.5\%$ of $4\pi\text{ sr}$).

\subsubsection{Electromagnetic Calorimeter (ECAL)}

The electromagnetic calorimeter (see Fig.~\ref{fig:inl3}) 
is used to measure the direction and energy of photons 
and electrons. It is made of 11360 bismuth germanate crystals 
($\text{Bi}_4\text{Ge}_3\text{O}_{12}$ abbreviated as BGO).
It covers the range in polar angle of $42^\circ-138^\circ$
for the barrel region and of $10^\circ-35^\circ$ and 
$145^\circ-170^\circ$ for the end-cap regions.
It must be noted that there is a gap in the coverage 
of $7^\circ$ between the end cap and the barrel regions.
An upgrade of the detector in 1995 has partially solved 
the problem by adding scintillator to the detector gap.

\subsubsection{Time Expansion Chamber (TEC)}

\begin{figure}[htbp]
  \begin{center}
    \includegraphics[width=12cm]{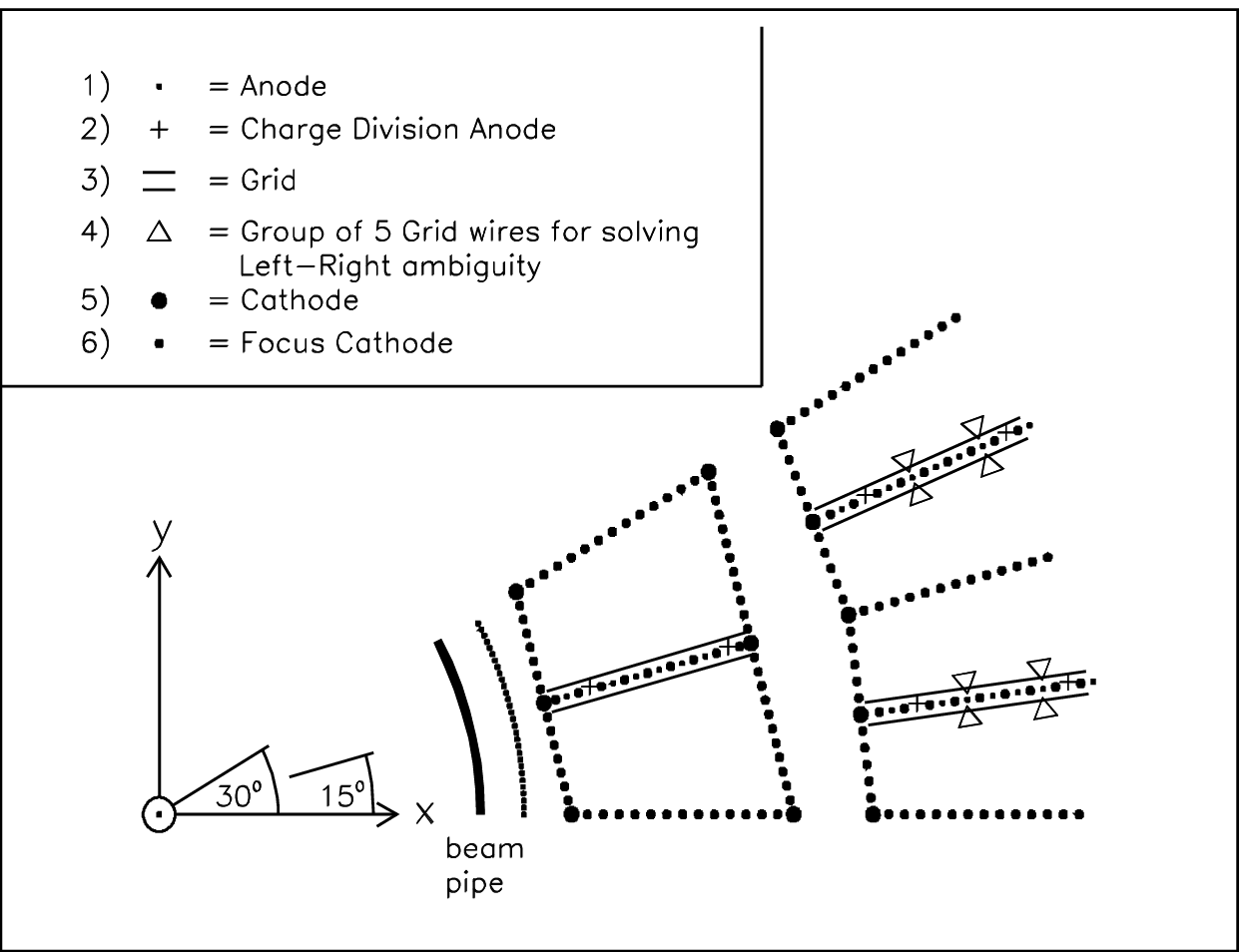}
  \end{center}
\scaption{Schematic view of a segment of TEC, showing an inner sector and 
part of two outer sectors.}
  \label{fig:tecsec}  
\end{figure}

\begin{figure}
  \begin{center}
    \includegraphics[width=12cm]{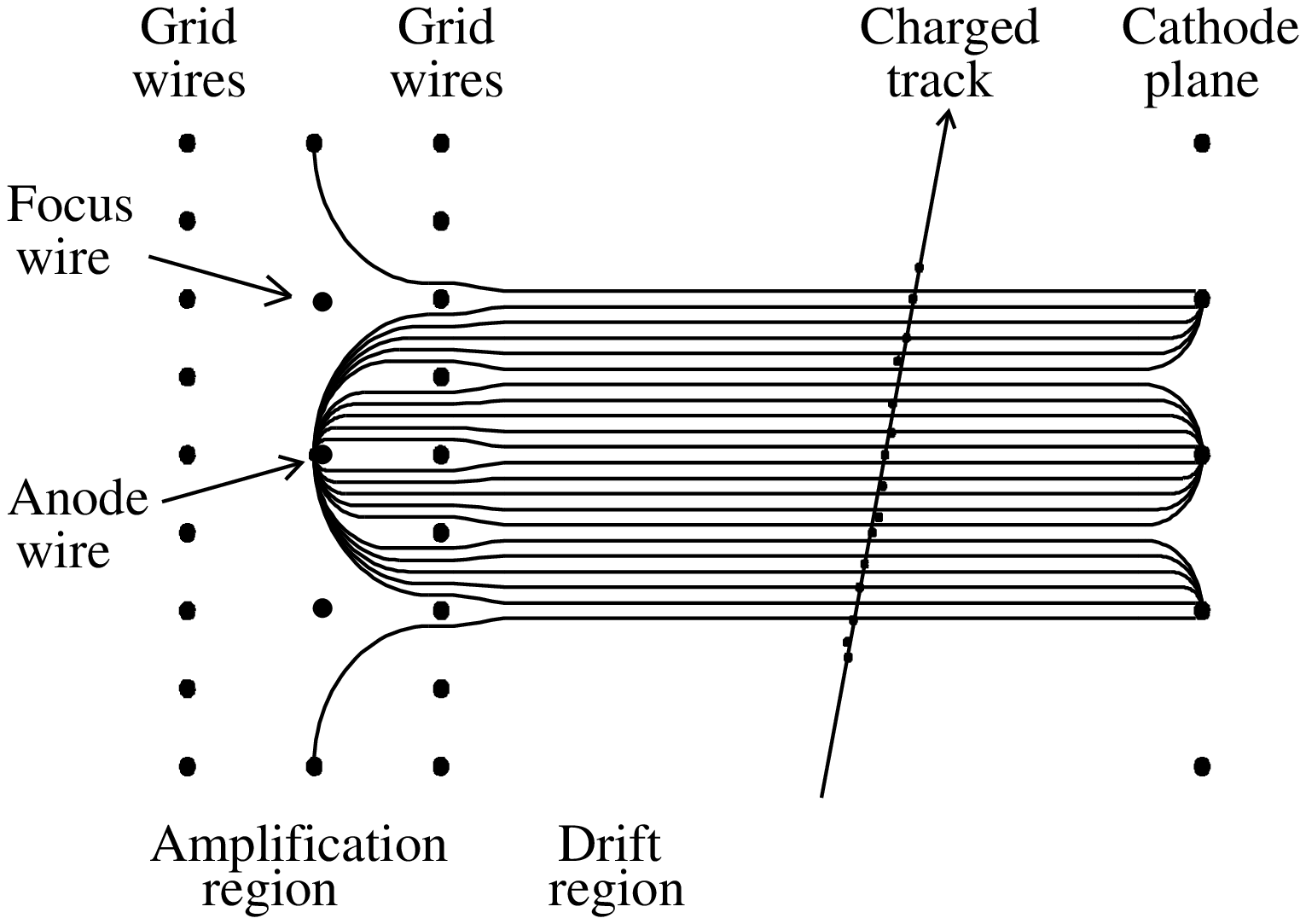}
  \end{center}
\scaption{The time expansion chamber principle of detection.}
  \label{fig:tecdrift}  
\end{figure}

The central tracking chamber is designed to measure 
the direction and curvature (hence, the transverse momentum,
which is calculated from the curvature) of
charged particles.
It consists of a cylindrical drift chamber placed 
along the beam axis. 
The chamber is filled with a mixture of $80\%$ carbon dioxide and 
$20\%$ iso-buthane at a temperature of 291K and a pressure of 1.2 bar.
The TEC is divided in two parts, the inner chamber starting 
at $8.5\text{ cm}$ from the interaction point and extending 
 to $14.3\text{ cm}$, and the outer chamber surrounding 
the inner one and extending to $46.9\text{ cm}$.
The inner part of the TEC is subdivided into 12 identical 
inner sectors each covering $30^\circ$ of the $xy$ plane. 
The outer part is subdivided into 24 identical outer sectors 
covering $15^\circ$ of the $xy$ plane (Fig.~\ref{fig:tecsec}).
Each inner and outer sector contains 8 and 54 sense wires (anodes), respectively.
These anode wires, stretched along the beam pipe, are the active part of 
the detection method used in the time expansion 
chamber principle (described in Fig.~\ref{fig:tecdrift}).  
A charged particle passing through the TEC chamber in the presence 
of a high homogeneous electric field ($0.9\text{kV}/\text{cm}$) 
causes a local ionization of the gas. The electrons 
produced by this ionization drift toward the nearest anode. After  
amplification, the signal produced on the anode by the electron 
flow is recorded as a hit.
This allows 
to record the path of the charged particle (track) in the $xy$ plane, from 
which its curvature and, consequently, its momentum can be reconstructed.  
The $z$ coordinate of a charged particle is measured  
by  two cylindrical systems of proportional wire chambers (Z chambers) 
placed around the outer TEC.
TEC and Z chambers allow the precise measurement of track parameters in the 
barrel region ($45^\circ < \theta < 135^\circ$).

The endcap regions are covered by proportional wire chambers, 
FTC (Forward Tracking Chambers), allowing the $z$ coordinate measurement 
in this region, but the TEC is hardly effective in measuring 
track parameters in this region, since the anode wires are parallel 
to the beam pipe. 
Therefore, a forward charged particle will traverse fewer wires and 
its curvature (and hence its momentum) will lack precision.
Therefore, tracks in the barrel region are more precisely 
measured than those in the endcap region.

\subsubsection{Silicon Micro-Vertex Detector (SMD)}  

\begin{figure}[htbp]
  \begin{center}
    \includegraphics[width=10cm]{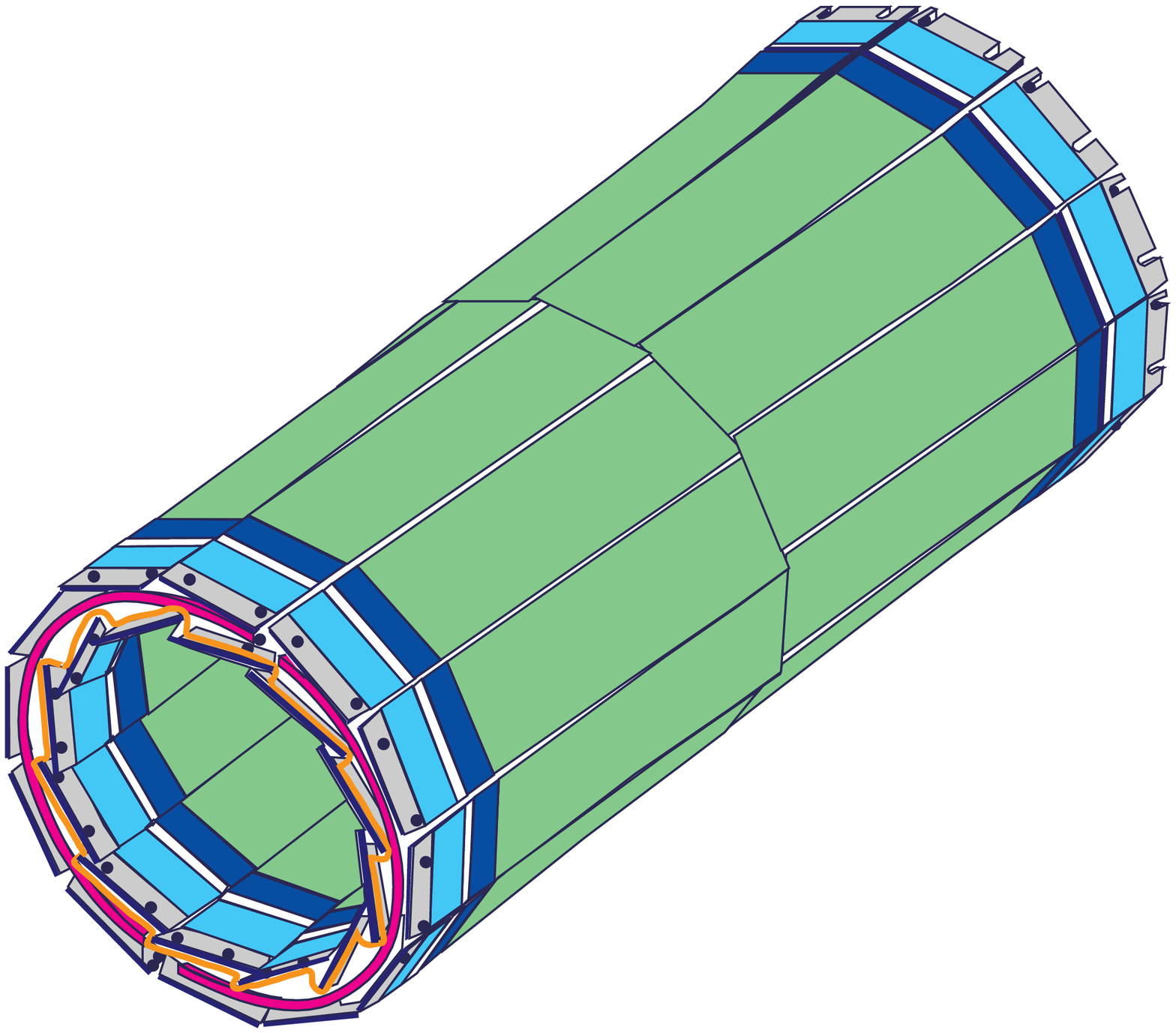}
  \end{center}
\scaption{Perspective view of the SMD.}
  \label{fig:smd}  
\end{figure}

The SMD (Fig.~\ref{fig:smd}) is located between the beam pipe and the TEC. 
It is the detector closest to the interaction point. It is aimed at measuring 
very precisely track parameters in order to pinpoint 
the impact parameter, which makes it perfectly designed for b-quark 
 identification. It provides  good $r\phi$ and $rz$ coordinate resolution for a  
polar angle range of $21.5^\circ-158.5^\circ$. Its high resolution improves the 
performance of the TEC (\ie{} better transverse momentum resolution).

The SMD consists of two radial layers supporting 12 ladders each.
The ladders are the basic element of the SMD. Each contains 
4 silicon microstrip sensors made of high-purity n-type silicon. 
On its junction side, each sensor carries strips 
designed to measure the $r\phi$ coordinate, 
while its ohmic side has strips perpendicular 
to those of the junction side, in order to measure the 
$rz$ coordinate.

\begin{figure}[htbp]
  \begin{center}
    \includegraphics[width=10cm]{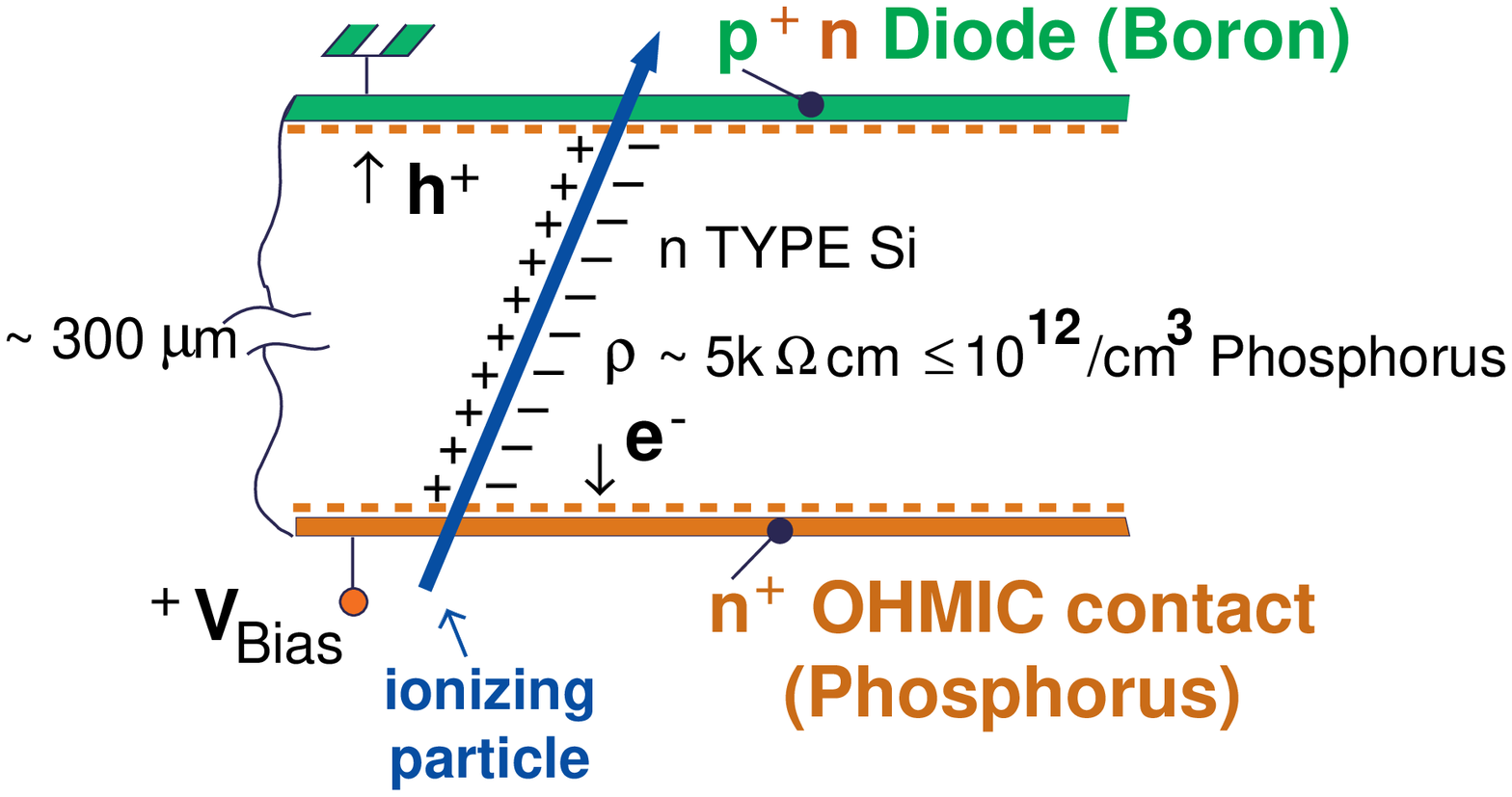}
  \end{center}
\scaption{The principle of detection of the SMD.}
  \label{fig:pnjunc}  
\end{figure}

The principle of detection of the silicon microstrip
(described in Fig.~\ref{fig:pnjunc}) is somewhat similar
to that of the time expansion chamber, but, of course, 
the medium is different. The detection, here, benefits from the 
semi-conductor properties of the material.
A particle passing through the silicon sensor will
produce electron-hole pairs. By applying a voltage 
bias between the two sides of the sensor, holes and 
electrons will drift to the nearest strips on both 
surfaces, allowing a simultaneous measurement  
of the $r\phi$ and $rz$ coordinates. 

\section{Data processing}

The way in which particles are detected and recorded 
by the various subdetectors can sometimes be  
very far from the physical quantities we 
want to measure. Therefore, it is necessary to 
translate the information coming from the detector 
into more convenient variables, which can later be 
used to perform  physics analyses such as those 
described in this thesis.

The treatment of the data requires three main steps. 
The first one is, of course, to decide if the signals
emitted by a detector are relevant to a physical 
process of interest. 
This role is attributed to the trigger systems.

The next step is the reconstruction of the data, 
which translates detector response into physical 
quantities and stores this information on storage 
media for later use.

The third step which is needed, even if it does not 
directly concern the data themselves, is very 
important, it is the detector simulation.
It enables us to understand (and to reproduce)
the response of the detector components to 
the passage of particles, from a given  
physics process. Therefore, it helps us to 
understand the data and how particles interact 
with the detector.

\subsection{Trigger system}

No matter how relevant the information coming from
the detector, the complete readout sequence of an event 
takes about $500\mu\text{s}$ (corresponding to the time of  
22 \ee beam crossings), during which the 
detector cannot process any new events.
Therefore, a multi-level trigger system is implemented to reduce 
this dead time by allowing abortion of the readout 
sequence as soon as possible if a piece of 
information is found to be faulty. This  
enables the detector to start sooner a new readout sequence, 
and hence the recording of a good event.

The goal of the trigger system is to 
improve the interface between the detector 
and the data acquisition system (DAQ). 
Furthermore, it decides what can be trusted  
as relevant for physics analysis and should be 
written onto tape, thus reducing both the amount 
of memory space and the time lost in recording 
useless information.
From the $45\text{kHz}$ collision only 
a few $\text{Hz}$ can be stored on tape.  
The trigger ensures a certain 
level of quality of the information 
written on tape.
 
The trigger system proceeds in a small number of steps: 
At first, it has to decide, from  signals emitted 
by a detector component, if these signals are
compatible with an \ee{} interaction. 
Then, if the signal is relevant to an \ee{} collision, 
it decides, depending on the quality of the response of 
the subdetectors, to allow 
or to veto the storage on tape of these signals
as an event.

There are three levels of trigger. 
The difference between the three levels
is the complexity of the operations treated and
the time needed to perform these calculations.
Such a  configuration allows a lower-level 
trigger to abort the readout sequence before 
higher-level triggers have completed more time
consuming operations, and hence  allows to save time 
in the readout process.

\subsubsection{Level-1}

The first-level trigger is a fast-response 
trigger.
It consists of 5 independent sub-triggers, 
the energy trigger (which analyses the response
from the calorimeters), the TEC trigger, 
the muon trigger (dedicated to the response
of the muon chambers), the scintillator trigger,
and the beamgate triggers.
The role of the first-level trigger is 
to initiate the readout sequence if an \ee{} collision 
is detected 
and to perform simple tests on an 
event in order to decide to keep it or 
not for further processing. 
Its decision time is about $20\mu\text{s}$.
In order to pass the level-1 trigger,
the event has to be selected by at least 1 
of the 5 sub-triggers.

\subsubsection{Level-2}

The level-2 trigger works in parallel to the 
first-level trigger and has access to the same 
information. It has more time to proceed 
to a decision, however.   
Only events which were selected by only one 
level-1 sub-trigger are considered by the 
level-2 trigger. Events which were selected by more
than one level-1 sub-trigger are automatically selected.
The level-2 trigger is aimed at rejecting the most obvious 
background events, such as cosmic events (a muon produced by cosmic rays), 
detector noise, and interaction of the beam with residual gas (beam-gas) or 
with the wall of the beam pipe (beam-wall). 

\subsubsection{Level-3}

The level-3 trigger uses the full data information and is 
able to perform full reconstruction of the data. 
It also has more time to perform more complex calculations,
correlating several level-1 sub-triggers relevant to a 
subdetector. 
A level-3 trigger is implemented for each subdetector. 
For example, events which are selected by the TEC trigger
are required to have tracks correlated with energy deposits 
in the calorimeter. 
Finally, if an event is selected by a level-3 trigger, 
it is written onto tape. This takes $500\mu\text{s}$.

The accepted events are grouped into runs of about 5000 events,
corresponding (originally) to the tape capacity but
also to a constant state of activity of the detector,
a new run being initialized in case of change 
of status of the detector.

For our analysis, the events are required to originate 
from runs where TEC, SMD, HCAL and ECAL triggers 
were active. This ensures a uniform level of 
quality for the full data sample, which decreases  
the systematic uncertainties.

\subsection{Event reconstruction}

The information written onto tape during the data acquisition 
consists essentially of the recording of the various  
signals emitted by the detector. This information
cannot be used directly in physics analysis. Further
processing is, therefore, needed in order to extract
the physical quantities relevant 
to particle physics analysis. 

The reconstruction proceeds in two steps. 
In the first step, using the program package REL3,
the various signals coming from one
subdetector are combined into a primitive object, 
characteristic of this subdetector.
The second step, by means of the subprogram
AXL3, processes these objects correlating the information 
from various subdetectors, to obtain a class 
of objects relevant to physics analysis.
There are several objects related to the main detector
components.
The following describes briefly the objects used in the  
present analysis.

\begin{itemize}

\item ASRC's (AXL3 Smallest Resolvable Clusters), or simply clusters:
These objects are obtained by combining the information from
the calorimeters (both hadronic and electromagnetic). 
They correspond to the smallest energy deposit which can be resolved. 
The ASRC's are used in this analysis to select hadronic events
(Sect.~\ref{sec:selcal}).

\item ATRK's (AXL3 TRacKs), also called tracks:
These objects are obtained by combining TEC, SMD and Z chamber 
information. They are also required to be matched with a 
calorimeter object. The ATRK's correspond to charged 
particles detected in the inner part of the L3 detector.
They are the main objects used in this analysis.

\end{itemize}

\subsection{Event simulation}

Natural ways to understand the data include their comparison  
to theoretical expectation or to try to reproduce their signature 
using a Monte Carlo generator which incorporates the present knowledge 
we already have 
of a reaction. The generation of Monte Carlo events is thus very 
important for our understanding of the underlying physics.

However, it must be kept in mind, that the detector is not 
$100\%$ efficient and part of the information can be  
lost or distorted by the various materials used for 
the detection. 
It is mandatory to understand the interaction 
between particles and the detector material, as well as the 
effect induced by the various parts of the detector.
This understanding is incorporated into a Monte Carlo  
program which simulates the perturbation 
induced by the detector and the detection itself.

Therefore, the events generated by the 
Monte Carlo event generator are also processed by SIL3~\cite{sil3}, 
a program based on the GEANT program 
(a general program package designed to simulate 
interactions between particles and detector 
materials) and aimed at simulating the whole
chain of detection (from the detection  
itself to the DAQ) of the L3 detector.

There are two levels of simulation in the L3 collaboration. 
The first is called ideal simulation.
It corresponds to the simulation of an ``ideal'' 
L3 detector for which all the various detector channels 
work at their maximum efficiency.

The second level of simulation is called realistic
simulation. This simulation is time dependent and 
the major changes in the detector during 
a period of data taking are incorporated.
This can be the permanent loss of detection channels, 
such as a dead crystal of BGO, noisy electronic channels,  
or eventual problems in a subdetector causing 
its inactivity. 
As for the TEC, the high voltage is permanently 
monitored ( a status is recorded every five minutes) 
during data taking in order to incorporate in the realistic 
simulation the loss of power (and consequently the partial  
or total loss of data) in one or all sectors. 
 The same runs used for the data are used in the 
Monte Carlo simulation.

The analysis reported here uses realistic Monte Carlo simulation.

\chapter{Event selection}\label{chap:sel}

   This analysis is based on data collected by the 
   L3 detector in 1994 and 1995 at an energy equal to the mass 
   of the \Z{} boson. The data sample corresponds 
   to approximately two million hadronic \Z{} decays. 
   Since the analysis makes extensive use of the reconstructed 
   charged-particle multiplicity distribution, 
   not only a good purity in hadronic events is needed, but also 
   a well understood selection of the charged tracks.
   This understanding cannot be achieved without a precise simulation of the 
   Central Tracker of L3.

   In order to fulfill the requirements of purity and track selection, 
   the events are selected in a two-step procedure.
   The first step selects hadronic events and removes most of 
   the background, using the energy
   measured in the electro-magnetic and hadronic calorimeters.
   The second step of the selection, more specific to this analysis, 
   is aimed at selecting 
   good tracks measured with the Central Tracking Detector,
   in order to obtain the best 
   reliability of data and Monte Carlo simulation while keeping  
   the number of tracks in the event as large as possible. Good
   agreement between data and simulation is essential in order to 
   reconstruct the charged-particle multiplicity distribution, 
   since this reconstruction is strongly 
   dependent on the description of the inefficiencies of the 
   L3 detector, which are obtained from simulations.

   Also used in our analysis are samples obtained from light- 
   and b-quark events, separately.  
   The procedure used to extract the charged-particle multiplicity distributions
   from these particular types of events is described in the third section.

\section{Calorimeter based selection}\label{sec:selcal}   
    
    The selection of hadronic events is based on the energy measured 
    in the hadronic and electro-magnetic calorimeters. 
    Its main purpose is to remove as large as possible a fraction of 
    the background in such a way
    that this does not affect the measurement of the charged tracks in TEC.  
    Of course, the background could
    be eliminated using the Tracking Chamber only. However, the cost 
    to pay in terms of efficiency would be rather large, since this would
    prevent any measurement of the low-multiplicity 
    events, which are highly contaminated by many background 
    sources (described later). 
    
   The background sources can be divided into two main categories~\cite{herve}:
   The first category consists of events originating from leptonic \Z{} channels 
   (\ee{}, \mumu{}, \toto{}). 
   The second category, called non-resonant background, contains sources such as 
   two-photon interactions, as well as beam-wall and beam-gas events.
    
    A preliminary cut on the 
    calorimeter cluster energy removes calorimeter clusters with an energy deposit 
    smaller than $100\MeV$, which are highly contaminated by electronic noise.
    Once these clusters are removed, we can proceed to the event selection.
    For that, we need to define a set of useful variables.
    First of all we define the visible energy, $\Ecal$ of an event as  
    the sum of all (remaining) cluster energies $E^\text{cal}_i$. 
    In a similar manner 
    we define the vectorial energy sum $\vec{E}^\text{cal}$, obtained by 
    summing the cluster energy along the particle direction as seen from 
    the interaction point, $\vec{n}_i$:

\begin{equation}
    \Ecal={\underset{i}{\sum}}E^\mathrm{cal}_i\phantom{2}\text{ and }
    \phantom{2}
    \vec{E}^\text{cal}={\underset{i}{\sum}}E^\mathrm{cal}_i\cdot\vec{n}_i.
\end{equation}
    We also define the longitudinal and the transverse energy imbalance,
    $\Ecpar\text{ and }\Ecperp$ as the projection along the $z$ axis 
    and in the plane perpendicular to the $z$ axis 
    of $\vec{E}^\text{cal}$ normalized to the visible energy $\Ecal$, 
    respectively:
    
\begin{equation}
\Ecpar=\frac{|E^\mathrm{cal}_{z}|}{\Ecal}\phantom{2}\text{ and }
\phantom{2}
       \Ecperp=
       \frac{
       \sqrt{(E^\mathrm{cal}_{x})^2+(E^\mathrm{cal}_{y})^2}}{\Ecal}.
\end{equation}
    
\subsubsection{Cuts on the rescaled visible energy}

\begin{figure}[htbp]
  \begin{center}
  \includegraphics[width=10cm]{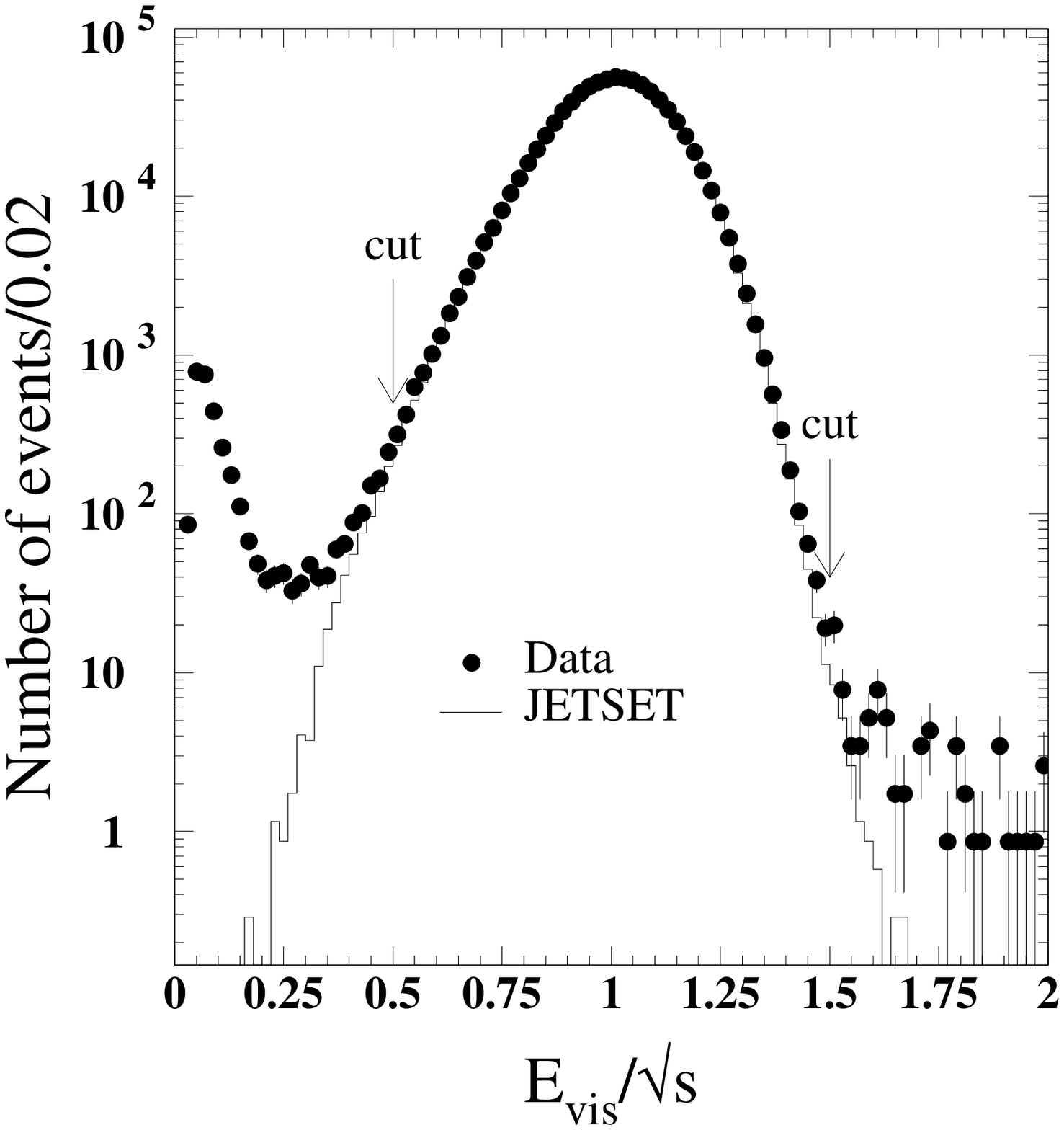}
  \scaption{Distribution of the visible energy 
   rescaled to $\sqrt{s}$ for the data  (dots) 
  compared to events simulated with JETSET 7.4 PS (line).
  As for all the following  selection plots, all cuts are applied, 
  except that on the variable plotted, the latter cut 
  being represented by arrows.}
  \label{fig:evis}  
  \end{center}
\end{figure}

   Hadronic \Z{} events are characterized by a visible energy
   centered around the center of mass energy, $\sqrt{s}$. 
   Non-resonant background, in particular beam-wall, beam-gas and 
   two-photon events, which typically have a much lower visible energy, are  
   easily discarded by a cut on $\Ecal{}$ (Fig.~\ref{fig:evis}).
   Selected events are required to satisfy  
\begin{equation}
   0.5<\frac{\Ecal}{\sqrt{s}}<1.5.  
\end{equation}
The role of the upper cut is to remove Bhabha events which are located 
at scaled energy higher than 1.5  because of the 
scaling factors (G-factors), which are used to take into account 
a shift to lower value of the energy detected for hadrons. Since the 
energy of electrons is fully detected, they should not be subjected 
to this shift consequently end up with an higher scaled energy. 
\subsubsection{Cut on the number of ASRC clusters} 
    
\begin{figure}[htbp]
  \begin{center}
  \includegraphics[width=10cm]{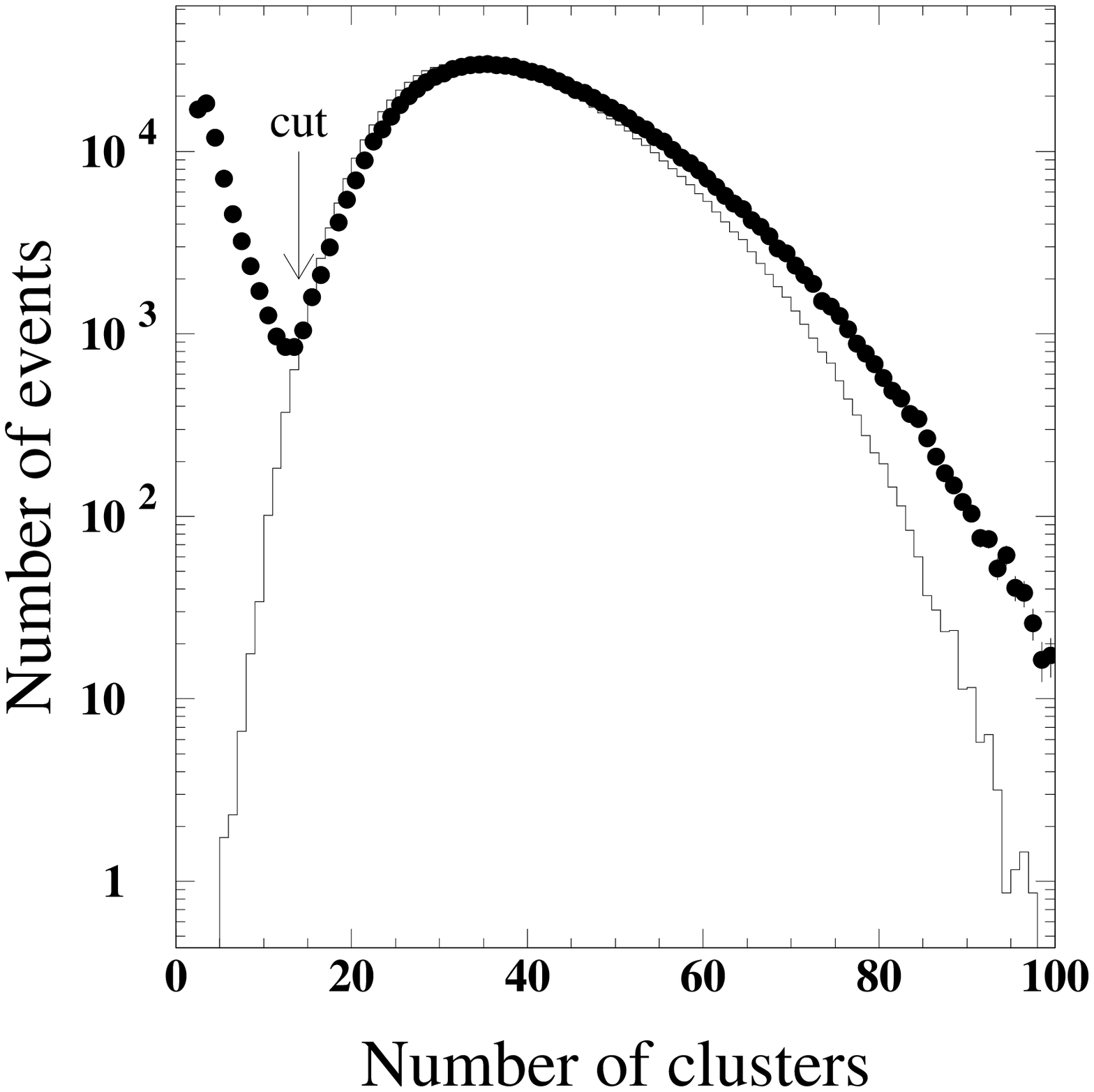}
  \scaption{Distribution of the number of ASRC clusters for the data 
   (dots) compared to JETSET 7.4 PS (line) predictions.}
  \label{fig:nclus}  
  \end{center}
\end{figure}

    Hadronic events usually have a larger particle multiplicity 
    than other processes. Hence, a 
    way to reduce the background contamination is to cut on 
    low-multiplicity events.
    By requiring that events have at least 14 ASRC clusters, most of the 
    \ee, \mumu~and \toto~background is eliminated (Fig.~\ref{fig:nclus}). 
    It must be noted that 
    the large discrepancy between Monte-Carlo and data for large 
    multiplicities is due to an incorrect description 
    of hadronic showers in the BGO crystals of the ECAL and not to 
    some kind of background contamination. Therefore, there is no reason in 
    this analysis, which uses only charged tracks, to cut on large 
    cluster multiplicities.  
\subsubsection{Cuts on the energy imbalance}

\begin{figure}[htbp]
  \begin{center}
  \includegraphics[width=8.4cm]{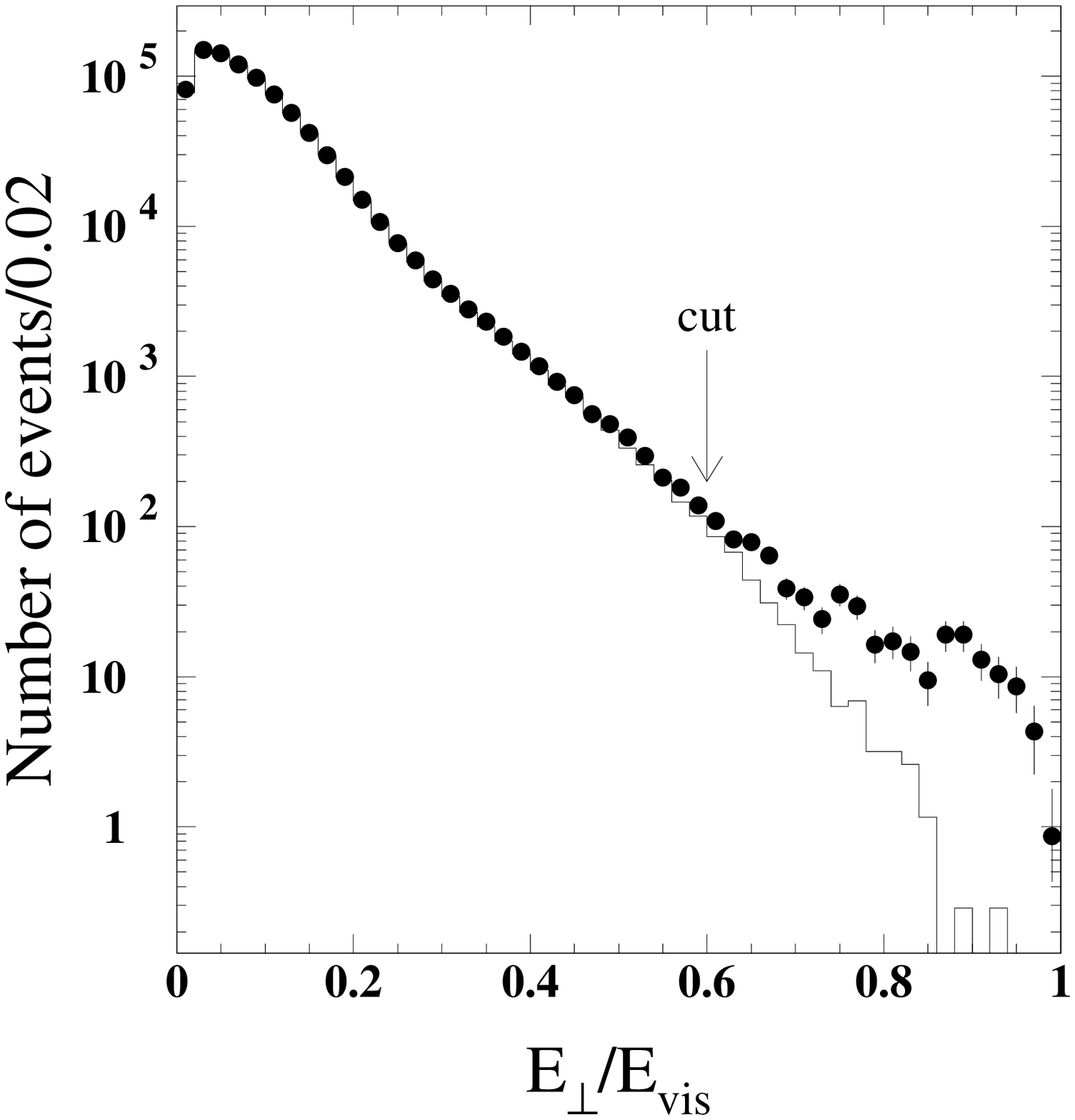}
  \includegraphics[width=8.4cm]{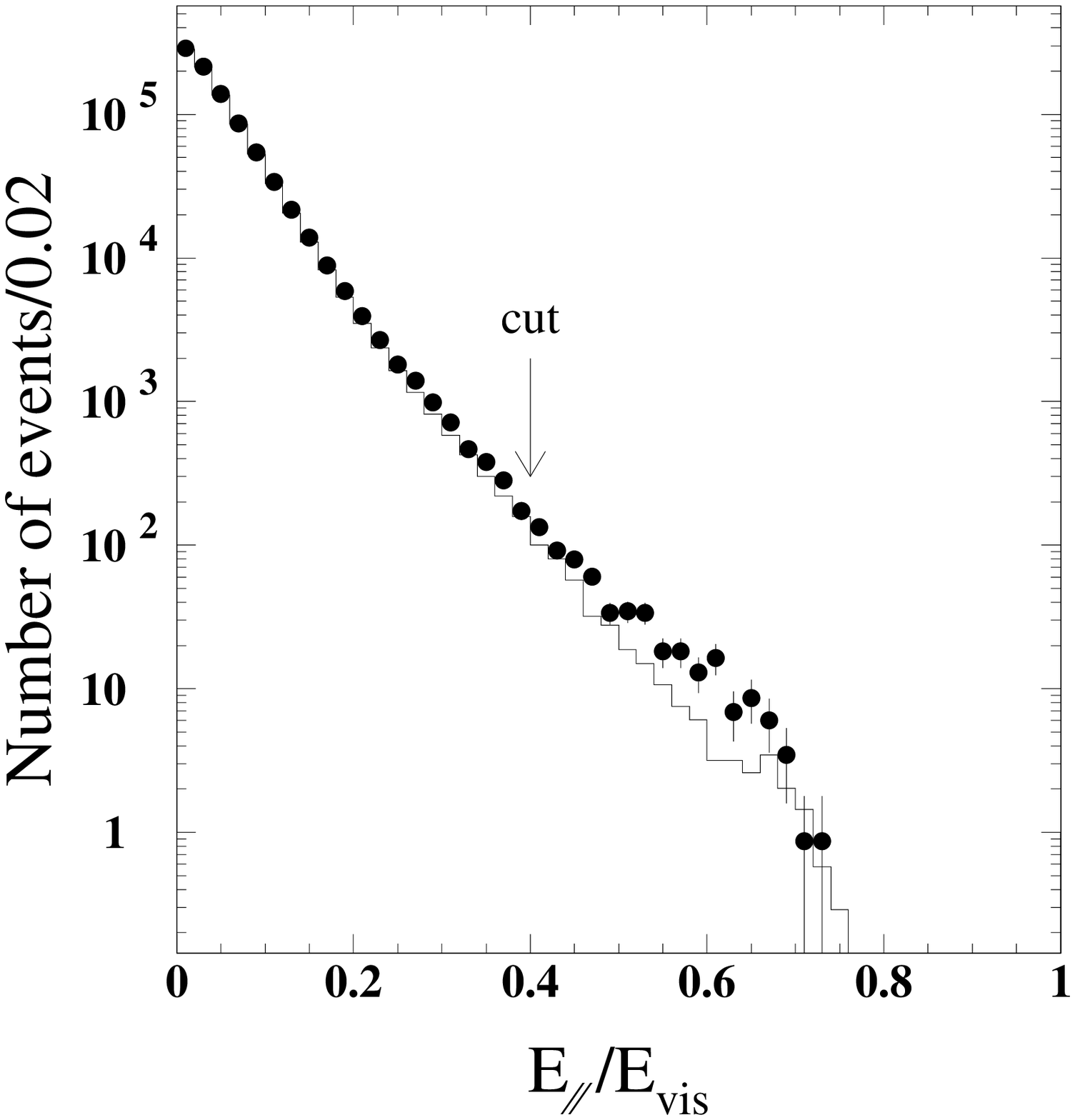}
  \scaption{Transverse and longitudinal energy imbalance for the data (dots) 
  and for the Monte-Carlo (line).}
  \label{fig:imb}  
  \end{center}
\end{figure}

 Since at LEP, the laboratory frame for \ee{} collisions coincides 
 with the center of mass frame, hadronic events
 are well balanced in energy flow.
 This is not the case for the non-resonant background, which
 usually has a large longitudinal energy imbalance. 
 Furthermore, due to $\tau$  
 decay into a quark or a lepton via the emission of neutrinos, 
 the \toto~background  has a larger energy imbalance  
 than the other \ee~channels.
 As shown in Fig.~\ref{fig:imb}, we require the longitudinal energy imbalance 
 $\Ecpar$ to be smaller than  0.4 and the transverse energy imbalance 
 $\Ecperp$ to be smaller than 0.6.
\subsubsection{Cut on the direction of the thrust axis}
\begin{figure}[htbp]
  \begin{center}
  \includegraphics[width=10cm]{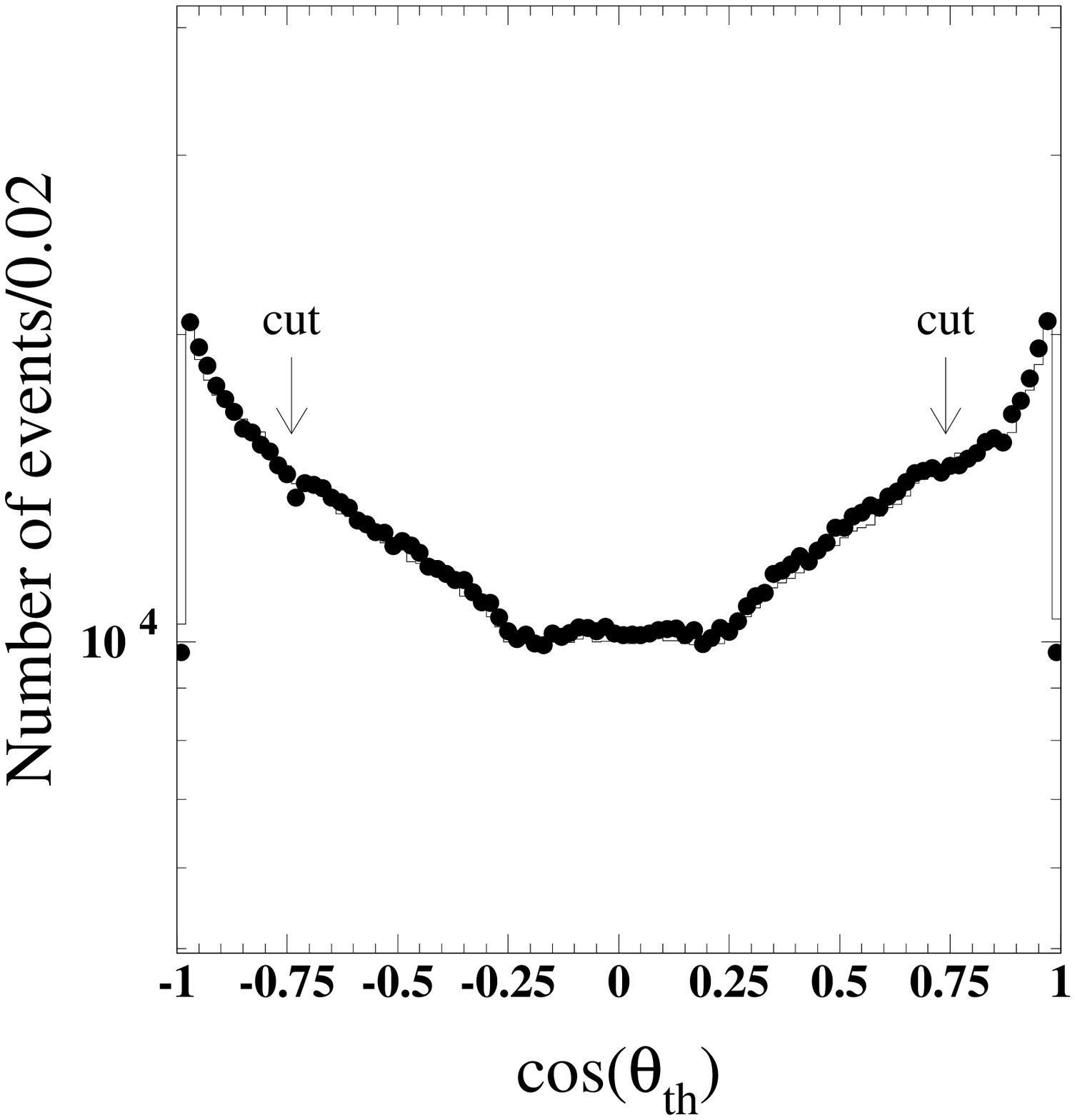}
  \scaption{Distribution of $\cos(\theta^{\mathrm{cal}}_{\mathrm{th}})$ for the data (dots) and
  for the Monte-Carlo (line).}
  \label{fig:cthr}  
  \end{center}
\end{figure}

   To ensure that the event lies within the full acceptance of the TEC and 
   since the TEC only poorly covers the end-cap region of the detector,
   we use only events which have the direction of the 
   thrust axis\footnote{The thrust axis $\vec{t}_1$ is defined  
   as the axis which maximizes the quantity 
   \begin{equation*}
   \frac{\sum_i|\vec{p}_i\cdot \vec{t}_1|}
   {\sum_i|\vec{p}_i|}.
   \end{equation*}
   The maximum value of this quantity is called the thrust.}
   within the barrel of the detector.
   Barrel events are selected by requiring 
   $|\cos(\theta^{\mathrm{cal}}_{\mathrm{th}})|< 0.74$ (Fig.~\ref{fig:cthr}), where 
   $\theta^{\mathrm{cal}}_{\mathrm{th}}$ is the polar angle 
   of the event thrust axis determined from calorimeter clusters.

  To summarize, the calorimeter selection criteria of hadronic 
  events are:
  
  \begin{itemize}
    \item $0.5<\Ecal/\sqrt(s)<1.5$
    \item $N_\mathrm{clus}>14$
    \item $\Ecpar<0.4 \text{ and } \Ecperp<0.6$
    \item $|\cos(\theta^{\mathrm{cal}}_{\mathrm{th}})|< 0.74.$
  \end{itemize} 
After having applied these cuts, approximately one million hadronic events remain, 
with a purity around $98\%$~\cite{herve}. All the non-hadronic background 
(\ie{} background which does not decay hadronically) 
has been removed, with the exception of $1.3\%$ of the \toto{} 
background, $0.23\%$ of the 
$\ee \text{q}\bar{\text{q}}$ background and $1.1\%$ of the \ee{}\toto{} background. 
This calorimeter pre-selection has the advantage of eliminating the background 
while being largely decoupled from the track selection, allowing 
relatively weak cuts on track momenta. 
The next step is the selection of charged particles using 
the central tracking detector.   

\section{TEC based selection}
   
   The main goals of the TEC track selection are 

   1. to remove badly reconstructed tracks and 

   2. to improve the simulation of track inefficiencies, 

   \noindent in order to have the best possible 
   reliability of the central tracking detector simulation, on which the reconstruction 
   of the charged-particle multiplicity and momentum distributions strongly depends.

   Since the event selection by means of the calorimeter clusters has rejected most of the 
   background, only a few cuts will be applied on an event basis.

\subsection{Track quality criteria}\label{sec:trksel}
   
\subsubsection{Transverse momentum}

\begin{figure}[htbp]
  \begin{center}
  \includegraphics[width=8.4cm]{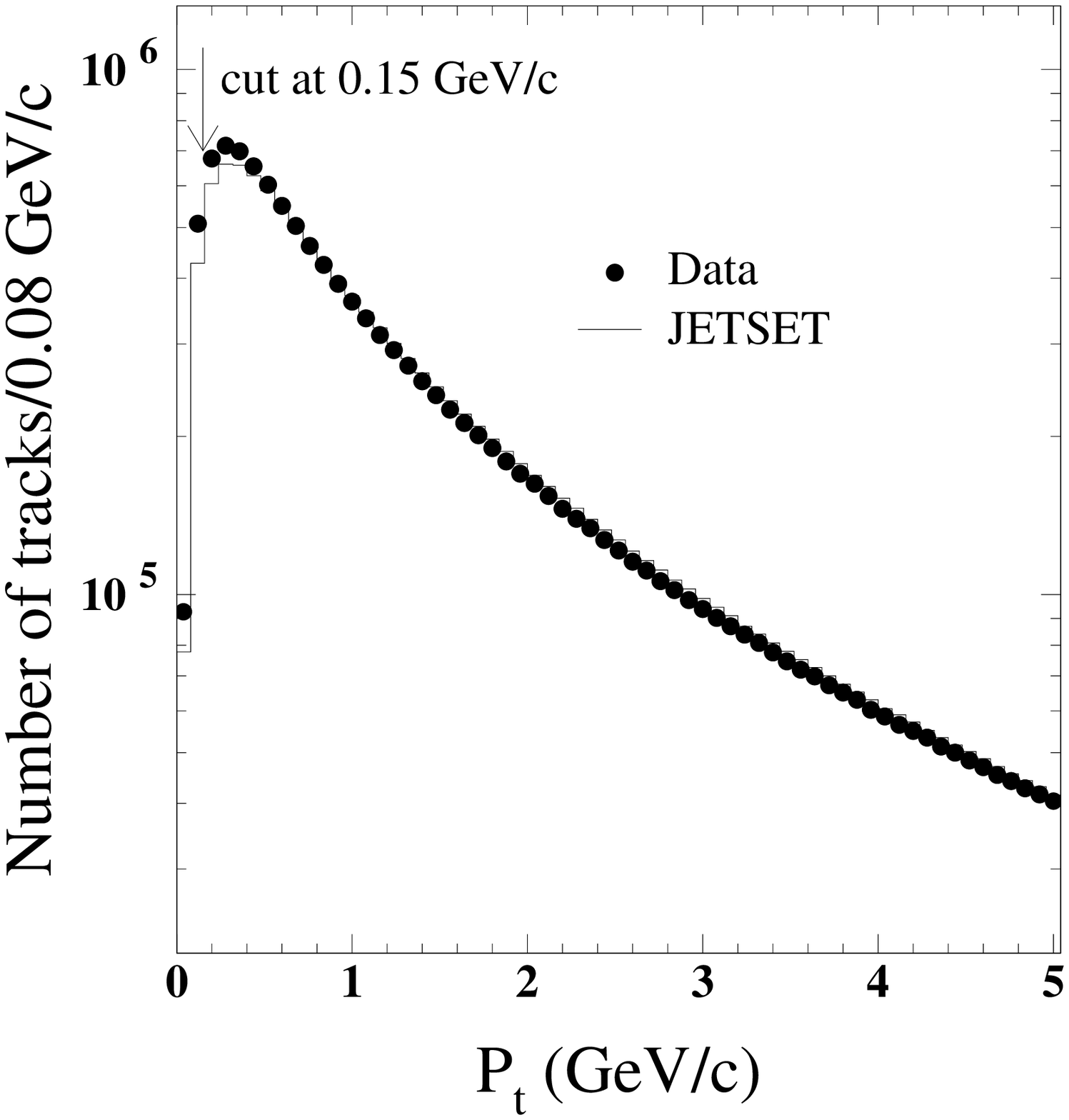}
  \scaption{Distribution of track transverse momenta for the data (dots) and
  for the Monte-Carlo (line).} 
  \label{fig:pt_sel}  
  \end{center}
\end{figure}

  The transverse momentum of a track is calculated from its curvature 
  imposed by the magnetic field in the  
  plane perpendicular to the beam axis. Tracks having a low transverse momentum 
  are easily contaminated by noise and must be removed. 
  Hence, the transverse momentum is required to be larger than 
  $150\MeV/c$ (Fig.~\ref{fig:pt_sel}) 
\subsubsection{Number of hits}

  When a particle, originating from the interaction point 
  and flying across the TEC,
  passes near a wire of TEC, it causes a local ionization of the gas leading 
  to an electric discharge on the wire. This is called a hit.
  There are 62 wires in the TEC, 8 wires in the inner TEC and 54 in the outer TEC.
  The larger the number of hits, the better is the resolution of the transverse 
  momentum, since the curvature is calculated from the path formed by the 
  subsequent hits.
  
\begin{figure}[htbp]
  \begin{center}
  \includegraphics[width=8.4cm]{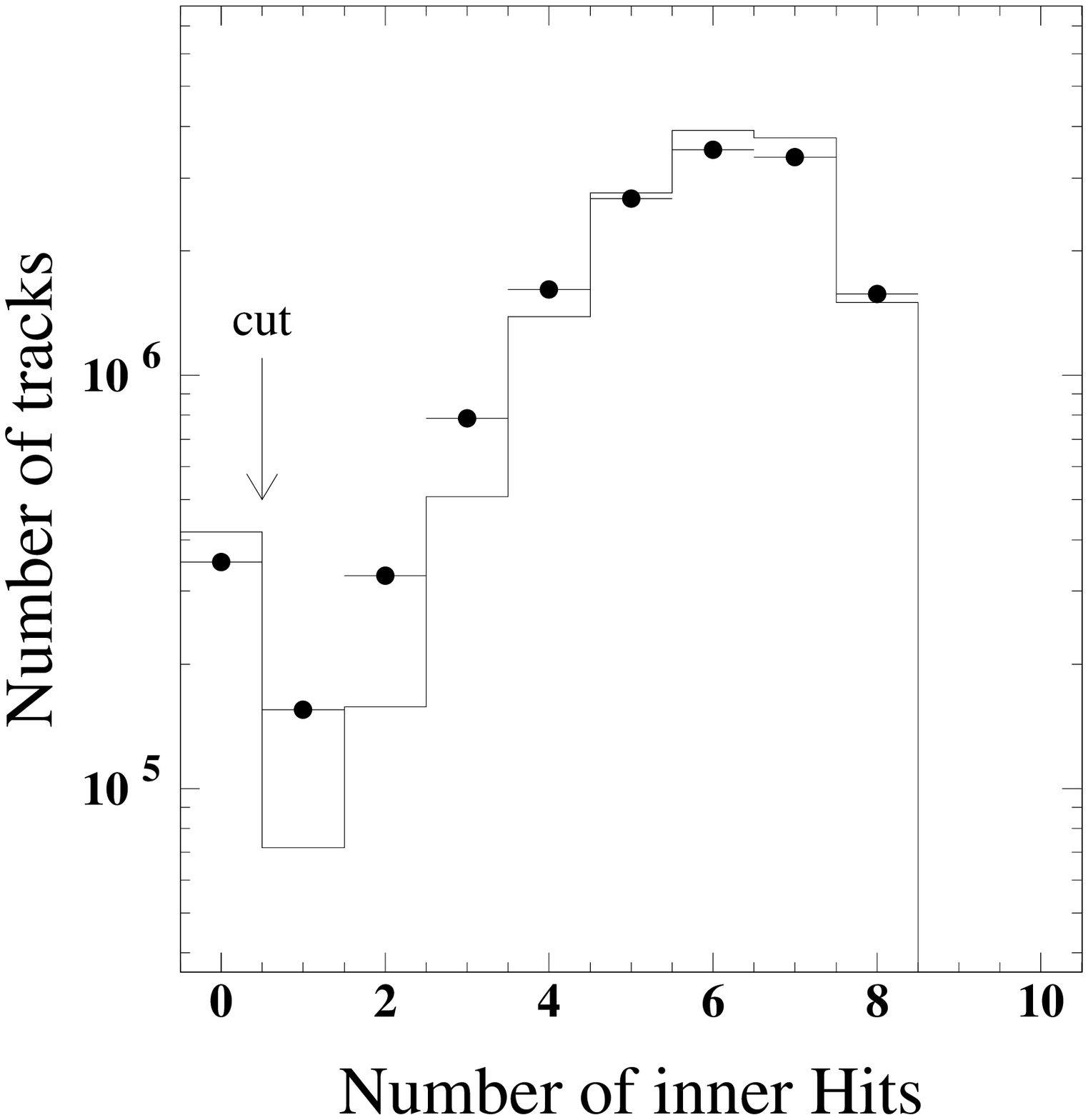}
  \scaption{Distribution of the number of hits in the inner TEC for the data (dots) and
  for the Monte Carlo (lines).} 
  \label{fig:inhits_sel}  
  \end{center}
\end{figure}

  Misreconstructed track segments usually have a small number of hits. 
  Furthermore, the absence of a hit in the inner TEC increases 
  strongly the chance 
  of misreconstruction of a track, since the track is not measured close
  to the interaction region. 
  Therefore, we require at least one hit in the inner TEC, which ensures
  that tracks come from the interaction region (Fig.~\ref{fig:inhits_sel}). 
  It will also help to solve the left-right ambiguity which occurs  
  when detecting charged particle with a wire chamber. 
  As the hits recorded from a charged particle passing near an 
  anode wire do not tell on which side of the wire, the charged 
  particle has been detected, two tracks can be reconstructed from
  a same set of hits.  
  One track corresponding to the real track of the charged particle, 
  the other, the mirror track, symmetric with respect to the wire
  of the track of the charged particle.
  
  Since the agreement between the data and 
  the Monte Carlo simulation is rather poor for the distribution of the 
  number of TEC hits (Fig.~\ref{fig:hits_sel} (a)), 
  the cut is chosen to lie in the middle of a region 
  of the distribution where the variation of the disagreement between data and 
  Monte Carlo is stable and no big change from bin to bin in this 
  disagreement is expected. 
  Therefore, the number of hits in the TEC is required to be at least 25.  
  Loss in track momentum resolution, which could result from the use of 
  such a low minimum requirement on the number of hits, is minimized by 
  the previous requirement of at least one hit in the inner TEC  
  and also by the choice of a rather large span (see below).
\subsubsection{Span of a track}
\begin{figure}[htbp]
  \begin{center}
  \includegraphics[width=8.4cm]{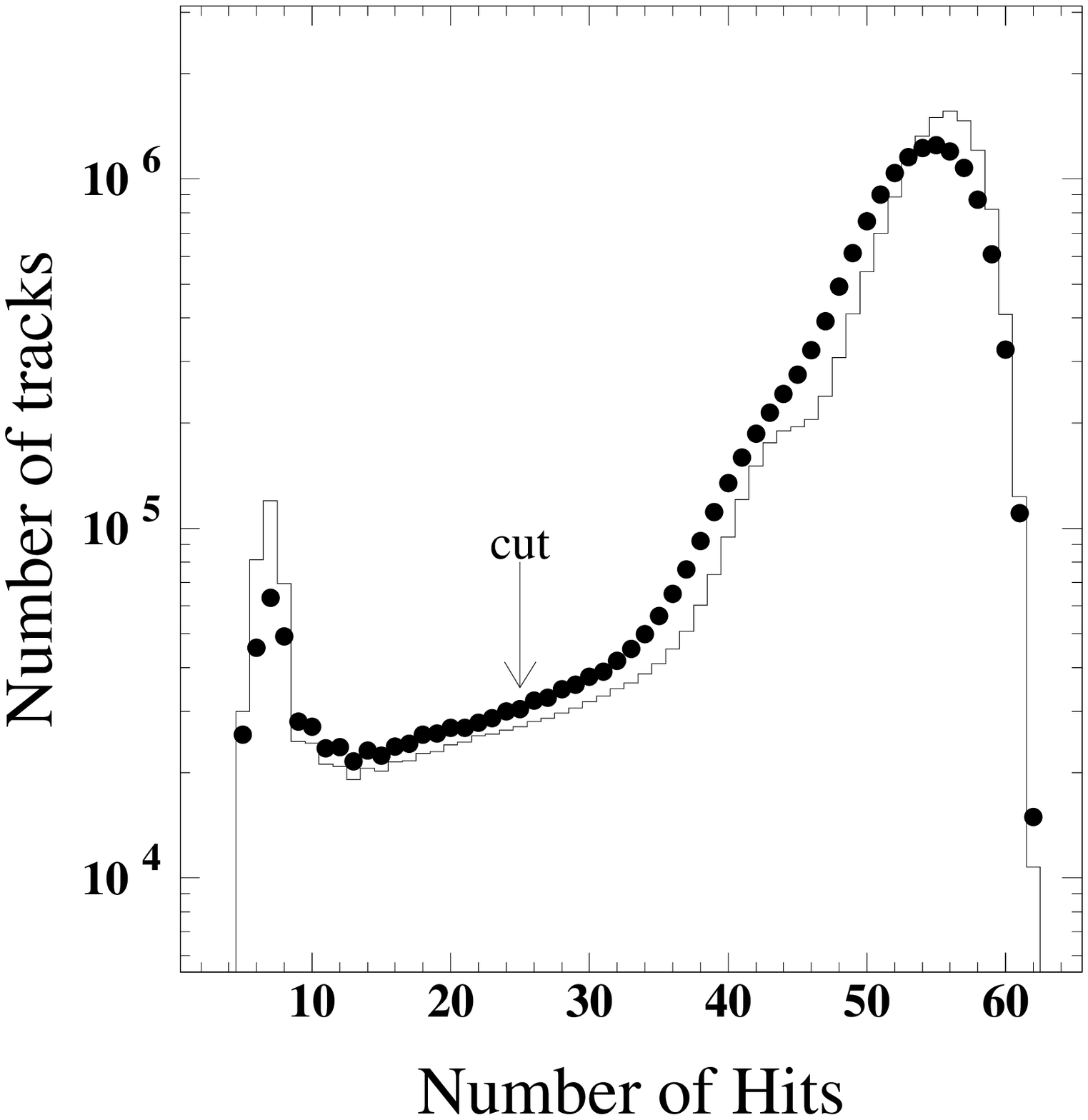}
  \includegraphics[width=8.4cm]{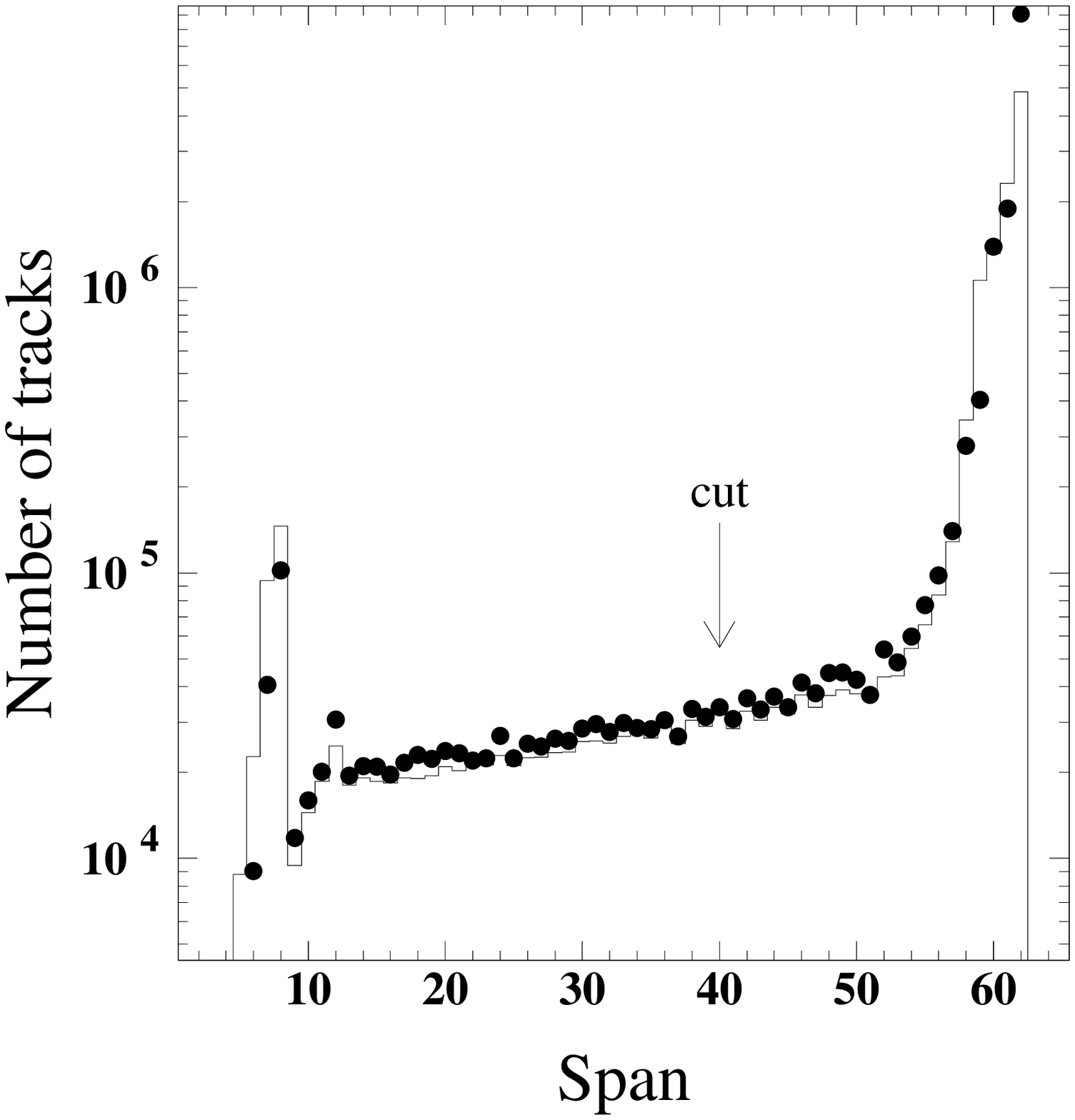}
  \scaption{Distribution of the number of hits in the whole TEC (left plot)
   and of the span of tracks (right plot) for the data (dots) and
  for the Monte-Carlo (line).} 
  \label{fig:hits_sel}  
  \end{center}
\end{figure}

  Tracks are reconstructed by combining hits. Sometimes, hits from different 
  tracks are mistakenly combined. Since these tracks usually have a smaller length than  
  well reconstructed tracks, it is possible to remove most of these tracks by requiring a 
  minimum length for each track. The length of a track is given by the span defined as:
  \begin{center}
  $\mathrm{Span}=\mathrm{W}_\mathrm{first}-\mathrm{W}_\mathrm{last}+1,$
  \end{center}
  where 
  $\mathrm{W}_\mathrm{first}$ and $\mathrm{W}_\mathrm{last}$ are the wire numbers of 
  the innermost hit (\ie{} the wire on which the first hit is left by the particle coming 
  from the interaction point when entering the TEC) and of 
  the outermost hit (\ie{} the wire on which the last hit has been recorded before 
  the particle leaves the TEC) recorded for a track. 
  All tracks are required to have a span of at least 40 
  (Fig.~\ref{fig:hits_sel} (b)).
\subsubsection{Distance of closest approach}
 
\begin{figure}[htbp]
  \begin{center}
  \includegraphics[width=8.4cm]{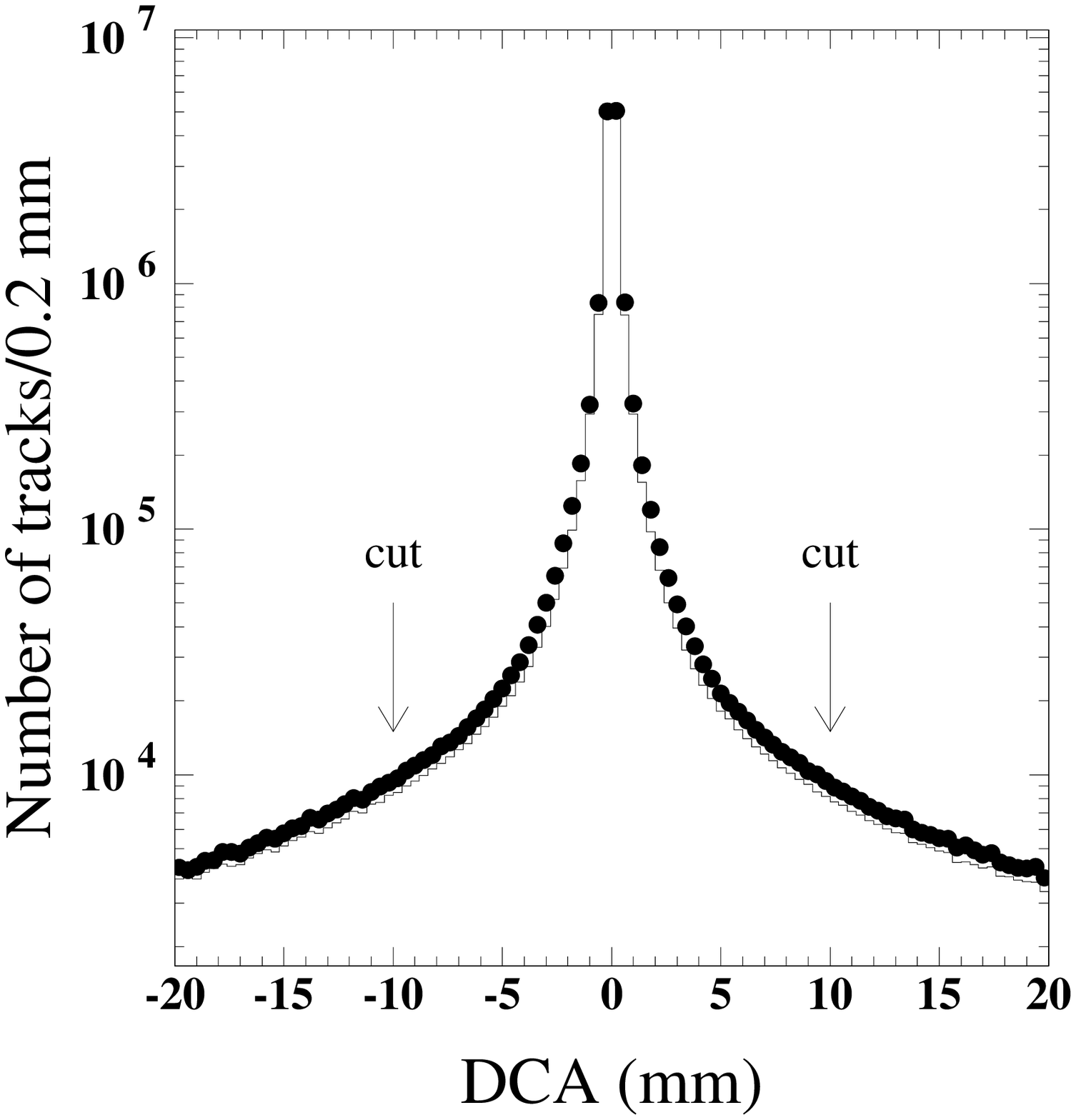}
  \scaption{Distance of closest approach of tracks for the data (dots) and
  for the Monte-Carlo (line).} 
  \label{fig:dca_sel}  
  \end{center}
\end{figure}

 To check if a track originates from the interaction vertex, each track is extrapolated 
 back to the interaction vertex. The distance of closest approach (DCA) to the interaction vertex
 is then calculated in the plane transverse to the beam direction. 
 In order to ensure that a track is coming from the interaction vertex, a DCA smaller than 10mm is
 required (Fig.~\ref{fig:dca_sel}).
\subsubsection{Azimuthal track angle \boldmath{$\phi$}}

\begin{figure}[htbp]
  \begin{center}
  \includegraphics[width=8.4cm]{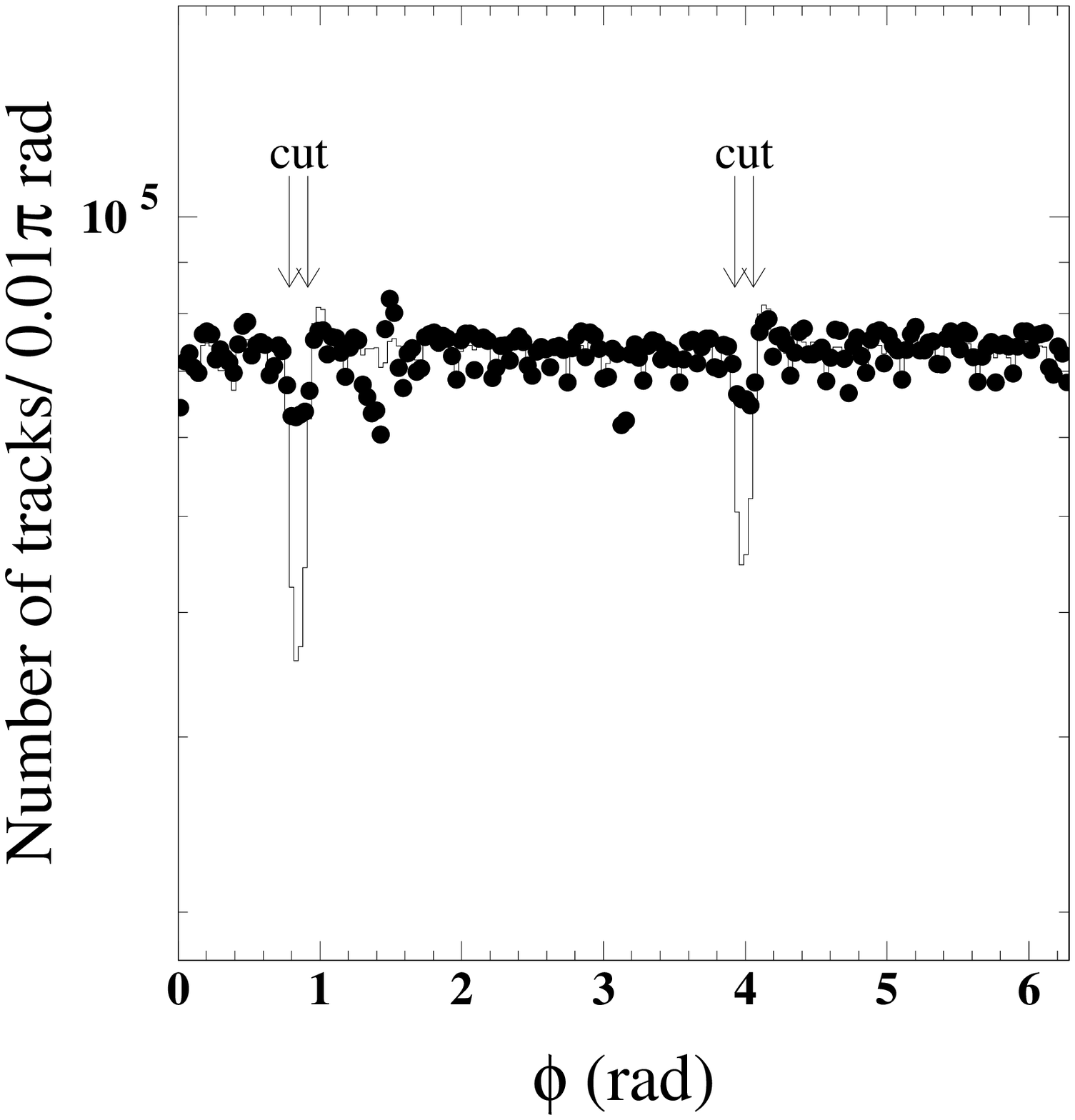}
  \scaption{Distribution of the azimuthal $\phi$ track angle for the data (dots) and
  for the Monte-Carlo (line).} 
  \label{fig:phi_sel}  
  \end{center}
\end{figure}

 Due to a wrong simulation of inefficiencies of two TEC sectors, 
 large discrepancies between data and 
 Monte Carlo are seen in the azimuthal angular distribution for the 
 two half-sectors located at 
 $45^\circ<\phi<52.5^\circ$ and $225^\circ<\phi<232.5^\circ$ 
 (Fig.~\ref{fig:phi_sel}).
 Therefore, tracks located in these two half sectors were 
 simply removed from the analysis. 
\subsection{TEC inefficiencies}\label{sec:tecinef} 
   
   During data taking, from time to time, high background levels, which are likely to 
   generate overcurrents or trips of anodes and cathodes, can cause the TEC to be 
   partly or totally turned off. 
   This leads to a temporary 
   loss of efficiency in certain TEC sectors or in the whole TEC. 

   The Monte Carlo simulation takes into account the major part of such 
   problems occurring during a data taking period 
   (rdvn format), but not if the problem has only occurred during a short 
   period of time.
   
\begin{figure}[htbp]
  \begin{center}
  \includegraphics[width=8.4cm]{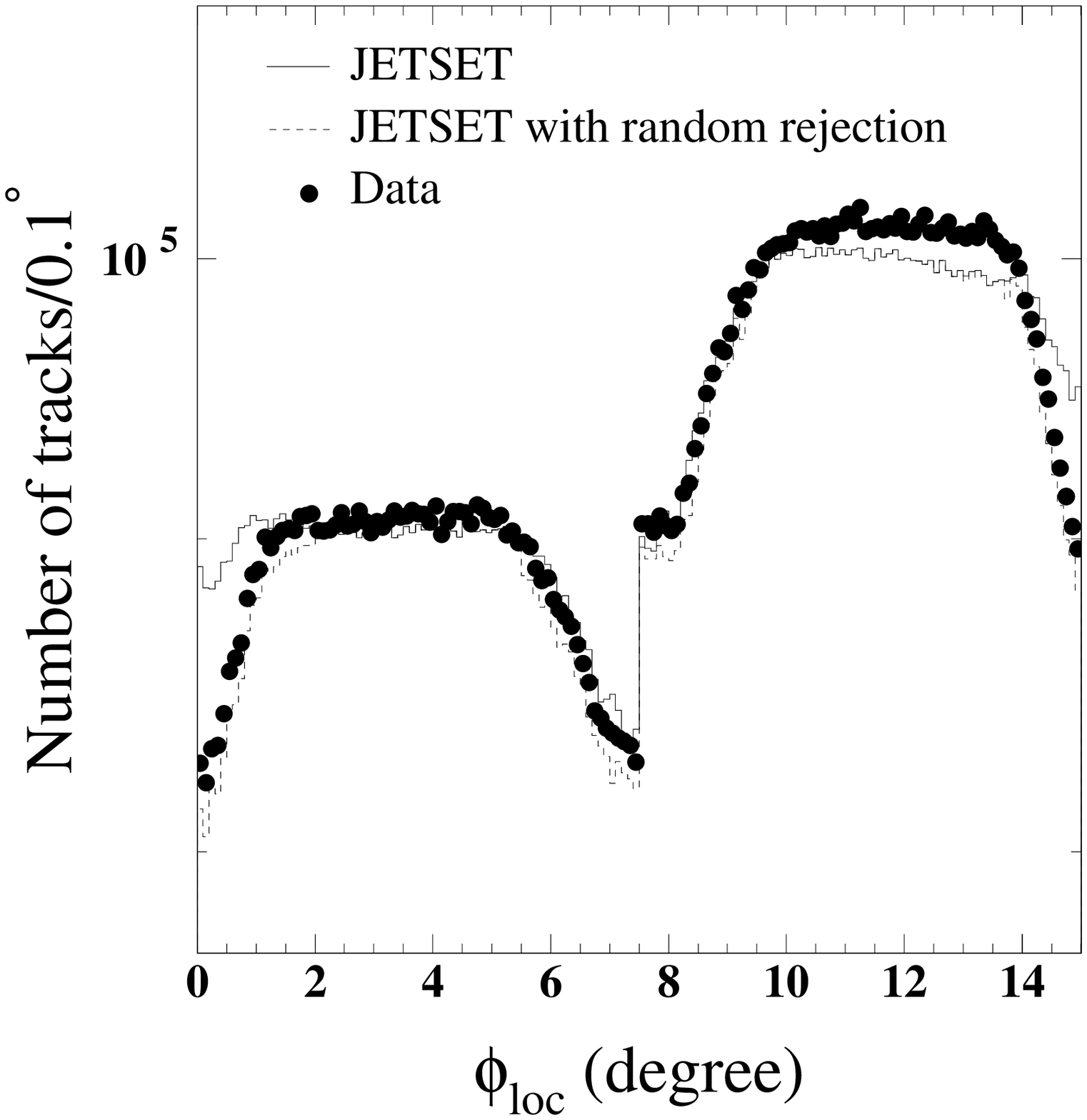}
  \scaption{Distribution of the $\phi$ local angle of tracks for the data (dots) and
  for the initial Monte Carlo (solid line) and for the Monte Carlo with  
  random rejection (dashed line).} 
  \label{fig:philoc_sel}  
  \end{center}
\end{figure}

   Finally, it appears that the Monte Carlo simulation underestimates the track losses close
   to the anodes and cathodes of the TEC. This discrepancy is 
   clearly seen in the distribution of outer $\phi$ local, 
   \ie{}, the distribution of the angle, $\phi_\text{loc}$, between the track 
   and the closest outer TEC anode (Fig.~\ref{fig:philoc_sel}).
  
   In order to improve the TEC inefficiency simulation, a random rejection of 
   Monte Carlo tracks is applied two degrees around anodes and cathodes.
   The random rejection leads to a better matching for the azimuthal 
   angular distribution between data and simulation and, therefore, to an overall 
   better agreement between data and Monte Carlo.
\subsection{Event selection}\label{sec:tecsel}

\begin{figure}[htbp]
\centering
  \includegraphics[width=8.4cm]{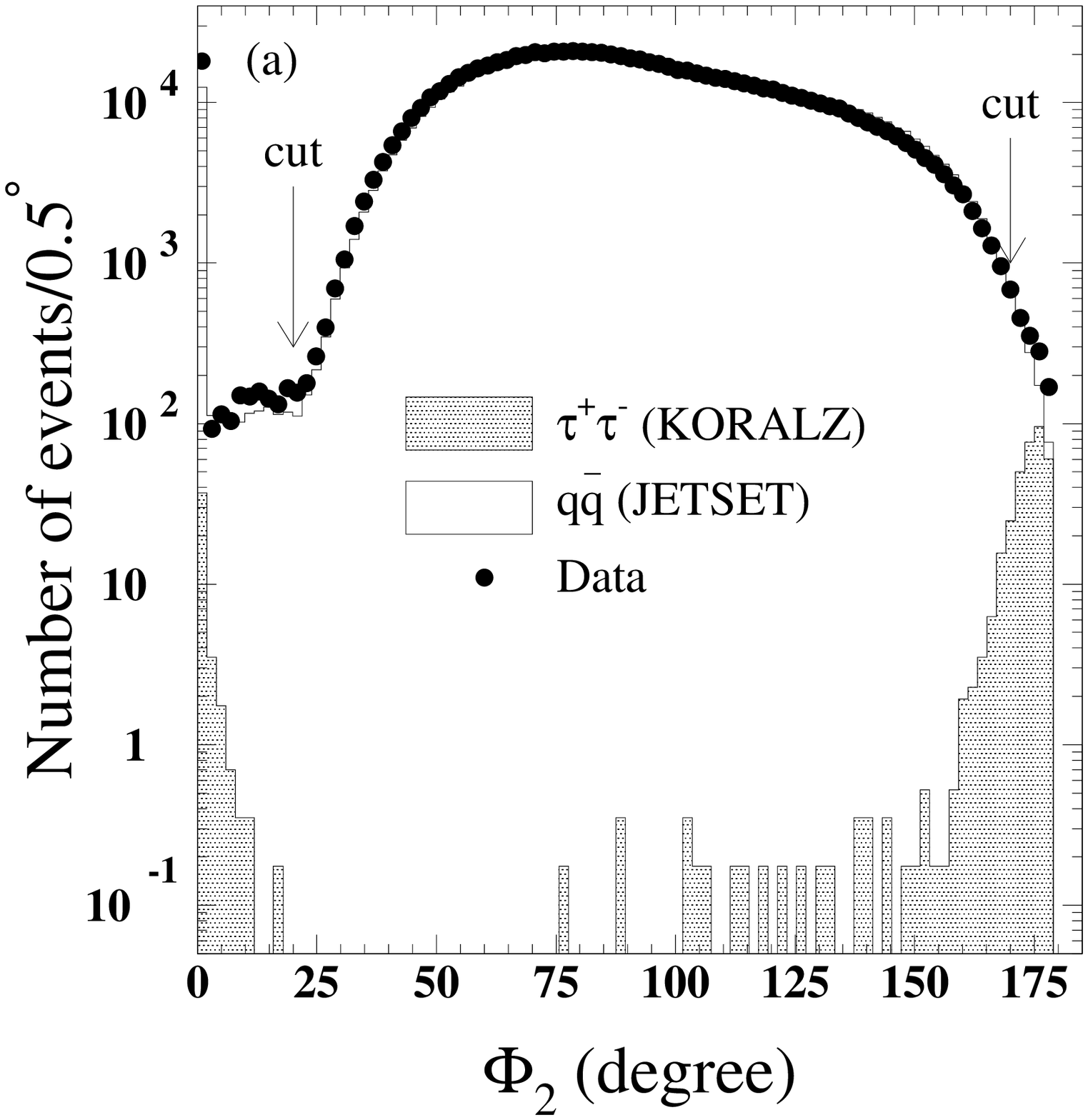}
  \includegraphics[width=8.4cm]{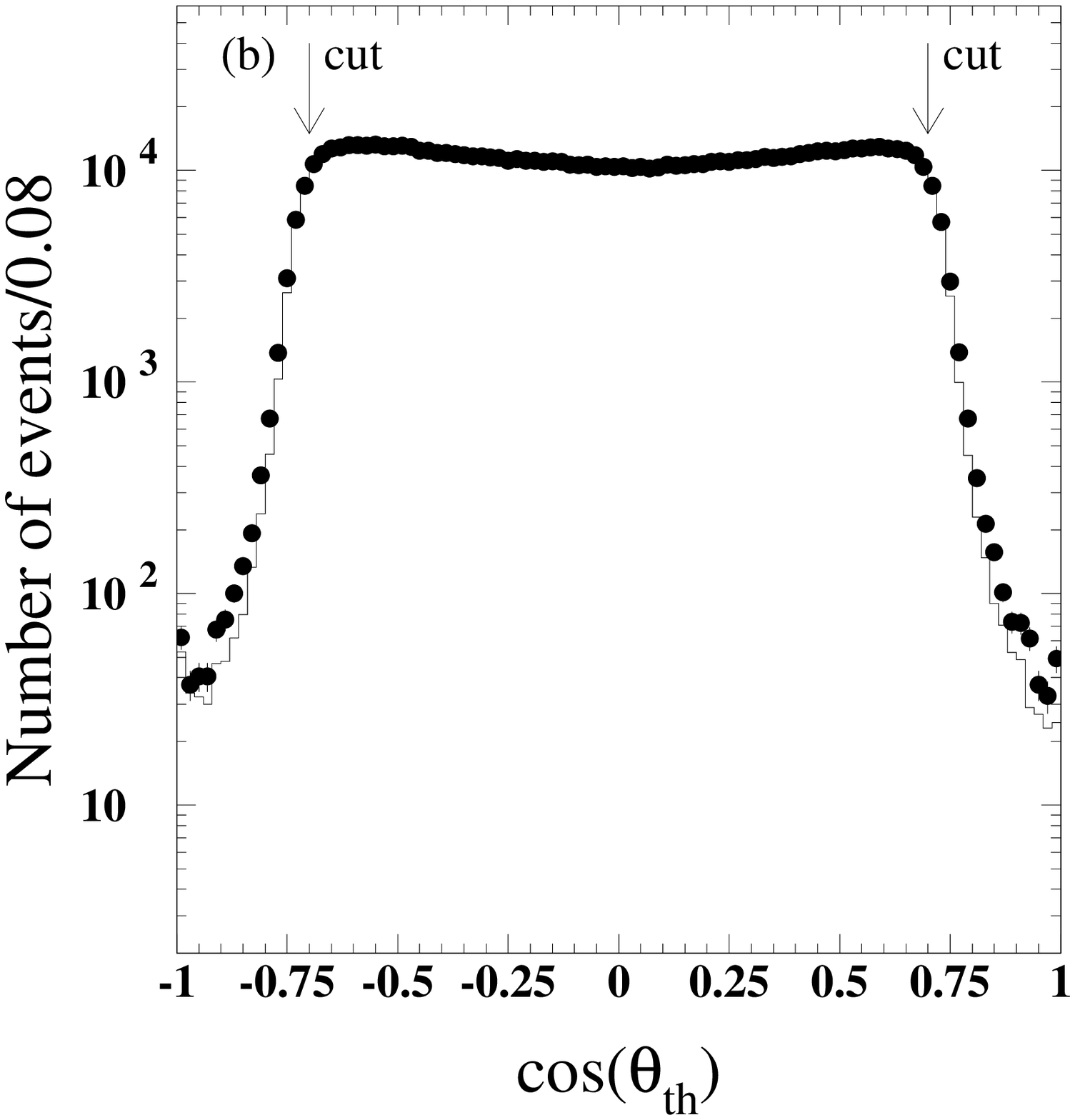}
 \scaption{Distribution of $\phi_2$ (a)
   and of $|\cos(\theta^{\mathrm{TEC}}_{\mathrm{th}})|$ (b) 
  for the data (dots) and for the Monte Carlo (line). The shaded area 
  in (a) indicates \toto{} background generated with 
  the KORALZ Monte Carlo and is normalized  
  to the relative production rate.} 
  \label{fig:angle_sel}  
\end{figure}

   Even though we have already applied a hadronic event selection using 
   calorimeter clusters, 
   a few additional cuts are needed. 
   In order to reject the remaining \toto{} background, we impose a cut 
   on the second largest angle $\phi_2$ between any two neighboring tracks
   in the $r\phi$ plane. Selected events are required to have their 
   $\phi_2$ angle between $20^\circ$ and $170^\circ$ (Fig.~\ref{fig:angle_sel}(a))
   which optimizes the rejection of \toto{} background without rejecting  
   too large a fraction of the hadronic \Z{} events. 

   Furthermore, events are required to have their thrust axis in the barrel 
   of the TEC. For that purpose, $|\cos(\theta^{\mathrm{TEC}}_{\mathrm{th}})|$  
   is required to be less than 0.7,  
   where $\theta^{\mathrm{TEC}}_{\mathrm{th}}$ 
   is the polar angle of the event thrust axis determined from charged tracks 
   (Fig.~\ref{fig:angle_sel}(b)).
   After selection, the purity in hadronic \Z{} 
   events is about $99\%$. What remains in the selected sample 
   is $0.19\%$ of the \ee\toto background and
   $0.07\%$ of both the \toto and the $\ee \text{q}\bar{\text{q}}$ background. 
 
   About 1 million events survive the selection.
\section{Light- and b-quark event selection}\label{sec:btag}

The selection of light-quark events  
(\Zto$\text{}\text{u}\bar{\text{u}}\text{, }
\text{d}\bar{\text{d}}\text{, }\text{s}\bar{\text{s}}
\text{ or }\text{c}\bar{\text{c}}$) 
and of b-quark events ($\Zbb$) 
proceeds in two steps. In the first step, a b-tag algorithm is used to 
define high purity samples of light- and b-quark events.
The second step applies the above-described general hadronic 
event selection procedures in order to obtain  samples from which 
charged-particle multiplicity distributions can be extracted.

The b-tag algorithm which is applied to 
discriminate between light-quark and b-quark events, relies on 
the full three-dimensional information on tracks recorded in the central tracking 
detector (TEC and SMD), in order to compute the probability that a track comes 
from the primary vertex. The method is fully described in~\cite{btag}, but the main
steps of the method are briefly summarized here.

The algorithm starts with the three-dimensional  reconstruction of the primary vertex  
by minimizing

\begin{equation}
\chi^2_{N_\text{trk}}=\overset{N_\text{trk}}{\underset{i=1}{\sum}}
(\vec{t}_i-\vec{f}(\vec{\nu},\vec{q}_i))^T
G^{-1}_i(\vec{t}_i-\vec{f}(\vec{\nu},\vec{q}_i))
+(\vec{\nu}-\vec{\nu}_\text{fill})^T
V^{-1}_\text{fill}(\vec{\nu}-\vec{\nu}_\text{fill}),
\end{equation}
where $N_\text{trk}$ is the number of tracks, $\vec{t}_i$ is the vector 
of measured parameters for the $i^\text{th}$ track, 
$G_i$ is the corresponding covariance matrix, $\vec{f}(\vec{\nu},\vec{q}_i)$
is the corresponding prediction assuming that the track originates at the vertex $\vec{\nu}$
with momentum $\vec{q}_i$, $\vec{\nu}_\text{fill}$ is the so-called fill vertex, \ie{} 
a measure of the position of the beam spot, and $V_\text{fill}$ is its covariance 
matrix. The $N_\text{trk}$ tracks involved in the $\chi^2$ 
have to satisfy the following criteria:

\begin{itemize}

\item being fitted with the Kalman filter~\cite{kalman},

\item $|d_{r\phi}| < \text{min} (10 \text{mm}, 
5\sigma(d_{r\phi}))$, where $d_{r\phi}$ and 
$\sigma(d_{r\phi})$ are the distance of closest approach 
in the $xy$ plane and its error,

\item $|d_{sz}|<100\text{mm}$, $d_{sz}$ being the distance of 
closest approach in the $sz$ plane.

\end{itemize}

If $P(\chi^2_{N_\text{trk}})$ is less than $4\%$ for a particular 
event, tracks are removed one by one and 
the $\chi^2$ is redetermined after having 
removed each track in turn. This results in the probability 
$P_i=P(\chi^2_{N_\text{trk}}-\chi^2_{N_\text{trk}-1,i})$ 
for the case that track $i$ 
is removed. 
Tracks, for which $P_i$ and $P(\chi^2_{N_\text{trk}})$ are both less than 
$2\%$ are definitely removed.
The fit procedure is repeated until no further track needs to be removed.
Primary vertices having less than three remaining tracks are rejected.

\begin{figure}[htbp]
  \begin{center}
    \includegraphics[width=9cm]{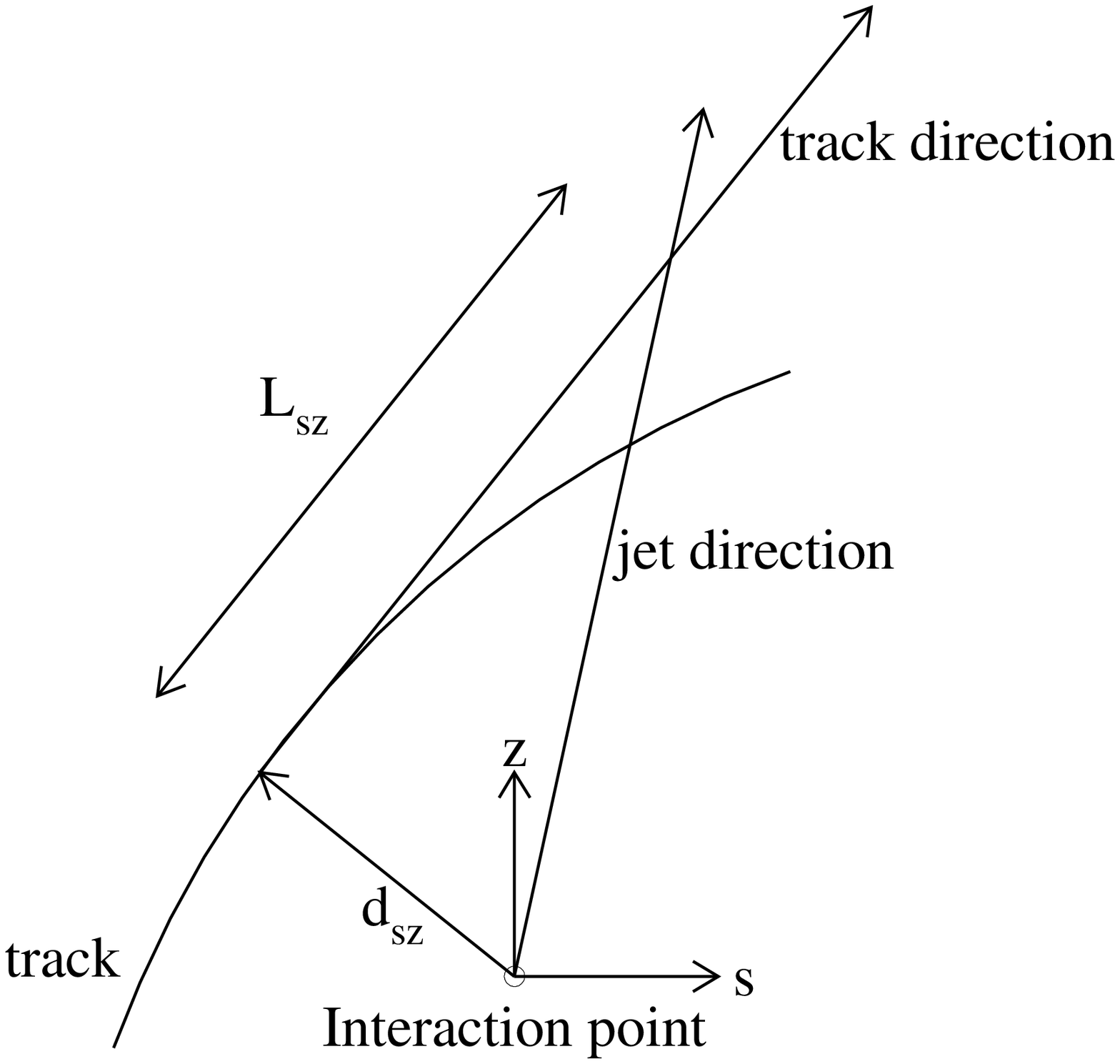}
  \scaption{Distance of closest approach of a track, $d_{sz}$,  
  together with its decay length, $L_{sz}$ in the $sz$ plane.}
  \label{fig:lsz}  
 \end{center}
\end{figure}

Once the primary vertex has been reconstructed, the decay lengths 
$L_{r\phi}$ and $L_{sz}$  measured in the 
$r\phi$ and $sz$ planes, respectively, can be estimated. 
They are defined as the distance  in the  $r\phi$ and 
$sz$ (see Fig.~\ref{fig:lsz}) planes between the impact point of a 
track and the primary vertex and correspond to independent measurements of 
the true decay length of the B hadron. 
They are used to compute an average decay length $L$.

From the significance defined as $S=L/\sigma_L$, 
the probability, $P(S_i)$, that a track with decay length, $L$, 
originates from the primary vertex, is computed.
The track probabilities $P(S_j)$ are combined into an event probability, 
$P_\text{event}$, which carries the sensitivity for an 
event to be a b-quark event, 
\begin{equation}
\label{eq:pevent}
P_\text{event}=\frac{\Pi}{2^N}\overset{N-1}{\underset{i=0}{\sum}}
\overset{N}{\underset{j=i+1}{\sum}}C^N_j\frac{(-\ln \Pi)^i}{i!}
\text{, where }
\Pi=\overset{N_+}{\underset{j=1}{\prod}}P(S_j),
\end{equation}
where $N_+$ is the number of tracks which have a 
positively signed decay length~\cite{btag}.

Due to the long life time and,  
hence, the long decay length of b hadrons, the 
probability $P_\text{event}$ is close to zero for a b-quark event, 
while for other types of events, $P_\text{event}$, is larger.
Therefore, to emphasize the low probability region, 
a discriminant $\delta$ is defined as

\begin{equation}
\label{eq:disc}
\delta=-\log P_\text{event}.
\end{equation}

\begin{figure}[htbp]
  \begin{center}
    \includegraphics[width=16.8cm]{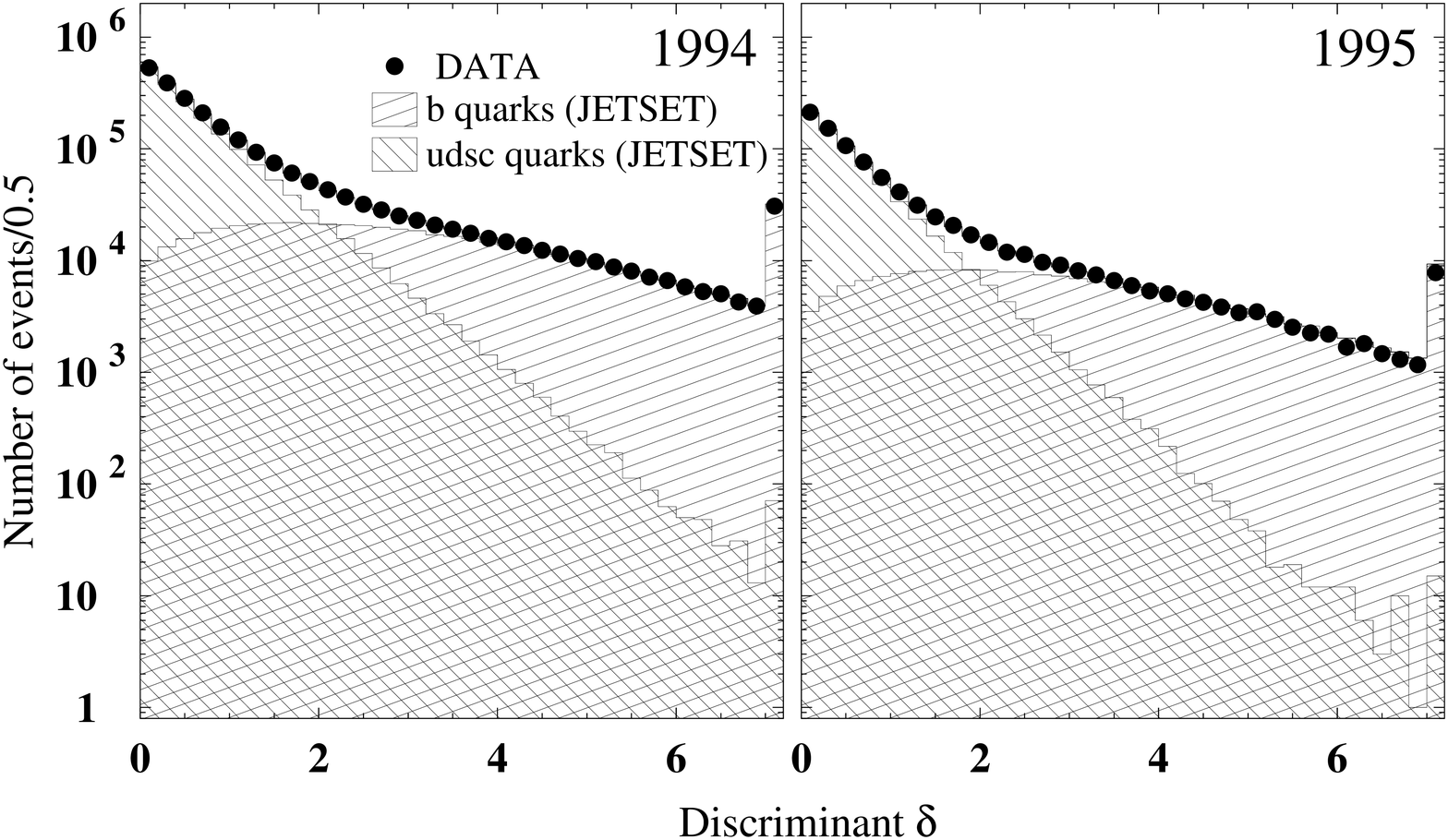}
\vspace{-0.5cm}
  \scaption{Event discriminant distribution for the 1994 data (left) and 
           for the 1995 data (right) compared
           with the discriminant distributions of the Monte Carlo for 
           all events, b-quark events and light-quark events.}
  \label{fig:disc94}  
\vspace{+1.cm}
    \includegraphics[width=16.8cm]{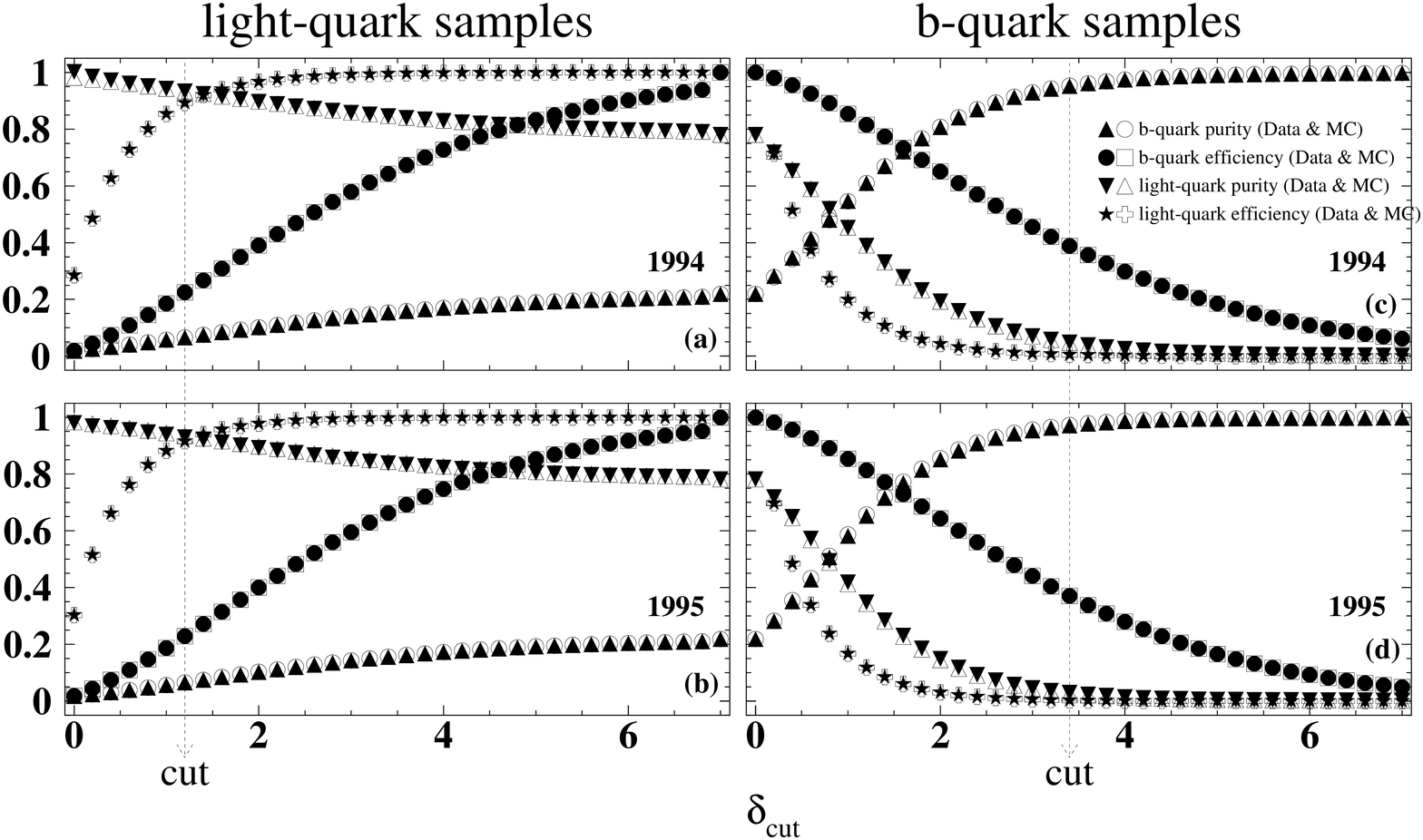}

\scaption{The efficiency and purity in the light-quark sample 
for the year 1994 (a) and 1995 (b) obtained by varying the cut on the 
discriminant value aimed to select light quarks ($\delta<\delta_\text{cut}$), 
and the efficiency and purity in the 
b-quark sample (right) obtained by varying the cut on the discriminant value aimed 
to select b quarks ($\delta>\delta_\text{cut}$) for the year 1994 (c) and 
for 1995 (d).}
  \label{fig:pur94}
\vspace{-0.5cm}  
\end{center}
\end{figure}

The distribution of this discriminant is shown in 
Fig.~\ref{fig:disc94} for the 1994 (left) and the 1995 (right) 
data taking periods, for both the data and JETSET 
for all events, together with the separate JETSET discriminant distributions 
for light-quark events and for b-quark events.
The selection or the rejection of b-quark events is based on the 
discriminant $\delta$.  
Light-quark events are selected by $\delta<\delta_\text{cut}$,
and b-quark events by $\delta>\delta_\text{cut}$.
The purity and efficiency of the light-quark sample
are shown in Fig.~\ref{fig:pur94}  as a function of 
$\delta_\text{cut}$ for the 1994 data sample (top left) 
and for the 1995 data sample (top right).
The purity and efficiency of the b-quark sample
are shown in Fig.~\ref{fig:pur94} as a function of 
$\delta_\text{cut}$ for the 1994 data sample (bottom left) 
and for the 1995 data sample (bottom right).

To minimize the size of the corrections which will have to be applied 
to the tagged samples in order to get pure light- and b-quark
samples, high purities in light and b quarks are required.
Therefore, the tagged light-quark event sample is selected by 
requiring a discriminant value $\delta<1.2$. For this cut, the purity 
and efficiency of the sample in light quarks for 1994 are $93.4\%$ and $89.2\%$, 
respectively, for 1995, $93.0\%$ and $91.6\%$.
The tagged b-quark event sample is selected by requiring a discriminant value
$\delta>3.4$, which leads to a purity and efficiency in 
b-quarks for 1994 of $95.2\%$ and $38.9\%$, respectively, and for 1995
of $97.0\%$ and $37.1\%$. 
These purities and efficiencies are not altered by the event selection.

Once the light- and b-quark samples are selected, we apply to 
them the same selection criteria which are applied to the full sample, 
as described previously in Sects.~\ref{sec:selcal} and~\ref{sec:trksel}.

\chapter{The charged-particle multiplicity distribution}\label{chap:cpmd}

Besides its theoretical interest discussed in Sect.~\ref{sec:tcpmd}, the 
interest in the \cpmd{} arises from the fact that 
the detection of charged particles 
is far more easy than the detection of neutral particles. To measure the full 
multiplicity distribution (charged and neutral particles), we would have 
to rely on the detection of energy deposits in the calorimeter, 
the calorimeter clusters. Since these clusters can represent 
from a fraction of the energy up to the whole energy of a particle, 
the correspondence between clusters and particles is rather 
difficult to establish.
Therefore, the extrapolation from energy clusters 
to particles would depend a lot on the simulation. As we have seen in the 
previous chapter, due to an underestimation of the noise in the calorimeter,
the agreement in terms of clusters between data and simulation is not 
good enough to perform such a measurement.

A charged particle is detected as a track in the
Central Tracking Chamber. Unlike a calorimeter cluster, a track and its 
kinematical content represents a particle and not a fragment of a particle 
(the track quality selections eliminate most of the badly measured or split tracks).
This increases both the reliability and the traceability of the final result 
since we can keep track of the charged particles from their detection 
to their reconstruction.
However, an accurate treatment is still needed to reconstruct the \cpmd{}. 

In the first section of this chapter, we discuss the steps needed to reconstruct
the \cpmd{} for the full, light- and b-quark samples, 
starting from the measured raw-data \cpmd{}. 
The next two sections introduce the calculation used to estimate their 
statistical errors and their systematic uncertainties.
The resulting \cpmd{s} and their principal moments 
are presented and discussed in the final section of this chapter.
 
\section{Reconstruction of the multiplicity distribution}\label{sec:rec}

\begin{figure}[htbp]
  \begin{center}
    \includegraphics[width=10cm]{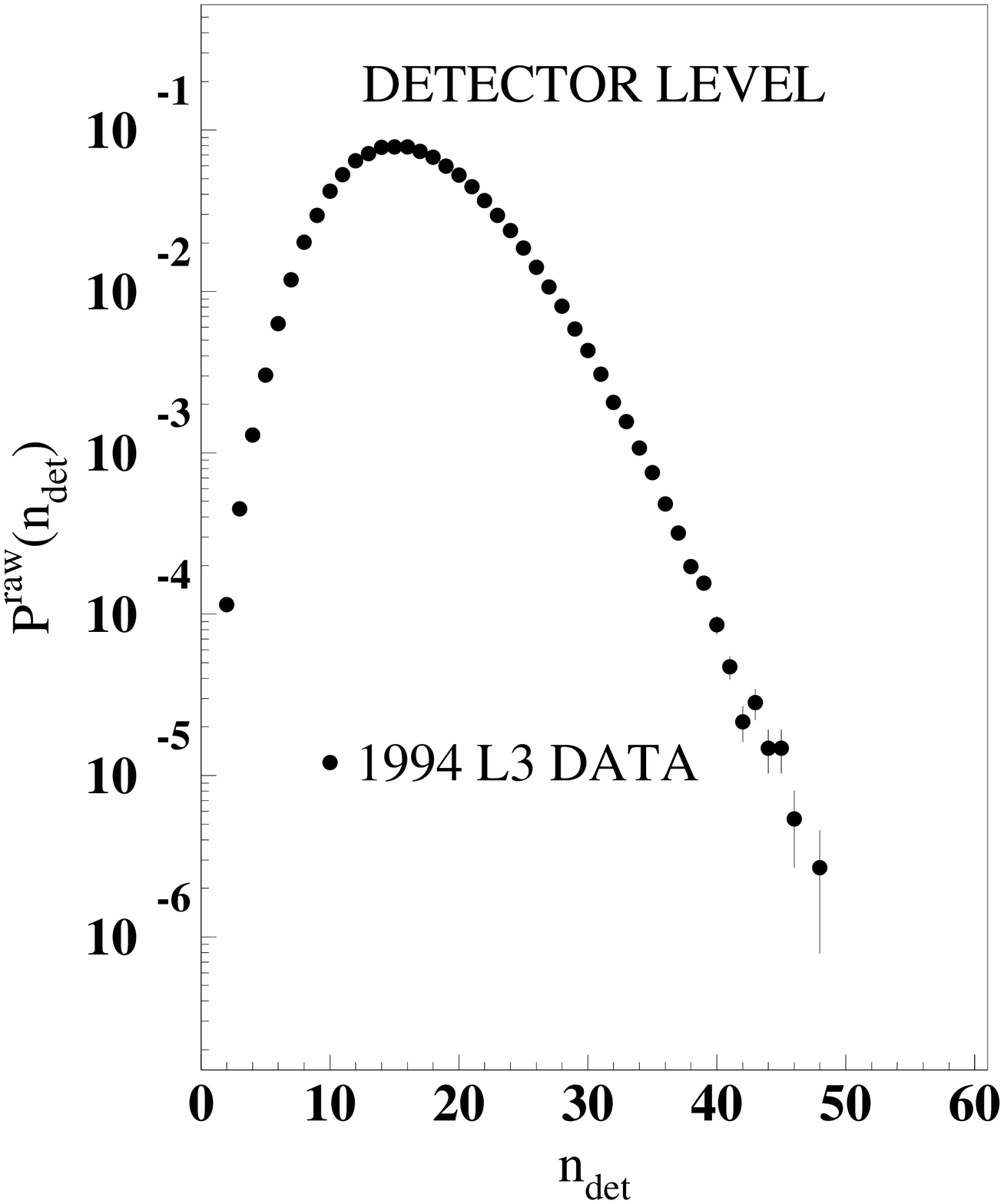}
  \end{center}
\scaption{The detected charged-particle multiplicity distribution of the 1994 data.}
  \label{fig:rawpn}  
\end{figure}

Because of the limited acceptance of the detector and of the selection procedures
which were used to obtain a pure sample of hadronic decays, not only do 
events escape both the detection and selection processes, but also do 
the detected events usually contain fewer particles 
than were produced. 
The most dramatic example of this effect is given in Fig.~\ref{fig:rawpn} by the 
charged-particle multiplicity distribution itself. 
If all charged particles in an event were detected, charge
conservation implies that their number would always be even. However, 
as shown here, we find both even and odd multiplicities.
Therefore, the treatment needed to reconstruct the charged-particle multiplicity has to 
take into account not only the undetected events, but also the undetected particles 
within an event. 
We therefore proceed in two steps. The first step uses an unfolding method which  
corrects the number of particles in an event. The second step 
corrects for event selection, including light- 
or b-quark selection and initial-state radiation.
An additional correction is applied to take into 
account the charged $\text{K}^0_\text{s}\text{ and }\Lambda$ decay products.

As a convention, and unless otherwise stated, we refer 
by $N(n)$ to the distribution of the number of events with a 
particle multiplicity $n$, and by $P(n)$ to the distribution of our 
estimate of the probability to obtain an event with a particle multiplicity of $n$,
\begin{equation}
P(n)=\frac{N(n)}{\underset{i}{\sum}N(i)}.
\end{equation}
The same convention will also apply to matrices.

\subsubsection{Correction for inefficiencies and limited acceptance of the detector}

\begin{figure}[htbp]
  \begin{center}
    \includegraphics[width=10cm]{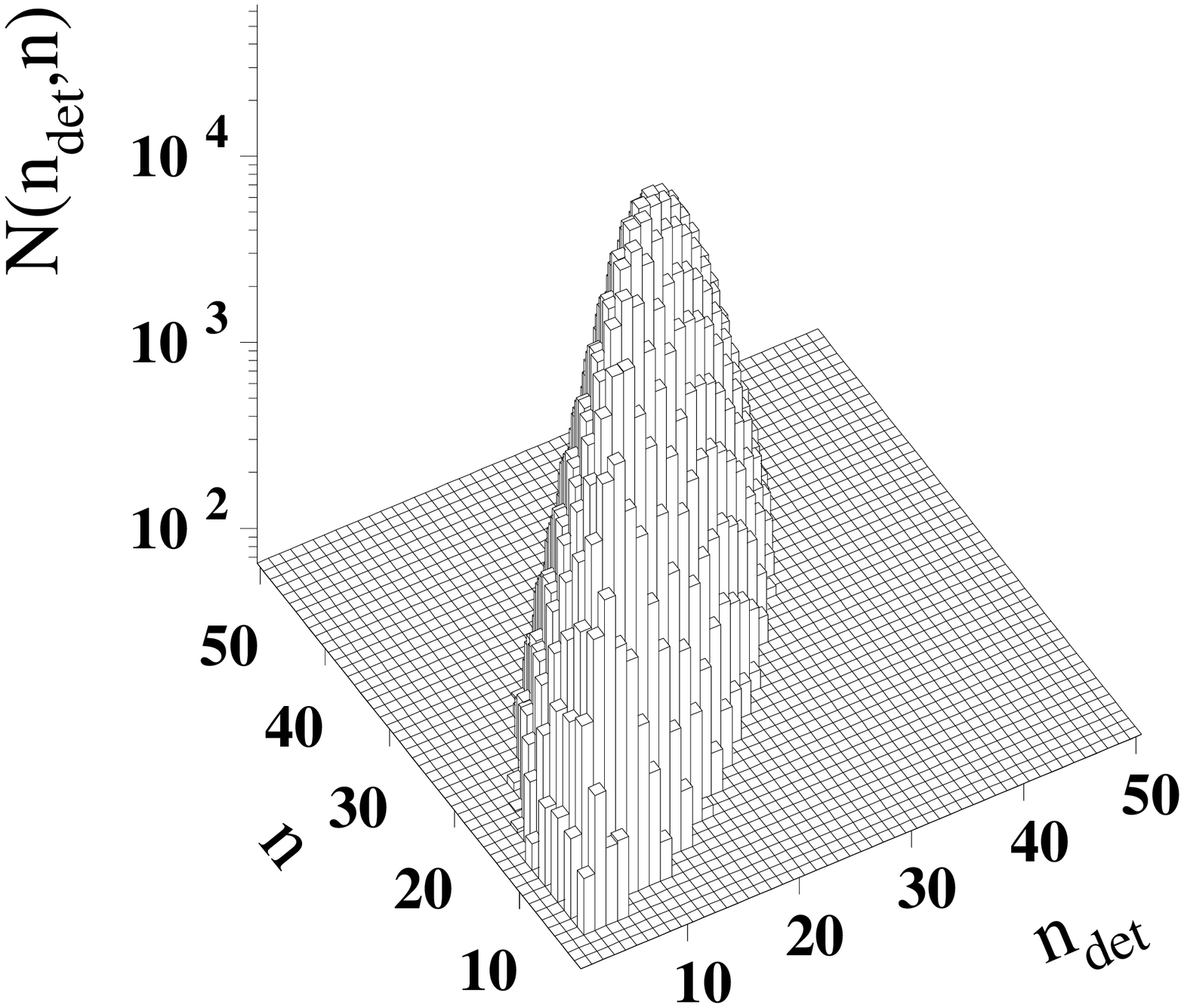}
  \end{center}
\scaption{The response matrix of the detector obtained with JETSET Monte Carlo.}
  \label{fig:mat}  
\end{figure}

The most common way to correct for detector inefficiencies consists of
multiplying bin-by-bin the raw data distribution
of a given variable, $V^\text{data}_\text{raw}$, 
by a correction factor which is the ratio of the distribution of the variable,  
$V^\text{MC}_\text{prod}$, generated by Monte Carlo
in the case of a detector working at $100\%$ efficiency 
and the distribution of the variable, 
$V^\text{MC}_\text{det}$, generated by Monte Carlo, passed 
through the simulation of the detector, reconstructed and selected 
as the data, 
\begin{equation}
\label{eq:var}
V^\text{data}_\text{cor}=V^\text{data}_\text{raw}
\frac{V^\text{MC}_\text{prod}}{V^\text{MC}_\text{det}}.
\end{equation}
However, this type of correction can only be used when the simulated and generated 
distributions are consistent with each other. (\eg{} have the same number of bins). 
Because the method intrinsically assumes the independence 
of each bin, it is better suited to correct for a global effect 
such as, \eg{}, a loss of events due to the selection procedure. 
Although this method will be used for that purpose  
later, it is inappropriate here because of bin-to-bin migration.  
This is not necessarily localized to adjacent bins and, therefore, cannot
be treated by simply changing the width of the bins in such way that 
the bin-to-bin migration would be taken into account.
Due to the imperfection of the detection process (not only the detector itself,
but also the track reconstruction and track quality cuts), 
the detected multiplicity is very often different from the original multiplicity. 
One can have detected multiplicities  
smaller or larger than the ones produced (as shown in Fig.~\ref{fig:mat}). 
Furthermore, this effect of the bin-to-bin migration depends  
only in a non-trivial way to the number of particle produced 
and cannot be corrected without a full simulation of the detector. 
Therefore, another method is needed to take into account, 
as properly as possible, the bin migrations between produced and detected 
multiplicities. This is done by a so-called unfolding method.

The unfolding method makes use of the detector response matrix, $\mathcal{N}$, 
which takes the bin migrations into account.
For each Monte Carlo event, this matrix $\mathcal{N}$ keeps track of the 
number of produced charged particles, $n$, 
and its associated number of detected tracks,  $n_{\text{det}}$, which have been 
processed and selected in the same way as the data tracks.
Each matrix element $\mathcal{N}(n_{\mathrm{det}},n)$ of 
$\mathcal{N}$ corresponds to the number of Monte Carlo events 
which have $n$ produced charged particles and 
$n_{\text{det}}$ detected tracks. The matrix found using events generated 
with JETSET is shown in Fig.~\ref{fig:mat}. 
The probability of detecting $n_\text{det}$ tracks, $P(n_\text{det})$, 
is related to the probability distribution of produced charged particles, 
$P(n)$ by
\begin{equation}
\label{eq:mat}
P(n_\text{det})={\underset{n}{\sum}}
                          \frac{\mathcal{N}(n_{\mathrm{det}},n)}
                           {{\underset{n_{\mathrm{det}}}{\sum}}
                           \mathcal{N}(n_{\mathrm{det}},n)}
                           P(n),
\end{equation}
Defining the migration matrix $\mathcal{M}$ by 
\begin{equation}
\label{eq:matm}
\mathcal{M}(n_\text{det},n)=
\frac{\mathcal{N}(n_{\mathrm{det}},n)}
                           {{\underset{n_{\mathrm{det}}}{\sum}}
                    \mathcal{N}(n_{\mathrm{det}},n)}.
\end{equation}
Eq.~(\ref{eq:mat}) can be rewritten as 
\begin{equation}
\label{eq:vecmat}
P_\text{det}=\mathcal{M}P_\text{prod},
\end{equation}
where $P_\text{det}$ and $P_\text{prod}$ are vectors whose elements are 
$P(n_\text{det})$ and $P(n)$, respectively. 

To estimate the produced multiplicity distribution of the data, 
$P^\text{data}_\text{prod}$, we can invert Eq.~(\ref{eq:vecmat}),
\begin{equation}
\label{eq:ivecmat}
P^\text{data}_\text{prod}=\mathcal{I}P^\text{raw}_\text{det},
\end{equation}
where $\mathcal{I}$ is obtained in a same manner as $\mathcal{M}$
\begin{equation}
\label{eq:imatm}
\mathcal{I}(n,n_\text{det})=
\frac{\mathcal{N}^\text{T}(n,n_{\mathrm{det}})}
{{\underset{n}{\sum}}
\mathcal{N}^\text{T}(n,n_{\mathrm{det}})}.
\end{equation}
Note that $\mathcal{I}\neq\mathcal{M}^\text{T}$ because of different normalizations.
The normalization of Eq.~(\ref{eq:matm}) makes $\mathcal{M}$ independent of the 
multiplicity distribution of the event generator. But this is not the case 
for Eq.~(\ref{eq:imatm}). Consequently, the use of the matrix $\mathcal{I}$ in 
Eq.~(\ref{eq:ivecmat}) could bias the result towards the distribution of the 
event generator. 

Therefore, we use a more elaborate method, known as Bayesian unfolding, 
in which Eq.~(\ref{eq:vecmat}) is used iteratively~\cite{agostini},\cite{susinno}. 
The probabilities of producing $n$ particles and of detecting 
$n_\text{det}$ tracks are related by Bayes' theorem:
\begin{equation}
\label{eq:bayes}
\mathcal{M}(n_\text{det},n)P(n)=
\mathcal{I}(n,n_\text{det})P(n_\text{det}).
\end{equation}
Hence
\begin{equation}
\label{eq:ptrans}
\mathcal{I}(n,n_\text{det})=
\mathcal{M}(n_\text{det},n)
\frac{P(n)}{P(n_\text{det})}.
\end{equation}
Taking $P(n_\text{det})$ from Equation (\ref{eq:mat}), this becomes 
\begin{equation}
\label{eq:invmat}
\mathcal{I}(n,n_\text{det})=
\frac{\mathcal{M}(n_\text{det},n)P(n)}
{{\underset{n}{\sum}}
\mathcal{M}(n_{\mathrm{det}},n)P(n)
}.
\end{equation}
Inserting this in Eq.~(\ref{eq:ivecmat}) gives an estimate of the 
produced multiplicity distribution:
\begin{equation}
\label{eq:basis}
P^\text{data}(n)={\underset{n_{\mathrm{det}}}{\sum}}
\frac{\mathcal{M}(n_\text{det},n)P(n)}
{{\underset{n}{\sum}}
\mathcal{M}(n_{\mathrm{det}},n)P(n)
}P^\text{raw}(n_\text{det}).
\end{equation}
This equation is the basis for the Bayesian unfolding. 
But instead of using directly the result of this equation,
and in order to remove the bias due to the use of a limited statistics
sample for the construction of the matrix $\mathcal{M}$, 
we will use Eq.~(\ref{eq:basis}) iteratively.

We start the iterative unfolding procedure by comparing the detected 
charged-particle multiplicity distribution of the data, 
$P^\text{raw}(n_\text{det})$ to 
$\sum \mathcal{M}(n_{\mathrm{det}},n)P^{(0)}(n)$, 
which, according to Eq.~(\ref{eq:mat}), corresponds to the 
detected distribution $P^{(0)}(n_\text{det})$. In principle, 
the initially produced
distribution, $P^{(0)}(n)$, can be anything, but 
here we use the charged-particle
multiplicity distribution produced from JETSET Monte Carlo, 
since we found that its fully simulated multiplicity distribution 
agrees rather well with the raw data. 
The $\chi^2/\text{dof}$ between the detected distributions, 
$P^\text{raw}(n_\text{det})$ and $P^{(0)}(n_\text{det})$  
is calculated, and if it is larger than 1, the ratio 
$C^{(1)}(n_\mathrm{det})$ is calculated,
\begin{equation}
\label{eq:first}
C^{(1)}(n_\mathrm{det})=\frac{P^{\mathrm{raw}}(n_\mathrm{det})}{
                          {\underset{n}{\sum}}
                           \mathcal{M}(n_{\mathrm{det}},n)
                          P^{(0)}(n)}.
\end{equation}
Using $C^{(1)}$ in Eq.~(\ref{eq:basis}), we can write
\begin{equation}
\label{eq:sec}
C^{(1)}(n)=
   {\underset{n_\mathrm{det}}{\sum}}
\mathcal{M}(n_{\mathrm{det}},n)
   C^{(1)}(n_\mathrm{det}),
\end{equation} 
with
\begin{equation}
\label{eq:cprod}
C^{(1)}(n)=\frac{P^{(1)}(n)}{P^{(0)}(n)},
\end{equation}
where $P^{(1)}(n)$ is the first-iteration estimate of the produced 
charged-particle multiplicity distribution of the data.

Using now $P^{(1)}(n)$ in Eq.~(\ref{eq:mat}), we can 
compare the detected charged-particle multiplicity distribution of the data 
$P^\text{raw}(n_\text{det})$  
to the estimate of the detected charged-particle multiplicity 
distribution $P^1(n_\mathrm{det})$ at detected level, 
\begin{equation}
\label{eq:mat1}
P^\text{(1)}(n_\text{det})={\underset{n}{\sum}}
                   \mathcal{M}(n_{\mathrm{det}},n)
                          P^{\text{(1)}}(n).
\end{equation}
Depending on the value of the $\chi^2$ between $P^\text{raw}(n_\text{det})$ and
$P^\text{(1)}(n_\text{det})$, we proceed to the next iteration by repeating 
with $P^{(1)}(n)$ instead of $P^{(0)}(n)$
the whole procedure described in Eqs.~(\ref{eq:first}) and~(\ref{eq:cprod}),   
leading finally to the estimate of the charged-particle multiplicity 
distribution of the data for the second iteration, 
\begin{equation}
\label{eq:pn2}
P^{(2)}(n)= 
P^\mathrm{(1)}(n)  
 C^{(2)}(n).
\end{equation}
By generalizing this result to the $q^\text{th}$ iteration we have:
\begin{equation}
\label{eq:pnn}
P^{(q)}(n)= 
P^{(q-1)}(n)  
C^{(q)}(n)=
P^{(0)}(n)\overset{q}{\underset{i=1}{\prod}}
C^{(i)}(n),
\end{equation}

\begin{equation}
\label{eq:cni}
      \text{where }
      C^{(i)}(n)=
      {\underset{n_\text{det}}{\sum}}\mathcal{M}(n_\text{det},n)
      \frac{P^\text{raw}(n_\text{det})}{
      {\underset{n}{\sum}}
       \mathcal{M}(n_\text{det},n)
            P^{(i-1)}(n)}.
\end{equation}

The iterative process is stopped when the $\chi^2/\text{dof}$ has become 
sufficiently small, \ie{,} smaller than 1. This occurs after the second
iteration. Therefore, $P^{(2)}(n)$ is taken as our estimate 
of the reconstructed \cpmd{} of the data, 
\begin{equation}
\label{eq:rec}
P^\text{rec}(n)=P^{(2)}(n).
\end{equation}

The unfolding method corrects only for detector inefficiencies, changing the 
multiplicity within a given event. In particular, 
it does not correct for events which were rejected by the event selection procedure.
To obtain a fully corrected charged-particle 
multiplicity distribution, we, therefore, need additional factors.

\subsubsection{Correction for event selection}

\begin{figure}[htbp]
  \begin{center}
    \includegraphics[width=8.4cm]{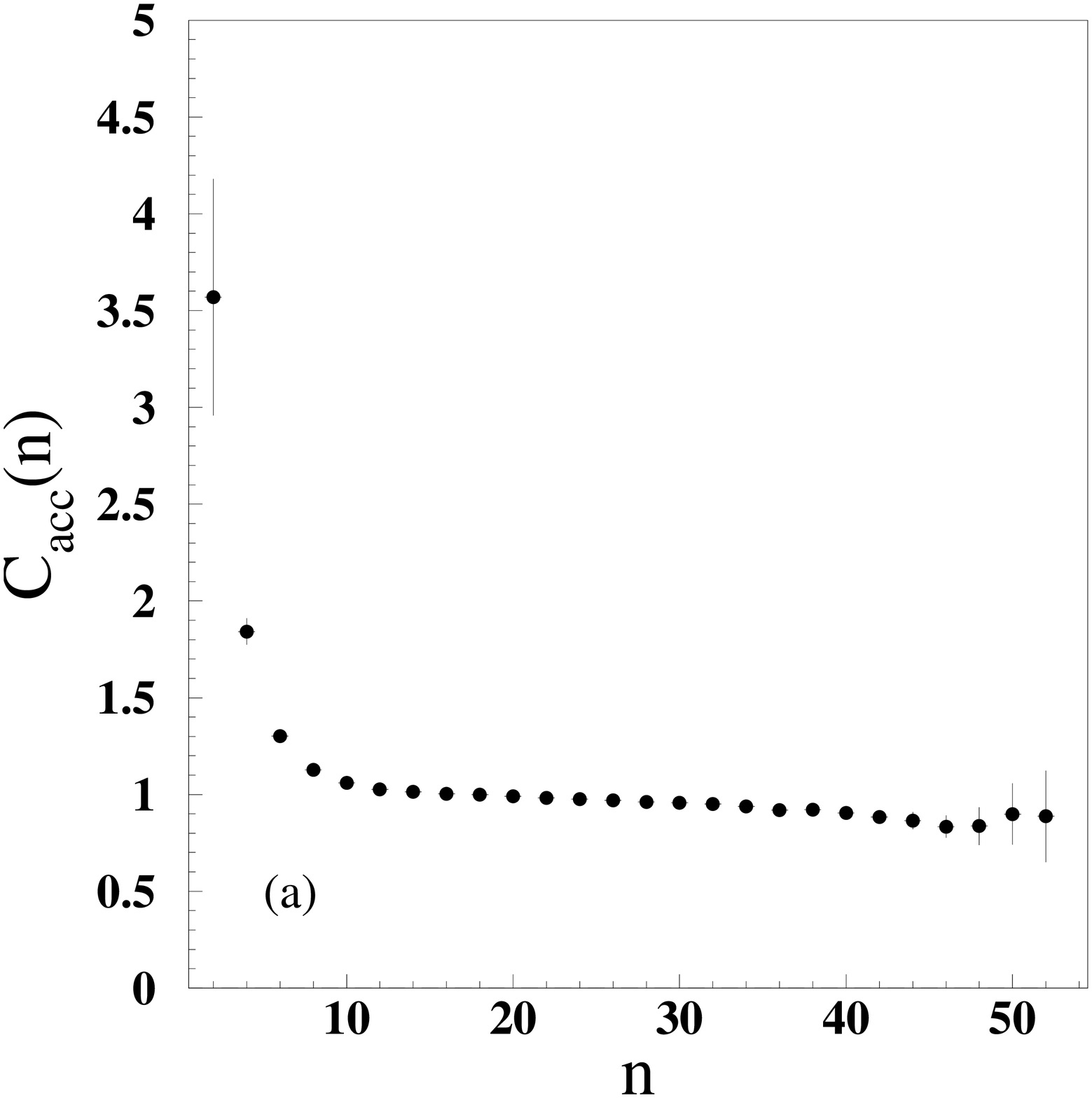}
    \includegraphics[width=8.4cm]{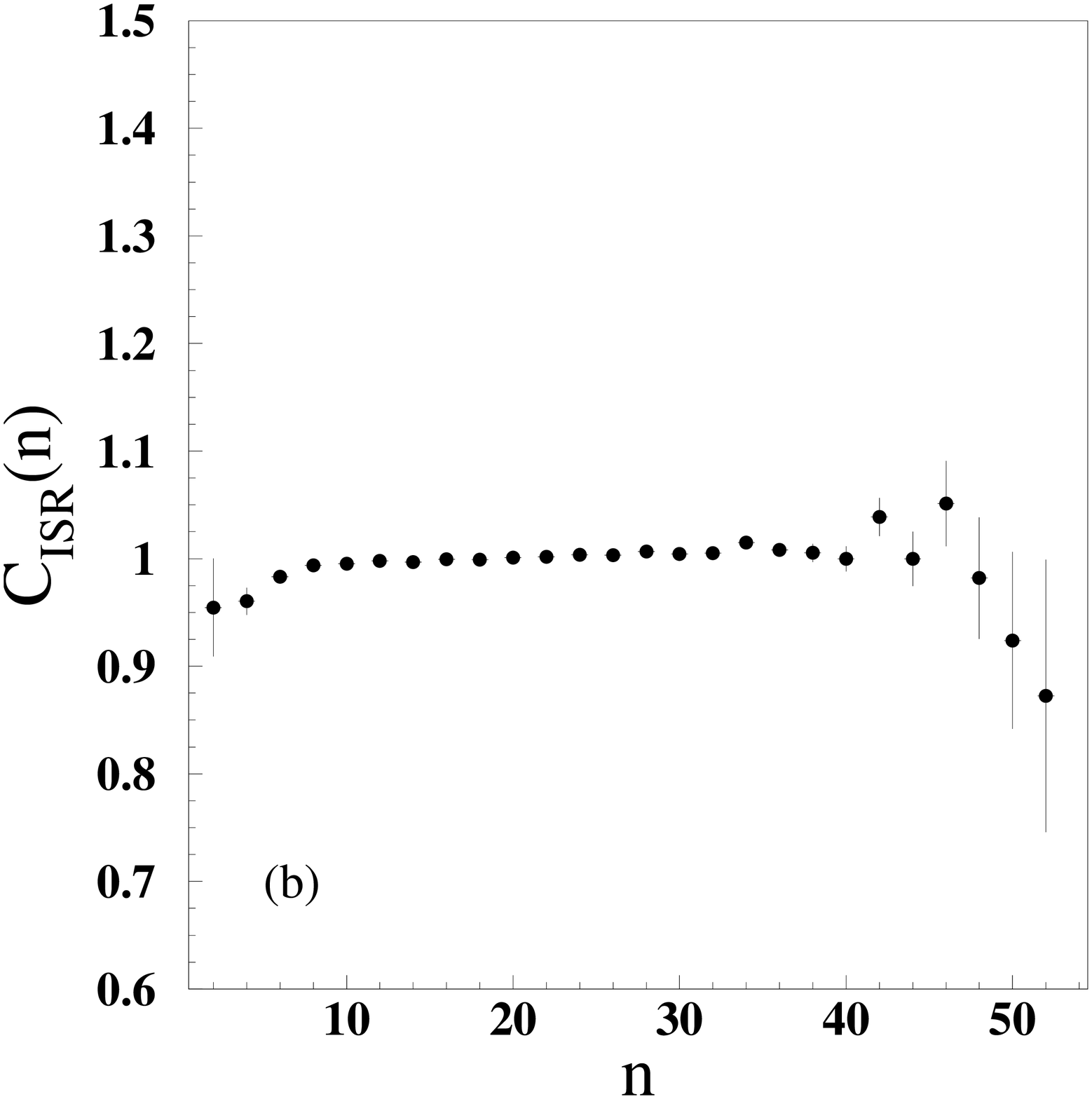}
  \end{center}
\scaption{ratios $C_\text{acc}(n)$, (a), 
and $C_\text{ISR}(n)$, (b), 
to correct for event selection and for ISR, respectively.}
  \label{fig:cor}  
\end{figure}

We first correct for events which were removed by the event selection. 
This is done by applying correction factors (Fig.~\ref{fig:cor} (a)),
 which are obtained by taking the ratio of the \cpmd{} 
of all Monte Carlo events at generator level,  
$P^\text{all}_\text{prod}(n)$, to that of the generator level 
which, when fully simulated, pass the selection procedure,  
$P^\text{acc}_\text{prod}(n)$,
\begin{equation}
\label{eq:acc}
C_\text{acc}(n)=\frac{P^\text{all}_\text{prod}(n)}
{P^\text{acc}_\text{prod}(n)}.
\end{equation}
\subsubsection{Correction for Initial-State Radiation}

Since we are interested in a pure sample of hadronic events at
the \Z{} energy, we also need to correct for Initial-State Radiation (ISR). 
This ISR refers to the emission of one or more photons by the 
electron or the positron, which shifts the energy of the event 
to a lower value. The ISR events have 
a signature similar to that of normal hadronic events, 
except that they have lower center of mass energies and, hence, 
lower charged-particle multiplicities. At the \Z{} energy, these 
events represent only a small fraction of the 
hadronic events and are well simulated by Monte Carlo. Therefore, their 
contribution may be easily corrected by a simple correction factor,
 (Fig.~\ref{fig:cor} (b)),  
corresponding to the ratio of the charged-particle multiplicity distributions 
of Monte Carlo
events generated with, $P^\text{ISR}(n)$, and without, 
$P^\text{noISR}(n)$, initial state radiation,
\begin{equation}
\label{eq:isr}
C_\text{ISR}(n)=\frac{P^\text{noISR}(n)}
                                  {P^\text{ISR}(n)}.
\end{equation}
These two corrections are applied to the reconstructed charged-particle 
multiplicity distribution of the data, $P(n)$, to obtain 
finally a fully corrected charged-particle multiplicity distribution,
\begin{equation}
\label{eq:cor}
P(n)=C_\text{acc}(n)
C_\text{ISR}(n)P^\text{rec}(n).
\end{equation}

\subsubsection{Correction for $\mathrm{K}^0_\mathrm{s}\text{ and }\Lambda$ decay products}

In the context of the strong interaction, particles which decay weakly or
electromagnetically are considered stable. However, experimentally 
it is more appropriate to consider particles which are detected.
This leads to a problem in the case of $\mathrm{K}^0_\mathrm{s}$ and 
$\Lambda$, both of which can decay weakly to two charged particles  
which can be detected in L3. Therefore, the detected charged-particle 
multiplicity distribution contains charged particles produced in these decays.
However, to obtain a charged-particle multiplicity distribution relevant to QCD, 
$\mathrm{K}^0_\mathrm{s}$ and $\Lambda$ should be considered stable. The 
necessary correction introduces the uncertainties on 
$\mathrm{K}^0_\mathrm{s}$ and $\Lambda$ production into the charged-particle 
multiplicity distribution. Particularly in the early days of LEP, these 
uncertainties were large, which led experiments to refrain from applying these 
corrections. At present, these uncertainties are much smaller.
In the unfolding procedure, we have, therefore, corrected the data to obtain 
the multiplicity distribution that does not include $\mathrm{K}^0_\mathrm{s}$ 
and $\Lambda$ decay products.
However, for comparison with other experiments, we also present the multiplicity 
distribution containing these decay products.

Inclusion of the $\mathrm{K}^0_\mathrm{s}$ and $\Lambda$ decay products corresponds 
to an increase of the number of charged particles in an event.
The problem we are facing is then similar to the track migration problem 
caused by detector acceptance and inefficiencies. Therefore, 
$\mathrm{K}^0_\mathrm{s}\text{ and }\Lambda$ decay processes 
are treated in a similar way, \ie{} by the use of a probability matrix.

Using the JETSET 7.4 PS generator, we build the matrix 
$k(n_{\text{K}^0},n_{\text{no}\text{K}^0})$ which represents the number of 
events which have $n_{\mathrm{K}^0}$ charged particles in the case that 
$\mathrm{K}^0_\mathrm{s}\text{ and }\Lambda$ decay products are 
included and $n_{\mathrm{no}\mathrm{K}^0}$ charged particles 
in the case that they are not.  
%
The probability matrix is then obtained by normalizing 
$k(n_{\mathrm{K}^0},n_{\text{no }\mathrm{K}^0})$ by the
distribution of the number of events having 
$n_{\text{no}\mathrm{K}^0}$ charged particles, assuming 
stable $\mathrm{K}^0_\mathrm{s}\text{ and }\Lambda$,
$N^\text{MC}(n_{\text{no}\text{K}^0})$,
\begin{equation}
\label{eq:normk}
K(n_{\mathrm{K}^0},n_{\text{no}\text{K}^0})=
\frac{k(n_{\text{K}^0},n_{\text{no}\text{K}^0})}
{N^\text{MC}(n_{\text{no}\text{K}^0})}.
\end{equation}
Since the process involved is far more simple than the 
detector response, and due to the fact that we can use as much statistics 
as we want, we do not need to use the iterative procedure described previously. 
The simple use of the probability matrix as described in Eq.~(\ref{eq:mat})
is sufficient to get a reliable result.
The charged-particle multiplicity distribution including 
$\mathrm{K}^0_\mathrm{s}\text{ and }\Lambda$ decay products, 
$P^{\mathrm{data}}(n_{\mathrm{K}^0})$, 
is then given by 
\begin{equation}
\label{eq:k0}
P^{\text{data}}(n_{\text{K}^0})=
   {\underset{n_{\text{no}\text{K}^0}}{\sum}} 
   K(n_{\text{K}^0},n_{\text{noK}^0})
   P^{\mathrm{data}}(n_{\mathrm{no }\mathrm{K}^0}).
\end{equation}

\subsubsection{Correction for light- or b-quark purities}

We need additional corrections specifically for the 
charged-particle multiplicity distributions of the light- and b-quark samples. 
These multiplicity distributions are corrected in the same way as the 
full sample described previously, but with an additional flavor dependent
 correction factor $C^\text{fl}_\text{purity}(n)$ to correct 
the light- or b-quark samples for b- or light-quark contaminations, 
respectively.  
The latter are obtained by taking the ratio of the charged-particle 
multiplicity distribution produced from Monte Carlo 
events for a given flavor, fl, $P^\text{fl}_\text{prod}$ and of the 
charged-particle multiplicity distribution of generated Monte Carlo events 
which have passed the flavor tagging, $P^\text{fl-tagged}_\text{prod}$,
\begin{equation}
\label{eq:cfl}
   C^{\mathrm{fl}}_{\mathrm{purity}}(n)=
\frac{P^{\mathrm{fl}}(n)}
     {P^{\mathrm{fl-tagged}}(n)}.
\end{equation}
The charged-particle multiplicity distributions of the 
light- and  b-quark samples,  
$P^{\mathrm{udsc}}_{\mathrm{data}}(n)$ and
$P^{\mathrm{b}}_{\mathrm{data}}(n)$, respectively, 
are given by 
\begin{equation}
\label{eq:fl}
            P^{\mathrm{fl}}_{\mathrm{data}}(n)=
          C^{\mathrm{fl}}_{\mathrm{purity}}(n)
 P^\mathrm{fl-tagged}_{\mathrm{data}}(n).
\end{equation}

\subsubsection{Combining 1994 and 1995 data samples}

There are many ways to combine data from different years, especially when 
they represent the same process at the same energy.
In our analysis, we choose to combine the charged-particle 
multiplicity distributions obtained from the 1994 and 1995 data samples, 
after these have been fully corrected, allowing to take into 
account the specificities of the corrections between the 2 years,
\begin{equation}
\label{eq:9495pn}
P^\text{94+95}_\text{data}(n)=
\frac{N^\text{94}_\text{data}(n)+N^\text{95}_\text{data}(n)}
{{\underset{n}{\sum}}(N^\text{94}_\text{data}(n)+
N^\text{95}_\text{data}(n))}, 
\end{equation}
where $N^\text{94}_\text{data}(n)$ and $N^\text{94}_\text{data}(n)$ are 
the fully corrected distributions of the number of events with 
a multiplicity $n$ for 1994 and 1995, respectively. 
However, it is not mandatory, here, to combine the 
multiplicity distribution after 
they have been corrected since we used exactly the same 
selection for the 1994 and 1995 data samples and that the 
corrections are also similar, but it 
allows us to perform consistency checks of the two samples. 
We find a $\chi^2$ of $15.4$ for 27 degrees of freedom 
between the \cpmd{} of the 1994 and 1995 data samples.

\section{Statistical errors}

In order to calculate the statistical errors on the charged-particle multiplicity
distributions and to estimate the errors on their moments, we need to calculate  
the covariance matrix of the charged-particle multiplicity distribution.
The covariance matrix takes into account the correlations
which exist between the multiplicities as well as the correlations which are 
introduced by the corrections applied on the data,
each term described in the previous section giving a contribution. 

For a given total number of events,  
the number of events 
with $n$ charged particles, $N(n)$, is 
distributed according to a multinomial distribution. 
Therefore, the errors are described in terms 
of a covariance matrix of the form:
\begin{equation}
\label{eq:covn}
\text{CoV}(N(n),N(m))=\frac{1}{\sum N(i)}
\begin{cases}
\text{ }N(n)[\sum N(n)-N(n)] &n=m\\
       -N(n)N(m)            &n\neq m.
\end{cases}
\end{equation}
In the limit of infinite statistics the correlations vanish and 
the diagonal terms becomes $[N(n)]^2$ as for the independent Poisson case.
The covariance matrix of the normalized charged-particle multiplicity
distribution $P(n)$ differs from $\text{CoV}(N(n),N(m))$ 
by a normalization factor:

\begin{equation}
\label{eq:covpn}
\text{CoV}(P(n),P(m))=\frac{1}{[\sum N(i)]^2}\text{CoV}(N(n),N(m))=
\frac{1}{\sum N(n)}
\begin{cases}
        P(n)[1-P(n)]        &n=m\\
       -P(n)P(m)            &n\neq m.
\end{cases}
\end{equation}
All corrections which are described in the previous section gives rise to 
their own contribution to the covariance matrix of the corrected data.
They are briefly described in the following.

\subsubsection{Contribution from the unfolding method}

The covariance matrix of $P^{(q)}(n)$, 
calculated from Eq.~(\ref{eq:pnn}), is given by
\begin{equation}
\label{eq:nscov}
\text{CoV}(P^{(q)}(i),P^{(q)}(j))=P^{(q)}(i) P^{(q)}(j)[
                                 \frac{\text{CoV}(P^{(0)}(i),P{(0)}(j))}
                                 {P^{(0)}(i) P^{(0)}(j)}
                                 +\overset{q}{\underset{m=1}{\sum}}
                        \frac{\text{CoV}(C^{(m)}(i),C^{(m)}(j))}
                             {C^{(m)}(i)C^{(m)}(j)}].
\end{equation}
Two simplifications are used in the calculation~\cite{agostini}. 
Firstly, $P^{(0)}(n)$ is assumed to be without statistical errors,
since it affects the result in a systematic way. (This effect will be examined 
later as a systematic contribution). Secondly, 
in the covariance matrix, we only take into account 
the correction factor $C^{(q)}(n)$ used in the last 
iteration only, neglecting the contribution of the 
$q-1$ other  $C^{(i)}(n)$. This is equivalent to replacing 
$P^{(0)}(n)$ by $P^{(q-1)}(n)$ 
as the starting multiplicity distribution.
The covariance matrix of $P^{(q)}(n)$ is then simplified to
\begin{equation}
\label{eq:scov}
\text{CoV}(P^{(q)}(i),P^{(q)}(j))=
P^{(q)}(i)
P^{(q)}(j)
\frac{\text{CoV}(C^{(q)}(i),C^{(q)}(j))}
{C^{(q)}(i) C^{(q)}(j)}.
\end{equation}
From the definition of $C^{(q)}_\text{prod}(n)$, given in 
Eq.~(\ref{eq:cni}), it is straightforward to obtain 
its covariance matrix:
\begin{equation}
\label{eq:covcn}
\begin{split}
\text{CoV}(C^{(q)}(i),C^{(q)}(j))
&={\underset{k}{\sum}}{\underset{l}{\sum}}
\mathcal{I}(k,i)\mathcal{I}(l,j)
\frac{\text{CoV}(P^\text{raw}(k),P^\text{raw}(l))}
{P^{(q-1)}_\text{sim}(k)P^{(q-1)}_\text{sim}(l)}\\
&+{\underset{k}{\sum}}{\underset{l}{\sum}}\mathcal{I}(k,i)\mathcal{I}(l,j)
\frac{P^\text{raw}(k)P^\text{raw}(l)}
{(P^{(q-1)}_\text{sim}(k)P^{(q-1)}_\text{sim}(l))^2}
\text{CoV}(P^{(q-1)}_\text{sim}(k),P^{(q-1)}_\text{sim}(l))\\
&+{\underset{k}{\sum}}{\underset{l}{\sum}}
\frac{\text{CoV}(\mathcal{N}(k,i),\mathcal{N}(l,j))}
{N_\text{gen}(k)N_\text{gen}(l)}
\frac{P^\text{raw}(k)P^\text{raw}(l)}
{P^{(q-1)}_\text{sim}(k)P^{(q-1)}_\text{sim}(l)}\\
&+{\underset{k}{\sum}}{\underset{l}{\sum}}\mathcal{M}(k,i)\mathcal{M}(l,j)
\frac{P^\text{raw}(k)P^\text{raw}(l)}
{P^{(q-1)}_\text{sim}(k)P^{(q-1)}_\text{sim}(l)}
\frac{\text{CoV}(N_\text{gen}(k),N_\text{gen}(l))}
{N_\text{gen}(k)N_\text{gen}(l)},
\end{split}
\end{equation}
\begin{equation}
\label{eq:ndet}
\text{where }N_\text{gen}(n)=
{\underset{n_\text{det}}{\sum}}\mathcal{N}(n_\text{det},n)\text{, }
\end{equation}
\begin{equation}
\label{eq:psim}
P^{(q-1)}_\text{sim}(n_\text{det})=
{\underset{n}{\sum}}\mathcal{M}(n_\text{det},n)
P^{(q-1)}(n),
\end{equation}
\begin{equation}
\label{eq:covsim}
\begin{split}
\text{and }
\text{CoV}(P^{(q-1)}_\text{sim}(k),P^{(q-1)}_\text{sim}(l))
&={\underset{o}{\sum}}{\underset{p}{\sum}}\mathcal{M}(o,k)
\mathcal{M}(p,l)
P^{(q-1)}_\text{prod}(o)P^{(q-1)}_\text{prod}(p)
\frac{\text{CoV}(N_\text{gen}(o),N_\text{gen}(p))}
{N_\text{gen}(o)N_\text{gen}(p)}\\
&+{\underset{o}{\sum}}{\underset{p}{\sum}}
\frac{\text{CoV}(\mathcal{N}(o,k),\mathcal{N}(p,l))}
{N_\text{gen}(o)N_\text{gen}(p)}
P^{(q-1)}(o)P^{(q-1)}(p).
\end{split}
\end{equation}
$N^\text{raw}(n_\text{det})$ and
$N_\text{gen}(n)$ being also distributed according a multinomial 
distribution, their covariance matrices are obtained by Eqs.~(\ref{eq:covn}) 
and~(\ref{eq:covpn}).
For $\mathcal{N}(n_\text{det},n)$,  
we exclude the possibility of correlation between generated multiplicities 
assuming that each Monte Carlo event is generated independently from 
each other. Hence, 
\begin{equation}
\label{eq:covmat}
\text{CoV}(\mathcal{N}(k,i),\mathcal{N}(l,j))=
\frac{1}{{\underset{i}{\sum}}\mathcal{N}(k,i)}
\begin{cases}
\mathcal{N}(k,i)[{\underset{i}{\sum}}\mathcal{N}(k,i)-\mathcal{N}(k,i)]  
&i=j\text{ and }k=l\\
-\mathcal{N}(k,i)\mathcal{N}(k,j)                              
&i\neq j\text{ and }k=l\\
0                                          & k\neq l.
\end{cases}
\end{equation}

\subsubsection{Contribution from event selection and ISR corrections} 

The covariance matrix of the event selection correction factors 
$C_\text{acc}(n)$ is given by
\begin{equation}
\label{eq:covacc}
\frac{\text{CoV}(C_\text{acc}(i),C_\text{acc}(j))}{C_\text{acc}(i)C_\text{acc}(j)}=
\frac{\text{CoV}(P^\text{all}_\text{prod}(i),P^\text{all}_\text{prod}(j))}
{P^\text{all}_\text{prod}(i)P^\text{all}_\text{prod}(j)}+
\frac{\text{CoV}(P^\text{acc}_\text{prod}(i),P^\text{acc}_\text{prod}(j))}
{P^\text{acc}_\text{prod}(i)P^\text{acc}_\text{prod}(j)}.
\end{equation}
We assume a multinomial distribution for the charged-particle
multiplicity distributions (see Eq.~(\ref{eq:covpn})) involved in both 
correction factors.

In the same way, the covariance matrix of the ISR correction factors 
$C_\text{ISR}(n)$ is given by 
\begin{equation}
\label{eq:covisr}
\frac{\text{CoV}(C_\text{ISR}(i),C_\text{ISR}(j))}{C_\text{ISR}(i)C_\text{ISR}(j)}=
\frac{\text{CoV}(P^\text{noISR}_\text{prod}(i),P^\text{noISR}_\text{prod}(j))}
{P^\text{noISR}_\text{prod}(i)P^\text{noISR}_\text{prod}(j)}+
\frac{\text{CoV}(P^\text{ISR}_\text{prod}(i),P^\text{ISR}_\text{prod}(j))}
{P^\text{ISR}_\text{prod}(i)P^\text{ISR}_\text{prod}(j)}.
\end{equation}
The covariance matrix of $P^\text{rec}(n_\text{rec})$ resulting from the 
unfolding procedure being given by Eq.~(\ref{eq:scov}),
the covariance matrix of the corrected charged-particle 
multiplicity distribution $P^\text{cor}(n)$  
given in Eq.~(\ref{eq:cor}) is simply
\begin{equation}
\label{eq:covcor}
\begin{split}
\frac{\text{CoV}(P^\text{cor}(i),P^\text{cor}(j))}{P^\text{cor}(i)P^\text{cor}(j)}
&=\frac{\text{CoV}(C^\text{acc}(i),C^\text{acc}(j))}{C^\text{acc}(i)C^\text{acc}(j)}\\
&+\frac{\text{CoV}(C^\text{ISR}(i),C^\text{ISR}(j))}{C^\text{ISR}(i)C^\text{ISR}(j)}\\
&+\frac{\text{CoV}(P^\text{rec}(i),P^\text{rec}(j))}{P^\text{rec}(i)P^\text{rec}(j)}.
\end{split}
\end{equation}
\subsubsection{Contribution from the addition of 
\boldmath{$\mathrm{K}^0_\mathrm{s}\text{ and }\Lambda$} decay products}

The covariance matrix of the charged-particle multiplicity distribution, 
to which the charged $\mathrm{K}^0_\mathrm{s}\text{ and }\Lambda$ decay products
have been added, $P^{\text{K}^0}_\text{data}(n_{\text{K}^0})$ 
(Eq.~(\ref{eq:k0})) is given by
\begin{equation}
\label{eq:covk0}
\begin{split}
\text{CoV}(P^{\text{K}^0}_\text{data}(i),P^{\text{K}^0}_\text{data}(j))
&={\underset{m}{\sum}}{\underset{n}{\sum}}K(m,i)K(n,j)
P^{\text{no}\text{K}^0}_\text{data}(m)P^{\text{no}\text{K}^0}_\text{data}(n)
\frac{\text{CoV}(N^{\text{no}\text{K}^0}_\text{MC}(m),
N^{\text{no}\text{K}^0}_\text{MC}(n))}
{N^{\text{no}\text{K}^0}_\text{MC}(m)N^{\text{no}\text{K}^0}_\text{MC}(n)}\\
&+{\underset{m}{\sum}}{\underset{n}{\sum}}
\frac{\text{CoV}(k(m,i),k(n,i))}{N^{\text{no}\text{K}^0}_\text{MC}(m)
N^{\text{no}\text{K}^0}_\text{MC}(n)}
P^{\text{no}\text{K}^0}_\text{data}(m)P^{\text{no}\text{K}^0}_\text{data}(n)\\
&+{\underset{m}{\sum}}{\underset{n}{\sum}}K(m,i)K(n,j)
\frac{\text{CoV}(P^{\text{no}\text{K}^0}_\text{data}(m),P^{\text{no}\text{K}^0}_\text{data}(n))}
{P^{\text{no}\text{K}^0}_\text{data}(m)P^{\text{no}\text{K}^0}_\text{data}(n)},
\end{split}
\end{equation}
where $P^{\text{no}\text{K}^0}_\text{data}(n_{\text{no}\text{K}^0})$ 
is the reconstructed,
corrected charged-particle multiplicity distribution of the data. 
Therefore, its covariance matrix incorporates all the contributions 
previously encountered.
Multinomial distributions are assumed for  
$N^{\text{no}\text{K}^0}_\text{MC}(n_{\text{no}\text{K}^0})$ 
(see Eq.~(\ref{eq:covn})).
We also assume multinomial distributions for the matrix 
$k(n_{\text{K}^0},n_{\text{no}\text{K}^0})$ and that
there is no correlation between $n_{\text{no}\text{K}^0}$ multiplicities. This 
is reasonable, since in the procedure we use to incorporate 
$\mathrm{K}^0_\mathrm{s}\text{ and }\Lambda$ decay products, the 
$n_{\text{no}\text{K}^0}$ multiplicities are generated (independently) first
and only then, $\mathrm{K}^0_\mathrm{s}\text{ and }\Lambda$ decay products are obtained
by decaying $\mathrm{K}^0_\mathrm{s}\text{ and }\Lambda$. 
The covariance matrix of $k(n_{\text{K}^0},n_{\text{no}\text{K}^0})$ 
is then given by 
\begin{equation}
\label{eq:covmk}
\text{CoV}(k(m,i),k(n,j))=\frac{1}{{\underset{i}{\sum}}k(m,i)}
\begin{cases}
k(m,i)[{\underset{i}{\sum}}k(m,i)-k(m,i)]  &i=j\text{ and }m=n\\
-k(m,i)k(m,j)                              &i\neq j\text{ and }m=n\\
0                                          &m\neq n.
\end{cases}
\end{equation}

\subsubsection{Contribution from the light- or b-quark purity correction}

This contribution is similar in its form to that for the event selection.  
Therefore, the covariance matrix of $P^\text{fl}_\text{data}(n)$
is given by
\begin{equation}
\label{eq:covpfl}
\begin{split}
\frac{\text{CoV}(P^\text{fl}_\text{data}(i),P^\text{fl}_\text{data}(j))}
{P^\text{fl}_\text{data}(i)P^\text{fl}_\text{data}(j)}
&=\frac{\text{CoV}(P^\text{fl}_\text{prod}(i),P^\text{fl}_\text{prod}(j))}
{P^\text{fl}_\text{prod}(i)P^\text{fl}_\text{prod}(j)}\\
&+\frac{\text{CoV}(P^\text{fl-tagged}_\text{prod}(i),P^\text{fl-tagged}_\text{prod}(j))}
{P^\text{fl-tagged}_\text{prod}(i)P^\text{fl-tagged}_\text{prod}(j)}\\
&+\frac{\text{CoV}(P^\text{fl-tagged}_\text{data}(i),P^\text{fl-tagged}_\text{data}(j))}
{P^\text{fl-tagged}_\text{data}(i)P^\text{fl-tagged}_\text{data}(j)}.
\end{split}
\end{equation}

\subsubsection{Combining covariance matrices of 1994 and 1995 data samples}

The covariance matrices of the charged-particle multiplicity distribution of 
the 1994 and 1995 data samples are combined,  after all 
contributions from the various corrections applied to the data have been 
taken into account. This yields to  
\begin{equation}
\label{eq:9495cov}
\text{CoV}(P^\text{94+95}_\text{data}(i),P^\text{94+95}_\text{data}(j))=
\overset{95}{\underset{\text{year}=94}{\sum}}
\frac{({\underset{i}{\sum}}N^\text{year}_\text{data}(i))^2
\text{CoV}(P^\text{year}_\text{data}(i),P^\text{year}_\text{data}(j))}
{({\underset{i}{\sum}}(N^\text{94}_\text{data}(i)+N^\text{95}_\text{data}(i)))^2}.
\end{equation}

\subsubsection{Statistical errors on moments}
Statistical errors on the moments are found by propagating  
the errors on the corresponding corrected charged-particle multiplicity 
distribution of the data, making use of the covariance matrix 
previously calculated.

\begin{itemize}

\item {\bf Variance of the mean multiplicity \boldmath{$\mu_1$}}

The mean multiplicity is defined as 
\begin{equation}
\label{eq:mu1}
\mu_1={\underset{n}{\sum}} n P(n),
\end{equation}
with variance 
\begin{equation}
\label{eq:varn}
\text{Var}(\mu_1)={\underset{n,m}{\sum}}n\cdot m \text{CoV}(P(n),P(m)).
\end{equation} 

\item {\bf Variance of \boldmath{$\mu_k$}}

In a similar way, a moment of order $k$ is defined by 
\begin{equation}
\label{eq:muk} 
\mu_k={\underset{n}{\sum}} n^k P(n),
\end{equation}
and its variance by 
\begin{equation}
\label{eq:varn2}
\text{Var}(\mu_k)={\underset{n,m}{\sum}}n^k\cdot m^k \text{CoV}(P(n),P(m)).
\end{equation}

\item {\bf Variance of the dispersion 
\boldmath{$D=\sqrt{\langle(n-\mu_1 )^2\rangle}$}}

\begin{equation}
\label{eq:vards}
\text{Var}(D)=\frac{1}{4D^2}{\underset{n,m}{\sum}}n\cdot m(n-2\mu_1)
(m-2\mu_1) \text{CoV}(P(n),P(m)).
\end{equation}

\item {\bf Variance of \boldmath{$\frac{\mu_1}{D}$}}
\begin{equation}
\label{eq:varnd}
\begin{split}
\text{Var}\left (\frac{\mu_1}{D}\right )=\left(\frac{\mu_1}{D}\right)^2
{\underset{n,m}{\sum}}n\cdot &m \left(\frac{1}{\mu_1}+
\frac{1}{2D^2}(n-2\mu_1)\right )\\ 
&\cdot \left (\frac{1}{\mu_1}+\frac{1}{2D^2}(m-2\mu_1)\right)
\text{CoV}(P(n),P(m)).
\end{split}
\end{equation}

\item {\bf Variance of the skewness \boldmath{$S=\frac{\langle(n-\mu_1 )^3\rangle}{D^3}$}}

\begin{equation}
\label{eq:vsk}
\begin{split}
\text{Var}(S)=\frac{1}{D^3}{\underset{n,m}{\sum}}n\cdot&m
\left (n^2-3n(\mu_1+\frac{SD}{2})+3(2\mu^2_1-\mu_2-SD\mu_1)\right )\\
&\cdot \left (m^2-3m(\mu_1+\frac{SD}{2})+
3(2\mu^2_1-\mu_2-SD\mu_1)\right )\\
&\cdot \text{CoV}(P(n),P(m)).
\end{split}
\end{equation}

\item {\bf Variance of the kurtosis 
\boldmath{$K=\frac{\langle(n-\mu_1)^4\rangle}{D^4}-3$}}
\begin{equation}
\label{eq:vkurt}
\begin{split}
\text{Var}(K)&=\frac{1}{D^3}{\underset{n,m}{\sum}}n\cdot m \text{CoV}(P(n),P(m))\\
&\cdot \{ n^3-4\mu_1n^2+[6\mu_1^2-2D^2(K+3)]n
-4[\mu_3+(D^2(K+3)-3\mu_2)\mu_1+3\mu_1^3] \}\\
&\cdot \{ m^3-4\mu_1m^2+[6\mu_1^2-2D^2(K+3)]m
-4[\mu_3+(D^2(K+3)-3\mu_2)\mu_1+3\mu_1^3] \}.\\
\end{split}
\end{equation}
\end{itemize}

\section{Systematic uncertainties}\label{sec:sys}

The systematic errors presented here are estimated for 
the combined 1994 and 1995 data samples. 
Each variation of the analysis procedure is performed 
separately for the two years, and the years are combined 
using Eq.~(\ref{eq:9495pn}). The resulting difference 
with the value obtained in the standard analysis is used 
to determine the systematic error as described below.
The systematic contributions to the errors of the charged-particle 
multiplicity distributions and their moments are classified into six
main categories:

\subsubsection{The track quality cuts}

The influence of the track quality cuts is investigated by varying independently each 
cut parameter (Sect.~\ref{sec:trksel}) using the values given in Table~\ref{tab:systsel}.
used to define a good track. For each cut parameter, $p$, starting from the original 
cut value $C^p_0$, we measure and reconstruct the charged-particle multiplicity 
distributions
and their moments using both smaller, $C^p_\alpha$, and larger, $C^p_\omega$,  
values of the cut parameter. 
The systematic contribution to the error from this cut parameter is obtained 
by taking half of the difference between the results ($P(n)$ or moments)
obtained from the two cut values.
This operation is repeated for all cut parameters, and the systematic contribution
to the error from the track quality is then taken to be the quadratic sum from
all the contributions:
\begin{equation}
\label{eq:trksys}
\Delta^\text{sys}_\text{track}P(i)=\frac{1}{2}\sqrt{\overset{\text{all cuts}}
{\underset{p}{\sum}}\text{max}(P^{C^p_\alpha}(i)-P^{C^p_\omega}(i))^2}
\text{, where }C^p_0\in [C^p_\alpha;C^p_\omega].
\end{equation}
This contribution is the dominant part of the systematic error. 
It contributes more than  $60\%$ of the systematic error 
on the mean charged-particle multiplicity (first row of Table~\ref{tab:sys} which 
shows the relative contribution expressed in terms of 
the square of the systematic error, $(\Delta^\text{sys}\langle n \rangle)^2$).
\begin{table}[htbp]
\begin{center}
\begin{tabular}{|c|c|c|}\hline
cuts                            & $C^p_\alpha$       & $C^p_\omega$      \\ \hline
$P_\text{t}>C^p$               &    $0.1\MeV/c$     &     $0.2\MeV/c$    \\
$\text{Span}\geq C^p$          &    32              &      48            \\
$N_\text{inner Hits}\geq C^p$  &    0               &     2              \\
$N_\text{Hits}\geq C^p$        &    20              &     30             \\
$|\text{DCA}|< C^p$            &    5mm             &     15mm           \\  \hline
\end{tabular}
\end{center}
\scaption{Alternative values used to determine the  
systematic error coming from the choice of track quality cuts.}
\label{tab:systsel}
\end{table}
\subsubsection{The event selection}

The same technique as for the track quality cuts is used for the  
parameters of the event selection (described in Sects.~\ref{sec:selcal}  
and~\ref{sec:tecsel}) using the alternative cut values given in 
Table~\ref{tab:sysvsel}.
It does not include the contribution from the light- or b-quark tagging.
This is the smallest 
contribution to the systematic error (these values are summarized in the 
second row of Table~\ref{tab:sysvsel}).
\begin{table}[htbp]
\begin{center}
\begin{tabular}{|c|c|c|}\hline
cut                            & $C^p_\alpha$  & $C^p_\omega$  \\ \hline
$C^p<\Ecal/\sqrt(s)$       &    0.44       &     0.56      \\
$\Ecal/\sqrt(s)<C^p$       &    1.42       &     1.55      \\
$N_\mathrm{clus}>C^p$      &    13         &     16        \\
$\Ecpar< C^p$              &    0.35       &     0.45      \\
$\Ecperp< C^p$             &    0.55       &     0.64      \\
$|\cos(\theta^{\text{cal}}_{\text{th}})|< C^p$ &  0.64 & 0.8   \\  
$|\cos(\theta^{\text{trk}}_{\text{th}})|< C^p$ &  0.6  & 0.8 \\ 
$\phi_2< C^p$              & $165^\circ$ & $175^\circ$       \\ 
\hline
\end{tabular}
\end{center}
\scaption{Alternative values used to determine the systematic error 
contribution due to the event selections.}
\label{tab:sysvsel}
\end{table}
\subsubsection{The light- and b-quark tagging method}

The contributions from light- and b-tagging are obtained by 
varying the values of the discriminants used for the selection. 
Applying values above and below the nominal discriminant value,  
the systematic uncertainty is taken as half of the  
difference between the corresponding charged-particle multiplicity 
distributions. Its contribution to the \mcpm{}  
is given in the third row of Table~\ref{tab:sys}.

\subsubsection{Monte Carlo model uncertainties}

Another important source of systematic error is 
the influence of the model used to correct the data. This error is estimated 
by varying the parameters in the Monte Carlo generator and by comparing the 
result of different Monte Carlo generators.
To investigate the influence of the parton shower algorithm, ARIADNE is 
used instead of JETSET to reconstruct the charged-particle multiplicity 
distribution of the data. 
The difference between the two reconstructed multiplicity distribution 
data sets is taken as the systematic uncertainty. 

The influence of the  
modeling of heavy-quark decays and its 
implementation in the Monte Carlo model is estimated by generating 
events with JETSET, for different values of the fragmentation parameter 
$\epsilon_\text{b}$.  The value used in JETSET by the L3 collaboration is 
$\epsilon_\text{b}=0.0035$. This value is varied by $\pm 0.0015$. As systematic 
uncertainty, we take half of the difference between the 
multiplicity distributions obtained using the 
larger and the smaller values of $\epsilon_\text{b}$.
The influence of the change in the value of other hadronization parameters 
such as the strangeness suppression were found to be negligible. 

The two contributions from the modeling are added in quadrature. 
While the contribution
of $\epsilon_\text{b}$ is small for the full sample and negligible for 
the light-quark sample, it is the largest theoretical contribution 
for the charged-particle multiplicity distribution of the b-quark 
sample. The total contribution of Monte Carlo uncertainties are given 
in the fourth row of Table~\ref{tab:sys}.

\subsubsection{The influence of the unfolding method}

The model independence of the unfolding method and its  
overall reliability is tested in several ways. The results of the tests 
are used to contribute to the systematic error.

First, in order to check the consistency of the method, 
we change the charged-particle multiplicity
distribution used to start the unfolding method, 
$P^{(0)}(n)$ (Eq.~(\ref{eq:pnn})), 
using instead the \cpmd{} generated independently 
with the ARIADNE generator. It must be noted that the starting 
distribution should not matter and, in principle, 
a uniform distribution could also be used.  
However, since we use a small number of iterations (only two, 
the variation in the number of iteration is investigated 
independently), it is preferable to start with a 
distribution which is close to the data already as it is 
the case for ARIADNE.

Also the number of iterations used to obtain the final result
in the unfolding procedure is changed to 4 iterations instead of 2. 
The difference between the charged-particle multiplicity distributions obtained 
after 2 and after 4 iterations represents the next systematic uncertainty. 

Since the unfolding method relies strongly on Monte Carlo, 
we investigate the dependence of the method on a 
given Monte Carlo sample. This is done    
by comparing the produced charged-particle multiplicity distribution of events 
generated with JETSET 
to that obtained by unfolding the simulated distribution of the 
same events using a response matrix of the detector 
determined using ARIADNE, and by a similar comparison with the roles of 
JETSET and ARIADNE exchanged. 
The difference between the unfolded distributions and 
the corresponding produced one (or between their moments) 
are taken as a systematic error and added in quadrature.

These contributions to the systematic error on the mean 
charged-particle multiplicity  are summarized in the 
fifth row of Table~\ref{tab:sys}.

\subsubsection{\boldmath{$\gamma$} conversion}

A photon, passing through the material of the detector, 
may convert into an \ee pair. 
This changes the number of detected 
charged particles, adding particles which do not intrinsically originate from the 
decay of the \Z{}. While this phenomenon is well known 
and treated by the simulation of the detector, a difference between data and 
simulation in the number of pairs of charged particles produced in this way  
constitutes a source of uncertainty. 
Therefore, we compare the rate of $\gamma$ conversions produced 
by the simulation in the central tracking chamber to the number of 
$\gamma$ converted in the data, identified using a simple secondary vertex 
reconstruction algorithm. The simulated rate is found to be slightly smaller 
(the difference does not exceed $15\%$) than the rate obtained from the data. 
The difference is taken as a systematic uncertainty. 
It contributes for $13.2\%$, $14.7\%$ and $8.0\%$ to the total systematic
error on the \mcpm{} for the full, light- 
and b-quark samples, respectively (sixth row of Table~\ref{tab:sys}).

All the above-mentioned contributions to the systematic error are estimated
for all the measurements (\ie{} charged-particle multiplicity distributions 
and moments) and added in quadrature.

In our analysis, uncertainties due to background processes (\eg{} \Z{} leptonic decays)
are found to be negligible or covered in the systematic contribution 
due to the event selection. 

\begin{table}[htbp]
\begin{center}
\begin{tabular}{|l|r|r|r|}\hline
systematic contribution   & full sample & light-quark sample & b-quark sample \\ \hline
track quality cuts  & $67.5\%$   &  $61.8\%$  & $71.3\%$      \\
event selection     & $0.2\%$    &  $0.3\%$   & $0.1\%$       \\
tagging             &            &  $3.0\%$   & $2.2\%$       \\
MC modelling         & $8.9\%$    &  $9.3\%$   & $8.5\%$      \\
unfolding method    & $10.2\%$   &  $10.9\%$  & $9.9\%$       \\
$\gamma$ conversion & $13.2\%$   &  $14.7\%$  & $8.0\%$       \\ \hline
\end{tabular}
\end{center}
\scaption{Relative contribution,  
$(\Delta^\text{sys}_\text{source}/\Delta^\text{sys}_\text{tot})^2$,  
of the various sources of systematic error expressed in terms 
of the square of the systematic error to the measurement of the \mcpm{}.}
\label{tab:sys}
\end{table}

\section{The charged-particle multiplicity distributions}

The charged-particle multiplicity distributions for the full, light- and 
b-quark samples have been measured together with their low-order moments and 
are presented separately in the following.

\subsection{All events}

\begin{figure}[H]
  \begin{center}
    \includegraphics[width=8.4cm]{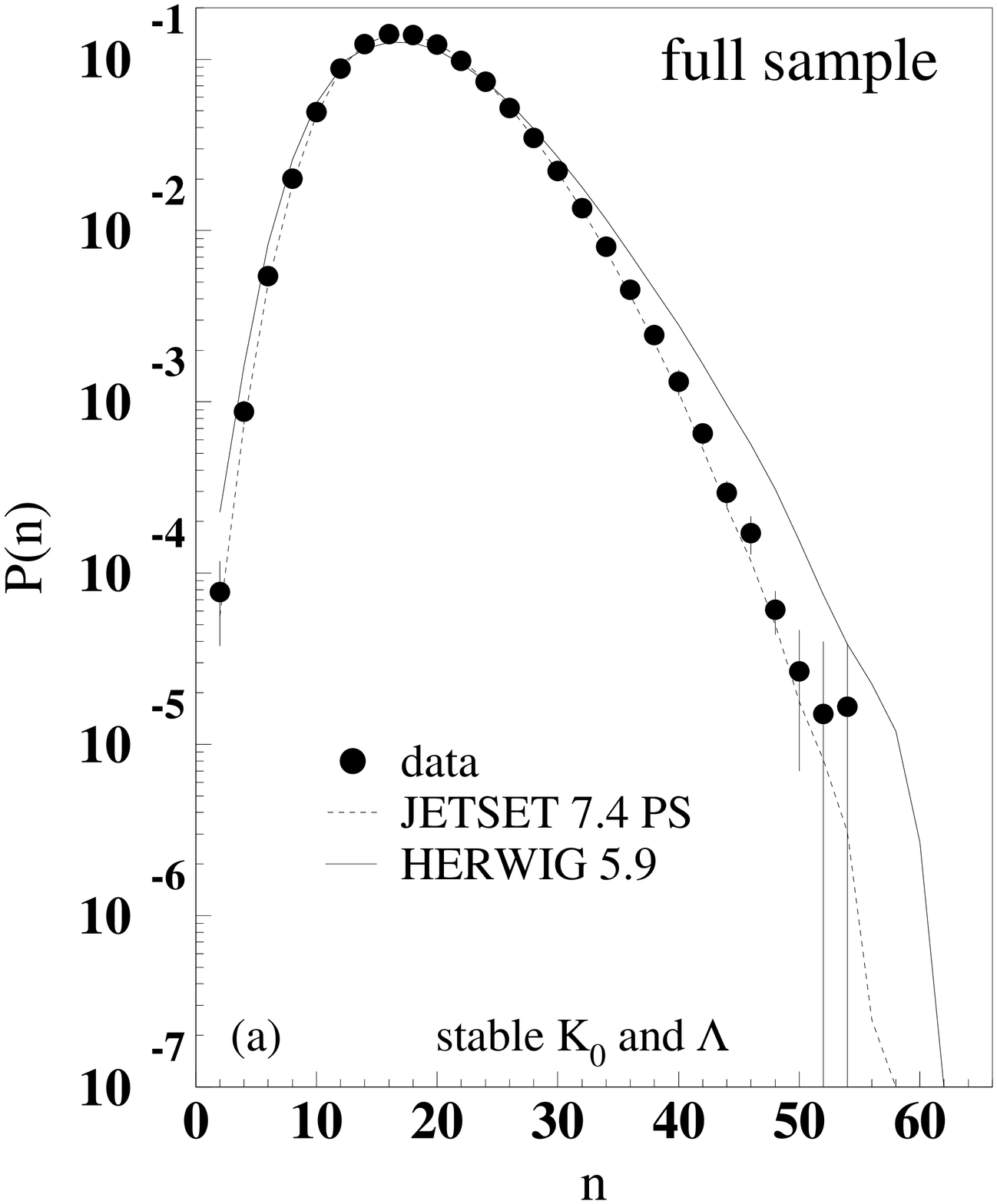}
    \includegraphics[width=8.4cm]{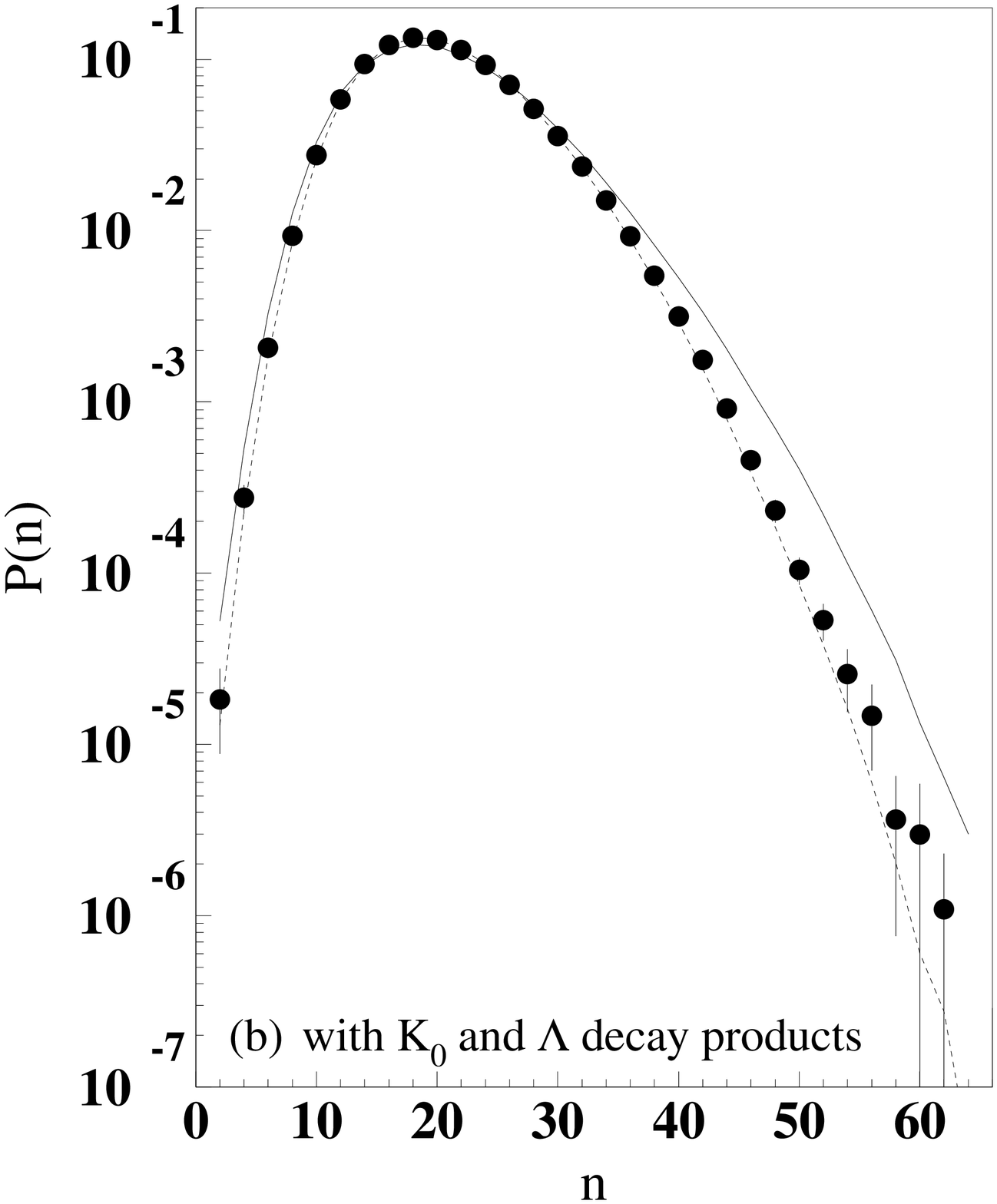}
  \end{center}
\scaption{The charged-particle multiplicity distribution of the full sample 
without \kl{} decay products (a) and including the
 \kl{} decay products (b), compared with JETSET and HERWIG. 
Errors include both statistical and systematic contributions.}
  \label{fig:mult}  
\end{figure}

The charged-particle multiplicity distribution of the full sample
is displayed in 
Fig.~\ref{fig:mult}(a), where 
$\mathrm{K}^0_\mathrm{s}\text{ and }\Lambda$ are assumed to 
be stable, and in Fig.~\ref{fig:mult}(b), where charged particles 
from the decay of $\mathrm{K}^0_\mathrm{s} \text{ and }\Lambda$ 
are included (see also Table~\ref{tab:pnall}). 
The main difference between the two distributions is 
an overall shift of the multiplicity distribution by about 2 charged particles. 
The distribution including charged particles from \kl{} decay
is also broader. 
Both distributions agree quite well with JETSET and ARIADNE, but HERWIG 
overestimates the data for both low and high multiplicities.

The high statistics at the \Z{} mass allows us to measure with high accuracy the 
low-order moments of the charged-particle multiplicity distribution, including
the skewness, $S$, and the kurtosis, $K$. 
They are summarized in Table~\ref{tab:momall} with and without
charged particles resulting from $\mathrm{K}^0\text{ and }\Lambda$
decay. 

The \mcpm{} including 
the decay products 
of $\mathrm{K}^0\text{ and }\Lambda$ is 
$\langle  n^{\mathrm{K}^0}\rangle = 20.46 \pm 0.01 \pm 0.11$.
This result is lower than the previous L3 measurement~\cite{91mult} 
($\langle n^{\mathrm{K}^0}\rangle= 20.79 \pm 0.03 \pm 0.52$), 
but agrees within the systematic error on the previous result.
\begin{table}[H]
\begin{center}
\begin{tabular}{|l|l|l|l|l|l|l|}\hline  
   & \multicolumn{3}{c|}{without K$^0$ and $\Lambda$ decay} &
   \multicolumn{3}{c|}{with K$^0$ and $\Lambda$ decay} \\ \cline{2-7}
\multicolumn{1}{|c|}{Variable} & \multicolumn{1}{c|}{Value} &
\multicolumn{2}{c|}{Error} 
& \multicolumn{1}{c|}{Value} & \multicolumn{2}{c|}{Error} \\ \cline{3-4} 
\cline{5-7}
 & &\multicolumn{1}{c|}{Stat.} & \multicolumn{1}{c|}{Syst.} &
 &\multicolumn{1}{c|}{Stat.} & \multicolumn{1}{c|}{Syst.} \\ \hline
$\langle  n\rangle $                                                       & 18.63   & $\pm$ 0.01   & $\pm$  0.11  & 20.46   & $\pm$ 0.01   & $\pm$  0.11   \\
$\langle  n^{2}\rangle $                                                   & 381.7   & $\pm$ 0.3    & $\pm$  4.4   & 457.7   & $\pm$ 0.3    & $\pm$  4.9    \\
$\langle  n^{3}\rangle $                                                   & 8524    & $\pm$ 10     & $\pm$  154   & 11108   & $\pm$ 12     & $\pm$  181    \\
$\langle  n^{4}\rangle $                                                   & 205918  & $\pm$ 362    & $\pm$  5137  & 290551  & $\pm$ 475    & $\pm$  6529   \\
$D$  & 5.888   & $\pm$ 0.005  & $\pm$  0.051 & 6.244   & $\pm$ 0.005  & $\pm$  0.051  \\
$\langle n\rangle/D$                                             & 3.164   & $\pm$ 0.002  & $\pm$  0.016 & 3.277   & $\pm$ 0.002  & $\pm$  0.016  \\
$S$    & 0.596   & $\pm$ 0.004  & $\pm$  0.010 & 0.600   & $\pm$ 0.004  & $\pm$  0.010  \\
$K$ & 0.51    & $\pm$ 0.01   & $\pm$  0.04  & 0.49    & $\pm$ 0.01   & $\pm$  0.03   \\
\hline\end{tabular}\end{center}
\scaption{Moments of the charged-particle multiplicity distribution for the full sample.}\label{tab:momall}
\end{table}

\subsection{Light-quark events}

Analytical pQCD calculations assume massless quarks. They do not take 
into account mass effects or the weak decay of heavy quarks.
It is therefore more meaningful to measure the 
\cpmd{} 
and its moments for light quarks only, thus allowing better 
comparison with analytical QCD calculations.
 
\begin{figure}[H]
  \begin{center}
    \includegraphics[width=8.4cm]{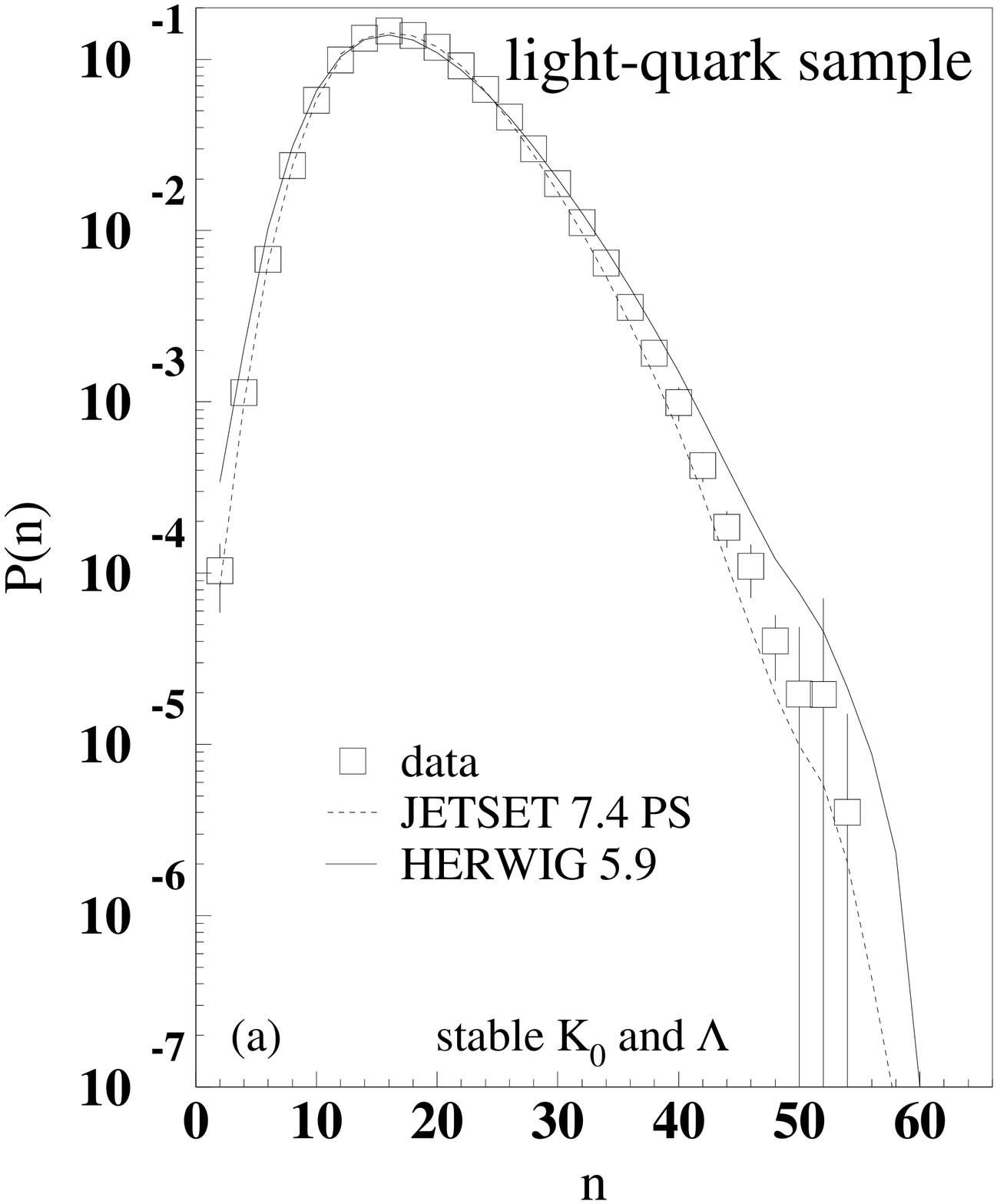}
    \includegraphics[width=8.4cm]{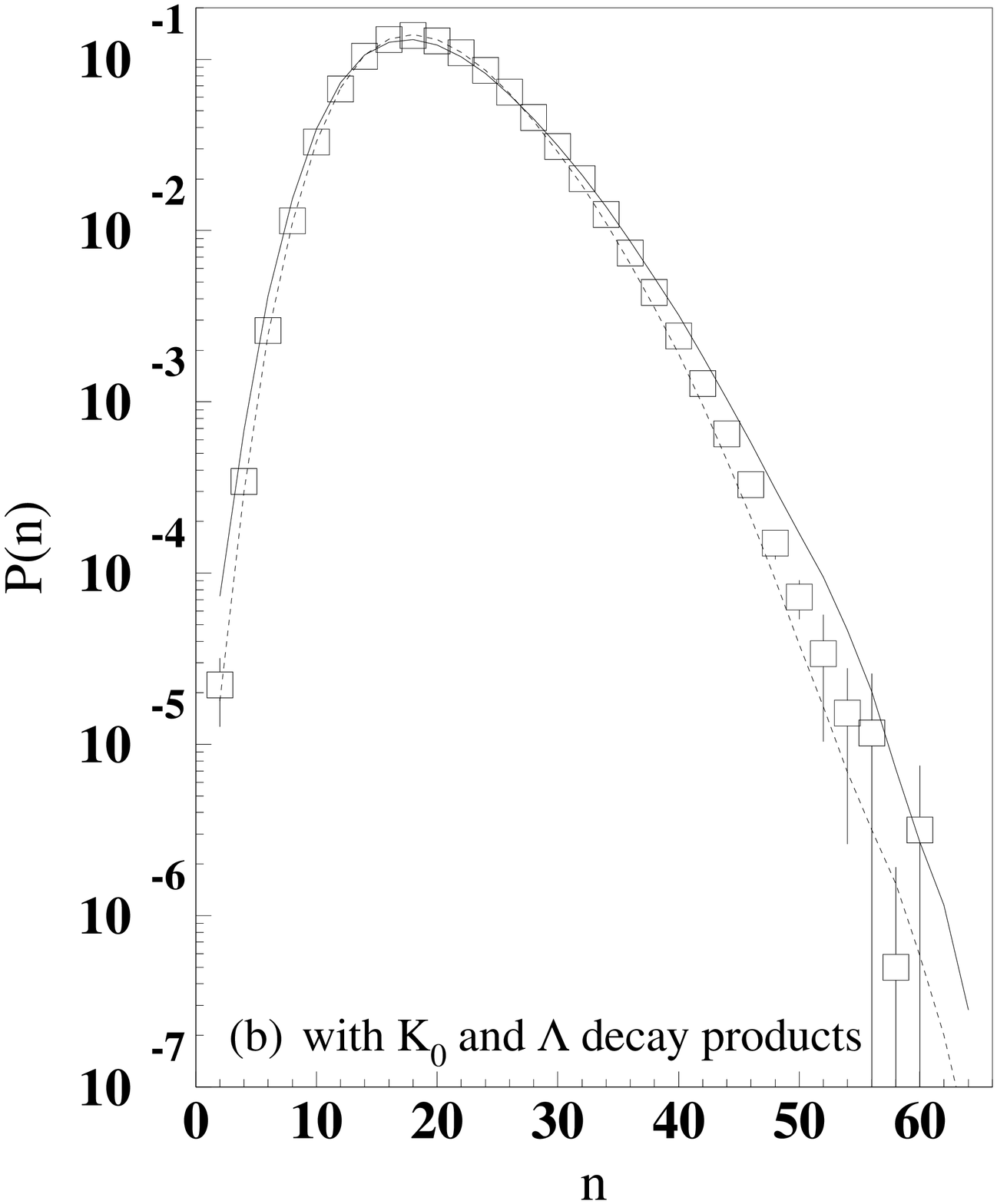}
  \end{center}
\scaption{The charged-particle multiplicity distribution for the 
light-quark sample without 
$\mathrm{K}^0\text{ and }\Lambda$ decay products (a) and including the
 $\mathrm{K}^0\text{ and }\Lambda$ decay products (b), compared with 
JETSET and HERWIG. 
Errors include both statistical and systematic contributions.}
  \label{fig:lightmult}  
\end{figure}

The charged-particle multiplicity distribution for the light-quark sample 
is shown in Fig.~\ref{fig:lightmult} (a)  and (b) assuming 
stable $\mathrm{K}^0\text{ and }\Lambda$ and including 
$\mathrm{K}^0\text{ and }\Lambda$ decay products, respectively 
(see also Table~\ref{tab:pnlight}). As for the 
full sample, JETSET and ARIADNE are found to agree well with the data, 
while HERWIG overestimates both low and high multiplicities.
The principal moments of the charged-particle multiplicity distribution of the 
light-quark samples are summarized in Table~\ref{tab:momltag}.
The \mcpm{} including the decay of 
K$^0$ and $\Lambda$ is found to be 
$\langle n^{\text{K}^0}\rangle =19.88 \pm 0.01 \pm 0.10$.

\begin{table}[H]
\begin{center}
\begin{tabular}{|l|l|l|l|l|l|l|}\hline  
   & \multicolumn{3}{c|}{without K$^0$ and $\Lambda$ decay} &
   \multicolumn{3}{c|}{with K$^0$ and $\Lambda$ decay} \\ \cline{2-7}
\multicolumn{1}{|c|}{Variable} & \multicolumn{1}{c|}{Value} &
\multicolumn{2}{c|}{Error} 
& \multicolumn{1}{c|}{Value} & \multicolumn{2}{c|}{Error} \\ \cline{3-4} 
\cline{5-7}
 & &\multicolumn{1}{c|}{Stat.} & \multicolumn{1}{c|}{Syst.} &
 &\multicolumn{1}{c|}{Stat.} & \multicolumn{1}{c|}{Syst.} \\ \hline
$\langle  n\rangle $                                                       & 18.07   & $\pm$ 0.01   & $\pm$  0.10  & 19.88   & $\pm$ 0.01   & $\pm$  0.10   \\
$\langle  n^{2}\rangle $                                                   & 360.0   & $\pm$ 0.3    & $\pm$  4.1   & 432.4   & $\pm$ 0.4    & $\pm$  4.5    \\
$\langle  n^{3}\rangle $                                                   & 7827    & $\pm$ 11     & $\pm$  142   & 10220   & $\pm$ 14     & $\pm$  166    \\
$\langle  n^{4}\rangle $                                                   & 184370  & $\pm$ 390    & $\pm$  4705  & 260701  & $\pm$ 531    & $\pm$  5852   \\
$D$  & 5.769   & $\pm$ 0.007  & $\pm$  0.054 & 6.111   & $\pm$ 0.007  & $\pm$  0.053  \\
$\langle n\rangle/D$                                             & 3.133   & $\pm$ 0.003  & $\pm$  0.019 & 3.252   & $\pm$ 0.003  & $\pm$  0.019  \\
$S$    & 0.613   & $\pm$ 0.005  & $\pm$  0.013 & 0.617   & $\pm$ 0.005  & $\pm$  0.011  \\
$K$ & 0.54    & $\pm$ 0.02   & $\pm$  0.06  & 0.53    & $\pm$ 0.02   & $\pm$  0.05   \\
\hline\end{tabular}\end{center}
\scaption{Moments of the charged-particle multiplicity distribution for the light-quark sample.}
\label{tab:momltag}
\end{table}
\subsection{b-quark events}

\begin{figure}[H]
  \begin{center}
    \includegraphics[width=8.4cm]{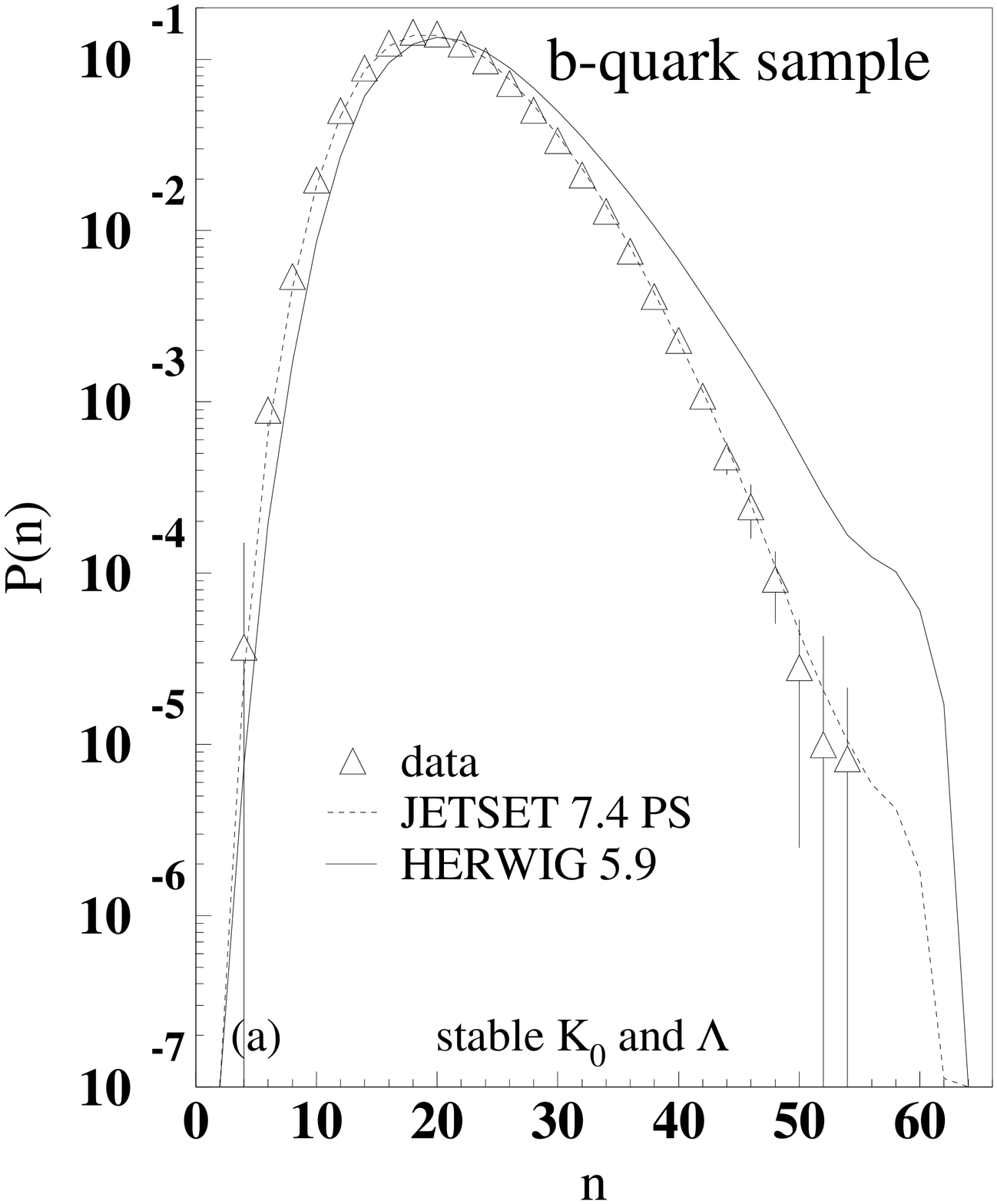}
    \includegraphics[width=8.4cm]{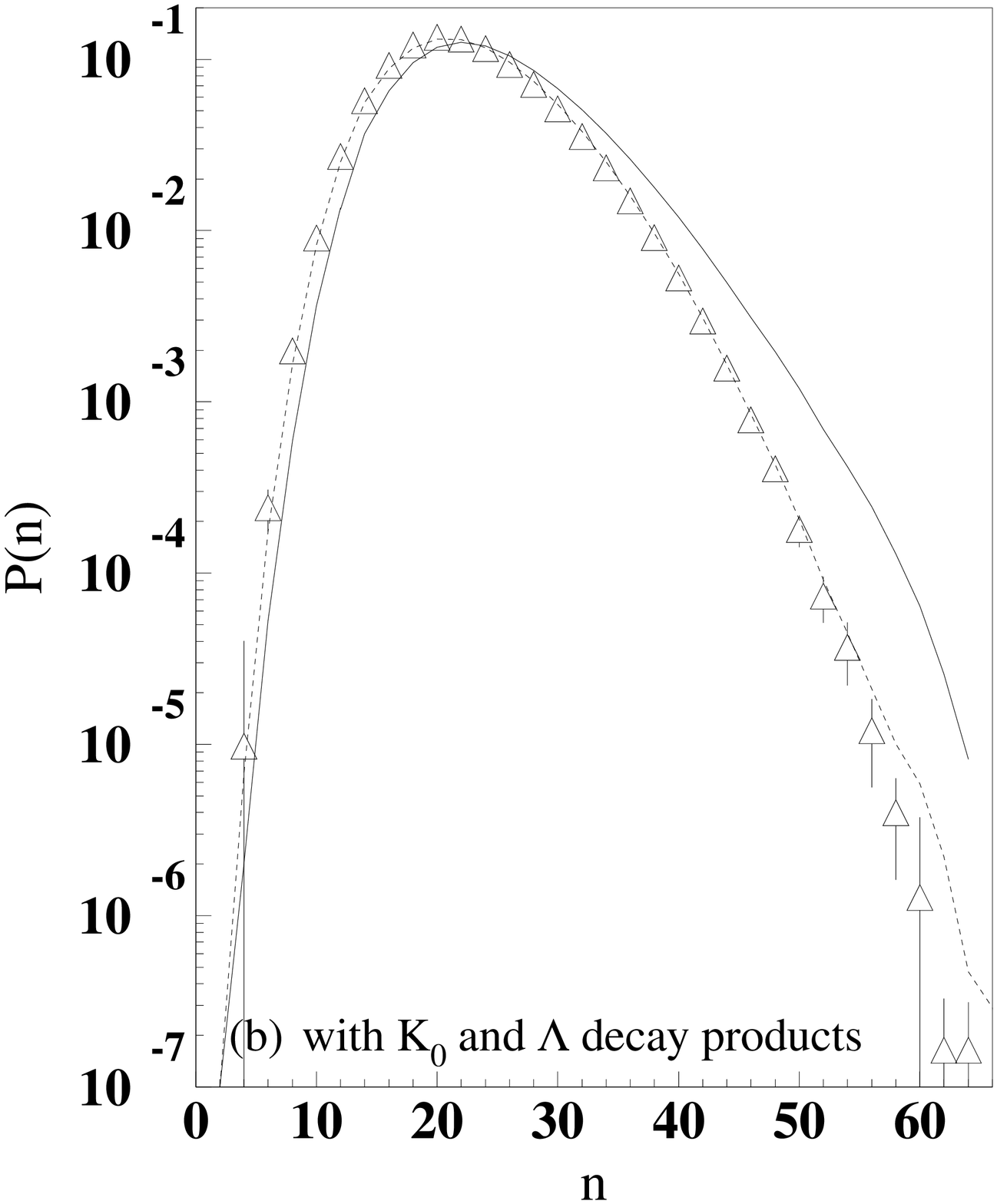}
  \end{center}
\scaption{The charged-particle multiplicity distribution for 
the b-quark sample without
$\mathrm{K}^0\text{ and }\Lambda$ decay products (a) and including the
 $\mathrm{K}^0\text{ and }\Lambda$ decay products (b), compared with 
JETSET and HERWIG. 
Errors include both statistical and systematic contributions.}
  \label{fig:bmult}  
\end{figure}

In order to investigate the influence of the heavy-quark mass or its weak decay, 
and to check the difference between the light- and heavy-quark charged-particle 
multiplicity distributions, the charged-particle multiplicity 
distribution and its moments are also measured for the b-quark sample.
The result without \kl{} decay products is given in 
Fig.~\ref{fig:bmult} (a) and that including charged 
particles produced in the decays of \kl{} 
in Fig.~\ref{fig:bmult} (b) and also in Table~\ref{tab:pnb}. 
As for the two other samples, 
the \cpmd{s} are found to agree well with JETSET and ARIADNE. 
The disagreement for high multiplicities is bigger for HERWIG  
than in the case of the light-quark sample. 
HERWIG, furthermore, underestimates the low multiplicities.

The moments of the charged-particle multiplicity distribution are 
summarized in Table~\ref{tab:mombtag}.
As an effect of the weak decay of the b-quark, the \mcpm{} of the b-quark 
sample is larger than that of the light-quark sample and is found to be 
$\langle n^{\text{K}^0}\rangle =22.45 \pm 0.03 \pm 0.14$
when the charged decay products of the \kl{} are included.
Furthermore, we find the difference between the \mcpm{} 
of the b-quark sample and of the light-quark sample to be
$\langle  n\rangle_\text{b-quark}-\langle  
n^{\mathrm{K}^0}\rangle_\text{light-quark} 
= 2.43 \pm 0.03 \pm 0.05  $ when \kl{} are considered 
as stable and $\langle  n^{\text{K}^0}\rangle_\text{b-quark}-\langle  
n^{\mathrm{K}^0}\rangle_\text{light-quark} 
= 2.58 \pm 0.03 \pm 0.05$ when \kl{} charged decay 
products are added.

\begin{table}[H]
\begin{center}
\begin{tabular}{|l|l|l|l|l|l|l|}\hline  
   & \multicolumn{3}{c|}{without K$^0$ and $\Lambda$ decay} &
   \multicolumn{3}{c|}{with K$^0$ and $\Lambda$ decay} \\ \cline{2-7}
\multicolumn{1}{|c|}{Variable} & \multicolumn{1}{c|}{Value} &
\multicolumn{2}{c|}{Error} 
& \multicolumn{1}{c|}{Value} & \multicolumn{2}{c|}{Error} \\ \cline{3-4} 
\cline{5-7}
 & &\multicolumn{1}{c|}{Stat.} & \multicolumn{1}{c|}{Syst.} &
 &\multicolumn{1}{c|}{Stat.} & \multicolumn{1}{c|}{Syst.} \\ \hline
$\langle  n\rangle $                                                       & 20.51    & $\pm$ 0.03   & $\pm$  0.14  & 22.45    & $\pm$ 0.03   & $\pm$  0.14   \\
$\langle  n^{2}\rangle $                                                   & 454      & $\pm$ 1      & $\pm$  6     & 542      & $\pm$ 1      & $\pm$  7      \\
$\langle  n^{3}\rangle $                                                   & 10787    & $\pm$ 40     & $\pm$  214   & 14006    & $\pm$ 48     & $\pm$  254    \\
$\langle  n^{4}\rangle $                                                   & $273.9\cdot 10^{3}$     & $\pm$ $1.5\cdot 10^{3}$ & $\pm$ $7.3\cdot 10^{3}$  
& $385.8\cdot 10^{3}$ & $\pm$ $1.9\cdot 10^{3}$   & $\pm$  $9.4\cdot 10^{3}$   \\
$D$  & 5.78     & $\pm$ 0.01   & $\pm$  0.05  & 6.16    & $\pm$ 0.01   & $\pm$  0.05   \\
$\langle n\rangle/D$                                             & 3.551    & $\pm$ 0.006  & $\pm$  0.016  & 3.645  & $\pm$ 0.005  & $\pm$  0.015  \\
$S$    & 0.574    & $\pm$ 0.017  & $\pm$  0.008 & 0.573   & $\pm$ 0.017  & $\pm$  0.007  \\
$K$ & 0.43     & $\pm$ 0.04   & $\pm$  0.04  & 0.42    & $\pm$ 0.04   & $\pm$  0.03   \\
\hline\end{tabular}\end{center}
\scaption{Moments of the charged-particle multiplicity distribution for the b-quark sample.}
\label{tab:mombtag}
\end{table}

\begin{table}[htbp]
\begin{center}
\begin{tabular}{|l|l|l|l|l|l|l|}\hline  
   & \multicolumn{3}{c|}{without K$^0$ and $\Lambda$ decay} &
   \multicolumn{3}{c|}{with K$^0$ and $\Lambda$ decay} \\ \cline{2-7}
\multicolumn{1}{|c|}{Multiplicity} & \multicolumn{1}{c|}{Value} &
\multicolumn{2}{c|}{Error} 
& \multicolumn{1}{c|}{Value} & \multicolumn{2}{c|}{Error} \\ \cline{3-4} 
\cline{5-7}
 & &\multicolumn{1}{c|}{Stat.} & \multicolumn{1}{c|}{Syst.} &
 &\multicolumn{1}{c|}{Stat.} & \multicolumn{1}{c|}{Syst.} \\ \hline
$\phantom{1}2$   &0.000075      & $\pm$0.000012    & $\pm$0.000038   &0.000018   & $\pm$0.000003    & $\pm$0.000009  \\
$\phantom{1}4$   &0.000854      & $\pm$0.000031    & $\pm$0.000137   &0.000268   & $\pm$0.000010    & $\pm$0.000051  \\
$\phantom{1}6$   &0.005415      & $\pm$0.000073    & $\pm$0.000203   &0.002054   & $\pm$0.000028    & $\pm$0.000104  \\
$\phantom{1}8$   &0.020121      & $\pm$0.00014     & $\pm$0.000779   &0.009328   & $\pm$0.000063    & $\pm$0.000368  \\
$10$             &0.049286      & $\pm$0.000205    & $\pm$0.001147   &0.027621   & $\pm$0.000108    & $\pm$0.000755  \\
$12$             &0.088697      & $\pm$0.000262    & $\pm$0.001408   &0.058426   & $\pm$0.000150    & $\pm$0.001115  \\
$14$             &0.122932      & $\pm$0.000291    & $\pm$0.001645   &0.093836   & $\pm$0.000173    & $\pm$0.001263  \\
$16$             &0.140991      & $\pm$0.000300    & $\pm$0.001268   &0.121719   & $\pm$0.000176    & $\pm$0.001274  \\
$18$             &0.138721      & $\pm$0.000292    & $\pm$0.000734   &0.133779   & $\pm$0.000168    & $\pm$0.000978  \\
$20$             &0.122072      & $\pm$0.000275    & $\pm$0.001286   &0.129539   & $\pm$0.000158    & $\pm$0.000883  \\
$22$             &0.097885      & $\pm$0.000250    & $\pm$0.001157   &0.113598   & $\pm$0.000148    & $\pm$0.000419  \\
$24$             &0.073501      & $\pm$0.000221    & $\pm$0.000874   &0.092586   & $\pm$0.000138    & $\pm$0.000639  \\
$26$             &0.051795      & $\pm$0.000188    & $\pm$0.000933   &0.070832   & $\pm$0.000125    & $\pm$0.000804  \\
$28$             &0.034650      & $\pm$0.000156    & $\pm$0.001353   &0.051225   & $\pm$0.000109    & $\pm$0.000998  \\
$30$             &0.022146      & $\pm$0.000126    & $\pm$0.000722   &0.035490   & $\pm$0.000093    & $\pm$0.000908  \\
$32$             &0.013439      & $\pm$0.000098    & $\pm$0.000835   &0.023541   & $\pm$0.000076    & $\pm$0.000822  \\
$34$             &0.008025      & $\pm$0.000077    & $\pm$0.000427   &0.014928   & $\pm$0.000061    & $\pm$0.000634  \\
$36$             &0.004460      & $\pm$0.000057    & $\pm$0.000307   &0.009190   & $\pm$0.000047    & $\pm$0.000461  \\
$38$             &0.002420      & $\pm$0.000043    & $\pm$0.000225   &0.005395   & $\pm$0.000036    & $\pm$0.000306  \\
$40$             &0.001302      & $\pm$0.000031    & $\pm$0.000222   &0.003119   & $\pm$0.000027    & $\pm$0.000221  \\
$42$             &0.000644      & $\pm$0.000022    & $\pm$0.000096   &0.001735   & $\pm$0.000020    & $\pm$0.000146  \\
$44$             &0.000293      & $\pm$0.000015    & $\pm$0.000049   &0.000903   & $\pm$0.000014    & $\pm$0.000090  \\
$46$             &0.000168      & $\pm$0.000012    & $\pm$0.000041   &0.000452   & $\pm$0.000010    & $\pm$0.000056  \\
$48$             &0.000058      & $\pm$0.000007    & $\pm$0.000016   &0.000228   & $\pm$0.000007    & $\pm$0.000035  \\
$50$             &0.000025      & $\pm$0.000005    & $\pm$0.000019   &0.000102   & $\pm$0.000004    & $\pm$0.000018  \\
$52$             &0.000013      & $\pm$0.000004    & $\pm$0.000025   &0.000051   & $\pm$0.000003    & $\pm$0.000013  \\
$54$             &0.000013      & $\pm$0.000008    & $\pm$0.000021   &0.000023   & $\pm$0.000003    & $\pm$0.000010  \\
$56$     &$0.2\cdot10^{-6}$& $\pm$$0.4\cdot10^{-6}$    & $\pm$0.000001   &0.000013   & $\pm$0.000003    & $\pm$0.000007  \\
$58$             &              &                 &                  &0.000003   & $\pm$0.000001    & $\pm$0.000003  \\
$60$             &              &                 &                  &0.000003   & $\pm$0.000001    & $\pm$0.000003  \\
$62$             &              &                 &                  &0.000001   & $\pm$0.000001    & $\pm$0.000001  \\
\hline\end{tabular}\end{center}
\scaption{Charged-particle multiplicity distributions with and without K$^0$ and $\Lambda$ decay products for
the full sample.}
\label{tab:pnall}
\end{table}
\begin{table}[htbp]
\begin{center}
\begin{tabular}{|l|l|l|l|l|l|l|}\hline  
   & \multicolumn{3}{c|}{without K$^0$ and $\Lambda$ decay} &
   \multicolumn{3}{c|}{with K$^0$ and $\Lambda$ decay} \\ \cline{2-7}
\multicolumn{1}{|c|}{Multiplicity} & \multicolumn{1}{c|}{Value} &
\multicolumn{2}{c|}{Error} 
& \multicolumn{1}{c|}{Value} & \multicolumn{2}{c|}{Error} \\ \cline{3-4} 
\cline{5-7}
 & &\multicolumn{1}{c|}{Stat.} & \multicolumn{1}{c|}{Syst.} &
 &\multicolumn{1}{c|}{Stat.} & \multicolumn{1}{c|}{Syst.} \\ \hline
 $\phantom{1}2$  &   0.000098   & $\pm$   0.000017  & $\pm$   0.000042  &   0.000021  & $\pm$   0.000004  & $\pm$   0.000009  \\
 $\phantom{1}4$  &   0.001089   & $\pm$   0.000044  & $\pm$   0.000152  &   0.000331  & $\pm$   0.000014  & $\pm$   0.000055  \\
 $\phantom{1}6$  &   0.006742   & $\pm$   0.000102  & $\pm$   0.000263  &   0.002582  & $\pm$   0.000041  & $\pm$   0.000132 \\
 $\phantom{1}8$  &   0.023975   & $\pm$   0.000184  & $\pm$   0.000806  &   0.011335  & $\pm$   0.000087  & $\pm$   0.000402 \\
 $10$            &   0.057501   & $\pm$   0.000273  & $\pm$   0.001198  &   0.032821  & $\pm$   0.000148  & $\pm$   0.000803 \\
 $12$            &   0.099218   & $\pm$   0.000339  & $\pm$   0.001419  &   0.066995  & $\pm$   0.000200  & $\pm$   0.001156 \\
 $14$            &   0.133099   & $\pm$   0.000373  & $\pm$   0.001520  &   0.104305  & $\pm$   0.000226  & $\pm$   0.001224 \\
 $16$            &   0.146113   & $\pm$   0.000377  & $\pm$   0.001090  &   0.130215  & $\pm$   0.000226  & $\pm$   0.001127 \\
 $18$            &   0.138063   & $\pm$   0.000363  & $\pm$   0.000791  &   0.137874  & $\pm$   0.000213  & $\pm$   0.000863 \\
 $20$            &   0.117316   & $\pm$   0.000338  & $\pm$   0.001396  &   0.128385  & $\pm$   0.000199  & $\pm$   0.000939 \\
 $22$            &   0.091121   & $\pm$   0.000302  & $\pm$   0.001183  &   0.109303  & $\pm$   0.000187  & $\pm$   0.000584 \\
 $24$            &   0.066498   & $\pm$   0.000262  & $\pm$   0.000846  &   0.086274  & $\pm$   0.000171  & $\pm$   0.000679 \\
 $26$            &   0.045660   & $\pm$   0.000221  & $\pm$   0.000903  &   0.064172  & $\pm$   0.000152  & $\pm$   0.000788 \\
 $28$            &   0.029891   & $\pm$   0.000180  & $\pm$   0.001320  &   0.045495  & $\pm$   0.000132  & $\pm$   0.000974 \\
 $30$            &   0.018742   & $\pm$   0.000144  & $\pm$   0.000650  &   0.030750  & $\pm$   0.000110  & $\pm$   0.000857 \\
 $32$            &   0.011102   & $\pm$   0.000111  & $\pm$   0.000799  &   0.019974  & $\pm$   0.000089  & $\pm$   0.000774 \\
 $34$            &   0.006490   & $\pm$   0.000086  & $\pm$   0.000383  &   0.012434  & $\pm$   0.000070  & $\pm$   0.000590 \\
 $36$            &   0.003574   & $\pm$   0.000064  & $\pm$   0.000282  &   0.007410  & $\pm$   0.000054  & $\pm$   0.000416 \\
 $38$            &   0.001907   & $\pm$   0.000048  & $\pm$   0.000206  &   0.004343  & $\pm$   0.000041  & $\pm$   0.000275 \\
 $40$            &   0.001000   & $\pm$   0.000035  & $\pm$   0.000219  &   0.002434  & $\pm$   0.000031  & $\pm$   0.000199 \\
 $42$            &   0.000426   & $\pm$   0.000022  & $\pm$   0.000080  &   0.001284  & $\pm$   0.000021  & $\pm$   0.000124 \\
 $44$            &   0.000187   & $\pm$   0.000015  & $\pm$   0.000042  &   0.000651  & $\pm$   0.000014  & $\pm$   0.000076  \\
 $46$            &   0.000109   & $\pm$   0.000012  & $\pm$   0.000036  &   0.000330  & $\pm$   0.000011  & $\pm$   0.000051  \\
 $48$            &   0.000038   & $\pm$   0.000008  & $\pm$   0.000015  &   0.000149  & $\pm$   0.000007  & $\pm$   0.000029  \\
 $50$            &   0.000020   & $\pm$   0.000006  & $\pm$   0.000028  &   0.000072  & $\pm$   0.000006  & $\pm$   0.000018  \\
 $52$            &   0.000018   & $\pm$   0.000009  & $\pm$   0.000051  &   0.000033  & $\pm$   0.000005  & $\pm$   0.000023  \\
 $54$            &   0.000003   & $\pm$   0.000004  & $\pm$   0.000010  &   0.000014  & $\pm$   0.000003  & $\pm$   0.000012  \\
 $56$            &              &                   &                   &   0.000011  & $\pm$   0.000004  & $\pm$   0.000014  \\
 $58$            &              &                   &                   &   0.000001  & $\pm$   0.000001  & $\pm$   0.000001  \\
 $60$            &              &                   &                   &   0.000003  & $\pm$   0.000002  & $\pm$   0.000004  \\
\hline\end{tabular}\end{center}
\scaption{Charged-particle multiplicity distributions with and without K$^0$ and $\Lambda$ decay products for
the light-quark sample.}
\label{tab:pnlight}
\end{table}
\begin{table}[htbp]
\begin{center}
\begin{tabular}{|l|l|l|l|l|l|l|}\hline  
   & \multicolumn{3}{c|}{without K$^0$ and $\Lambda$ decay} &
   \multicolumn{3}{c|}{with K$^0$ and $\Lambda$ decay} \\ \cline{2-7}
\multicolumn{1}{|c|}{Multiplicity} & \multicolumn{1}{c|}{Value} &
\multicolumn{2}{c|}{Error} 
& \multicolumn{1}{c|}{Value} & \multicolumn{2}{c|}{Error} \\ \cline{3-4} 
\cline{5-7}
 & &\multicolumn{1}{c|}{Stat.} & \multicolumn{1}{c|}{Syst.} &
 &\multicolumn{1}{c|}{Stat.} & \multicolumn{1}{c|}{Syst.} \\ \hline
$\phantom{1}2$  &             &                  &                  &                      &                  &          \\
$\phantom{1}4$  &   0.000034  & $\pm$   0.000020 & $\pm$   0.000112 &   0.000009           & $\pm$   0.000005 & $\pm$   0.000030  \\
$\phantom{1}6$  &   0.000842  & $\pm$   0.000103 & $\pm$   0.000143 &   0.000227           & $\pm$   0.000028 & $\pm$   0.000060 \\
$\phantom{1}8$  &   0.005317  & $\pm$   0.000228 & $\pm$   0.000695 &   0.001944           & $\pm$   0.000085 & $\pm$   0.000270 \\
$10$            &   0.019552  & $\pm$   0.000416 & $\pm$   0.001111 &   0.008961           & $\pm$   0.000188 & $\pm$   0.000663 \\
$12$            &   0.049975  & $\pm$   0.000629 & $\pm$   0.001634 &   0.026978           & $\pm$   0.000324 & $\pm$   0.001120 \\
$14$            &   0.089681  & $\pm$   0.000785 & $\pm$   0.002247 &   0.057434           & $\pm$   0.000447 & $\pm$   0.001571 \\
$16$            &   0.125570  & $\pm$   0.000859 & $\pm$   0.002153 &   0.092830           & $\pm$   0.000508 & $\pm$   0.001910 \\
$18$            &   0.143850  & $\pm$   0.000869 & $\pm$   0.001663 &   0.121985           & $\pm$   0.000507 & $\pm$   0.001818 \\
$20$            &   0.140402  & $\pm$   0.000835 & $\pm$   0.001421 &   0.135052           & $\pm$   0.000471 & $\pm$   0.001417 \\
$22$            &   0.122014  & $\pm$   0.000785 & $\pm$   0.001342 &   0.130966           & $\pm$   0.000439 & $\pm$   0.000563 \\
$24$            &   0.097834  & $\pm$   0.000726 & $\pm$   0.001264 &   0.115071           & $\pm$   0.000421 & $\pm$   0.000747 \\
$26$            &   0.072607  & $\pm$   0.000648 & $\pm$   0.001298 &   0.093411           & $\pm$   0.000405 & $\pm$   0.001048 \\
$28$            &   0.050486  & $\pm$   0.000554 & $\pm$   0.001741 &   0.071269           & $\pm$   0.000377 & $\pm$   0.001320 \\
$30$            &   0.033232  & $\pm$   0.000462 & $\pm$   0.001200 &   0.051408           & $\pm$   0.000338 & $\pm$   0.001316 \\
$32$            &   0.020745  & $\pm$   0.000370 & $\pm$   0.001116 &   0.035160           & $\pm$   0.000289 & $\pm$   0.001193 \\
$34$            &   0.012608  & $\pm$   0.000296 & $\pm$   0.000680 &   0.023016           & $\pm$   0.000238 & $\pm$   0.000956 \\
$36$            &   0.007315  & $\pm$   0.000225 & $\pm$   0.000468 &   0.014572           & $\pm$   0.000192 & $\pm$   0.000704 \\
$38$            &   0.003959  & $\pm$   0.000170 & $\pm$   0.000356 &   0.008841           & $\pm$   0.000150 & $\pm$   0.000493 \\
$40$            &   0.002179  & $\pm$   0.000130 & $\pm$   0.000292 &   0.005114           & $\pm$   0.000114 & $\pm$   0.000356 \\
$42$            &   0.001010  & $\pm$   0.000088 & $\pm$   0.000106 &   0.002853           & $\pm$   0.000084 & $\pm$   0.000214 \\
$44$            &   0.000449  & $\pm$   0.000057 & $\pm$   0.000084 &   0.001503           & $\pm$   0.000058 & $\pm$   0.000120 \\
$46$            &   0.000218  & $\pm$   0.000042 & $\pm$   0.000073 &   0.000741           & $\pm$   0.000038 & $\pm$   0.000080  \\
$48$            &   0.000084  & $\pm$   0.000024 & $\pm$   0.000034 &   0.000378           & $\pm$   0.000027 & $\pm$   0.000059  \\
$50$            &   0.000024  & $\pm$   0.000012 & $\pm$   0.000023 &   0.000165           & $\pm$   0.000016 & $\pm$   0.000035  \\
$52$            &   0.000008  & $\pm$   0.000007 & $\pm$   0.000032 &   0.000066           & $\pm$   0.000008 & $\pm$   0.000020  \\
$54$            &   0.000006  & $\pm$   0.000007 & $\pm$   0.000011 &   0.000033           & $\pm$   0.000005 & $\pm$   0.000014  \\
$56$            &             &                  &                  &   0.000010           & $\pm$   0.000004 & $\pm$   0.000005  \\
$58$            &             &                  &                  &   0.000003           & $\pm$   0.000002 & $\pm$   0.000002  \\
$60$            &             &                  &                  &   0.000001           & $\pm$   0.000001 & $\pm$   0.000002  \\
$62$            &             &                  &   &   $0.2\cdot10^{-6}$  & $\pm$   $0.1\cdot10^{-6}$& $\pm$ $0.1\cdot10^{-6}$   \\
$64$            &             &                  &   &   $0.2\cdot10^{-6}$  & $\pm$   $0.1\cdot10^{-6}$& $\pm$ $0.1\cdot10^{-6}$   \\
\hline\end{tabular}\end{center}
\scaption{Charged-particle multiplicity distributions with and without K$^0$ and $\Lambda$ decay products for
the b-quark sample.}
\label{tab:pnb}
\end{table}

\chapter
[Inclusive charged-particle \texorpdfstring{\boldmath{$\xi$}}{$\xi$} spectrum]
{Inclusive charged-particle \texorpdfstring{\boldmath{$\xi$}}{$\xi$} spectrum}
\markboth{\large{Inclusive charged-particle \unboldmath{$\xi$} spectrum}}{}

This chapter is dedicated to the measurement 
of the inclusive charged-particle spectrum in
$\xi=-\text{ln}(2p/\sqrt{s})$, $p$ being the momentum of the
particle and $\sqrt{s}$ the center of mass energy. 
In the first section, we describe briefly, the reconstruction of 
the $\xi$ spectrum. This is followed by the 
description and the estimation of both statistical and systematic
errors.  
In the last section, we present the measurement of the $\xi$ spectrum  
and its peak position for the full, light-quark 
and b-quark samples. The resulting spectra are compared to the analytical 
QCD expectation in the framework of the Local Parton-Hadron
Duality.

\section{Reconstruction of the inclusive spectrum}

The method used to reconstruct the $\xi$ spectrum is very similar 
to that used for the charged-particle multiplicity 
distribution (Sect.~\ref{sec:rec}). Therefore, we present 
here just the major steps which enable us to access the fully 
reconstructed $\xi$ spectrum. 

\subsubsection{Correction for inefficiencies and limited acceptance 
of the detector}

The correction of the  $\xi$ spectrum for inefficiencies and limited 
acceptance of the detector uses the same Bayesian unfolding method 
as we already used for the charged-particle multiplicity distribution. 
The variables are, of course, different. 
We define, here, $n_\text{t}$ as the total number of detected tracks in 
the Monte Carlo sample. Any track of this sample has been generated with 
a certain momentum $p$ (and hence $\xi$ value). 
Due to detector resolution, the measured value is shifted with 
respect to the generated one. 
Therefore, the main purpose of the Bayesian unfolding is, in this case, 
to correct for this shift

The number of tracks $n_\xi(i_\text{g})$ produced with a 
$\xi$ value between $\delta\xi(i_\text{g}-1)$ and $\delta\xi i_\text{g}$. 
is given by
\begin{equation}
\label{eq:nxig}
n_\xi(i_\text{g})=\int_{0}^{\infty}\text{d}\xi_\text{det}
\int_{\delta\xi(i_\text{g}-1)}^{\delta\xi i_\text{g}}
\frac{\partial^2 n_\text{t}}{\partial\xi_\text{gen}\partial\xi_\text{det}}
\text{d}\xi_\text{gen}.
\end{equation}
Similarly, the number of tracks $n_\xi(j_\text{d})$ with a  
$\xi$ value measured between $\delta\xi(j_\text{d}-1)$ and $\delta\xi j_\text{d}$ 
is given by
\begin{equation}
\label{eq:nxid}
n_\xi(j_\text{d})=\int_{0}^{\infty}\text{d}\xi_\text{gen}
\int_{\delta\xi(j_\text{d}-1)}^{\delta\xi j_\text{d}}
\frac{\partial^2 n_\text{t}}{\partial\xi_\text{gen}\partial\xi_\text{det}}
\text{d}\xi_\text{det}.
\end{equation}

We build the matrix $m(j_\text{d},i_\text{g})$ representing  
the number of tracks with $\xi$ values generated, $\xi_\text{gen}$, between 
$\delta\xi(i_\text{g}-1)$ and $\delta\xi i_\text{g}$ and measured, $\xi_\text{det}$,
 between $\delta\xi(j_\text{d}-1)$ and $\delta\xi i_\text{d}$, 
where $\delta \xi$ is the size of 
the interval, which is chosen in this analysis to be 0.2,
\begin{equation}
\label{eq:mxi}
m_\xi(j_\text{d},i_\text{g})=\int_{\delta\xi(i_\text{g}-1)}^
{\delta\xi i_\text{g}}\text{d}\xi_\text{gen}
\int_{\delta\xi(j_\text{d}-1)}^{\delta\xi j_\text{d}}
\frac{\partial^2 n_\text{t}}{\partial\xi_\text{gen}\partial\xi_\text{det}}
\text{d}\xi_\text{det}.
\end{equation}

In addition, this matrix contains a 
``0-particle'' bin,  $m(j_\text{d},0)$, which represents detected tracks 
which do not have any corresponding generated particle. We find in this category 
mainly the charged particles coming from the decay of 
 $\text{K}^0_\text{s}$ and $\Lambda$, 
since, as discussed in the previous chapter,  
$\text{K}^0_\text{s}$ and $\Lambda$ are considered stable 
at generator level.
There is also a small contribution from mis-reconstructed tracks.
In the Bayesian unfolding, 
this ``0-particle'' bin acts by rejecting on a statistical 
basis a certain number of tracks in each $\xi$ interval  
$[\delta\xi(j_\text{d}-1),\delta\xi j_\text{d}]$.
The probability matrix $M_\xi(j_\text{d},i_\text{g})$ of detecting a particle 
with $\xi$ generated within the interval $\delta\xi(i_\text{g}-1)$ and $\delta\xi i_\text{g}$
and measured between $\delta\xi(j_\text{d}-1)$ and $\delta\xi j_\text{d}$ is obtained by 
\begin{equation}
\label{eq:nmxi}
M_\xi(j_\text{d},i_\text{g})=\frac{m_\xi(j_\text{d},i_\text{g})}
{\underset{j_\text{d}}{\sum}m_\xi(j_\text{d},i_\text{g})}.
\end{equation}
Using the Bayesian unfolding with the variables defined above, 
in the same manner as for the case of the charged-particle multiplicity 
distribution (see Sect~\ref{sec:rec}), we can  
correct the number of detected data tracks, 
$n^\text{data}_\xi(j_\text{d})$ having 
$\xi$ measured within the interval $[\delta\xi(j_\text{d}-1),\delta\xi j_\text{d}]$.
After unfolding, we obtain $n^\text{rec}_\xi(i_\text{g})$, the 
reconstructed number of tracks having $\xi$ produced 
within the interval $[\delta\xi(i_\text{g}-1),\delta\xi i_\text{g}]$. 
It corresponds to the result obtained from the Bayesian unfolding 
method after five iterations.
While only two steps were needed to obtain a stable result 
for the charged-particle multiplicity distribution, a stable result 
for the $\xi$ distribution cannot be achieved with less than five steps. 
This may be explained by the fact that the JETSET Monte Carlo $\xi$ distribution 
used to start the iteration procedure does not agree well with the data 
(see Sect.~\ref{sec:resxi}), whereas the agreement was rather good for the 
multiplicity distribution.  

\subsubsection{Other corrections}

In addition to the correction for detector inefficiencies, the data are further 
corrected for event selection, Initial State Radiation, 
taking into account of \kl{} charged decay products as 
well as b- or light-quark contamination in light- or b-quark samples.
These corrections are all applied bin-by-bin using multiplicative 
correction factors,
\begin{equation}
\label{eq:effect}
C_\text{effect}(i_\text{g})=
\frac{n^\text{no effect}_\xi(i_\text{g})}
{n^\text{effect}_\xi(i_\text{g})},
\end{equation}
where $n^\text{effect}_\xi(i_\text{g})$ and 
$n^\text{no effect}_\xi(i_\text{g})$ are the number of    
charged particles produced with $\xi$ generated in the interval   
$[\delta\xi i_\text{g},\delta\xi(i_\text{g}-1)]$. They are obtained from 
Monte Carlo samples which do or do not include the effect we want to correct for.  
The elements $n_\xi(i_\text{g})$ of the 
$\xi$ spectra are then given by 
\begin{equation}
\label{eq:xispec}
n_\xi(i_\text{g})=n^\text{rec}_\xi(i_\text{g})
C_\text{acc}(i_\text{g})C_\text{ISR}(i_\text{g})
C_{\text{K}^0}(i_\text{g}),
\end{equation}
where $C_\text{acc}(i_\text{g})$, $C_\text{ISR}(i_\text{g})$ and 
$C_{\text{K}^0}(i_\text{g})$ are the multiplicative factors 
correcting for event selection, initial-state radiation and for the 
inclusion of charged particles coming from the decay of
$\text{K}^0_\text{s}$ and $\Lambda$, respectively.
An additional correction factor $C^\text{purity}_\text{fl}(i_\text{g})$ is used 
on light- and b-quark tagged samples to correct for b- and light-quark 
contaminations,
\begin{equation}
\label{eq:flpur}
n^\text{fl}_\xi(i_\text{g})=
n^\text{fl-tagged}_\xi(i_\text{g})C^\text{purity}_\text{fl}(i_\text{g}).
\end{equation}
In order to simplify the calculation used to reconstruct and correct the data 
(in particular for the unfolding method), we did not normalize the result  
until now. The $\xi$ distribution is normalized in such a way that 
the integral over the $\xi$ distribution corresponds to the \mcpm{}.
We find a $\chi^2$ of  $31.7$ for $40$ degree of freedom between the 1994 and 1995 data samples. 
The combined 1994 and 1995 normalized 
number of charged particles $p^\text{94+95}_\xi(i_\text{g})$ having $\xi$ 
in the interval $[\delta\xi(i_\text{g}-1),\delta\xi i_\text{g}]$ is then given by
\begin{equation}
\label{eq:renorm}
p^\text{94+95}_\xi(i_\text{g})=
\frac{n^{94}_\xi(i_\text{g})+n^{95}_\xi(i_\text{g})}
{\delta \xi{\underset{n}{\sum}}(N^\text{94}_\text{data}(n)+
N^\text{95}_\text{data}(n))},
\end{equation}
where $N^\text{94}_\text{data}(n)$ and $N^\text{95}_\text{data}(n)$ are 
the numbers of events having $n$ charged particles in 1994 and 1995, 
respectively, as defined in the previous chapter. 
\section{Estimation of the errors}

\subsection{Statistical errors}

Because of the very large statistics of the sample 
(there are about 30 million of charged particles which are produced), 
we can ignore the correlation in the covariance matrix 
$\text{CoV}(n_\xi(i),n_\xi(j))$ and take into account of only the 
diagonal terms. 
This constitutes a slight overestimation of the errors, which is still 
reasonable in view of the sample size.
Therefore, the statistical error on $p^\text{94+95}_\xi(i_\text{g})$, 
$\Delta p^\text{94+95}_\xi(i_\text{g})$ is given by
\begin{equation}
\label{eq:xistat}
\Delta p^\text{94+95}_\xi(i_\text{g})=
\frac{\sqrt{(\Delta n^{94}_\xi(i_\text{g}))^2+(\Delta n^{95}_\xi(i_\text{g}))^2}}
{\delta \xi{\underset{n}{\sum}}(N^\text{94}_\text{data}(n)+
N^\text{95}_\text{data}(n))},
\end{equation}
where $\Delta n^{y}_\xi(i_\text{g})$, $y=94\text{ or }95$, 
is the statistical error for 1994 or 1995
on the number of tracks with $\xi$ produced within $[\delta\xi(i_\text{g}-1),\delta\xi 
i_\text{g}]$.
The error $\Delta n^{y}_\xi(i_\text{g})$ takes into account all the terms we use 
to correct the data and is for each year of the form 
\begin{equation}
\label{eq:xistaty}
\Delta n_\xi(i_\text{g})=n_\xi(i_\text{g})
\sqrt{
(\frac{\Delta n^\text{rec}_\xi(i_\text{g})}{n^\text{rec}_\xi(i_\text{g})})^2
+{\underset{\text{effect}}{\sum}}\left \{
(\frac{\Delta n^\text{effect}_\xi(i_\text{g})}{n^\text{effect}_\xi(i_\text{g})})^2+
(\frac{\Delta n^\text{no effect}_\xi(i_\text{g})}{n^\text{no effect}_\xi(i_\text{g})})^2
\right \}},
\end{equation}
where any of the $\Delta n_\xi(i_\text{g})$ corresponds to $\sqrt{n_\xi(i_\text{g})}$.

\subsection{Systematic errors}

The sources of systematic errors investigated here are the same, with the 
exception of $\gamma$ conversion,   
as those we investigated for the charged-particle multiplicity distribution, 
namely event selection, track quality cuts, Monte Carlo modelling,  
unfolding method, and b-tagging. 
The systematic error for the combined 1994 and 1995 samples 
is estimated in the same way as it was  
for the charged-particle multiplicity distribution.
Table~\ref{tab:xisys} presents the contributions to the total systematic
error from the various sources. This result is obtained from the 
integral of the $\xi$ distribution which corresponds to the \mcpm{}. 
It expresses the relative contribution of the various 
source of systematic error of an average $\delta \xi$ bin.   
It allows a direct comparison with Table~\ref{tab:sys}.
As for the \mcpm{} calculated in the previous chapter, 
the main source of systematic error is the track quality cuts.
It is, also, to be noted that the contribution 
from the unfolding method is larger here than for the \cpmd{} 
(Table~\ref{tab:sys}).
\begin{table}[htbp]
\begin{center}
\begin{tabular}{|l|r|r|r|}\hline
systematic contribution   & full sample & light-quark sample & b-quark sample \\ \hline
track quality cuts  & $63.3\%$    &  $64.8\%$           & $48.9\%$       \\
event selection     & $0.3\%$     &  $0.3\%$            & $2.0\%$        \\
tagging             &             &  $0.9\%$            & $8.1\%$        \\
modelling           & $19.5\%$    &  $17.6\%$           & $23.3\%$       \\
unfolding method    & $16.9\%$    &  $16.4\%$           & $17.7\%$       \\ \hline
\end{tabular}
\end{center}
\scaption{Relative contribution of the various sources of systematic 
error to the measurement of the integral over the 
inclusive charged-particle spectrum.}
\label{tab:xisys}
\end{table}

\section[Inclusive charged-particle $\xi$ spectrum]
{Inclusive charged-particle \boldmath{$\xi$} spectrum}\label{sec:resxi}

\begin{figure}[htbp]
\begin{center}
    \includegraphics[width=8.4cm]{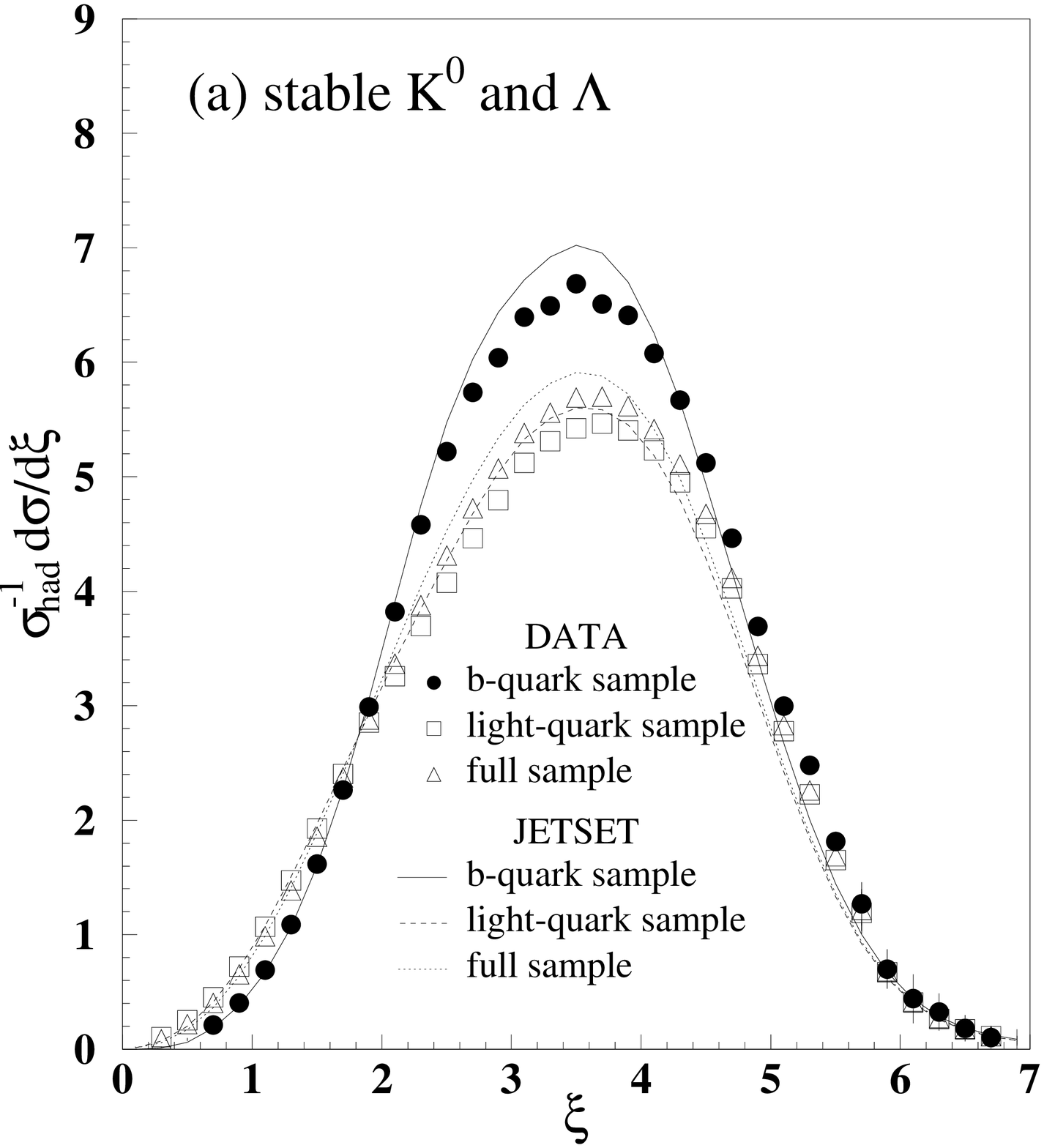}
    \includegraphics[width=8.4cm]{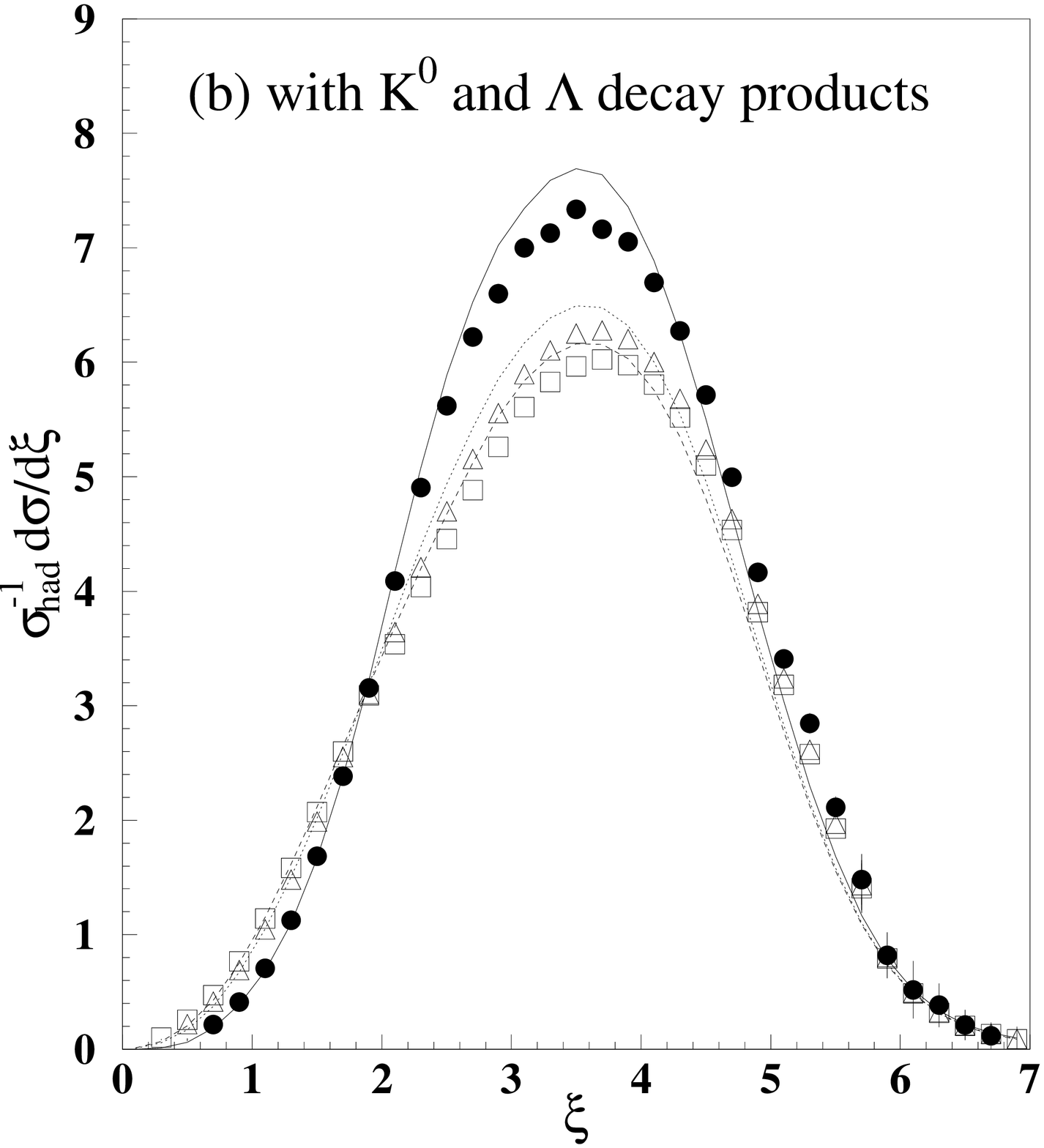}
\scaption{$\xi$ spectrum for the full, light- and b-quark samples 
compared to JETSET expectations.}
  \label{fig:xialljt}  
    \includegraphics[width=8.4cm]{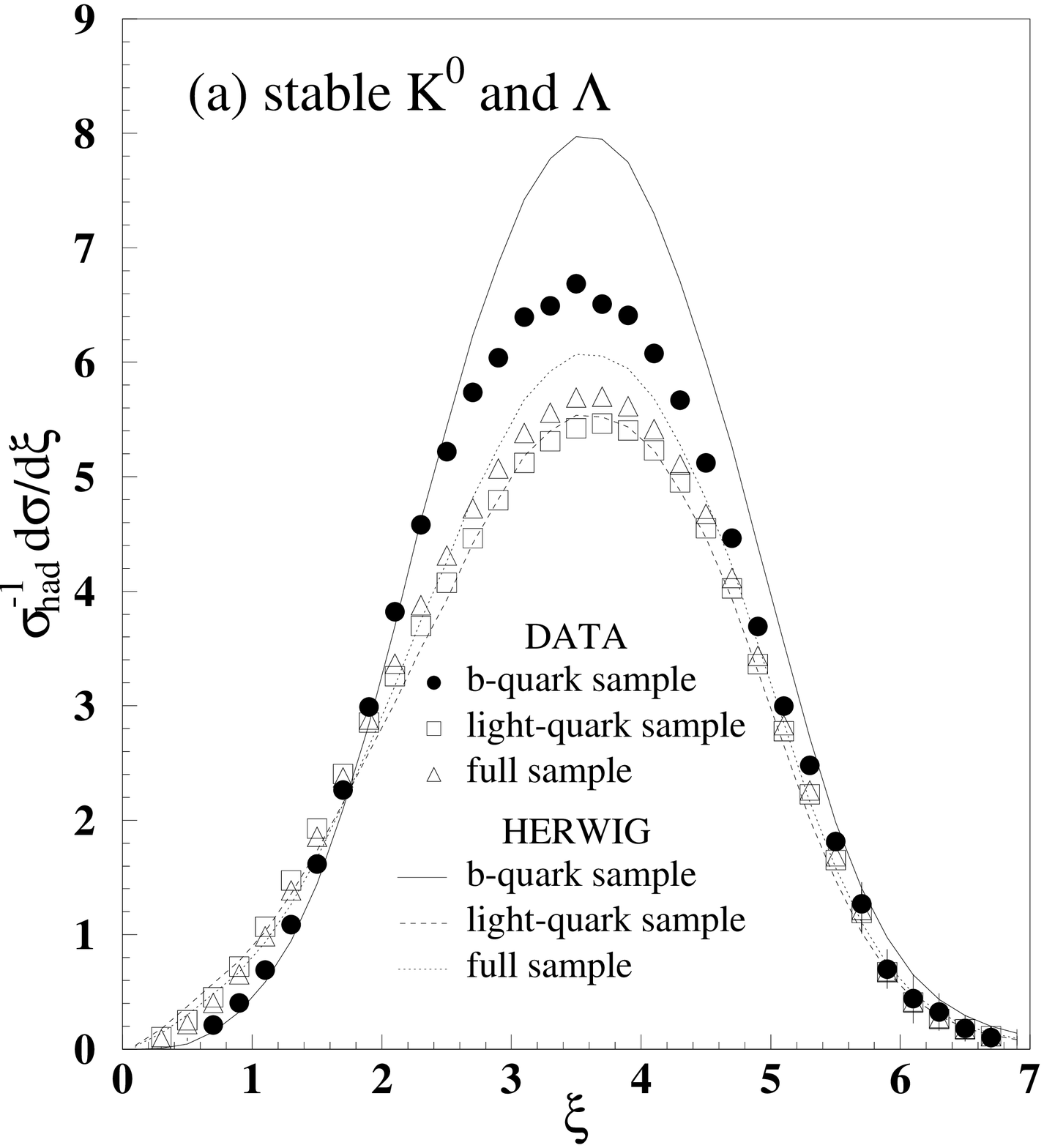}
    \includegraphics[width=8.4cm]{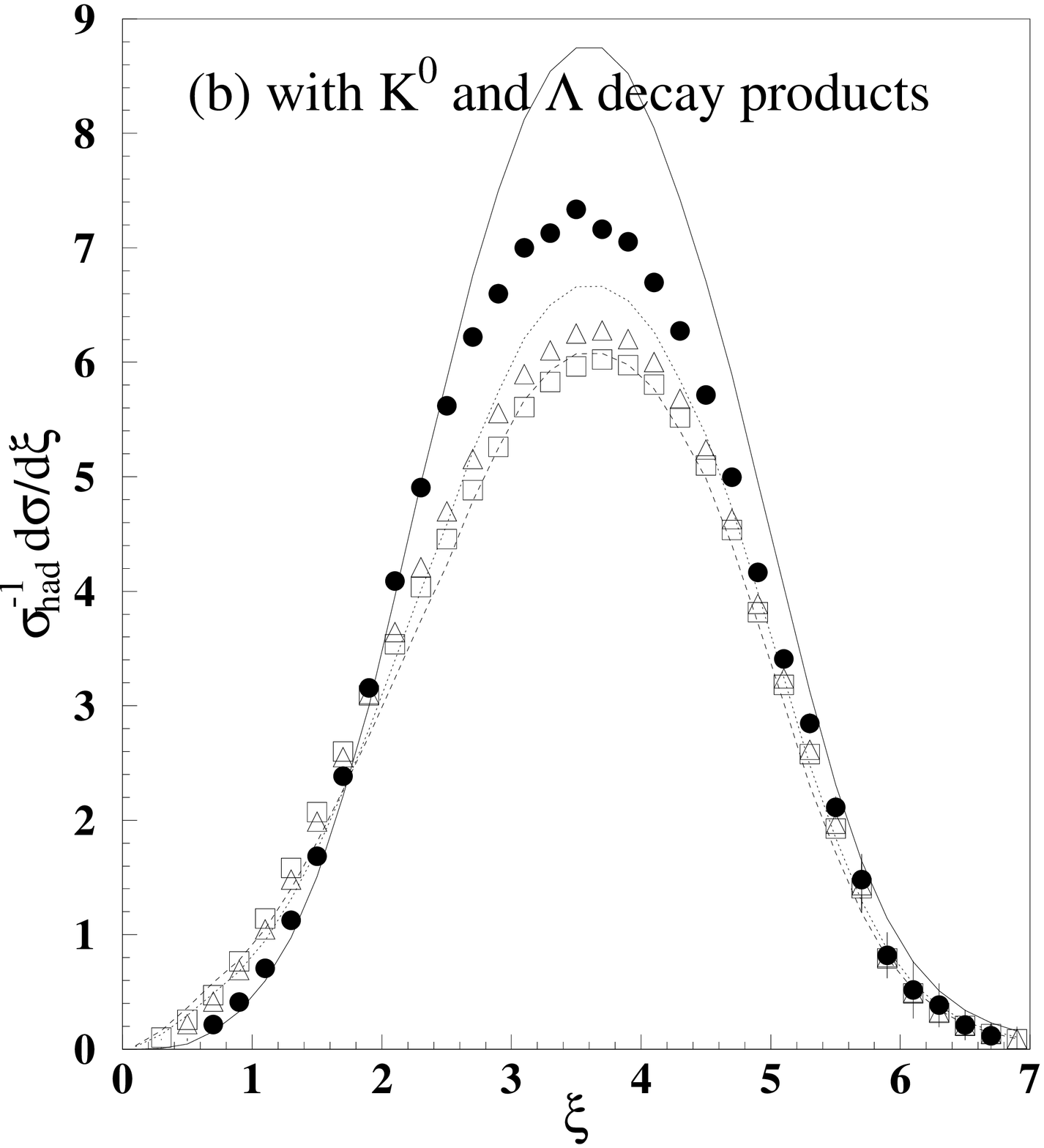}
\scaption{$\xi$ spectrum for the full, light- and b-quark samples 
compared to HERWIG expectations.}
\end{center}
  \label{fig:xiallhw}  
\end{figure}

The inclusive charged-particle $\xi$ spectrum is measured for the full, 
light- and b-quark samples.
The resulting distributions for the three samples are shown together with 
JETSET in Fig.~\ref{fig:xialljt} and 
with HERWIG in Fig.~\ref{fig:xiallhw}, assuming stable $\text{K}^0_\text{s}$ and 
$\Lambda$ for the left plots and including $\text{K}^0_\text{s}$ and 
$\Lambda$ charged decay products for the right plots.
For both light- and b-quark samples, as well as for the full sample, 
it is seen that, JETSET overestimates $\xi$ 
around the peak region, while high $\xi$ are underestimated.  
HERWIG shows even larger disagreement 
for the b-quark and the full samples. 
However, it gives a relatively good description
of the light-quark sample. 
The disagreement in HERWIG is  caused by
a poor implementation (or tuning) of b-quark fragmentation. 
It must be noted that the disagreement of HERWIG for the charged-particle 
multiplicity distribution cannot be related entirely to the b-quark 
fragmentation implementation, even if it has some effects, as seen in 
Fig.~\ref{fig:bmult}, where a very large shift of the b-quark 
charged-particle multiplicity distribution is displayed. Differences are also 
seen for the light-quark samples (Fig.~\ref{fig:lightmult}).

\subsection{Mean charged-particle multiplicity}
By integrating over the whole $\xi$ spectrum, it is possible to obtain 
a measurement of the \mcpm{}.
This result provides a cross-check of the direct measurement 
performed in the previous chapter.
The average numbers of charged particles obtained from the $\xi$ spectra are 
summarized in Table~\ref{tab:xcheck} for the $\xi$ distribution of the full, light- 
and b-quark samples, including and excluding charged particles from 
\kl{} decay.
\begin{table}[htbp]
\begin{center}
\begin{tabular}{|c|c||c|}\cline{2-3}
  \multicolumn{1}{c|}{} &
  \multicolumn{2}{c|}{Mean charged-particle multiplicity}\\
\cline{2-3}
  \multicolumn{1}{c|}{} &
  \multicolumn{1}{c||}{without $\text{K}^0_\text{s}$ and $\Lambda$ decays} & 
  \multicolumn{1}{c|}{with $\text{K}^0_\text{s}$ and $\Lambda$ decays} \\
\hline

full sample           & $18.56\pm 0.01\pm 0.10$ & $20.50\pm 0.01 \pm 0.11$   \\
\hline
light-quark sample    & $18.02\pm 0.01\pm 0.10$ & $19.95\pm 0.01 \pm 0.12$   \\
\hline
b-quark sample        & $20.55\pm 0.03\pm 0.11$ & $22.56\pm 0.03 \pm 0.12$   \\
\hline
\end{tabular}\end{center}
\vspace{-0.2cm}
\scaption{Average number of charged particles obtained by integrating over 
the $\xi$ spectrum. The first error quoted is statistical, the second systematic.}
\label{tab:xcheck}
\end{table}
The average numbers of charged particles measured from the $\xi$ distribution 
are found to be in good agreement with the results obtained from the 
direct measurement (see Table~\ref{tab:momall}, ~\ref{tab:momltag} 
and~\ref{tab:mombtag}), thus reconfirming the consistency 
of both our measurements and the methods used to obtain them.
\subsection[$\xi^\star$ measurement]{$\xi^\star$ measurement}

An important parameter which can be extracted from 
the $\xi$ spectrum is its peak position, $\xi^\star$. 
Both the shape of the $\xi$ distribution
 and the evolution of $\xi^\star$ 
with the center-of-mass energy are predicted by analytical QCD 
assuming Local Parton-Hadron Duality. 
This is usually seen as an important test of 
pQCD and of the importance of the coherence effect. 
In the Double Leading Logarithm Approximation (DLLA),  
analytical QCD calculations predict 
the shape of the $\xi$ spectrum to be a Gaussian.
With next to leading order corrections, taking into account gluon interferences 
responsible for coherence effects, the Modified Leading Logarithm Approximation 
(MLLA) skews and flattens the shape of the $\xi$ spectrum, thereby shifting the 
peak position, $\xi^\star$, to a higher value.
 
\begin{figure}[htbp]
\centering
    \includegraphics[width=8.4cm]{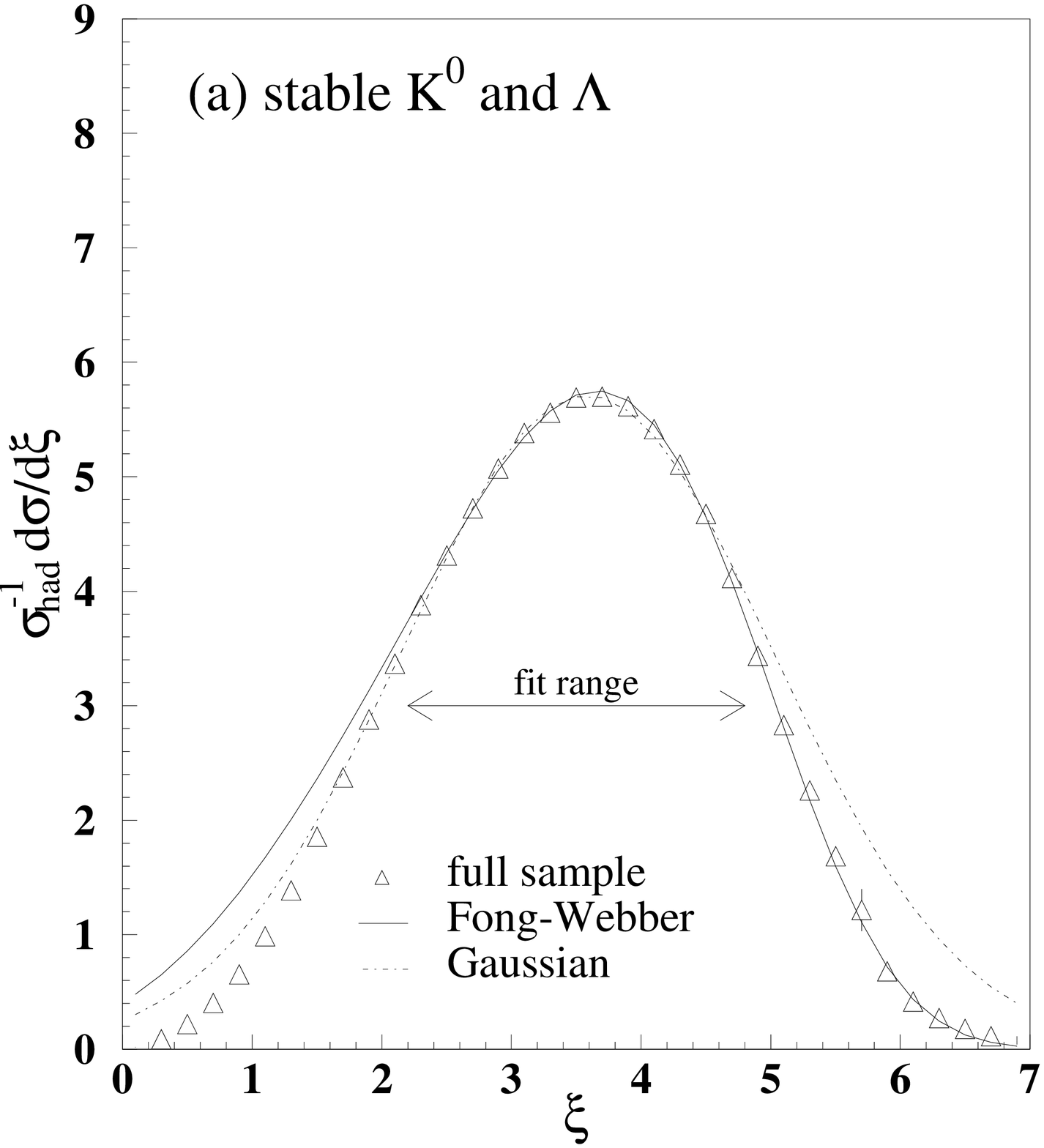}
    \includegraphics[width=8.4cm]{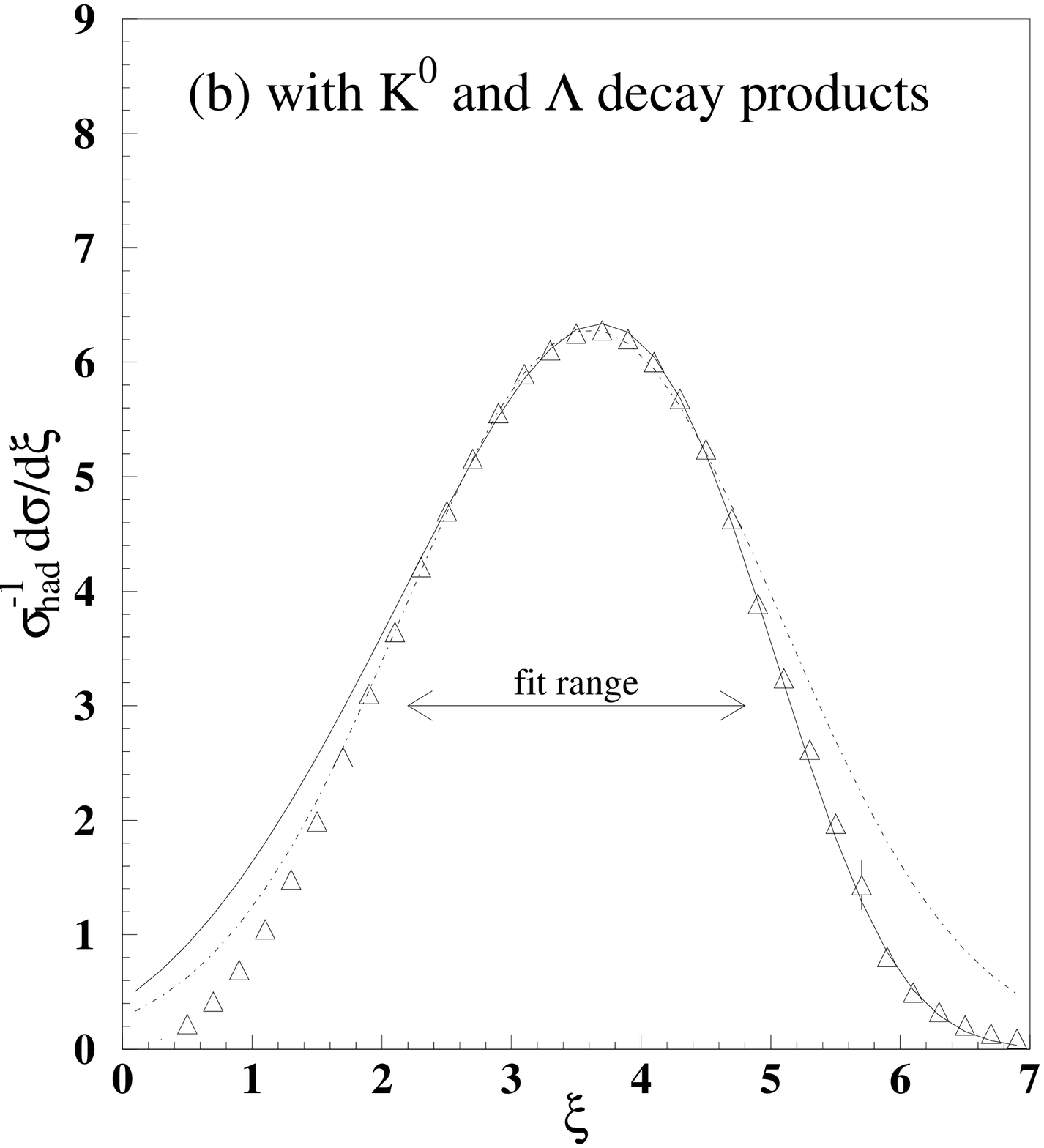}
\scaption{$\xi$ spectrum with Gaussian and Fong-Webber parametrizations 
for the full sample without (a) and with (b) 
$\text{K}^0_\text{s}$ and $\Lambda$ charged decay products}
  \label{fig:xifit_full}  
\end{figure}

\begin{figure}[htbp]
\centering
    \includegraphics[width=8.4cm]{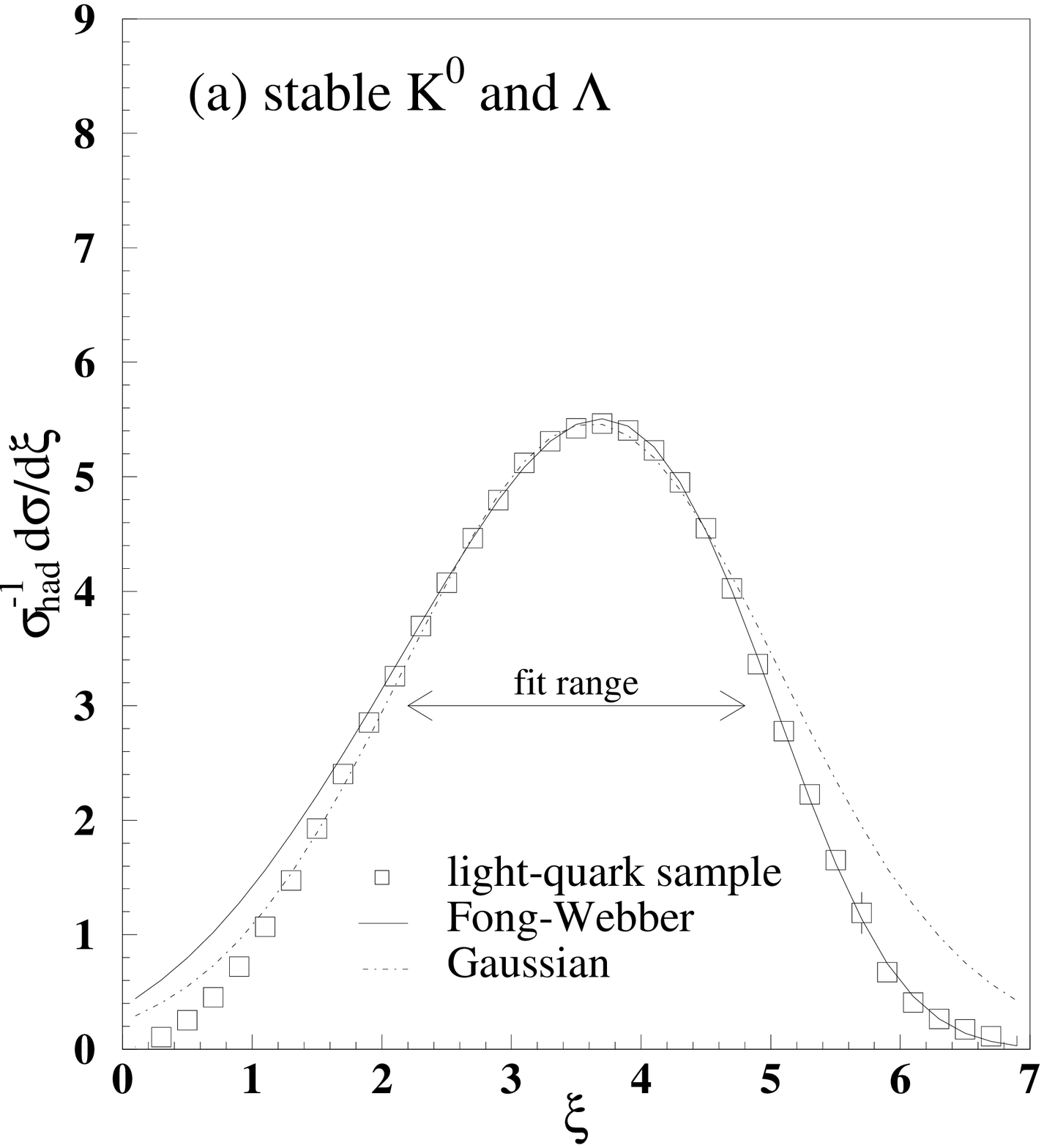}
    \includegraphics[width=8.4cm]{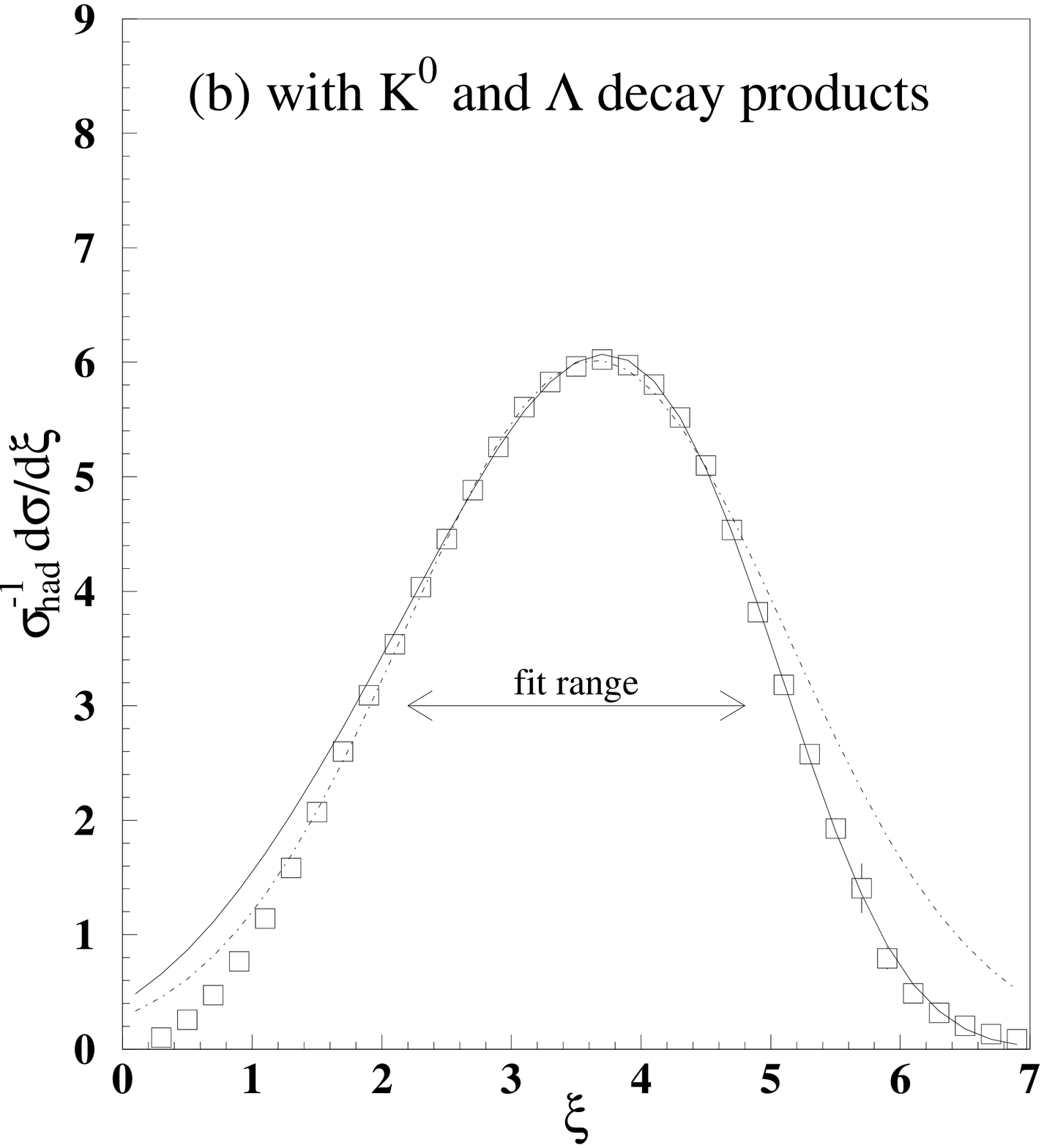}
\scaption{$\xi$ spectrum with Gaussian and Fong-Webber parametrizations 
for the light-quark sample without (a) and with (b) 
$\text{K}^0_\text{s}$ and $\Lambda$ charged decay products.}
  \label{fig:xifit_ltag}  
\end{figure}

We performed fits to the $\xi$ spectra over the interval $[2.3,4.7]$
\footnote{This region corresponding to $60\%$ of the $\xi$ distribution 
is commonly used to compare to other LEP experiment.} with 
the Gaussian parametrization as expected by DLLA, and also with a skewed Gaussian  
using the Fong-Webber parametrization, which reproduces the MLLA expectation 
around the peak position $\xi^\star$. 
In the fitting procedure both statistical and systematic 
errors are included.
We found good agreement for both parametrizations around the peak value for 
the full and the light-quark sample, with or without $\text{K}^{0}_\text{s}$ 
and $\Lambda$ decay products. The Fong-Webber fits have $\chi^2$ 
confidence levels of $49\%$ for the full sample and $88\%$ for the 
light-quark sample. (These $\chi^2$ confidence levels are for 
distributions which exclude \kl{} decay products, but similar 
values are found for the other distributions.) 
The Gaussian fits give somewhat lower $\chi^2$ confidence levels, $20\%$ 
for the full sample and $11\%$ for the light-quark sample, respectively.  
The fitted distributions are shown in 
Fig.~\ref{fig:xifit_full} for the full sample and in Fig.~\ref{fig:xifit_ltag} 
for the light-quark sample.

\begin{figure}[htbp]
\centering
    \includegraphics[width=8.4cm]{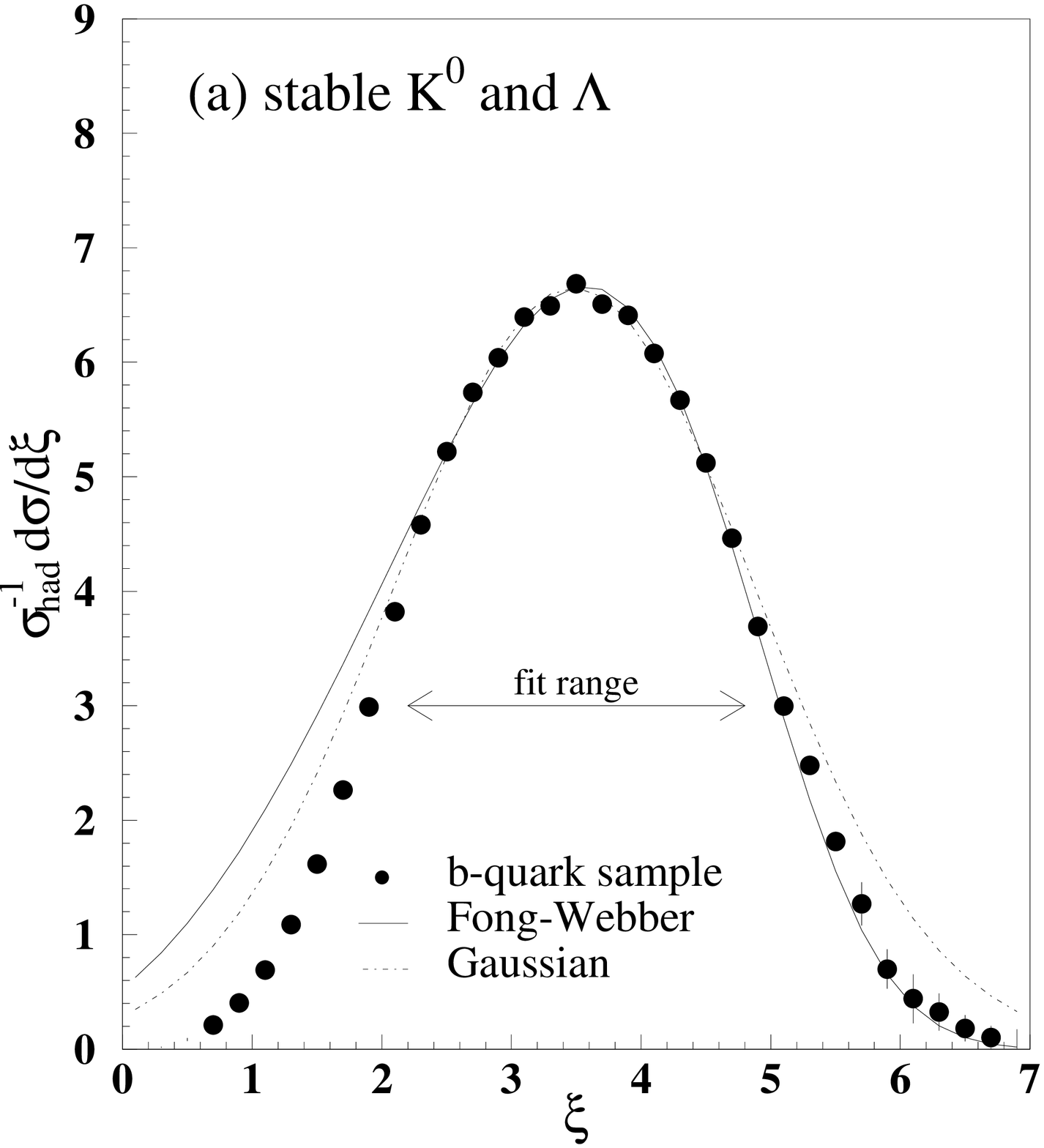}
    \includegraphics[width=8.4cm]{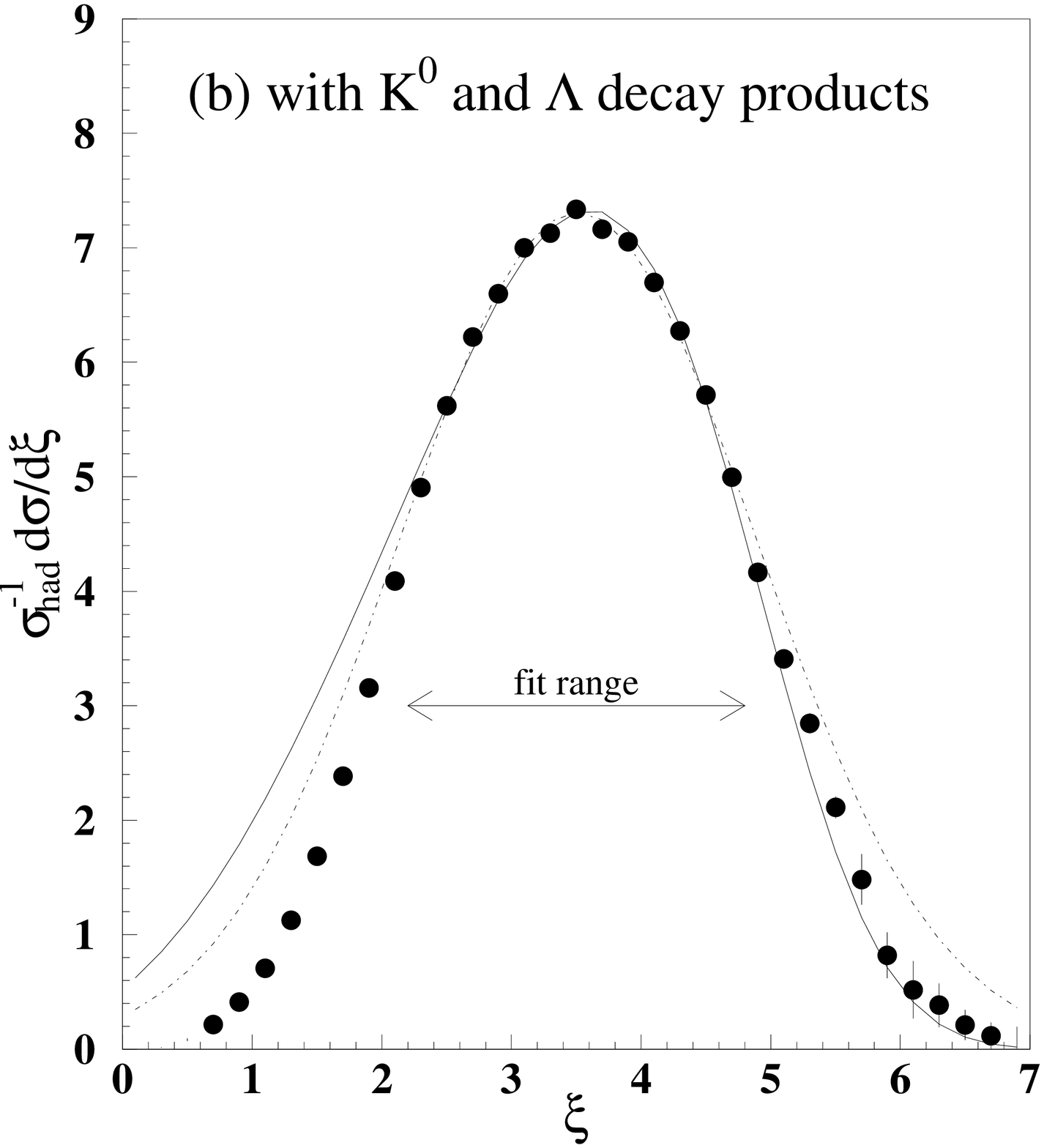}
\scaption{$\xi$ spectrum with Gaussian and Fong-Webber parametrizations 
for the b-quark sample without (a) and with (b) 
$\text{K}^0_\text{s}$ and $\Lambda$ charged decay products.}
  \label{fig:xifit_btag}  
\end{figure}

For the b-quark sample, the agreement with the Fong-Webber 
parametrization is poor with $\chi^2$ confidence level of $5 \cdot 10^{-4}\%$.  
The agreement with the Gaussian parametrization for this sample is 
acceptable with a $\chi^2$ confidence level of $11\%$.
The fitted distributions of the b-quark samples are shown in 
Fig.~\ref{fig:xifit_btag}.
This poor agreement of the MLLA fit (besides the fact that massless 
quarks are assumed) may be due to the fact that some particles, originating from
the weak decay rather than from the partonic shower, would mask part of the 
coherence effect induced by the gluon interference in the parton shower.

From the fits performed on the $\xi$ distributions, we extract 
the peak position, $\xi^\star$. 
In Table~\ref{tab:xival_gs}, we present the peak positions obtained from the 
Gaussian parametrization for the full, light- and b-quark samples 
with and without \kl{} decay products, in Table~\ref{tab:xival_fw},  
those obtained from the Fong-Webber parametrizations.

An additional contribution to the systematic
error is obtained by changing the fit range, using 
both a larger and a smaller fit range.
This systematic contribution is quadratically 
added to the systematic errors obtained from our 
usual systematic sources (\eg{} track quality cuts, 
event selection, unfolding method, theoretical 
uncertainties, tagging).
The systematic error of $\xi^\star$ is largely dominated 
by the change of the fit range, which represents more than $90\%$ 
of the total systematic error, as shown in Table~\ref{tab:sxistar}. 

We also measure the ratio $\xi^\star_\text{ltag}/\xi^\star_\text{full}$ and 
$\xi^\star_\text{btag}/\xi^\star_\text{full}$, combining results from 
Gaussian and Fong-Webber fits. Assuming stable \kl{} we find:
$$
\xi^\star_\text{ltag}/\xi^\star_\text{full}=1.008\pm 0.003 \pm 0.003
\text{ and }
\xi^\star_\text{btag}/\xi^\star_\text{full}=0.974\pm 0.003 \pm 0.004.
$$
When the \kl{} decay products are included,
$$
\xi^\star_\text{ltag}/\xi^\star_\text{full}=1.008\pm 0.003 \pm 0.001
\text{ and }
\xi^\star_\text{btag}/\xi^\star_\text{full}=0.975\pm 0.003 \pm 0.004.
$$
So, these results are found not to be sensitive to the inclusion  
of the \kl{} decay products. 
These ratios also show a clear flavor dependence of the $\xi$ spectrum.
$\xi^\star_\text{btag}/\xi^\star_\text{full}$ is found to be in good agreement 
with a previous measurement performed by the OPAL collaboration~\cite{ksifl_opal}.

\begin{table}[htbp]
\begin{center}
\begin{tabular}{|c|c||c|}\cline{2-3}
  \multicolumn{1}{c|}{} &
  \multicolumn{2}{c|}{$\xi^\star$ from Gaussian fit}\\
\cline{2-3}
  \multicolumn{1}{c|}{} &
  \multicolumn{1}{c||}{without \kl{} decays} & 
  \multicolumn{1}{c|}{with \kl{} decays} \\
\hline
full sample           & $3.684\pm 0.007\pm 0.018$  & $3.712\pm 0.008\pm 0.018$   \\
\hline
light-quark sample    & $3.714\pm 0.008\pm 0.020$  & $3.743\pm 0.009\pm 0.021$   \\
\hline
b-quark sample        & $3.584\pm 0.007\pm 0.009$  & $3.613\pm 0.007\pm 0.009$   \\
\hline
\end{tabular}\end{center}
\vspace{-0.2cm}
\scaption{Peak position, $\xi^\star$, of the $\xi$ spectra,  
obtained from a Gaussian parametrization.
The first error quoted is statistical, the second systematic.}
\label{tab:xival_gs}

%
\vspace{0.5cm}
%

\begin{center}
\begin{tabular}{|c|c||c|}\cline{2-3}
  \multicolumn{1}{c|}{} &
  \multicolumn{2}{c|}{$\xi^\star$ from Fong-Webber fit}\\
\cline{2-3}
  \multicolumn{1}{c|}{} &
  \multicolumn{1}{c||}{without \kl{} decays} & 
  \multicolumn{1}{c|}{with \kl{} decays} \\
\hline
full sample           & $3.715\pm 0.007\pm 0.011$  & $3.741\pm 0.007\pm 0.011$   \\
\hline
light-quark sample    & $3.745\pm 0.007\pm 0.009$  & $3.770\pm 0.008\pm 0.009$   \\
\hline
b-quark sample        & $3.625\pm 0.007\pm 0.026$  & $3.656\pm 0.007\pm 0.026$   \\
\hline
\end{tabular}\end{center}
\vspace{-0.2cm}
\scaption{Peak position, $\xi^\star$, of the $\xi$ spectra, 
obtained from the Fong-Webber parametrization of a skewed Gaussian. 
The first error quoted is statistical, the second systematic.}
\label{tab:xival_fw}
%
%
\vspace{0.5cm}
%

\begin{center}
\begin{tabular}{|l|l|l|l|}\hline
systematic contribution   & full sample & light-quark sample & b-quark sample \\ \hline
track quality cuts  & $77.8\%$     &  $78.3\%$           & $72.9\%$       \\
event selection     & \phantom{1}$0.8\%$     &  \phantom{1}$1.8\%$  & \phantom{1}$0.5\%$        \\
tagging             &              &  \phantom{1}$0.5\%$           & \phantom{1}$5.0\%$        \\
MC modelling        & $17.8\%$     &  $16.8\%$           & $18.0\%$       \\
unfolding method    & $3.6\%$     &  \phantom{1}$2.6\%$           & \phantom{1}$3.6\%$       \\ 
\hspace{1.0cm}Total & \hspace{1.0cm}$29.1\%$       & \hspace{1.0cm}$21.7\%$    
& \hspace{1.0cm}$96.0\%$             \\
\hline
\hspace{1.0cm}fit range & \hspace{1.0cm}$70.9\%$    
& \hspace{1.0cm}$78.3\%$ & \hspace{1.0cm}\phantom{1}$4.0\%$       \\
\hline
\end{tabular}
\end{center}
\scaption{Relative contribution of the various sources of systematic
error to the measurement of $\xi^\star$ obtained from the Gaussian fit.
In the first 5 rows, the square of the contribution from the   
various sources we have used for all the analysis
are  expressed relative to the quadratic sum of all 5 sources. Then, 
in the last two rows, the contribution 
of the sum of these sources and the contribution due to the fit range 
are expressed relative to the quadratic sum of these two contribution.}
\label{tab:sxistar}
\end{table}

\chapter
[\texorpdfstring{\boldmath{$H_q$}}{$H_q$} moments of the \cpmd{}]
{\texorpdfstring{\boldmath{$H_q$}}{$H_q$} moments of the \cpmd{}}\label{chap:hq}
\markboth{\large{$H_q$ moments of the \cpmd{}}}{}

This is the first chapter 
dedicated to the detailed study of the shape of the 
charged-particle multiplicity distribution, 
which is the starting point of the analysis. 
In order to understand the origin of the 
shape of the \cpmd{}, in the next chapters, 
the full sample events will be classified  
into several categories and their \cpmd{s} measured. 

In this chapter, the focus is kept on the entire 
charged-particle multiplicity distribution.
Its shape is analyzed using the ratios of cumulant factorial 
moments to factorial moments, $H_q$, which resolve the relative 
weight of a single $q$-particle correlation function 
on the shape of the distribution.
This study not only uses the charged-particle multiplicity distribution
of the full sample but also those of the light- and b-quark samples.

In the first section of this chapter, we describe the various 
steps needed to obtain a reliable measurement of the $H_q$. 
This includes 
the evaluation of both statistical and systematic errors, 
followed by a short study of the truncation in the tail of the 
charged-particle multiplicity distribution.
The next section presents the measurement of the $H_q$ moment 
for the charged-particle multiplicity distribution of the 
full, light- and b-quark samples. Results are compared to 
the numerous analytical QCD predictions which exist up to 
the Next to Next to Leading Logarithm Approximation (NNLLA).
The last section of the chapter describes an attempt at finding 
an answer on the origin of the $H_q$ behavior (or at least finding 
a way to reproduce it) based on the study of various Monte Carlo models.

\section[Measurement of the {$H_q$} moments]
{Measurement of the \boldmath{$H_q$} moments}

In this section, we present the steps  needed to obtain
a reliable $H_q$ measurement. This includes, of course, 
the estimation of both the statistical and systematic errors, but also more 
specific problems as the influence of the statistics on the measurement, 
the sensitivity of the $H_q$ moments 
to the truncation of the tail of the charged-particle 
multiplicity distribution and its influence on the 
result.

\subsection[{$H_q$} correlation]
{\boldmath{$H_q$} correlation}

\begin{figure}[htbp]
\centering
    \includegraphics[width=8.4cm]{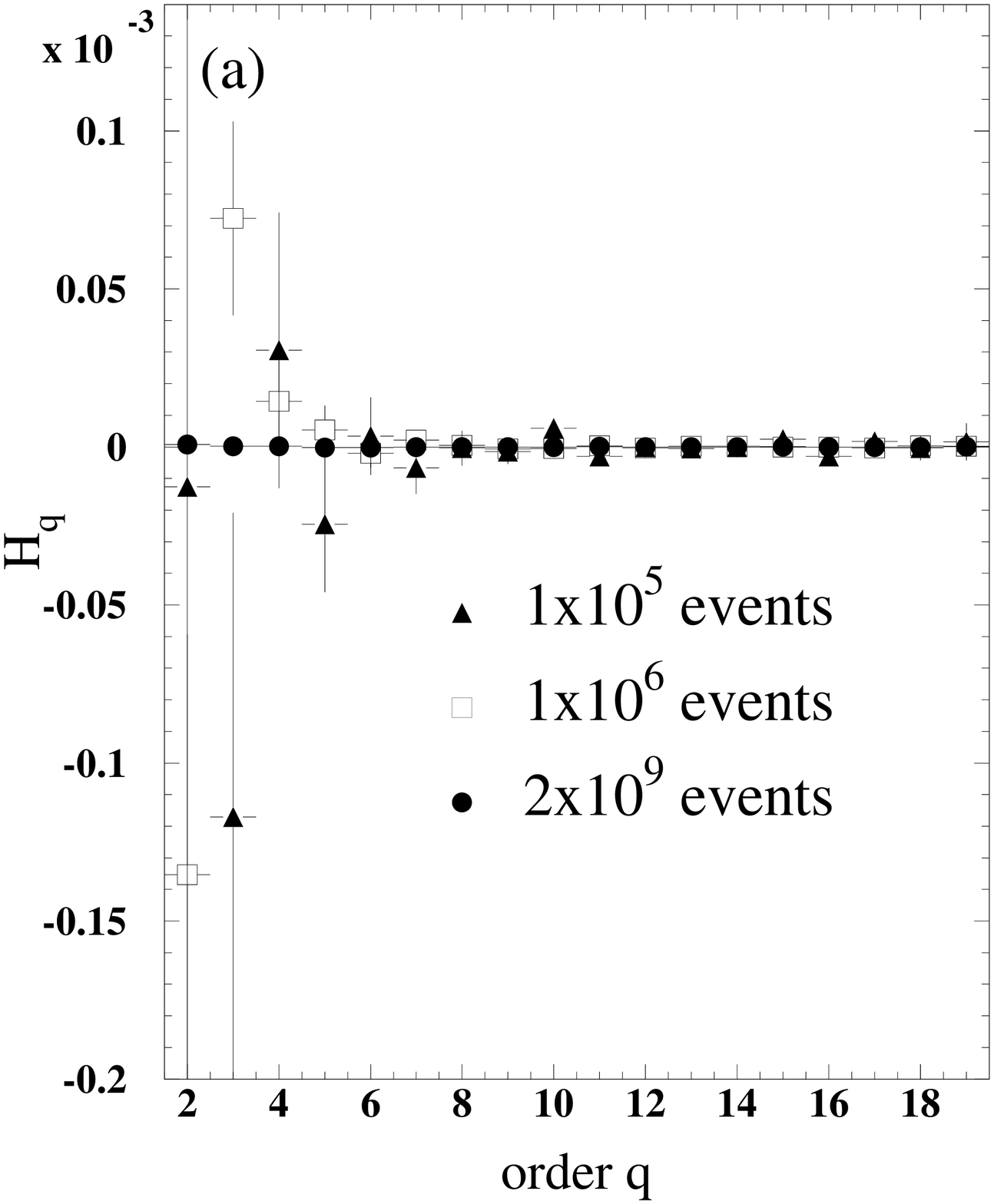}
    \includegraphics[width=8.4cm]{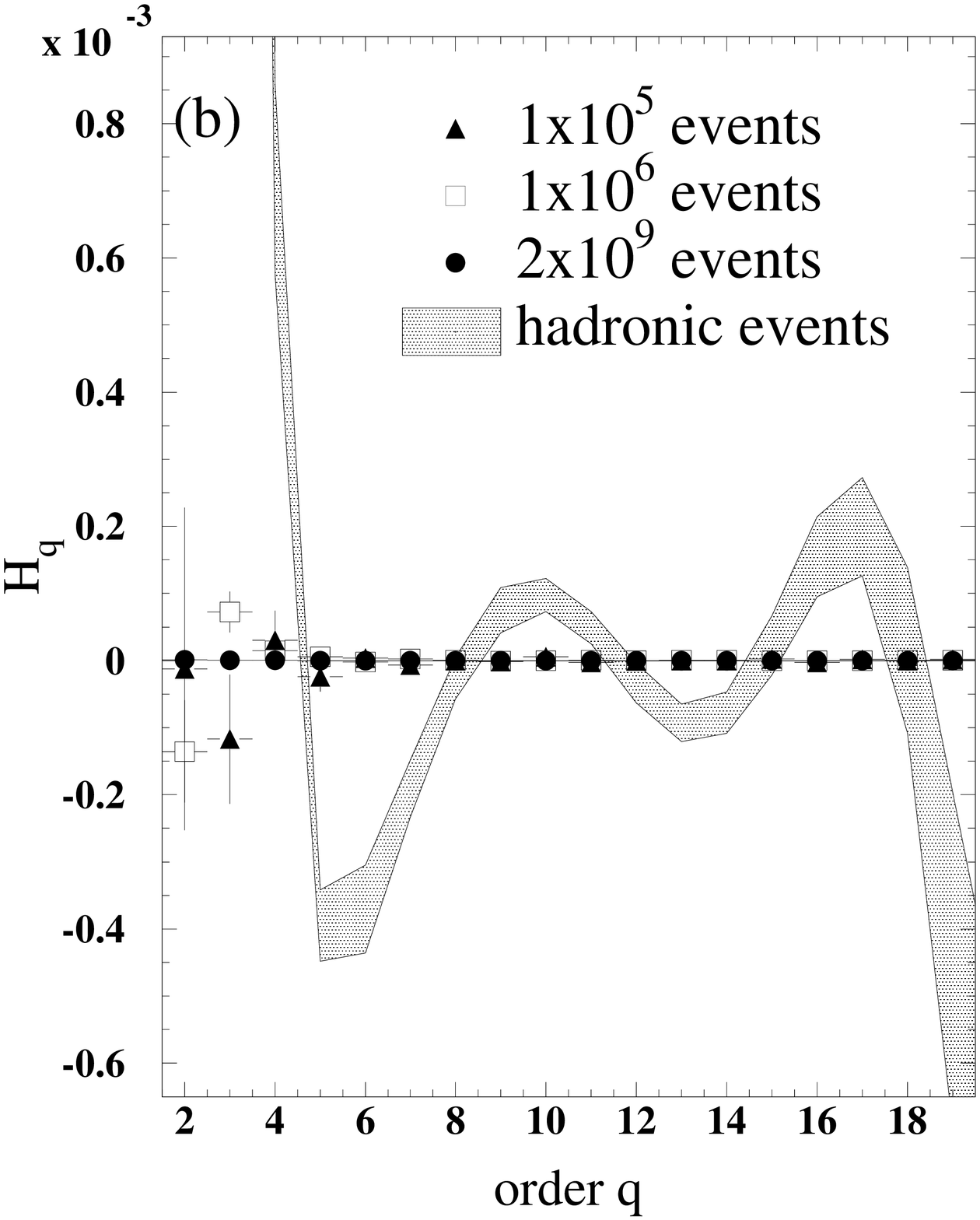}
\scaption{Comparison between the $H_q$ moments obtained from
events generated randomly according a Poisson distribution 
for various size samples (a). The results are further compared in 
(b) to the data.}
\label{fig:hqpois}  
\end{figure}

In order to test the consistency of the measurements of the 
$H_q$ moments, we determine the $H_q$ obtained from 
distributions of various statistics and study the 
resulting correlation between $H_q$. It has been shown 
in a previous analysis that this type of correlation 
is small at low $q$~\cite{opal_hq}.  
Here, we extend this analysis to values of $q$  
up to the value of the \mcpm{}.
We perform this study using events 
generated according to a Poisson distribution.
Knowing that, mathematically, all $H_q$ moments of 
a Poisson distribution are zero, we first 
try to evaluate how much this result is affected by 
the statistics. Therefore, we generate randomly events
according to a Poisson distribution having the same  
mean value as the experimental \cpmd{} of the full sample. 
We generate several samples containing up to 
$2\cdot 10^9$  events. A few examples are given in Fig.~\ref{fig:hqpois}.

In Fig.~\ref{fig:hqpois}(a) the effect on the $H_q$ of limited  
statistics is shown for a Poisson distribution, which 
expects zero for all $q$.
We see large deviations from zero. 
However, these deviations are seen only for $q$ smaller than 8 and 
decrease with the increase of the statistics.  
For larger $q$ they are all very close to 0. 
The $2\cdot 10^9$ sample does not show any significant 
deviation from the mathematical Poisson distribution.
However, this effect of the limited statistic is seen 
to be negligible compared to the size of the $H_q$ oscillation seen 
in the data (Fig.~\ref{fig:hqpois}(b)) and hence smaller than the  
size of the effect we want to study.
This shows us that the measurement of the $H_q$ moments 
should be rather independent of the size of the data samples.

Because of the mathematical expression of the $H_q$ (Eq. (\ref{eq:hqt})) 
by which an $H_q$ moment is calculated iteratively from the previous ones,  
one can worry about correlations which may exist between the $H_q$. Since 
the $H_q$ are supposed to measure the relative weight of the 
genuine $q$-particle correlation function, we investigate 
the reliability of this result.
Therefore, we determine from the previously generated Poisson 
distributions also their correlation matrices. 

\begin{figure}[htbp]
\centering
    \includegraphics[width=8.4cm]{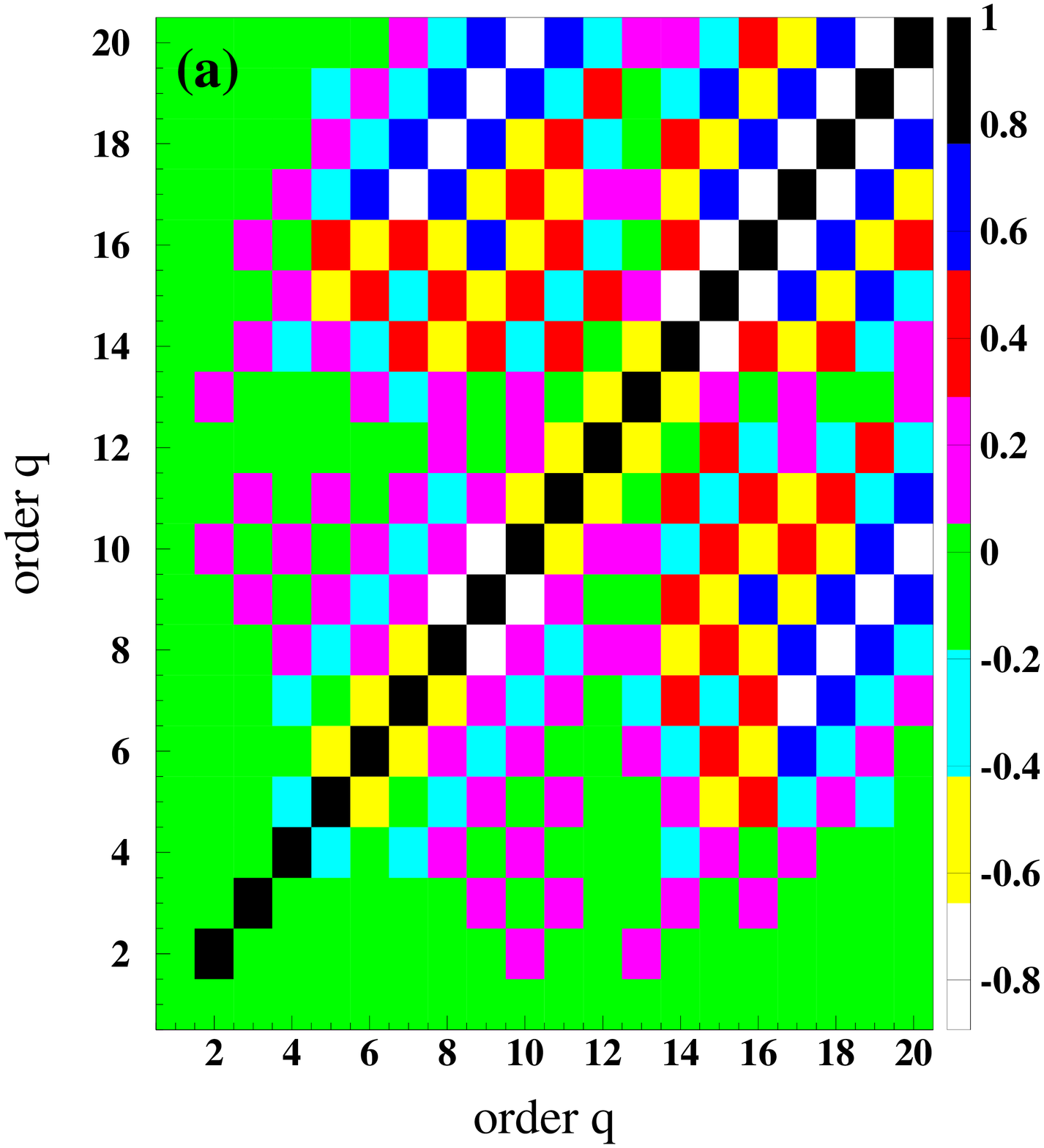}
    \includegraphics[width=8.4cm]{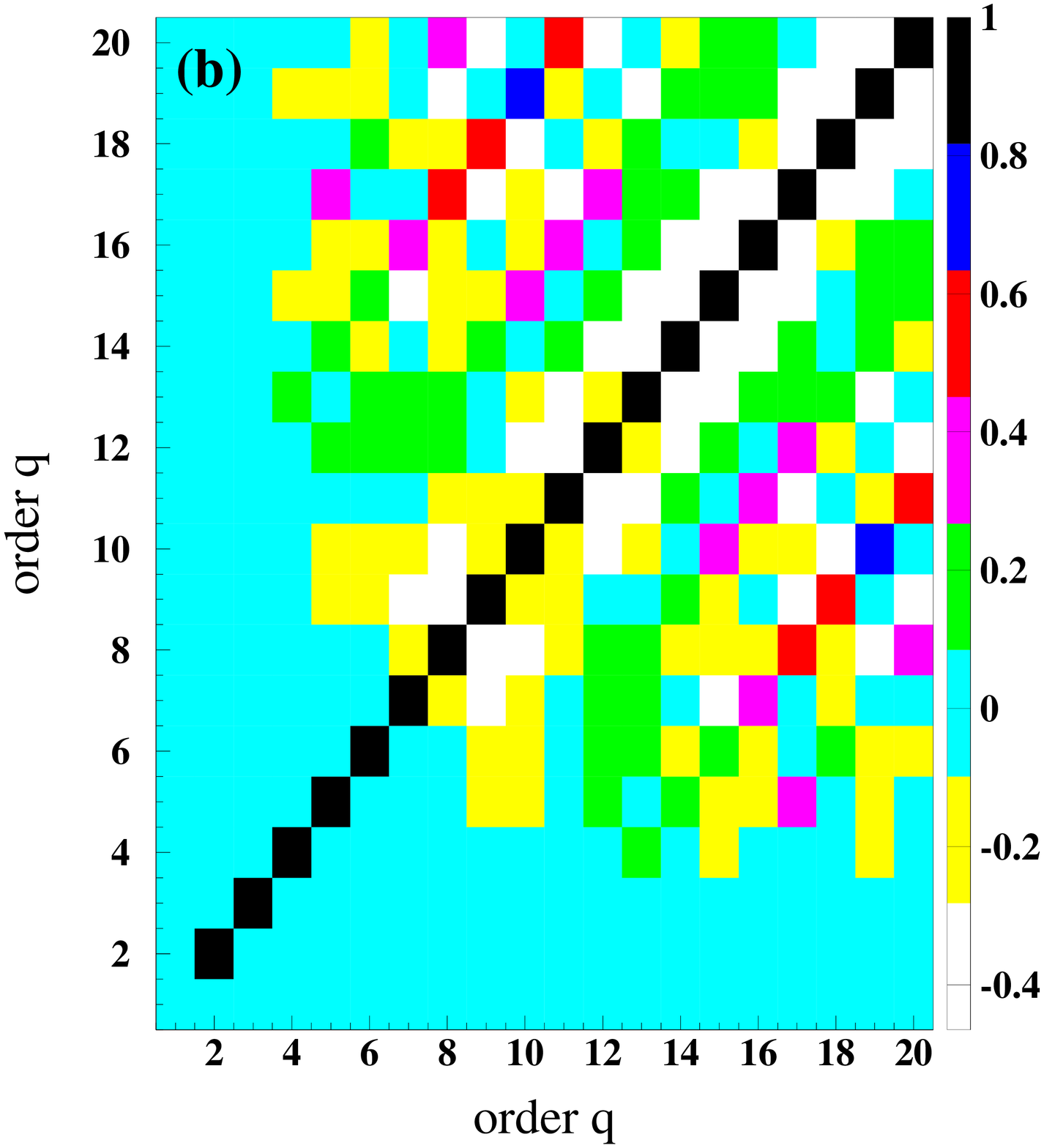}

    \includegraphics[width=8.4cm]{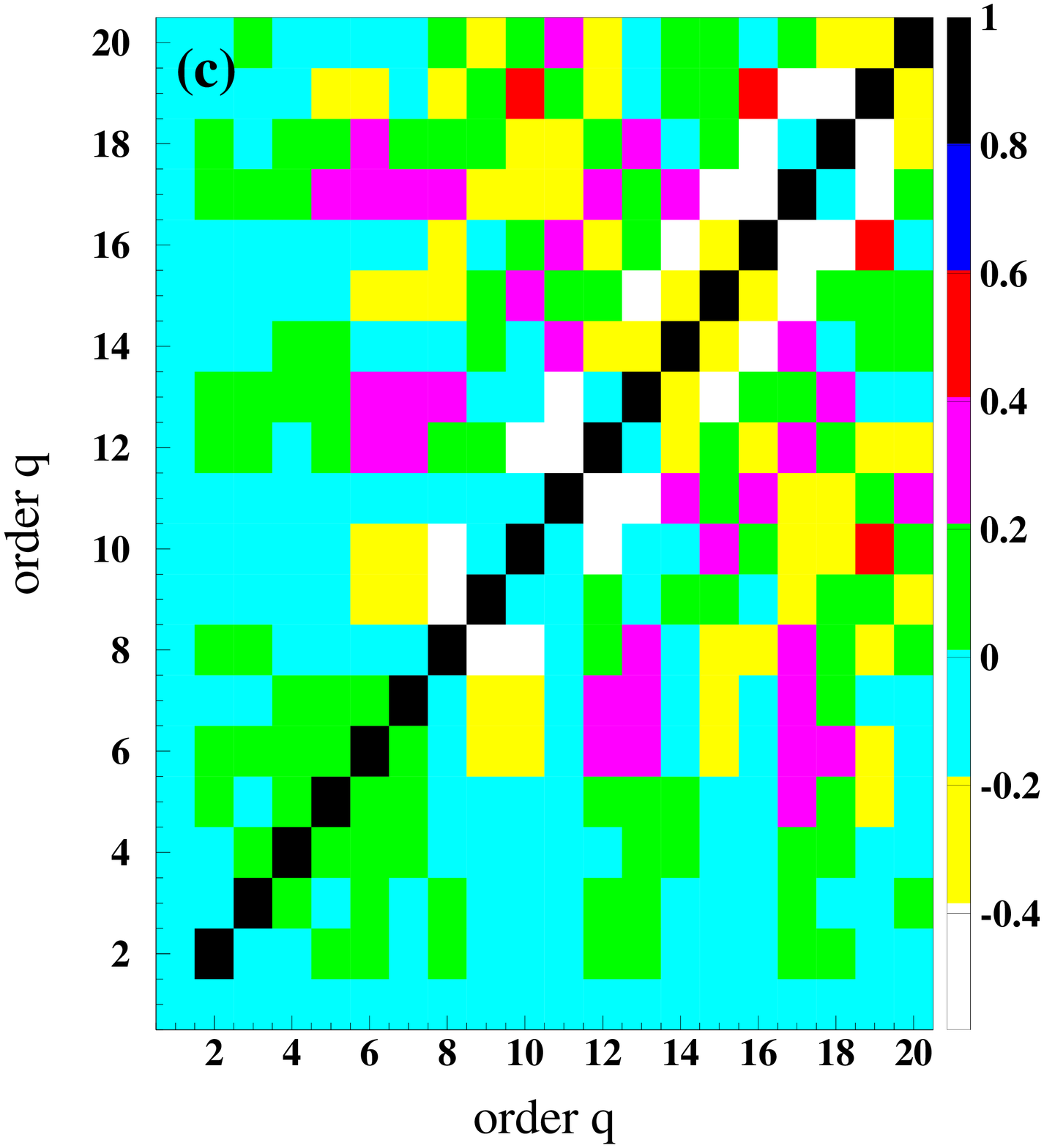}
    \includegraphics[width=8.4cm]{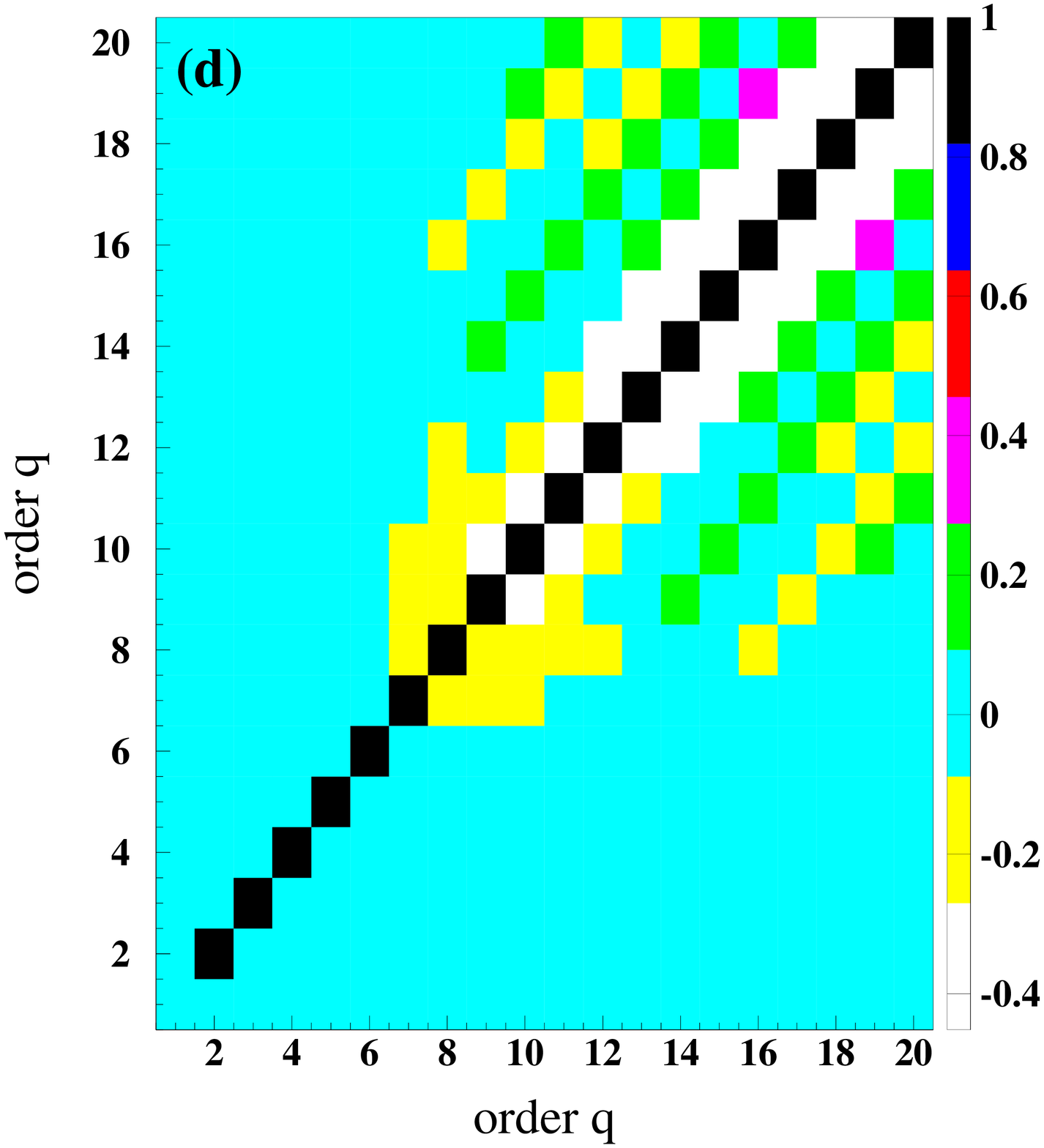}
\scaption{Correlation matrices of $H_q$ obtained from samples generated 
according to a Poisson distributions with $1\cdot 10^5$ events (a), 
$5\cdot 10^5$ events (b), $1\cdot 10^6$ events (c) 
and $2\cdot 10^9$ events (d).}
\label{fig:hqcor}  
\end{figure}

In Fig.~\ref{fig:hqcor} we can see a clear dependence of the
correlation on the statistics.
The sample which has the smallest number of events  
(Fig.~\ref{fig:hqcor}(a)) 
exhibits a large number of correlation peaks at large $q$, but most of 
the correlations are relatively weak. 
At low $q$, for example $H_3$ appears to be predominantly correlated  
to $H_{9}$, $H_{11}$, $H_{14}$ and $H_{16}$,  
but these correlations are only about $20\%$. 
By increasing $q$ the overall correlation pattern is  
increased, showing alternation of peaks of correlation and of 
anti-correlation,  
but most of these peak values stay in an acceptable 
range of less than $40\%$. 
Nevertheless, hard anti-correlation 
(about $80\%$) occurs between the $H_q$ ranges of $H_{7}-H_{10}$ 
and $H_{17}-H_{20}$. Furthermore, at large $q$ values 
large anti-correlation (about $80\%$) exists with adjacent $H_q$. 
For example, $H_{16}$ is highly correlated with $H_{15}$ and 
$H_{14}$, but it shows rather acceptable correlation with 
other $H_q$.
When the statistics of the sample is increased, 
the shape of the correlation pattern remains about the same,  
but it is weakened. Also, the size of the overall 
correlation is decreased. 
With a statistics of half a million of events (Fig.~\ref{fig:hqcor}(b)), 
the correlations do not exceed  
$40\%$, with only one exception of about $70\%$ between 
$H_{10}$ and $H_{19}$.
By increasing further the statistics (Fig.~\ref{fig:hqcor}(c)), 
the correlation pattern is further weakened and the 
size of the correlation is further decreased.

For a sample of 1 million events (Fig.~\ref{fig:hqcor}(c)),
similar to  our  data samples,  
the main correlation is the correlation between adjacent 
$H_q$ for $q$ larger than 12. 
The size of these correlations is only $40\%$. This is 
perfectly acceptable to carry out the analysis on the whole 
range of $H_q$.
For the $2\cdot 10^9$ event sample (Fig.~\ref{fig:hqcor}(d)), 
only the (anti-)correlations remain between adjacent $H_q$ 
for large $q$, which are in their maximum 
about the same size as for the 1 million sample ($40\%$). 
Other correlations have almost completely disappeared.
Therefore, we can conclude in view of the size of most our 
data samples (about 1 million), that correlations are rather small
and lie within an acceptable range for the whole range of $q$ 
on which the analysis is carried out, for 
$q$ between $2$ and $\langle n \rangle(\pm D$, 
where $\langle n \rangle$  is the \mcpm{} and $D$ the dispersion). 
The b-quark sample, which has lower statistics (about $100.000$ events) 
should be more influenced by correlations for large $q$ ($q>16$). 
Nevertheless, since this sample also has a larger \mcpm{} than the 
other samples, it should be safe to present the $H_q$ moments  
on the same range of $q$ as used for the other samples. 

\begin{figure}[htbp]
\centering
    \includegraphics[width=8.4cm]{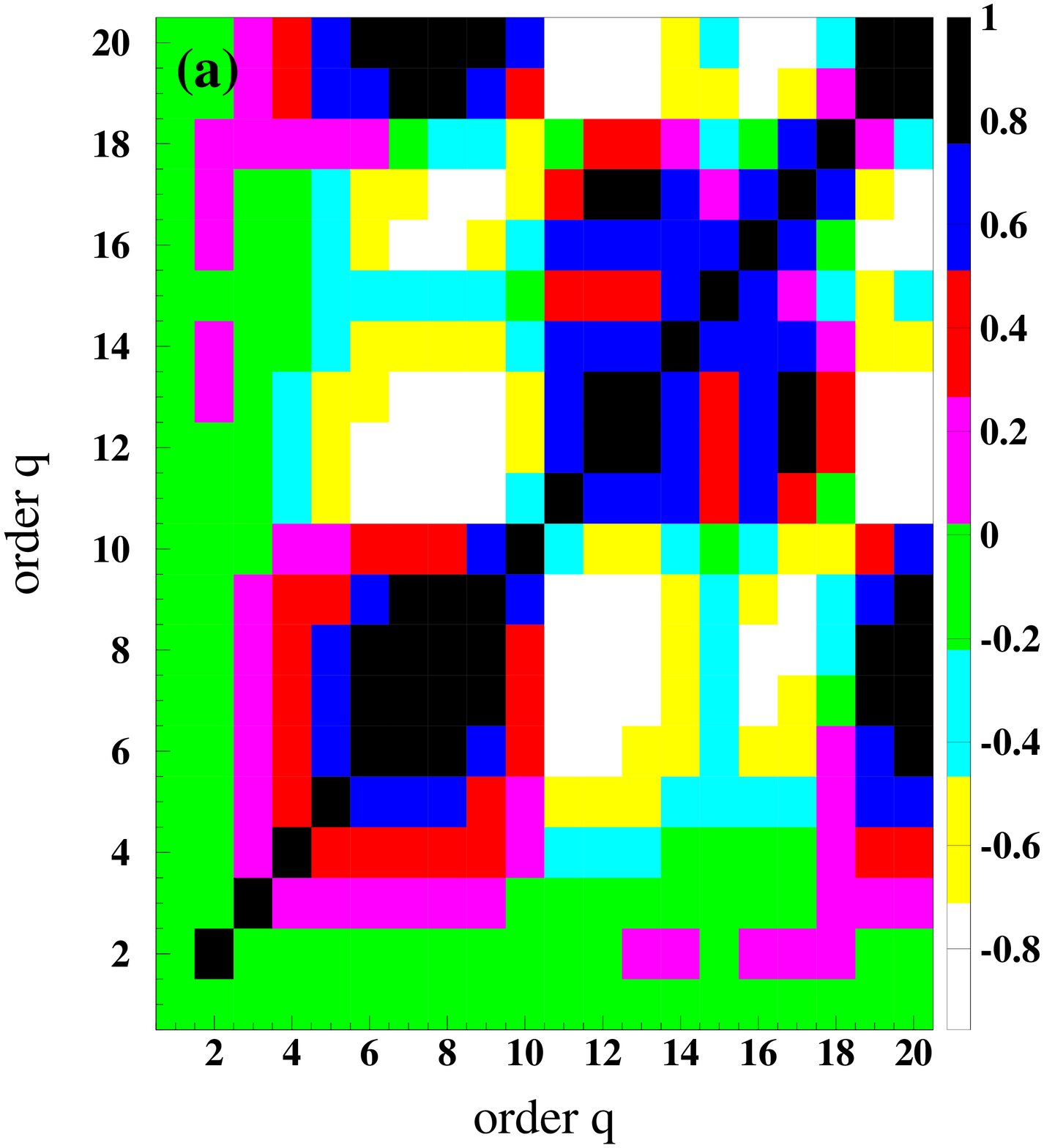}
    \includegraphics[width=8.4cm]{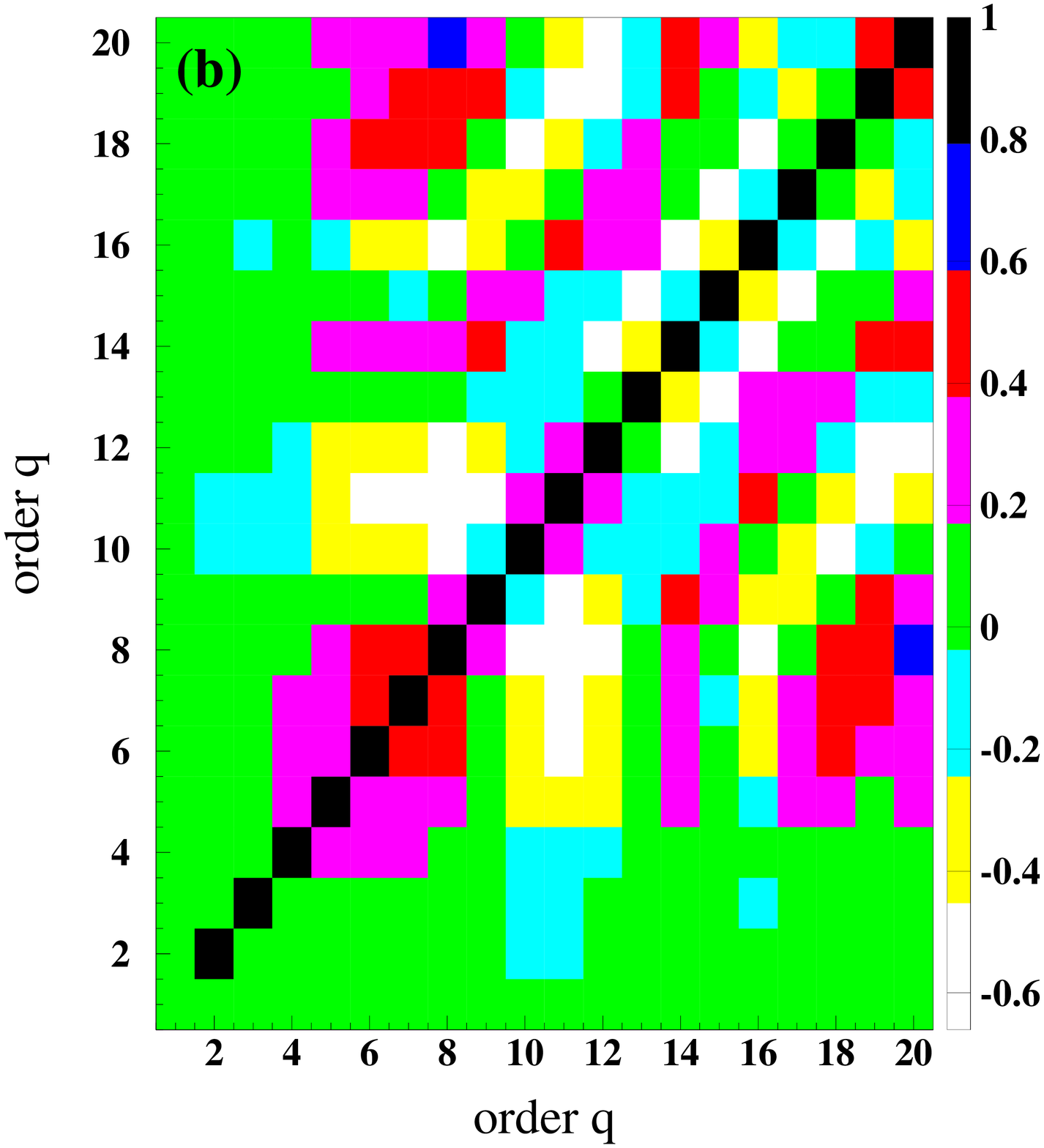}

    \includegraphics[width=8.4cm]{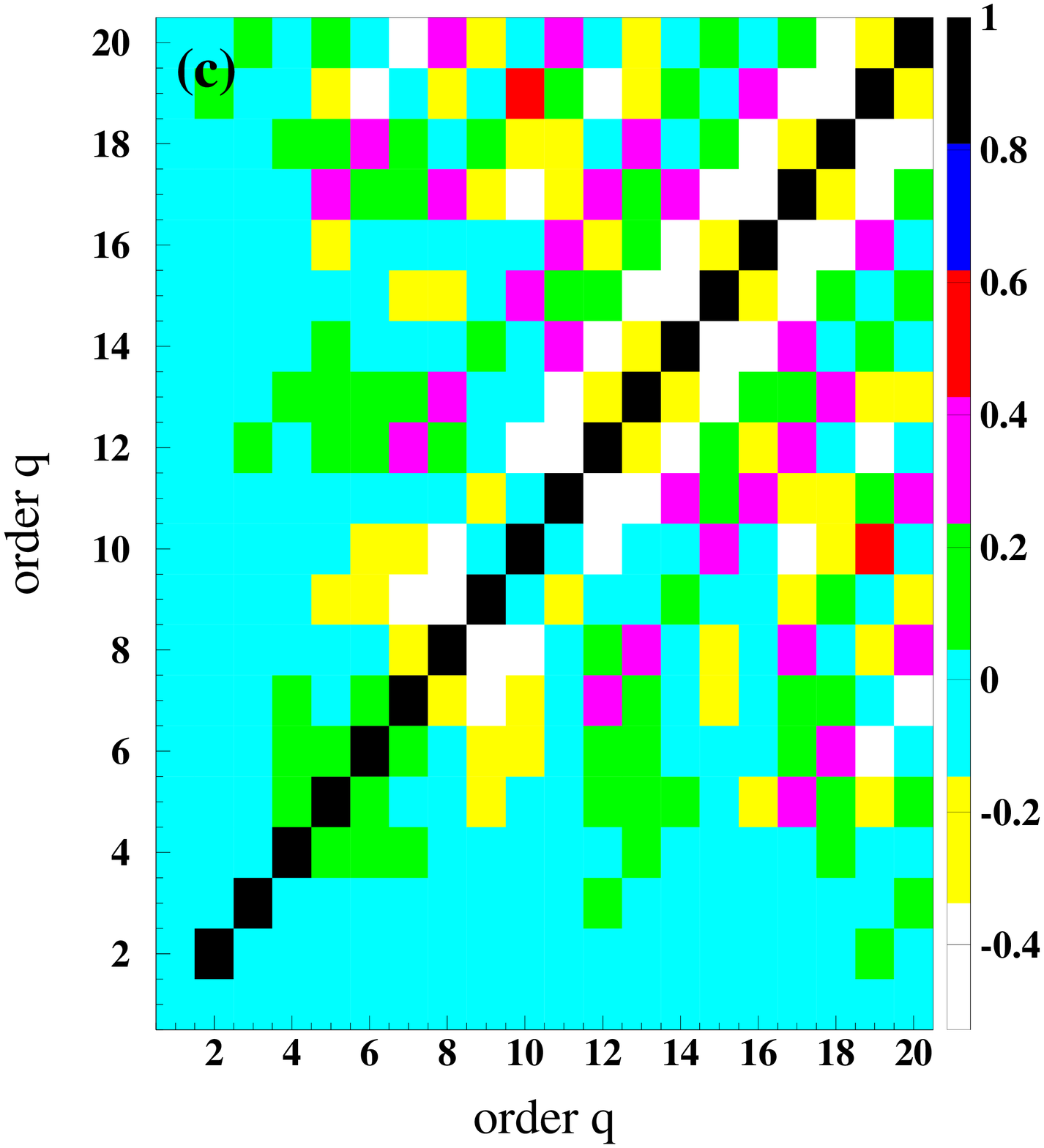}
    \includegraphics[width=8.4cm]{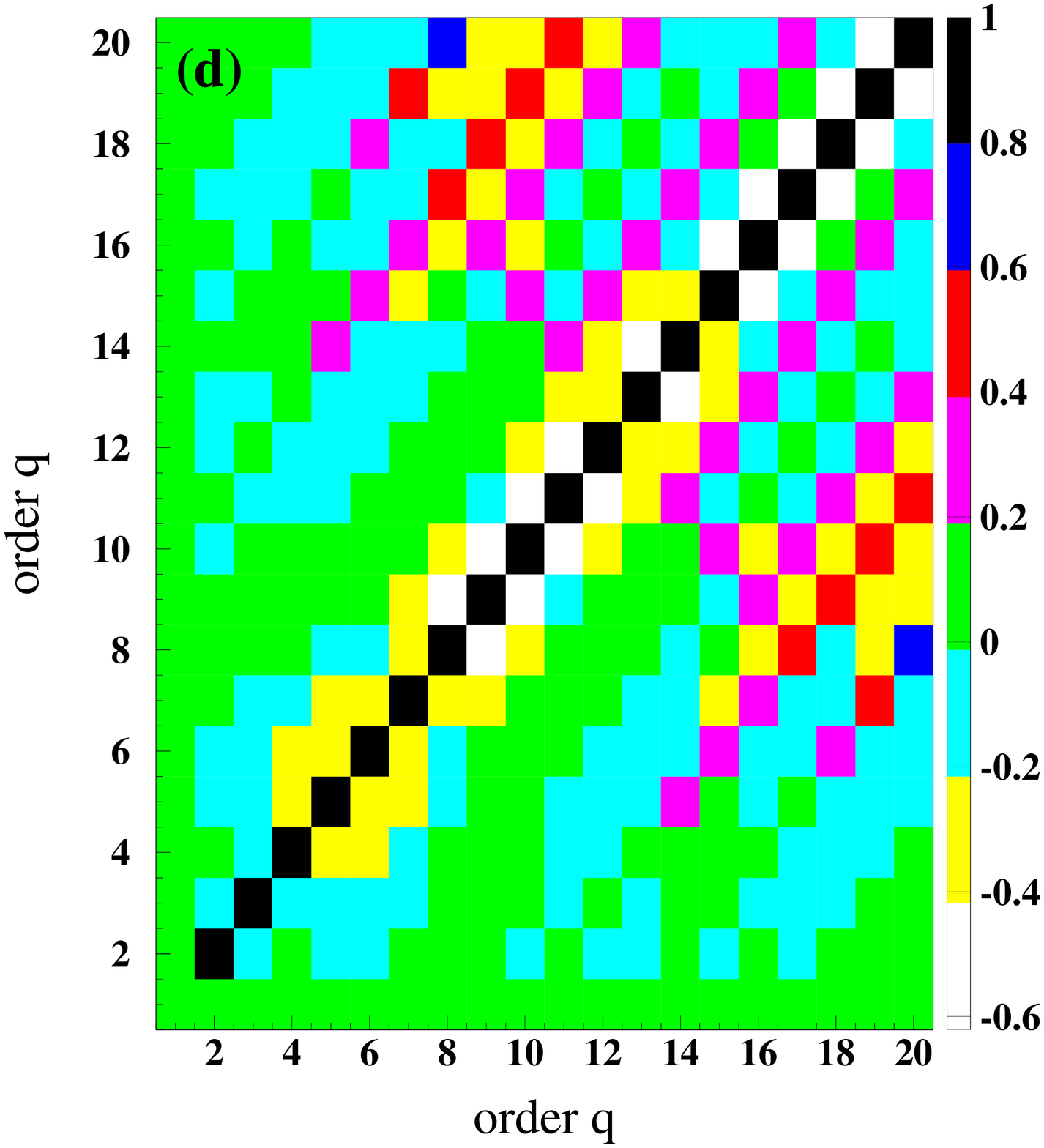}
\scaption{Effect of statistical fluctuation on the $H_q$ 
correlation matrices obtained from various 1-million event 
samples generated according to a Poisson distribution. In (a) 2 events have 
been added at large multiplicities, in (b) only one, (c) 
none and (d) the distribution of (c) is truncated.}
\label{fig:poistrunc}  
\end{figure}

An important feature we will discuss in the next sub-section 
is the importance of statistical fluctuations. Using the 
Poisson distribution of the 
million event sample, we determine 
the effect of statistical fluctuation 
by adding one or two events in the tail of the distribution.
In Fig.~\ref{fig:poistrunc}(a), is shown that by adding two events in the 
tail of the distribution, the whole correlation pattern is completely 
destroyed. It suddenly shows very strong correlations. Also with only one event
added in the tail of the distribution (Fig.~\ref{fig:poistrunc}(b)),
we have relatively strong correlations. The correlation pattern 
is minimal in the original distribution (Fig.~\ref{fig:poistrunc}(c)). 
In Fig.~\ref{fig:poistrunc}(d) 
we apply a truncation in the tail of the distribution. This causes  
a slight increase of the correlation between adjacent $H_q$, but gives 
practically no correlation elsewhere. This is a trend, somehow, similar 
to the $2\cdot 10^9$ events sample. Therefore, the truncation has the 
advantage to decrease the sensitivity of the $H_q$ to the statistics 
of the sample, which will make easier the comparison of 
samples of different statistics.

In conclusion, from these studies, we find that our $H_q$ analysis 
can be carried out up to a $q$ value about equal to the \mcpm{}.
Furthermore, in order to restore the original correlation pattern and 
subsequently the $H_q$ behavior, as well as  uniformizing the correlation
 pattern between samples of different statistics, it is important  
to get rid of the statistical fluctuation. 
The whole procedure of truncation, as well as other 
aspects of the truncation are discussed in the next section.

\newpage

\subsection[{$H_q$} and truncation]
{\boldmath{$H_q$} and truncation}

The first truncation we should talk about, as an experimentalist, 
is the truncation in the tail of the charged-particle multiplicity
distribution as a consequence of the finite statistics of our samples. 
In any experimental multiplicity distribution, there is a maximum 
multiplicity $n_\text{max}$ above which no events are seen.  
Therefore, we should rewrite Eq.~(\ref{eq:fq}), the mathematical 
definition of the factorial moment $F_q$ in a more experimental way:
\begin{equation}
\label{eq:fqexp}
F_q=\frac{\overset{n_\text{max}}{\underset{n=q}{\sum}}n(n-1)....(n-q+1)P(n)}
{\overset{n_\text{max}}{\underset{n=q}{\sum}}nP(n)},
\end{equation}
where $n_\text{max}$ is the largest multiplicity obtained for a given sample.
But we should also worry about the significance of  
high multiplicities which lack precision, as seen in Fig.~\ref{fig:mult} and 
Table~\ref{tab:pnall} for the full sample where 
$\text{K}^0_\text{s}$ and $\Lambda$ are assumed to be stable.
Multiplicities larger than 48 have relative errors larger
than $78\%$, 
which clearly means that the knowledge we have of these multiplicities 
is almost non-existent. They are statistical fluctuations,
their presence or their absence 
does not have any real physical meaning, and hence they should be removed.
Although their impact on the measurement of the mean of the charged-particle 
multiplicity distribution is negligible, this is not the case for 
the measurement of the $H_q$ moments.
Previous analyses have shown~\cite{trunc} that the $H_q$ moments are very 
sensitive to the truncation of the tail of the charged-particle 
multiplicity distribution. This sensitivity comes from the definition  
of the factorial moment, $F_q$ (Eq.~(\ref{eq:fqexp})). 
For example, if we impose 
a truncation at $n_\text{max}-1$, 
we can write for the non-normalized factorial moments, 
\begin{equation}
\label{eq:fqtrunc}
\tilde{F}_q=\overset{n_\text{max}-1}{\underset{n=q}{\sum}}\frac{n!}{(n-q)!}P(n)+
            \frac{n_\text{max}!}{(n_\text{max}-q)!}P(n_\text{max}),
\end{equation}
and we see that the importance of the last term rises with the order $q$.
Thus, we see that a statistical fluctuation  
at very high multiplicity can have an
important influence on the final result. Since the $H_q$ measurement 
is expected to   
perform a detailed analysis of the shape of the charged-particle 
multiplicity distribution, such a high multiplicity statistical fluctuation 
can destroy or mask part of the information we want to gather.
Therefore, multiplicities which can be considered as dominated by statistical 
fluctuations must be removed from the sample for the $H_q$ analysis.
These multiplicities are characterized by the instability of their
values and, hence, have very large relative error. 
In Fig.~\ref{fig:mult}, we noticed a large change in the relative error between 
$n=48$ ($29\%$) and $n=50$ ($78\%$). 
Therefore, it seems reasonable to remove multiplicities larger than 48.
These relative errors, which take into account both statistical and systematic
errors, allow to take into account both aspects of the statistical fluctuation 
at high multiplicities, one being due to, {\it{stricto sensus}}, the statistics, 
the other due to the sensitivity to the choice of selection criteria of 
high multiplicity ``boundary'' events. 

The problem of sensitivity of the $H_q$ to the truncation of the 
tail of the multiplicity distribution is not limited to multiplicities 
which are largely dominated by statistical 
fluctuations, they are only the most astonishing example of this 
effect. 
As we know from the above preliminary study of the $H_q$,  
truncating statistical fluctuation is 
fully justified by the fact that it removes correlations 
introduced by such events, 
but we also know that truncating meaningful 
information will introduce even further correlations. 
Anyhow, the truncation will always have an influence on the result.
Contrary to the study in the previous sub-section, which was 
performed on a Poisson distribution, 
 we cannot predict the theoretical behavior of the experimental 
\cpmd{}. Therefore, we cannot precisely quantify the bias of the 
$H_q$ introduced by the use of the truncated distribution.
To limit this bias, the truncation must be limited to 
the elimination of multiplicities which are sensitive to 
large statistical fluctuations.
 
\begin{figure}[htbp]
\centering
    \includegraphics[width=8.4cm]{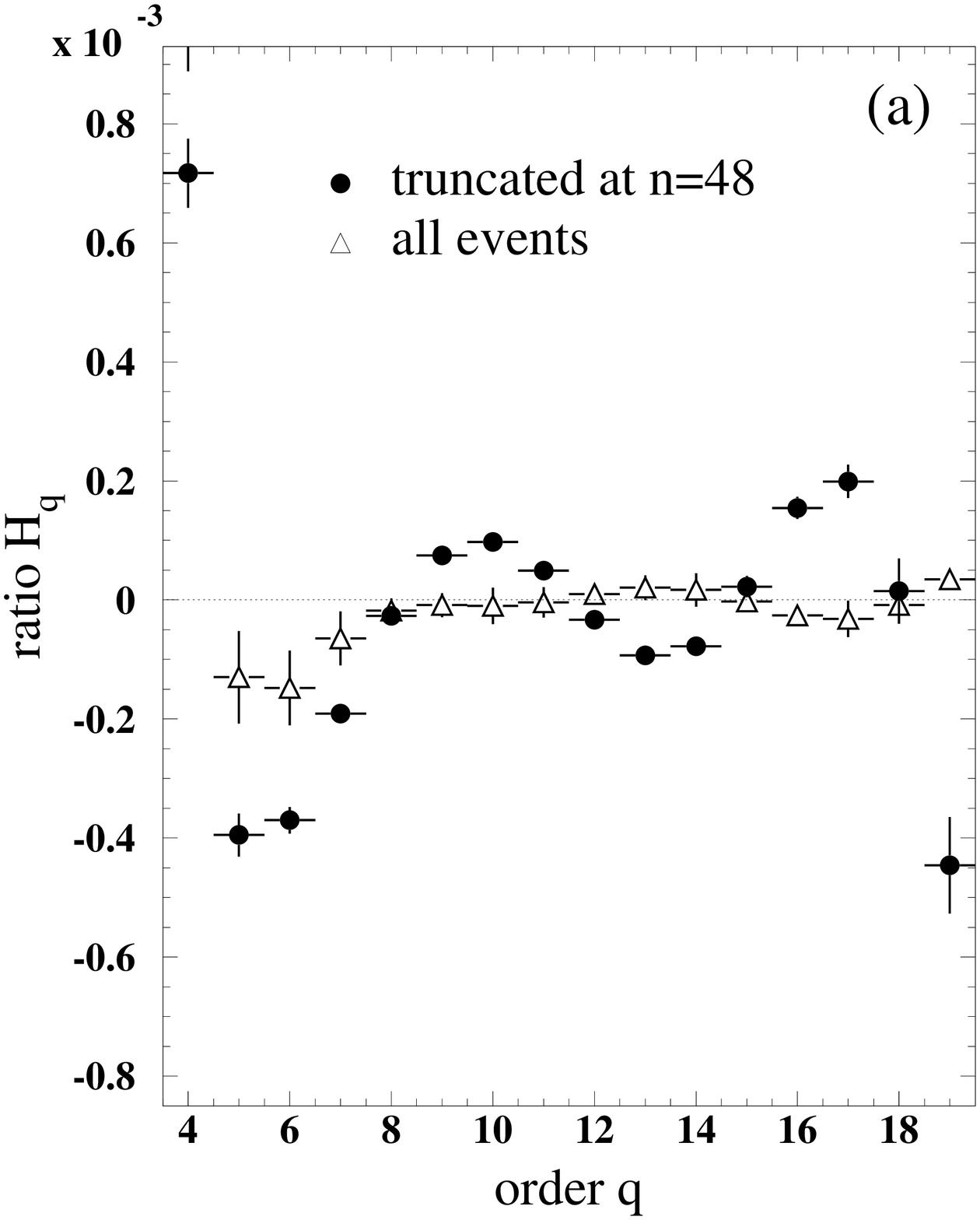}
    \includegraphics[width=8.4cm]{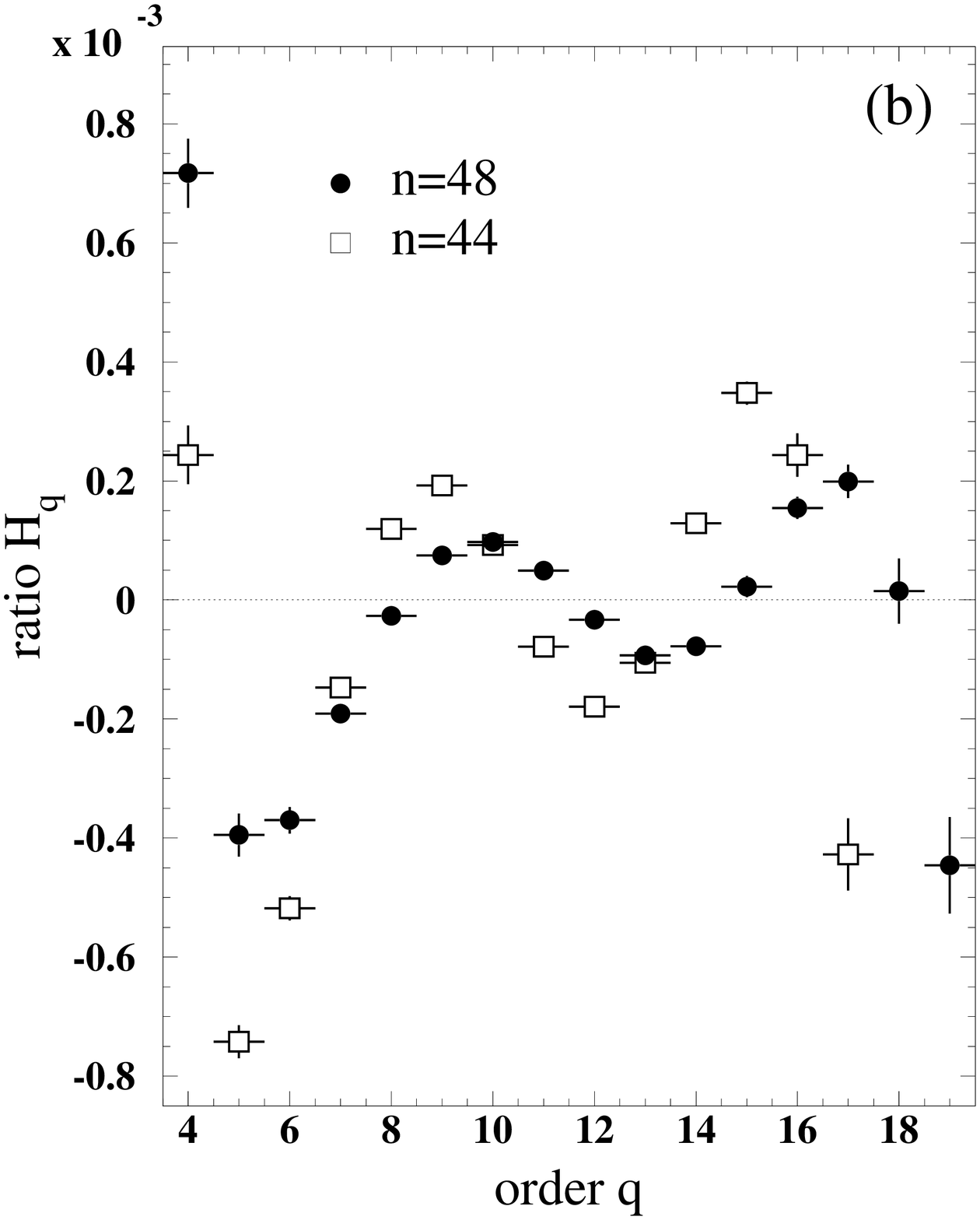}
\scaption{(a) Comparison between the reference $H_q$ moment 
(coming from the full sample truncated at $n=48$) and $H_q$ moment 
obtained from the distribution before applying the truncation, (b) 
for a strongly truncated (at $n=44$) distribution. 
Only statistical errors are shown.}
\label{fig:trunc}  
\end{figure}

The effects of the truncation are shown in Fig.~\ref{fig:trunc}, 
where the reference $H_q$ sample (in which multiplicities
influenced by large statistical fluctuations are removed) 
for the full sample without $\text{K}^0_\text{s}$ and $\Lambda$ 
decay products is compared 
to the same distribution where no truncation has been applied 
(Fig.~\ref{fig:trunc}(a)) and 
where the distribution has been exaggeratedly truncated 
(Fig.~\ref{fig:trunc}(b)), removing 
multiplicities known with a rather good accuracy (and therefore 
assumed to be statistically significant). At low $q$, 
we can also note an improvement of the statistical error, obtained 
by removing multiplicities having large statistical error (and hence
being influenced by statistical fluctuation).

The effect of the truncation being part 
of our measurement, we have to make sure that 
all distributions  we want to compare are affected 
by the truncation in the same way.
Therefore, it is not possible to express the truncation as a fixed 
cut on the multiplicity distribution. It is very unlikely that 
the sensitivity to statistical fluctuation of the multiplicities larger than 48 
is the same for the full sample without and with $\text{K}^0_\text{s}$ and 
$\Lambda$ decay products. The second distribution contains the 
same events, for which the \kl{} have been allowed to decay.  
These events are found to have on average two more 
charged particles than in the first place. In other words, the 
multiplicity distribution can be described, in first approximation, 
as shifted by about 2 particles. So are the statistical fluctuations.
Therefore, in order to remove these events in both distributions, 
we remove the tail of the second \cpmd{} which 
corresponds to the fraction of events removed by a 
truncation at 48 in the first. 

Because of rather long lifetimes of \kl{'s}, their 
decays take place at the end of the chain of processes
leading to the final state. Therefore, the $q$-particle 
correlations in which their decay products (not the 
\kl{} themselves, which are already taken into account) 
are involved are certainly limited to their closest neighbors 
in momentum space. 
Therefore, it is reasonable to expect correlations only between 
a small number of particles. In terms of $q$-particle correlation 
functions, this means that the $H_q$ moments obtained from the two 
distributions, with and without \kl{} decay products, 
should have rather similar values, except at low $q$ 
(small changes are nevertheless expected as a consequence of the 
correlation between the $H_q$ moments, which were found to be 
rather small in the previous section).

\begin{figure}[htbp]
\centering
    \includegraphics[width=8.cm]{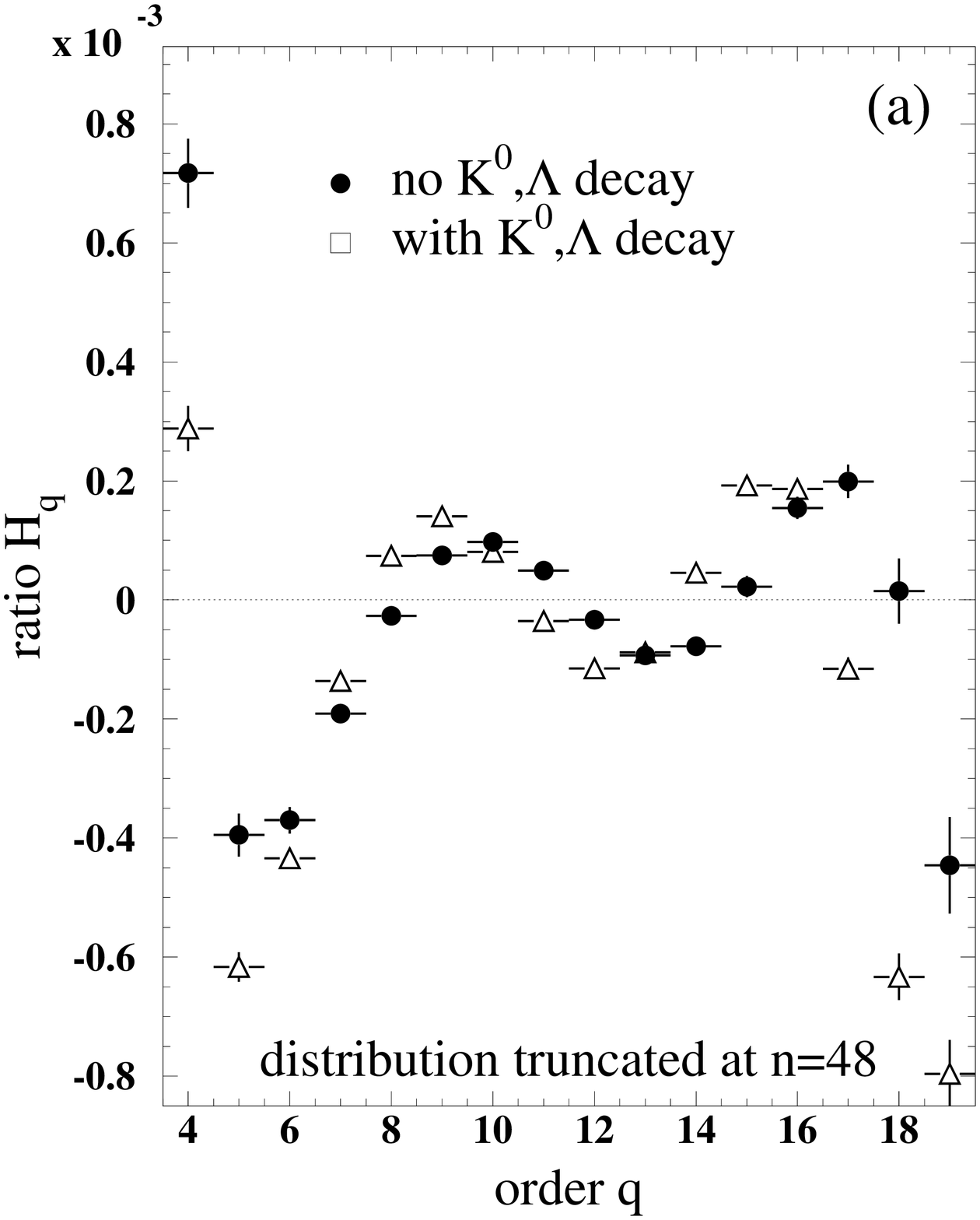}
    \includegraphics[width=8.cm]{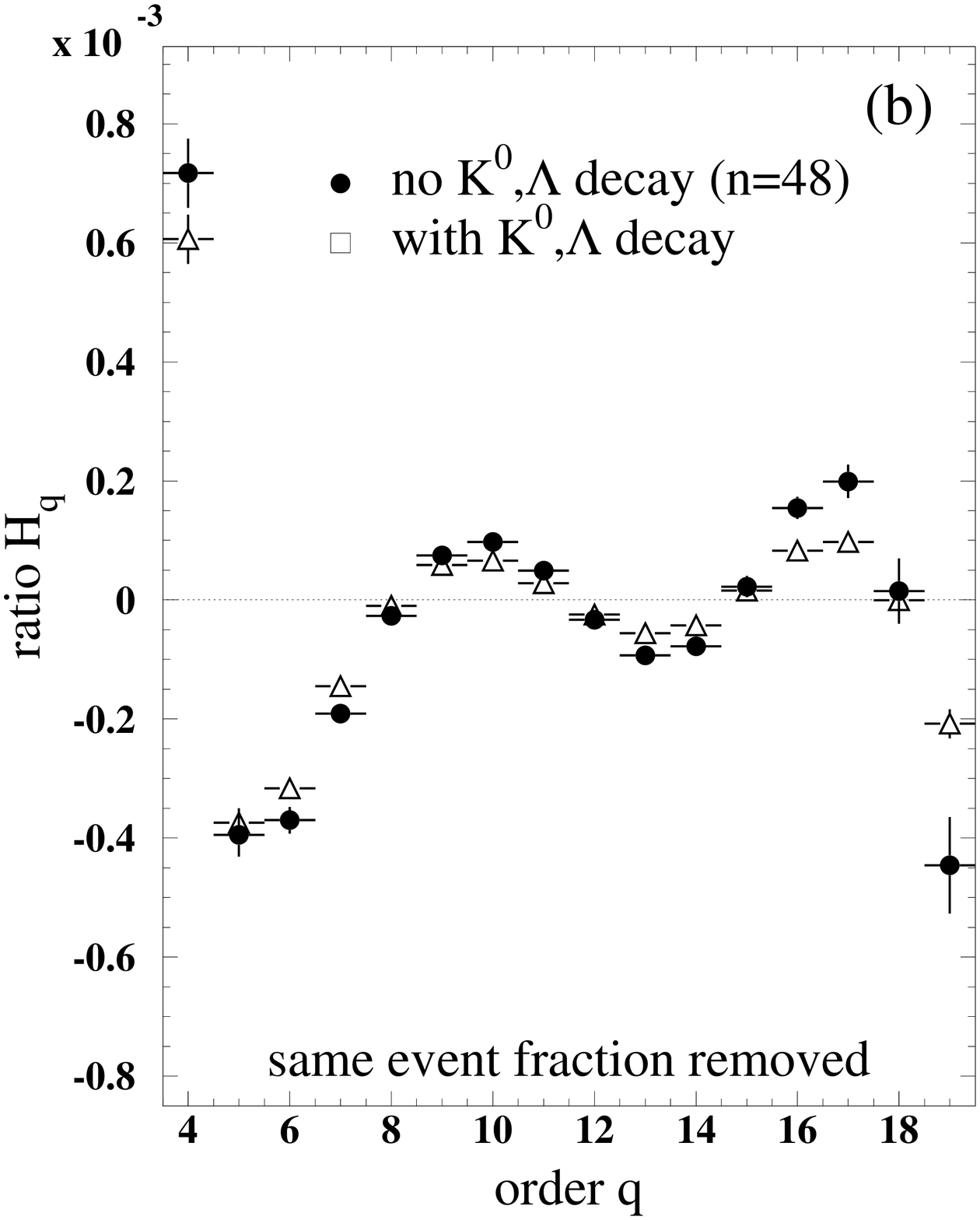}
\vspace{-0.2cm}
\scaption{Comparison between the full samples without and with 
$\text{K}^0_\text{s}$ and $\Lambda$ decay products, (a) where the 
truncation has removed multiplicities larger than 48 
and (b) where the truncation has removed the fraction of events 
defined by a truncation on multiplicities larger than 48 on 
the full sample without $\text{K}^0_\text{s}$ and $\Lambda$ 
decay products. Only statistical errors are shown.}
\label{fig:tr_fr1}  
%
%
%
  \begin{center}
    \includegraphics[width=8.cm]{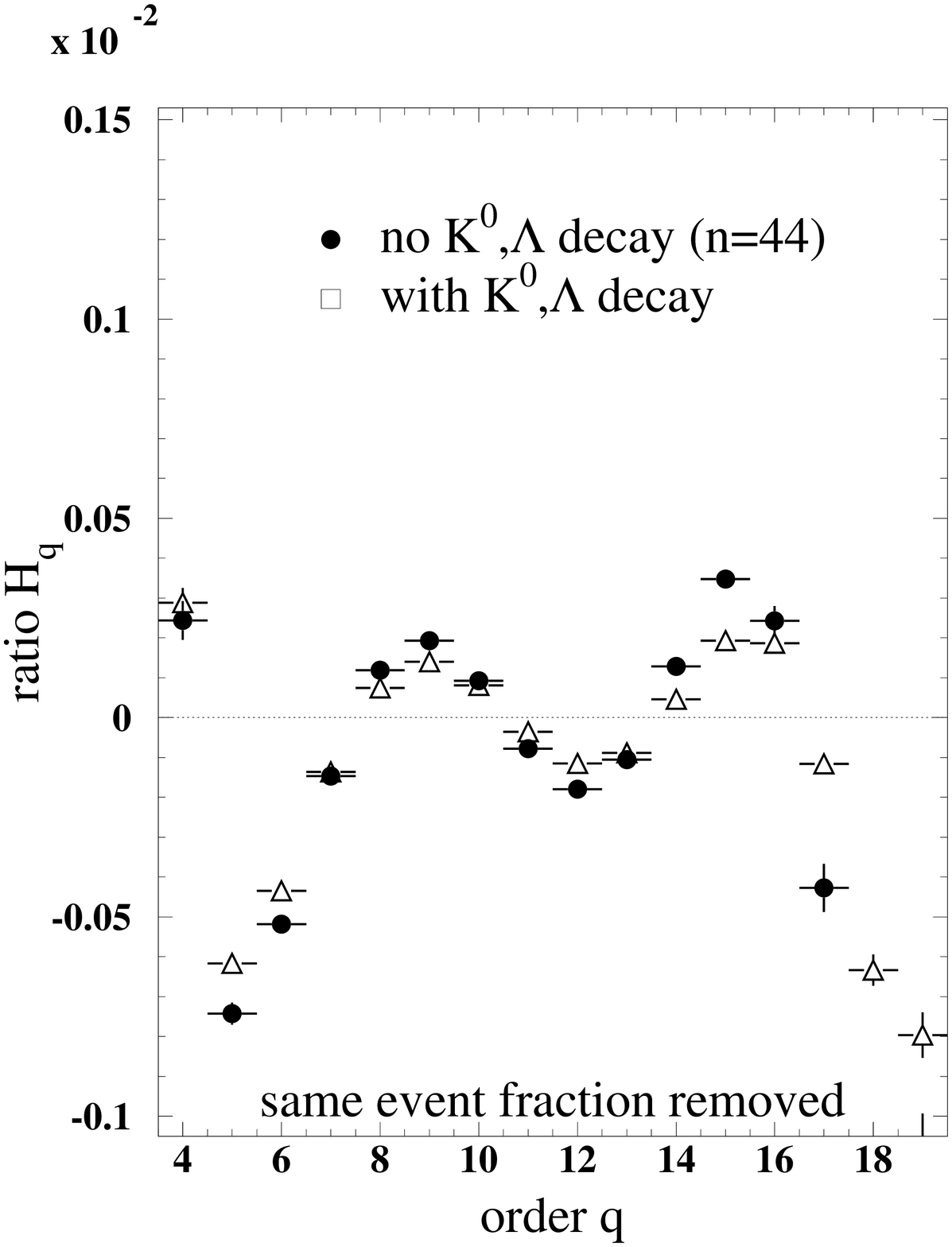}
  \end{center}
\vspace{-0.7cm}
\scaption{Comparison between the full samples without and with 
$\text{K}^0_\text{s}$ and $\Lambda$ decay products 
where the truncation has removed the fraction of events 
defined by a truncation on multiplicities larger than 44 on 
the full sample without $\text{K}^0_\text{s}$ and $\Lambda$ 
decay products, corresponding to the truncation of multiplicities
larger than 48 for the full sample including decay products of 
$\text{K}^0_\text{s}$ and $\Lambda$. Only statistical errors 
are shown.}
\label{fig:tr_fr2}  
\end{figure}

This is illustrated in Figs.~\ref{fig:tr_fr1}
and~\ref{fig:tr_fr2}, where we compare the full sample with 
and without $\text{K}^0_\text{s}$ and $\Lambda$ decay products. 
When we apply a truncation at $n=48$ for both distributions, 
the two distributions disagree. However, if the fraction of events  
corresponding to a truncation on multiplicities larger than 48 
for the full sample where $\text{K}^0_\text{s}$ and $\Lambda$ 
are stable is removed from the full sample including 
$\text{K}^0_\text{s}$ and $\Lambda$ (which corresponds to 
a truncation on multiplicities larger than 52) we have 
good agreement between the two samples. 

Now, if we apply the inverse, \ie{} we fix the truncation on the 
multiplicity at $n=48$ for the full sample where the 
$\text{K}^0_\text{s}$ and $\Lambda$ decay products are taken 
into account and remove the corresponding fraction in the 
other sample (which corresponds to a truncation on multiplicities
larger than 44) we also have a good agreement between the two 
distributions. This shows us (apart from the fact that the 
weakly decaying short life-time particles don't have any 
influence on the shape of the charged-particle multiplicity 
distribution) that we have to define the truncation in terms 
of an equal fraction of events to be removed in the tail of the 
multiplicity distribution.

Only $0.005\%$ of the events are in fact removed
by this truncation.

We have seen in this sub-section that the amplitude of the 
oscillation is increased by the truncation, thereby 
amplifying an already existing behavior. This is due to 
an increase in the size of the correlations between $H_q$, which 
then amplifies the original $H_q$. 
On the other hand, statistical fluctuation destroying 
the original correlation pattern will partly mask the 
original $H_q$ behavior as seen in the previous sub-section.
This is illustrated in Fig.~\ref{fig:trunc}(a), where 
the original data sample, having large statistical fluctuation 
in its tail, has much smaller oscillations than the truncated one.
Since the statistical fluctuation masks the oscillatory 
behavior, the fact that we have small oscillation strongly suggests that 
the oscillatory behavior is not due to the truncation.
Nevertheless, too strong truncation may increase the oscillation size. 
This is illustrated in Fig.~\ref{fig:trunc} (b), 
where we have applied a truncation removing well measured 
and hence statistically significant multiplicities and the size 
of the oscillation is increased.

\subsection{Statistical errors}

The statistical errors on the $H_q$ are obtained by two different 
methods:
The first one, an analytical method, which we prefer, is 
error propagation, making use of the 
covariance matrix of the charged-particle multiplicity distribution. 
The dependence of our estimate of the 
covariance matrix on the statistics, together with the form of 
the derivative of the $H_q$
moments, which can emphasize even small changes in the covariance matrix, 
causes this method to give reliable results only for high-statistics samples. 
At low statistics, the approximations and 
assumptions  used to estimate the covariance matrix 
are not valid anymore.
Since it would be rather difficult to obtain a better estimate 
of the covariance matrix, a Monte Carlo based method is found to be 
a rather advantageous alternative of treating the error,
all terms being naturally taken into account by this method.
Therefore, the analytical method is used for the full and 
the light-quark samples where large statistics makes it suitable, 
but the second method will be used for the b-quark sample, 
which has smaller statistics.

\subsubsection{Analytical method}

In this method, the statistical errors are obtained by 
error propagation, which makes use of the covariance matrix
of the \cpmd{}, 
$\text{CoV}(P(i),P(j))$. 
The variance of the $H_q$, $\text{Var}(H_q)$, is given by
\begin{equation}
\label{eq:varhq}
\text{Var}(H_q)={\underset{i,j}{\sum}}\frac{\partial H_q}{\partial P(i)}
\frac{\partial H_q}{\partial P(j)}\text{CoV}(P(i),P(j)),
\end{equation}
where $\frac{\partial H_q}{\partial P(i)}$ is a partial derivative  
of $H_q$ as a function of the multiplicity $P(i)$.
These terms can be easily obtained by writing $H_q$ as a function 
of the non-normalized factorial moment $\tilde{F_q}$ 
corresponding to:
\begin{equation}
\label{eq:ufq}
\tilde{F_q}=\overset{n_\text{max}}{\underset{n=q}{\sum}}n(n-1)....(n-q+1)P(n)
=F_q\tilde{F_1}^q.
\end{equation}
It must be noted that $\tilde{F_1}$ is nothing but the mean of the multiplicity 
distribution.
Dividing the cumulant factorial moment, $K_q$, (Eq.~(\ref{eq:kq})) by $F_q$,
we can express $H_q$ as a sum of factorial moments:
\begin{equation}
\label{eq:hqfq}
H_q=1-\overset{q-1}{\underset{m=1}{\sum}}\frac{(q-1)!}{m!(q-m-1)!}H_{q-m}
\frac{F_{q-m}F_{m}}{F_q}\text{ with }H_1=F_1=1.
\end{equation}
We see also that the normalization factor of the factorial moments, 
corresponding to the mean of the multiplicity distribution cancels, leading to 
\begin{equation}
\label{eq:hqfqtilde}
H_q=1-\overset{q-1}{\underset{m=1}{\sum}}\frac{(q-1)!}{m!(q-m-1)!}H_{q-m}
\frac{\tilde{F}_{q-m}\tilde{F}_{m}}{\tilde{F}_q}.
\end{equation} 
Therefore, the derivative of $H_q$ is given by 
\begin{equation}
\label{eq:dhq}
\text{d}H_q=-\overset{q-1}{\underset{m=1}{\sum}}
\frac{(q-1)!}{m!(q-m-1)!}H_{q-m}\frac{\tilde{F}_{q-m}\tilde{F}_{m}}{\tilde{F}_q}
(\frac{\text{d}H_{q-m}}{H_{q-m}}+\frac{\text{d}\tilde{F}_{q-m}}{\tilde{F}_{q-m}}+
\frac{\text{d}\tilde{F}_{m}}{\tilde{F}_{m}}-
\frac{\text{d}\tilde{F}_{q}}{\tilde{F}_{q}}),
\end{equation}
\begin{equation}
\label{eq:dfqtilde}
\text{where }
\text{d}\tilde{F}_{q}=
\begin{cases}
n(n-1)....(n-q+1)\text{d}P(n) &n\geq q\\
0                             &n<q.
\end{cases}
\end{equation}
The first non-zero partial derivative, $\frac{\partial H_2}{\partial P(n)}$, 
obtained from Eqs.~(\ref{eq:dhq}) and~(\ref{eq:dfqtilde}) is given by:
\begin{equation}
\label{eq:dh2}
\frac{\partial H_2}{\partial P(n)}=(H_2-1)
\begin{large}
\begin{cases}
\frac{2n}{\tilde{F}_1}-\frac{n(n-1)}{\tilde{F}_2}  &n\geq 2\\
\frac{2n}{\tilde{F}_1}                             &n=1.
\end{cases}
\end{large}
\end{equation}
The partial derivatives of the higher $H_q$ moments are found iteratively, 
starting with the $H_2$ partial derivative, $\frac{\partial H_2}{\partial n}$ 
in Eq.~(\ref{eq:dhq}).
This way of calculating statistical errors works well, unless the sample size 
becomes too small. Samples which have less than 500.000 events cannot 
be processed by this method. At that point, the combination 
of the deterioration in the precision of the covariance matrix,  
together with the higher-order terms of the $H_q$ derivative,  
cause the resulting statistical error to significantly deviate 
from its real value and become larger. 
In order to obtain a reliable estimation, we need to take 
into account all orders of correlation. It would be very difficult
to do this analytically. Therefore, we use a Monte Carlo 
based method.

\subsubsection{Monte Carlo based error calculation}

The principle of this method is very simple. 
A large number of  
multiplicity distributions are generated from the experimental one by  
allowing random variation of the multiplicities which compose it.
Each of these generated multiplicity
distributions gives an $H_q$ value. The error is then extracted  
from the distribution of the generated $H_q$.

In order to generate a multiplicity distribution as close as possible 
to the experimental result, the generation process has to follow the 
same treatment and reconstruction process as used for the experimental data.
Since events are produced independently of each other, the random variation
of the multiplicity distribution is obtained by imposing Poisson fluctuation 
of the number of events produced with a given multiplicity.

The generation process starts by imposing a Poisson fluctuation on the number 
of events for each detected multiplicity $N^\text{raw}(i)$ of the data. 
This allows us to calculate a completely new (but statistically 
consistent with the original distribution) probability distribution, 
$P^\text{raw}_1(i)$.

Next, making use of the detector response matrix, 
$\mathcal{M}$ , we impose a Poisson fluctuation on the number of events 
$N(n_\text{det},n_\text{prod})$ with $n_\text{det}$ 
detected particles and $n_\text{prod}$ produced tracks.
We then obtain not only a completely new detector response 
matrix $\mathcal{M}_1$ but also new Monte Carlo produced and detected 
 multiplicity distributions, $P_1(n_\text{prod})$ and 
$P_1(n_\text{det})$.

\begin{figure}[htbp]
\centering
    \includegraphics[width=8.4cm]{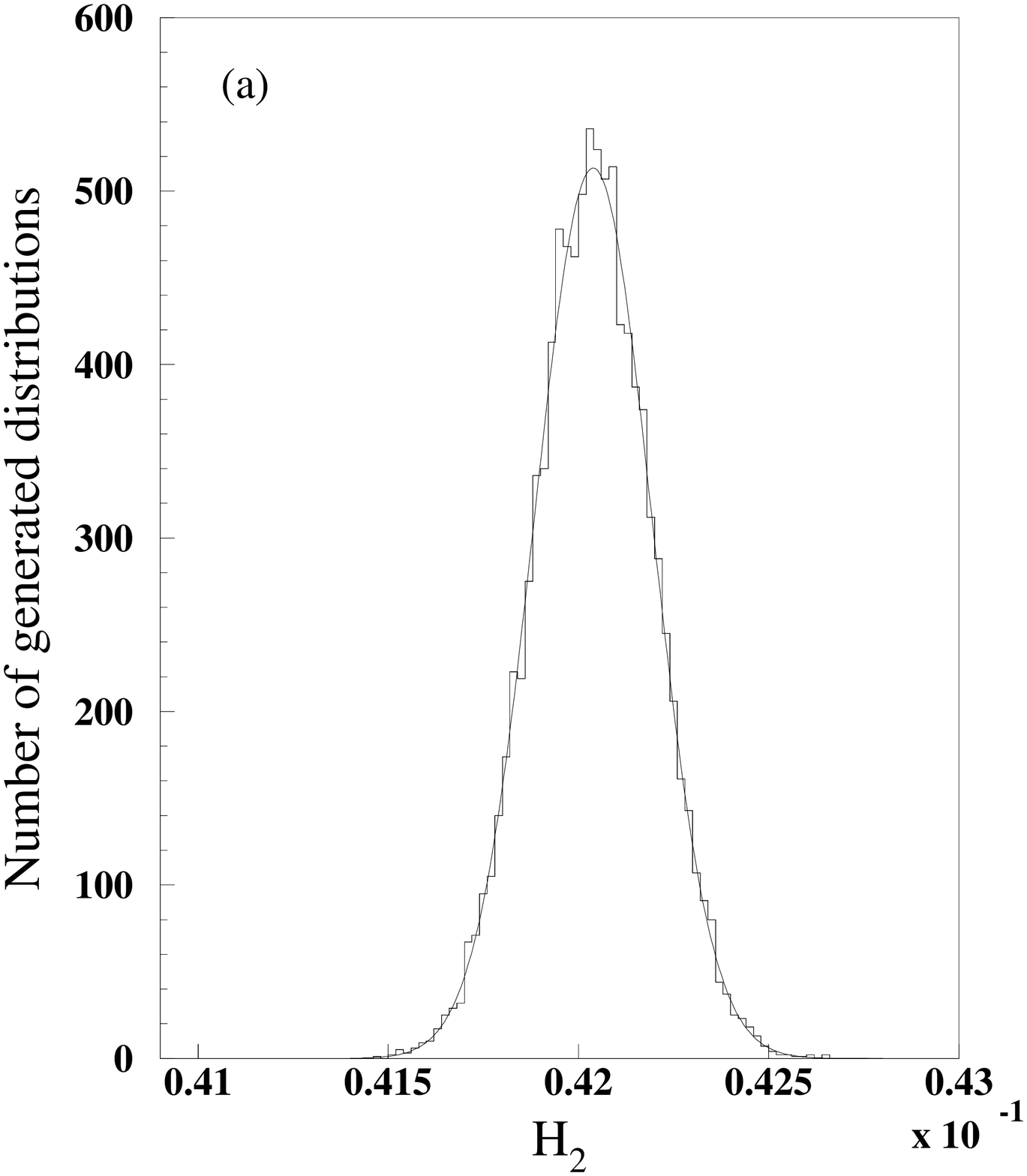}
    \includegraphics[width=8.4cm]{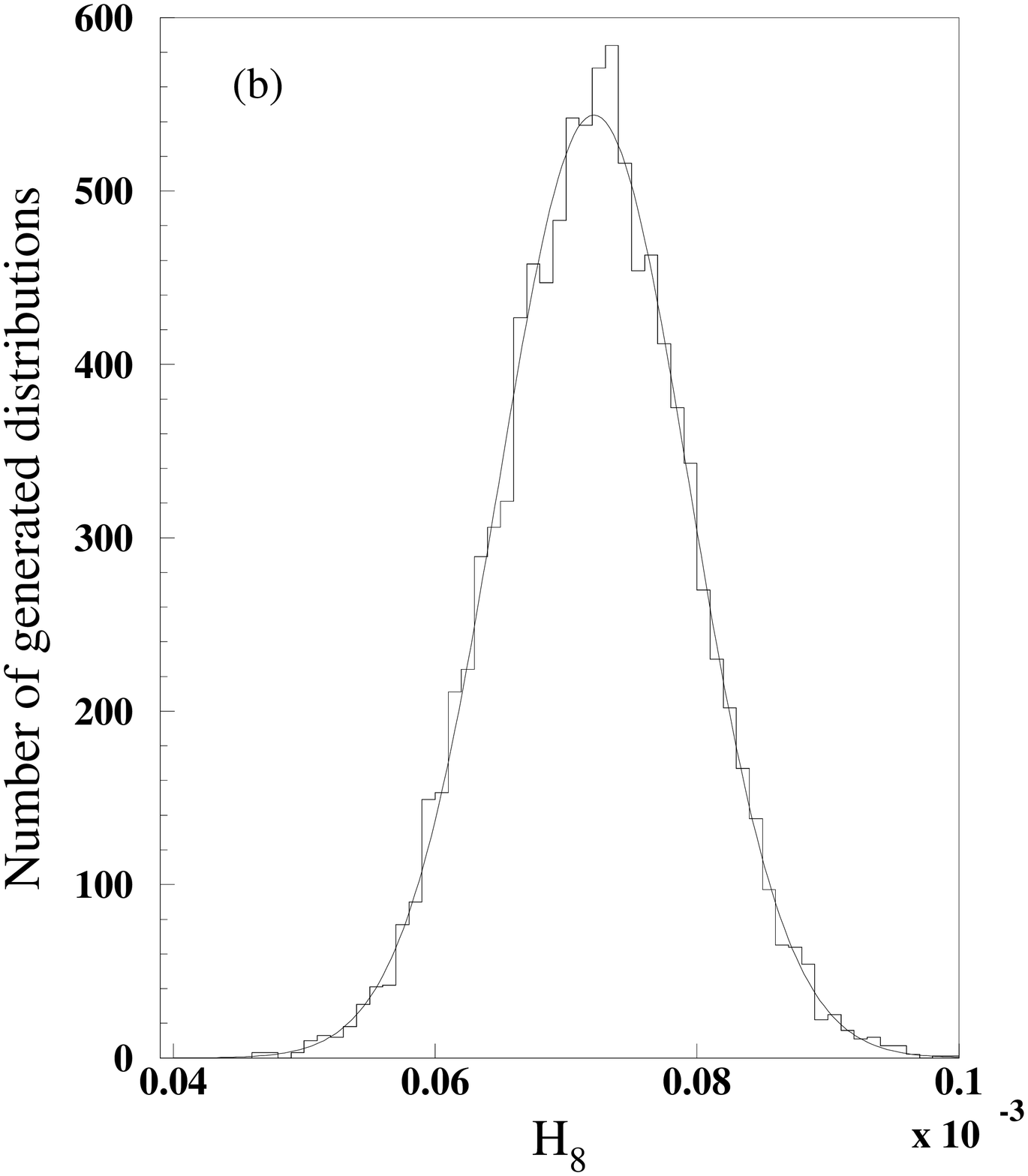}
\scaption{Distributions of the $H_2$, (a), and $H_8$, (b), obtained after 
the generation of 10.000 distributions.}
\label{fig:errdist}  
\end{figure}

We then perform a reconstruction of the new data distribution, 
$P^\text{raw}_1(i)$, by the Bayesian unfolding method making use
of the newly obtained probability matrix $\mathcal{M}_1$. 
A reconstructed charged-particle multiplicity distribution 
of the data, $P^\text{data}_1(n_\text{prod})$, is obtained.
This distribution has, of course, to be corrected, 
like the normal data sample, for event selection, initial-state  
radiation, $\text{K}^0_\text{s}$ and $\Lambda$ decay products, 
light- or b-quark purities. Since all these corrective factors
have a statistical influence on the final result, Poisson 
fluctuations of the number of events of each multiplicity are 
also imposed on all the multiplicity distributions which 
compose them. We finally get a completely new fully reconstructed 
and corrected multiplicity distribution, 
$P^\text{data}_1(n_\text{prod})$, from which a new set of $H^{(1)}_q$ 
moments is calculated. 
The whole operation is then repeated 1000 times, in order to 
obtain $H^{(i)}_q$ distributions.
After having checked that the distributions of $H^{(i)}_q$ are 
close to Gaussian (see Fig.~\ref{fig:errdist} as examples) centered 
around the original $H_q$ measurement, the statistical
error on the $H_q$ is taken to be the half-width of the distribution.
\begin{figure}[htbp]
\centering
    \includegraphics[width=8.4cm]{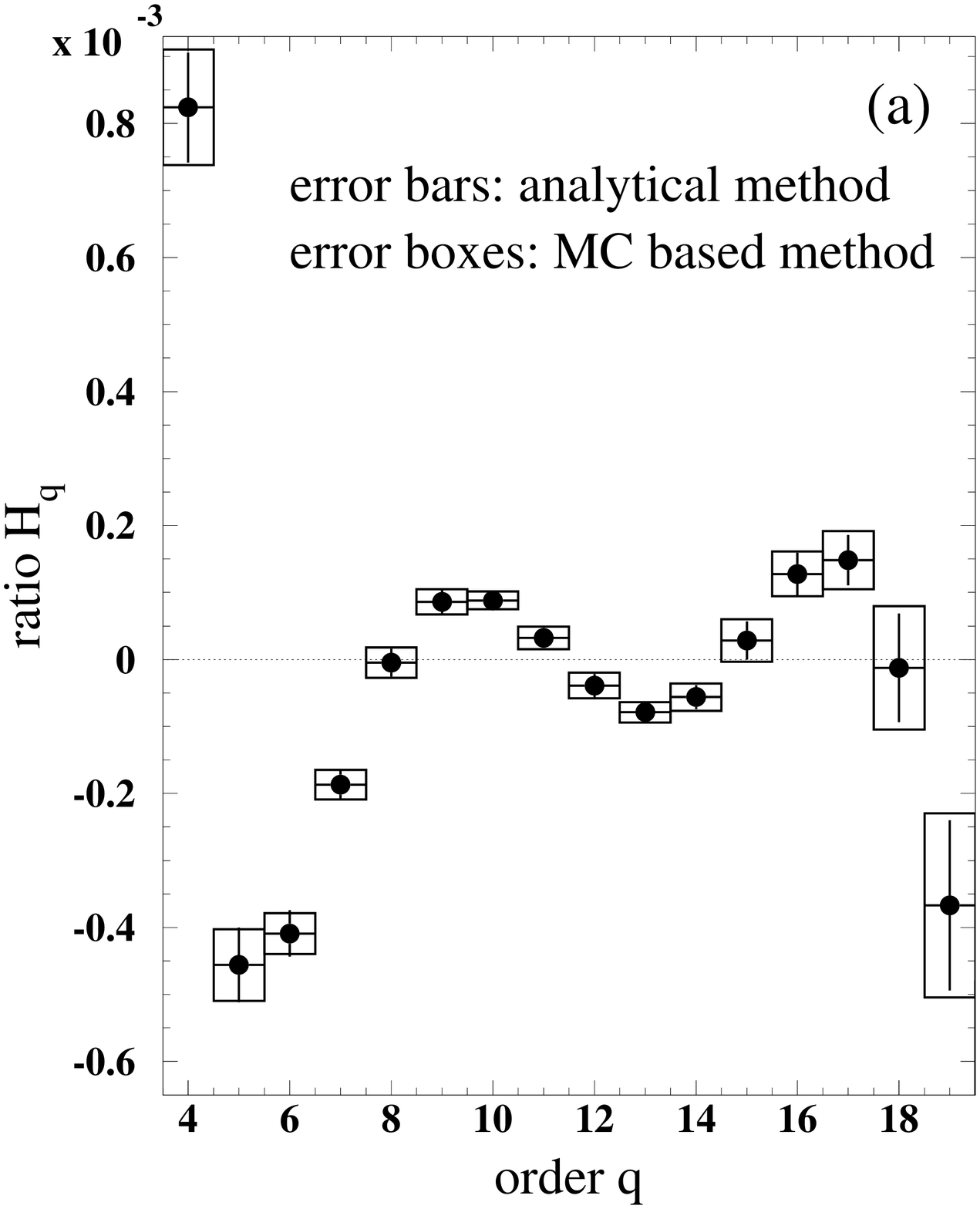}
    \includegraphics[width=8.4cm]{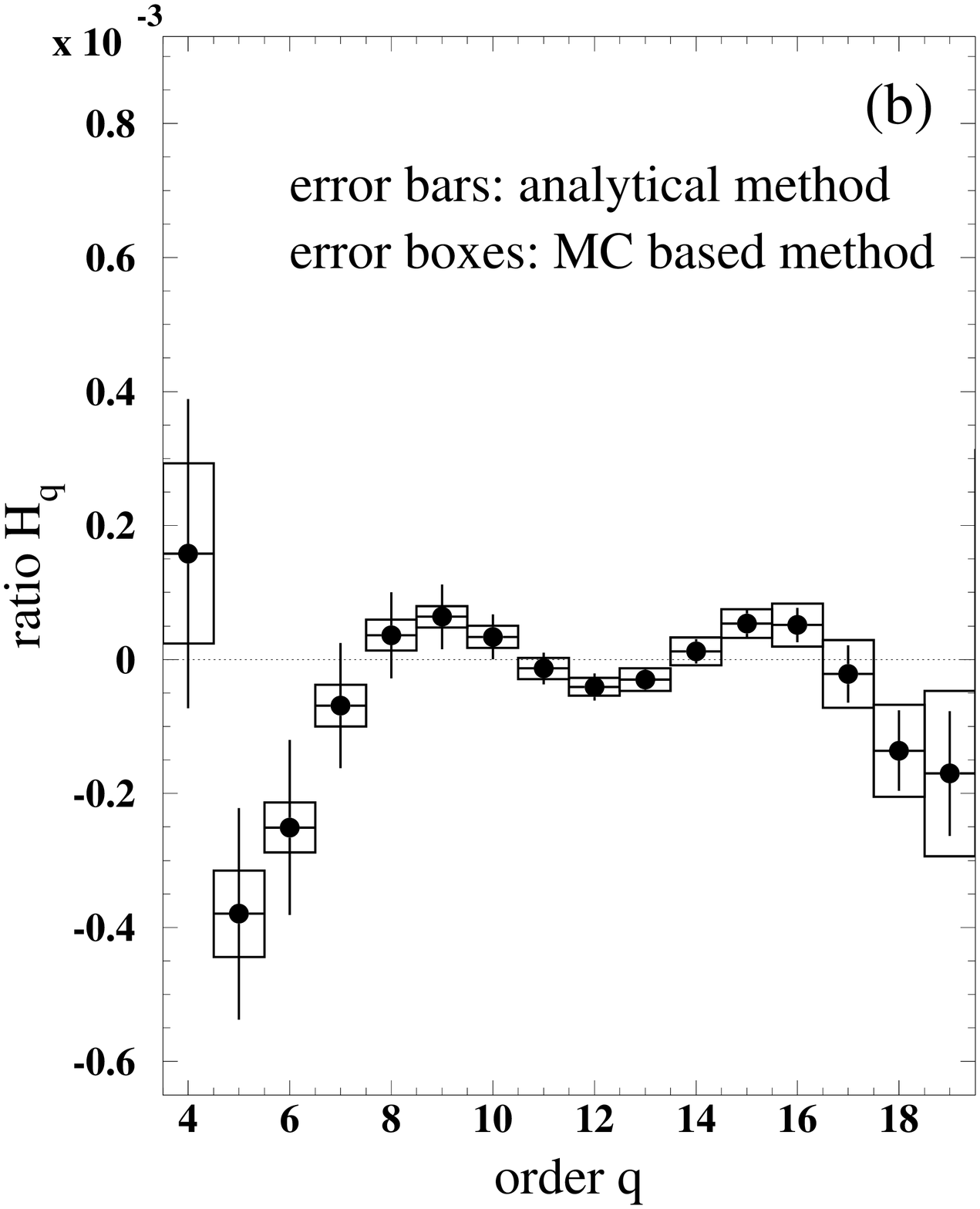}
\scaption{$H_q$ moments for light-quark sample (a) and 
b-quark sample (b). The error lines is for the 
statistical error calculated by the analytical method, the 
error boxes for the statistical errors calculated with 
the Monte Carlo based method.}
\label{fig:errcomp}  
\end{figure}
Since this procedure is rather time consuming, the number of generated 
multiplicity distributions is chosen in such a way that the precision of the error 
(\ie{} the error on the error) is sufficiently accurate.
For 1000 multiplicity distributions, the error is known within an accuracy of $3.2\%$, 
which is sufficiently small.  
(For comparison, the error committed by giving the  
value of the \mcpm{} of the 
b-quark sample with a precision of 0.01 is $16\%$). 
Furthermore, in view of the size of 
the systematic error, which is always the major contribution to 
the error in this analysis, the accuracy of $3.2\%$ for 
the statistical error is fine.
A comparison between the analytical method and this method given 
in Fig.~\ref{fig:errcomp}, shows both the limitation of the 
analytical method at low statistics and the advantage of the use  
of the analytical method. For the light-quark sample 
(Fig.~\ref{fig:errcomp}(a)) where the statistics is high (about 1,200.000 
events after corrections), the two methods give the same results, 
but for the b-quark sample, where the statistics is smaller 
(about 300.000 events after corrections), the analytical method, 
which is not able to insure a proper 
cancellation of the correlations introduced by the reduction of the statistics, 
largely overestimates the statistical errors on the $H_q$.

\subsection{Systematic errors}

The estimation of the systematic errors is based on the 
systematic study of the \cpmd{} described in Sect.~\ref{sec:sys}, where 
we have determined \cpmd{s} corresponding 
to the various checks of systematic effects. Here, we determine 
the $H_q$ moments from these distributions. 
The resulting $H_q$ moments are then compared to  
the $H_q$ moments of the reference sample in the same way 
as  was described in Sect.~\ref{sec:sys}, replacing 
the charged-particle multiplicity distributions 
by the $H_q$ moments.
Contributions to the systematic errors are then due to 
the track quality cuts, which are again the largest 
contribution, the event selection,  
uncertainties in Monte Carlo modelling, the unfolding method used to 
reconstruct the charged-particle multiplicity distribution, and the 
light- or b-tagging method.

\section{Results for the full, light- and b-quark samples}

\begin{figure}[htbp]
\centering
    \includegraphics[width=8.4cm]{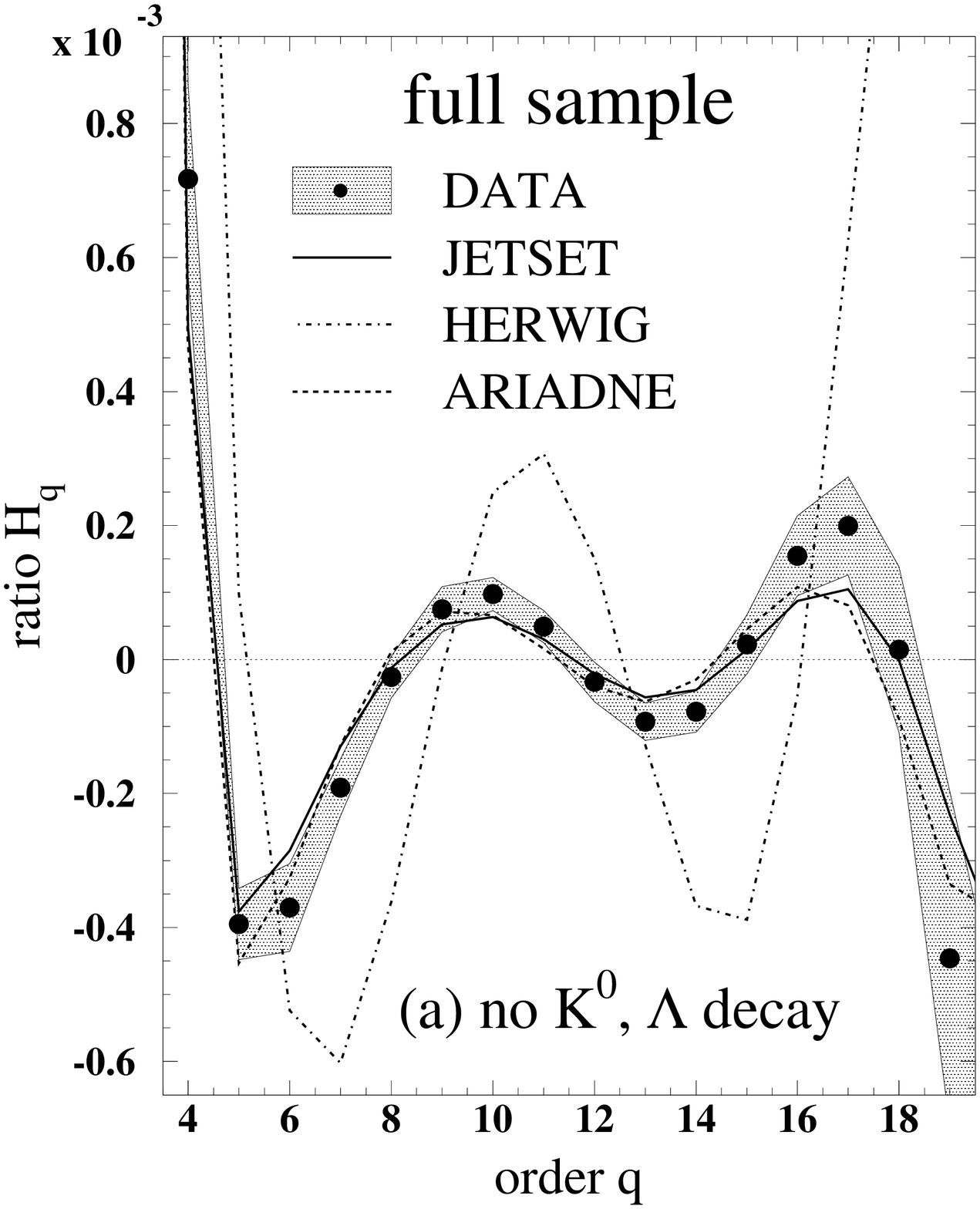}
    \includegraphics[width=8.4cm]{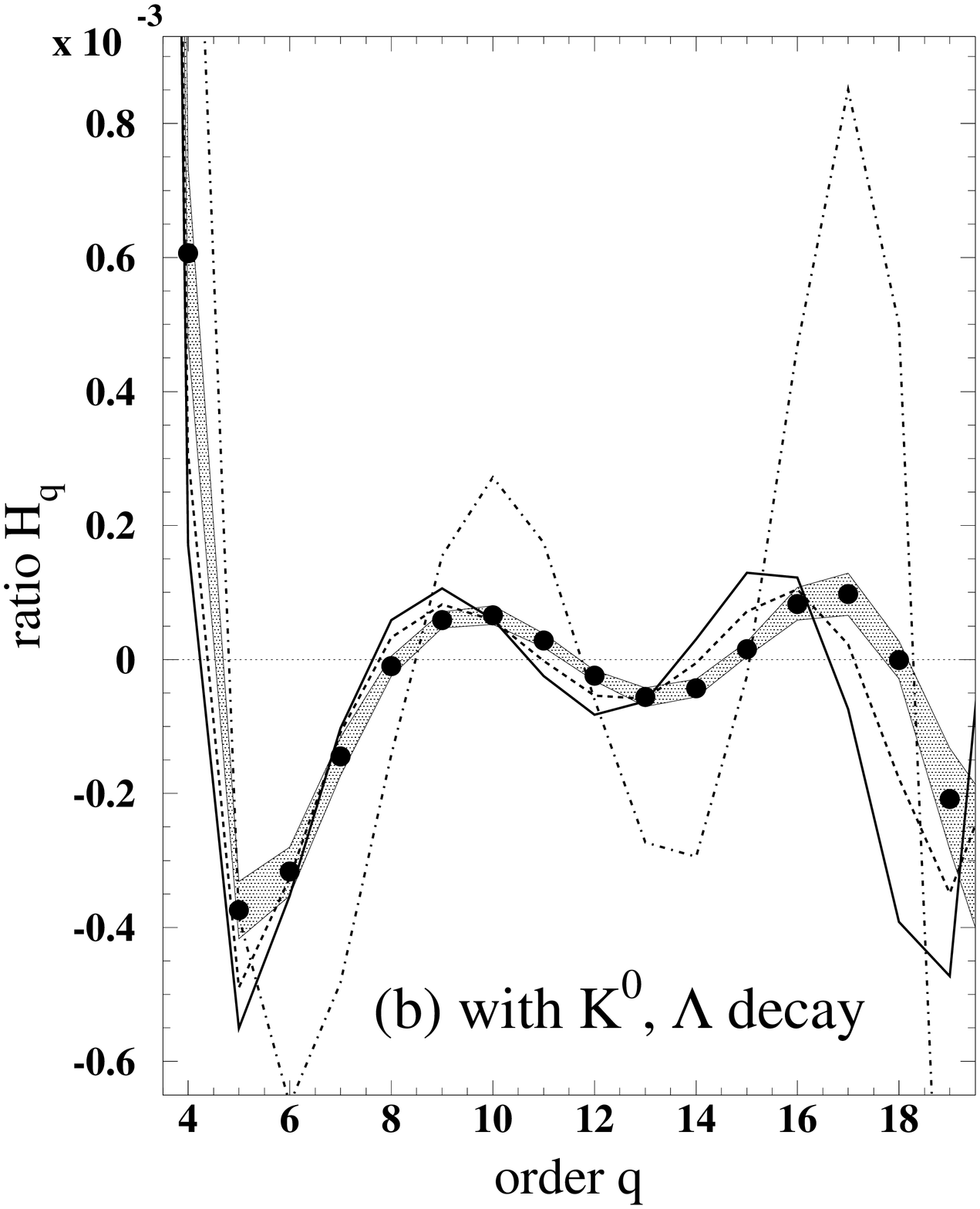}
\scaption{$H_q$ as a function of $q$ for the full samples 
(a) without and (b) with \kl{} decay products,
together with JETSET, HERWIG and ARIADNE.}
\label{fig:hqfull}  
\end{figure}

\begin{figure}[htbp]
\centering
    \includegraphics[width=8.4cm]{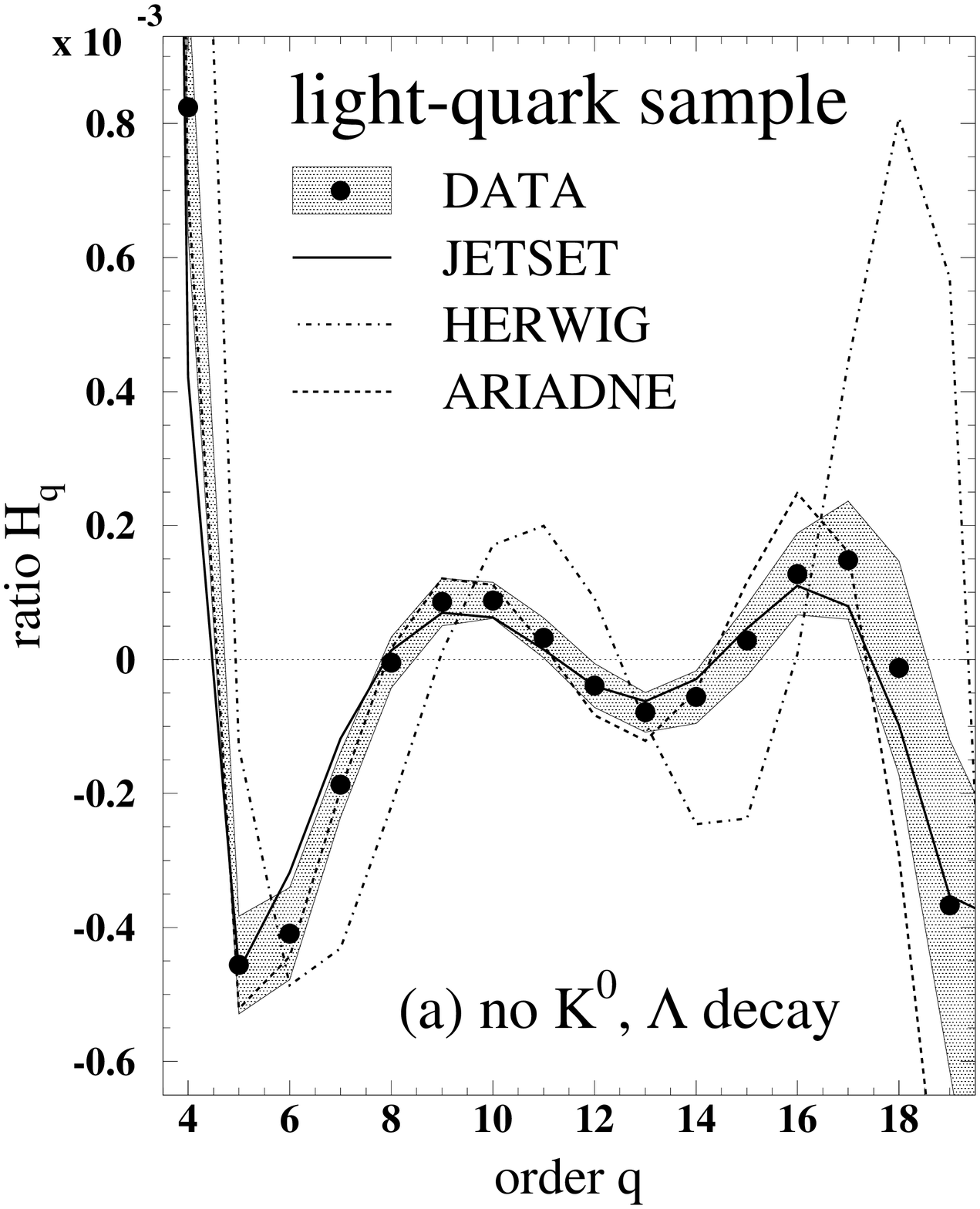}
    \includegraphics[width=8.4cm]{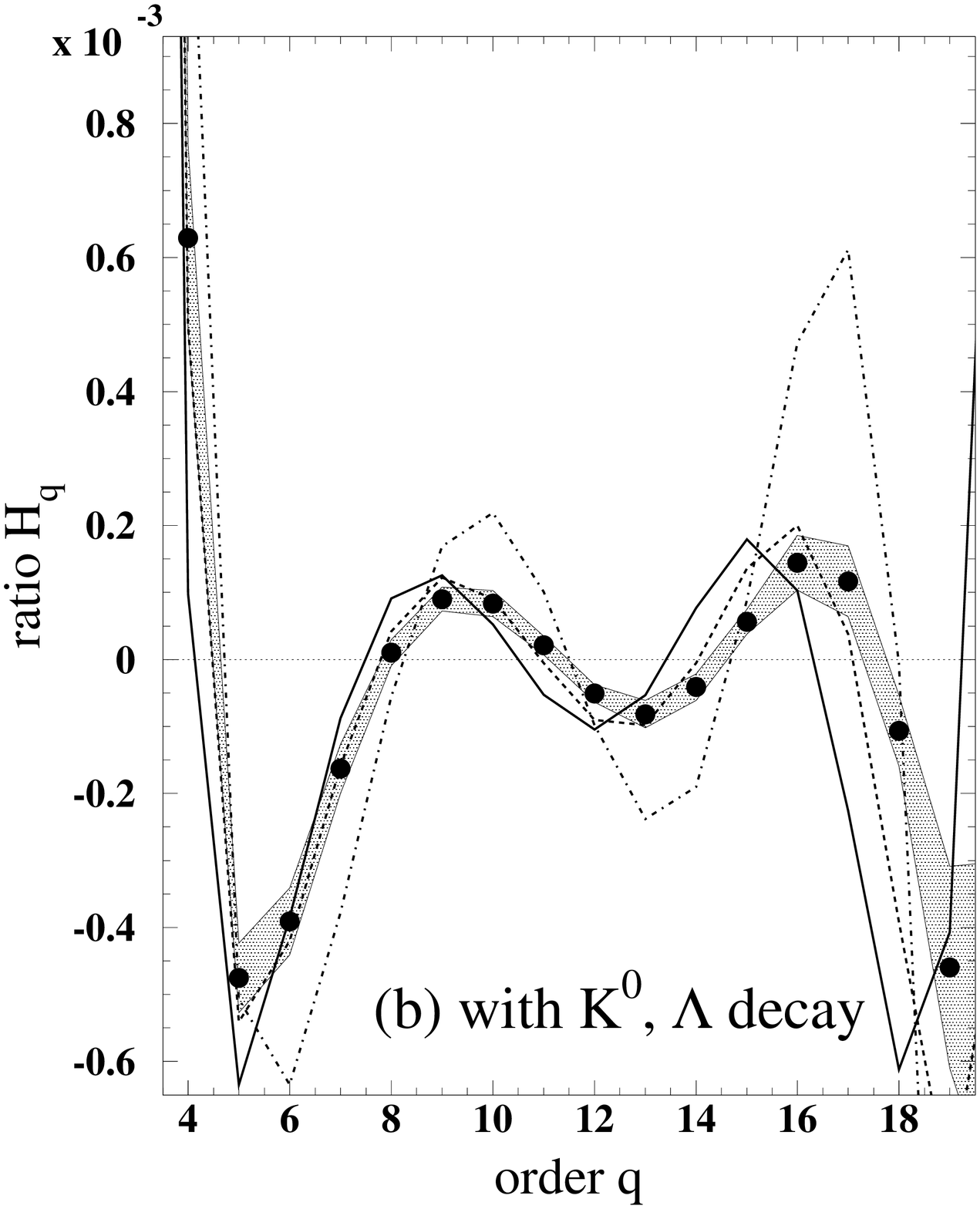}
\scaption{$H_q$ as a function of $q$ for the light-quark samples 
(a) without and (b) with \kl{} decay products,
together with JETSET, HERWIG and ARIADNE.}
\label{fig:hqlight}  
\end{figure}

\begin{figure}[htbp]
\centering
    \includegraphics[width=8.4cm]{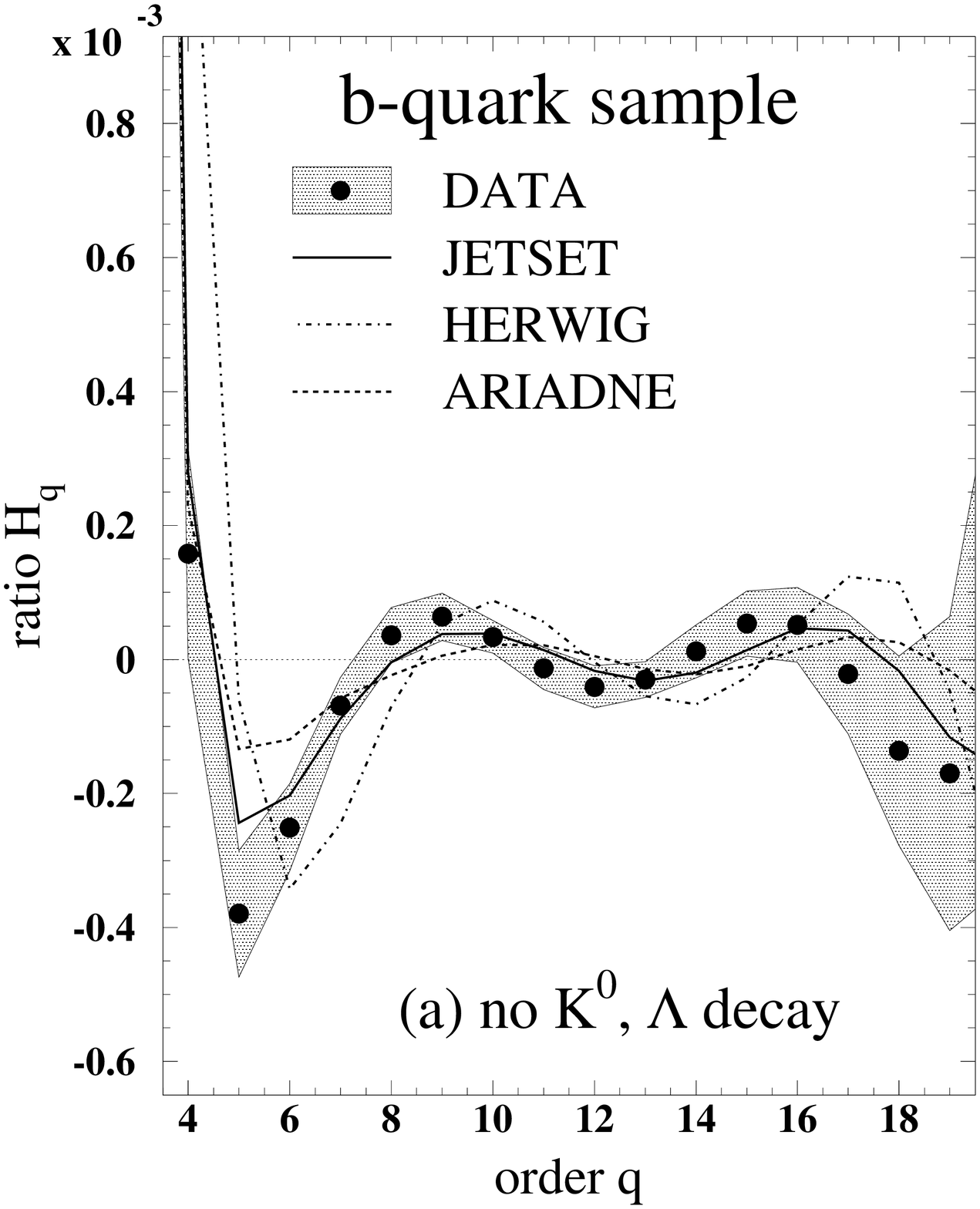}
    \includegraphics[width=8.4cm]{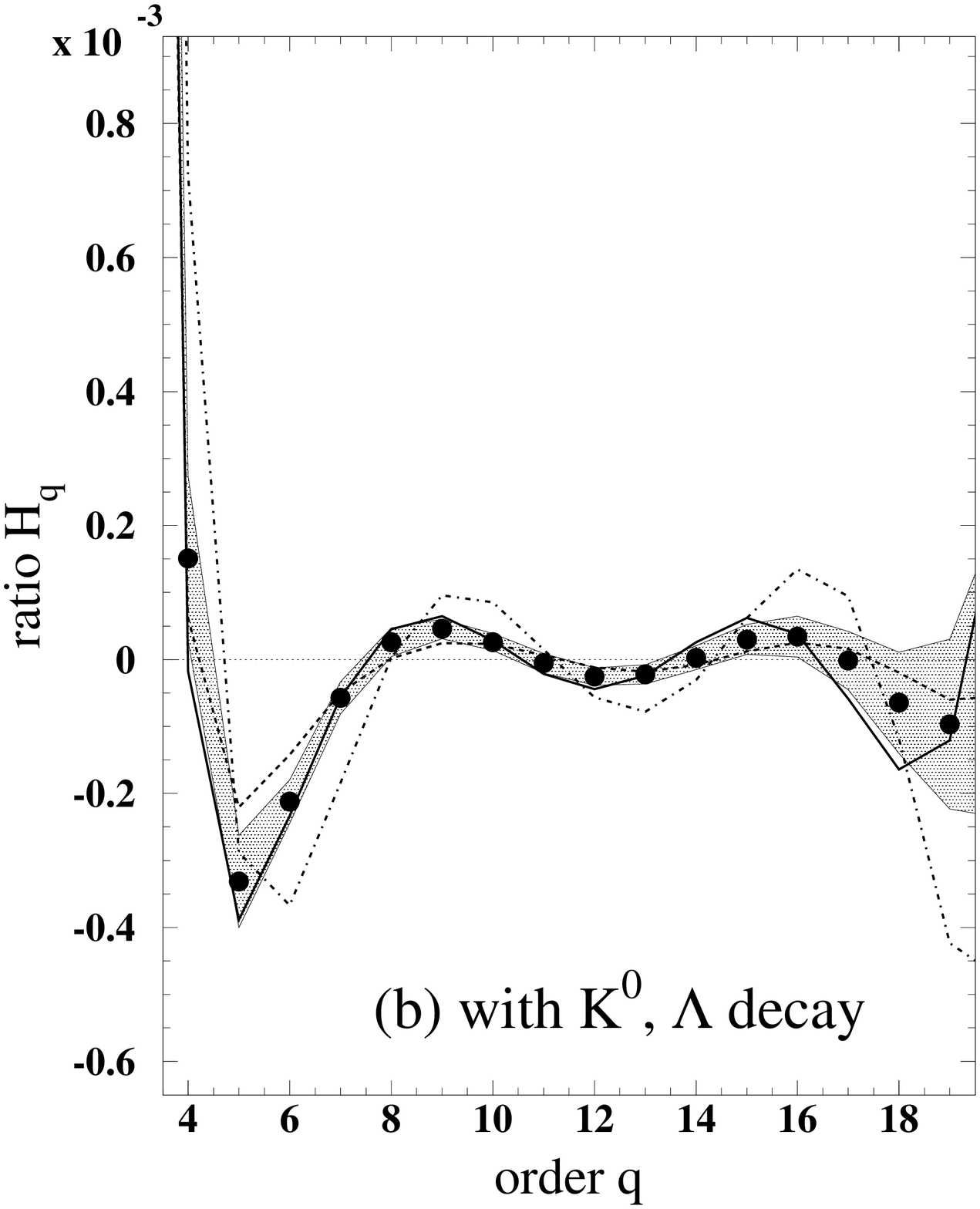}
\scaption{$H_q$ as a function of $q$ for the b-quark samples 
(a) without and (b) with \kl{} decay products,
together with JETSET, HERWIG and ARIADNE.}
\label{fig:hqb}  
\end{figure}

The $H_q$ moments are measured for the full, light-quark and 
b-quark samples without and with \kl{} decay 
products. They are shown as a function of the order $q$ 
in Figs.~\ref{fig:hqfull}, 
\ref{fig:hqlight} and~\ref{fig:hqb}, together with $H_q$ moments calculated 
from \cpmd{s} obtained from events generated with JETSET, ARIADNE 
and HERWIG. The size of the Monte Carlo samples is similar to that of the 
data, except for JETSET which contains 3 times more events. 
For all the data samples, the $H_q$ moments 
 exhibit a first negative minimum 
at $q=5$ and quasi-oscillation for higher $q$.
While JETSET and ARIADNE show relatively  good agreement 
for all the samples, the $H_q$ moments calculated from events generated 
with HERWIG do not agree with any of them.
For all samples, they show a shift of at least one order for 
all extrema. Furthermore, the $H_q$ obtained for the HERWIG full and 
light-quark samples, have amplitudes of oscillation which are much 
larger than those found in the data.
For the b-quark sample, apart from the one-order shift of the 
extrema, the amplitudes of the oscillation have about the size of 
those of the data.
\begin{figure}[htbp]
\centering
    \includegraphics[width=8.4cm]{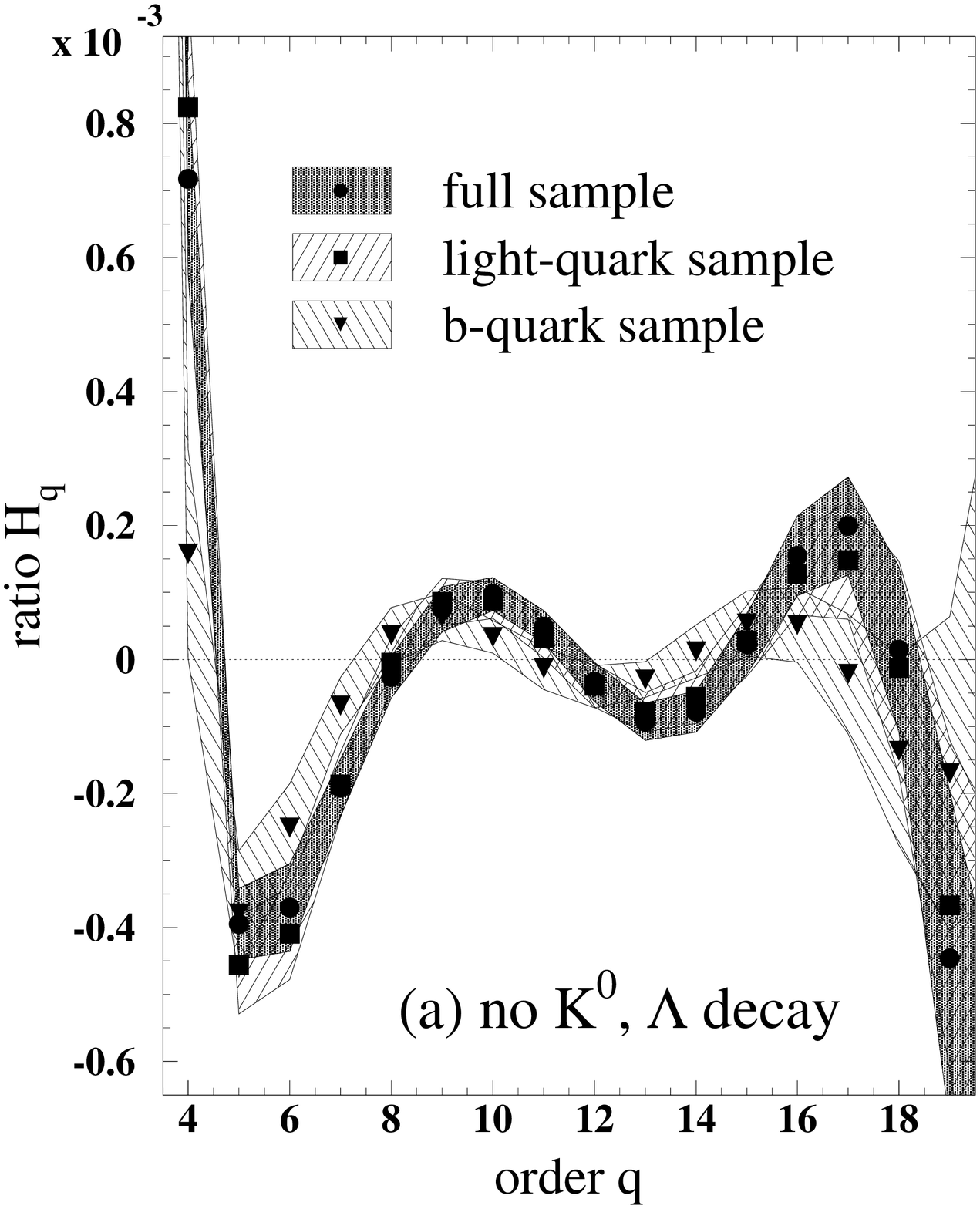}
    \includegraphics[width=8.4cm]{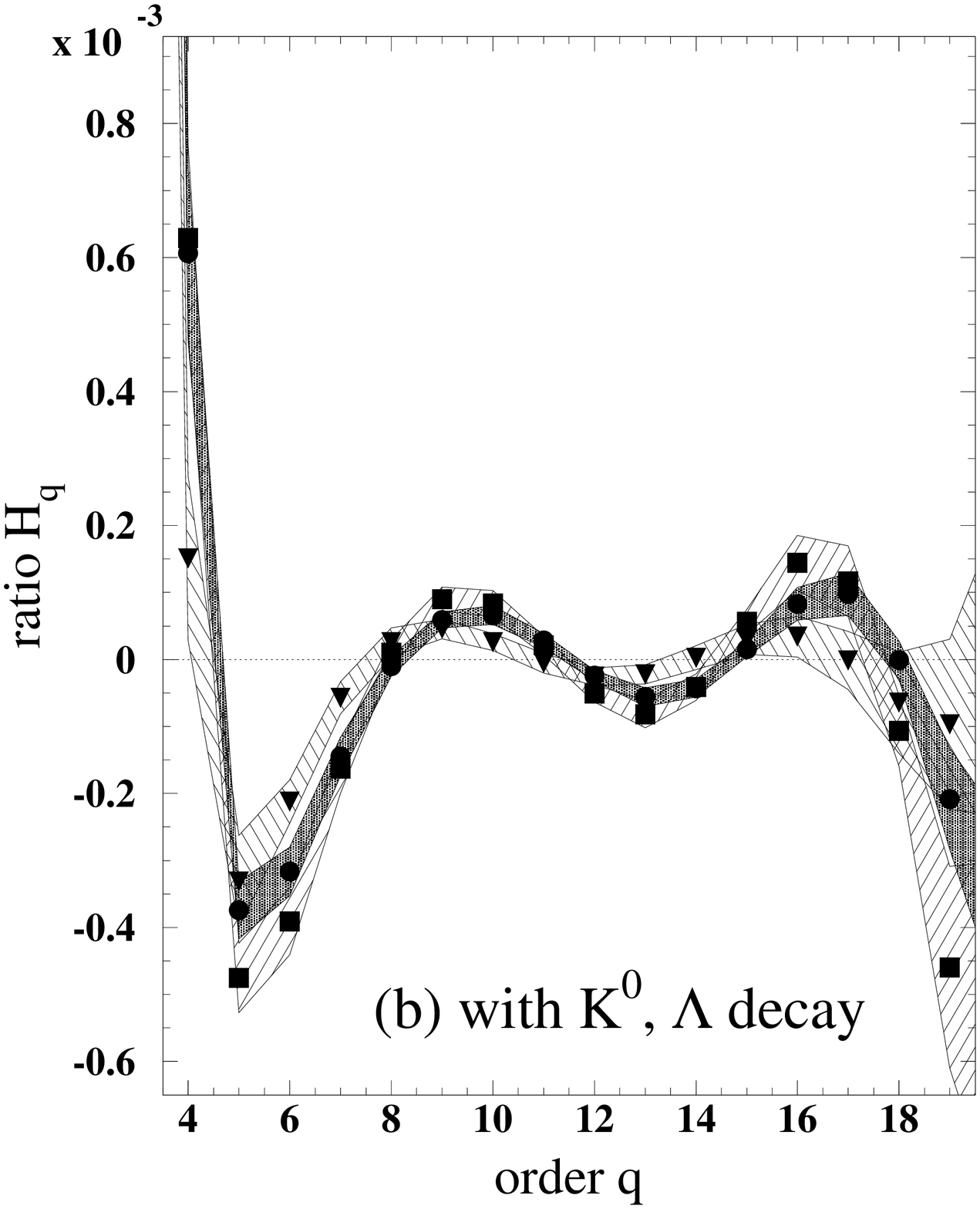}
\scaption{$H_q$ as a function of $q$ for the full, light- and 
b-quark samples (a) without and (b) with  \kl{} 
decay products.}
\label{fig:hqflb}  
\end{figure}

In Fig.~\ref{fig:hqflb}, we can see that the full, light- and 
b-quark samples agree 
well with each other, with only a small difference
for small $q$ ($q<5$), between the b-quark sample and the other samples. 
This indicates that the weak decay of the b quark does 
not have much influence on the shape of the \cpmd{}. 
It must be noted that neither does the weak decay of \kl{} have
 much influence on the shape of the multiplicity
distribution. 

The $H_q$ behavior observed for the data is qualitatively 
similar to that predicted by the NNLLA assuming 
Local Parton-Hadron Duality. However, also JETSET agrees 
well with all the data samples (and HERWIG, even if it does not
agree with the data, predicts the same kind of features,  
\ie{} a first negative minimum followed by quasi-oscillation). 
However, none of the parton showers used by those Monte Carlo models
have implemented NNLLA. Rather, they use parton showers which 
are close in form to the MLLA with, in addition,  
full energy-momentum conservation.

Therefore, we attempt to find which aspect of the 
Monte Carlo generation is responsible for this 
agreement.

\section[Monte Carlo analysis of the {$H_q$}]
{Monte Carlo analysis of the \boldmath{$H_q$}}

In view of the good agreement of JETSET with the data for 
all samples, we vary several options in JETSET and 
study their influence on the shape of the charged-particle 
multiplicity distribution and hence, on the $H_q$ behavior.

Since analytical QCD predicts the $H_q$ behavior for partons, 
we tried first to change in JETSET, options related to the parton 
production, keeping in all cases the Lund string fragmentation 
model for the hadronization.
We try the following:

\begin{itemize}

\item No angular ordering in the parton shower. This makes it 
essentially an LLA shower with, in addition, energy-momentum 
conservation.  
Removing this constraint allows more partons to be  
generated. Therefore, the multiplicity distribution 
generated that way should have a larger average number 
of particles and should be broader.

\item Partons are generated according  
to $\mathcal{O}(\alpha_\text{s})$ and 
$\mathcal{O}(\alpha^2_\text{s})$ matrix elements and even 
only $\text{q}\bar{\text{q}}$.
Since only a small number of partons are 
generated under these models, the charged-particle 
multiplicity distribution  
has smaller mean and dispersion.

\end{itemize}

\begin{figure}[htbp]
\centering
    \includegraphics[width=8.4cm]{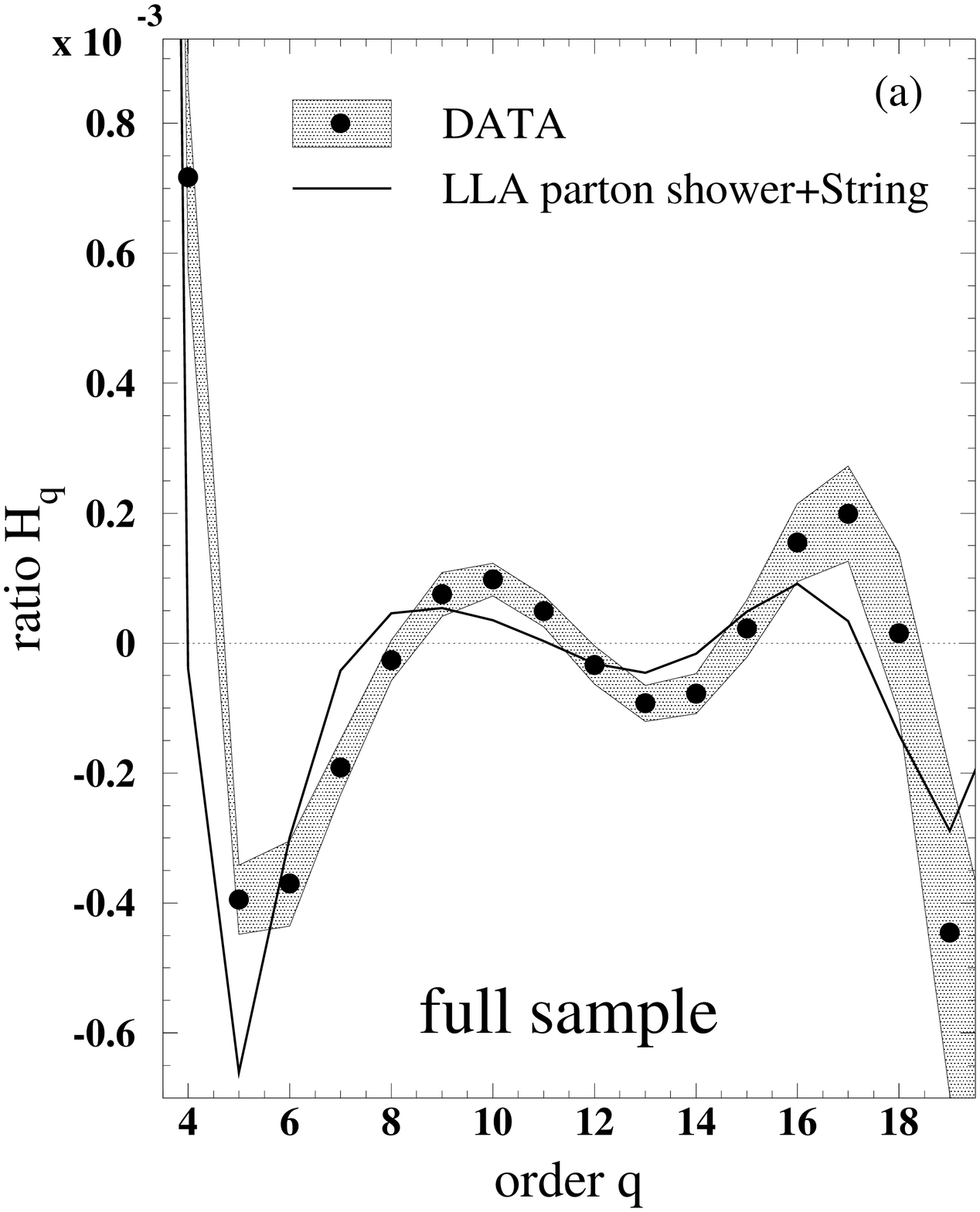}
    \includegraphics[width=8.4cm]{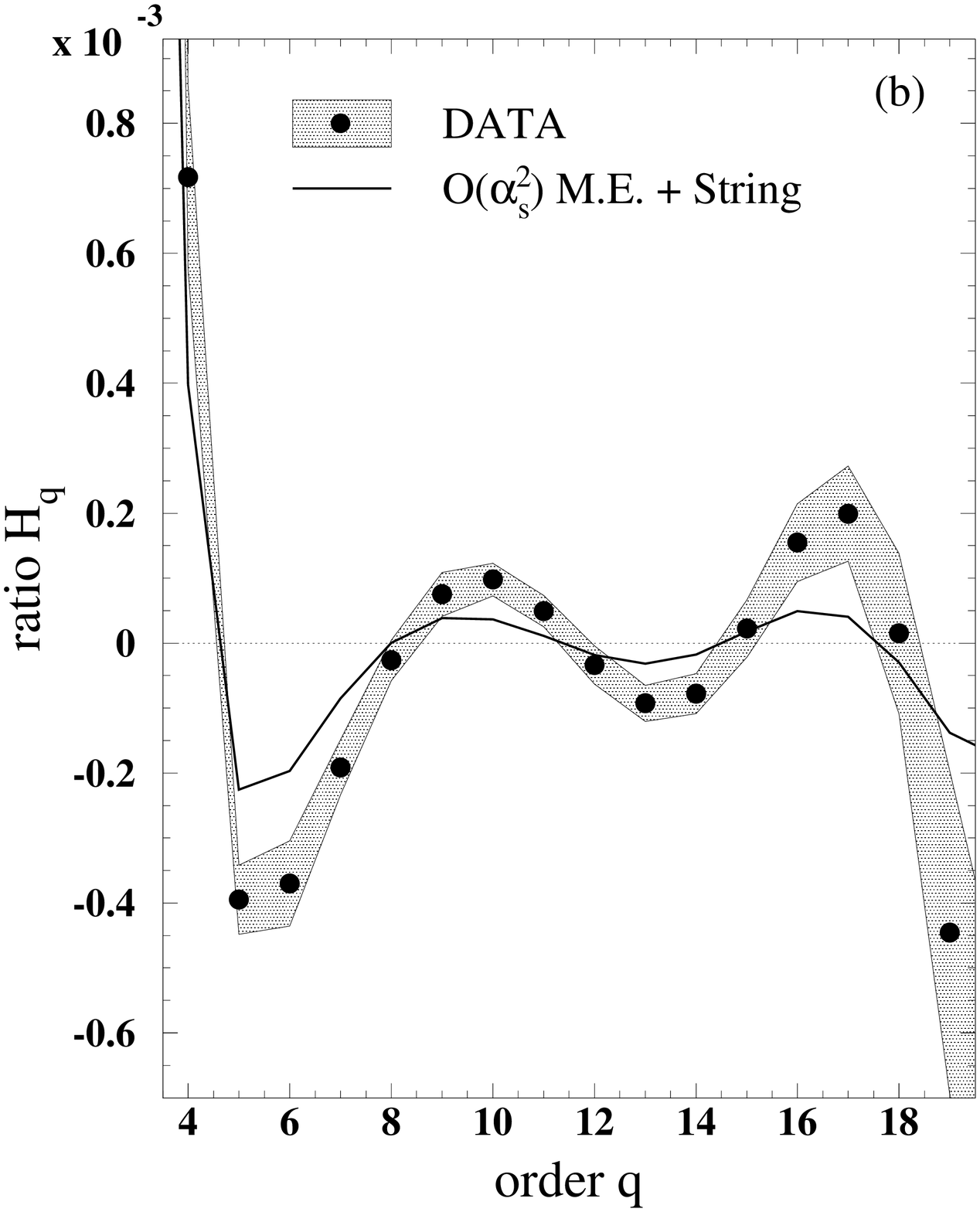}

    \includegraphics[width=8.4cm]{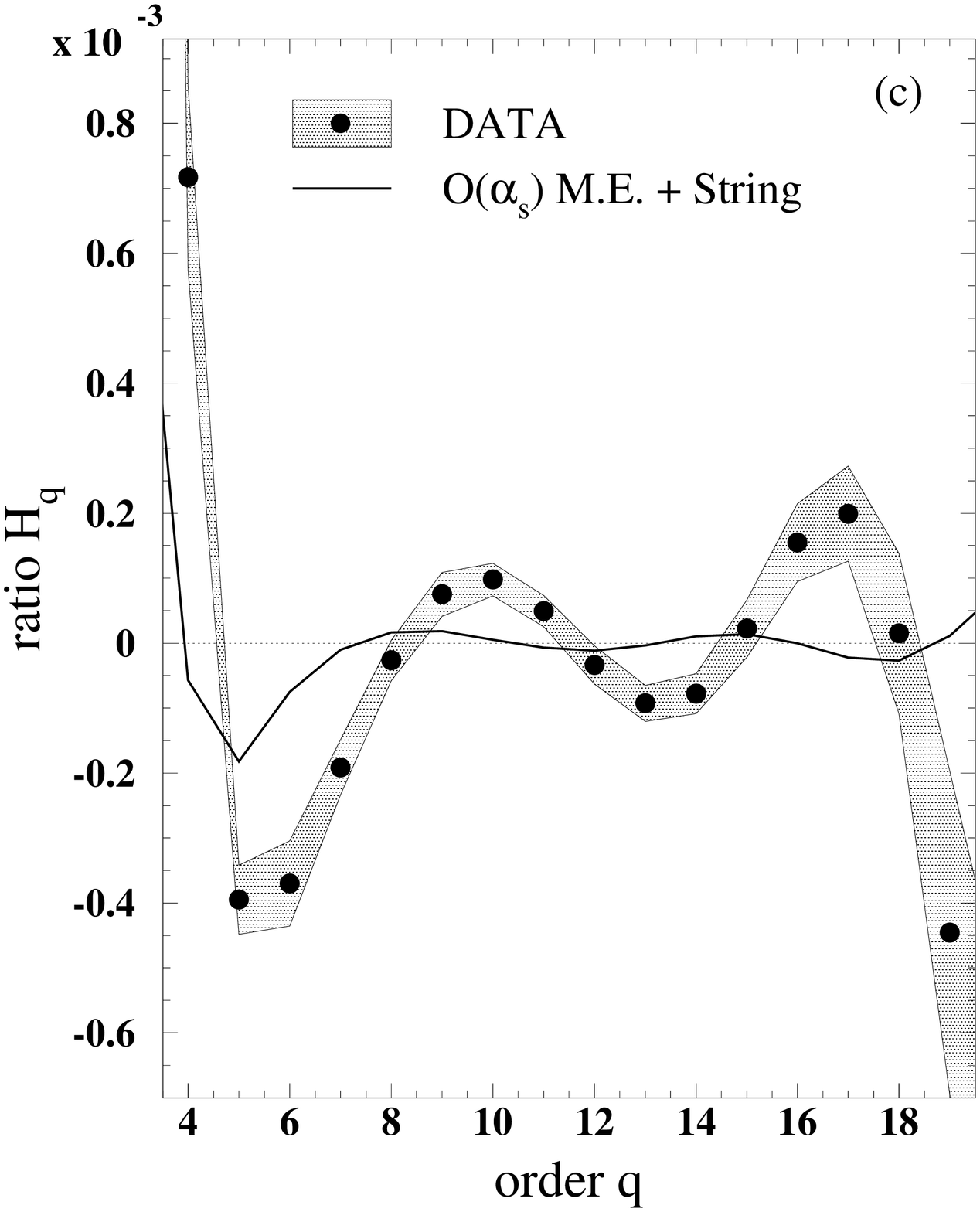}
    \includegraphics[width=8.4cm]{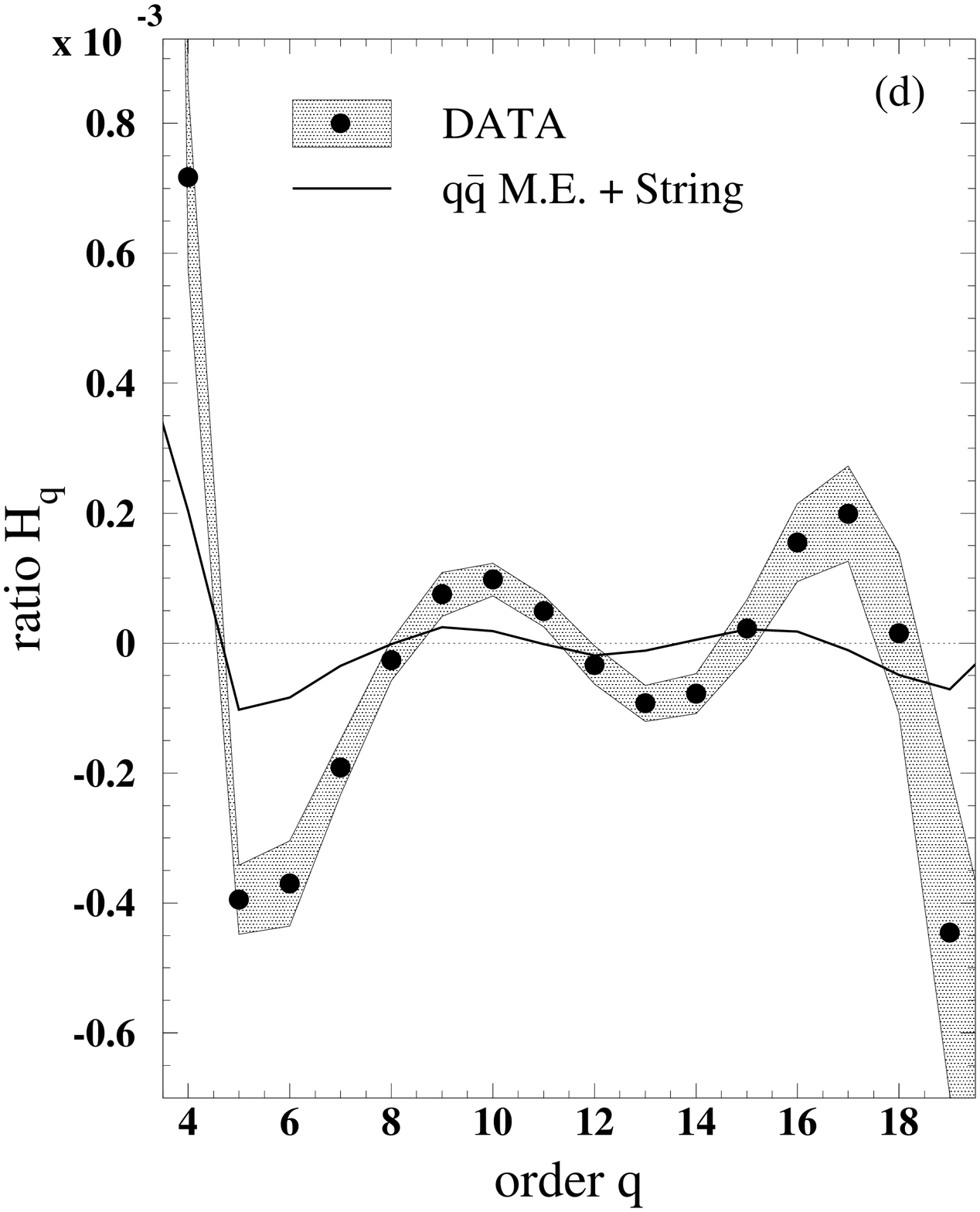}
\scaption{$H_q$ moment of the full sample, 
together with events generated with JETSET (a) using a LLA parton shower, 
(b) using $\mathcal{O}(\alpha^2_\text{s})$ matrix element, 
(c) using $\mathcal{O}(\alpha^1_\text{s})$ matrix element and (d) 
generating a $\text{q}\bar{\text{q}}$ pair only,
and in all cases the Lund string fragmentation.}
\label{fig:hq_mc1}  
\end{figure}

For all four cases (\ie{} multiplicity distribution generated with Lund string 
fragmentation and partons according LLA parton shower, 
$\mathcal{O}(\alpha^2_\text{s})$, $\mathcal{O}(\alpha^1_\text{s})$,  
and $\text{q}\bar{\text{q}}$ only), we find that the $H_q$ moments 
calculated from the 
corresponding charged-particle multiplicity distribution, have a first 
negative minimum around $q=5$ followed by quasi-oscillations as is 
seen in Fig.~\ref{fig:hq_mc1}.
Even though, these various models do not reproduce the $H_q$ behavior exactly, 
we find qualitative 
agreement with the $H_q$ oscillatory behavior seen in the data. 
The matrix element models have slightly 
smaller oscillation size than the parton shower models. 
We notice that the depth of the first 
negative minimum decreases with decreasing matrix element order. 
Also the amplitudes of the oscillation 
are smaller for the distributions generated with partons 
according matrix elements, 
but it must be noted that these distributions contain fewer   
particles than those generated according to the parton 
shower implementation.

This shows us that the overall features of the $H_q$
behavior do not depend on a particular model for 
the generation of partons. 

The next step is to consider that the oscillatory behavior may originate 
from the fragmentation, 
which could simulate some higher-order aspects
of pQCD. Therefore, we repeat the previous study, 
replacing the Lund string model by the independent fragmentation 
model. We also allow in both Lund string and independent 
fragmentation models, resonances to decay. As a further test, we 
split the samples into light- and b-quark samples.

\begin{figure}[htbp]
\centering
    \includegraphics[width=8.4cm]{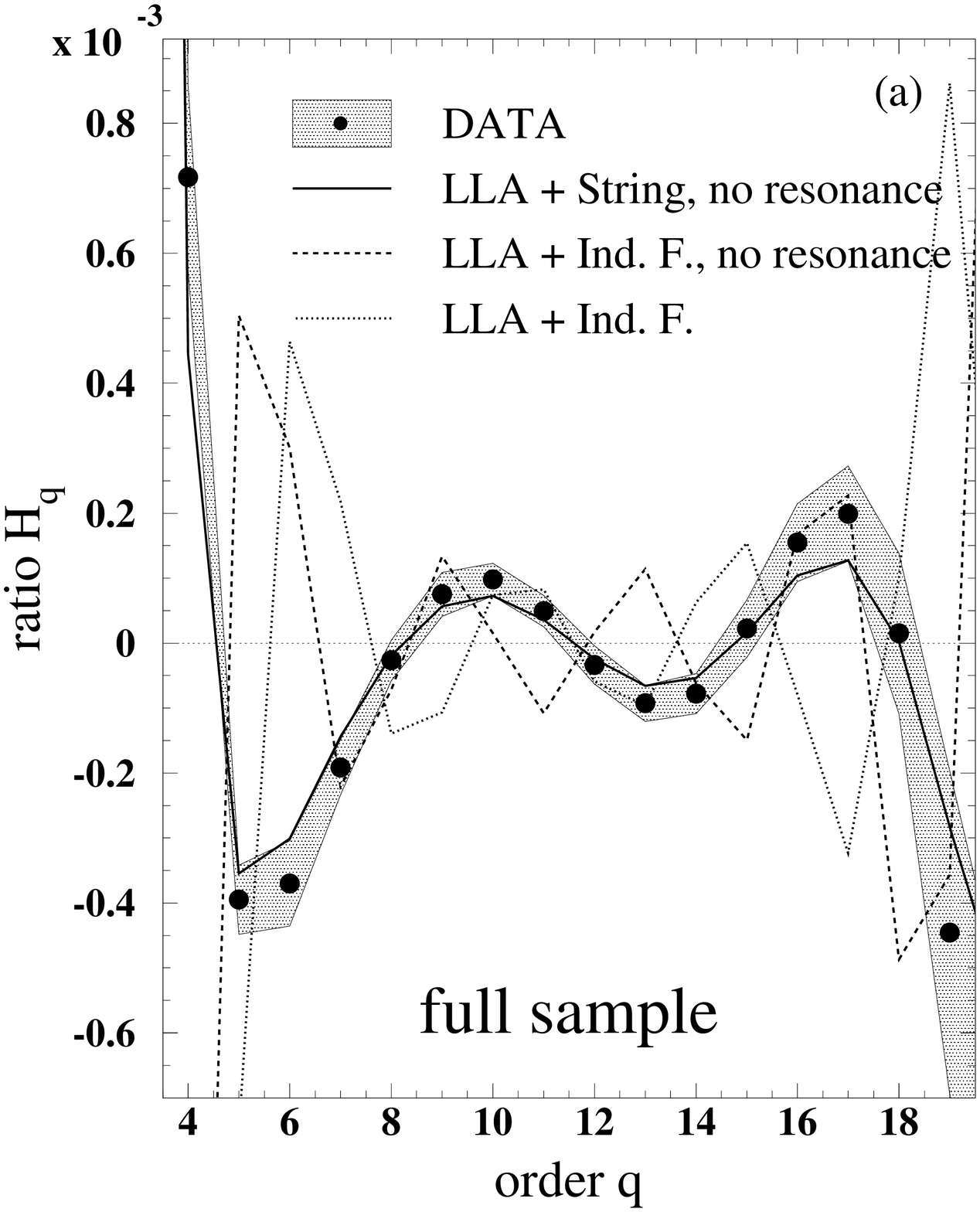}
    \includegraphics[width=8.4cm]{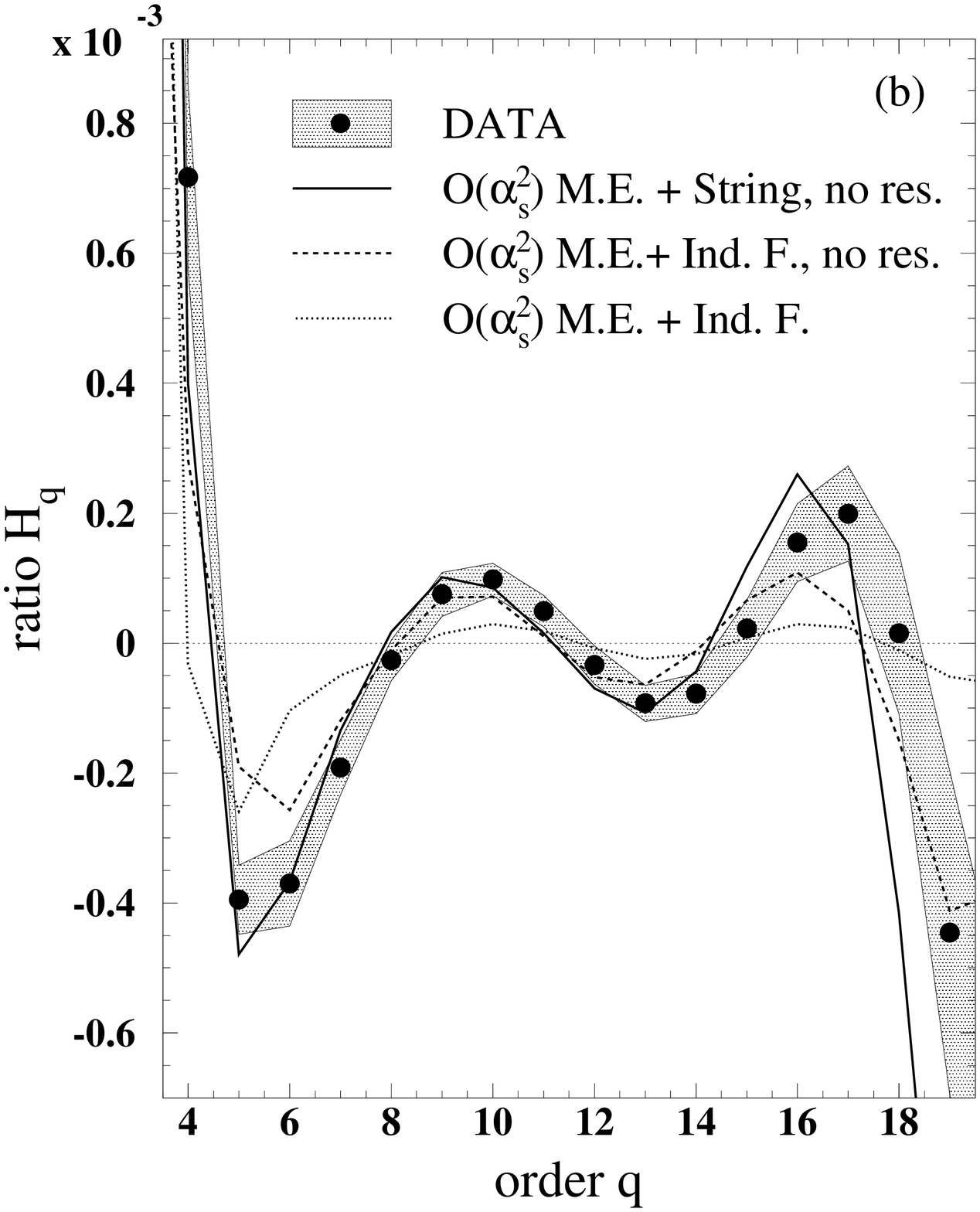}

    \includegraphics[width=8.4cm]{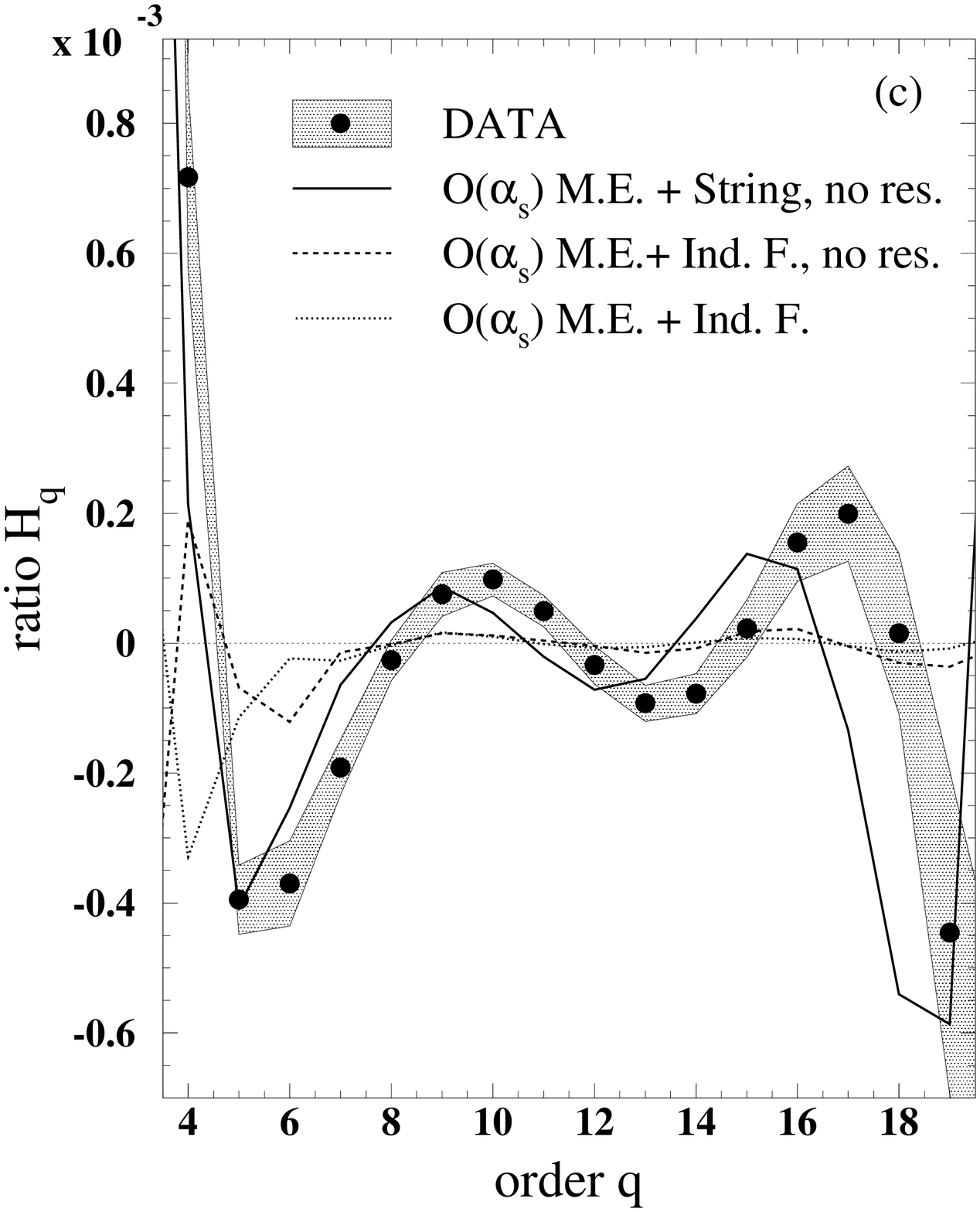}
    \includegraphics[width=8.4cm]{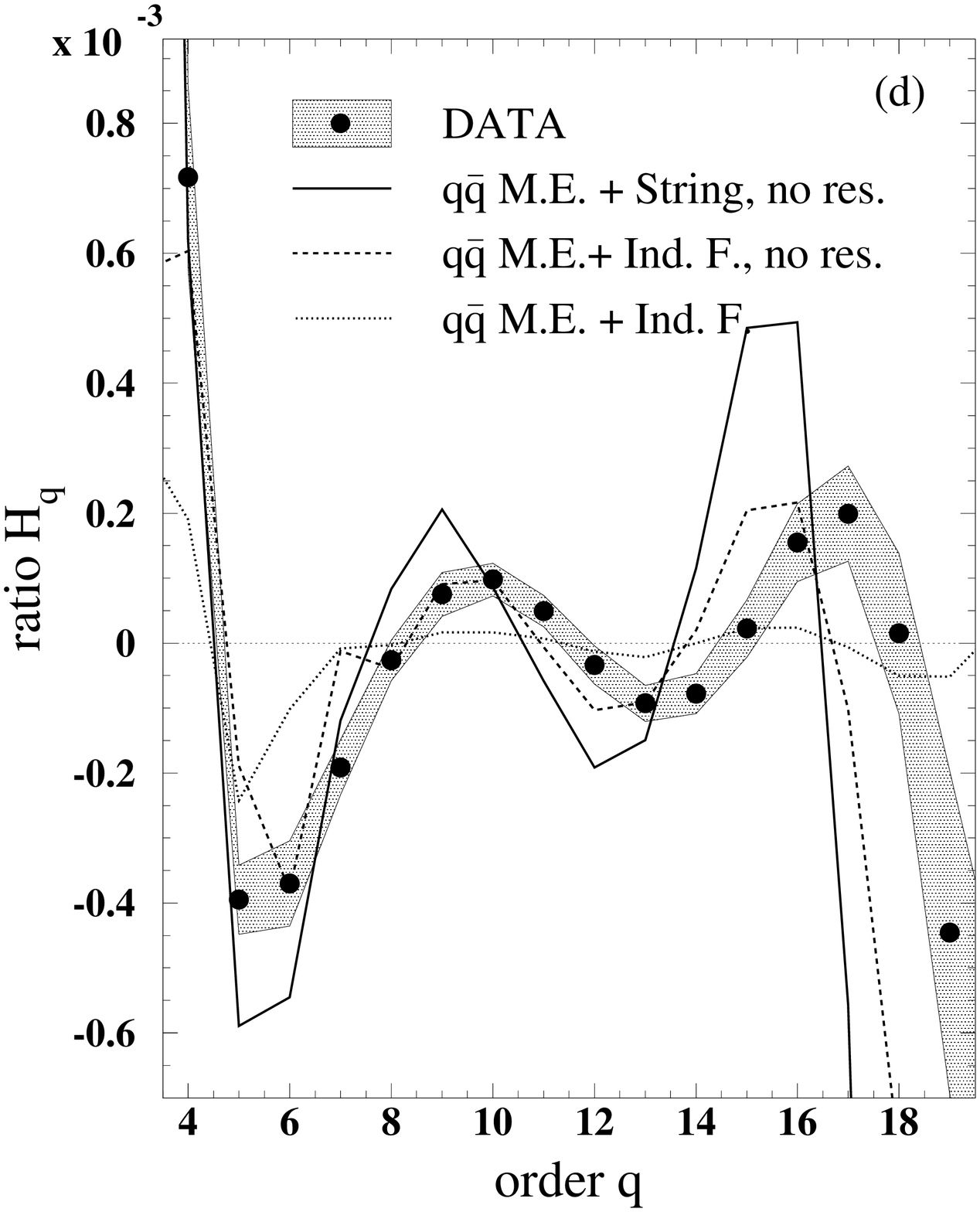}
\scaption{$H_q$ moment of the full sample compared to 
events generated (a) using LLA parton shower, (b) with 
$\mathcal{O}(\alpha^2_\text{s})$ matrix element,  
(c) $\mathcal{O}(\alpha_\text{s})$ matrix element  
and (d) only $\text{q}\bar{\text{q}}$  for 
various fragmentation options.}
\label{fig:hq_mc1_f}  
\end{figure}

For the full samples (Fig.~\ref{fig:hq_mc1_f}),  
most of what we try gives oscillations of about the size of the data.  
However, the $H_q$ moments obtained with a 
LLA parton shower and independent fragmentation 
(with and without resonances) shows a different behavior. 
It has a first negative minimum at $q=4$ and, instead of
smooth oscillation for higher $q$, the $H_q$ moment 
oscillates with a short period. 
In this particular case, the combination of the LLA parton shower and  
independent fragmentation models gives birth to a very 
wide distribution with a dispersion near 10 and a mean of the \cpmd{} 
of about 30.  

\begin{figure}[htbp]
\centering
    \includegraphics[width=8.4cm]{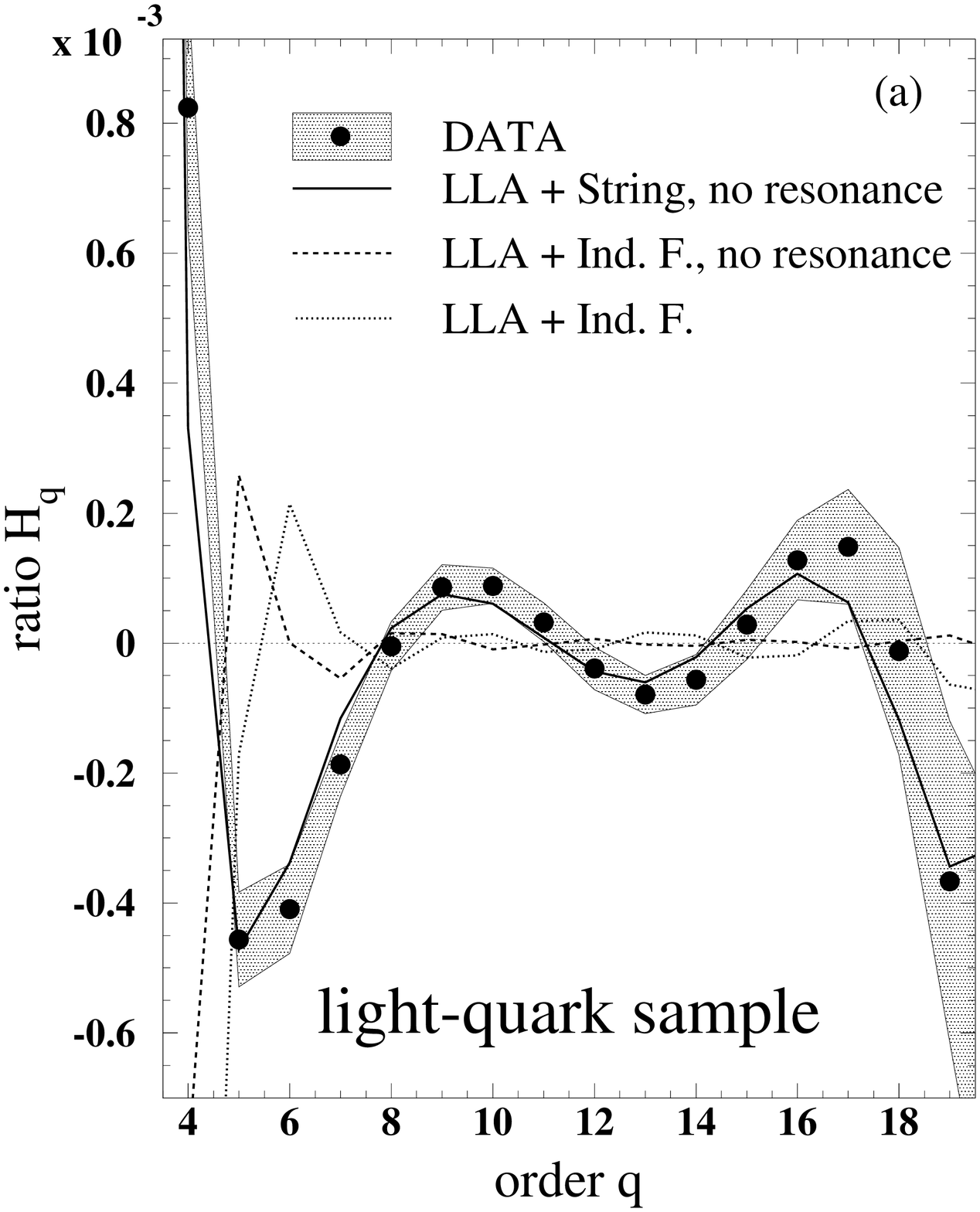}
    \includegraphics[width=8.4cm]{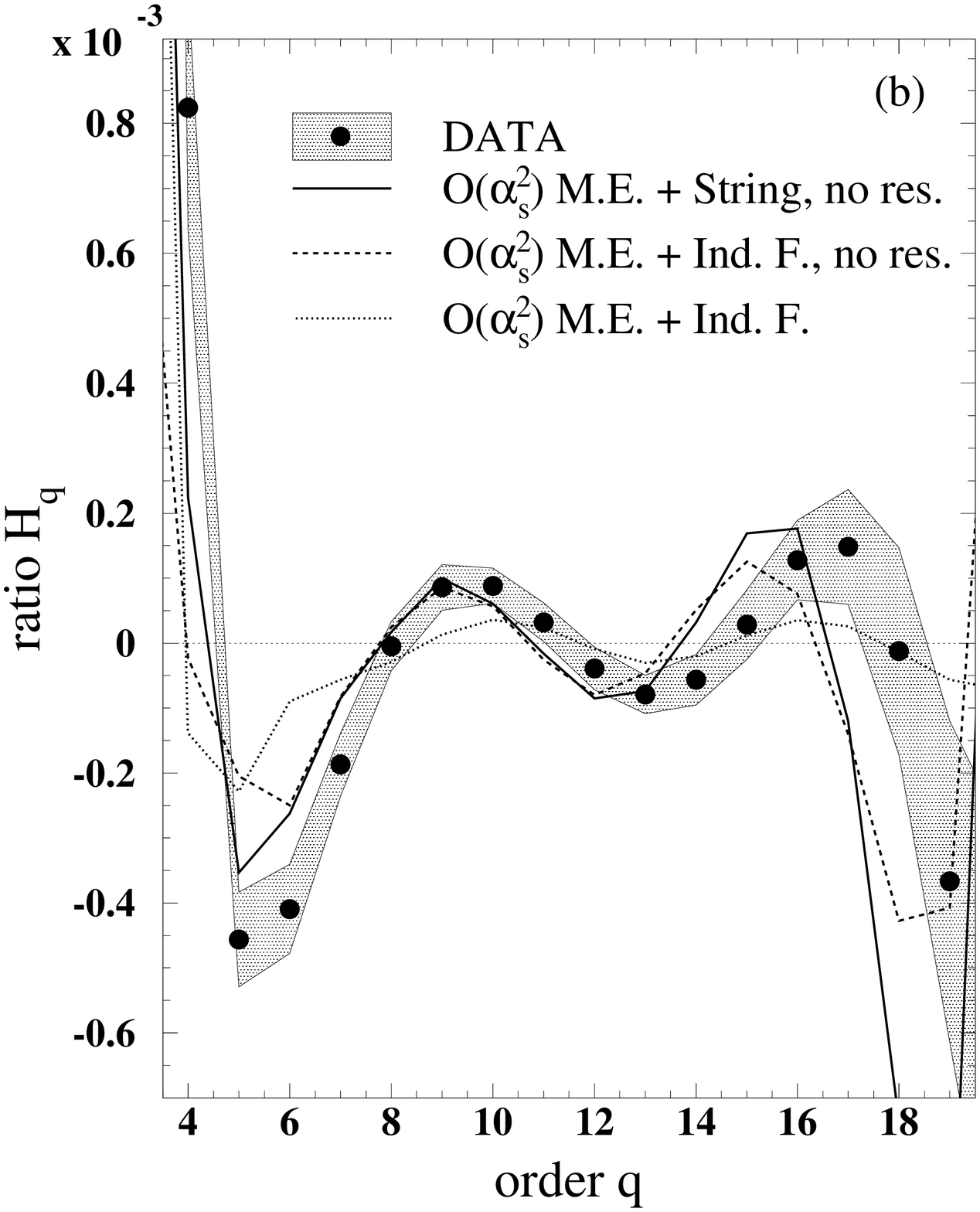}

    \includegraphics[width=8.4cm]{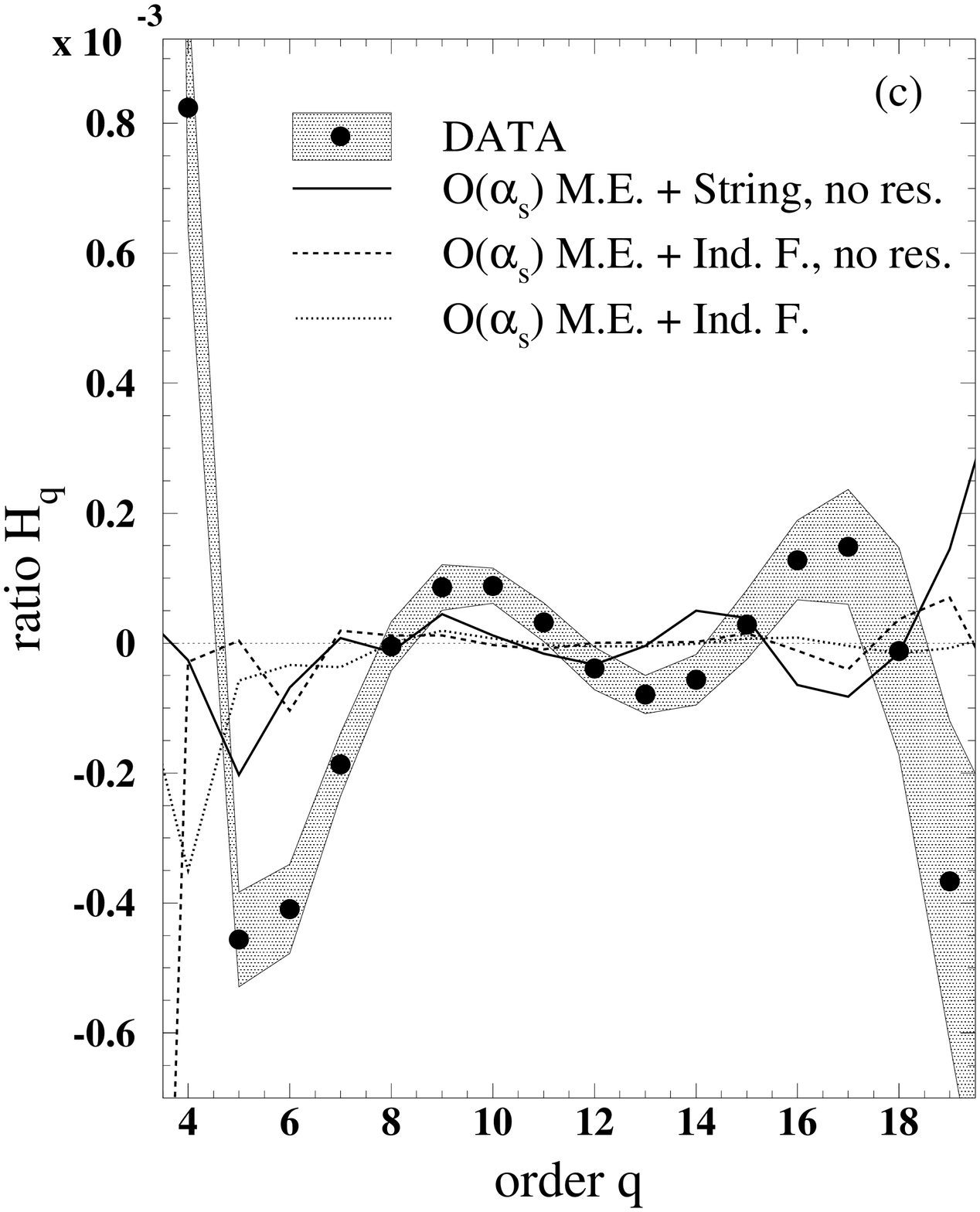}
    \includegraphics[width=8.4cm]{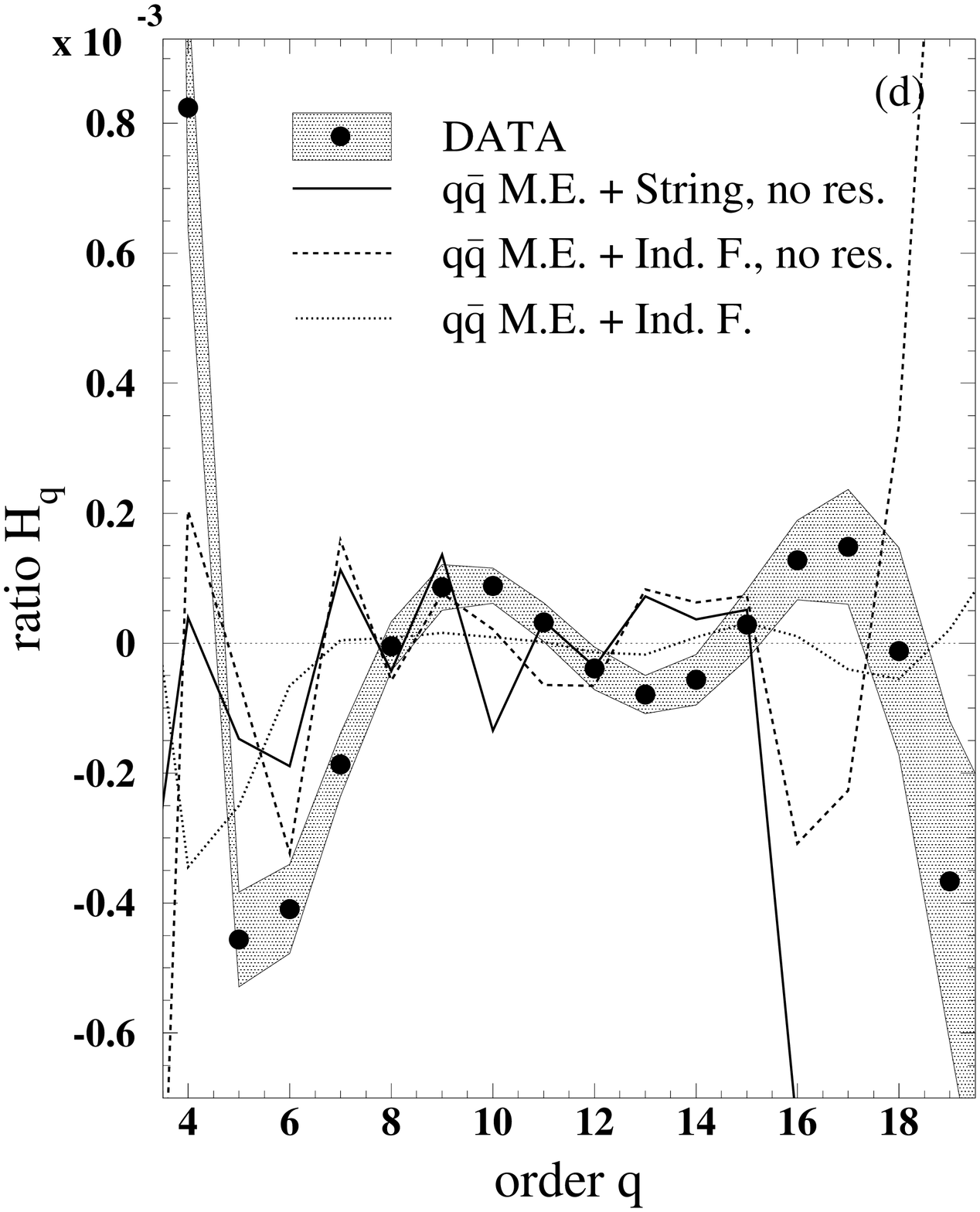}
\scaption{$H_q$ moment of the light-quark sample compared to 
events generated (a) using LLA parton shower, (b) with 
$\mathcal{O}(\alpha^2_\text{s})$ matrix element,  
(c) $\mathcal{O}(\alpha_\text{s})$ matrix element  
and (d) only $\text{q}\bar{\text{q}}$  for 
various fragmentation options.}
\label{fig:hq_mc1_l}  
\end{figure}

Also for the light-quark samples (Fig.~\ref{fig:hq_mc1_l})  
we obtain for most of the samples $H_q$ moments which have a first negative 
minimum for $q$ between 4 and 6 and oscillatory behavior for larger $q$, but 
these oscillations are usually smaller. More specifically, the $H_q$ moments 
generated from distributions obtained with LLA parton shower and 
independent fragmentation, 
without resonance decays shows a first negative minimum at 4
followed by a positive maximum at 5 and small oscillation. With resonance 
decays, we have a similar behavior but shifted by one order.
For only $\text{q}\bar{\text{q}}$ with both string and 
independent fragmentation without resonance decay, we cannot 
see any oscillation but some erratic behavior. This seems to be 
related to the absence of the resonances, since the model using 
independent fragmentation with resonances shows the oscillatory behavior.

\begin{figure}[htbp]
\centering
    \includegraphics[width=8.4cm]{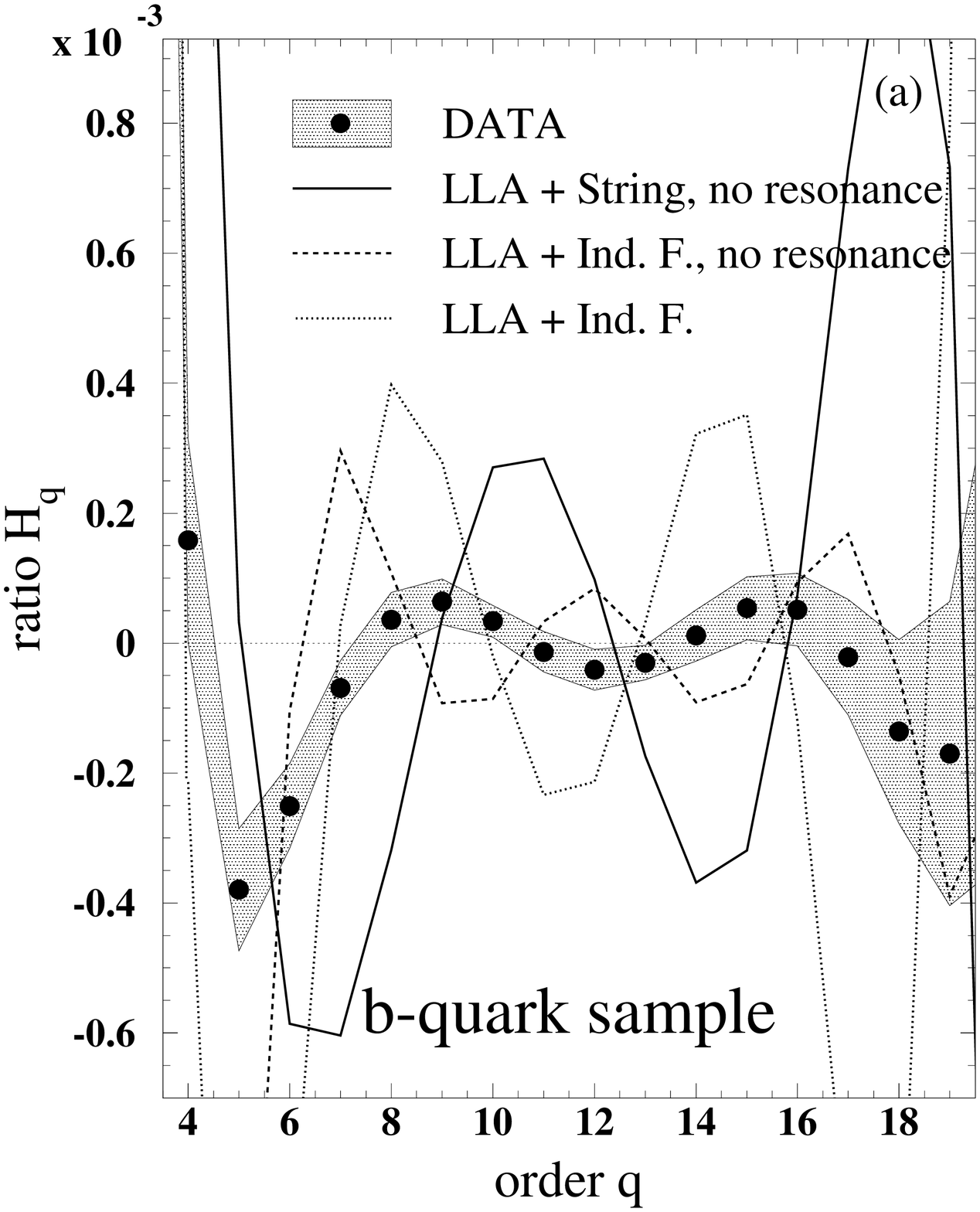}
    \includegraphics[width=8.4cm]{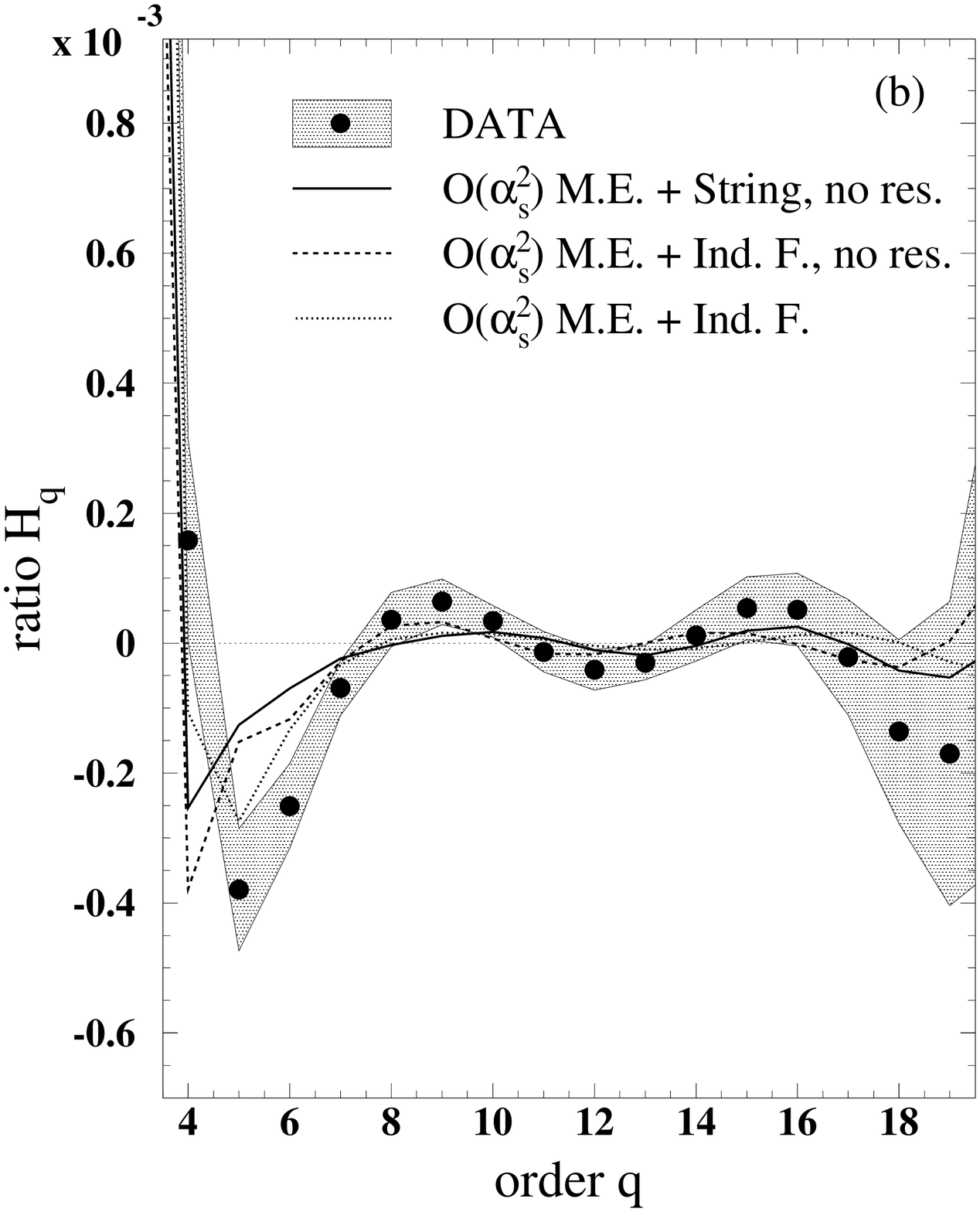}

    \includegraphics[width=8.4cm]{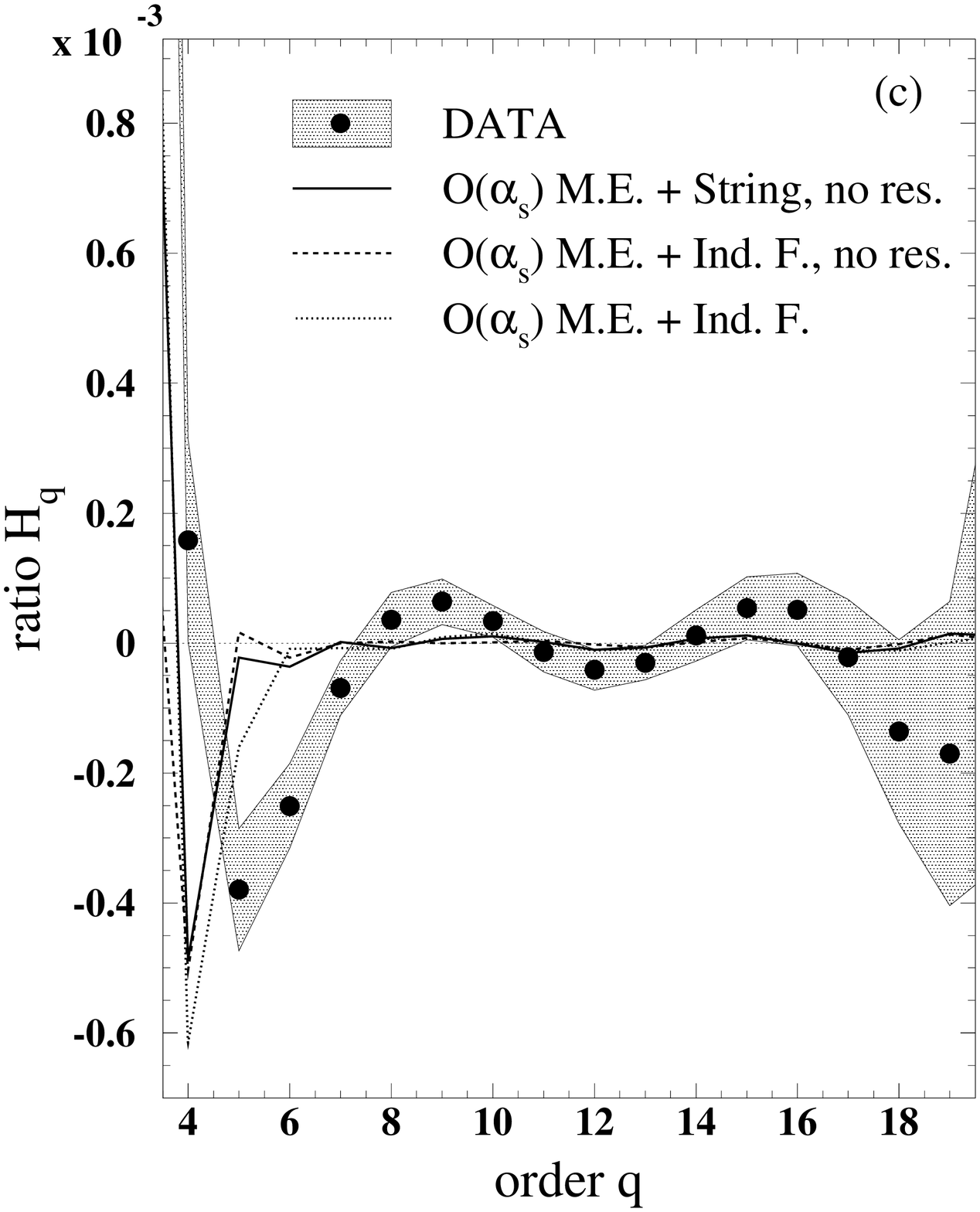}
    \includegraphics[width=8.4cm]{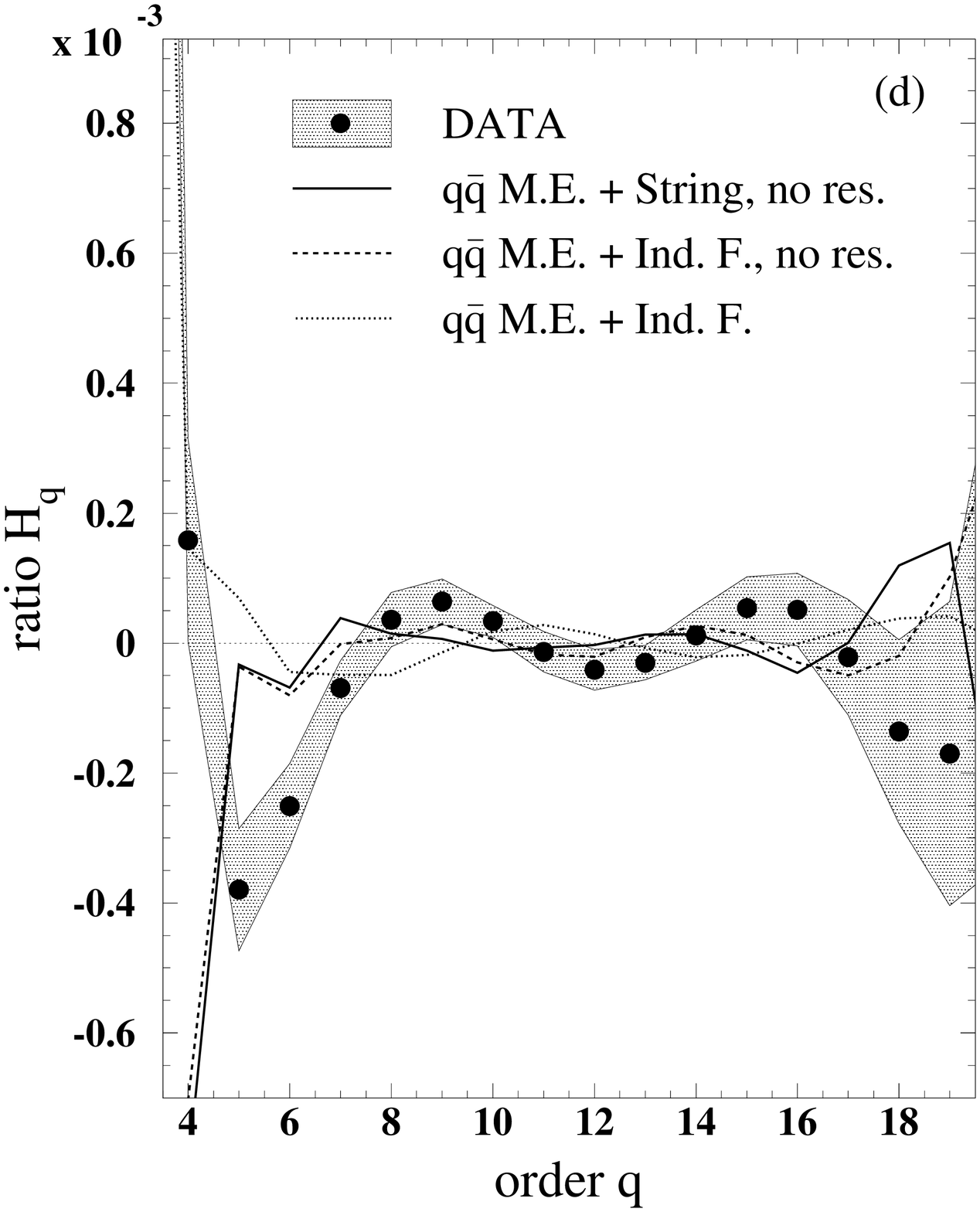}
\scaption{$H_q$ moment of the b-quark sample compared to 
events generated (a) using LLA parton shower, (b) with 
$\mathcal{O}(\alpha^2_\text{s})$ matrix element,  
(c) $\mathcal{O}(\alpha_\text{s})$ matrix element  
and (d) only $\text{q}\bar{\text{q}}$  for 
various fragmentation options.}
\label{fig:hq_mc1_b}  
\end{figure}

\begin{figure}[htbp]
\centering
    \includegraphics[width=8.4cm]{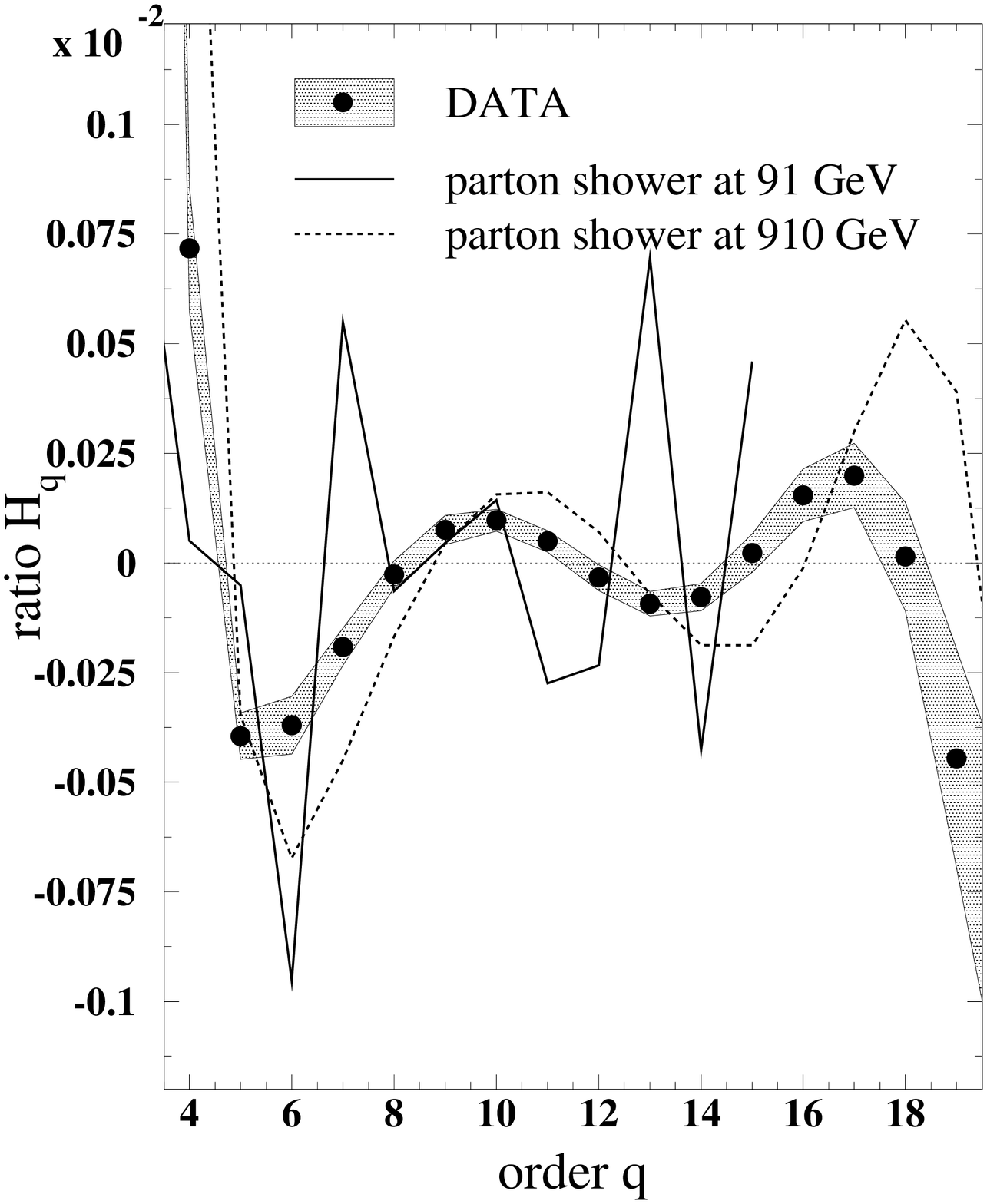}
\scaption{$H_q$ moment of the full sample compared to 
that of parton distributions obtained at 91~\GeV{} and 
at 910~\GeV{}.}
\label{fig:hq_parton}  
\end{figure}

For the b-quark sample (Figs.~\ref{fig:hq_mc1_b}),
all the \cpmd{s} give $H_q$ moments for which oscillatory behavior 
is seen. Except for the LLA parton shower, which has a large 
amplitude compared to the data, the \cpmd{s} for which partons have been 
generated using matrix elements have relatively small oscillations.

Most of the models we have checked show a first negative minimum at  
about $q=5$, followed by an oscillatory behavior 
qualitatively similar to the one seen in the data, no matter 
which model is used to generate the partons and for most of the 
options we tried for the fragmentation. 

Concerning the absence of $H_q$ oscillation in some samples, 
this cannot be related to a specific aspect of the fragmentation.  
The two \cpmd{s} in question concern completely different, even
opposite,  models. One generating partons according to LLA showers, the 
other using only $\text{q}\bar{\text{q}}$. Furthermore, in the first case the 
two models which do not have oscillation are those 
using independent fragmentation with and without resonance.  
But in the second case, it is both string and independent fragmentation 
without resonances which do not show the oscillation.

As an additional test, we also determined the $H_q$ moments of the 
parton multiplicity distribution instead of the \cpmd{}. 
This test was performed on the default parton shower of the JETSET 
model at the center-of-mass energy of 91~\GeV and of 910~\GeV.
The resulting $H_q$ moments are shown in Fig.~\ref{fig:hq_parton} together 
with that of the data. We see that the $H_q$ moments obtained from 
the partons at the \Z{} energy do not present the usual 
oscillatory behavior but has an erratic behavior. However its 
first negative minimum is at $q=6$. At 910~\GeV{}, for which 
the parton multiplicity distribution has about the same mean 
as the \mcpm{} at the \Z{} energy, the $H_q$ moments display 
an oscillatory behavior having about the same amplitude as the data, 
but shifted by one order.

It seems that there is no particular aspect of the Monte Carlo   
responsible for the presence or the absence of the $H_q$ oscillatory behavior, 
but that it is due to a collective effect, various aspects of the Monte Carlo 
contributing to the oscillations in various models. From this Monte Carlo 
study, even if we fail in finding a unique origin for these oscillations, 
we can nevertheless conclude that these oscillations can be reproduced 
without the need of NNLLA of perturbative QCD. 

Therefore, in the next chapter, we will attempt to challenge 
more directly perturbative QCD, by measuring the $H_q$ moments  
for the jet multiplicity distribution at energy scales where
pQCD is the dominant mechanism.

\chapter[Analysis of jet multiplicity distributions with the 
\texorpdfstring{\boldmath{$H_q$}}{$H_q$}]
{Analysis of jet multiplicity distributions with the 
\texorpdfstring{\boldmath{$H_q$}}{$H_q$}}
\label{chap:hqjet}
\markboth{\large{Analysis of jet multiplicity distributions with the $H_q$}}{}

In the previous chapter, we compared the $H_q$ moments of the 
\cpmd{} with various analytical QCD predictions and 
in particular with the NNLLA predictions. 
Originally applying to partons, these predictions are made valid 
for final-state particles by the use of 
the Local Parton-Hadron Duality (LPHD) assumption. This assumption 
may be summarized from our point of interest as  
the hypothesis that the shape of the partonic distribution
is not distorted by the hadronization. Therefore, the 
$H_q$ moments obtained for the partonic distribution should be similar 
to those of the final-state particle multiplicity distribution and, 
by extension, to those of the \cpmd{}. 
Therefore, the comparison of the $H_q$ moments 
of the \cpmd{} to the pQCD prediction rests strongly on the validity 
of LPHD.

In order to remove the dependence on LPHD, 
we use the jet multiplicity distribution instead of the \cpmd{}. 
Since jets obtained for an energy scale  above 1-2~\GeV{} fall 
into the domain of validity of pQCD, they correspond 
to partons. 

In the first section, the various steps needed to 
measure the jet multiplicity distributions are briefly 
described, as well as the estimation of statistical and 
systematic errors. The next section is dedicated 
to the measurement of the $H_q$ moments for a wide range 
of energy scales above and below the 1~\GeV{} limit of
the perturbative QCD region. The results are compared 
to analytical QCD expectations.

\section{Experimental procedures}

The jet multiplicity distribution is defined as 
the distribution of the number of jets reconstructed from an event 
at a given energy scale.
These jets are built with the Durham algorithm~\cite{dur} 
using charged particles only. 
This has the advantage that the jet multiplicity distribution 
obtained with \Ycut{} near zero  corresponds to the \cpmd{}. Since most 
of the comparison is done relative to the \cpmd{}, we 
know that for very small \Ycut{} values, the jet multiplicity 
distribution will approach the \cpmd{}.
  
The main disadvantage is that the method does not use 
the full event, but only charged particles. 
This may affect the number of reconstructed 
jets. The effect will increase with decreasing value of \Ycut{}
 (the effect will be  maximal for \Ycut{=0} where each particle 
is resolved as a jet). However, particles 
are distributed in such a way that, 
at energy scales close to the domain of validity of pQCD, 
the number of reconstructed jets does not depend too 
much on the fact that we use only the charged particles.
Furthermore, since our main interest
is the shape of the distribution, it does not really matter 
that we use the \cpmd{} instead of the multiplicity distribution 
of all particles. Both distributions have equivalent shape. 
Therefore, even for low \Ycut{} where 
the bias is maximum, the difference will not be large. 

Using the Durham algorithm, the cut-off parameter value \Ycut{} 
defines a jet energy scale which 
is closely related to the transverse energy \Ejet{} of the jet,  
\begin{equation}
\label{Escale}
E^\text{jet}_\text{t}=E_\text{cm}\sqrt{y_\text{cut}},
\end{equation}
where $E_\text{cm}$ is the center-of-mass energy of the collision.
Therefore, in this chapter we will preferably refer to 
the jet energy scale, \Ejet{}, rather than to the cut-off 
parameter \Ycut{}. Nevertheless, Table~\ref{tab:ycet} 
shows the correspondence between the two parameters.

\begin{table}[htbp]
\begin{center}
\begin{tabular}{|c|c|c|c|c|c|c|}\hline
\Ycut{}  & $1.2\cdot10^{-6}$ & $2.7\cdot10^{-6}$ & $4.8\cdot10^{-6}$ 
         & $7.5\cdot10^{-6}$ & $1.1\cdot10^{-5}$ & $1.9\cdot10^{-5}$\\ \hline
$E^\text{jet}_\text{t}$ (\GeV) & 0.1 & 0.15 & 0.2 & 0.25 & 0.3 & 0.4\\ 
\hline\hline
\Ycut{}  & $3\cdot10^{-5}$   & $4.3\cdot10^{-5}$ & $1.2\cdot10^{-4}$ 
         & $3.9\cdot10^{-4}$ & $8.2\cdot10^{-4}$ & $1.1\cdot10^{-3}$\\ \hline
$E^\text{jet}_\text{t}$(\GeV)  & 0.5 & 0.6  & 1   & 1.8  & 2.6 & 3 \\ \hline
\end{tabular}
\end{center}
\scaption{Correspondence between the jet energy scale 
\Ejet{} and the cut-off parameter \Ycut{}.}
\label{tab:ycet}
\end{table}
From the range of energy resolution given in Table~\ref{tab:ycet},
we are able to follow in great detail  
the evolution of partons into hadrons, and hence we should 
be able to match any change in the behavior of the $H_q$ moments
to any phase which could occur in the evolution of partons
into hadrons.

The jet multiplicity distributions are reconstructed in  
exactly the same way as the 
\cpmd{}. The only difference is that we build the distribution 
of the number of events with $n_\text{jet}$ jets, $N(n_\text{jet})$, 
instead of the distribution of the number of events with 
$n$ charged-particles, $N(n)$. 
As for the \cpmd{}, these distributions are corrected for 
detector inefficiencies by the Bayesian unfolding method. 
They are also corrected for the selection procedures and initial-state 
radiation using bin-by-bin correction factors. 
Since the $H_q$ of the light- and b-quark samples are found to agree 
with the $H_q$ of the full sample, only results for the full sample 
are shown. Further, we limit our 
study to the case where \kl{} are assumed to be stable.

The estimation of both statistical and systematic 
errors follows the same procedures as already used for the \cpmd{}. 
In the estimation of the systematic errors we include contributions
from changing the track quality criteria  and 
the event selection, from the 
influence of the unfolding method and also a contribution 
coming from Monte Carlo modelling uncertainties.

Furthermore, in the determination of the $H_q$ moments from 
the \jmd{s}, we use the same criteria for the truncation 
as used for the \cpmd{} (\ie{}, we remove in the \jmd{s} 
the same fraction of events as removed by a cut
on multiplicities larger than 48 in the \cpmd{} of the 
full sample). In this way we are able to directly compare 
the $H_q$ moments obtained from the \jmd{s} to  
those obtained from the \cpmd{} of the full sample.
To estimate the statistical errors on the 
$H_q$ moments, we use the Monte Carlo based method 
described in the previous chapter. The systematic 
errors of the $H_q$ moments are obtained in the usual way.

\section[$H_q$ moments of the jet multiplicity distributions]
{\boldmath{$H_q$} moments of the jet multiplicity distributions}

\begin{figure}[htbp]
\centering
    \includegraphics[width=8.4cm]{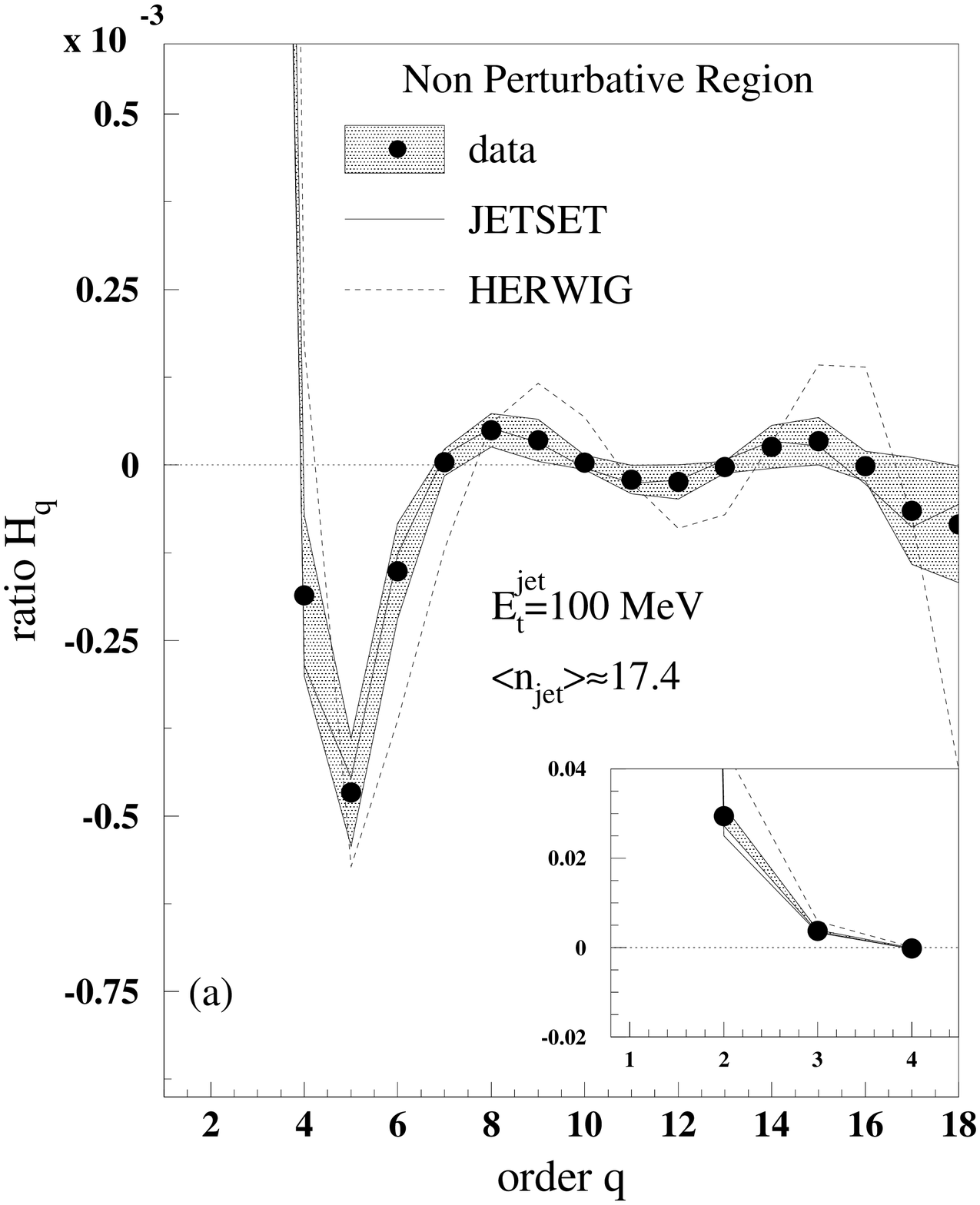}
    \includegraphics[width=8.4cm]{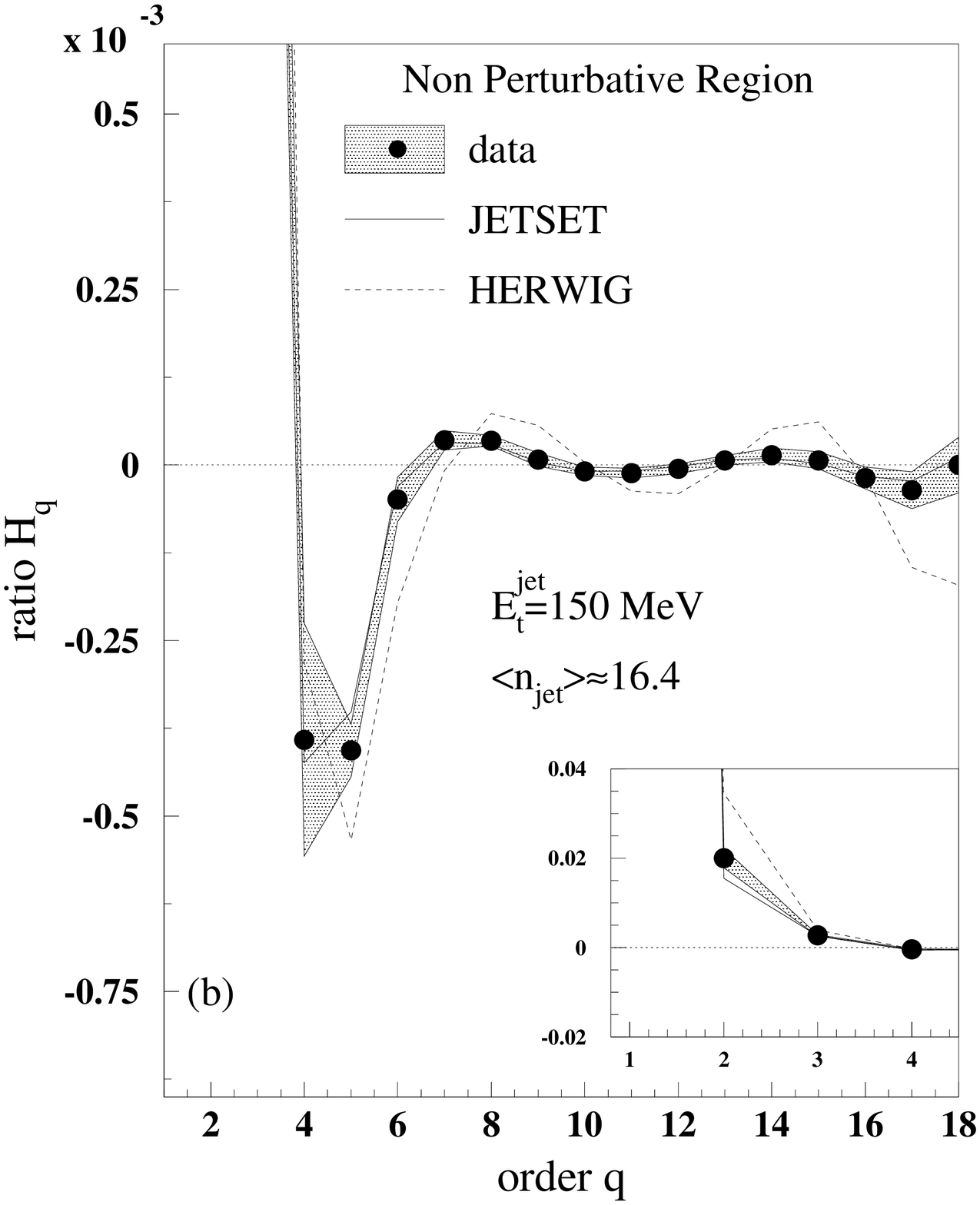}

    \includegraphics[width=8.4cm]{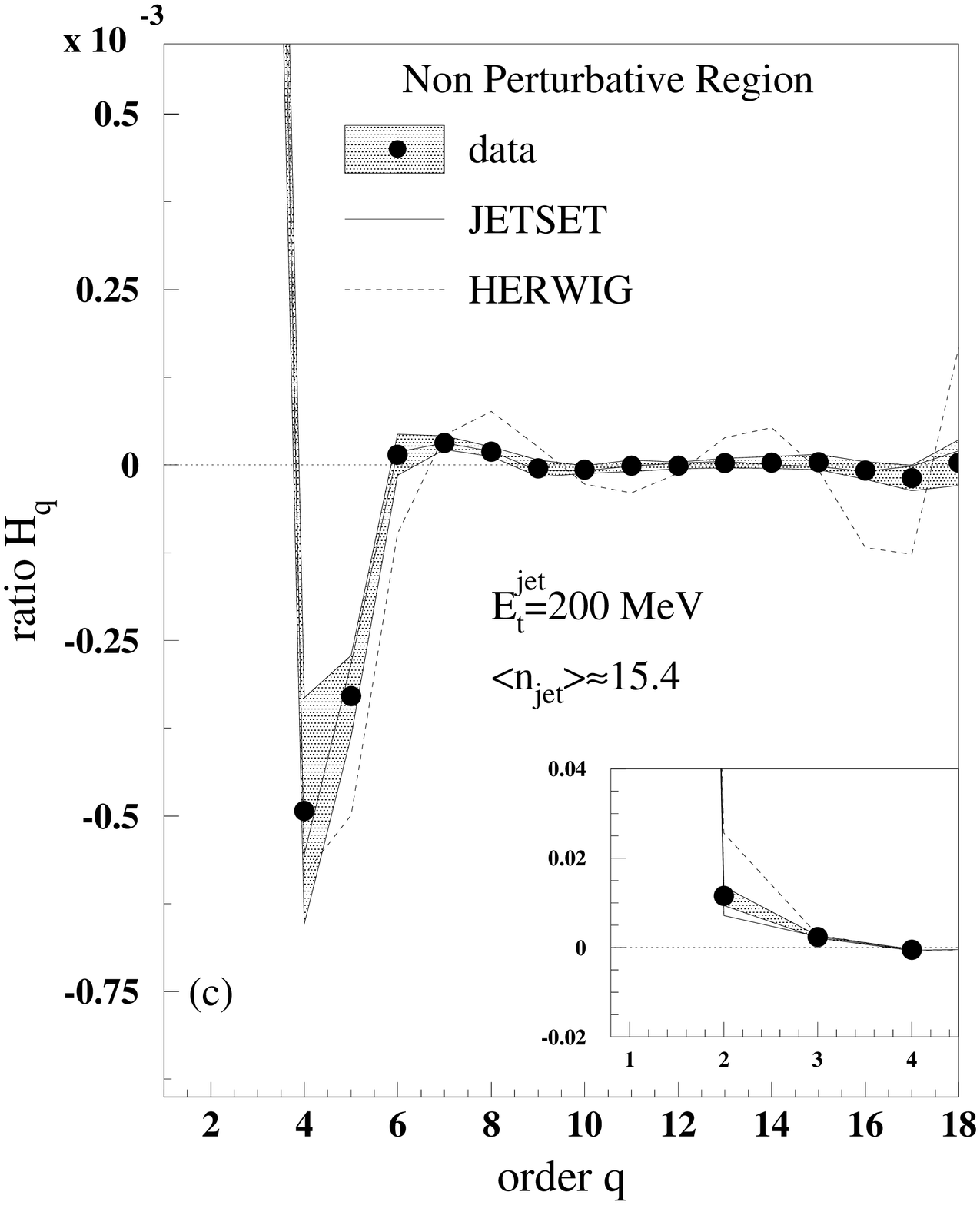}
    \includegraphics[width=8.4cm]{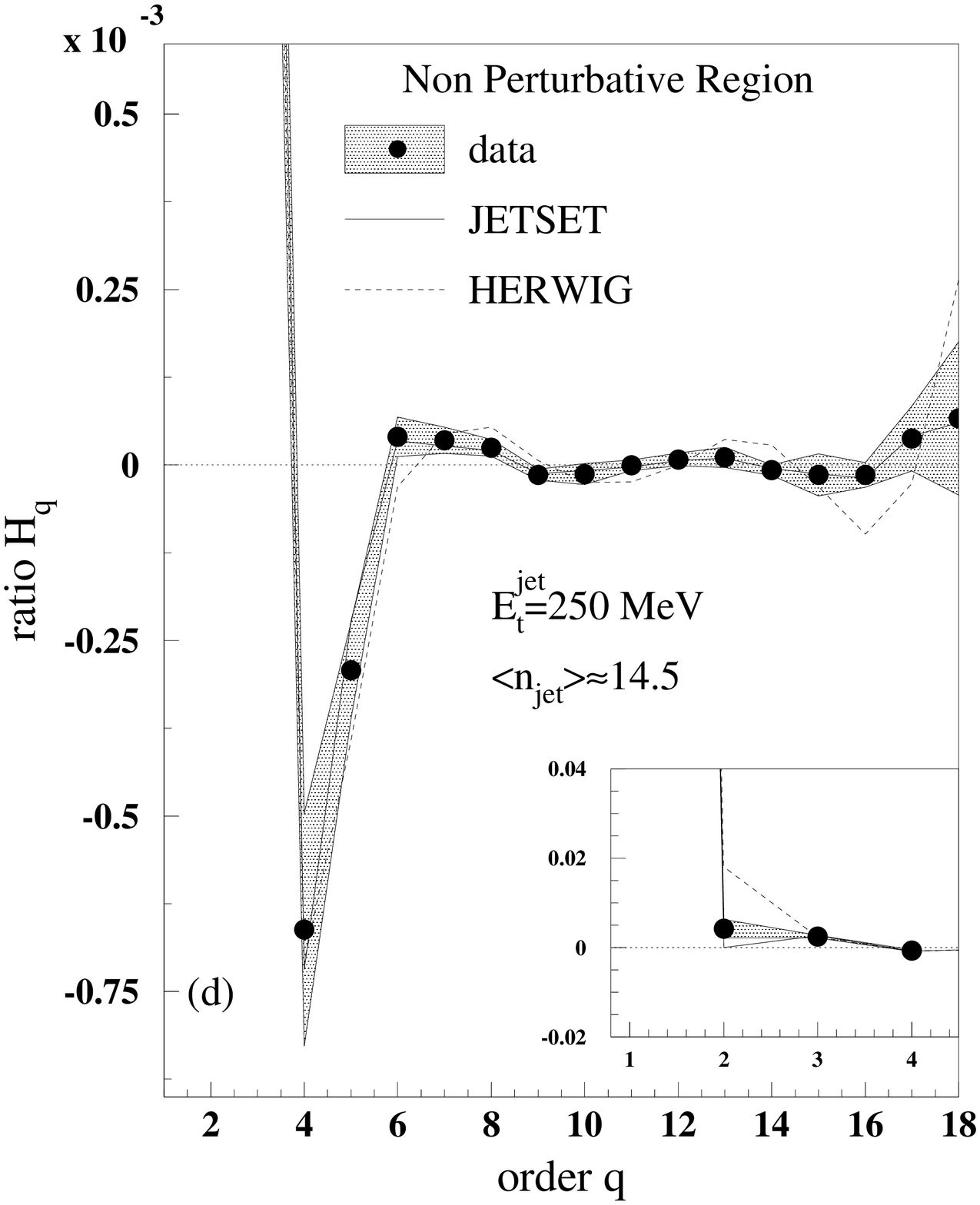}
\scaption{$H_q$ moments of \jmd{} obtained with (a) \Ejet{=100~\MeV{}}, 
(b) \Ejet{=150~\MeV{}}, (c) \Ejet{=200~\MeV{}} and 
(d) \Ejet{=250~\MeV{}}.}
\label{fig:hq_j1234}  
\end{figure}

\begin{figure}[htbp]
\centering
    \includegraphics[width=8.4cm]{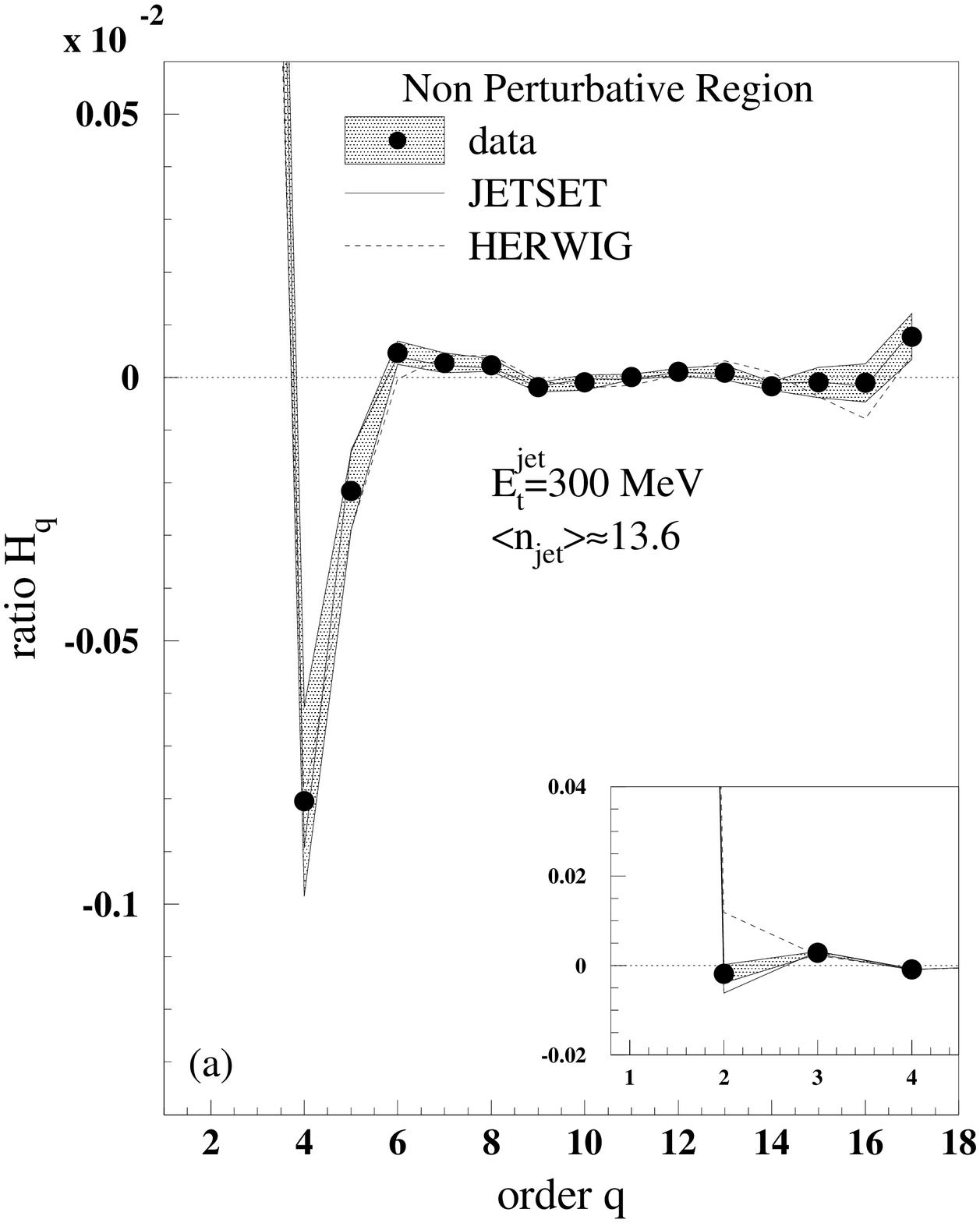}
    \includegraphics[width=8.4cm]{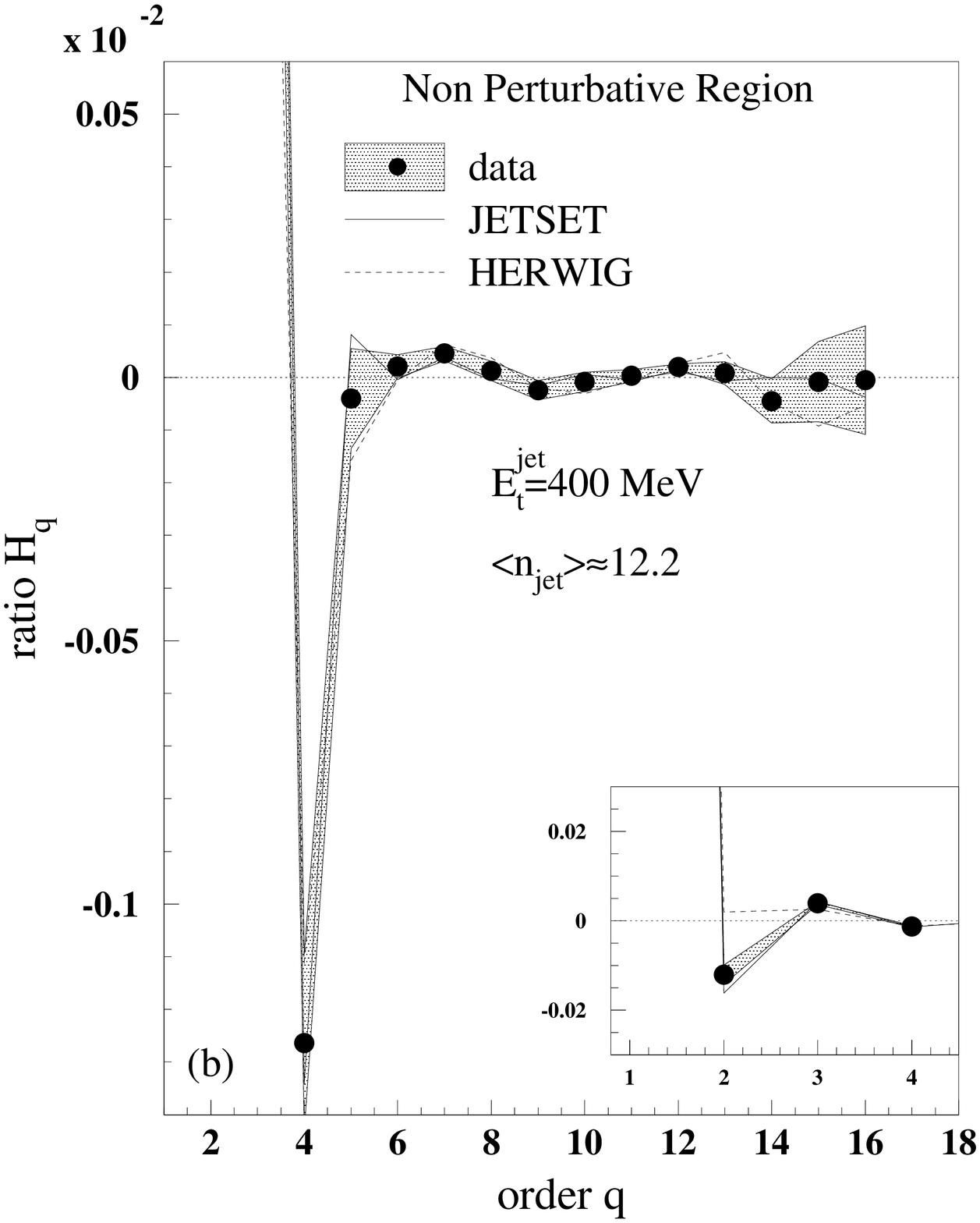}

    \includegraphics[width=8.4cm]{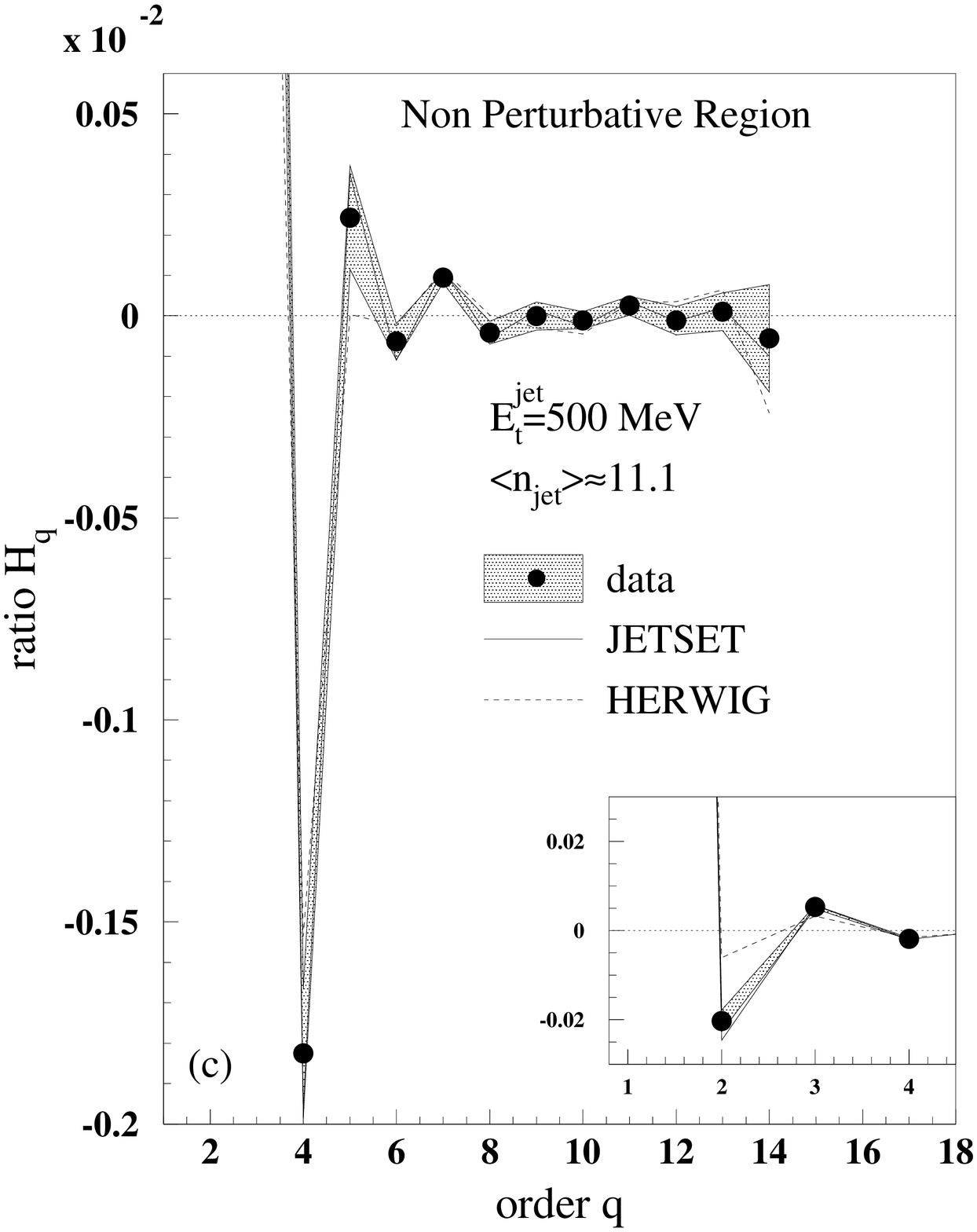}
    \includegraphics[width=8.4cm]{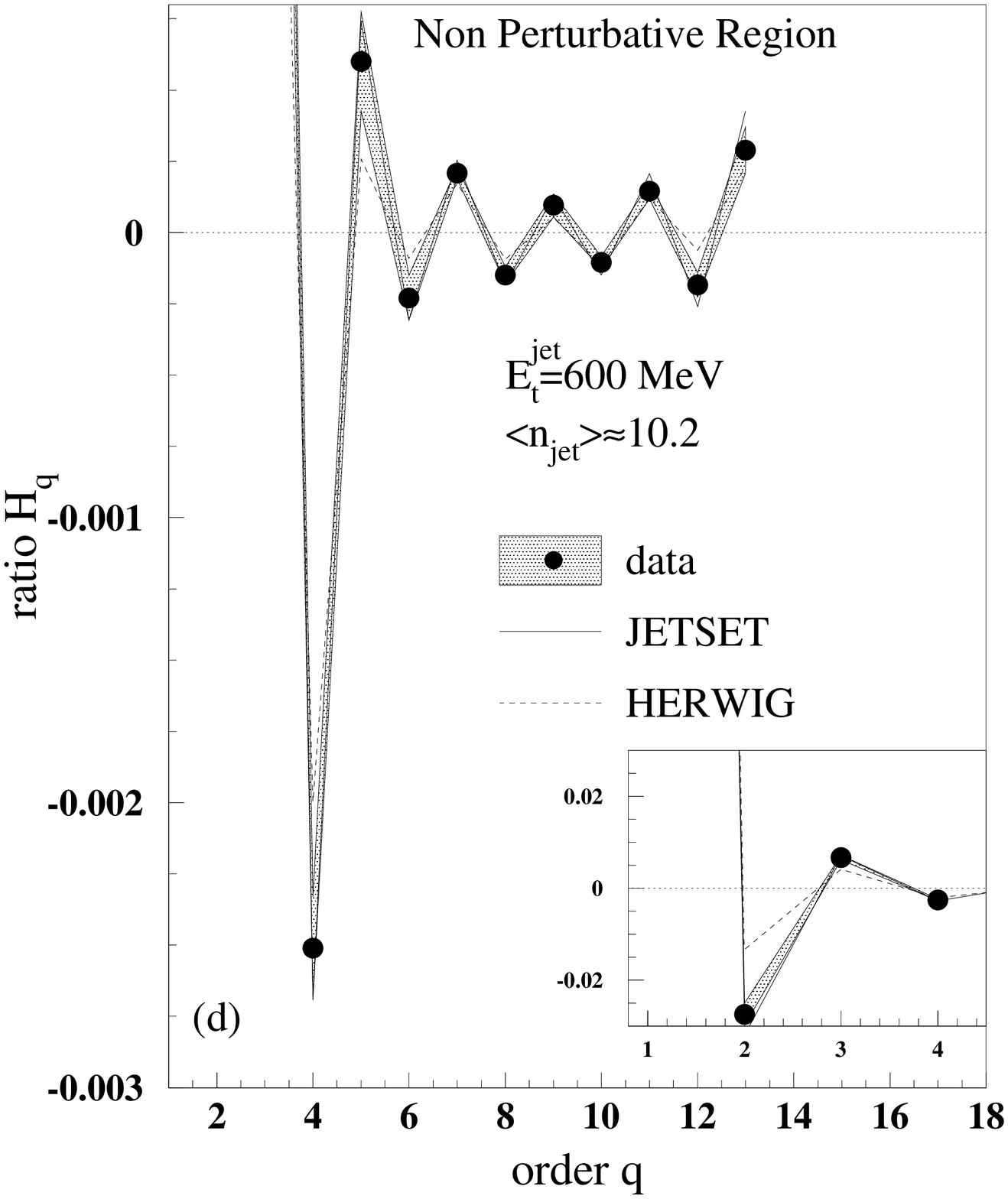}
\scaption{$H_q$ moments of \jmd{} obtained with (a) \Ejet{=300~\MeV{}}, 
(b) \Ejet{=400~\MeV{}}, (c) \Ejet{=500~\MeV{}} and  
(d) \Ejet{=600~\MeV{}}.}
\label{fig:hq_j5678}  
\end{figure}

\begin{figure}[htbp]
\centering
    \includegraphics[width=8.4cm]{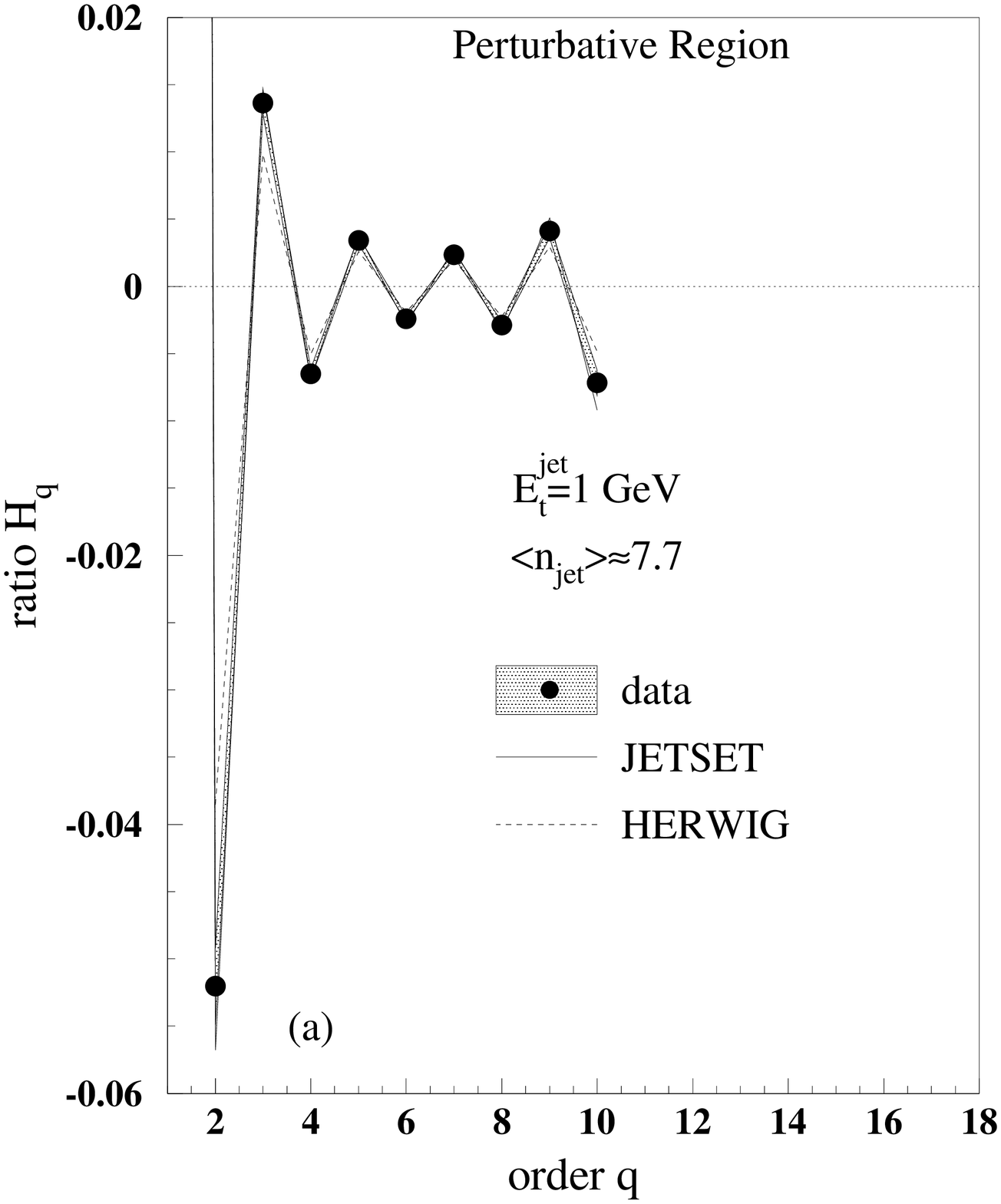}
    \includegraphics[width=8.4cm]{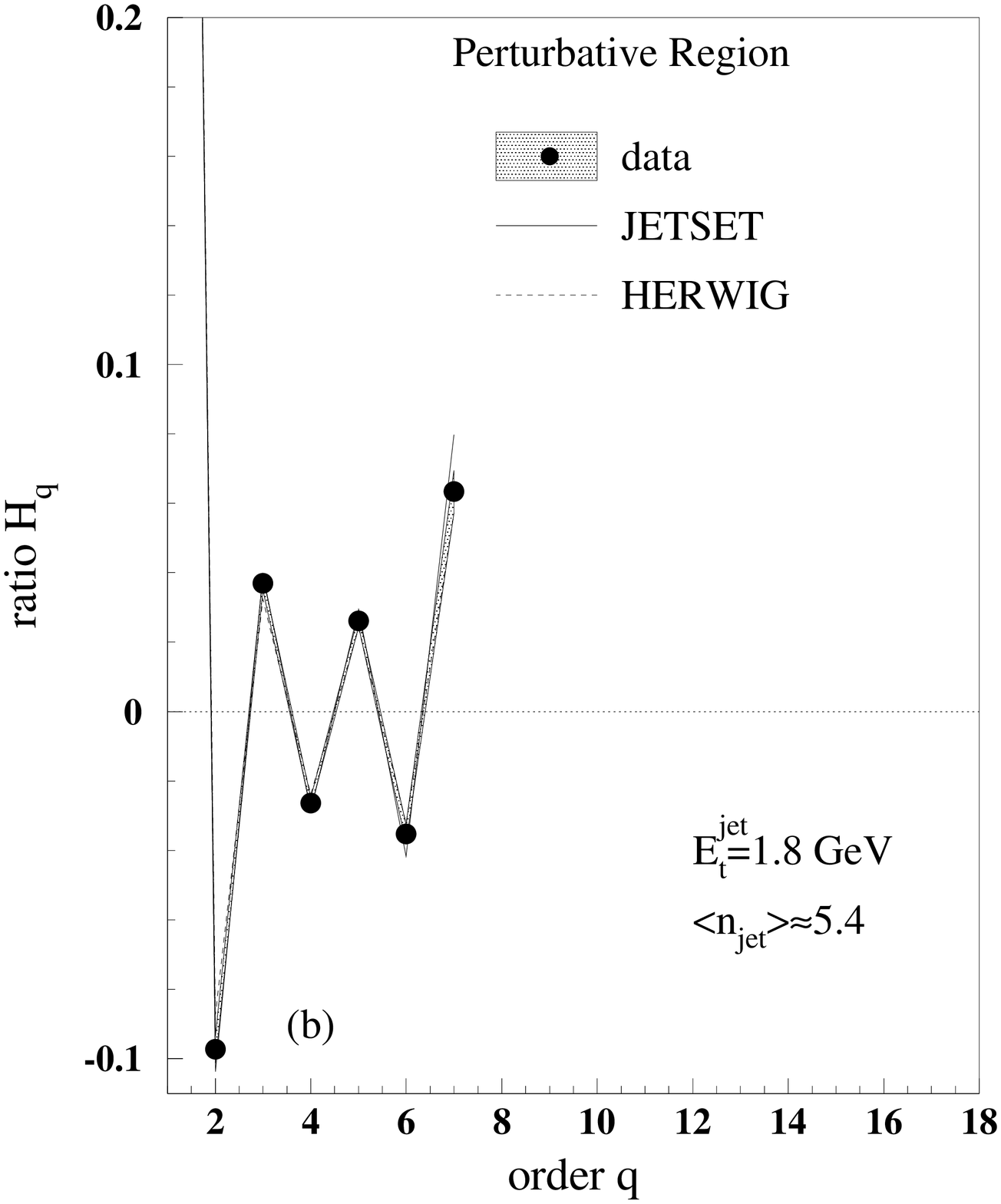}

    \includegraphics[width=8.4cm]{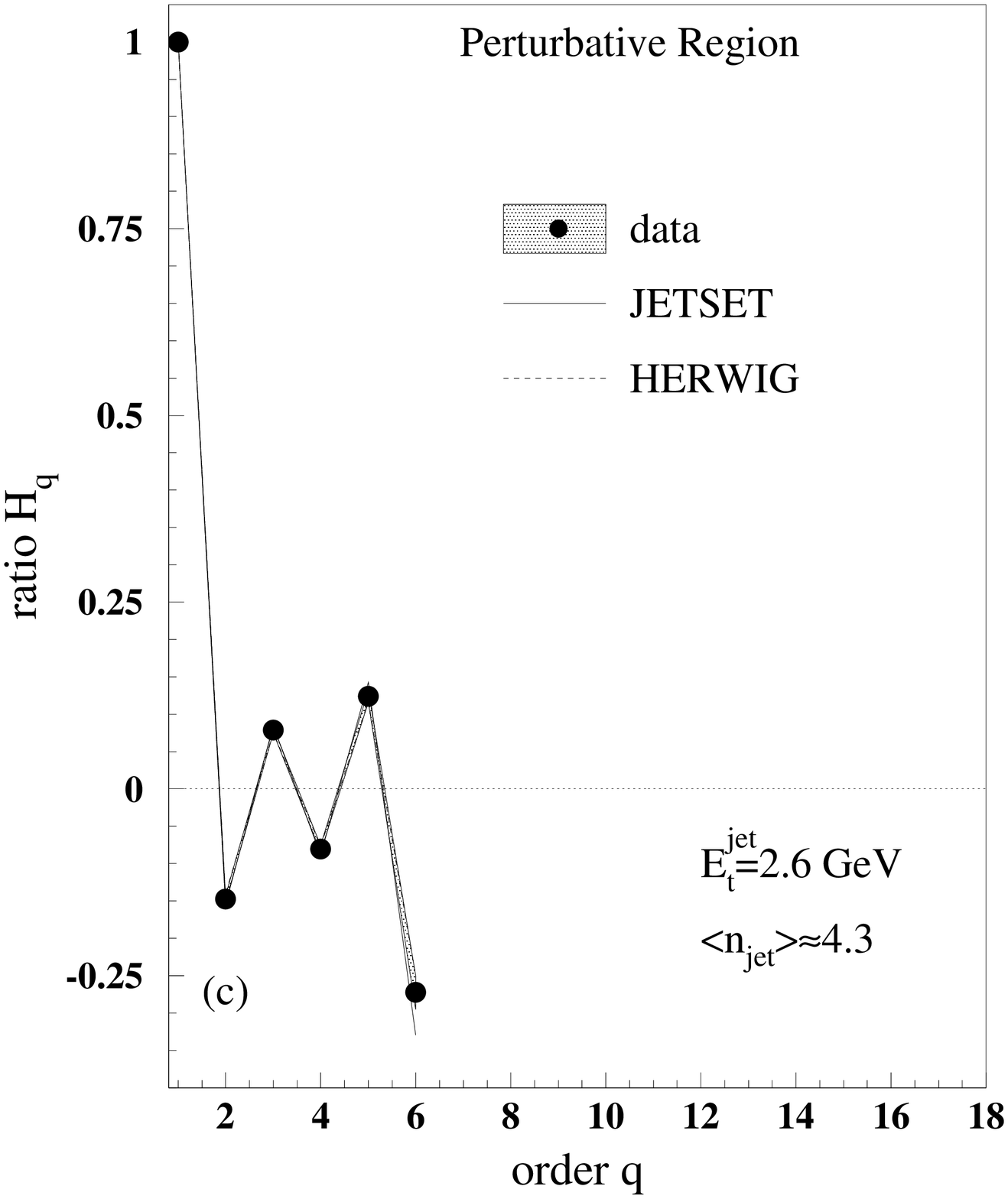}
    \includegraphics[width=8.4cm]{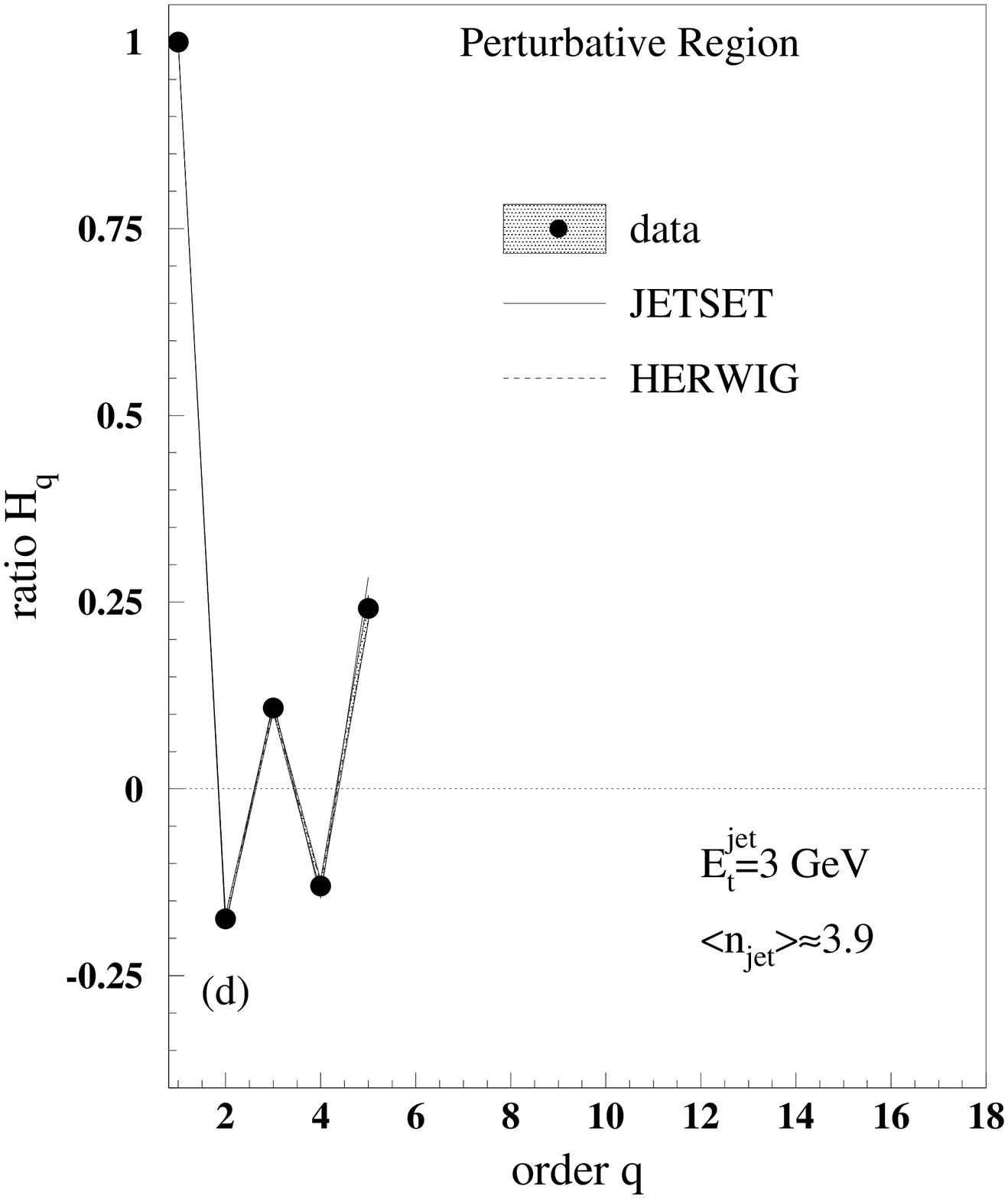}
\scaption{$H_q$ moments of \jmd{} obtained with (a) \Ejet{=1~\GeV{}}, 
(b) \Ejet{=1.8~\GeV{}},(c) \Ejet{=2.6~\GeV{}} and  
(d) \Ejet{=3~\GeV{}}.}
\label{fig:hq_j9112}  
\end{figure}

The $H_q$ moments determined from the \jmd{s} are shown in 
Figs.~\ref{fig:hq_j1234} to~\ref{fig:hq_j9112}, for  
the \Ycut{} values given in Table~\ref{tab:ycet}. 
For \Ejet{=100~\MeV{}} (Fig.~\ref{fig:hq_j1234}(a)), 
the $H_q$ moments 
show a first negative minimum at $q=5$ and oscillatory 
behavior for larger vales of $q$. This behavior is similar 
to that observed for the \cpmd{}, but the oscillations 
are smaller in amplitude. By increasing the energy resolution 
of the jet algorithm, the amplitude of the oscillation decreases 
further. At \Ejet{=200~\MeV{}} (Fig.~\ref{fig:hq_j1234}(c)), 
the first negative minimum has shifted to $q=4$, 
the oscillation has disappeared for $q>8$ and is much 
reduced between $q=6$ and $q=8$. Between \Ejet{=200~\MeV{}} 
and \Ejet{=300~\MeV{}} (Fig.~\ref{fig:hq_j5678}(a)), 
the first negative minimum 
at $q=4$ deepens sharply.  
The same sharp decrease is also 
seen for $q=2$ in the insert. For $q>6$, $H_q$ remains as it was 
at lower energy scales. At \Ejet{=400~\MeV{}} (Fig.~\ref{fig:hq_j5678}(b)), 
a new first minimum appears at $q=2$ 
with a much deeper value than for $q=4$, and for $q<5$ 
$H_q$ alternates between positive and negative values.
For larger $q$ values, the $H_q$ remain roughly constant about 0.

At \Ejet{=500~\MeV{}} (Fig.~\ref{fig:hq_j5678}(c)), 
we have a complete change of the $H_q$: 
the smooth quasi-oscillations we see   
for the lowest energy scales have disappeared and $H_q$ 
alternates between positive and negative values for all $q$. 
We notice also that this sign-changing behavior has an amplitude 
almost twice as large as than at \Ejet{=400~\MeV{}}. 
As we increase the energy scale further, there is no noticeable 
change in the $H_q$ with respect to \Ejet{=500~\MeV{}}. 
The only difference is that the absolute value of each $H_q$ has increased.
At \Ejet{=1~\GeV{}} (Fig.~\ref{fig:hq_j9112}(a)), 
we enter the domain of validity of the perturbative region and 
the behavior is similar to that already observed for 
$E^\text{jet}_{t}>500~\MeV{}$, 
but the amplitude is much larger. 

It seems that 
the increase in the scale of the $H_q$ values 
is due to the decrease of the mean jet multiplicity. 
Increasing the energy scale further does not bring anything new.  
We still have the sign-changing behavior as we already 
observed, but again the $H_q$ moments have greater values.

In these figures we also find an overall good agreement 
with JETSET. HERWIG is seen to disagree with the data 
at small energy scales ($E^\text{jet}_{t}<250~\MeV{}$). 
Nevertheless, this disagreement diminishes with the 
increase of the energy scale, and for larger energy 
scales and in particular in the perturbative region, 
HERWIG agrees well with the data.

\begin{figure}[htbp]
\centering
    \includegraphics[width=8.4cm]{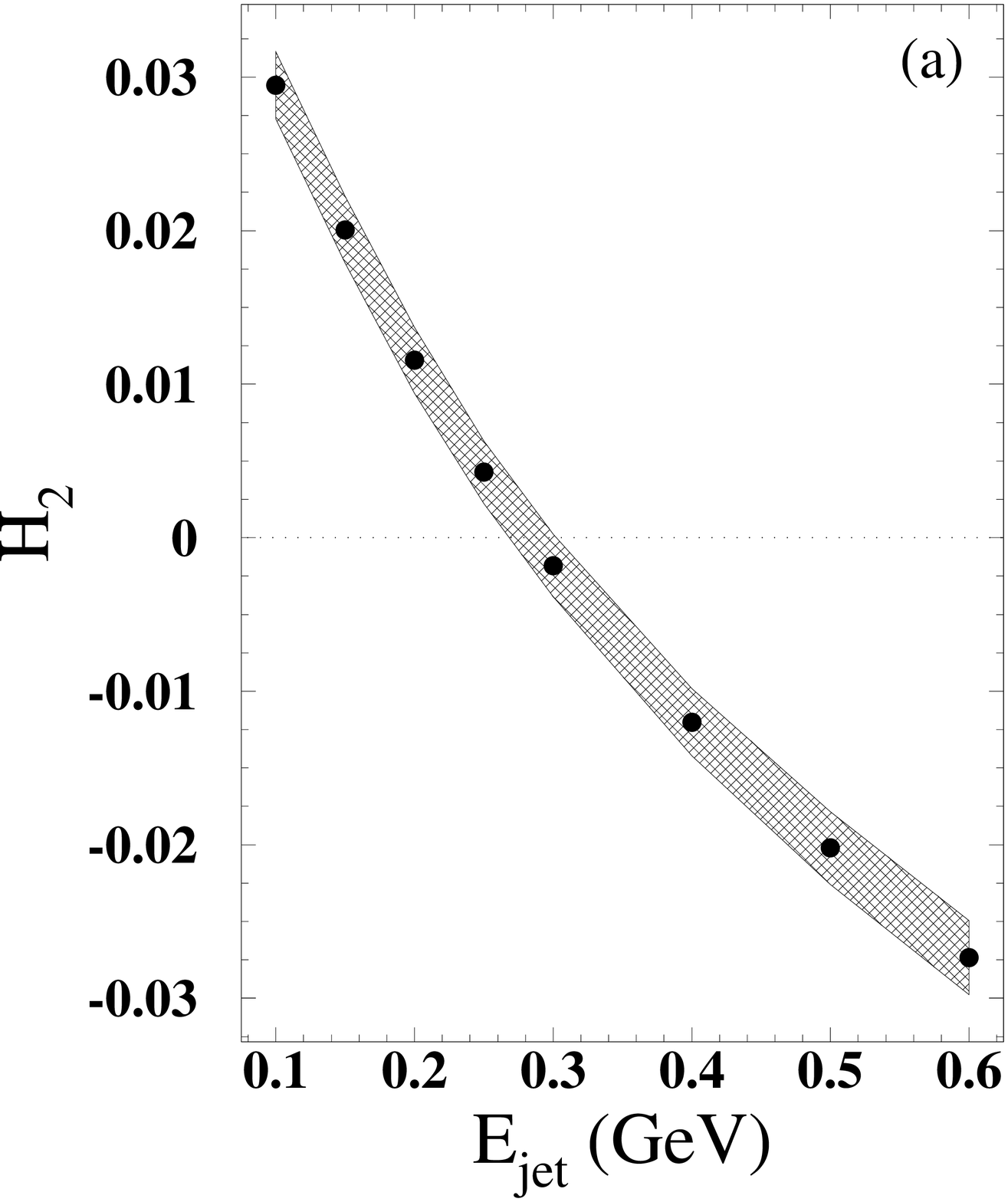}
    \includegraphics[width=8.4cm]{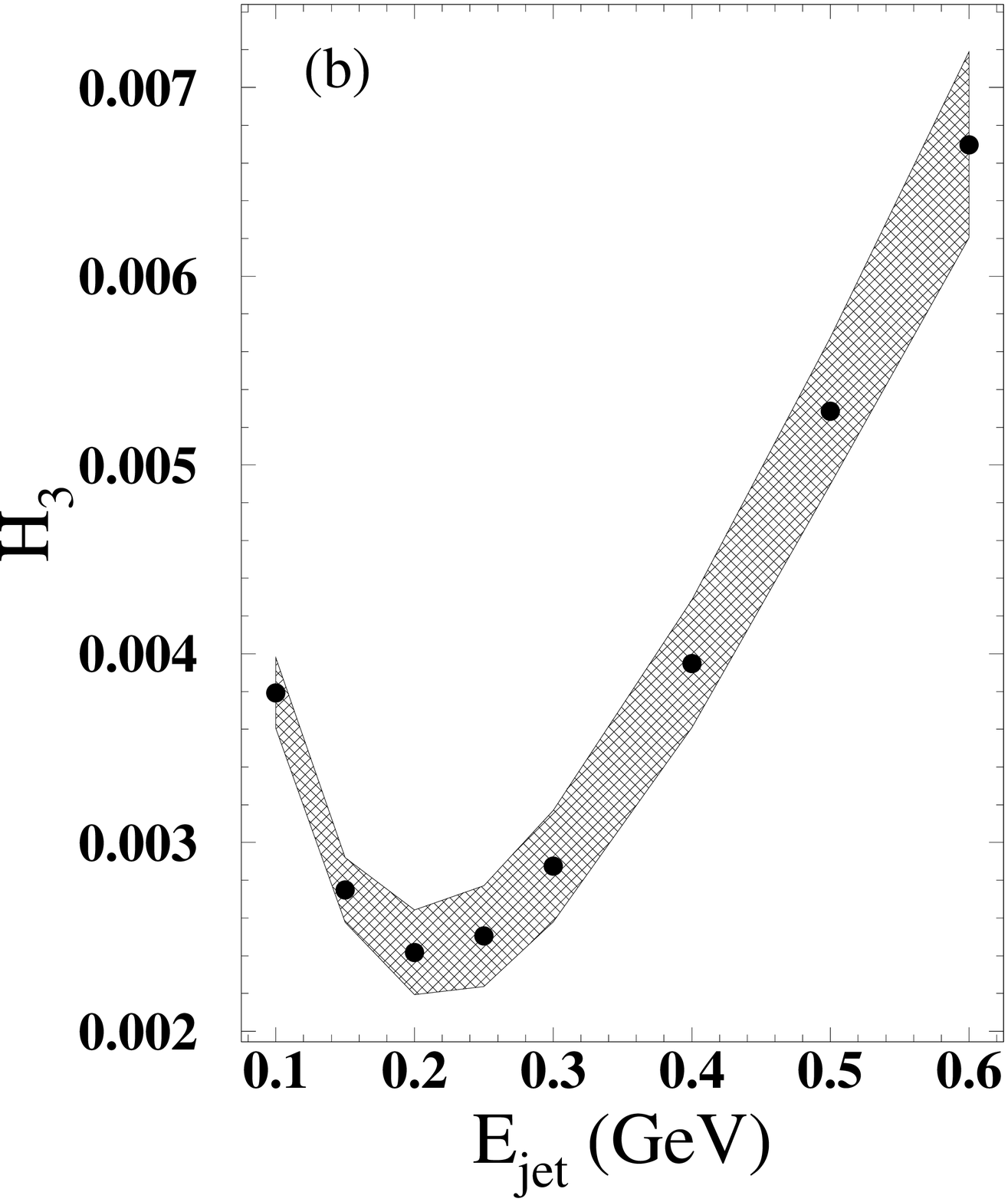}

    \includegraphics[width=8.4cm]{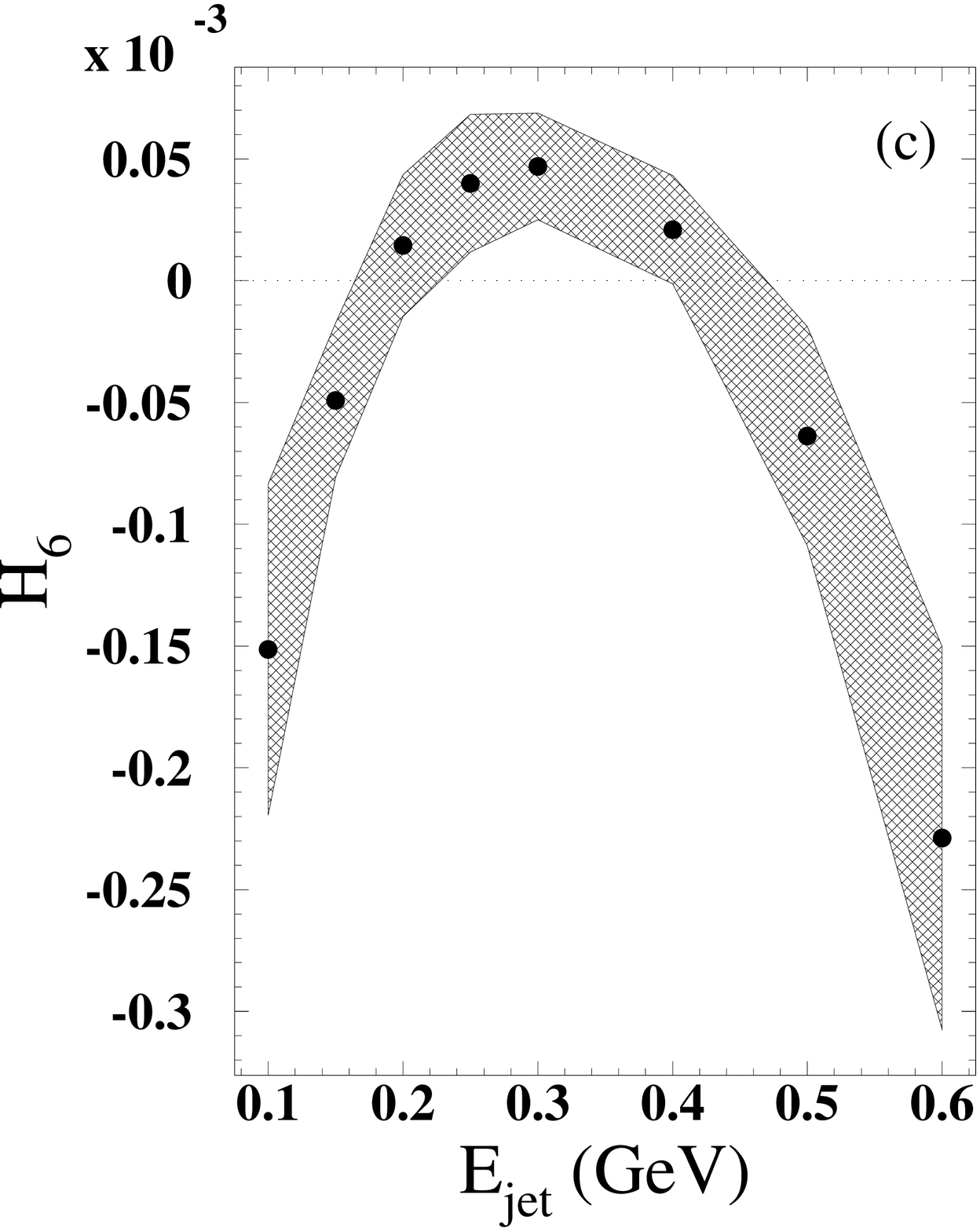}
    \includegraphics[width=8.4cm]{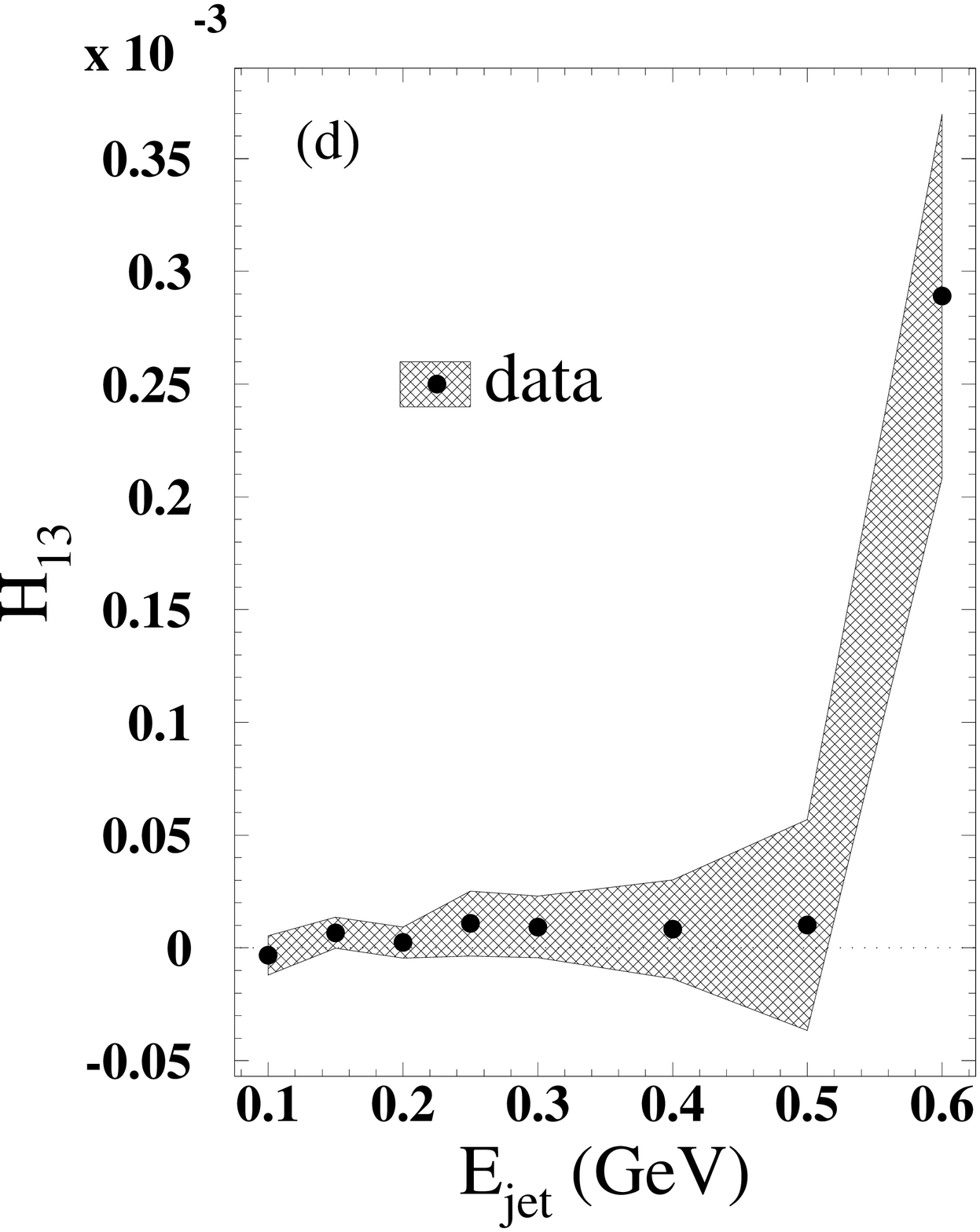}
\scaption{Evolution of $H_2$, (a), $H_3$, (b), 
$H_6$, (c), and $H_{13}$, (d),  with the jet energy scale.}
\label{fig:hvsej2}  
\end{figure}

We can summarize the evolution of the $H_q$ behavior 
with the energy scale in three main steps. 
At low energy scale,  
$E^\text{jet}_\text{t}<200~\MeV{}$,  
the $H_q$ behavior is qualitatively similar 
to that of the \cpmd{}, 
with a first negative minimum around $q=5$ and  
oscillatory behavior for larger values of $q$.
The second step,    
$200~\MeV{}<E^\text{jet}_\text{t}<500~\MeV{}$,   
can be described as a transition phase where the 
oscillation disappears and a first negative 
minimum appears at $q=2$. Finally, in the 
third step, $E^\text{jet}_\text{t}>500~\MeV{}$,  
a completely different behavior is observed 
where the $H_q$ alternates between negative 
and positive values.

The details of this evolution are easier to see, 
in particular for the transition phase,
when the $H_q$ is plotted as a function of \Ejet{}. 
Fig.~\ref{fig:hvsej2}(b) illustrates the 
beginning of the transition phase. 
It is characterized by the appearance of a 
minimum at \Ejet{=200~\MeV{}} in $H_3$, which 
also marks the end of the step of the $H_q$ 
oscillatory behavior similar to that of the  
\cpmd{}. Fig.~\ref{fig:hvsej2}(a) represents 
the evolution of $H_2$ with the energy scale 
and illustrates one of the major changes in 
the $H_q$ behavior which occurs during the 
transition phase. 
This is the appearance of the first negative 
minimum at $q=2$ and the first sign of the 
sign changing $H_q$ behavior which  
characterize the $H_q$ at large \Ejet and 
in the perturbative region.
Fig.~\ref{fig:hvsej2}(c) shows the 
evolution of $H_6$ with the energy scale.  
We can see that all the transition 
phase is located around the maximum value.
The end of this transition phase and the 
beginning of the full sign-changing $H_q$ 
behavior at large $q$ is easily seen, \eg{} $H_{13}$   
in Fig.~\ref{fig:hvsej2}(d).  
The whole transition phase is characterized by a
 plateau where $H_{13}$ has a stable value 
near 0, up to an energy scale of $500\MeV$. 

\section{Comparison with theoretical expectations}

Assuming that the $H_q$ behavior is described by 
analytical QCD, then we should expect to find this 
behavior for partons.
Assuming that the jets correspond to the partons, 
we should find the $H_q$ behavior predicted for partons 
for the jets produced at an energy scale of 1-2~\GeV.
But as we have discussed above, in the perturbative 
region we see a sign-changing $H_q$ behavior which is not 
described by any of the analytical QCD predictions, and 
in particular is not described by the NNLLA. 

The oscillatory behavior appears only 
in the final stages of the hadronization process, 
far away from the perturbative region. 
Therefore, it is not possible to link this oscillatory 
behavior to the NNLLA.
We can conclude that the $H_q$ oscillatory behavior 
observed for the \cpmd{} and for \jmd{s} obtained 
at very low energy scales has no relation with 
the $H_q$ behavior predicted by the NNLLA.

\chapter{2- and 3-jet event multiplicity distributions}\label{chap:hq23j}

In search of an alternative way of describing the shape of the \cpmd{}, 
we test, in this chapter, phenomenological approaches~\cite{23jet,lbquark} 
by making use of the \cpmd{s} of the 2-jet and 3-jet events taken from 
the full sample, as well as the light- and b-quark samples.
Two main approaches are tested experimentally in this chapter:

In the first one, we assume that the dominant influence on the shape 
of the \cpmd{} of the full sample comes from the jet 
configuration of the events. 
Assuming that the \cpmd{} of both the 2-jet and the 3-jet events can  
be described by a negative binomial distribution, the full sample 
would be described by a weighted sum of the negative binomial 
distributions of the 2-jet and 3-jet events~\cite{23jet}.
(We first assume here that we 
have only 2-jet and 3-jet events, the 3-jet events  
being the events which are not classified 
as 2-jet events). The parametrization describing the 
full sample would then be
\begin{equation}
\label{eq:2nbd_jet}
f_\text{full}(n)=
    R_2 f^\text{NB}_\text{2-jet}(n,\langle n_2\rangle,k_2)+
 (1-R_2)f^\text{NB}_\text{3-jet}(n,\langle n_3\rangle,k_3), 
\end{equation}
where the parameter of the NBDs are found from the means and dispersions 
measured from the experimental \cpmd{} of these two samples. 
A previous analysis~\cite{deljet} found it possible 
to describe simultaneously the \cpmd{s} 2-, 3- 
and 4-jet events by NBDs for certain \Ycut{} 
values of the JADE algorithm. 
This gave credit to this approach.

As an extension, we also investigate 
this approach for the \cpmd{s} of the light- and b-quark 
samples, assuming they can be related in the same way
as the full sample to the jet configurations of their events.

The second approach relies on the flavor composition 
of the full sample. It 
assumes that both the \cpmd{s} of the 2-jet and the 3-jet events 
can be described by a weighted sum of negative binomial distributions
related to the flavor composition of these 2-jet and 3-jet 
events~\cite{lbquark}. 

\begin{equation}
\label{eq:2nbd_fl}
f_\text{2,3-jet}(x)=R_\text{b} 
f^\text{NB}_{\text{2,3-jet light-quark}}
(x,\langle n_\text{light}\rangle,k_\text{light})+
(1-R_\text{b})f^\text{NB}_{\text{2,3-jet b-quark}}
(x,\langle n_\text{b}\rangle,k_\text{b}),
\end{equation}
where $R_\text{b}$ is the relative hadronic cross section of the b-quark system.

This hypothesis is also tested on the full sample, trying 
in this case to parametrize it using negative binomial distribution to  
describe individually the \cpmd{s} of the light- and b-quark samples.

In order to test all these hypotheses, we decompose the full, light- 
and b-quark samples into 2-jet and 3-jet events obtained for 
different values of the cut-off parameter \Ycut{} of the Durham algorithm.
Their corresponding \cpmd{s} are then reconstructed and used 
to calculate the means and dispersions which are used as 
parameters of the NBD parametrizations.

In the first section, we describe briefly the various steps 
needed to reconstruct the \cpmd{s} of the 2-jet and 3-jet events. 
In the next section, we present the determination of the 
moments (including the $H_q$ moments) of the 2-jet and 
3-jet events obtained from the full, light- and b-quark 
samples.
In the penultimate section, the two phenomenological approaches are  
confronted with the experimental results discussed.
Finally, in the last section, we present an extension 
of the phenomenological approaches, 
by decomposing the full sample into 
well defined jet topologies such as pencil-like 2-jet 
and Mercedes-like 3-jet events with, in addition,  
the remaining events coming from 3-jet events having
a softer gluon jet. 
This approach seems to explain 
the origin of the $H_q$ oscillatory behavior of the 
\cpmd{} of the full sample, as due to the diversity 
of jet topologies and hard gluon radiation which 
coexist at the \Z{} energy.

\section{Experimental procedures}

In this section are briefly summarized all the procedures needed for 
our measurements. Since most of the corrections and error estimations 
have already been discussed in previous chapters, the various steps
needed for the present measurements will be just indicated.

The classification into 2-jet and 3-jet events 
is achieved using the 
Durham jet algorithm. We define as 2-jet event, an event
which has been classified at a given value of the  cut-off parameter, \Ycut{} 
as 2-jet event. Events not classified as 2-jet are called 3-jet. 

Since the charged particles represent only a fraction 
of the particles produced (and of the energy radiated) during an \ee{} collision, 
it is preferable, as we do, to apply the jet algorithm to the 
whole event and not only to charged particles.
While most of the events are not affected by this change, some of the 
events, having $y_\text{23}$ values ($y_\text{23}$ is the value 
of \Ycut{} for which a 2-jet event becomes a 3-jet event)
close to the \Ycut{} value used for a particular 2-jet definition,
will be rather sensitive to whether the neutral particles 
are included in the jet algorithm.  

Once the event has been classified as a 2- or 3-jet event, its number of 
charged particles is extracted. The \cpmd{s} of the 2- or 3-jet events 
are then built from the distribution of the number of charged particles 
in the same way as in the previous 
chapters. Also reconstruction and corrections as well as the estimation 
of the statistical and systematic errors are identical 
(see Chapter~\ref{chap:cpmd} for more details). 

In order to have a more flexible definition of 2-jet events and to avoid 
any strong dependence, in our analysis, on a particular \Ycut{} value,
we use six different values of the cut-off parameter. 
This also enables us to study the evolution of the  
the \cpmd{} with the cut-off parameter.
The six \Ycut{} values have been chosen such that both the 2-jet and 
the 3-jet sub-samples always have enough statistics to allow  simultaneously 
both types analyses. 
The fraction of 2-jet events, $R_2$, obtained for the 
\Ycut{} values used in our analysis are given in Table~\ref{tab:R2} 
for the full, light- and b-quark samples, respectively.  
While, in general, we observe rather 
similar 2-jet fraction for the full, light- and b-quark samples for each 
\Ycut{}, we see a depletion in 2-jet events 
obtained with \Ycut{=0.002} for the b-quark sample. This may be explained 
by some mis-assignment between 2-jet and 3-jet events due to 
the weak decay of the b quark at such a low \Ycut{} value. 

\begin{table}[htbp]
\begin{center}
\begin{tabular}{|c|c|c|c|}\hline
\Ycut{}   & full sample & light-quark sample & b-quark sample \\ \hline
0.030     & $82.0\%$    &  $81.2\%$          & $83.0\%$       \\
0.015     & $71.5\%$    &  $71.1\%$          & $73.3\%$       \\
0.010     & $64.6\%$    &  $64.1\%$          & $66.7\%$       \\
0.006     & $55.1\%$    &  $54.6\%$          & $57.0\%$       \\
0.004     & $46.5\%$    &  $46.4\%$          & $46.8\%$       \\
0.002     & $29.3\%$    &  $31.3\%$          & $22.0\%$       \\ \hline
\end{tabular}
\end{center}
\scaption{Fraction of 2-jet events, $R_2$, for the 
full, light- and b-quark samples obtained for the 6 \Ycut{} values used in this analysis.}
\label{tab:R2}
\end{table}
The fully corrected \cpmd{s} of the 2- and 3-jet events obtained from the 
full, light- and b-quark samples are then used for our analysis.
We restrict the analysis to the \cpmd{s} where \kl{} are considered  
stable.

\begin{figure}[htbp]
\begin{center}
    \includegraphics[width=8.4cm]{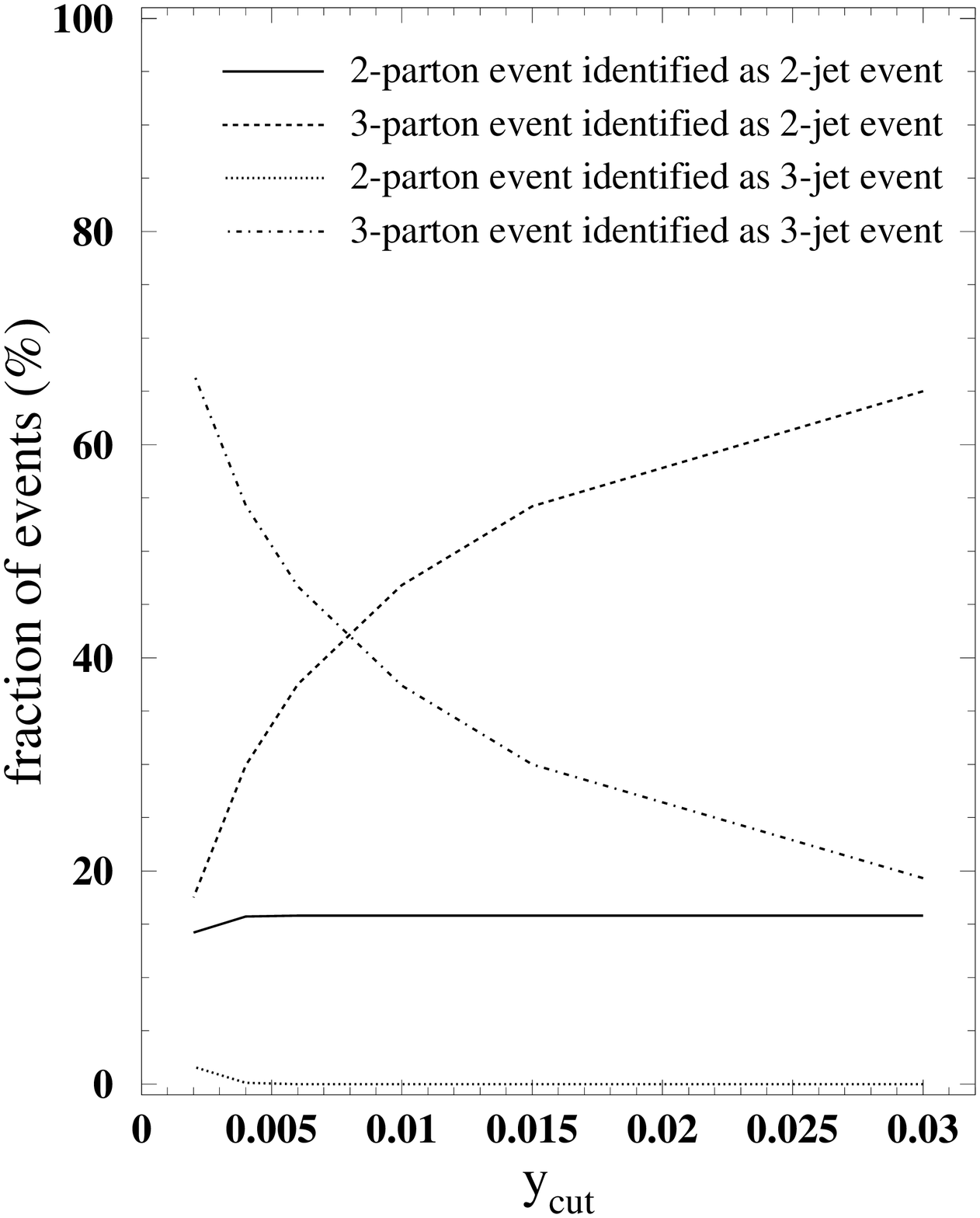}
    \includegraphics[width=8.4cm]{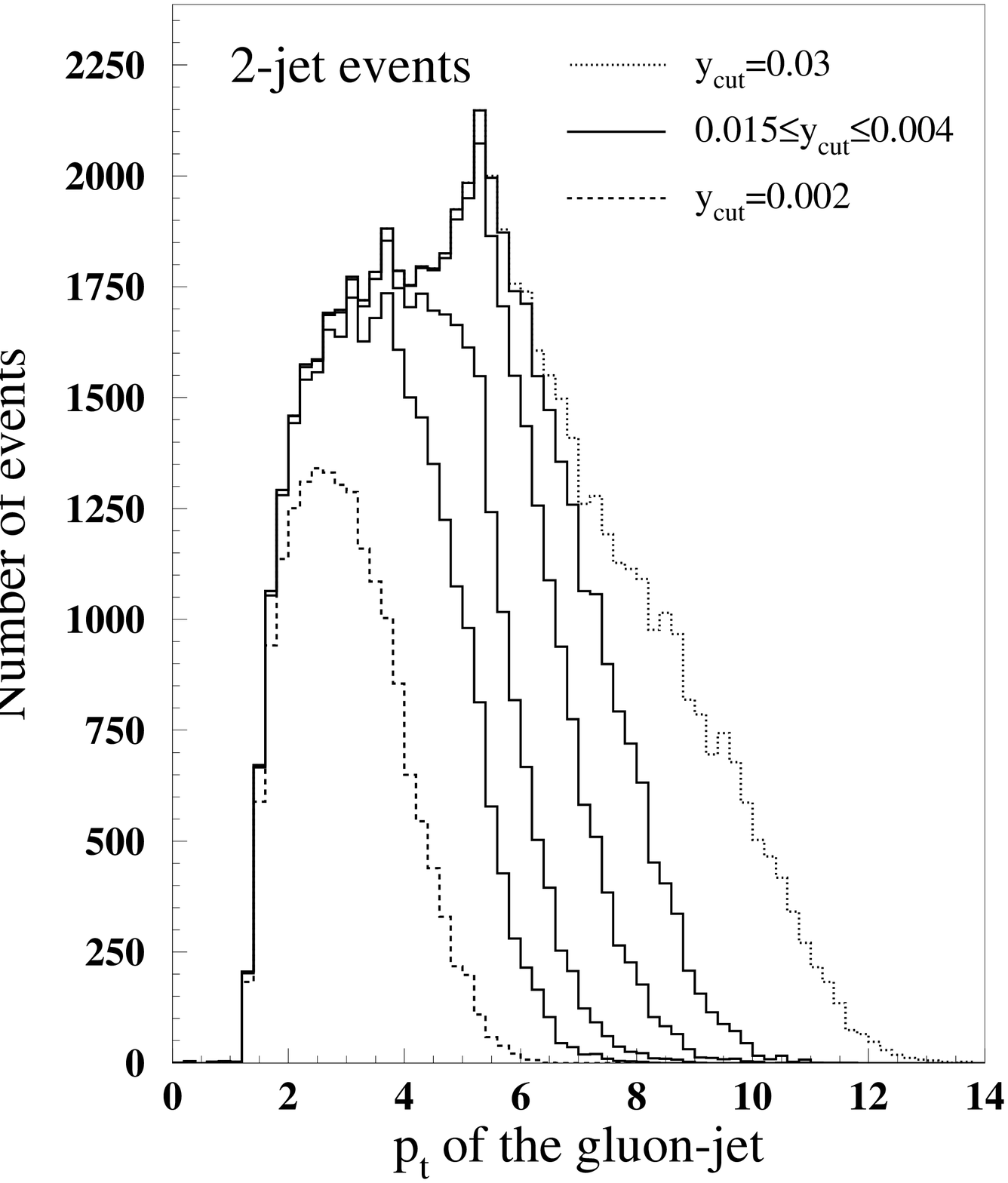}
\end{center}
\vspace{-1cm}
\begin{multicols}{2}
\scaption{Fraction of events generated with JETSET 
$\mathcal{O}(\alpha_\text{s})$ with 2 or 3 partons and 
classified as 2- or 3-jet.}
\label{fig:parton_jet}  
\scaption{Transverse momentum of the gluon jet in 
JETSET $\mathcal{O}(\alpha_\text{s})$ 3-parton events 
classified as 2-jet events for several \Ycut{} values.}
\label{fig:pt_gluon} 
\end{multicols}
\end{figure}

Since the fraction of identified 2- or 3-jet events varies with the 
cut-off parameter value, \Ycut{}, it is interesting to link
this number of reconstructed jets to the number of primary partons.  
To check that, we use Monte Carlo events generated in a relatively simple 
case. We use JETSET to generate events according to the 
$\mathcal{O}(\alpha_\text{s})$ 
matrix element followed by the Lund string fragmentation. 
In this simple case, the particles produced in the final state  
come from a maximum of 3 partons. 
We use the Durham algorithm with the same \Ycut{} values used in our 
analysis to reconstruct jets from the final-state particles and 
we compare the result to the number of initial partons  
generated by JETSET.
Fig.~\ref{fig:parton_jet} shows the fraction of events with 
2- and 3-parton final states which have been identified 
as 2- or 3-jet events by the jet algorithm.

We first note that the 2-parton events 
represent only a small fraction ($16\%$) of the events.
We find that these events are almost always 
reconstructed as 2-jet events by the jet-algorithm for the 
range of \Ycut{} used in the analysis even for small \Ycut{} 
values.   
At \Ycut{=0.002}, the fraction of events mistaken as 
3-jet events is only $1.6\%$. For this \Ycut{} value 
about $80\%$ of the 3-parton events are identified 
as 3-jet, however this fraction will decrease when 
the \Ycut{} value will be increased.   
The remaining 3-parton events correspond to 
pencil-like events accompanied by a very 
low transverse momentum gluon, collinear to the quark-jet 
direction. These events cannot be  
identified as 3-jet events by the jet algorithm even 
at very low \Ycut{} value. Using a value smaller than 
0.002 would increase the fraction of 2-parton events 
mistaken as 3-jet but would not have any effect 
in reducing the fraction of 3-parton events 
mistaken as 2-jet events.
Therefore, changing the \Ycut{} value will act only 
on the identification of 3-parton events as 2- or 
3-jets. 
By changing the \Ycut{} value, we mainly change 
the ``hardness'' criterion of the primary gluon. 
This is illustrated in Fig.~\ref{fig:pt_gluon}, which shows 
the transverse momentum of the gluon in 3-parton events which 
have been identified as 2-jets at the different \Ycut{} values.
Since the cut-off parameter of the Durham algorithm is linked 
to the transverse momentum of the jet, 
the gluon will be considered part of the quark jet  
if the transverse momentum of the gluon is smaller 
than $E_\text{cms}\sqrt{y_\text{cut}}$. 
Therefore, the jet configuration depends mainly on the 
hardness of the primary gluon.

\section{Moments of the 2- and 3-jet events}

\begin{figure}[htbp]
\centering
    \includegraphics[width=8.4cm]{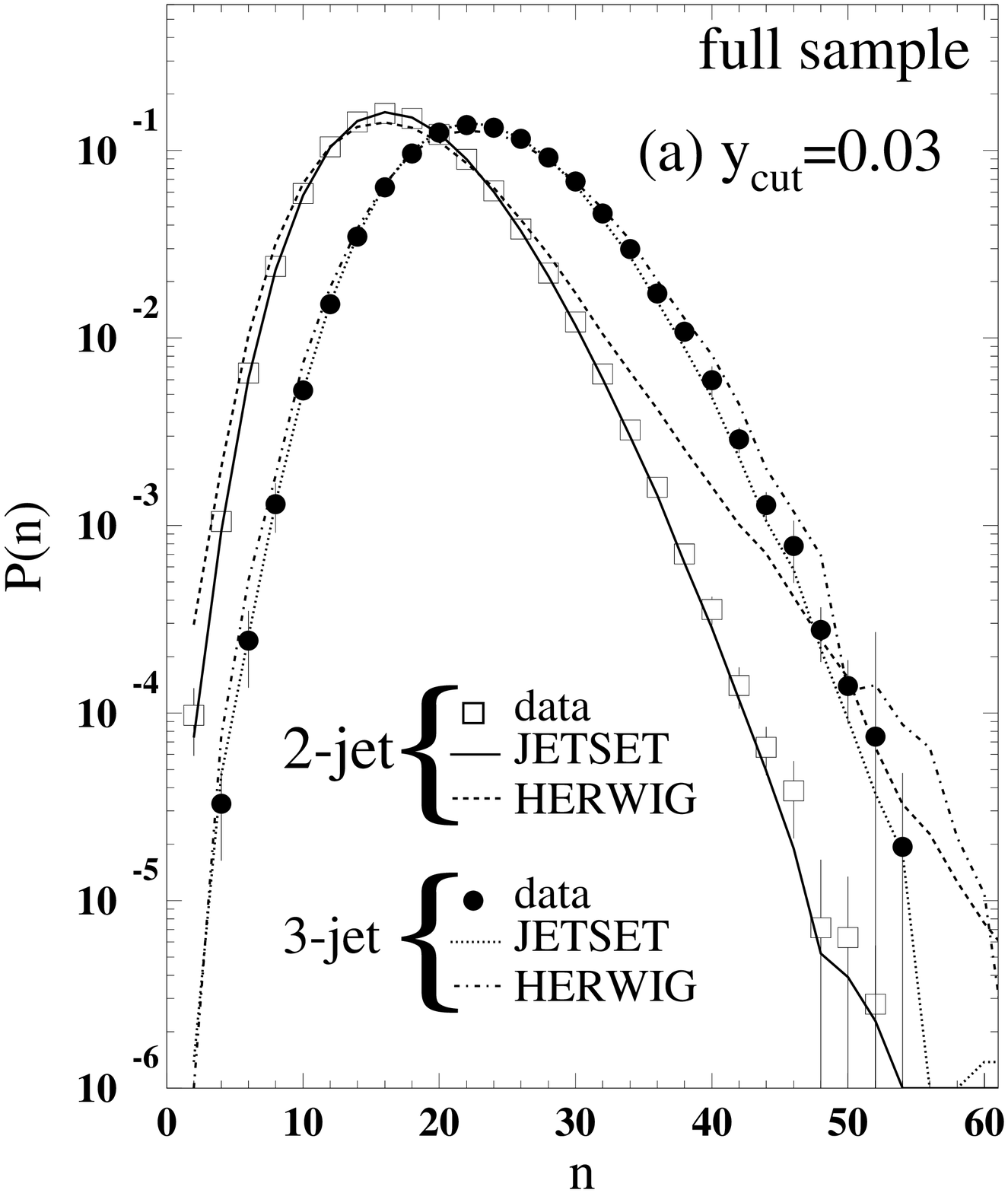}
    \includegraphics[width=8.4cm]{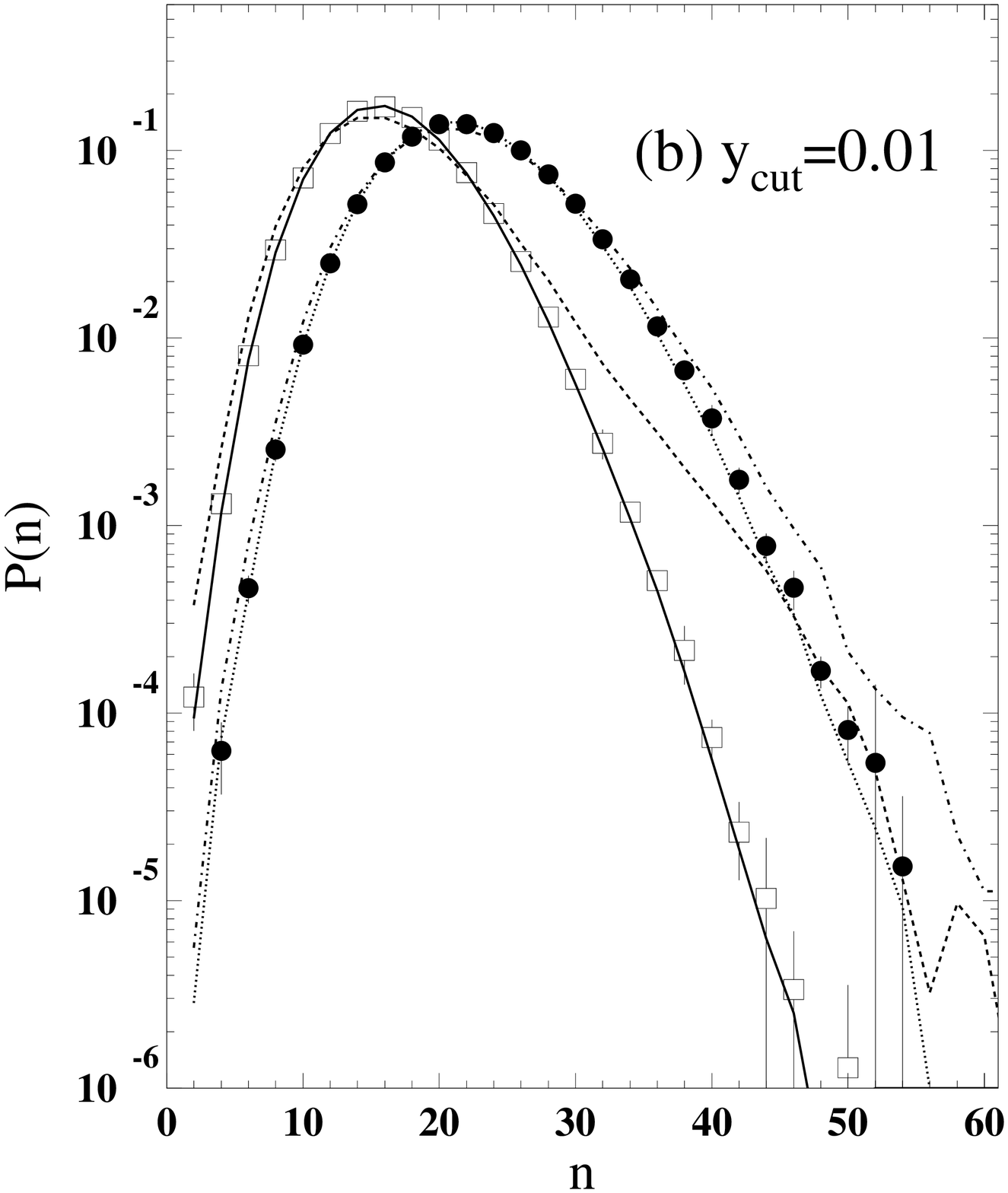}
\scaption{\Cpmd{s} of 2-jet and 3-jet events obtained from the 
full sample with (a) \Ycut{=0.03} and (b) \Ycut{=0.01}.}
\label{fig:pn23j_f}  
\end{figure}

\begin{figure}[htbp]
  \begin{center}
    \includegraphics[width=8.4cm]{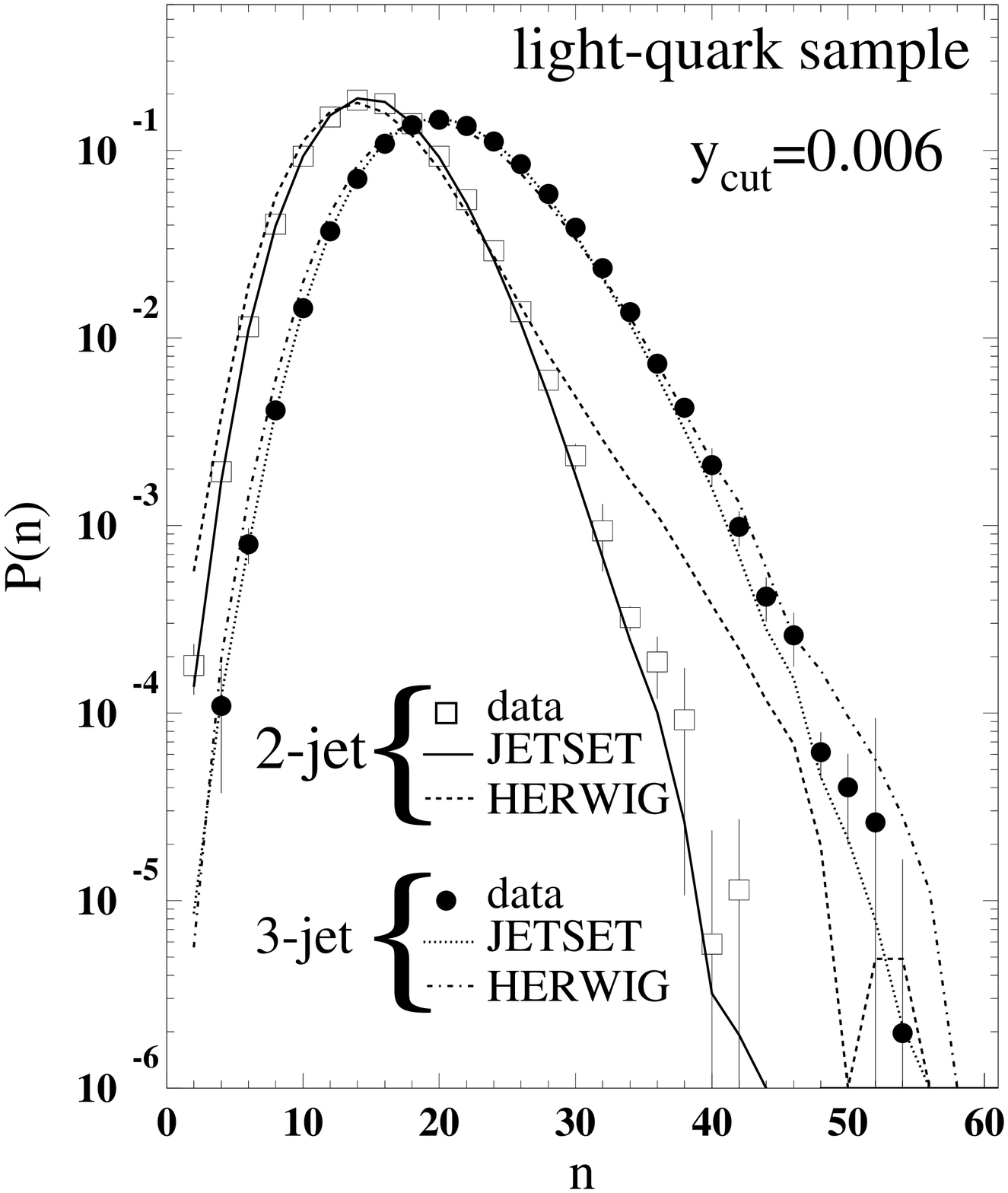}
    \includegraphics[width=8.4cm]{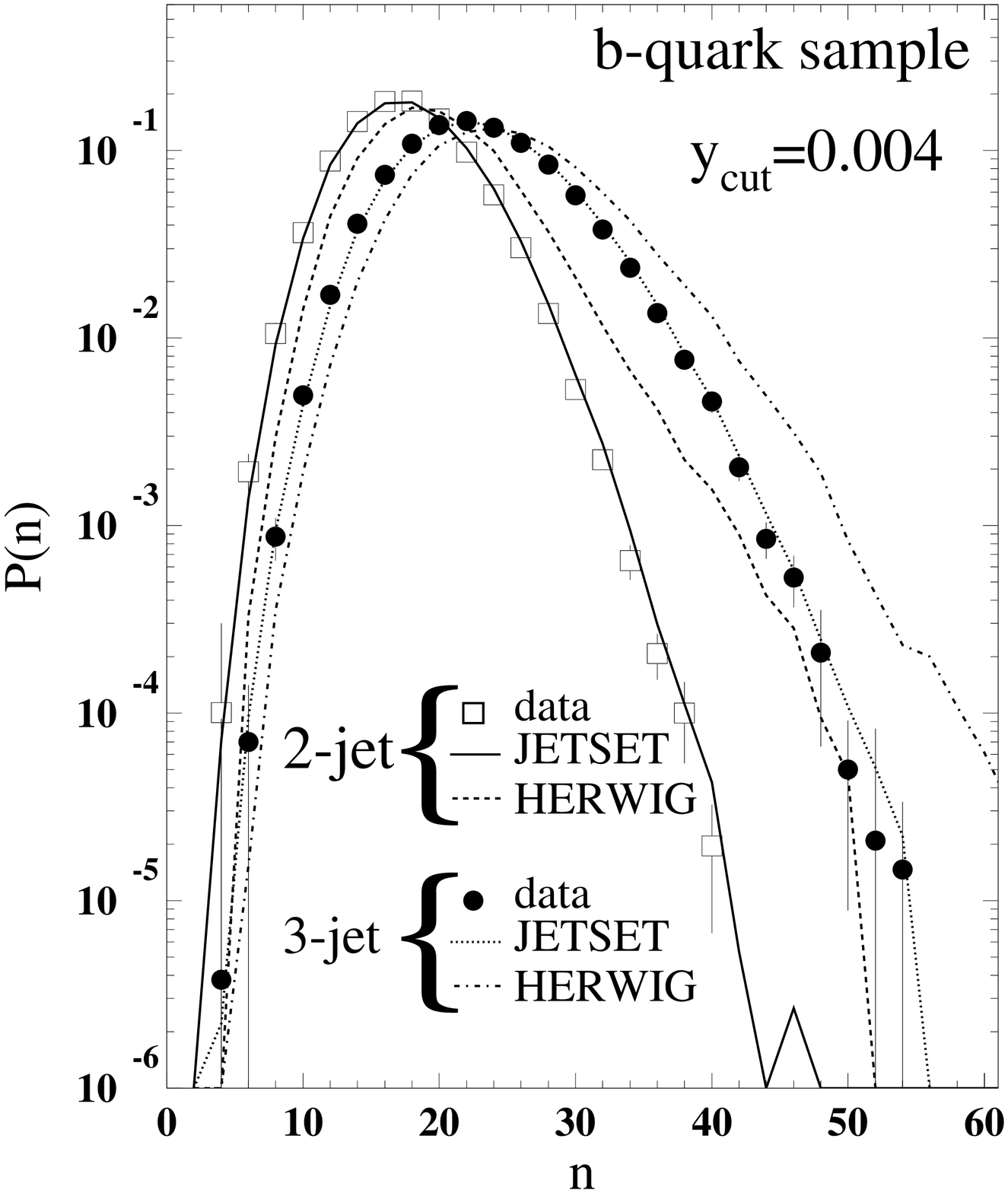}
  \end{center}
\begin{multicols}{2}
\scaption{\Cpmd{s} of 2-jet and 3-jet events obtained from the light-quark sample with \Ycut{=0.006}.}
\label{fig:pn23j_l}  
\scaption{\Cpmd{s} of 2-jet and 3-jet events obtained from the b-quark sample with \Ycut{=0.004}.}
\label{fig:pn23j_b} 
\end{multicols}
\end{figure}

Examples of \cpmd{s} measured for 2-jet and 3-jet events are shown 
in Fig.~\ref{fig:pn23j_f}(a) and (b) for the full sample with 
\Ycut{=0.03} and \Ycut{=0.01}, in Fig.~\ref{fig:pn23j_l} for the 
light-quark sample with \Ycut{=0.006} and in Fig.~\ref{fig:pn23j_b}  
for the b-quark sample with \Ycut{=0.004}. 
We note that JETSET agrees with the data in all cases, but 
HERWIG shows large disagreement for high multiplicities 
in the 2-jet events of all samples. 
The agreement is rather good for the 3-jet events of the full sample  
and of the light-quark sample, but it is bad for the b-quark sample.

We determined the mean, $\langle n\rangle$, and the dispersion, $D$, 
for all the 2- and 3-jet event \cpmd{s}. 
They are given in Table~\ref{tab:par2j} for the 2-jet 
events and in Table~\ref{tab:par3j} for the 3-jet events. 

\begin{figure}[htbp]
  \begin{center}
    \includegraphics[width=8.4cm]{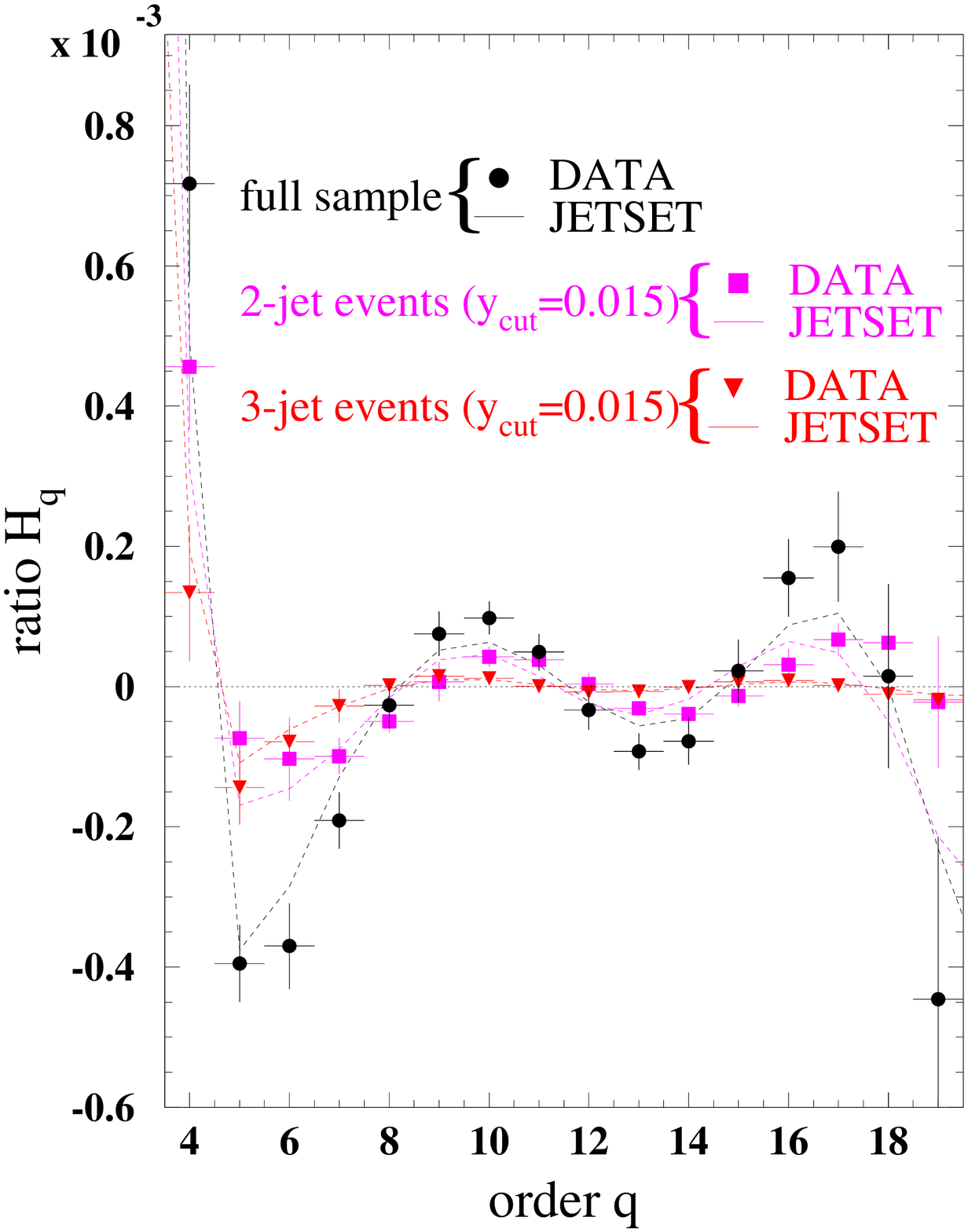}
    \includegraphics[width=8.4cm]{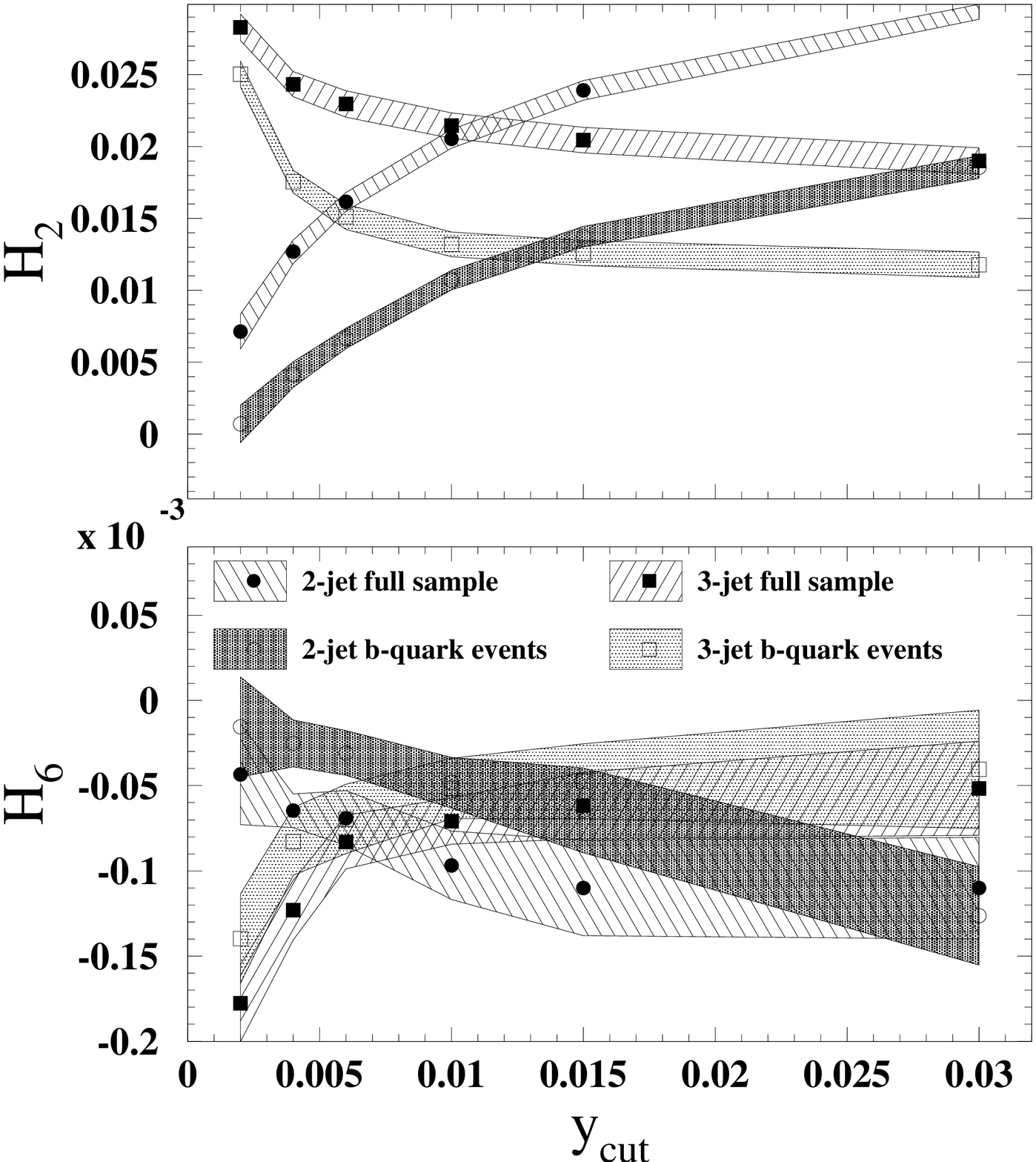}
  \end{center}
\begin{multicols}{2}
\scaption{$H_q$ moments of the full sample and 
and of 2- and 3-jet events obtained with \Ycut{=0.015}.}
\label{fig:hqf23j}  
\scaption{$H_2$ (top plot) and $H_6$ (bottom plot) plotted versus \Ycut{} 
for 2- and 3-jet obtained from both light- and b-quark samples.}
\label{fig:h26} 
\end{multicols}
\end{figure}

\begin{table}[htbp]
\begin{center}
\begin{tabular}{|c|c|c|c|c|}\cline{3-5}
  \multicolumn{2}{c|}{} &
  \multicolumn{1}{c|}{full sample} &
  \multicolumn{1}{c|}{light-quark sample} &
  \multicolumn{1}{c|}{b-quark sample}\\
\hline
$0.03$   &  $\langle n\rangle$   &  $17.51\pm0.01\pm0.08 $  & $16.95\pm0.01\pm0.07  $  &  $19.44\pm0.02\pm0.11$   \\
         &  $D$   & \phantom{1}$5.24\pm0.01\pm0.04  $  & \phantom{1}$5.10\pm0.01\pm0.04   $  &  \phantom{1}$5.15\pm0.01\pm0.04 $   \\
\hline
$0.015$  &  $\langle n\rangle$   &  $16.94\pm0.01\pm0.07 $  & $16.36\pm0.01\pm0.07  $  &  $18.90\pm0.02\pm0.10$   \\
         &  $D$   & \phantom{1}$4.95\pm0.01\pm0.04  $  & \phantom{1}$4.80\pm0.01\pm0.04   $  &  \phantom{1}$4.88\pm0.02\pm0.03$    \\   
\hline
$0.01$   &  $\langle n\rangle$   &  $16.57\pm0.01\pm0.07 $  & $15.99\pm0.01\pm0.06  $  &  $18.55\pm0.02\pm0.10$   \\
         &  $D$   & \phantom{1}$4.78\pm0.01\pm0.03  $  & \phantom{1}$4.62\pm0.01\pm0.04   $  &  \phantom{1}$4.71\pm0.02\pm0.03$    \\ 
\hline
$0.006$  &  $\langle n\rangle$   &  $16.07\pm0.01\pm0.07 $  & $15.48\pm0.01\pm0.06  $  &  $18.07\pm0.01\pm0.09$   \\
         &  $D$   & \phantom{1}$4.56\pm0.01\pm0.03  $  & \phantom{1}$4.39\pm0.01\pm0.03   $  &  \phantom{1}$4.48\pm0.03\pm0.03$    \\   
\hline
$0.004$  &  $\langle n\rangle$   &  $15.62\pm0.01\pm0.06 $  & $15.04\pm0.01\pm0.06  $  &  $17.64\pm0.01\pm0.09$   \\
         &  $D$   & \phantom{1}$4.38\pm0.01\pm0.03  $  & \phantom{1}$4.20\pm0.01\pm0.04   $  &  \phantom{1}$4.33\pm0.03\pm0.04$    \\  
\hline
$0.002$  &  $\langle n\rangle$   &  $14.65\pm0.01\pm0.06 $  & $14.21\pm0.01\pm0.05  $  &  $16.90\pm0.01\pm0.08$   \\
         &  $D$   & \phantom{1}$4.07\pm0.01\pm0.04  $  & \phantom{1}$3.90\pm0.01\pm0.05   $  &  \phantom{1}$4.12\pm0.06\pm0.05$    \\  
\hline
\end{tabular}\end{center}
\vspace{-0.2cm}
\scaption{Mean $\langle n\rangle$ and dispersions $D$ of the 2-jet events.}
\label{tab:par2j}
\end{table}
\begin{table}[htbp]
\begin{center}
\begin{tabular}{|c|c|c|c|c|}\cline{3-5}
\cline{3-5}
  \multicolumn{2}{c|}{} &
  \multicolumn{1}{c|}{full sample} &
  \multicolumn{1}{c|}{light-quark sample} &
  \multicolumn{1}{c|}{b-quark sample}\\
\hline
$0.03$   &  $\langle n \rangle$   &  $23.74\pm0.02 \pm0.13$ & $23.14\pm0.02\pm0.12  $  &$25.80\pm0.09\pm0.15  $     \\
         &  $D$   &  \phantom{1}$5.96 \pm0.01 \pm0.07$ & \phantom{1}$5.83\pm0.02\pm0.08   $  &\phantom{1}$5.81\pm0.05\pm0.05   $     \\
\hline
$0.015$  &  $\langle n \rangle$   &  $22.91\pm0.01\pm0.12$  & $22.32\pm0.01\pm0.12  $ & $24.97\pm0.06\pm0.15  $     \\
         &  $D$   &  \phantom{1}$5.89 \pm0.01\pm0.07$  & \phantom{1}$5.75\pm0.01\pm0.07   $ & \phantom{1}$5.75\pm0.04\pm0.06   $     \\
\hline
$0.01$   &  $\langle n \rangle$   &  $22.41\pm0.01\pm0.13$  & $21.82\pm0.01\pm0.12  $ & $24.48\pm0.05\pm0.16  $     \\
         &  $D$   &  \phantom{1}$5.85\pm0.01\pm0.06 $  & \phantom{1}$5.71\pm0.01\pm0.07   $ & \phantom{1}$5.71\pm0.04\pm0.06   $     \\ 
\hline
$0.006$  &  $\langle n \rangle$   &  $21.79\pm0.01\pm0.13$  & $21.22\pm0.01\pm0.12  $ & $23.77\pm0.05\pm0.15  $     \\
         &  $D$   &  \phantom{1}$5.81 \pm0.01\pm0.06$  & \phantom{1}$5.67\pm0.01\pm0.07   $ & \phantom{1}$5.71\pm0.03\pm0.06   $     \\ 
\hline
$0.004$  &  $\langle n \rangle$   &  $21.28\pm0.01\pm0.14$  & $20.74\pm0.01\pm0.13  $ & $23.06\pm0.04\pm0.15  $     \\
         &  $D$   &  \phantom{1}$5.77 \pm0.01\pm0.06$  & \phantom{1}$5.64\pm0.01\pm0.06   $ & \phantom{1}$5.71\pm0.03\pm0.05   $     \\ 
\hline
$0.002$  &  $\langle n \rangle$   &  $20.31\pm0.01\pm0.14$  & $19.86\pm0.01\pm0.13  $ & $21.54\pm0.03\pm0.16  $     \\
         &  $D$   &  \phantom{1}$5.74\pm0.01\pm0.06 $  & \phantom{1}$5.62\pm0.01\pm0.06   $ & \phantom{1}$5.77\pm0.02\pm0.06   $     \\ 
\hline
\end{tabular}\end{center}
\vspace{-0.2cm}
\scaption{Mean $\langle n \rangle$ and dispersion $D$ of the 3-jet events.}
\label{tab:par3j}
\end{table}
We further determined the $H_q$ moments for the 2- and 3-jet events 
in the full, light- and b-quark samples. 
We used the same truncation criteria as in the 
previous chapters. A comparison between the $H_q$ for 
the full sample and the  $H_q$ for the 2-jet and 3-jet events obtained 
from the full sample with a \Ycut{=0.015} is given in Fig.~\ref{fig:hqf23j}. 
It shows that even the largest $H_q$ oscillations obtained 
from any 2- or 3-jet event samples (what we see in Fig.~\ref{fig:hqf23j} for 
\Ycut{=0.0015} is also true for any other \Ycut{}) are much 
smaller than those in the full sample.

\begin{figure}[htbp]
\centering
\vspace{-0.4cm}
    \includegraphics[width=7.4cm]{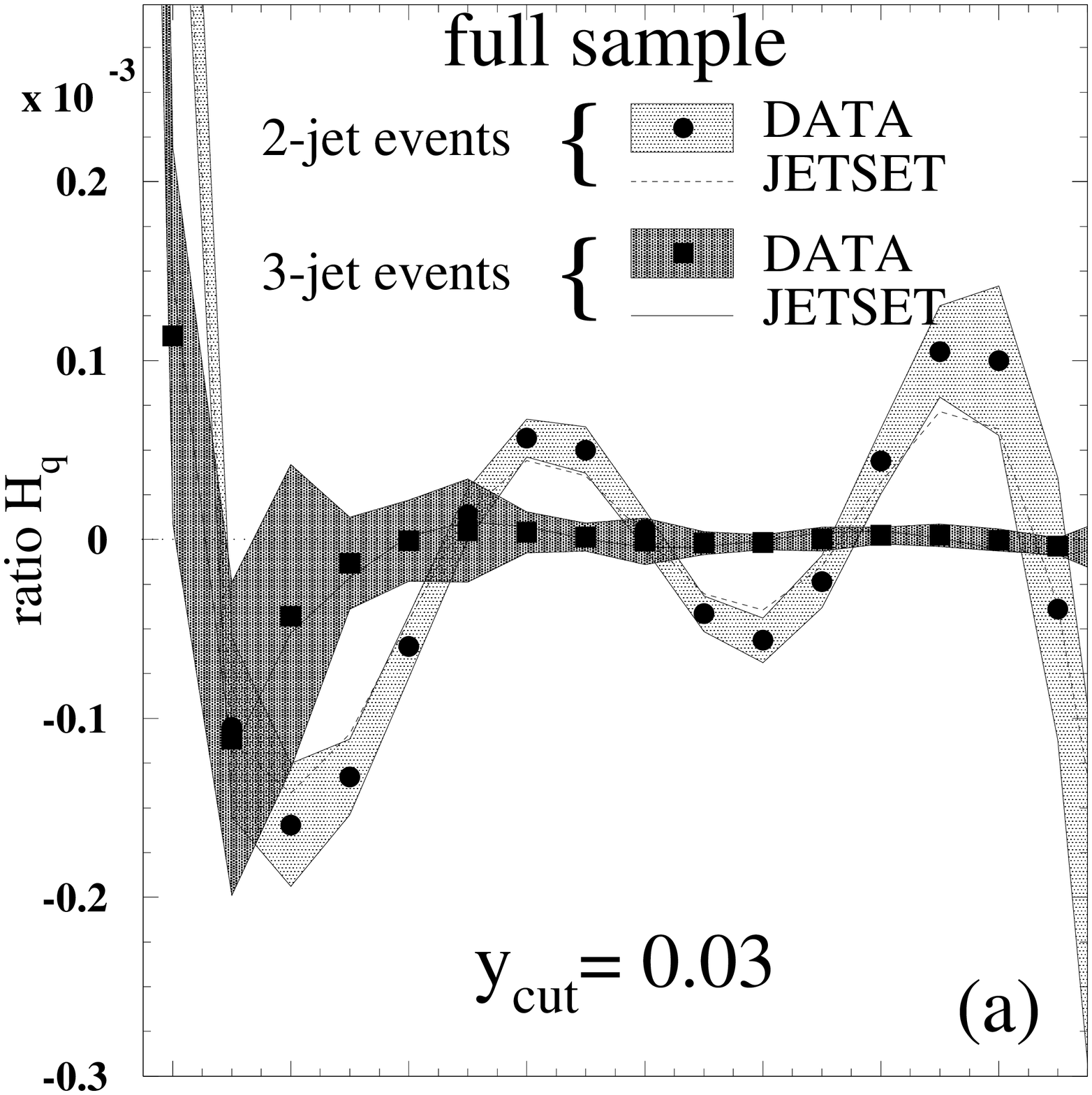}
    \includegraphics[width=7.4cm]{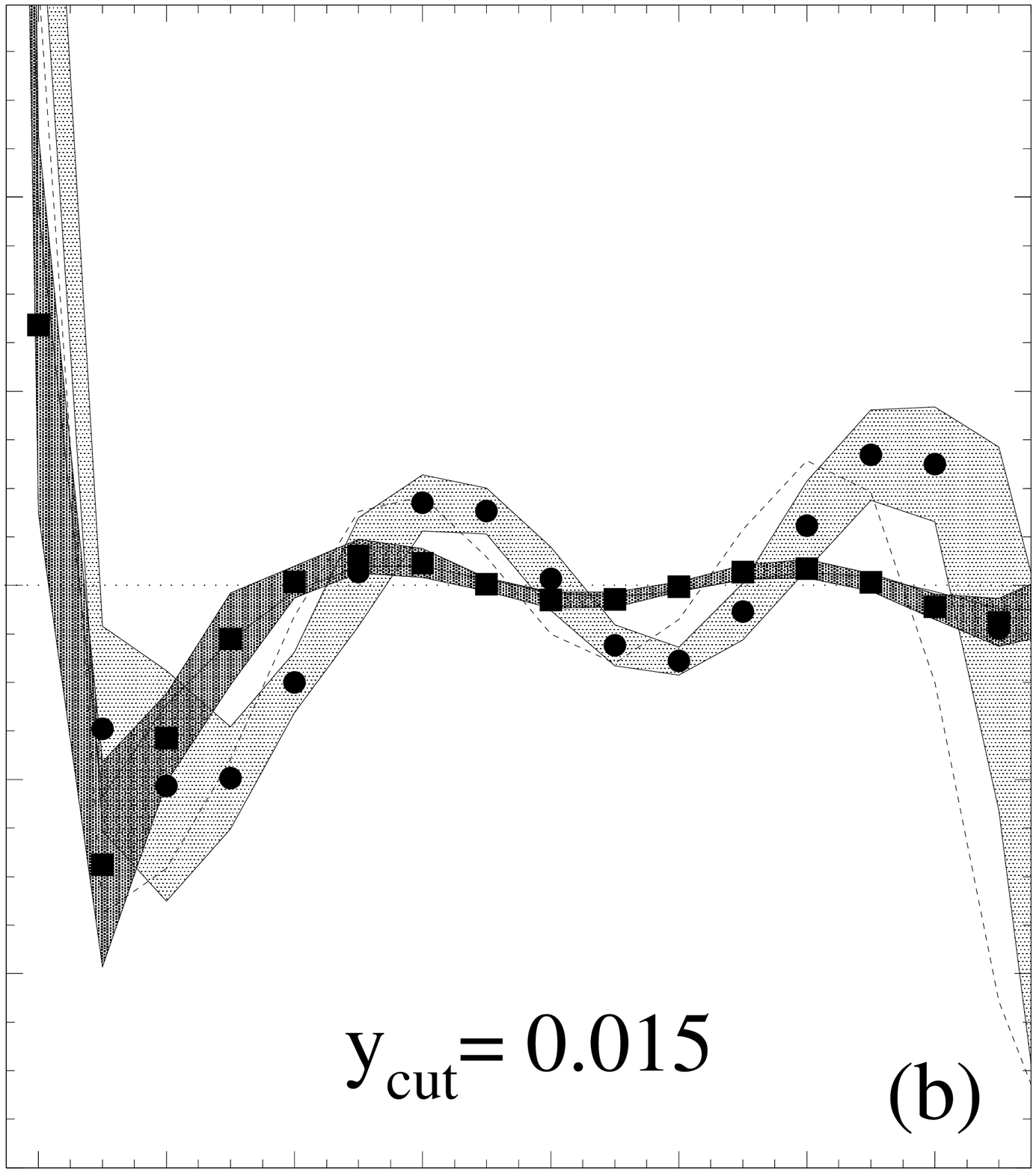}

\vspace{-1.4cm}

    \includegraphics[width=7.4cm]{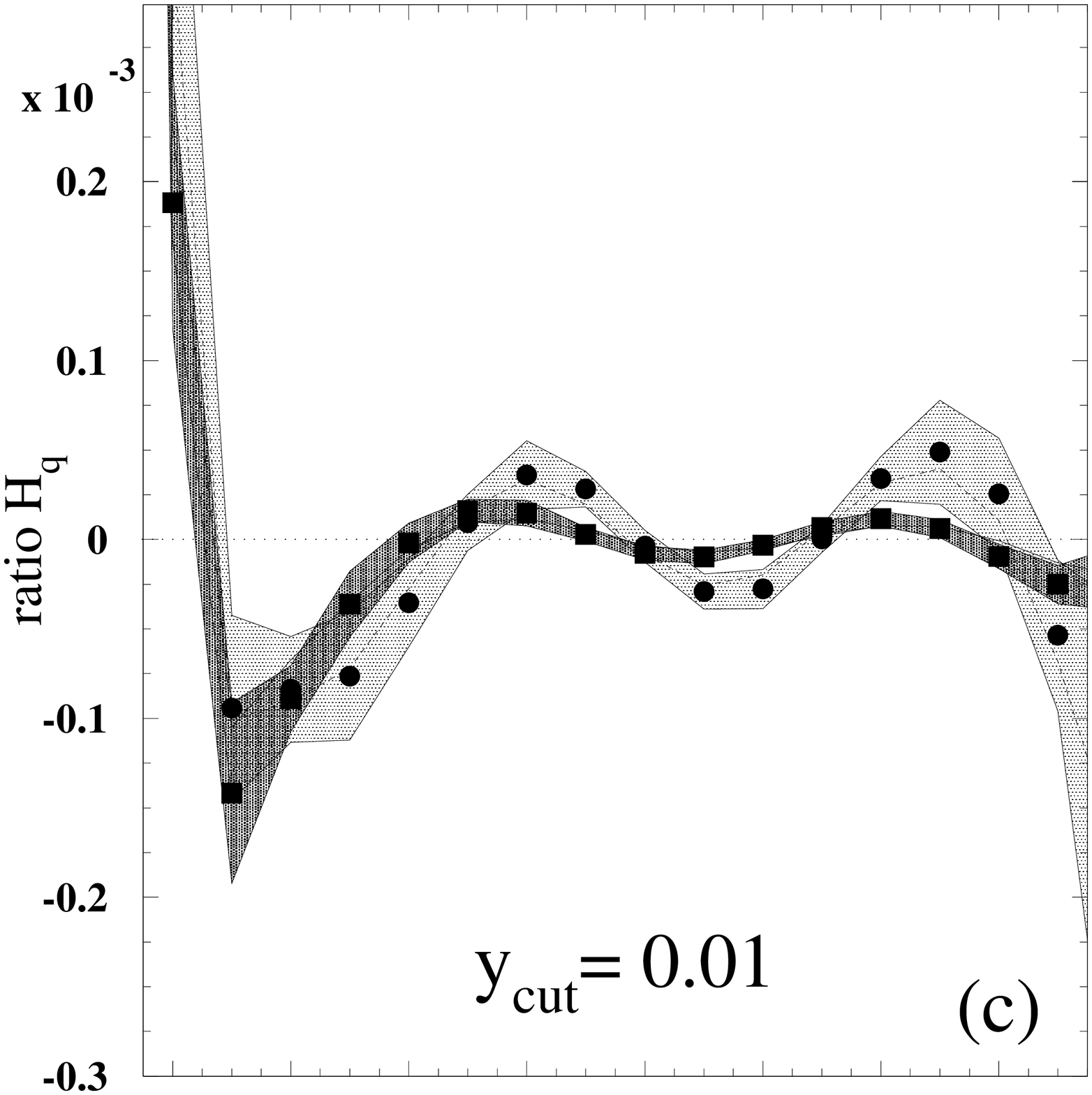}
    \includegraphics[width=7.4cm]{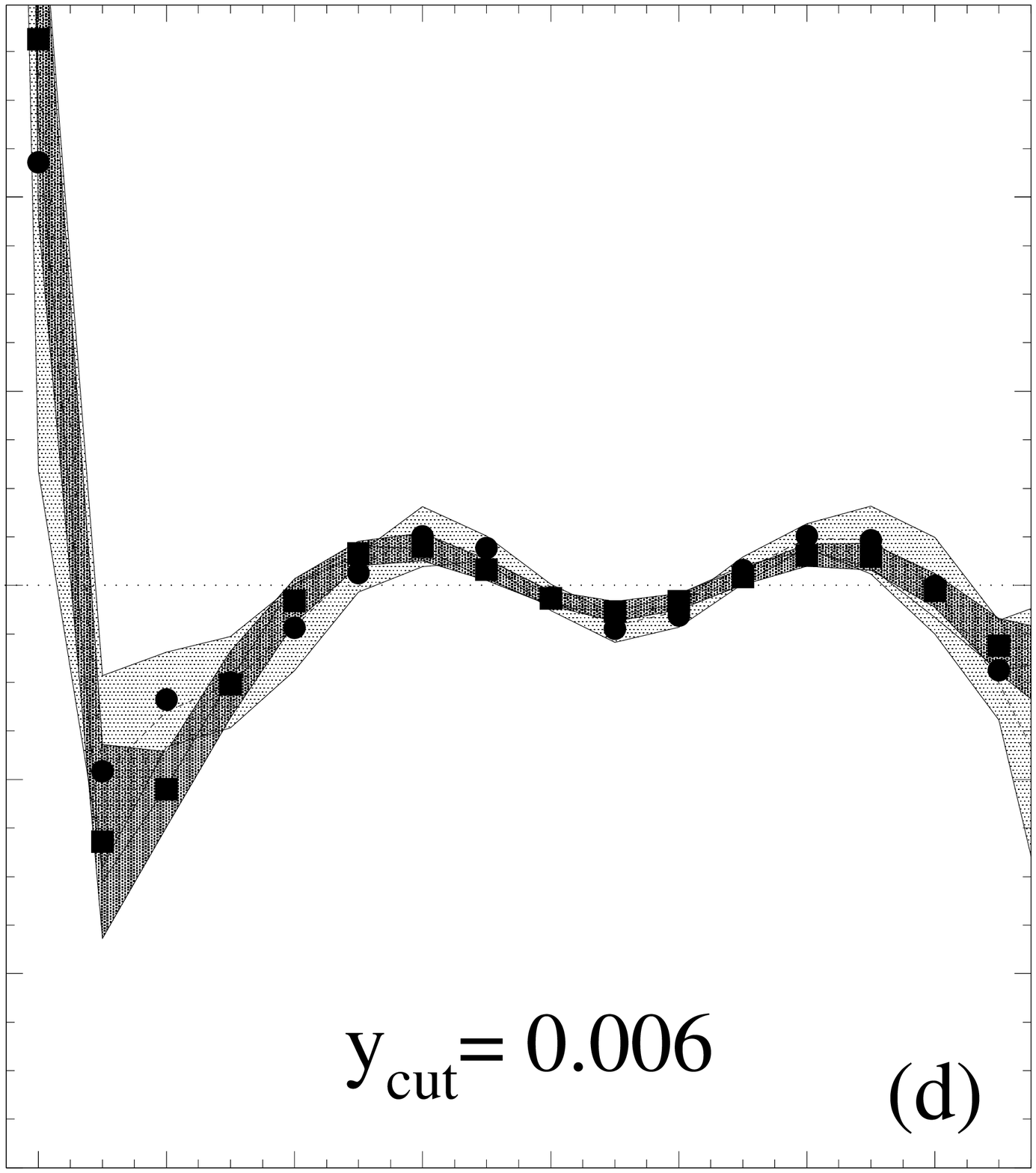}
\vspace{-1.4cm}

    \includegraphics[width=7.4cm]{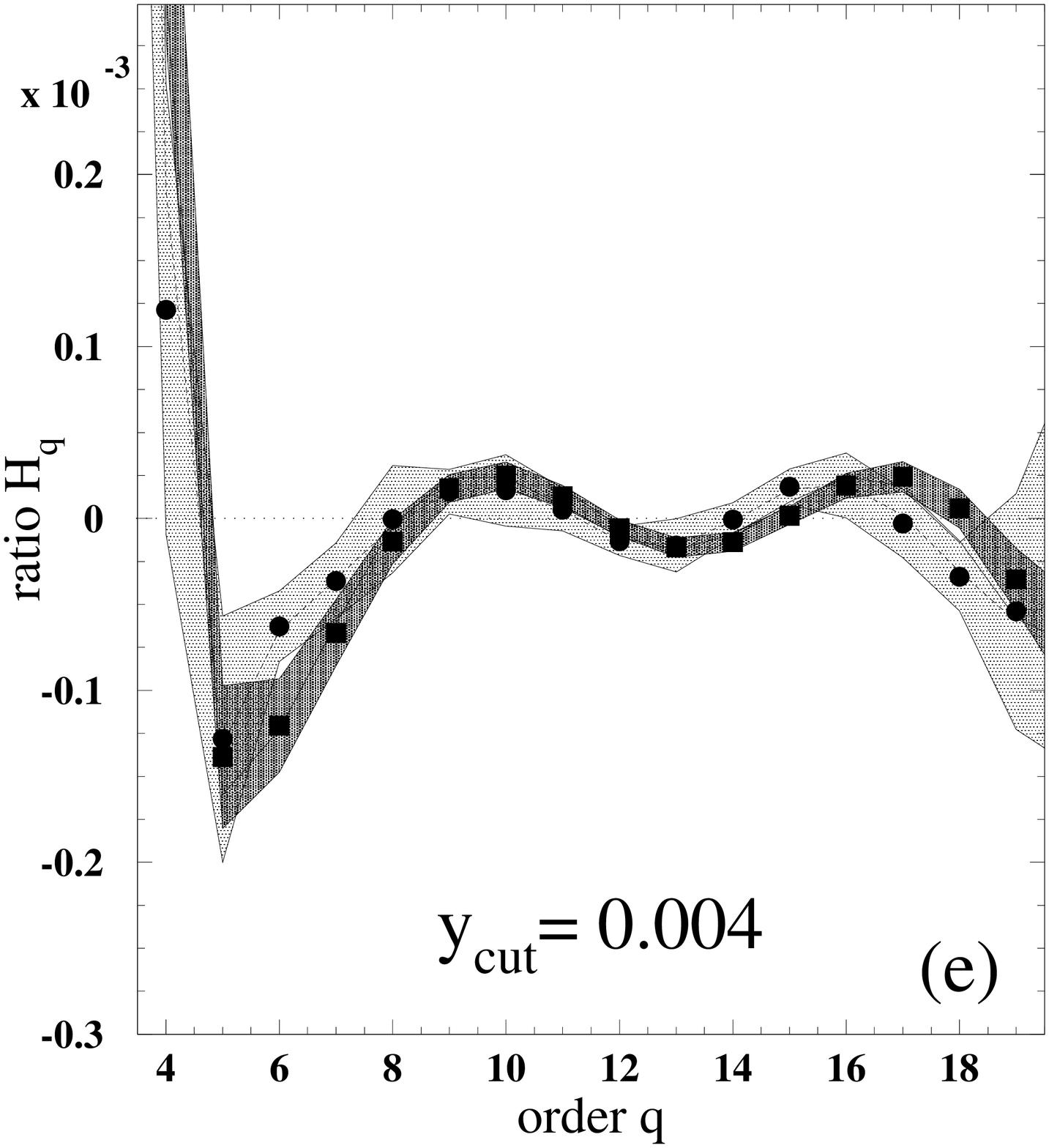}
    \includegraphics[width=7.4cm]{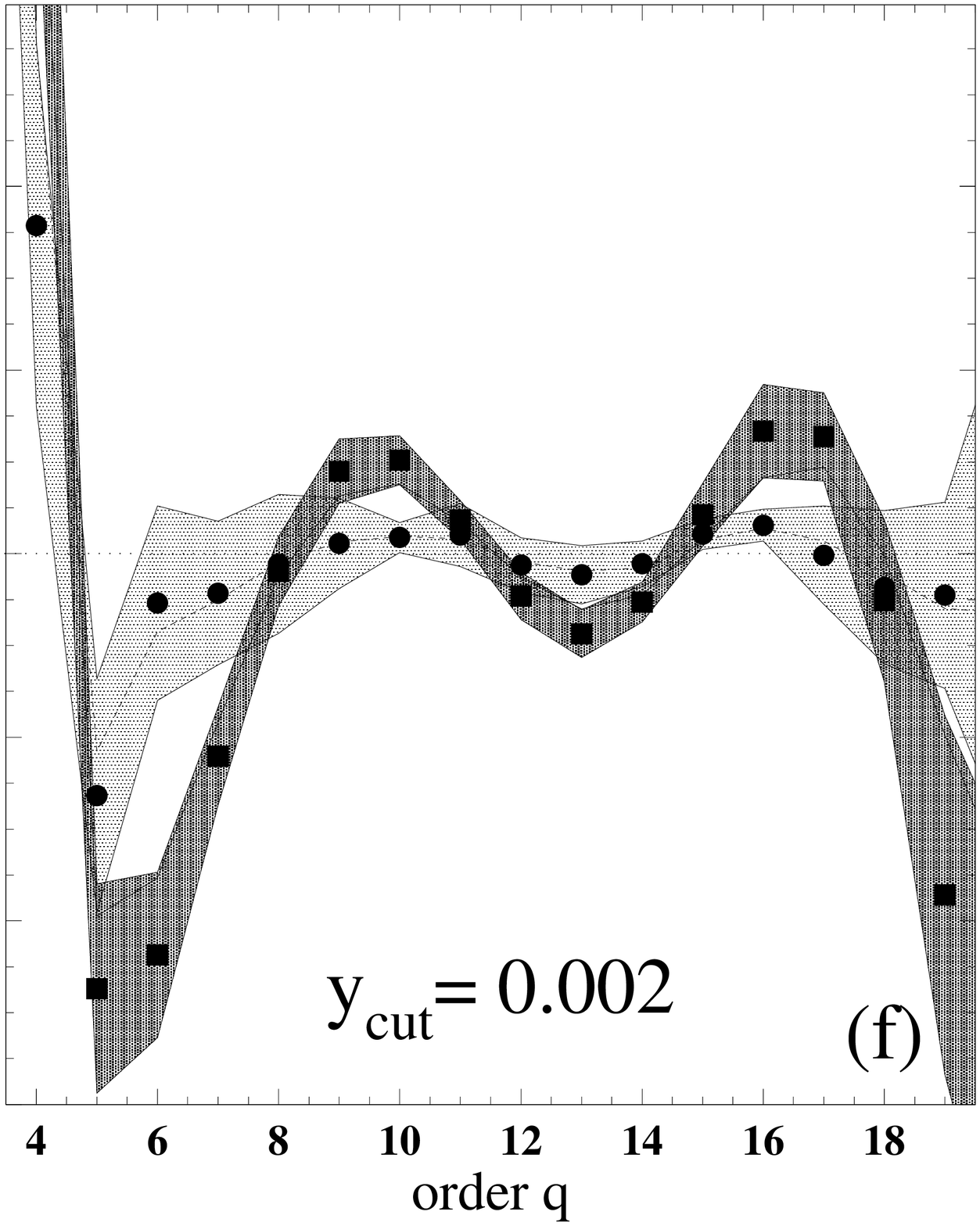}

\vspace{-0.2cm}
\scaption{$H_q$ moments measured from the \cpmd{s} of the 2- and 3-jet of the 
full sample for (a) \Ycut{=0.03}, (b) \Ycut{=0.015}, 
(c) \Ycut{=0.01}, (d) \Ycut{=0.006}, 
(e) \Ycut{=0.004} and (f) \Ycut{=0.002}.}
\label{fig:hq_j12f}  
\end{figure}

Fig.~\ref{fig:hq_j12f} shows the 
$H_q$ moments for 2- and 3-jet events obtained from the full sample 
for the 6 \Ycut{}  
values.
For the 2-jet events we find that the size of the oscillations decreases with decreasing \Ycut{}. 
For \Ycut{=0.030}, where the 2-jet fraction represents $82\%$ of the 
full sample and can include rather broad 2-jet events, the oscillations 
have about half the amplitude of those of the full sample.  
They further decrease gradually until \Ycut{=0.004}, where the oscillations
have almost completely disappeared.

For the 3-jet events, we have the opposite trend. 
For \Ycut{=0.03} and \Ycut{=0.015}, the $H_q$ moments do not show 
any oscillation. As \Ycut{} decreases, the amplitude of the 
oscillation increases to reach about half the size of those 
of the full sample at \Ycut{=0.002}. 

We must also note that the absence of oscillations in the 2-jet samples and 
in the 3-jet samples are obtained for completely different jet configurations. 
The low \Ycut{} values for which the absence of oscillation
occurs for 2-jet events, identify as 2-jet events only 
those events which have narrow jets, 
almost back to back (pencil-like 2-jet events). 
For the 3-jet events, it is for large \Ycut values that the oscillations
disappear. For these \Ycut{} values, the events selected as 3-jet can  
have a very broad jet configuration, and hence they have a configuration 
close to the Mercedes-like 
3-jet events. In both configurations, the energy is shared almost equally 
among  the jets.

\begin{figure}[htbp]
\centering
    \includegraphics[width=7.4cm]{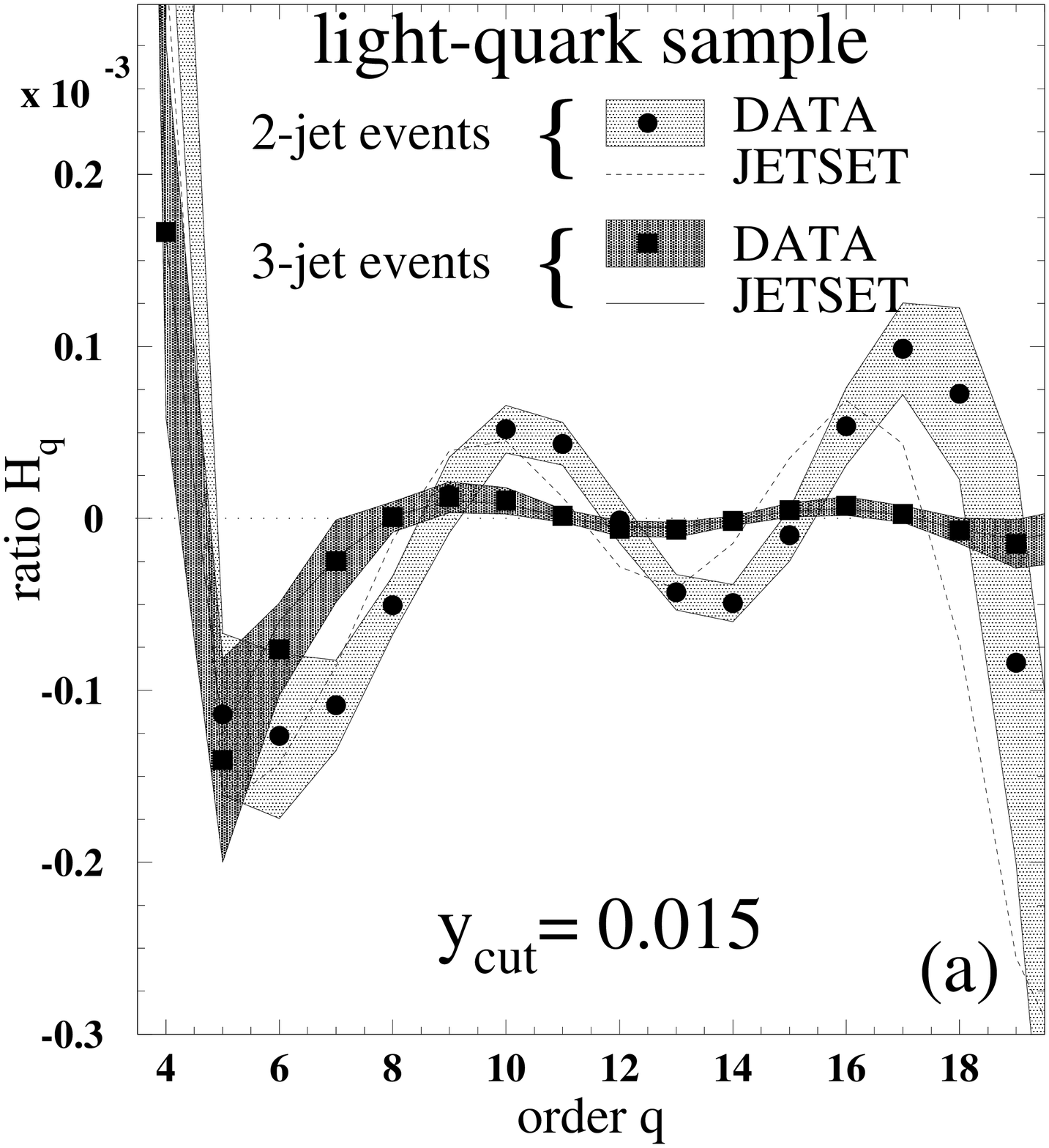}
    \includegraphics[width=7.4cm]{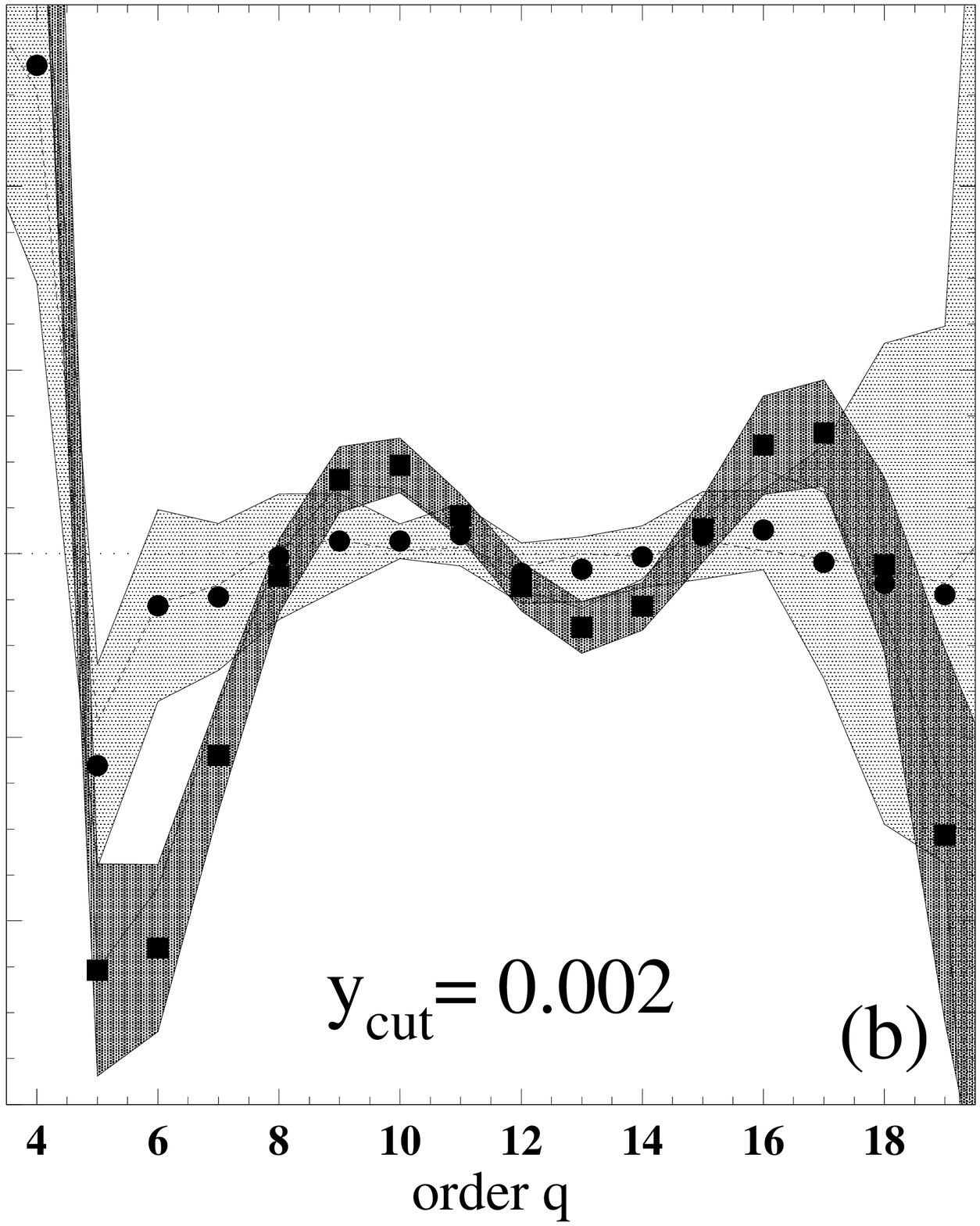}
\scaption{$H_q$ moments measured from the \cpmd{s} of the 2- and 3-jet in the 
light-quark sample for (a) \Ycut{=0.015} and (b) \Ycut{=0.002}.}
\label{fig:hq_j26l}  
\end{figure}

\begin{figure}[htbp]
\centering
    \includegraphics[width=7.4cm]{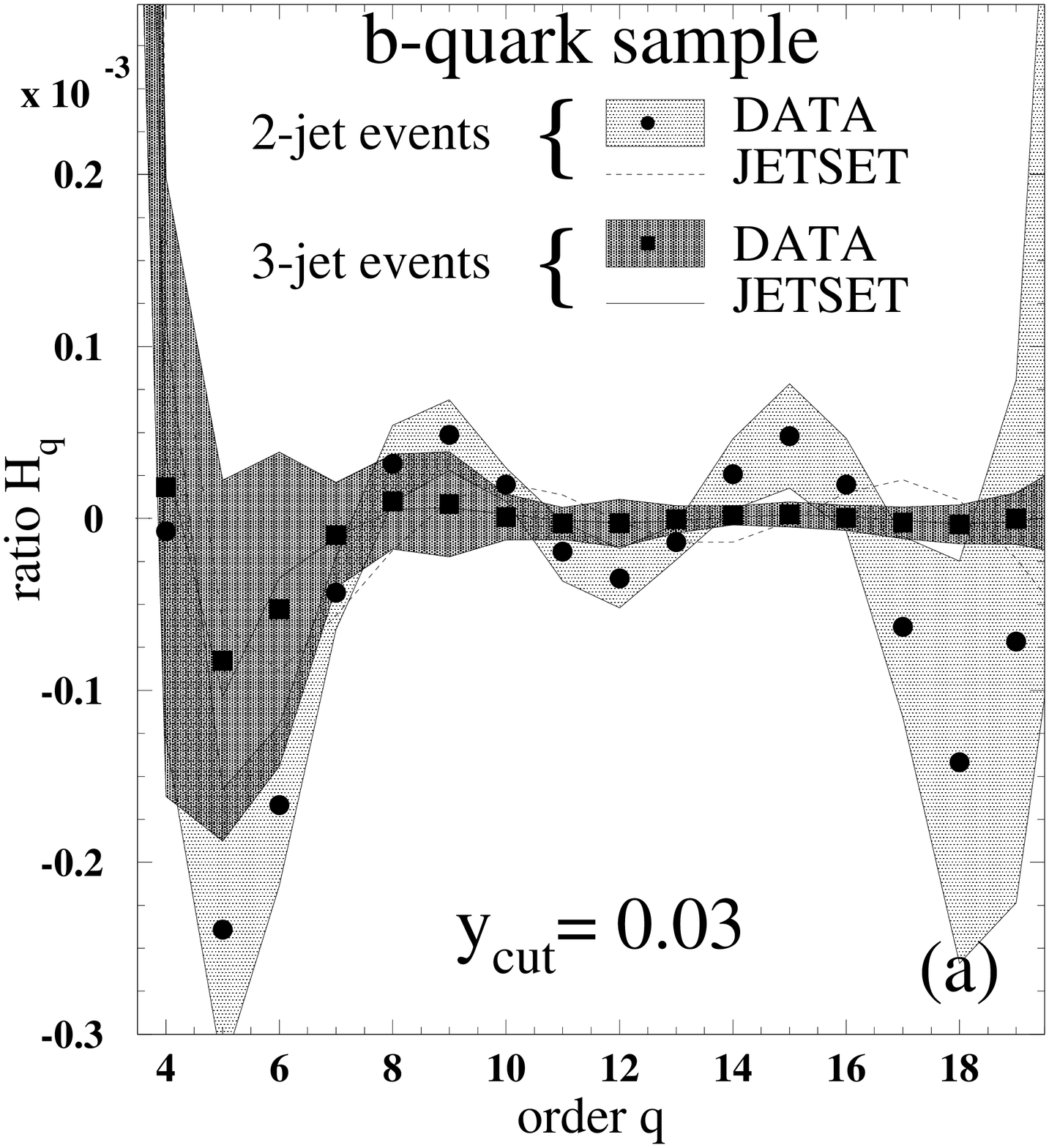}
    \includegraphics[width=7.4cm]{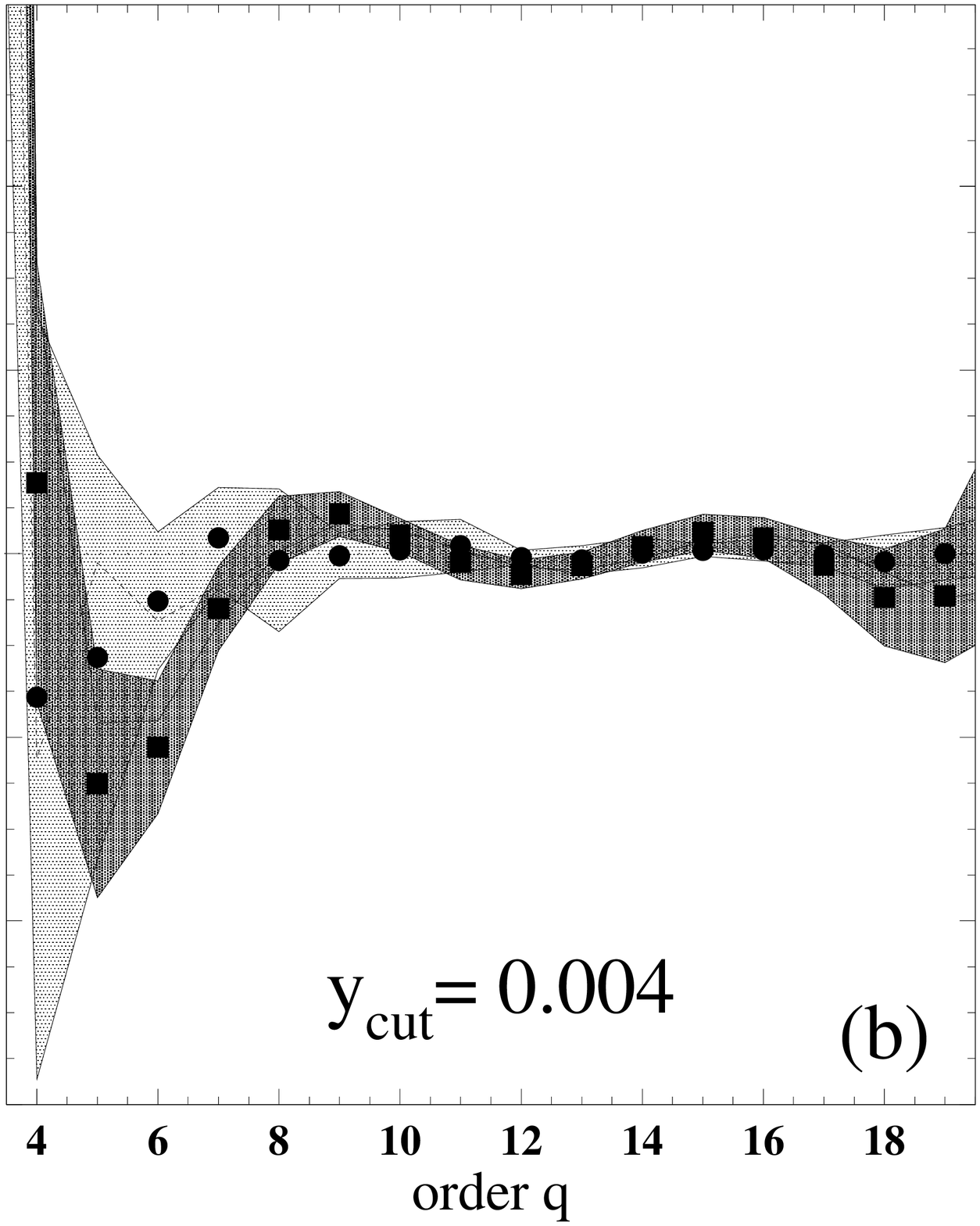}
\scaption{$H_q$ moments measured from the \cpmd{s} of the 2- and 3-jet in the 
b-quark sample for (a) \Ycut{=0.03} and (b) \Ycut{=0.004}.}
\label{fig:hq_j15b}  
\end{figure}

We also measure the $H_q$ moments of the \cpmd{s} of the 2-jet and 3-jet events 
obtained from the light- and b-quark samples. 
The $H_q$ behavior  does not exhibit any significant difference 
from that of the $H_q$ moments derived from the 2- and 
3-jet events obtained from the full sample. Two examples are given 
in Fig.~\ref{fig:hq_j26l} for the light-quark sample and two in 
Fig.~\ref{fig:hq_j15b} for the b-quark sample. 
The major differences, we found, are confined to low $q$ values ($q<5$). 
This is further illustrated 
in Fig.~\ref{fig:h26}, where $H_2$ and $H_6$ are plotted 
as a function of \Ycut{} for the 2- and 3-jet events obtained 
from the light- and b-quark samples. For $H_2$, we see a 
rather large difference for both 2- and 3-jet 
between the light- and b-quark samples, while for $H_6$, differences between 
light- and b-quark samples have almost disappeared.  


\section{The phenomenological approach}

This approach views the shape of \cpmd{} of the full sample as the 
result of the superposition of distributions originating from 
various processes related to the topology of the events, \eg{},  
2-jet and 3-jet or light-quark and heavy-quark events. 
Assuming that each of these processes can be described by a relatively 
simple parametrization such as the NBD, the \cpmd{} of the full sample would 
then be a weighted sum of all the contributions. Altogether, these various 
contributions would explain the shape of the \cpmd{} and ultimately its  
$H_q$ behavior. In the framework of this approach, we investigate  
two hypotheses.
In the first hypothesis we assume that the shape of the 
\cpmd{} of the full sample 
arises from the superposition of 2-jet and 3-jet events~\cite{23jet}.
In the second hypothesis we assume that the full sample and also those 
of the 2-jet and the 3-jet events can be parametrized by 
superposition of flavor related distributions~\cite{lbquark}.

\subsection{2- and 3-jet superposition}

In this hypothesis we assume that the main features of the shape of the 
\cpmd{} of full, light- and b-quark samples responsible for the 
oscillatory behavior of the $H_q$ moments, can be described by a 
weighted sum of two NBDs (Eq. (\ref{eq:2nbd_jet})), the two NBDs 
themselves describing the \cpmd{s} of the 2-jet and of the 3-jet 
events, respectively. 
The parameters of the 2 NBD's are obtained from the experimental \cpmd{s} of 
the 2-jet and 3-jet events (Tables~\ref{tab:par2j} and~\ref{tab:par3j} 
for the 2-jet and the 3-jet events, respectively), the weight between the 
2 NBD's is taken as   
the 2-jet fraction obtained for a given \Ycut{} value (Table~\ref{tab:R2}). 
This fraction can also be calculated from the means of the \cpmd{s}  
of the 2-jet, 3-jet events and of the full samples.

We first compare the data and the single NBDs, 
$f_\text{2-jet}^\text{NB}(n,\langle n_2\rangle,k_2)$ and 
$f_\text{3-jet}^\text{NB}(n,\langle n_3\rangle,k_3)$
where the parameters $\langle n_{2,3}\rangle$ and $k_{2,3}$ are taken from the 
experimental 2- and 3-jet \cpmd{s} according to 
Eq.~(\ref{eq:park}).
The corresponding $\chi^2$ confidence levels are given in Table~\ref{tab:NBD}.
We find good agreement for the 2-jet events obtained 
for \Ycut{} smaller than 0.01. 
We note also that the agreement increases gradually with  
decreasing \Ycut{} value.

For the 3-jet events, \ie{}, non-2-jet events,
we observe the opposite of what is 
observed for the 2-jet events. The agreement with the 
single-NBD parametrization improves with increasing \Ycut{} value. 
We have good agreement with  the 
\cpmd{} of the 3-jet events for \Ycut{} values larger than 0.06.

\begin{figure}[htbp]
\begin{center}
    \includegraphics[width=8.4cm]{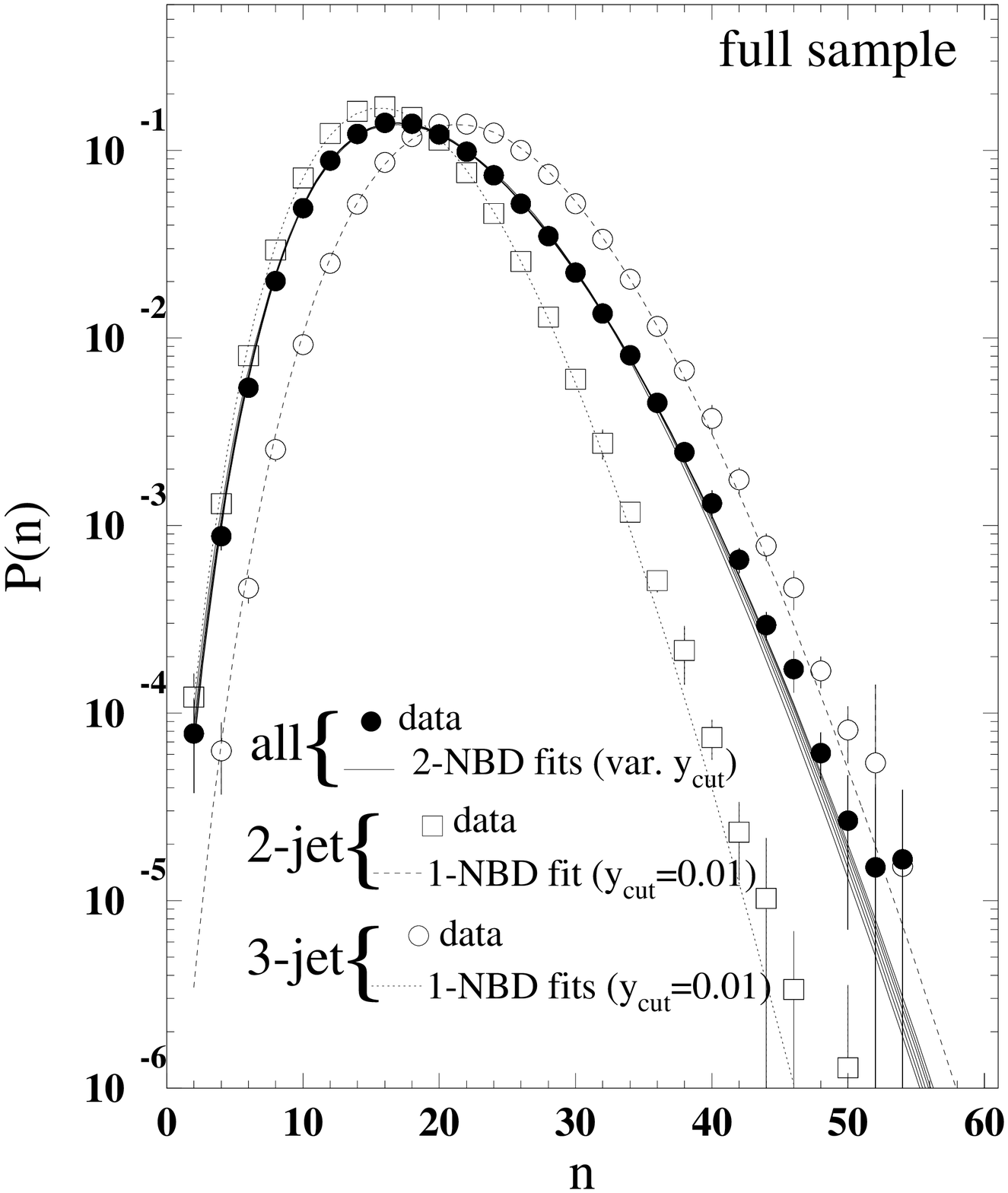}
    \includegraphics[width=8.4cm]{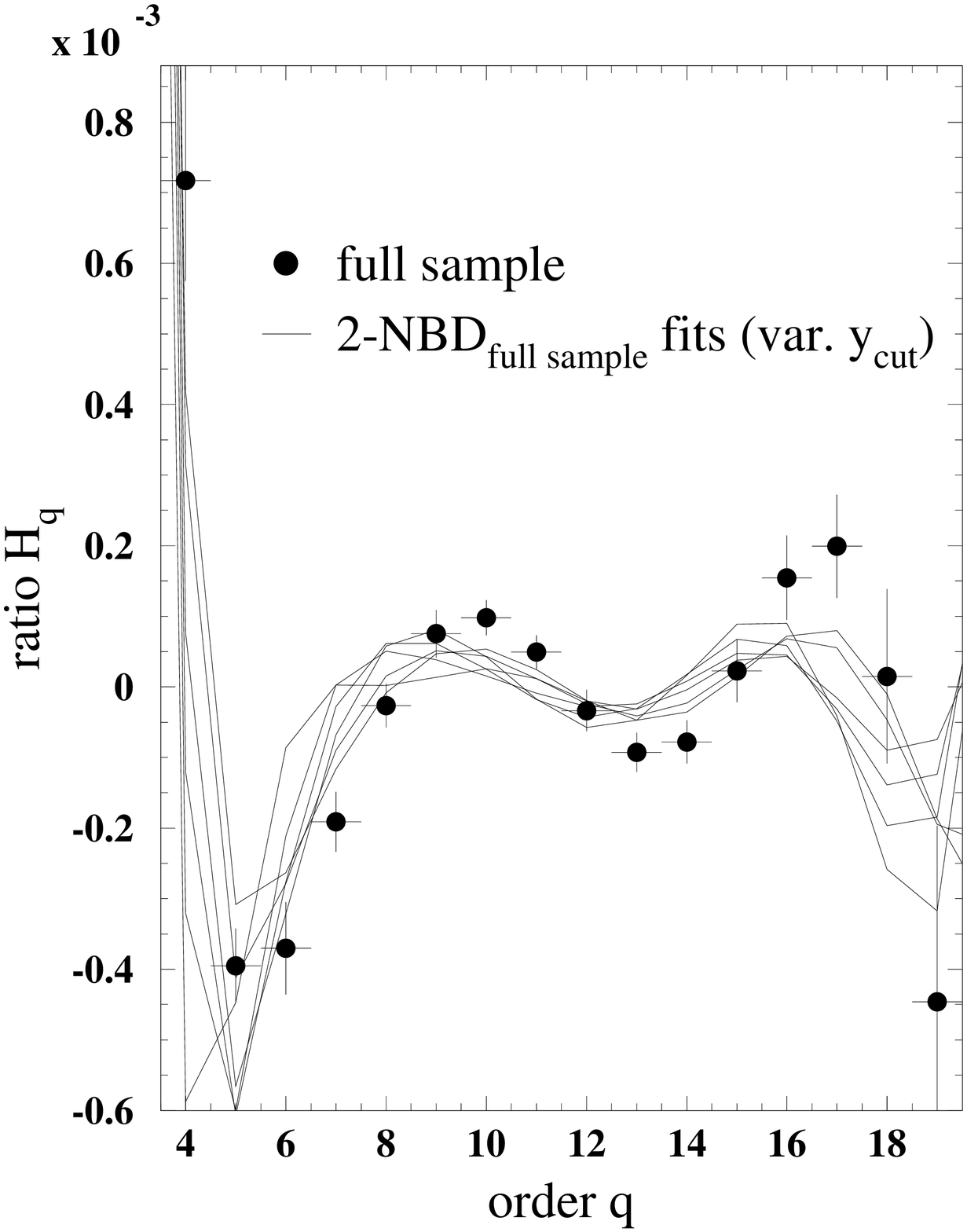}
\end{center}
\vspace{-1cm}
\begin{multicols}{2}
\scaption{\Cpmd{s} of the full sample and of 2-jet and 3-jet events obtained 
with \Ycut{=0.01} fitted by various two-NBD and single-NBD parametrizations, 
respectively.}
\label{fig:2nbd_fa}  
\scaption{$H_q$ moments of the full sample, together with the $H_q$ moments 
calculated from the two-NBD parametrization.}
\label{fig:2nbd_fb} 
\end{multicols}
\begin{center}
    \includegraphics[width=8.4cm]{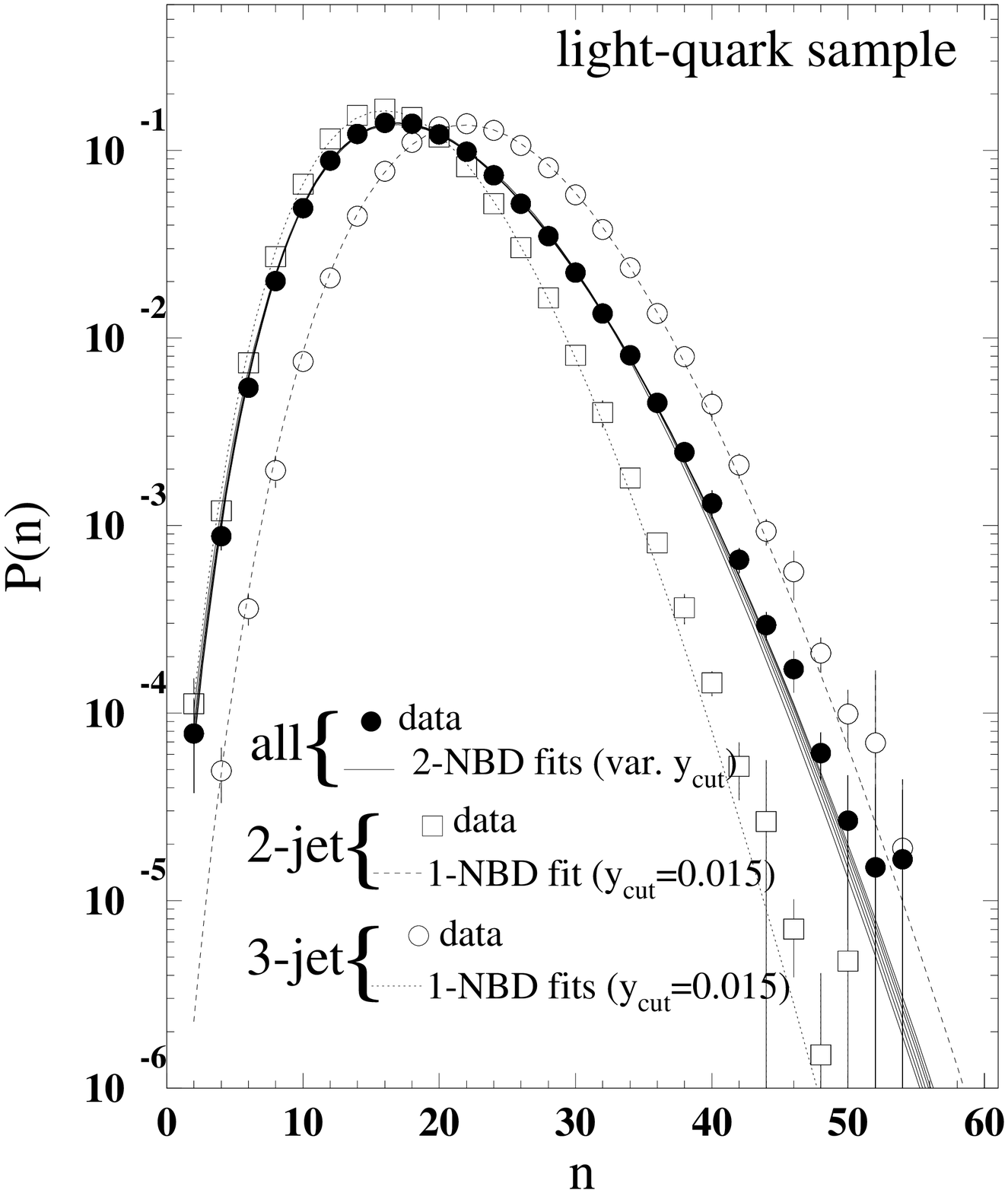}
    \includegraphics[width=8.4cm]{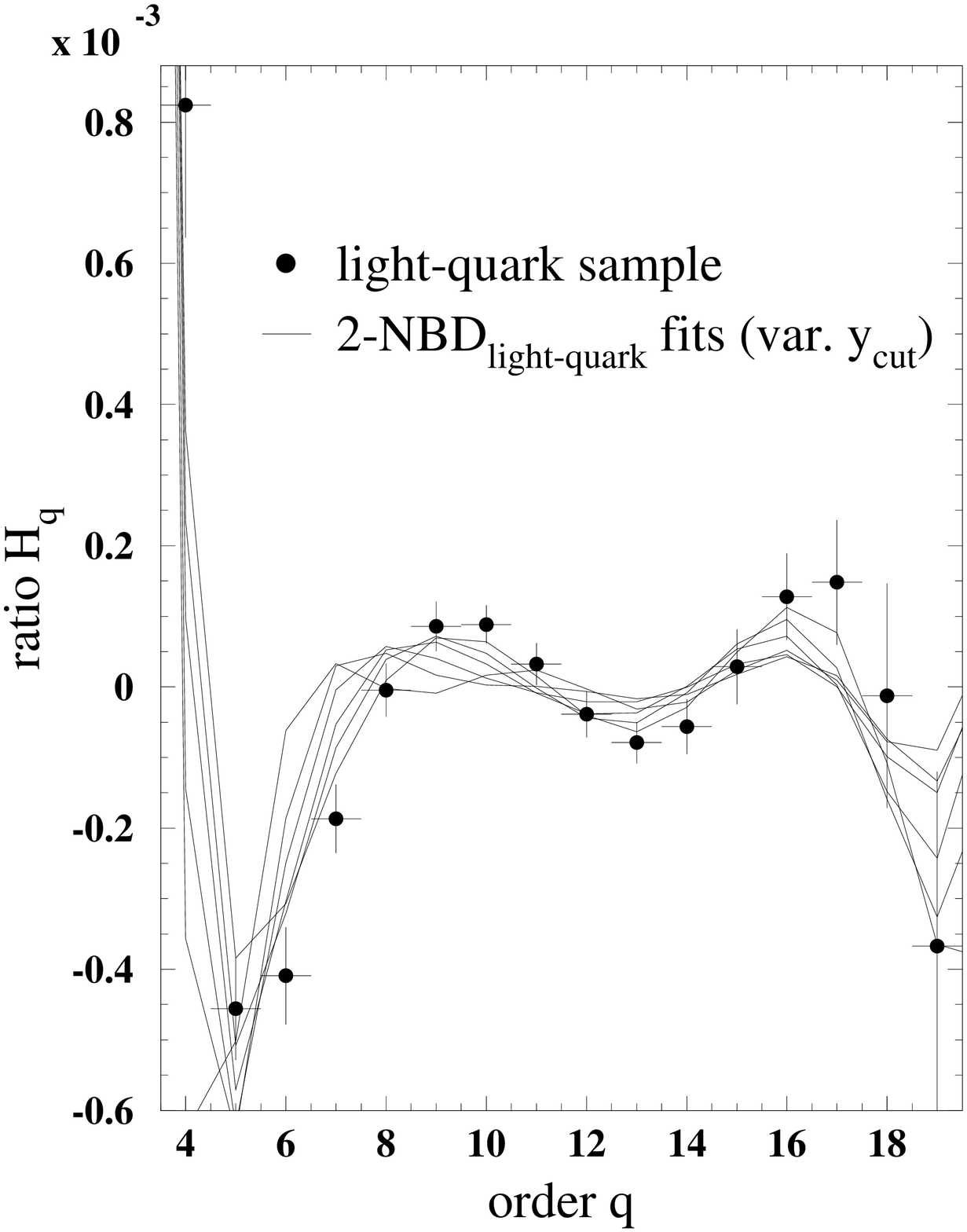}
\end{center}
\vspace{-1cm}
\begin{multicols}{2}
\scaption{\Cpmd{s} of the light-quark sample and of 2-jet and 3-jet events obtained 
with \Ycut{=0.01} fitted by various two-NBD and single-NBD parametrizations, respectively.}
\label{fig:2nbd_la}  
\scaption{$H_q$ moments of the light-quark sample, together with the $H_q$ moments 
calculated from the two-NBD parametrization.}
\label{fig:2nbd_lb} 
\end{multicols}
\end{figure}

\begin{figure}[htbp]
\begin{center}
    \includegraphics[width=8.4cm]{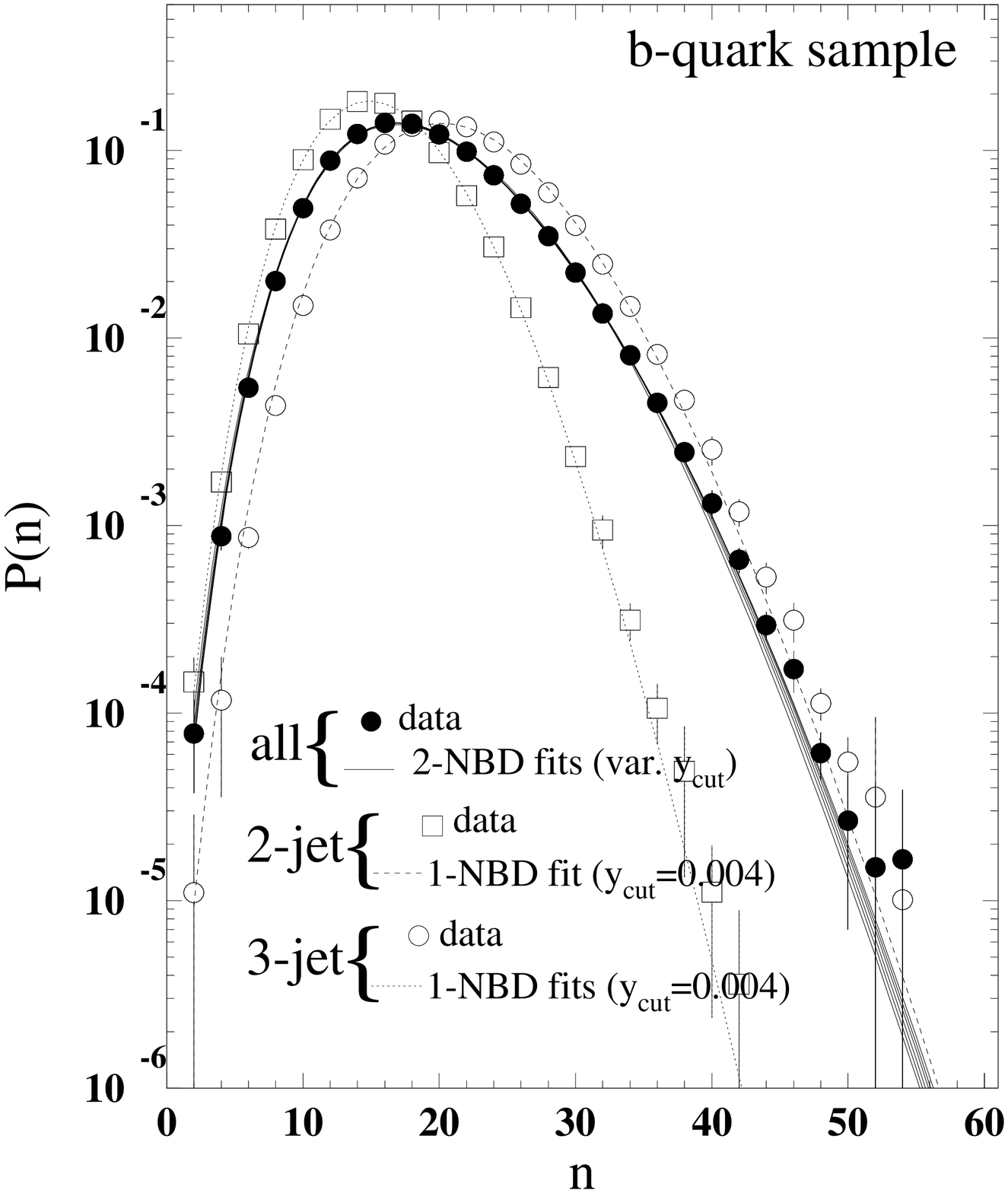}
    \includegraphics[width=8.4cm]{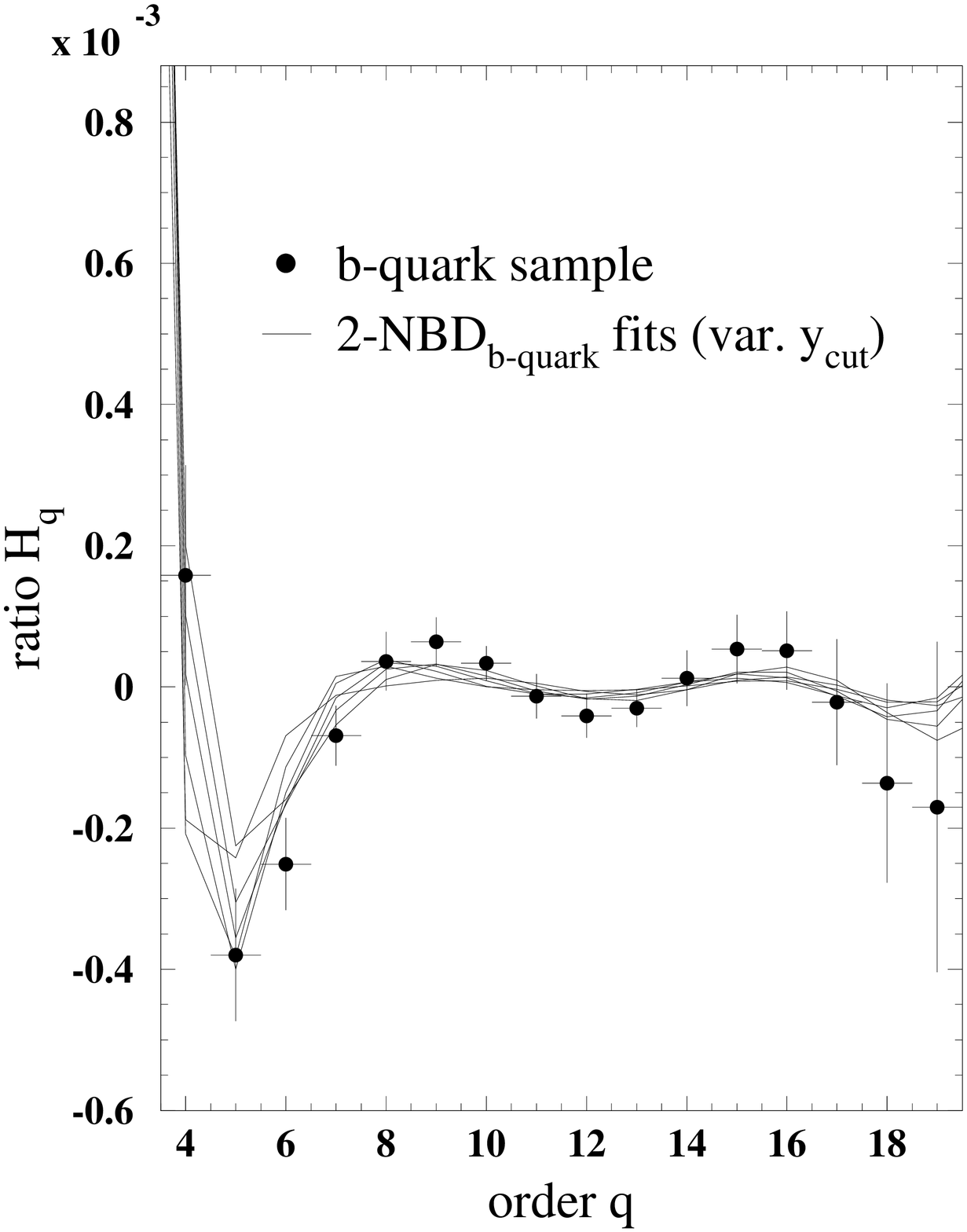}
\end{center}
\vspace{-1cm}
\begin{multicols}{2}
\scaption{\Cpmd{s} of the b-quark sample and of 2-jet and 3-jet events obtained 
with \Ycut{=0.01} fitted by various two-NBD and single-NBD parametrizations, respectively.}
\label{fig:2nbd_ba}  
\scaption{$H_q$ moments of the b-quark sample, together with the $H_q$ moments 
calculated from the two-NBD parametrization.}
\label{fig:2nbd_bb} 
\end{multicols}
\vspace{-0.5cm}
\end{figure}

A few examples of the single-NBD parametrization, together with the 
\cpmd{s} of 2-jet events and of 3-jet events are given in
Figs.~\ref{fig:2nbd_fa}, \ref{fig:2nbd_la}  and~\ref{fig:2nbd_ba} for the 
full, light- and b-quark samples, respectively. 
The same behavior is observed in the same proportion for the full, light- and 
b-quark samples. It seems there is a strong relation between the 
agreement with the single-NBD expectation and the jet configuration of 
the events. 

For the 2-jet events we see that agreement is obtained at low \Ycut{} 
values. For these values, the jet algorithm resolves fewer and fewer 
events as 2-jet events, which means that the remaining 2-jet events 
have a less ambiguous 2-jet status than those of the previous ones 
(these events have relatively narrow jets, and may be 
described as pencil-like 2-jet events).

For the 3-jet events, the agreement with the single-NBD is obtained at large 
\Ycut{} value. This means that the fraction of events not considered as 2-jet 
events is small and these events have topologies very different from the 
2-jet topologies (they are close to Mercedes-like 3-jet events). 
Thus, agreement with a single-NBD is obtained for events 
which have completely different jet configurations and also 
completely different \cpmd{s}, namely pencil-like 2-jet and 
Mercedes-like 3-jet events. This would suggest that the 
mixture of jet configurations plays an important role 
in the origin of the oscillatory behavior of the $H_q$, 
since the $H_q$ moments obtained from the \cpmd{s} of these 
jet configurations themselves (multiplicity distributions obtained from 
2-jet with \Ycut{} smaller than 0.006 and 3-jet with \Ycut{} larger than 
0.01) do not show this oscillatory behavior 
(Fig.~\ref{fig:hq_j12f}). This will be discussed
in more detail in the next section.

Although we both conclude that the jet configuration plays an important role, 
what we find is rather in contradiction with DELPHI analysis~\cite{deljet}. 
This collaboration claimed good agreement 
between \cpmd{s} and the single-NBD parametrization 
simultaneously for the 2-jet, 3-jet and 4-jet obtained 
at the same \Ycut. This is never the case in our analysis, since where 
there is good agreement with the 2-jet events, there is poor 
agreement for the 3-jet events (and {\it vice versa}).

Next, we compare the \cpmd{} of the full, light- and b-quark 
samples with the distribution $f_\text{full}(x,y_\text{cut})$ 
obtained from the weighted sum of the two NBDs (Eq.~(\ref{eq:2nbd_jet})), 
one corresponding to the \cpmd{} of the 2-jet events, the other to the 
3-jet events both obtained at the same \Ycut{} value. 
This procedure gives us a fully constrained two-NBD  parametrization
of the \cpmd{s}. The $\chi^2$ confidence levels are given in 
Table~\ref{tab:2NBD}.

We find an overall good agreement between these two-NBD parametrizations and 
the \cpmd{} of the full, light- and b-quark samples. The \cpmd{s} of the  
full, light- and b-quark samples are shown  
in Figs.~\ref{fig:2nbd_fa}, ~\ref{fig:2nbd_la} and~\ref{fig:2nbd_ba}, 
together with the parametrizations for various values of \Ycut{}. 
This agreement is also reflected in the $H_q$. 
The $H_q$ calculated from the two-NBD parametrization  
is seen to agree quite well with the data for the 
full sample as well as for both light- and b-quark samples, 
as seen in Figs.~\ref{fig:2nbd_fb}, ~\ref{fig:2nbd_lb} and~\ref{fig:2nbd_bb}, 
respectively.
This agreement is rather odd since, as we have seen previously, for a given 
\Ycut{} value, only one of the two distributions can be well described by a NBD.
Therefore, we cannot conclude that 
the two-NBD hypothesis in terms of 2-jet and non-2-jet events 
is a success.

However, as we cannot claim the success of this two-NBD parametrization, 
we cannot ignore that very singular jet configurations as the ones 
observed above are well described by a single-NBD 
parametrization. This gives some indication of where the  
aspect of the \cpmd{}, responsible for the oscillatory behavior of 
the $H_q$, might lie. We have already identified two (non overlapping) 
components of the \cpmd{} which do not have this oscillatory behavior.
As it will be discussed later in this chapter, the answer might lie in 
the multijet part of the \cpmd{} which is ignored in this parametrization, 
so far.

\begin{table}[htbp]
\begin{center}
\begin{tabular}{|c|c|c||c|c||c|c|}\cline{2-7}
  \multicolumn{1}{c|}{} &
  \multicolumn{2}{c||}{full sample} &
  \multicolumn{2}{c||}{light-quark sample} & 
  \multicolumn{2}{c|}{b-quark sample} \\
\hline
$y_\text{cut}$&
2-jet event &
3-jet event &
2-jet event & 
3-jet event &
2-jet event & 
3-jet event \\
\hline
$0.03$     & $10^{-6}$  & 0.89              & $10^{-9}$  & 0.90                & $10^{-8}$ & 0.94              \\
\hline
$0.015$    & $10^{-4}$  & 0.34              & $10^{-6}$  & 0.36                & $10^{-6}$ & 0.60              \\
\hline
$0.01$     & 0.002             & 0.02              & $10^{-3}$  & 0.04                & $10^{-4}$ & 0.21              \\
\hline
$0.006$    & 0.42              & $10^{-5}$  & 0.11              & $10^{-4}$    & 0.34             & 0.05              \\
\hline
$0.004$    & 0.83              & $10^{-9}$  & 0.65              & $10^{-8}$    & 0.49             & $10^{-4}$  \\
\hline
$0.002$    & 0.82              & $10^{-14}$ & 0.97              & $10^{-13}$   & 0.81             & $10^{-5}$  \\
\hline
\end{tabular}\end{center}
\vspace{-0.2cm}
\scaption{$\chi^2$ confidence level between the NBD parametrization and their 
experimental counterpart.}
\label{tab:NBD}
\end{table}
\begin{table}[htbp]
\begin{center}
\begin{tabular}{|c|c|c|c|}\cline{2-4}
\hline
$y_\text{cut}$&
full sample &
light-quark sample &
b-quark sample        \\
\hline
$0.03$     & 0.04               & $10^{-3}$         & $10^{-3}$     \\
\hline
$0.015$    & 0.57               & 0.19              & 0.15    \\
\hline
$0.01$     & 0.74               & 0.42              & 0.56     \\
\hline
$0.006$    & 0.48               & 0.19              & 0.86     \\
\hline
$0.004$    & 0.03               & 0.01              & 0.54     \\
\hline
$0.002$    & $10^{-12}$         & $10^{-11}$        & $10^{-4}$    \\
\hline
\end{tabular}\end{center}
\vspace{-0.2cm}
\scaption{$\chi^2$ confidence level between the 2-jet, 3-jet  
parametrizations and their experimental counterpart.}
\label{tab:2NBD}
\end{table}

\subsection{Light- and b-quark superposition}

Instead of viewing the origin of the main features of the shape 
and of the $H_q$ oscillatory behavior
of the \cpmd{} of the full sample as due to the superposition of 
the 2-jet and 3-jet events, the other hypothesis relates 
these features to the flavor content of the sample~\cite{lbquark}.

Therefore, with this hypothesis, we attempt to describe the 
\cpmd{s} of the 2-jet events and the 3-jet events of the full 
sample themselves as a weighted sum of two NBDs, using 
parameters taken from the \cpmd{} of the 2-jet (or 3-jet) 
events of the light- and b-quark samples (Eq.~(\ref{eq:2nbd_fl})).
Knowing that the $H_q$ moments of the full, light- and b-quark samples 
have oscillations of about the same size (see Fig.~\ref{fig:hqflb}), 
as it is the case for 2-jet and 3-jet obtained from the full, light- and 
b-quark samples, we can also test, 
as a consistency check, this hypothesis on the full sample.

Results are summarized in Table~\ref{tab:2nbdfl} for the \cpmd{s} 
of the 2-jet and 3-jet events.
\begin{table}[htbp]
\begin{center}
\begin{tabular}{|c|c|c|}\cline{2-3}
\hline
$y_\text{cut}$&
2-jet events &
3-jet events \\
\hline
$1$        & $1\cdot 10^{-18}$   &                    \\
\hline
$0.03$     & $8\cdot 10^{-12}$   & 0.56                \\
\hline
$0.015$    & $2\cdot 10^{-9}$    & 0.04                \\
\hline
$0.01$     & $1\cdot 10^{-5}$    & $2\cdot 10^{-4}$    \\
\hline
$0.006$    & 0.01                & $8\cdot 10^{-8}$    \\
\hline
$0.004$    & 0.24                & $3\cdot 10^{-12}$   \\
\hline
$0.002$    & $7\cdot 10^{-5}$    & $1\cdot 10^{-16}$   \\
\hline
\end{tabular}\end{center}
\vspace{-0.2cm}
\scaption{$\chi^2$ confidence level between the \cpmd{s} of the 2-jet, 3-jet 
events of the full sample and the light, b-quark parametrization.}
\label{tab:2nbdfl}
\end{table}
The parametrization of the \cpmd{s} of the 2-jet and 
3-jet events are mainly found to be  
in disagreement with the data, except 
for the 2-jet sample for \Ycut{=0.006} and 
\Ycut{=0.004}. But this seems to be more related to the fact 
that for these \Ycut{} values, the 2-jet or 3-jet events 
of the full sample are already described by single-NBDs 
(see Table~\ref{tab:NBD}).

The parametrization of the \cpmd{} of the full 
sample by two NBD's representing the light- and b-quark contributions 
is found not to describe the data ($\chi^2$ confidence level of 
$1\cdot 10^{-18}$). 
Therefore, the shape of the \cpmd{} of the full sample, or even 
of the 2- and 3-jet events cannot be described in terms of 
light- and b-quark superposition. Examples of that were already given 
in terms of the $H_q$ moments for the full, light- and b-quark 
samples (Fig.~\ref{fig:hqflb} in Chapter~\ref{chap:hq}) and  
for the 2-jet and 3-jet samples in Fig.~\ref{fig:h26}, which show    
that the oscillatory behavior is very similar for the three samples. 
We can deduce from that, that if there is a phenomenon responsible 
for the oscillations, it plays the same role in the full sample  
as in both light- and b-quark samples and consequently cannot 
be due to the flavor composition of the sample.

\section[Origin of the $H_q$ oscillatory behavior]
{Origin of the $H_q$ oscillatory behavior}

In the previous sections, we found that  
extreme 2- or 3-jet configurations 
did not show the $H_q$ oscillatory behavior seen in 
the full sample and that they were quite well 
described by single NBD's. 

Since both these extreme 2- and 3-jet configurations 
co-exist  in the full sample, in this section, we try  
to isolate simultaneously these 
configurations from the full sample. 
As seen in the Monte Carlo study  
at parton level in the previous section, the remaining events,  
which can neither be categorized, undoubtfully, as 2-jet nor as
3-jet events, can be identified as 3-jet events where the third 
jet is a gluon-jet with energy smaller than that of the two quark-jets.  
In the following, we will call these events, soft-jet events.

Following Eq. (\ref{eq:multi}), we can write 
the charged-particle multiplicity distribution in terms of these 
three components as

\begin{equation}
\label{eq:3pn}
P(n)=R_2(y_\text{2-jet}) P_\text{2-jet}(n)+
               R_3(y_\text{3-jet}) P_\text{3-jet}(n)+
               (1-R_2(y_\text{3-jet})-R_2(y_\text{2-jet})) 
               P_\text{soft-jet}(n),
\end{equation}
where the $P_\text{2-jet}(n)$, 
          $P_\text{3-jet}(n)$ and 
          $P_\text{soft-jet}(n)$ are the 
\cpmd{s} of the extreme 2- and 3-jet events and 
of the soft-jet events, respectively. $R_2(y_\text{2-jet})$ and 
$R_3(y_\text{3-jet})$ are the rate of the 2-jet and 3-jet 
events obtained for the \Ycut{} values $y_\text{2-jet}$ and 
$y_\text{3-jet}$, respectively.

The \cpmd{} of the soft-jet events is reconstructed from events remaining 
once we have identified the 2-jet and 3-jet events 
having \cpmd{s} which do not show the $H_q$ oscillatory 
behavior. From our choice of \Ycut{}, we use for the pencil-like 
2-jet events the \cpmd{s} obtained with \Ycut{=0.004} or 
\Ycut{=0.002}. For the Mercedes-like 3-jet events, 
the \cpmd{s} are obtained with \Ycut{=0.03} or \Ycut{=0.015}.
Therefore, we have four different possible \cpmd{s} for  
soft-jet events. These distributions are corrected and 
reconstructed in the same way as any other distributions 
used in this analysis. 
The production rates, means and dispersions of the \cpmd{s} of 
soft-jet events are given in Table~\ref{tab:soft}. 
We also include, the $\chi^2$ confidence 
levels of the comparison of the \cpmd{} with the single-NBD. 
The agreement is rather bad, except for the case 
with \Ycut{} values of 0.004 and 0.015, respectively. 
As an extension to the phenomenological approach tested in the 
previous section, we also test for this case 
the agreement between \cpmd{} of the full sample and 
its parametrization obtained by the use of 3 NBD's  
describing the 2-, 3- and soft-jet events, as given in 
Eq.~(\ref{eq:fits}). With a $\chi^2$ confidence level of 0.96, 
the three-NBD parametrization is found to be in very good agreement  
with the \cpmd{} of the full sample.

\begin{figure}[htbp]
\centering
    \includegraphics[width=8.4cm]{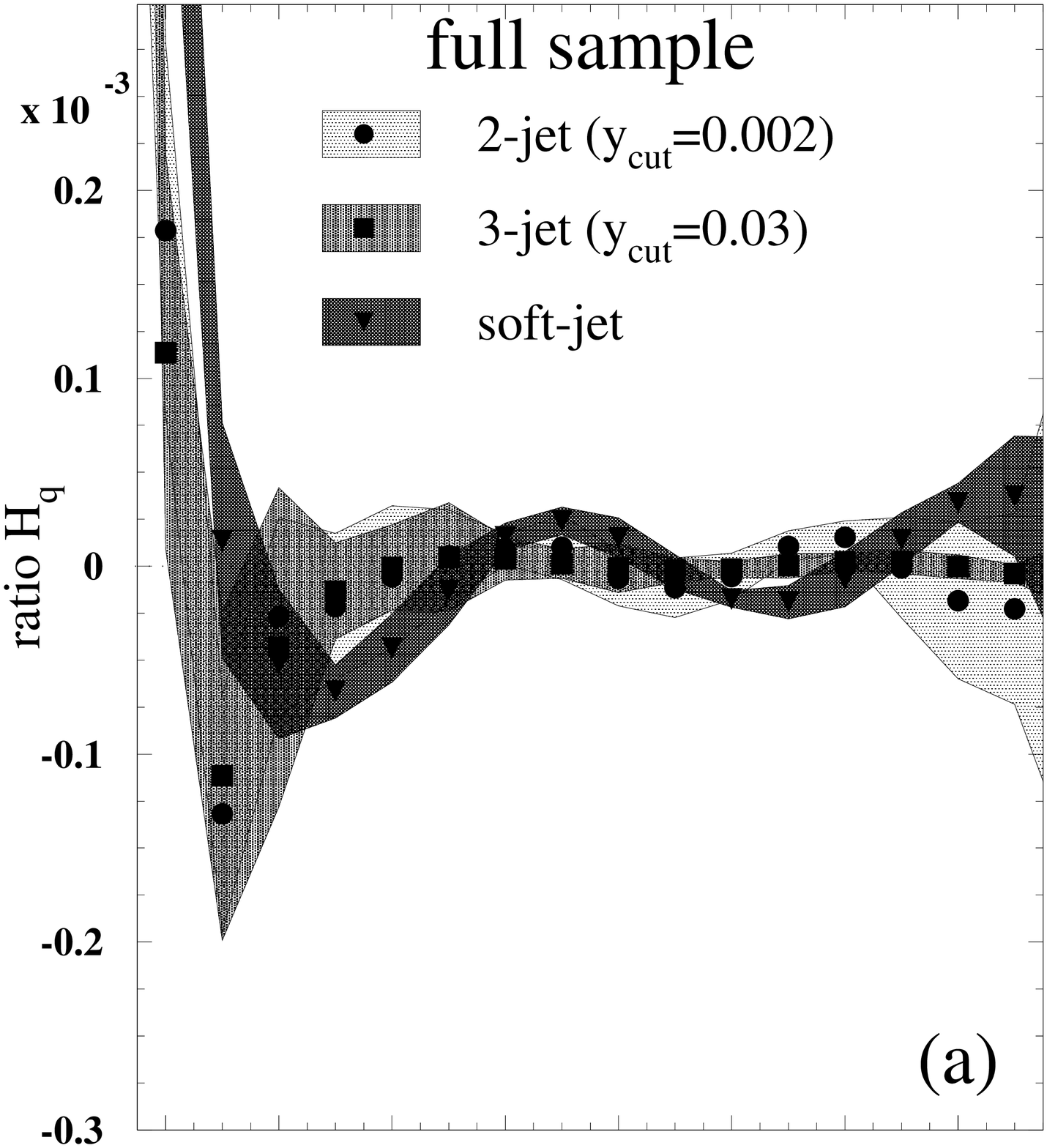}
    \includegraphics[width=8.4cm]{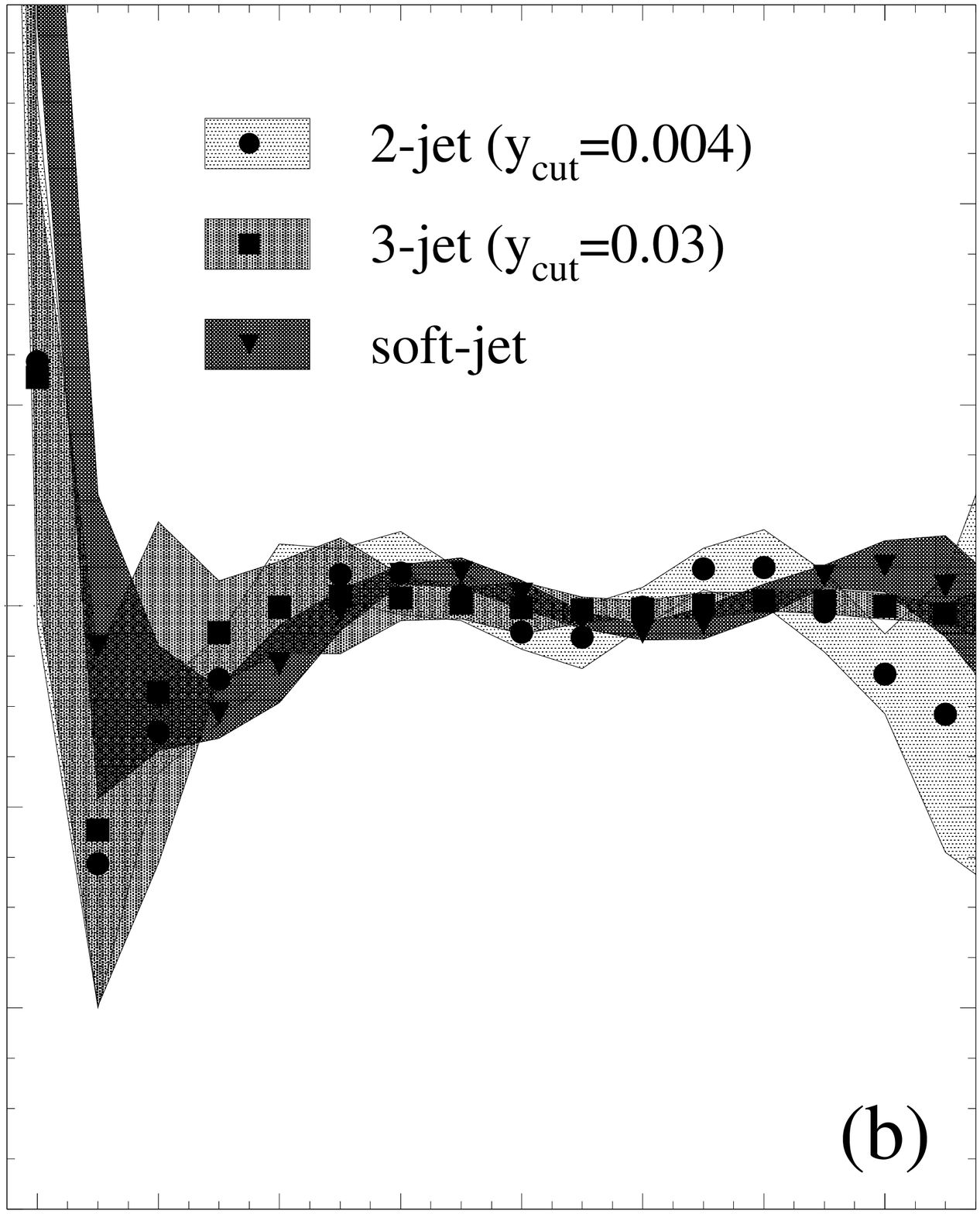}

\vspace{-1cm}

    \includegraphics[width=8.4cm]{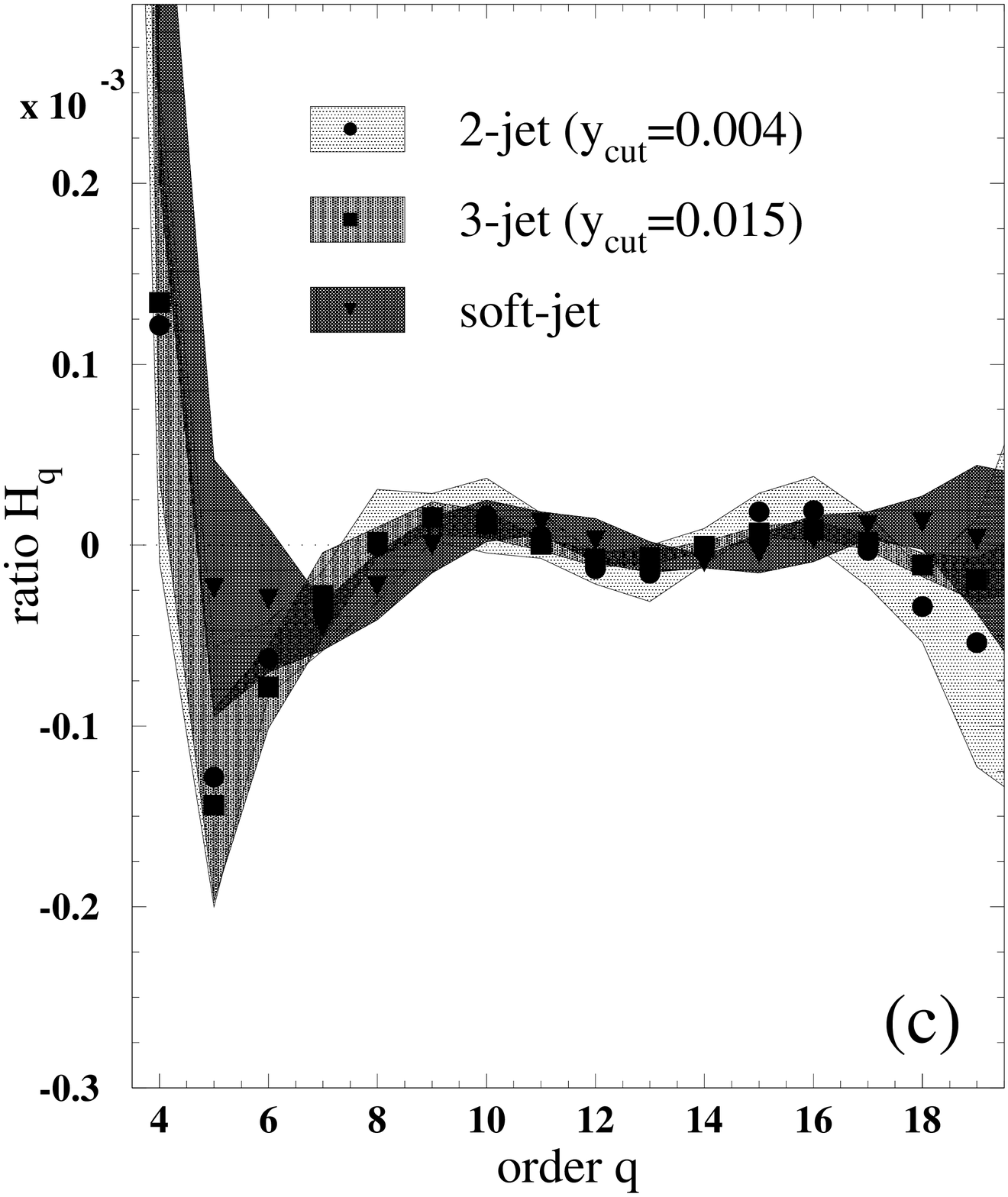}
    \includegraphics[width=8.4cm]{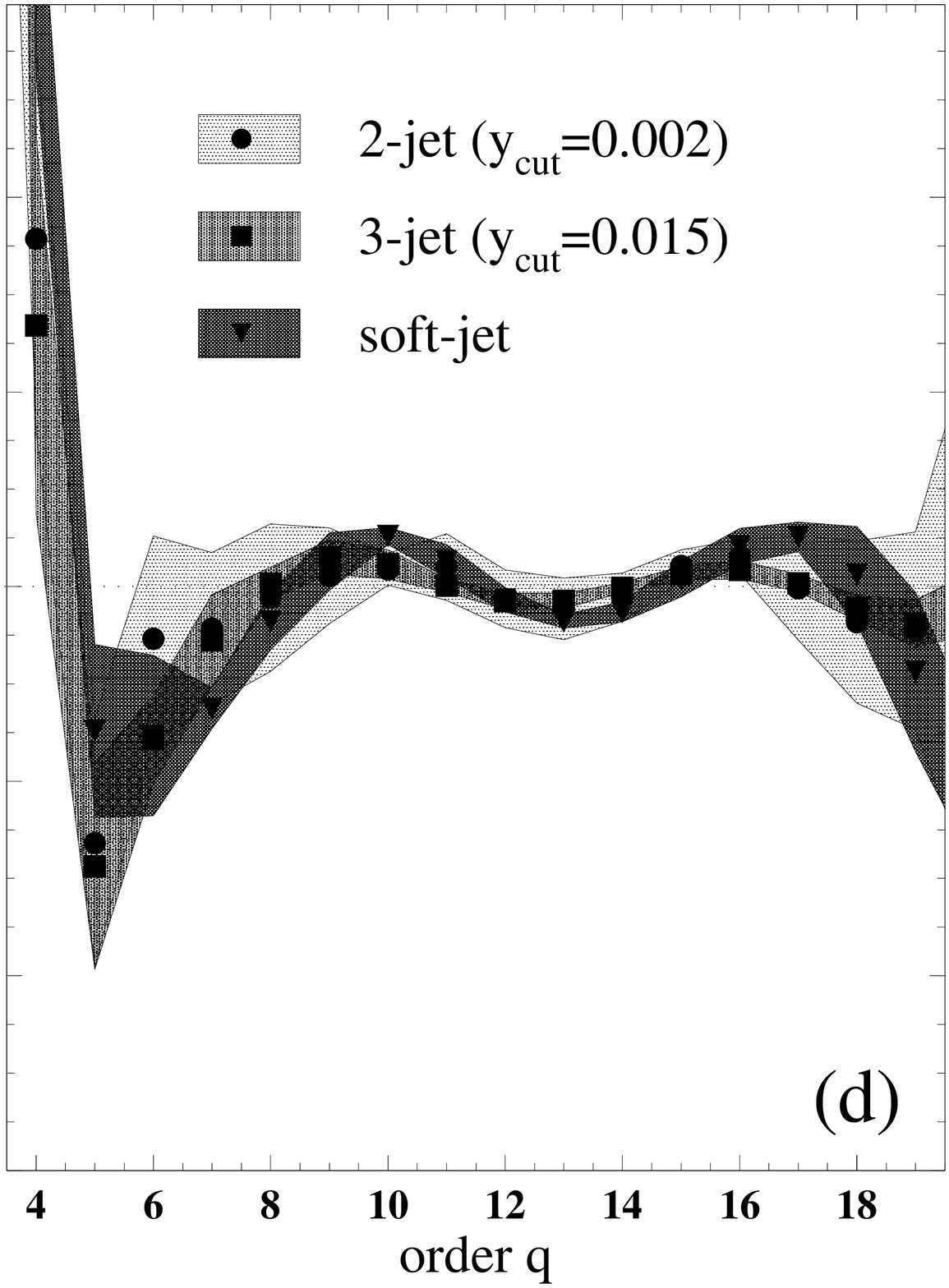}
\scaption{$H_q$ moments of the \cpmd{} of the 2-, 3- and soft-jet events obtained 
with (a) \Ycut{=0.002} for the 2-jet and \Ycut{=0.03} for the 3-jet, 
    (b) \Ycut{=0.004} for the 2-jet and \Ycut{=0.03} for the 3-jet,
    (c) \Ycut{=0.004} for the 2-jet and \Ycut{=0.015} for the 3-jet and 
    (d) \Ycut{=0.002} for the 2-jet and \Ycut{=0.015} for the 3-jet.}
\label{fig:hq_js1}  
\end{figure}

Also the $H_q$ moments of the soft-jet \cpmd{s} are determined. They are 
shown together with the $H_q$ moments of the 2-jet and 3-jet 
events in Fig.~\ref{fig:hq_js1}.
We see that the amplitudes of the oscillations are 
comparable to the residual oscillations seen in the 2-jet and 
3-jet samples.

\begin{table}[htbp]
\begin{center}
\begin{tabular}{|c|c|c|c|c|}\cline{2-5}
  \multicolumn{1}{c|}{} &
  \multicolumn{1}{c|}{$R_\text{soft-jet}$} &
  \multicolumn{1}{c|}{mean} &
  \multicolumn{1}{c|}{Dispersion} &
  \multicolumn{1}{c|}{$\chi^2$ C.L. with NBD}\\
\hline
$y_\text{2-jet}=0.002$  & $52.7\%$& $19.12\pm0.01\pm0.11 $  & $5.145\pm0.005\pm0.053$  &  $7\cdot10^{-8}$   \\
$y_\text{3-jet}=0.030$  &  &                        &                          &                    \\
\hline
$y_\text{2-jet}=0.004$  & $35.5\%$ &$20.01\pm0.01\pm0.11 $  & $5.236\pm0.006\pm0.052$  &  $1\cdot10^{-4}$   \\
$y_\text{3-jet}=0.030$  &  &                        &                          &                    \\   
\hline
$y_\text{2-jet}=0.004$  & $25\%$ &$19.41\pm0.01\pm0.10 $  & $5.014\pm0.006\pm0.047$  &  $0.02$            \\
$y_\text{3-jet}=0.015$  &  &                        &                          &                    \\ 
\hline
$y_\text{2-jet}=0.002$  & $42.2\%$ &$18.53\pm0.01\pm0.10 $  & $4.891\pm0.005\pm0.048$  &  $2\cdot10^{-5}$   \\
$y_\text{3-jet}=0.015$  &  &                        &                          &                    \\   
\hline
\end{tabular}\end{center}
\vspace{-0.2cm}
\scaption{Mean $\langle n\rangle$ and dispersion $D$ and $\chi^2$ confidence level with a 
NBD parametrization of the soft-jet events, for four different definitions}
\label{tab:soft}
\end{table}
For the \cpmd{} of the soft-jet events obtained by excluding 2-jet events 
with \Ycut{=0.004} and 3-jet events with \Ycut{=0.015} (Fig.~\ref{fig:hq_js1}(c)), 
the oscillation has almost completely disappeared, 
as for the 2- and 3-jet events. 
\begin{figure}[htbp]
  \begin{center}
    \includegraphics[width=8.4cm]{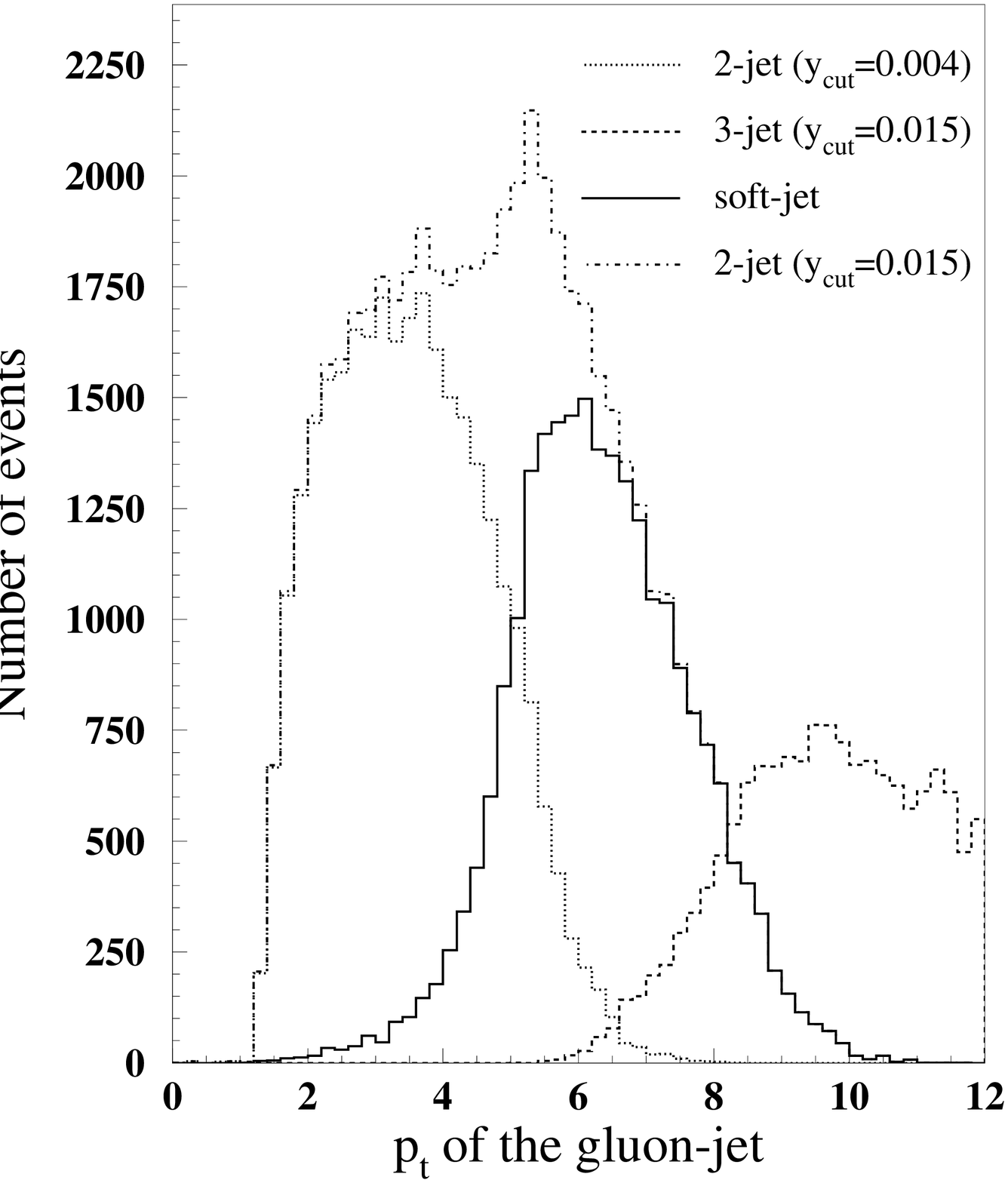}
    \includegraphics[width=8.4cm]{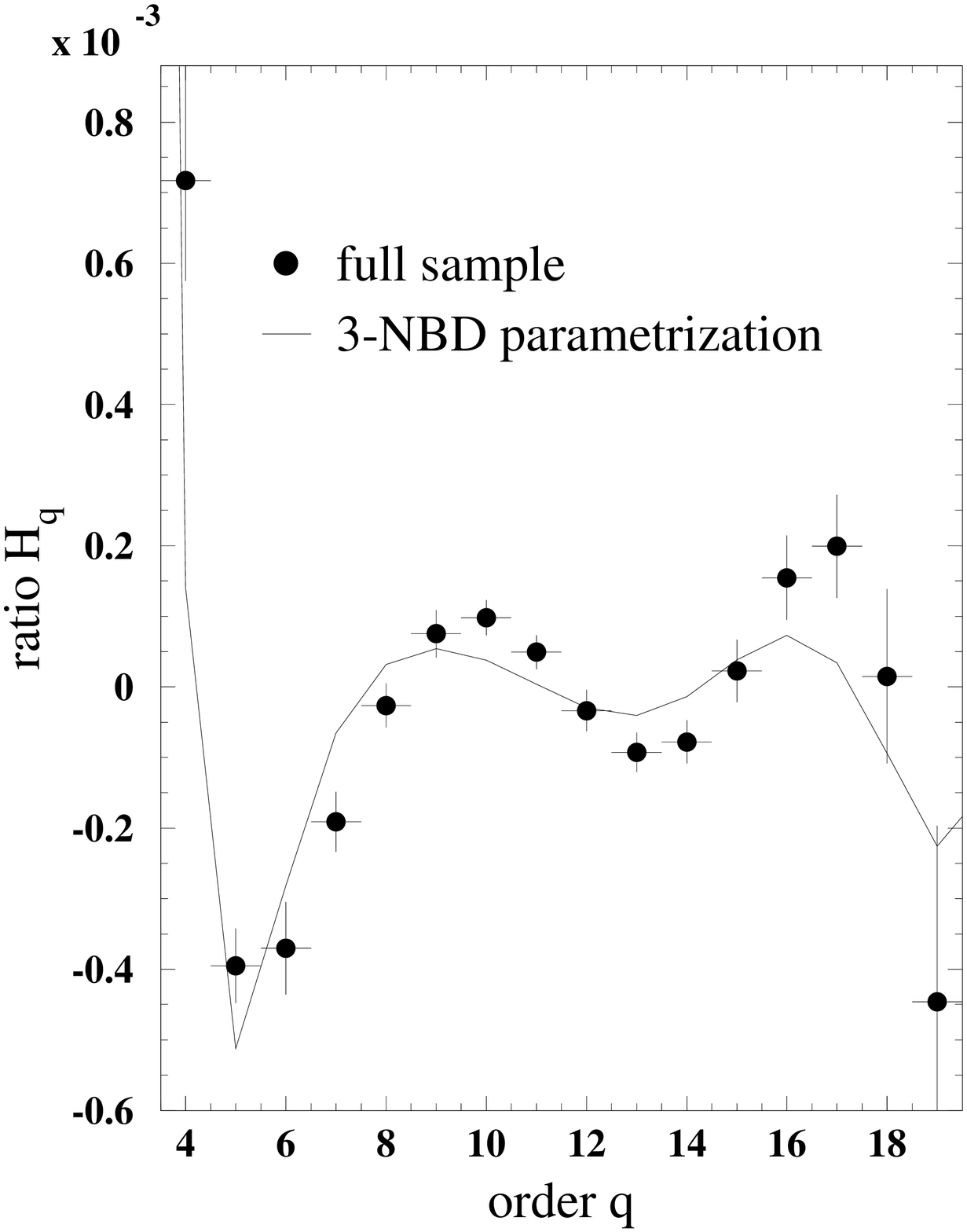}
  \end{center}
\vspace{-1cm}
\begin{multicols}{2}
\scaption{Transverse momentum of the gluon in the thrust frame  
for events which have been resolved as 2-jet and 3-jet with 
\Ycut{=0.015}, together with 2-jet events resolved with 
\Ycut{=0.004} and soft-jet events.}
\label{fig:3sample}  
\scaption{Comparison with the $H_q$ moments of the full sample and 
the $H_q$ moments calculated from the three-NBD parametrization.}
\label{fig:hq_3nbd} 
\end{multicols}
\vspace{-0.8cm}
\end{figure}
\vspace{1.5cm}
\vspace{-1.5cm}
Therefore, we are able to decompose 
\cpmd{} of the full sample into three \cpmd{s} 
characterizing different jet configurations, whose  
$H_q$ moments do not show oscillatory 
behavior (or show only very faint oscillations). 
It seems clear that the 
origin of the $H_q$ oscillatory behavior is  
the co-existence of the various jet configurations 
in the full sample. 

The effect of the co-existence of the various jet 
configurations is also illustrated by the gluon jet energy 
in the $\mathcal{O}(\alpha_\text{s})$ ME Monte Carlo
events we have studied previously. 
In Fig.~\ref{fig:3sample}, which shows  
transverse momentum of the gluon jets in the thrust frame, 
 for both 2-jet and 3-jet samples obtained with \Ycut{=0.015},  
where the 3-jet sample does not have $H_q$ oscillation, 
the 2-jet distribution of the gluon momentum has a completely 
asymmetric behavior showing a structure around 6~\GeV{}. 
Now, if this 2-jet sample is decomposed into 
2-jet events obtained with \Ycut{=0.004} whose $H_q$ 
do not show oscillatory behavior and the remaining events 
(\ie{}, the soft-jet events), we see that this structure  
in the gluon momentum has been resolved as the peak  
 of the momentum distribution of the gluon for 
soft-jet events. 
Furthermore, all three samples have now a rather symmetric
distribution of the gluon momentum with 
clearly differentiated peak positions.

Without arguing about the origin of the structure in the
transverse momentum distribution, since it 
is only a events generated with $\mathcal{O}(\alpha_\text{s})$ 
ME Monte Carlo, this illustrates that the 
$H_q$ oscillations  disappear only in samples composed of 
events representing relatively similar jet 
topologies (and hence of similar gluon energies). If we 
try to group together jet topologies which are 
rather different,  
as it is the case, \eg{}, in 2-jet events 
resolved with \Ycut{=0.015}, the $H_q$ oscillatory 
behavior does not disappear completely,  
even if it is smaller than in the full sample.
This may also explain the difference with the DELPHI 
analysis~\cite{deljet}. The decomposition into 2-, 3- and 
4-jet samples is rather different (besides the difference in 
jet algorithm) from our 2-jet, soft-jet and 3-jet event 
decomposition.    
This is because the 4-jet events with their high multiplicities  
would rather be classified by the jet algorithm as part of 
our 3-jet events than of our soft-jet events, which have 
a low multiplicity compared to 3-jet events. 
Therefore, the DELPHI collaboration tempts to 
concentrate their effort on the study of an upper tail 
of the 3-jet multiplicity, which,  in our analysis is 
already described in its whole by a single-NBD.  
In our analysis we found that the meaningful  
information is rather located at the 
boundary between 2-jet and 3-jet events, where 
a jet-topology different from that of the 2-jet and 3-jet events 
has an important role in the \cpmd{}.

Using the three-NBD parametrization, we calculate the 
$H_q$ moments and compare them to those 
of the full sample (Fig.~\ref{fig:hq_3nbd}). 
We find that it reproduces the oscillatory behavior 
of the measured $H_q$ moments.
Therefore, we can conclude that the 
phenomenological approach is successful when three distinct jet 
topologies are assumed to be responsible for 
the shape of the \cpmd{}.

Besides this conclusion, one can also make a statement on the 
physical origin of the oscillation of the $H_q$ moments.
As we have seen in the previous chapters, 
the oscillations by themselves are not related to  
pQCD and are mainly caused by the soft hadronization 
process. Here we see that 
we are able to decompose the full sample into three different
samples which do not have $H_q$ oscillatory 
behavior. Among all the different tests we tried in order 
to get rid of this oscillatory behavior, 
this is the only one which gives samples which do not 
have $H_q$ moments showing oscillations (even though 
they still show the first minimum).

Therefore, it seems clear that the origin of the $H_q$ oscillatory 
behavior is related to the co-existence of the various jet topologies 
in the full sample, in \ee{} related to the interplay of 
soft physics and hard gluon radiation.

\chapter{Multiplicity distributions in restricted rapidity windows}

In this chapter, the \cpmd{} is analysed in restricted phase space, 
namely in various central intervals of pseudo-rapidity 
and in their outside complement regions. 

As for the \cpmd{} in full phase space, several attempts 
have been made to describe the \cpmd{s} in restricted 
rapidity interval using the negative binomial distribution, in \ee{} 
annihilation~\cite{tassorap} as well as in hadron-hadron and 
lepton-nucleon~\cite{leptorap} collisions. However, in \ee{} the description 
by a single negative binomial distribution has been 
ruled out at LEP energies~\cite{delrap,alephrap}.
In the first section of this chapter we measure 
the \cpmd{s} and their basic moments
in both central and non-central rapidity intervals. After having 
verified that the negative binomial distribution cannot 
describe our data, in the second section, we concentrate our effort 
on the study of the shape of these \cpmd{s} using the $H_q$ moments.

\section{The charged-particle multiplicity distributions}

In this analysis we define six rapidity intervals. 
For each interval, we build two \cpmd{s}, the first one obtained 
by taking the charged particles which have a rapidity value  
inside intervals centered at 0, $[-|\eta_i|,|\eta_i|]$, 
the second one by taking the remaining charged particles. 
We will refer to the first type of intervals as central intervals, 
and the second as outside intervals.

\begin{figure}[htbp]
\centering
\includegraphics[width=9.5cm]{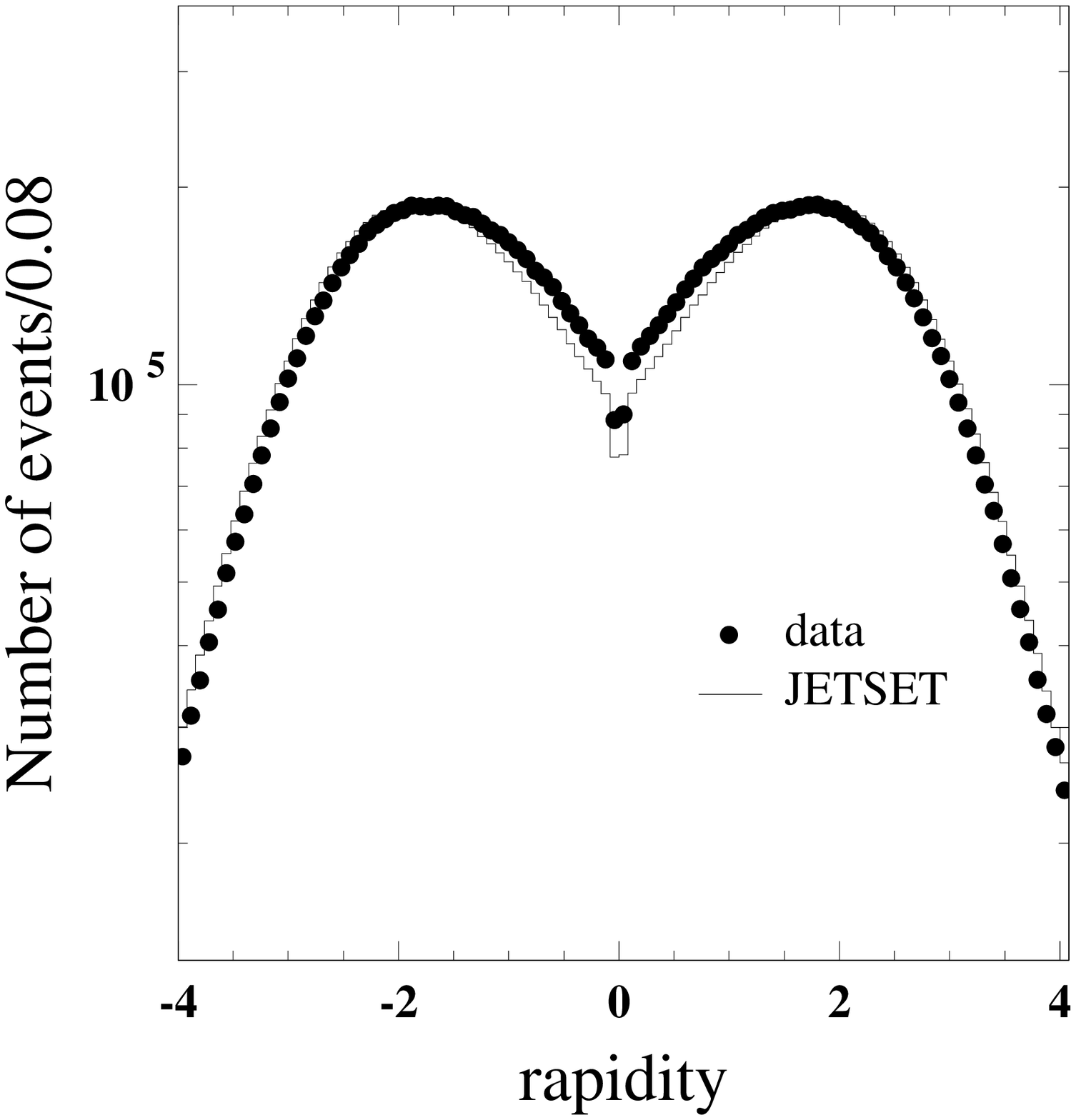}
\scaption{Rapidity distribution obtained  
for the raw data and JETSET at detector level.}
\label{fig:rapidity}
\end{figure}

For simplicity, by rapidity, we mean, in fact, pseudo-rapidity. 
The pseudo-rapidity is equivalent 
to the rapidity when massless particles are assumed  
and is more appropriate to our measurements since 
we do not determine the mass of the particle. It is defined as 
\begin{equation}
\label{eq:rapidity}
\eta=\frac{1}{2}\ln \left (\frac{p+p_\parallel}
{p-p_\parallel} \right ),
\end{equation}
where $p$ is the momentum of the particle and 
$p_\parallel$ its longitudinal component in the thrust frame. 
The rapidity distribution obtained 
for the raw data and JETSET at detector level 
are given in Fig.~\ref{fig:rapidity}.

The \cpmd{s} are reconstructed in the same way as in  
Chapter~\ref{chap:cpmd}, using a Bayesian unfolding method, and corrected 
for acceptance, event selection and initial-state radiation. 
Also the statistical and systematic errors on the \cpmd{s} are  
determined similarly to what we did in Chapter~\ref{chap:cpmd}. 
We limit this analysis to the full sample and we assume stable 
\kl{}.

\begin{figure}[htbp]
\centering
    \includegraphics[width=8.4cm]{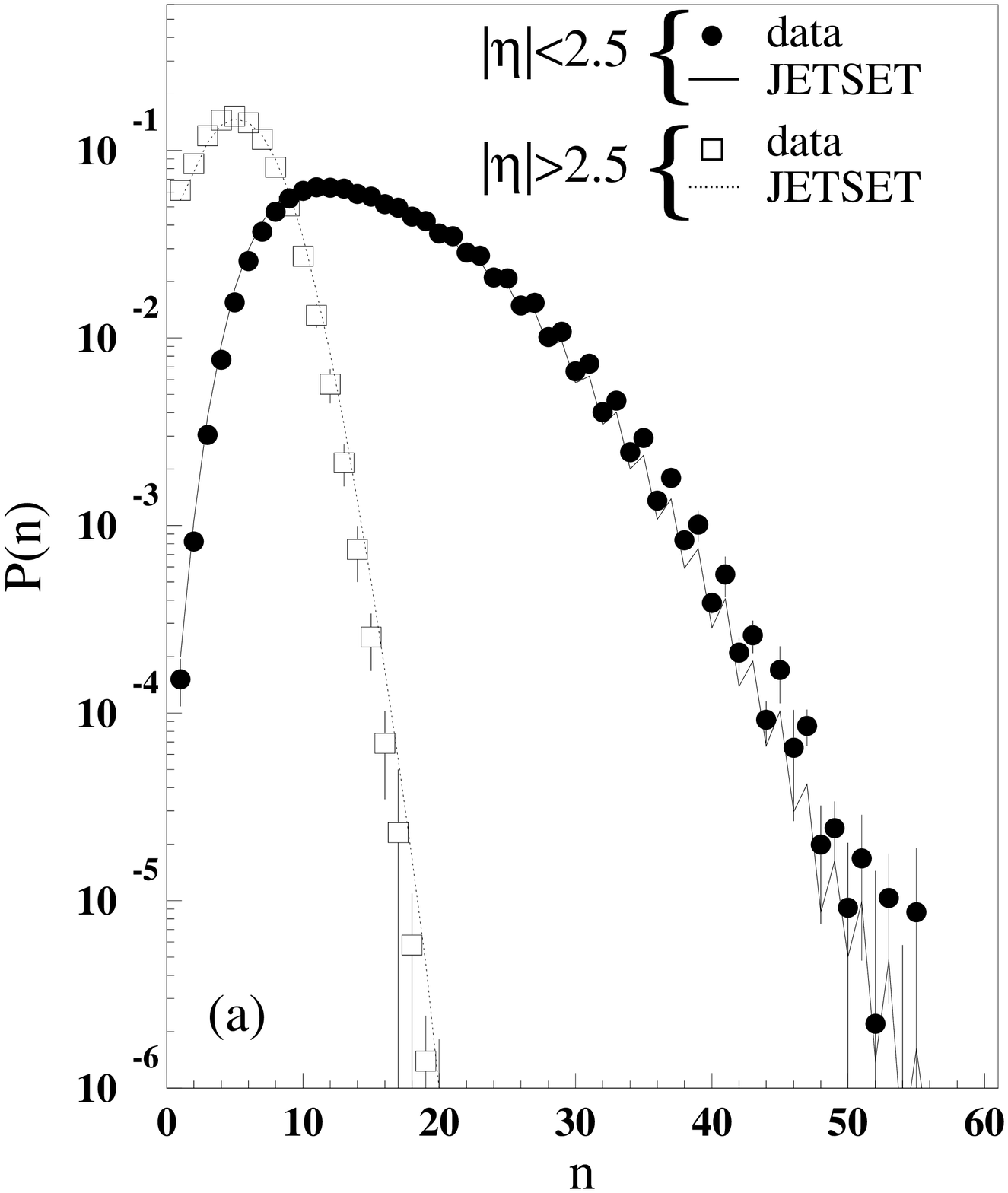}
    \includegraphics[width=8.4cm]{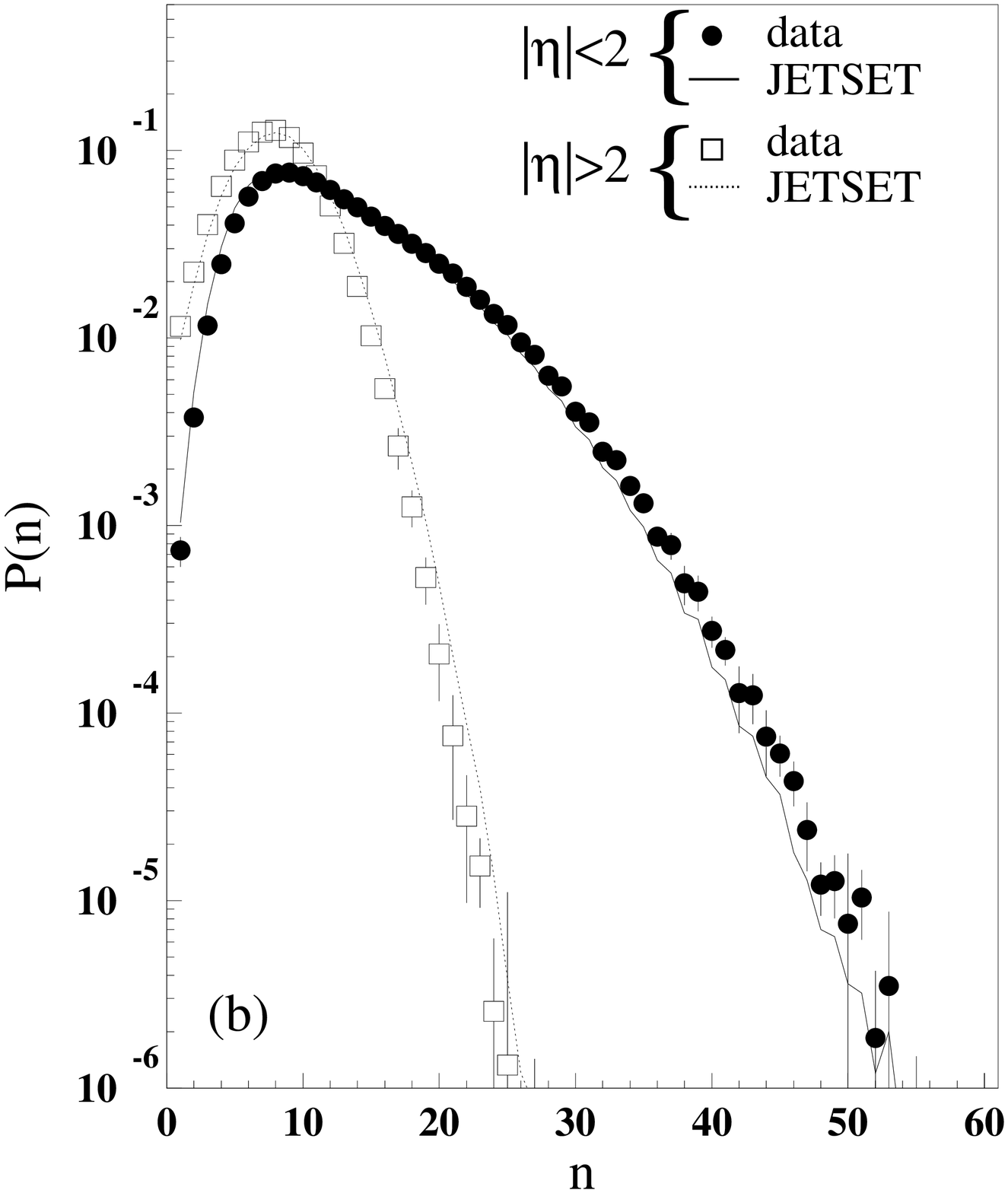}

    \includegraphics[width=8.4cm]{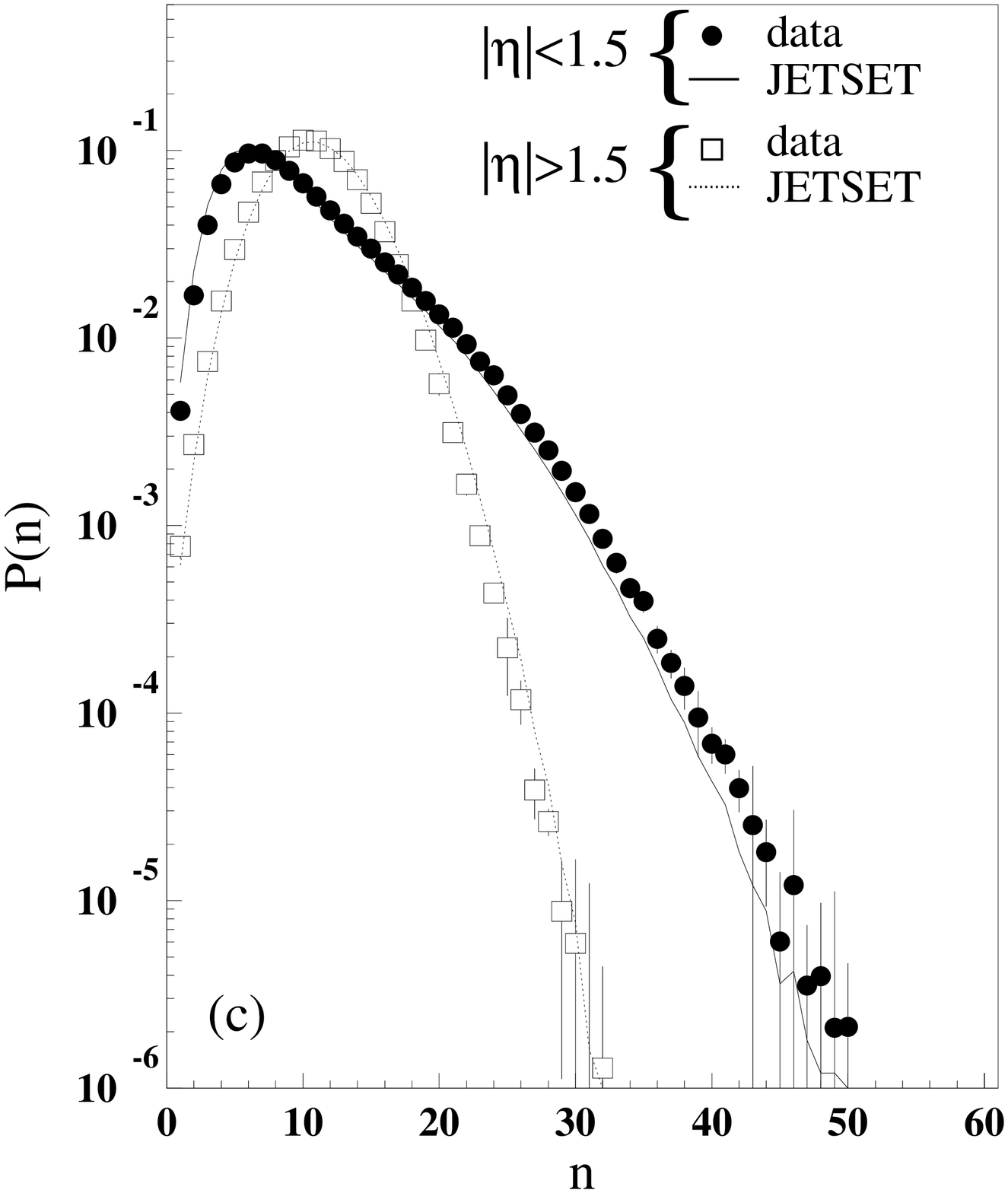}
    \includegraphics[width=8.4cm]{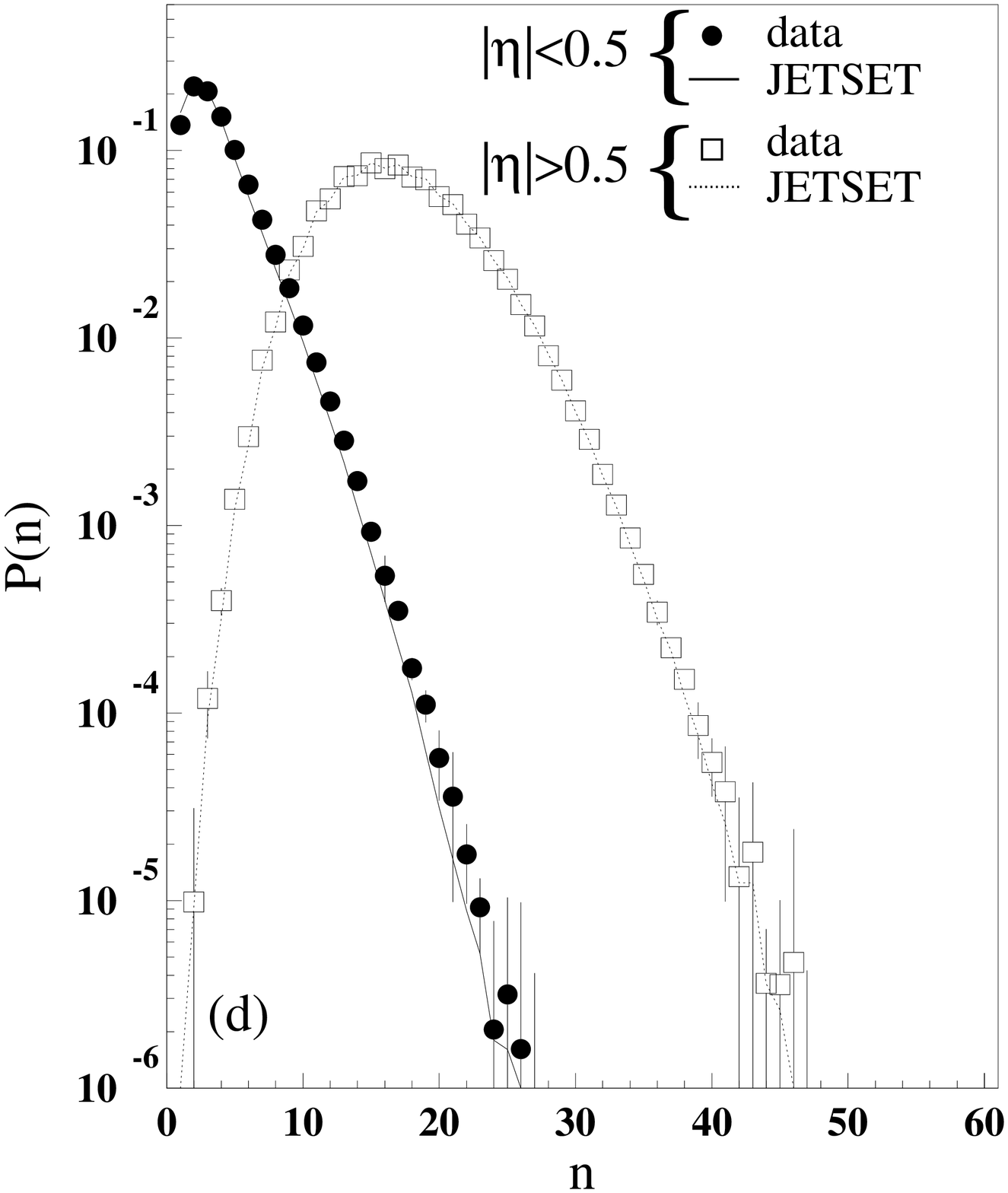}
\scaption{\Cpmd{s} for both central and non-central rapidity intervals.}
\label{fig:pn_rap}  
\end{figure}

A few examples of \cpmd{s} obtained in central rapidity intervals  
and in the outside regions are shown in Fig.~\ref{fig:pn_rap}. 
The odd-even fluctuation in Fig.~\ref{fig:pn_rap}(a) and (b) 
are due to the fact that the full sample only has even $n$. 
The case of Fig.~\ref{fig:pn_rap}(c) ($|\eta|<1.5$ and $|\eta|>1.5$)  
is particularly interesting since this rapidity value divides 
the \cpmd{} of the full sample into two distributions having 
rather similar means. But as we can see, the two 
distributions have completely different shape. 
In the central rapidity interval, the \cpmd{} is highly 
asymmetric and has a shoulder around $n=24$.  
In the outside rapidity region, the distribution is much more 
symmetric and has a slightly higher peak value. 
The basic moments,  such as 
the mean, dispersion, skewness and kurtosis in both central 
interval and outside region are summarized in 
Table~\ref{tab:mom_c_nc}.

For the central rapidity intervals, the skewness and kurtosis increase 
sharply when the size of the interval is decreased, 
characterizing the highly asymmetric distribution  
seen in small rapidity intervals. 

For the outside region, the kurtosis and skewness also  
reflect rather important changes in the \cpmd{} with an  
increasing size of the outside region. However, 
these changes are much smaller than those observed in the 
central rapidity intervals. 

As already observed in previous analyses~\cite{delrap,alephrap}, 
we see for medium  central rapidity intervals 
(Fig~\ref{fig:pn_rap}(b) and (c)) 
a peculiar shoulder structure in the upper tail 
(near $n=24$) of the \cpmd{}.  
This shoulder has been associated to a deviation from the 
shape of a negative binomial distribution~\cite{delrap}.  
We tried to parametrize these distributions with the 
 negative binomial distribution. Our 
results with confidence level close to 0, confirm the results 
of previous analyses, thus showing again that the single negative 
binomial distribution does not describe the \cpmd{s} 
in central or outside intervals.

\begin{table}[htbp]
\begin{center}
\begin{tabular}{|c|c|l|l|}\cline{3-4}
  \multicolumn{2}{c|}{} &
  \multicolumn{1}{c|}{central interval $|\eta|<\eta_i$} &
  \multicolumn{1}{c|}{outside region, $|\eta|>\eta_i$}\\

\hline
$\eta_i=3$    &  $\langle n \rangle$     &             $16.536\pm0.006\pm0.094   $  & \phantom{1}$2.158\pm0.001\pm0.071   $     \\
              &  $D$                     &  \phantom{1}$6.581\pm0.004\pm0.047    $  & \phantom{1}$1.458\pm0.001\pm0.036   $     \\
              &  $\langle n \rangle/D$   &  \phantom{1}$2.513\pm0.001\pm0.011    $  & \phantom{1}$1.481\pm0.001\pm0.025   $     \\
              &  $S$                     &  \phantom{1}$0.590\pm0.002\pm0.014    $  & \phantom{1}$1.881\pm0.003\pm0.055   $     \\
              &  $K$                     &  \phantom{1}$0.289\pm0.008\pm0.033    $  & \phantom{1}$2.53\phantom{1}\pm0.01\phantom{1}\pm0.17      $     \\   
\hline

$\eta_i=2.5$  &  $\langle n \rangle$     &  $14.50\phantom{1}\pm0.01\phantom{1}\pm0.10      $  & \phantom{1}$4.216\pm0.002\pm0.014   $     \\
              &  $D$                     &  \phantom{1}$6.720\pm0.004\pm0.048    $  & \phantom{1}$2.248\pm0.001\pm0.049   $     \\
              &  $\langle n \rangle/D$   &  \phantom{1}$2.157\pm0.001\pm0.013    $  & \phantom{1}$1.876\pm0.001\pm0.014   $     \\
              &  $S$                     &  \phantom{1}$0.751\pm0.003\pm0.016    $  & \phantom{1}$0.888\pm0.002\pm0.014   $     \\
              &  $K$                     &  \phantom{1}$0.46\phantom{1}\pm0.01\phantom{1}\pm0.05       $  & \phantom{1}$0.383\pm0.005\pm0.031   $     \\   
\hline

$\eta_i=2$    &  $\langle n \rangle$     &             $11.854\pm0.006\pm0.092   $  & \phantom{1}$6.879\pm0.002\pm0.094   $     \\
              &  $D$                     &  \phantom{1}$6.433\pm0.005\pm0.049    $  & \phantom{1}$2.993\pm0.002\pm0.047   $     \\
              &  $\langle n \rangle/D$   &  \phantom{1}$1.843\pm0.001\pm0.013    $  & \phantom{1}$2.299\pm0.002\pm0.082   $     \\
              &  $S$                     &  \phantom{1}$0.991\pm0.003\pm0.018    $  & \phantom{1}$0.402\pm0.002\pm0.014   $     \\
              &  $K$                     &  \phantom{1}$0.97\phantom{1}\pm0.01\phantom{1}\pm0.070      $  & \phantom{1}$0.014\pm0.004\pm0.028   $     \\   
\hline
$\eta_i=1.5$  &  $\langle n \rangle$     &  \phantom{1}$8.892\pm0.005\pm0.076    $  & \phantom{1}$9.867\pm0.002\pm0.074   $     \\
              &  $D$                     &  \phantom{1}$5.558\pm0.005\pm0.047    $  & \phantom{1}$3.532\pm0.002\pm0.026   $     \\
              &  $\langle n \rangle/D$   &  \phantom{1}$1.600\pm0.001\pm0.013    $  & \phantom{1}$2.794\pm0.002\pm0.008   $     \\
              &  $S$                     &  \phantom{1}$1.281\pm0.003\pm0.019    $  & \phantom{1}$0.310\pm0.002\pm0.015   $     \\
              &  $K$                     &  \phantom{1}$1.93\phantom{1}\pm0.02\phantom{1}\pm0.11       $  & \phantom{1}$0.110\pm0.005\pm0.015   $     \\   
\hline

$\eta_i=1$    &  $\langle n \rangle$     &  \phantom{1}$5.804\pm0.004\pm0.062    $  & $12.949\pm0.002\pm0.064   $     \\
              &  $D$                     &  \phantom{1}$4.087\pm0.004\pm0.041    $  & \phantom{1}$4.086\pm0.003\pm0.014   $     \\
              &  $\langle n \rangle/D$   &  \phantom{1}$1.420\pm0.001\pm0.014    $  & \phantom{1}$3.169\pm0.002\pm0.018   $     \\
              &  $S$                     &  \phantom{1}$1.607\pm0.004\pm0.042    $  & \phantom{1}$0.405\pm0.003\pm0.010   $     \\
              &  $K$                     &  \phantom{1}$3.27\phantom{1}\pm0.03\phantom{1}\pm0.27      $  & \phantom{1}$0.226\pm0.007\pm0.010   $     \\   
\hline

$\eta_i=0.5$  &  $\langle n \rangle$     &  \phantom{1}$2.731\pm0.002\pm0.051    $  & $15.961\pm0.003\pm0.074   $     \\
              &  $D$                     &  \phantom{1}$2.164\pm0.003\pm0.035    $  & \phantom{1}$4.920\pm0.003\pm0.030   $     \\
              &  $\langle n \rangle/D$   &  \phantom{1}$1.262\pm0.001\pm0.016    $  & \phantom{1}$3.244\pm0.002\pm0.021   $     \\
              &  $S$                     &  \phantom{1}$2.279\pm0.006\pm0.066    $  & \phantom{1}$0.472\pm0.003\pm0.006   $     \\
              &  $K$                     &  \phantom{1}$5.871\pm0.06\phantom{1}\pm0.60      $  & \phantom{1}$0.307\pm0.008\pm0.015   $     \\   
\hline

\end{tabular}\end{center}
\vspace{-0.2cm}
\scaption{Means and dispersions, skewness and kurtosis of the \cpmd{} 
in both central rapidity interval and outside regions.}
\label{tab:mom_c_nc}
\end{table}


\section[{$H_q$} moments]
{\boldmath{$H_q$} moments}

\begin{figure}[htbp]
\centering
\vspace{-0.4cm}
    \includegraphics[width=7.4cm]{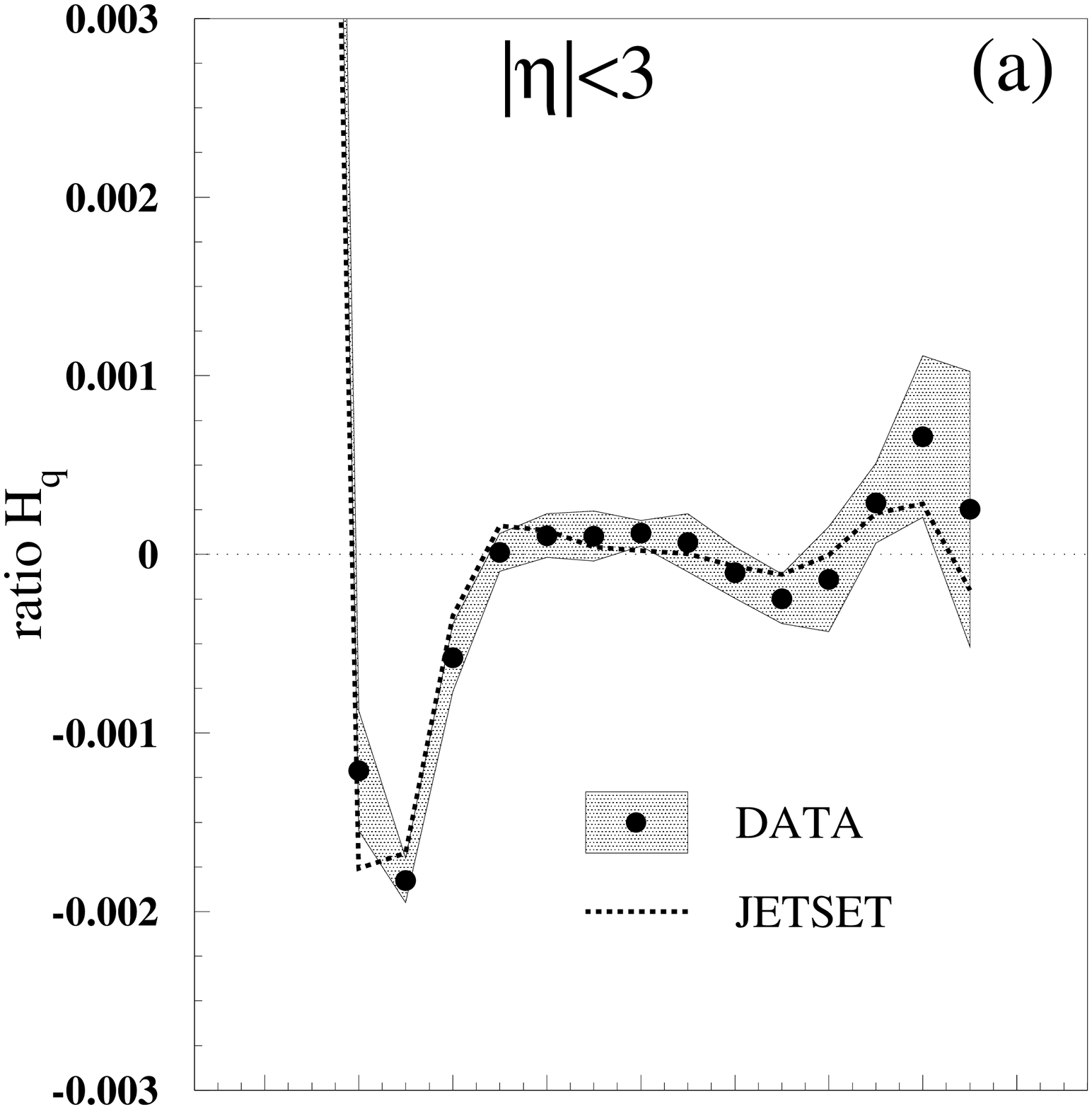}
    \includegraphics[width=7.4cm]{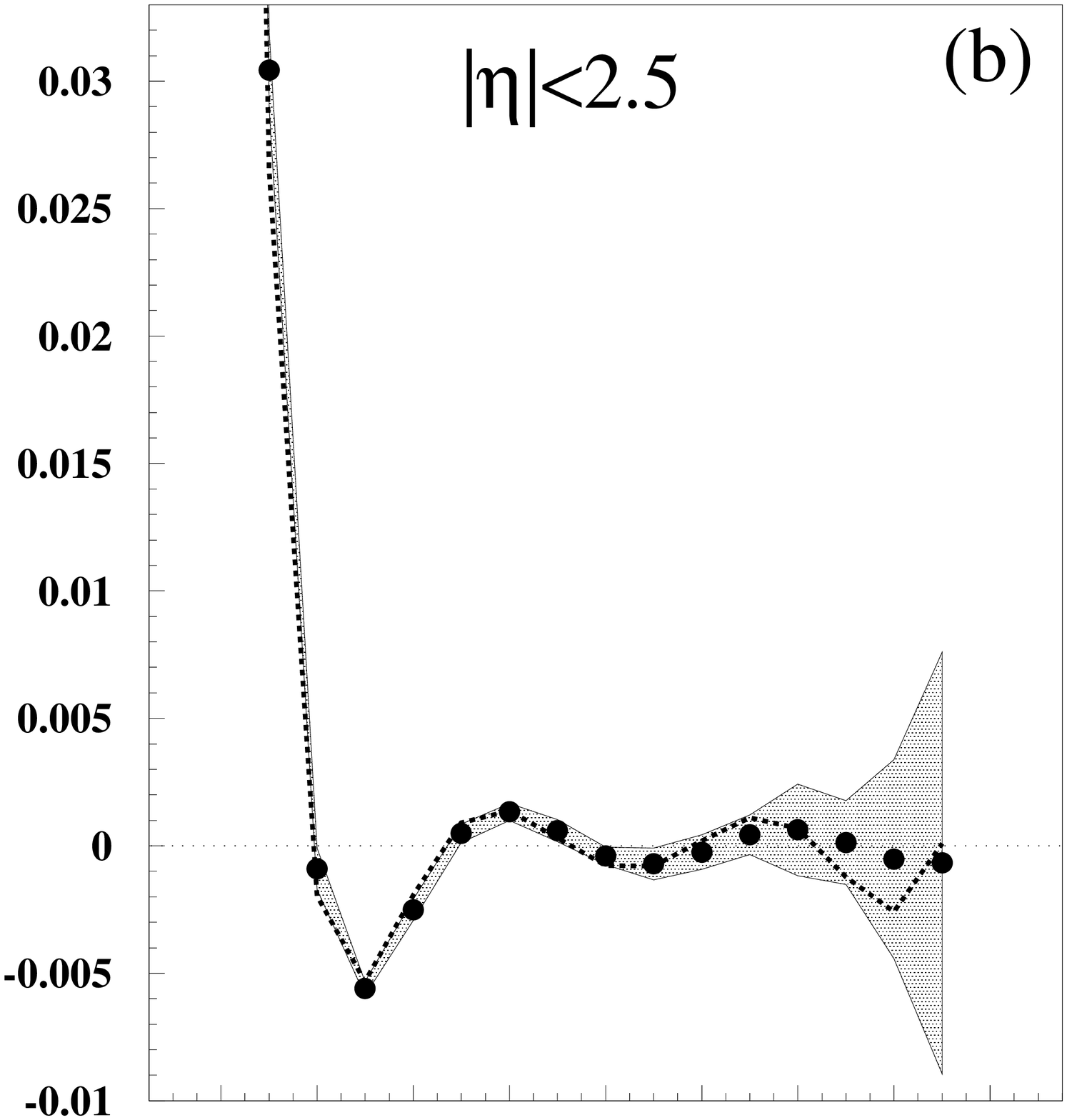}

\vspace{-1.4cm}

    \includegraphics[width=7.4cm]{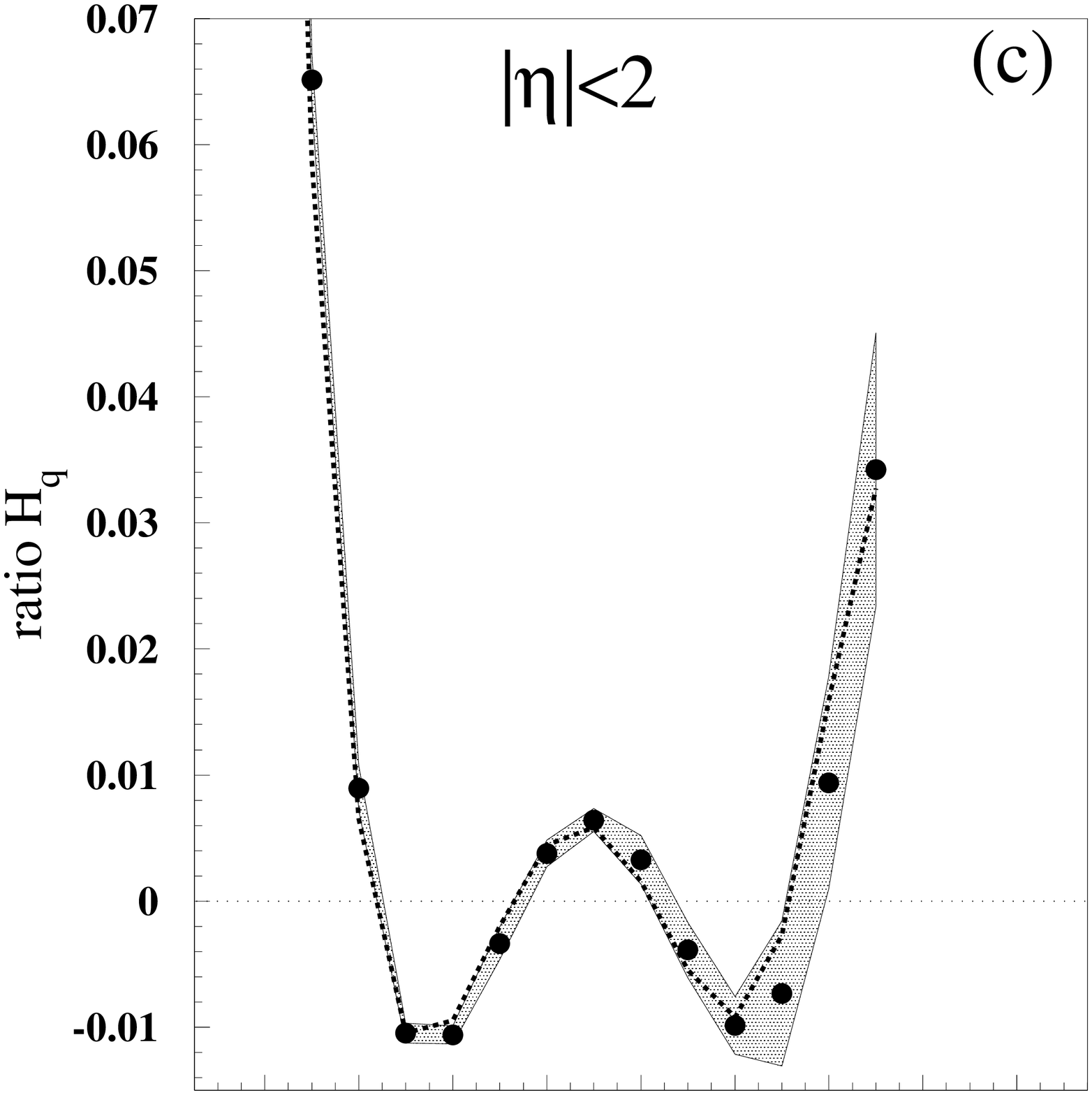}
    \includegraphics[width=7.4cm]{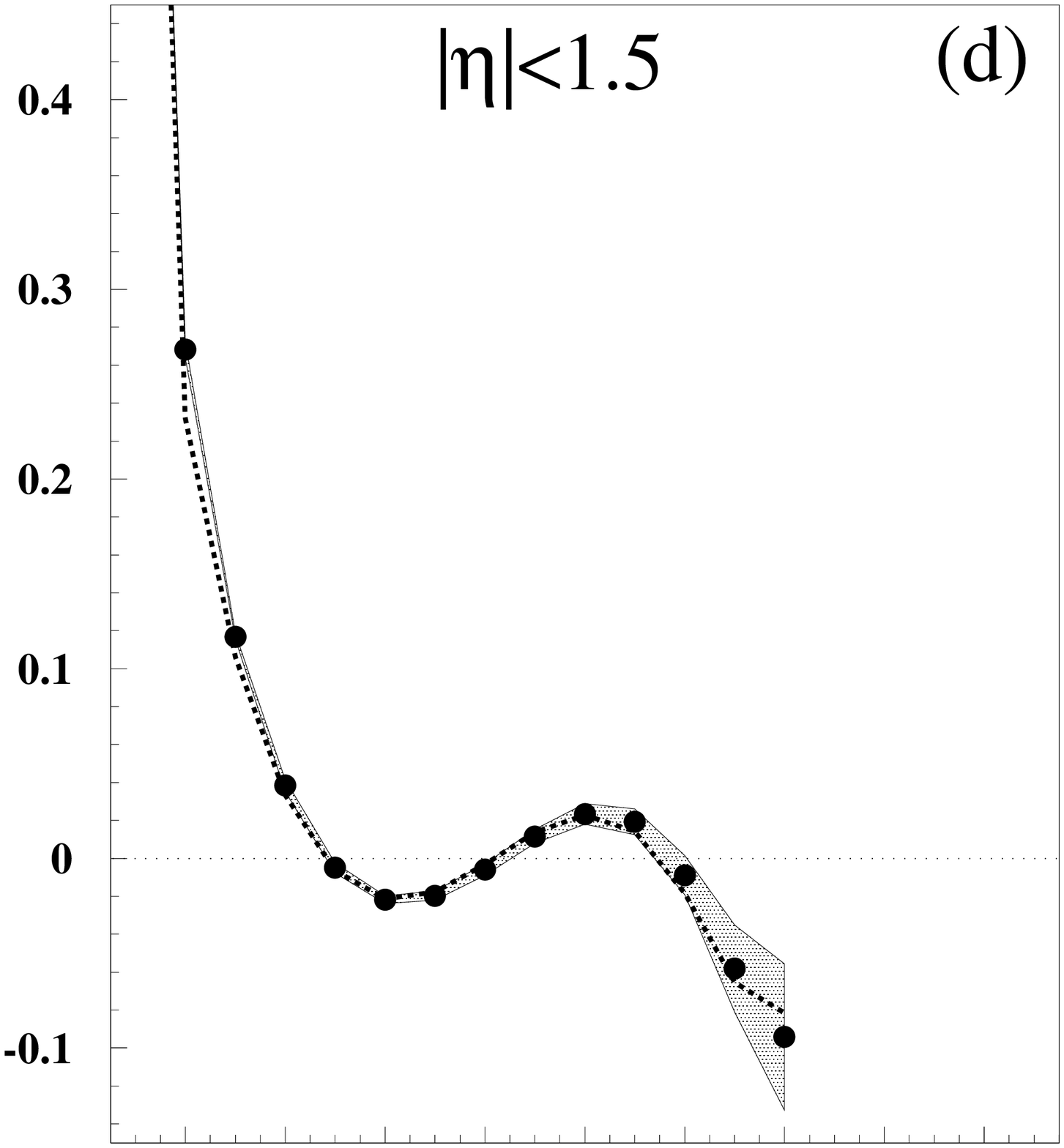}
\vspace{-1.4cm}

    \includegraphics[width=7.4cm]{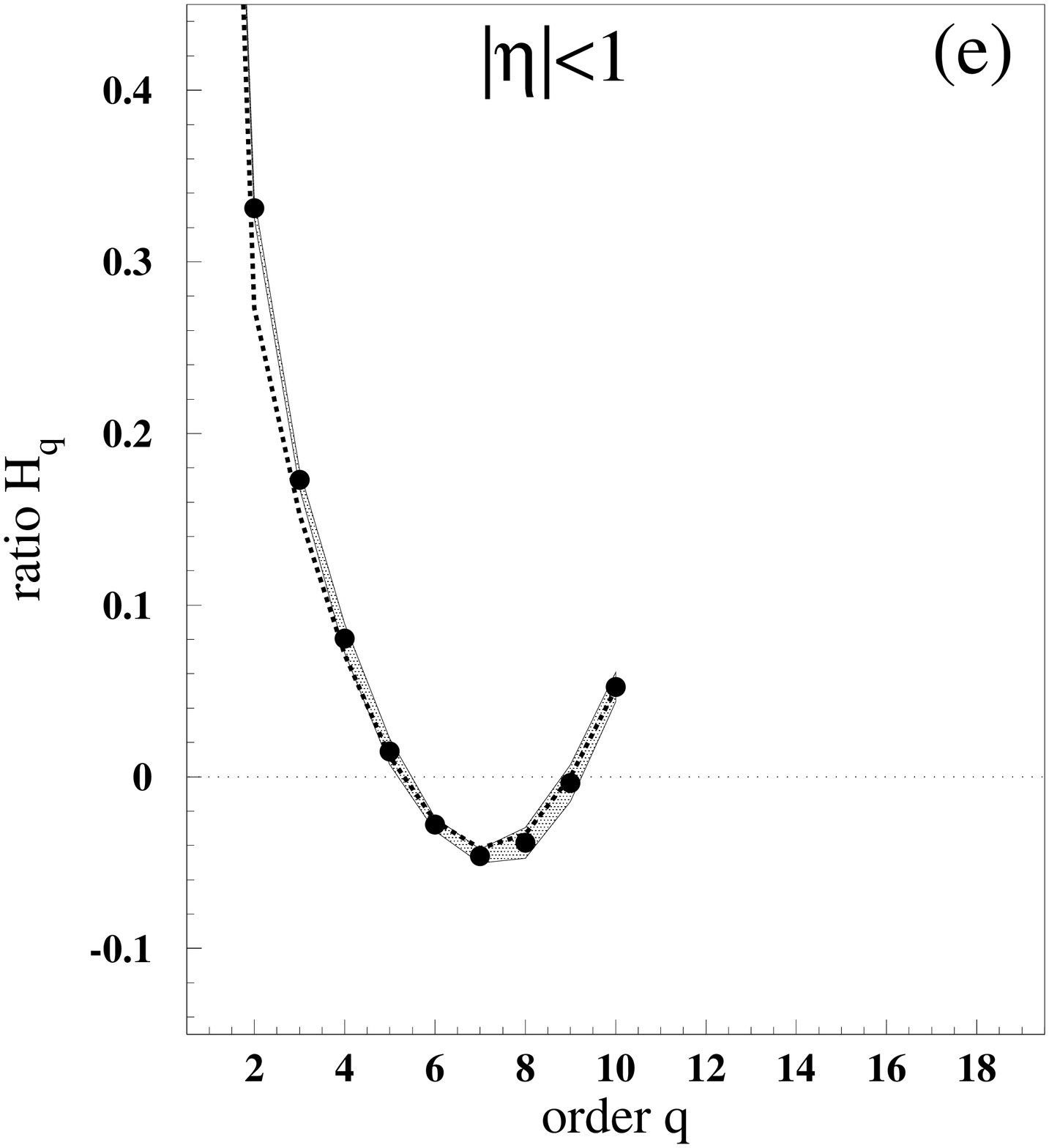}
    \includegraphics[width=7.4cm]{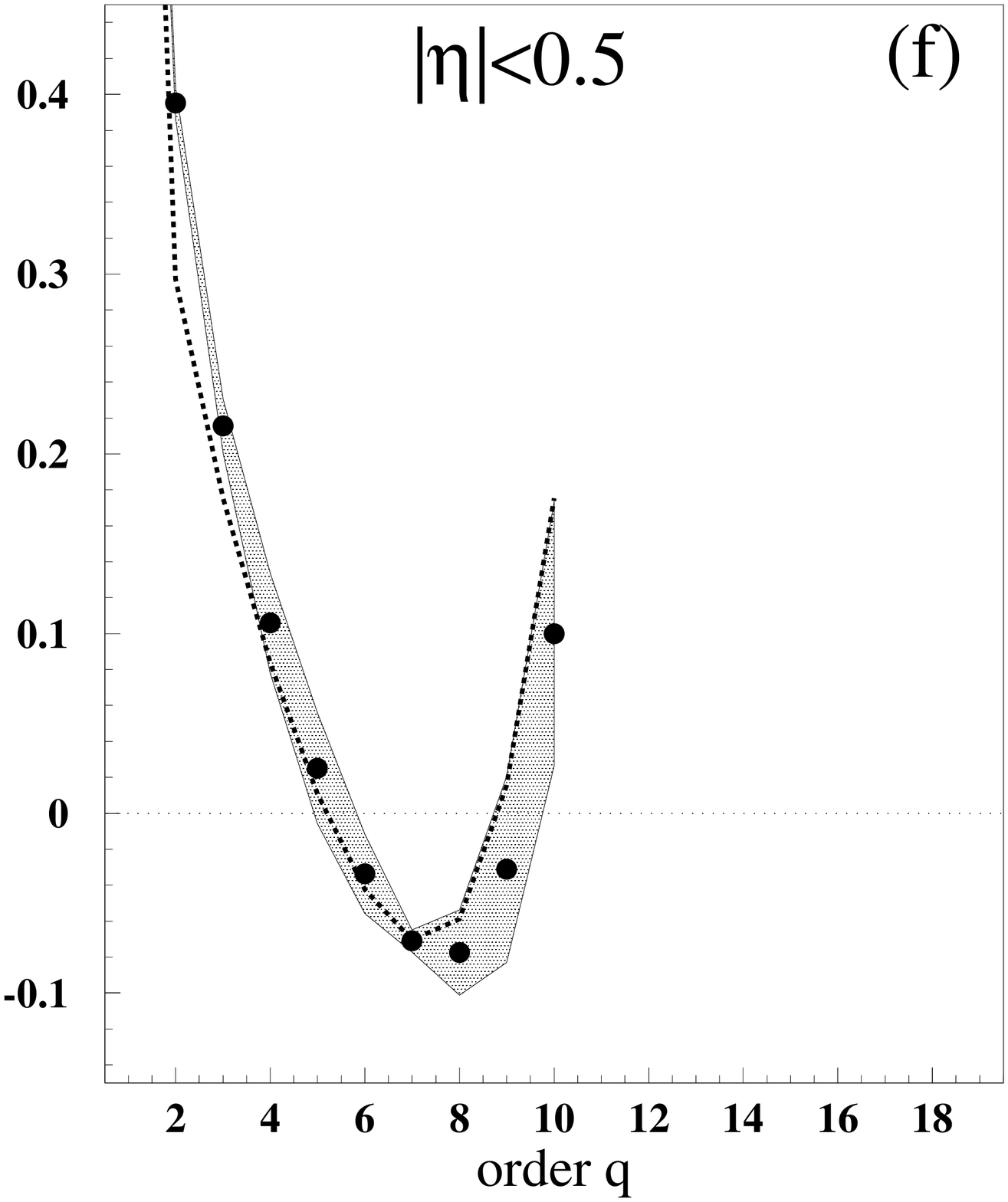}

\vspace{0.2cm}
\scaption{$H_q$ moments measured from \cpmd{s} obtained 
with (a) $|\eta|<3$, (b) $|\eta|<2.5$
     (c) $|\eta|<2$, (d) $|\eta|<1.5$,
     (e) $|\eta|<1$ and (f) $|\eta|<0.5$.}
\label{fig:rap_c1_f} 
\end{figure}

We also measure the $H_q$ moments the various rapidity intervals.  
The $H_q$ moments found in central rapidity intervals 
are shown in Fig.~\ref{fig:rap_c1_f}.
As for the sample in full phase space (Fig.~\ref{fig:hqfull}), 
we see an oscillatory 
behavior with a first negative minimum near $q=5$. This minimum
shifts to higher values of $q$ when the rapidity interval is decreased.
We also note a sharp increase in the amplitude of 
the oscillations, as well as of the depth of the first 
negative minimum.

\begin{figure}[htbp]
\centering

\vspace{-0.4cm}
    \includegraphics[width=7.4cm]{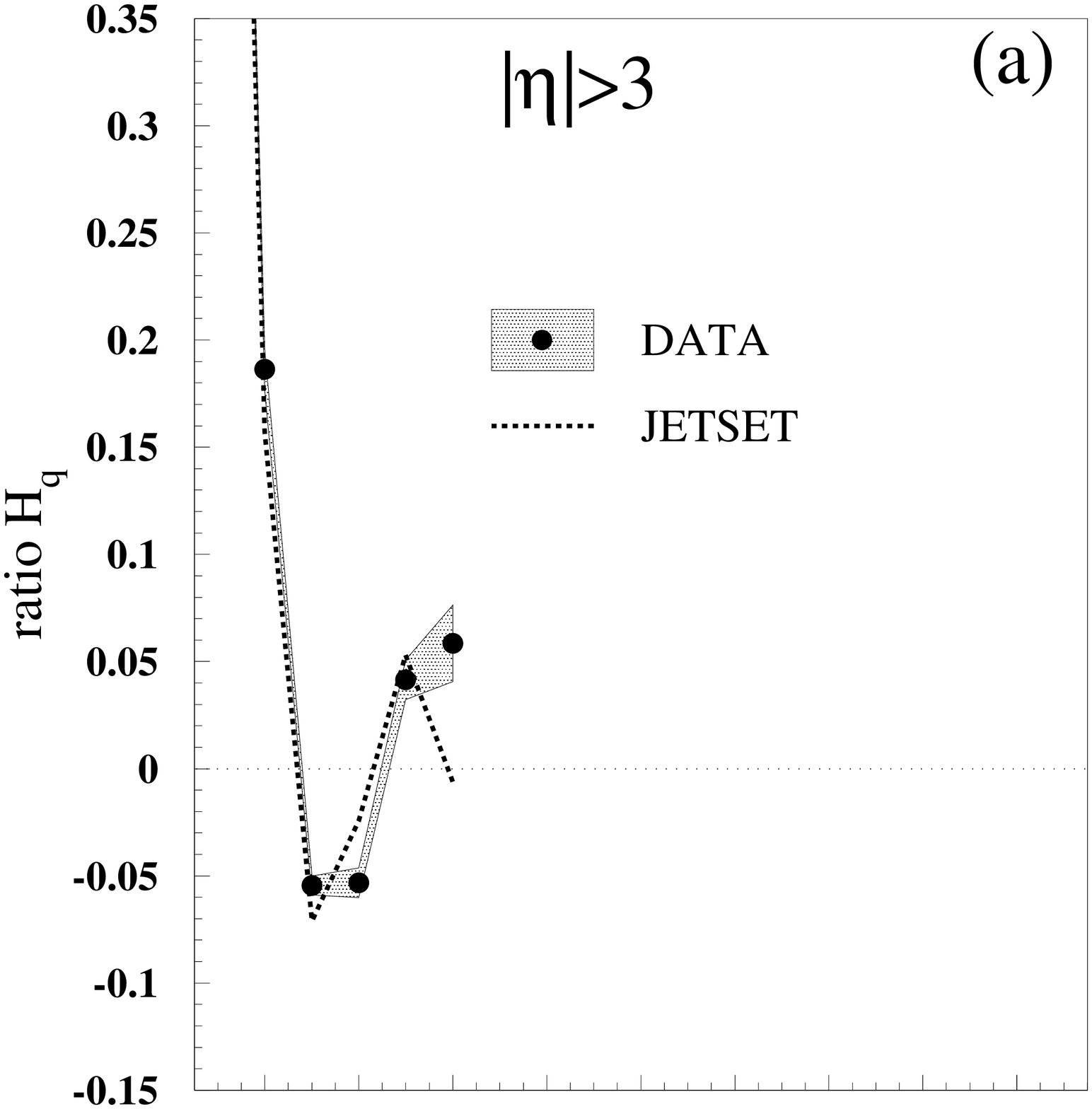}
    \includegraphics[width=7.4cm]{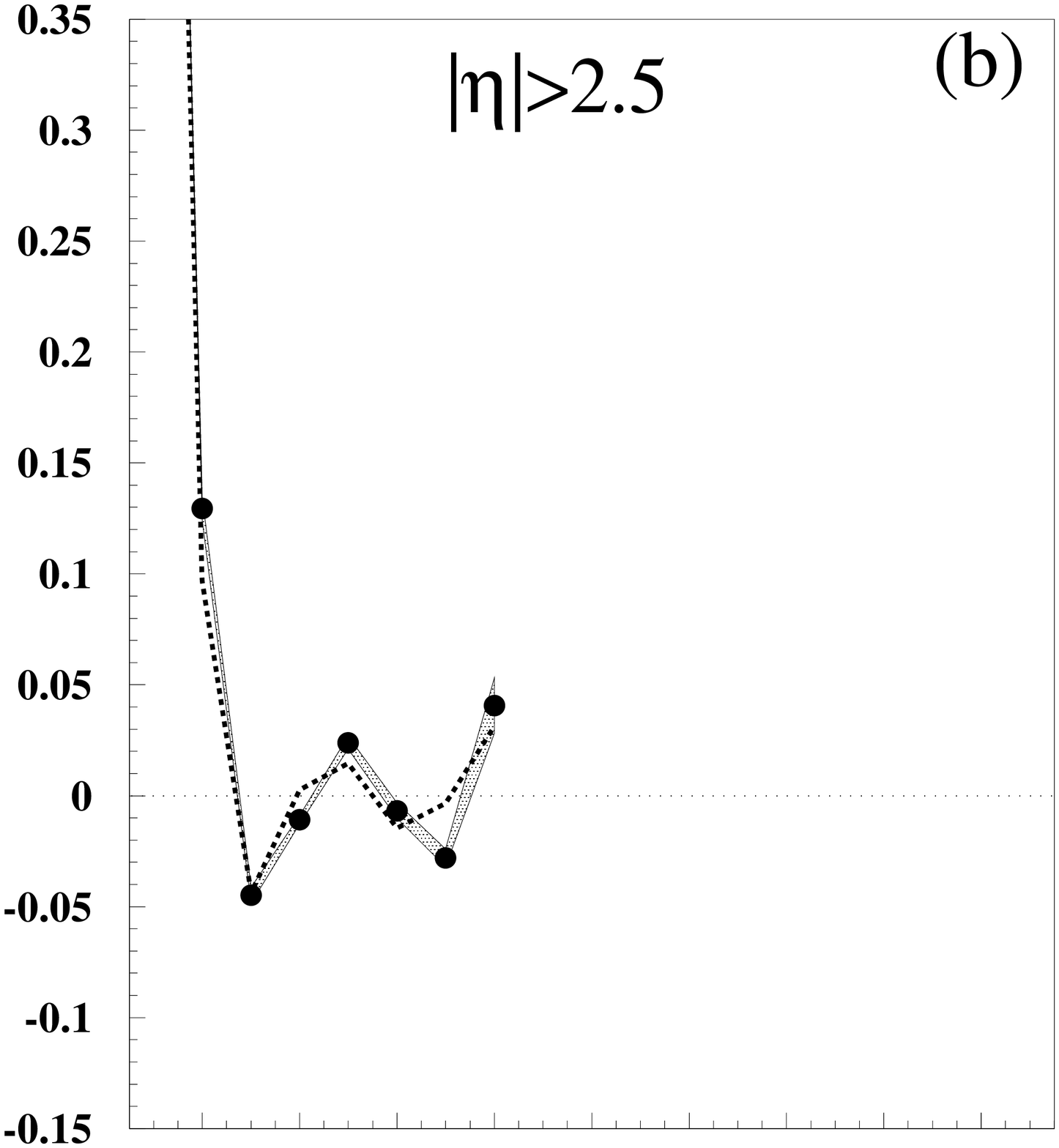}

\vspace{-1.4cm}

    \includegraphics[width=7.4cm]{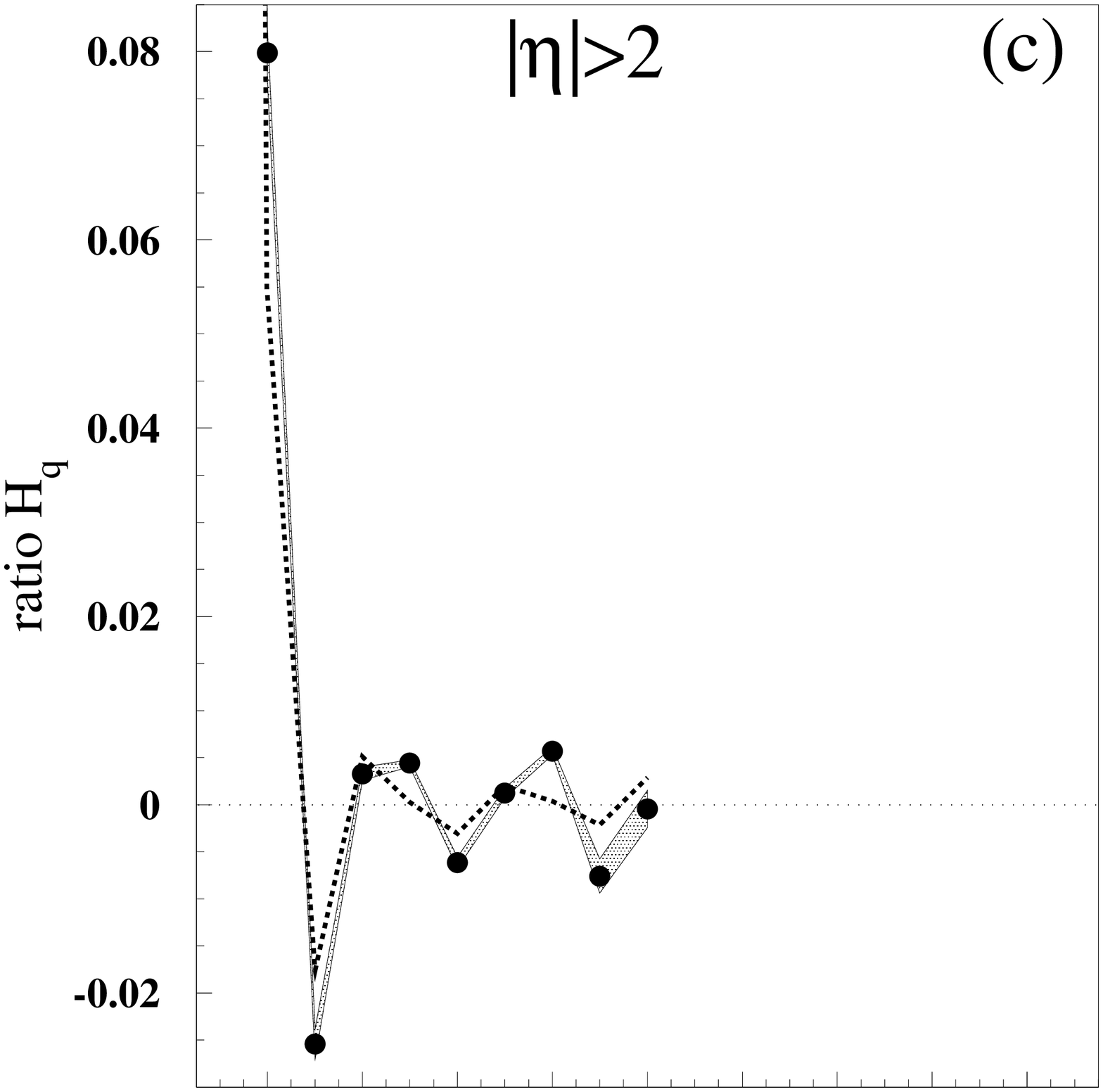}
    \includegraphics[width=7.4cm]{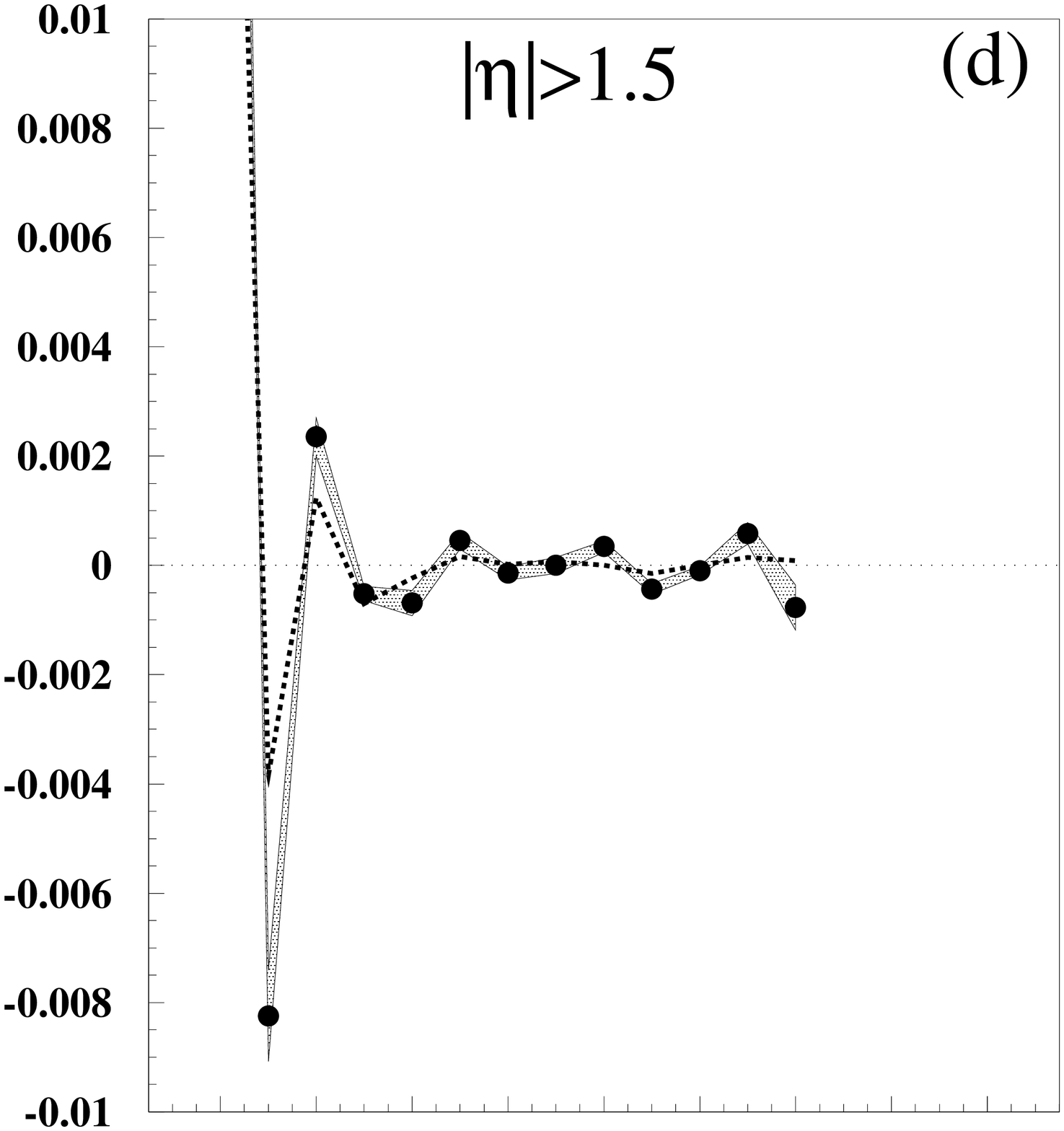}
\vspace{-1.4cm}

    \includegraphics[width=7.4cm]{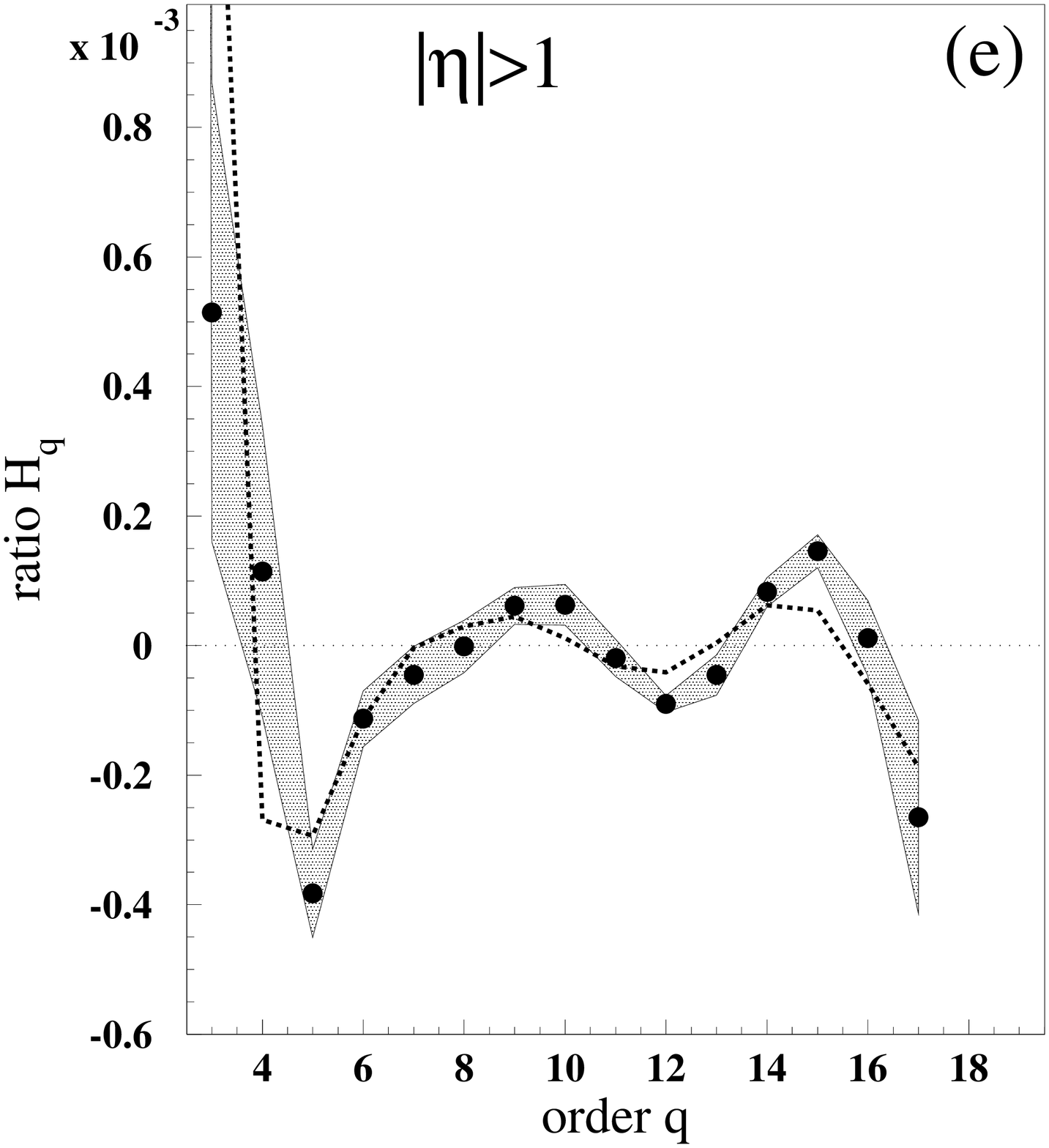}
    \includegraphics[width=7.4cm]{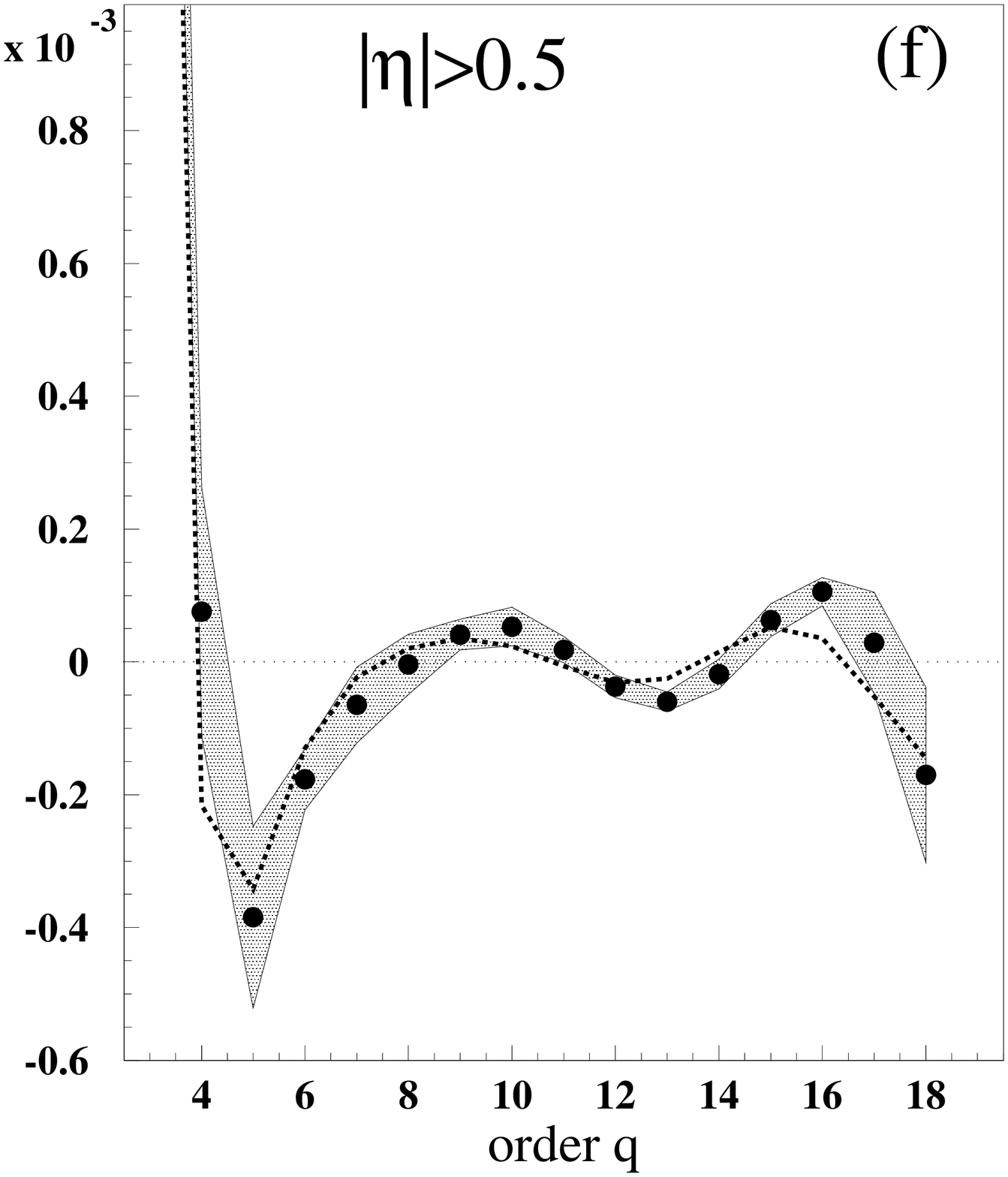}

\vspace{-0.2cm}
\scaption{$H_q$ moments measured from \cpmd{s} obtained 
with (a) $|\eta|>3$, (b) $|\eta|>2.5$
     (c) $|\eta|>2$, (d) $|\eta|>1.5$,
     (e) $|\eta|>1$ and (f) $|\eta|>0.5$.}
\label{fig:rap_nc1_f}  
\end{figure}

\begin{figure}[htbp]
  \begin{center}
    \includegraphics[width=8.4cm]{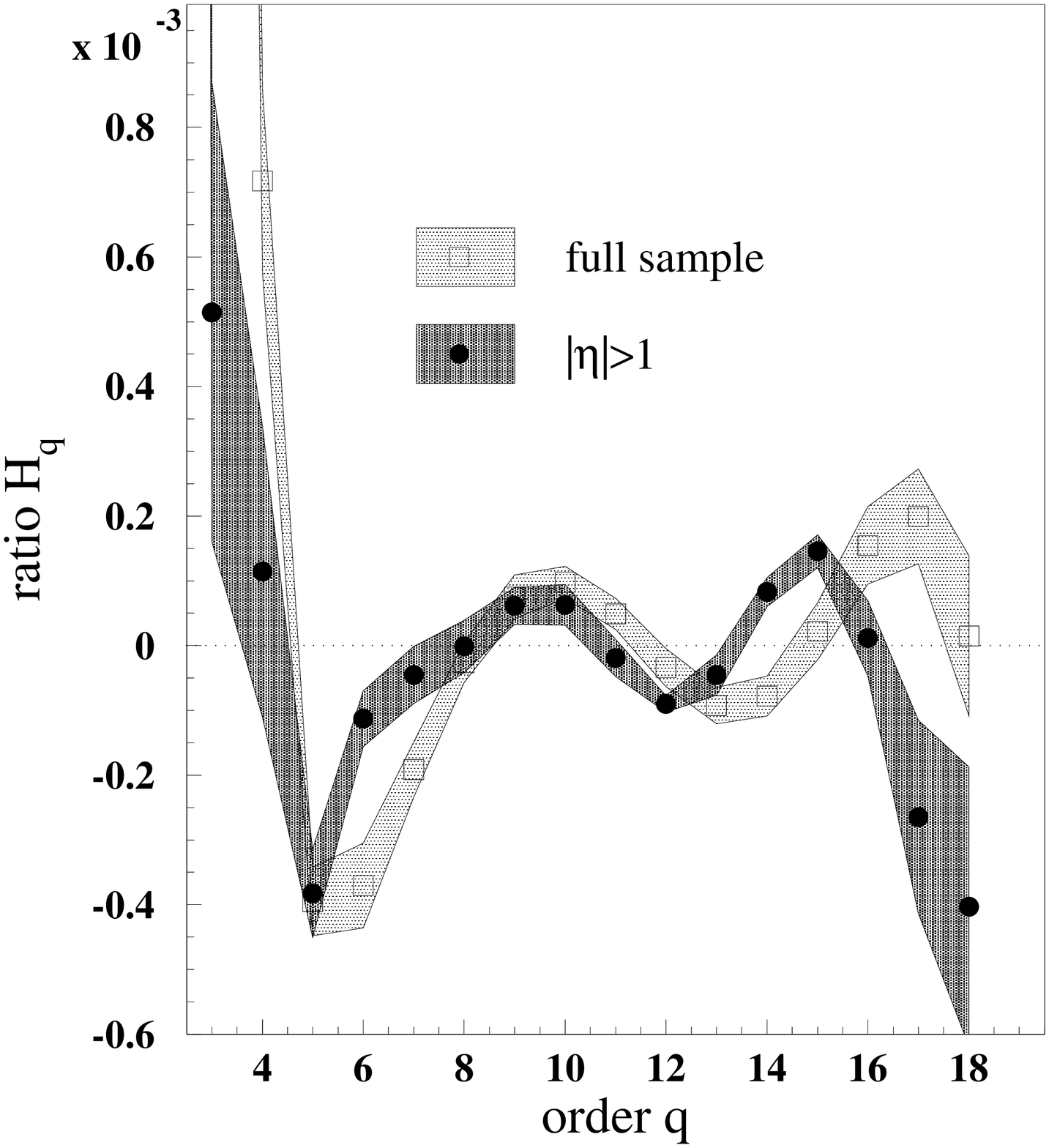}
    \includegraphics[width=8.4cm]{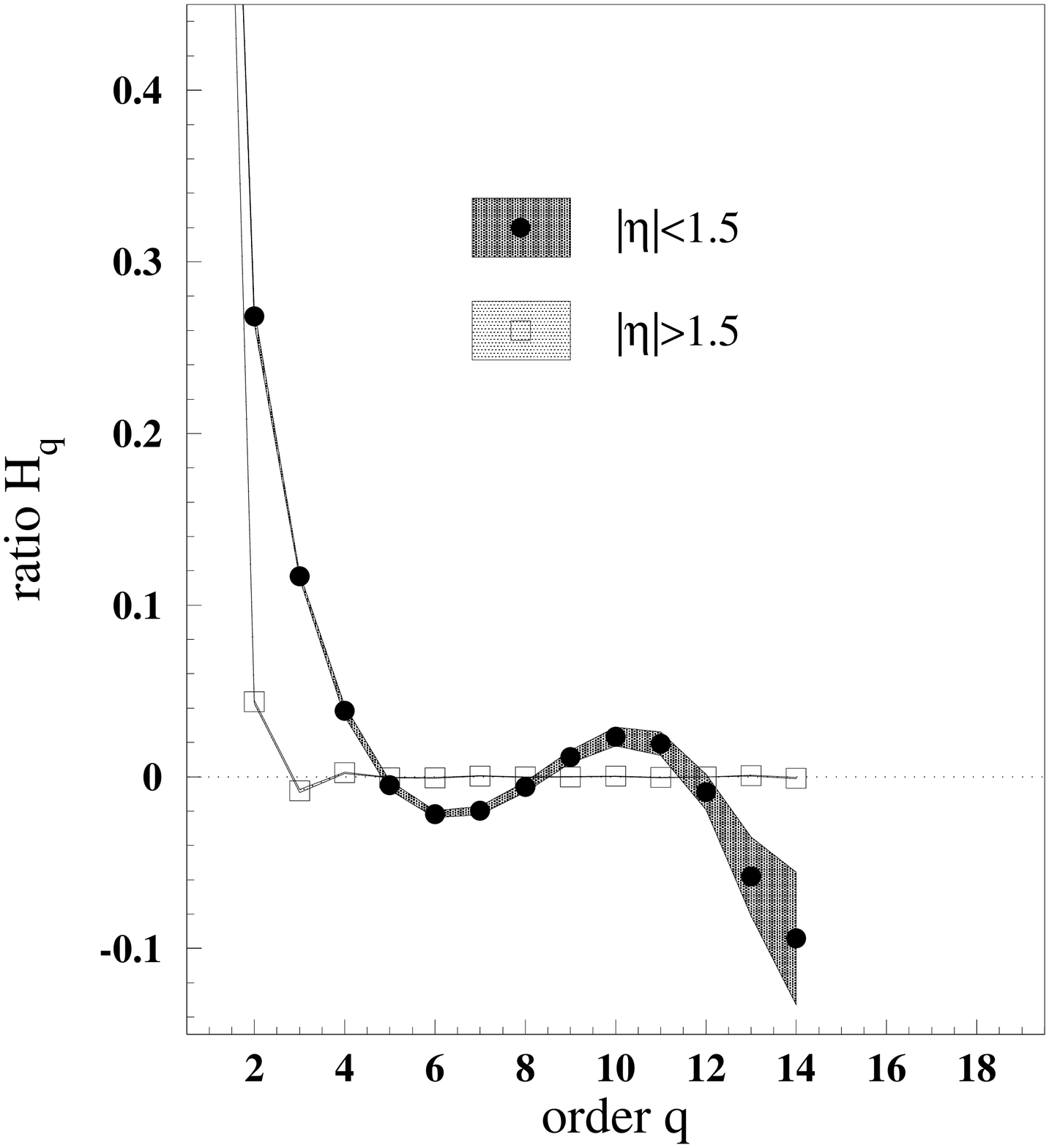}
  \end{center}
\begin{multicols}{2}
\scaption{$H_q$ moments obtained in the outside rapidity region  
($|\eta|>1$ compared to that obtained in full phase space.} 
\label{fig:nc1_full}  
\scaption{$H_q$ moments obtained in the central rapidity interval 
with $|\eta|<1.5$, compared to those obtained from the remaining charged 
particles with $|\eta|>1.5$.}
\label{fig:rap_c_nc} 
\end{multicols}
\end{figure}

For the outside rapidity regions shown in Fig.~\ref{fig:rap_nc1_f}, 
$H_q$ moments with a first negative minimum near $q=5$ and 
quasi-oscillations  are observed only for $|\eta|>1$ and $|\eta|>0.5$, 
which correspond to rapidity intervals where the majority of the 
charged particles are found. Furthermore, as shown in 
Fig.~\ref{fig:nc1_full},  the amplitude of these oscillations 
is about the size of those of the full sample 

For smaller outside rapidity regions, the $H_q$ behavior 
changes drastically, the first negative minimum is no longer 
at $q=5$ but at $q=3$ (visible only in Fig.~\ref{fig:rap_c_nc}). 
The smooth oscillatory behavior 
observed until now is replaced by a more 
erratic behavior. We still see quasi-oscillations but 
their period is somewhat shortened. There is also 
an increase of the amplitude as less and 
less particles are included.

For $\eta_i=1.5$, where central and outside intervals have 
comparable mean multiplicity, we find that the $H_q$ moments 
in the two regions are completely different both 
in scale and behavior as seen in Fig.~\ref{fig:rap_c_nc}.

In all cases, the JETSET generator agrees well with 
the data.

\begin{figure}[htbp]
  \begin{center}
    \includegraphics[width=7cm]{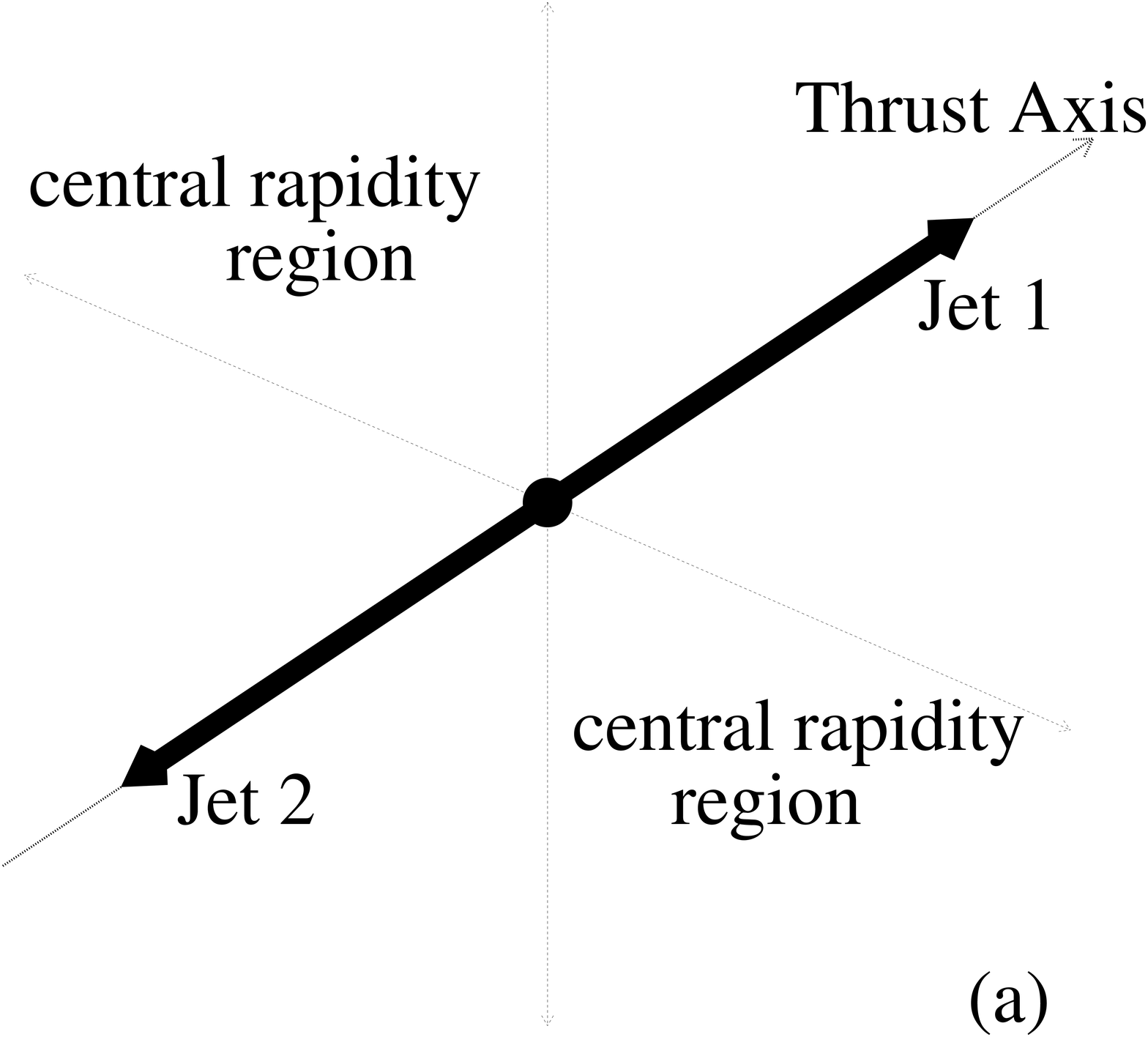}
    \includegraphics[width=7cm]{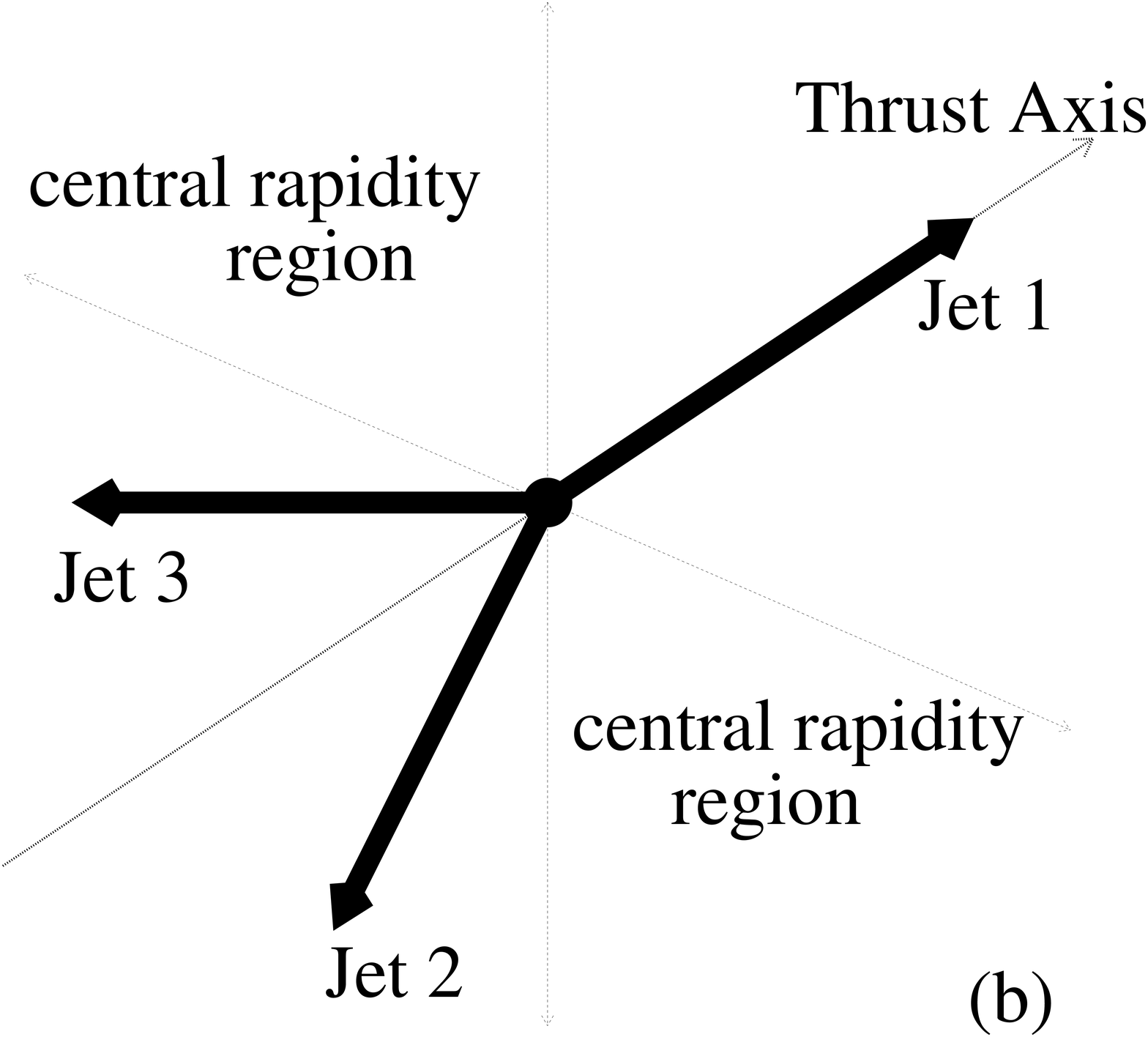}
\scaption{Schematic representation  of 
the central rapidity windows in (a) a 2-jet event and 
(b) in a 3-jet event.}
\label{fig:rap_jet}
  \end{center}
\end{figure}

\section{Discussion}

The so-called shoulder seen in medium central rapidity intervals 
has been associated in a previous analysis~\cite{delrap} to 
the presence of various jet topologies, thus 
reflecting the number of primary partons produced.  
The restriction imposed by the central rapidity interval 
enhances the difference between the \cpmd{s} 
of the different jet topologies, causing this 
shoulder to appear.
This may be explained in a very simple way. Since the 
rapidity is calculated in the thrust frame,  
the difference between the various topologies is enhanced, 
especially between 2-jet and non-2-jet events. 

In the case of a 2-jet event (Fig.~\ref{fig:rap_jet}(a)), 
the thrust axis is collinear 
with the jet axes. Therefore, selecting particles within a 
central rapidity interval, will select particles located 
in the phase space region between the two jets, which is  
depleted (see Fig.~\ref{fig:rapidity}).  
The \cpmd{} of 2-jet events will contain on average
a rather small number of charged particles. 
Since 2-jet events represent the majority of the events, 
the peak position will mainly be determined by the 2-jet events.

For the 3-jet events (Fig.~\ref{fig:rap_jet}(b)), 
the jet axes can  deviate from 
the thrust axis, depending on the energy taken by the third jet.
If the energy of the third jet is not too important, the thrust 
axis will still have the direction of the most energetic jet.  
However, depending on the energy of the third jet, the central 
rapidity region can overlap with the cone of one of the two 
other jets. The \cpmd{} of the 3-jet events will 
contain more particles than that of the 2-jet events, since it will not only
pick up particles in the intra-jet regions but also from 
inside the jet cone. 
This will be responsible for an enhancement of the difference between the 
different jet topologies. The more energetic the third jet,  
the larger will be the difference with the \cpmd{} of other jet topologies.
In the case of spherical events, with 4 or more jets, which somehow 
randomizes the direction of the thrust axis,  
the number of particles included in the \cpmd{} of small 
central rapidity intervals will be maximum.

In the previous chapter, we have seen that 
the appearance or the disappearance of the oscillatory 
behavior was due to the mixture of at least 3 different jet topologies. 
Resolving the various jet topologies into individual samples causes  
the $H_q$ oscillations to disappear in these individual samples. 
But, as we know, there is no visible shoulder 
structure in the \cpmd{} in full phase space. 
By increasing the difference in the general characteristics between 
the various jet topologies leads to the appearance 
of a clearly visible shoulder structure. The $H_q$ oscillations 
of these samples have much larger amplitude. The reason may 
well be associated to the shoulder effect and hence to the increase 
in the difference between the different jet topologies.

In order to verify that, one might 
want to find another experimental way of increasing the difference 
between the various final states composing the \cpmd{}. 
For example, one can look at $H_q$ moments 
for other processes such as hadronic or even heavy-ion collisions where 
both the diversity in final states and their number are tremendously 
increased. Such a review of $H_q$ moments has already been  
taken~\cite{hqhadron}. 
The $H_q$ moments measured in these samples clearly show an increase 
of the size of the oscillation with the number of possible 
final states. Going from \ee{} to heavy-ion collisions 
indeed increases the size of the oscillation.
Therefore, it is the same phenomenon which causes the oscillation to 
increase or to decrease. The $H_q$ moments obtained from \cpmd{s} 
of events which have relatively similar jet topologies 
(such as in the previous chapter) do not show 
the oscillatory behavior. Combining events from different jet topologies 
(as it is the case for the full sample) will lead the  $H_q$ moments 
to exhibit this oscillatory behavior. Combining events from 
jet topologies which are even more different, as it is the case 
in a restricted central rapidity interval, will increase the size of the 
$H_q$ oscillations.

In this chapter, we confirm the conclusion of the previous chapters
associating the origin of the $H_q$ oscillation in \ee{} at the \Z{} energy, 
as due to non-perturbative effects initiated by the diversity in jet topologies.

The origin of the $H_q$ oscillation  may be different 
at higher energies or for other process. 
In Fig.~\ref{fig:hq_parton}, we see that the $H_q$ moments 
of partons generated at $900\GeV$ by the parton shower 
of the JETSET Monte Carlo model show the oscillatory behavior 
but not at $90\GeV$.
Furthermore, in heavy-ion collision, 
the $H_q$ oscillation cannot be associated to jet topologies 
and neither to perturbative QCD. The large number of fragments  
produced during this type of collision won't 
take parts to the reactions and will evolve independently. 
therefore we will not have jets but 
many aggregates of particles to which will be associated the 
diversity in final states. Furthermore,  the size of these 
cluster prevent any perturbative QCD prediction.
 
The only property in common between these different 
type of reactions are the fact that they all have the ability to   
produced a large variety of final 
states, which may explain this $H_q$ oscillatory 
behavior.

\chapter*{Conclusions}
\addcontentsline{toc}{chapter}{\numberline{}Conclusions}
\markboth{\large{Conclusions}}{\large{Conclusions}}

The charged-particle multiplicity distribution and its 
moments as well as the inclusive charged particle 
momentum distribution have been measured for the full 
sample and  for the light- and b-quarks samples with an 
accuracy never reached before.

A detailed study of the shape of the \cpmd{} has been 
performed for the full, light- and b-quark samples 
using the $H_q$ moments. 
As a property of the full sample, the $H_q$ moments, when 
plotted as a function of the order $q$, exhibit an 
oscillatory behavior.  
This property is usually interpreted as a confirmation of NNLLA of pQCD.
However, we find that this oscillatory behavior, 
which is observed not only in \ee{} collisions but also for 
hadronic as well as heavy-ion collisions, is reproduced by a wide 
range of Monte Carlo models. Investigations performed 
on different models of parton 
generation with both parton shower and matrix elements and 
for different fragmentation models have displayed  
oscillatory behavior in all cases. But as there is no implementation 
of NNLLA in these Monte Carlo models, it appears that there is no need 
of NNLLA to produce \cpmd{s} having $H_q$ oscillatory behavior. 

In view of this rather inconclusive result, we 
question the validity of the prediction. Since 
 for charged particles the prediction relies strongly on the 
assumption of 
local parton-hadron duality, we extend the analysis to the jet 
multiplicity, assuming that jets obtained for energy scales above 
1--2~\GeV{} fall into the domain of validity of perturbative QCD. 
We thus avoid any assumption concerning the evolution of partons into 
hadrons such as local parton-hadron duality. 
The analysis of the $H_q$ moments of the jet multiplicity
distributions reveals that this oscillatory 
behavior appears only for very small 
$y_{\mathrm{cut}}$, corresponding to energy scales $\lesssim 100\MeV$, 
\ie{}, far from the perturbative region. 
At \Z{} energies, the $H_q$ oscillatory behavior appears only 
during hadronization. 
Therefore, we conclude that at the present energies 
the $H_q$ oscillatory behavior observed in the 
\cpmd{} is not related to the NNLLA of pQCD, but rather to the hadronization.

In search of an alternative origin of this $H_q$ behavior, we have 
investigated a more phenomenological approach which assumes 
that the shape of the multiplicity distribution results from a 
superposition of various types of events.
This was investigated, using a superposition of negative binomial 
distributions (NBDs). The \cpmd{} of each individual topology was  
parametrized using a NBD with parameters (the mean and 
dispersion) measured in the corresponding experimental \cpmd{}.

We found that it was possible to decompose the full sample into  
a minimum of three samples characterized by \cpmd{s} for which 
the $H_q$ moments do not exhibit oscillations.
These samples are characterized by the fact that they represent 
completely different jet topologies as pencil-like 2-jet events, 
Mercedes-like 3-jet events and what we call soft-jet events, 
\ie{}, 3-jet events with a low momentum gluon-jet.
Furthermore, each of these samples appears to be well described by a NBD, 
while the \cpmd{} of the full sample is 
found to be well described by a weighted sum of the three NBDs. 
Also the $H_q$ moments calculated from the three-NBD 
parametrization are 
found to agree rather well with those of the full sample.
Thus, we find that this phenomenological 
approach is successful 
in describing the full sample as a superposition of three NBDs.

Furthermore, the decomposition in terms of jet topology appears 
to be the only way to find samples which do not have $H_q$ oscillation. 
We also tried to separate the full as well as the 2-jet and 3-jet 
samples into light 
and b-quark samples, but neither the period of the oscillation nor 
its amplitude are influenced by these decompositions.

By studying the \cpmd{} in restricted central rapidity intervals, which have 
the property of enhancing the difference between jet topologies, we found that 
the size of the amplitude of the oscillation is linked to the compositeness 
(in jet topology) of the sample. In other words, by isolating 2-jet or 3-jet
events, we group together in these samples, events having rather similar 
jet topologies and hence without $H_q$ oscillation. On the other hand,  
restricting to central rapidity windows, which causes an enhancement 
of the difference between the jet topologies, increasing the difference 
between the events grouped into a same sample, will increase the size of the 
oscillation.

Therefore, we conclude that the origin of the oscillatory behavior 
is mainly an artifact appearing during the hadronization, but 
whose existence is linked to the diversity in jet topology and 
hence is related to the wide energy range available to the gluon.

At \Z{} energies, the oscillatory behavior is a non-perturbative phenomenon, 
but caused by the difference in topology initiated by hard gluon radiation.

It must also  be noted that since the $H_q$ moments are rather similar   
for extreme 2-jet and 3-jet events, the shape of the multiplicity 
distribution seems not influenced by the jet topology itself, but by the 
mixture of jet topologies.

Therefore, the main features in the shape of the charged-particle multiplicity
still visible in the final states is related to the number of primary partons, 
more precisely to the energy shared by the primary partons and to the hadronization.

\bibliographystyle{unsrtmod}
\bibliography{thesis}
\addcontentsline{toc}{chapter}{\numberline{}Bibliography}

\chapter*{Summary}
\addcontentsline{toc}{chapter}{\numberline{}Summary}
\markboth{\large{Summary}}{\large{Summary}}

In this thesis, we perform an analysis on \ee{} hadronic 
\Z{} decays recorded in 1994 and 1995 
by the L3 detector of LEP at center-of-mass energy
corresponding to the \Z{} mass.
The analysis is performed in parallel for all hadronic events,  
and for selected b-quark and light-quark events. 

The distribution of two variables,  
the charged-particle multiplicity $n$ and the inclusive charged-particle 
momentum $\xi$ are measured, from which all the analysis 
is carried out. 

Measuring the inclusive charged-particle momentum $\xi$ distribution 
is of interest since, under local parton-hadron duality, 
the $\xi$ distribution is assumed to be directly related to 
the amount of gluons produced at low momentum during the perturbative 
cascade as predicted by perturbative QCD.
Furthermore, measuring the $\xi$ distribution for the full, 
light- and b-quark samples allows to quantify the effect 
of the weak decay of the b-quark which is not accounted for 
by perturbative QCD prediction.

From the measurement of the \cpmd{}, which is the main 
study of the thesis, it is possible to extract information
concerning the dynamics of the interaction. 
By studying the moments of the \cpmd{}, one can 
obtain informations on particle correlation. For this purpose, 
we measure the $H_q$ moments of the \cpmd{} giving the 
relative amount of genuine $q$-particle correlation. 
These $H_q$ moments 
are known to display oscillations when plotted versus the 
order $q$. An explanation for these oscillations  
has been given by perturbative QCD. 
In the framework of the next-to-next to leading 
logarithm approximation (NNLLA), one predicts 
such an oscillatory behavior for the $H_q$ moments 
calculated from the {\it{parton}}  
multiplicity distribution. Under the assumption of the 
local parton-hadron duality, which claims that the shape 
of the multiplicity distribution is not distorted by  
hadronization, this behavior is also predicted for $H_q$ 
calculated from the final-state {\it{particle}} multiplicity 
distribution, such as the \cpmd{}.

However, further studies performed on Monte Carlo 
show that this oscillatory behavior can be reproduced without the need 
for the NNLLA, thus suggesting the need of an explanation different from that  
provided by perturbative QCD.
Therefore further tests are conducted in order to get a more 
definitive answer, this time using the jet multiplicity distribution.
When using jets, we assume that jets obtained at energy scales above 1--2~\GeV{} 
are described by perturbative QCD, thereby decreasing the role of  
local parton-hadron duality. 
However, we find that the oscillatory $H_q$ behavior is observed 
only for jets obtained at very low scales, far away from the 1--2~\GeV{} limit 
of validity of perturbative QCD.
This leads us to conclude that this oscillatory behavior observed in the 
$H_q$ moments measured from the \cpmd{} is not related to that  
predicted by NNLLA of perturbative QCD.

Therefore, in the absence of confirmation of pQCD, we search 
for an alternative explanation using a more phenomenological 
approach. 
We investigate the possibility that the features in the shape of the \cpmd{} responsible 
for oscillatory $H_q$ behavior could be due to the fact that the \cpmd{} derives 
from a superposition of final states of differing event topology, such as 
2-jet, 3-jet events or light- and b-quark events. Assuming we are able to parametrize the 
\cpmd{} of each of these samples, the \cpmd{} of the full sample would be 
parametrized by a weighted sum of the individual parametrizations. 
This is investigated using a negative binomial distribution (NBD), which 
has been used in the past to parametrize the \cpmd{} of a large number 
of processes at various energies. The \cpmd{s} of individual topologies are 
parametrized by a NBD with parameters measured from the 
experimental \cpmd{s}. We try two approaches, an approach which describes 
the \cpmd{} of the full sample as a mixture of various jet topologies  
and another one which describes it as a mixture of light- and b-quark events.

We find that the \cpmd{} of the full sample, as well as its oscillatory $H_q$ 
behavior, is successfully described by a mixture 
of 3 different jet topologies. Furthermore, $H_q$ measured from the \cpmd{} of these 
topologies, themselves, do not show the oscillatory behavior. 
This suggests that the oscillatory 
behavior in the $H_q$ of the full sample originates from the diversity 
of jet topologies. This lets us conclude, that the $H_q$ oscillatory behavior 
observed at the \Z{} energy is a non-perturbative 
phenomenon due to the diversity of topologies of 
hard gluon production and soft mechanisms. 
This conclusion is further supported by a study of the $H_q$ using particles 
in restricted rapidity intervals, which enables the difference of jet topologies 
to be enhanced.

\chapter*{Samenvatting}
\addcontentsline{toc}{chapter}{\numberline{}Samenvatting}
\markboth{\large{Samenvatting}}{\large{Samenvatting}}

In dit proefschrift wordt een analyse gedaan op de hadronische Z$^0$-vervallen
uit e$^+$e$^-$, opgenomen in 1994 en 1995 met de L3 detector van Lep bij een
massamiddelpuntsenergie overeenkomend met de Z$^0$ massa. De analyse is in
parallel uitgevoerd voor alle hadronische gevallen, en voor speciaal
geselecteerde b-quarks of lichte quarks. Gemeten zijn verdelingen van twee 
variabelen, de geladen multipliciteit $n$ en de inclusive impuls van de geladen
deeltjes $\xi$, die als uitgangspunt voor de verdere analyse zijn genomen.

Het meten van de inclusieve impuls van de geladen deeltjes $\xi$ is van belang,
omdat bij locale parton-hadron dualiteit wordt aangenomen, dat de 
$\xi$-verdeling direct gekoppeld is aan het aantal gluonen, dat wordt 
geproduceerd bij lage impuls, zoals door quantum chromo dynamica 
in storingsrekening (QCDS) wordt voorspeld. Verder staat het meten van de 
$\xi$-verdeling voor alle gevallen en voor de lichte- en b-quark gevallen 
toe om het effect te bepalen van het zwakke verval van het b-quark, dat niet 
in rekening wordt gebracht door QCDS.

Uitgaande van de geladen multipliciteitsverdeling is het mogelijk informatie
te verkrijgen over de dynamica van de wisselwerking. Dit is het belangrijkste
onderzoek in dit proefschrift. Door de momenten van de 
multipliciteitsverdeling van geladen deeltjes te bestuderen, kan men
informatie verkrijgen betreffende de deeltjescorrelatie. Hiertoe meten we de 
$H_q$-momenten van de verdeling van geladen deeltjes, die de relatieve
hoeveelheid correlatie van de werkelijke $q$-deeltjes geven.

Het is bekend, dat deze $H_q$ momenten van de verdeling van geladen deeltjes
oscillaties laten zien, als ze worden uitgezet tegen de orde $q$. Een 
verklaring voor deze oscillaties kan worden gegeven door QCDS. In het kader 
van de "next to next to leading logarithm" (NNLLA) benadering wordt een 
dergelijk oscillerend gedrag berekend uit de parton multipliciteitsverdeling. 
Met de veronderstelling van locale parton-hadron dualiteit, dat de vorm van de
multipliciteitsverdeling niet verandert door hadronisatie, wordt dit gedrag 
ook voorspeld voor  de $H_q$ berekend uit een multipliciteitsverdeling van de
deeltjes in de eindtoestand, zoals de multipliciteitsverdeling van de geladen
deeltjes.

Echter verdere Monte Carlo studies laten zien, dat dit oscillerend gedrag kan
worden gereproduceerd zonder dat de NNLLA nodig is, aangevend, dat er een
andere uitleg nodig is, dan die gegeven door QCDS. Om hierop een definitiever
antwoord te krijgen zijn verdere tests uitgevoerd, deze keer met hulp van de
multipliciteitsverdeling van jets (deeltjesbundels). Als we deze jets
gebruiken nemen we aan, dat ze verkregen bij energieschalen boven 1--2~\GeV{}
beschreven worden door QCDS, waarbij de rol van locale parton-hadron dualiteit
vermindert. Toch vinden we, dat het oscillerend gedrag voor $H_q$ alleen voor
jets bij zeer lage energieschalen wordt gevonden ver van de 1--2~\GeV{}
geldigheidlimiet van QCDS. Hieruit concluderen we, dat het oscillerend gedrag
waargenomen in de $H_q$ momenten en afgeleid uit de multipliciteitsverdeling 
van geladen deeltjes niets te maken heeft met datgene, wat door NNLLA in QCDS
wordt voorspeld.

Daarom zoeken we bij afwezigheid van de bevestiging van QCDS naar een
alternatieve verklaring met behulp van een fenomenologischer benadering.
We onderzoeken de mogelijkheid, dat de kenmerken in de vorm van de
multipliciteit van de geladen deeltjes, die het oscillatiegedrag van $H_q$
veroorzaken, zou kunnen komen door het feit, dat deze multipliciteitsverdeling
wordt afgeleid uit de superpositie van eindtoestanden van verschillende
topologie, zoals uit 2 jet en 3 jet gevallen of uit lichte quarks en b
quarks. Onder de aanname, dat we in staat zijn de multipliciteitsverdeling van
geladen deeltjes voor al deze types te parametriseren, wordt de verdeling voor
de gehele verzameling gevormd door een gewogen som van de individuele
parametrisaties. Dit is onderzocht met behulp van een negatieve
binomiaalverdeling (NBV), in het verleden gebruikt om de
multipliciteitsverdeling van geladen deeltjes van een groot aantal processen
bij verschillende energie\"en te beschrijven. De multipliciteitsverdelingen
van geladen deeltjes van individuele topologie\"en zijn geparametriseerd met
een NBV met parameters gemeten in de overeenkomstige experimentele
distributies. We proberen twee benaderingen, een eerste, die de
multipliciteitsverdeling van geladen deeltjes beschrijft als een mengsel van
verschillende jet topologie\"en en een andere, die het beschrijft als een
mengsel van lichte en b quark gevallen.

We vinden, dat de multipliciteitsverdeling van geladen deeltjes van de complete
verdeling tezamen met zijn oscillerend karakter in $H_q$ succesvol kan worden
beschreven met een mengsel van drie verschillende jet topologie\"en. Daarbij
vertonen de $H_q$'s gemeten met de multipliciteitsverdeling van geladen 
deeltjes van deze topologie\"en zelf geen oscillerend gedrag. Dit geeft aan, 
dat het oscillerend gedrag in de $H_q$ van de complete verzameling een gevolg 
kan zijn van de verschillende jet topologie\"en. Dit laat ons concluderen, 
dat het oscillerend gedrag bij de $Z^0$ energie een niet verstorings effect 
is ten gevolge van de diversiteit van de topologie\"en van harde gluonen 
productie en zachte mechanismen. Deze conclusie wordt ook nog ondersteund 
door een studie van de $H_q$'s met gebruik van deeltjes in beperkte 
rapiditeitsintervallen, die het mogelijk maken om de verschillen van de jet 
topologie\"en te versterken.

\chapter*{Acknowledgements}
\addcontentsline{toc}{chapter}{\numberline{}Acknowledgements}
\markboth{\large{Acknowledgements}}{\large{Acknowledgements}}
{\it 

\vspace{-0.35cm}

I wish to express my gratitude to all the persons, collegues, 
friends, familly who made, by their support and also by their 
patience, this thesis possible. 

I, first, would like to thank Wolfram Kittel, my supervisor, 
who gave me the opportunity to work in the Experimental 
High Energy Physics group of the University of Nijmegen, at a 
time when it was easier to understand me while speaking in 
French rather than in English. I would like also to thank him 
for his support and for introducing me to the exciting 
field of multiparticle dynamics. Even though ``exciting'' may sound 
a bit paradoxical when applied to multiparticle dynamics, 
while knowing people who would rather use words like 
``obscure'' and ``marginal'', I must say that multiparticle 
dymanics is one of those domains of particle physics where 
we have still a lot to learn. Therefore, 
it makes any investigation an exciting adventure in the domain 
of the unknown and of the highly uncertain. Although, it may not be 
Star Trek (it is very unlikely to be attacked by unfriendly 
particles even charged ones, and if computers hurt sometimes, 
they don't bite), it is nevertheless a very interesting field to explore, 
and I hope that this thesis as well as the overall effort 
made by our group have helped to make this field more exciting 
and not anymore obscure or marginal.

I would also like to thank Wes Metzger, my co-supervisor, 
for his help and support during all these years, for the many critical 
discussions we had which have helped to channel the boiling 
flow of ideas in a more navigable way which has finally 
resulted in this thesis. 

I would like to thank for the usefull discussions we had and guidance provided 
by their publications, Igor Dremin, Alberto Giovannini, S\'andor Hegyi, 
Wolfgang Ochs and Roberto Ugoccioni, the cartographers of this 
world of the unknown and of the highly uncertain which was explored in my thesis.

I am also grateful to Dominique Duchesneau, John Field, Edigio Longo, 
Swagato Banerjee, Pete Duinker for their help and support.

I thank also all the members past and present of the 
Experimental High Energy Physics department of Nijmegen whom  
I had the opportunity to meet. In the past members, I would like 
in particular to thank for his friendship Sergei Chekanov 
who made the beginning of my stay at Nijmegen very enjoyable, 
and also Ljubisa Drndarski. 
Thanks also to Jorn van Dalen, Silke Duensing, Wim Lavrijsen, Bert Petersen, 
Blandine Roux, Michiel Sanders and Henric Wilkens.  
I have also to give a special thanks to Raphael Hakobyan and Yuan Hu 
who, sharing an office with me, have endured every day both my good 
and bad moods (both of them may be very painfull sometimes...). 
I also would like to thank Jaap Schotanus for always being available. 
Thanks also to Peter Klok and Frans Rohde for helping me 
with computers.
A special thanks also to Marjo van Wees-Mobertz, Hanneke Vos-van 
der Lugt and Annelies Oosterhof-Meij for their kind help 
in the administrative domain which will always look to me much more 
obscure than physics and for helping me to keep my serenity on many 
occasions.

Very special thanks to Catherine Adloff, Patrice Pag\`es and 
Fran\c{c}oise Touhari.

Last but not least, I wish to thank my parents for their 
unconditional love and support during all my life.
}

\chapter*{Curriculum vitae}
\addcontentsline{toc}{chapter}{\numberline{}Curriculum vitae}
\markboth{\large{Curriculum vitae}}{\large{Curriculum vitae}}
\pagestyle{empty}

The author of this thesis was born on $23^\text{rd}$ of June 1971 at Epinal 
in France. Having obtained a {\it{Baccalaur\'eat}} in sciences at the 
Claude Mathy lyc\'ee at Luxeuil in 1990, he studied mathematics 
and physics at the University of Montpellier, where 
he graduated with an M.Sc. degree in physics in 1994. He moved to 
the Louis Pasteur University at Strasbourg to specialize in 
subatomic physics where he obtained a D.E.A. (Dipl\^ome d'Etudes 
Approfondies) in nuclear and particle physics in 1995. After 
the completion of his military duties in the French Air Force, 
he started to work as a graduate student at the Catholic University of 
Nijmegen in the Netherlands under the supervision of Prof. E.W.~Kittel 
in 1996. The completion of this work has lead in 2001 
to the realization of the present doctoral dissertation and of a 
publication. 

As a graduate student, he has participated in several graduate 
schools, namely the Joint Belgian-Dutch-German graduate schools 
at Rolduc (The Netherlands) in 1996 and at Monschau (Germany) 
in 1997, the European School of High-Energy Physics at 
Saint Andrews (Scotland) in 1998, the NATO Advanced Study 
Institute at Nijmegen (The Netherlands) in 1999 and the 
CERN school of computing at Marathon (Greece) in 2000. 
He has also participated and given talks 
in many international conferences and meetings including 
the $8^\text{th}$ International Workshop on Multiparticle 
Production at M\'atrah\'aza (Hungary) in 1998, the 
$6^\text{th}$ International High-Energy Physics Euroconference in 
Quantum ChromoDynamics at Montpellier (France) also in 1998, 
the April 2000 meeting of the American Physical Society 
at Long Beach (California) and the $30^\text{th}$ 
International Symposium on Multiparticle Dynamics at 
Tihany (Hungary) both in 2000. 

Having lost the meaning of the word ``modesty'' 
while writing his curriculum vitae, the author remembered that 
all this was only a beginning and he recovered immediately.

\end{document}